\documentclass{article}
\usepackage{ijcai23}

\usepackage{times}
\usepackage{soul}
\usepackage{url}
\usepackage[hidelinks]{hyperref}
\usepackage[utf8]{inputenc}
\usepackage[small]{caption}
\usepackage{graphicx}
\usepackage{amsmath}
\usepackage{amsthm}
\usepackage{booktabs}
\usepackage{algorithm}
\usepackage{algorithmic}
\usepackage[switch]{lineno}

\usepackage{balance} 
\usepackage{booktabs}
\usepackage{graphicx}
\usepackage{makecell}
\usepackage{array}
\usepackage[export]{adjustbox}
 \usepackage{tabularx}
 
\usepackage[all]{nowidow}

\newcommand{\mm}[1]{}
\newcommand{\lt}[1]{}
\newcommand{\sh}[1]{}

\newcommand\blfootnote[1]{%
  \begingroup
  \renewcommand\thefootnote{}\footnote{#1}%
  \addtocounter{footnote}{-1}%
  \endgroup
}

\title{Modeling Moral Choices in Social Dilemmas \\ with Multi-Agent Reinforcement Learning}

\author{
Elizaveta Tennant$^1$\and
Stephen Hailes$^1$\and
Mirco Musolesi$^{1, 2}$
\affiliations
$^1$University College London \\
$^2$University of Bologna\\
\emails
\{l.karmannaya.16, s.hailes, m.musolesi\}@ucl.ac.uk}

\begin{document}

\maketitle

\begin{abstract}
    Practical uses of Artificial Intelligence (AI) in the real world have demonstrated the importance of embedding \textit{moral choices} into intelligent agents. They have also highlighted that defining top-down ethical constraints on AI according to any one type of morality is extremely challenging and can pose risks. A bottom-up learning approach may be more appropriate for studying and developing ethical behavior in AI agents. In particular, we believe that an interesting and insightful starting point is the analysis of emergent behavior of Reinforcement Learning (RL) agents that act according to a predefined set of \textit{moral rewards} in social dilemmas.

    In this work, we present a systematic analysis of the choices made by intrinsically-motivated RL agents whose rewards are based on moral theories. We aim to design reward structures that are simplified yet representative of a set of key ethical systems. Therefore, we first define moral reward functions that distinguish between consequence- and norm-based agents, between morality based on societal norms or internal virtues, and between single- and mixed-virtue (e.g., multi-objective) methodologies. Then, we evaluate our approach by modeling repeated dyadic interactions between learning moral agents in three iterated social dilemma games (Prisoner's Dilemma, Volunteer's Dilemma and Stag Hunt). We analyze the impact of different types of morality on the emergence of cooperation, defection or exploitation, and the corresponding social outcomes. Finally, we discuss the implications of these findings for the development of moral agents in artificial and mixed human-AI societies. \blfootnote{Version with Appendix: \small{\url{https://arxiv.org/abs/2301.08491}}}
    \blfootnote{Code: \small{\url{https://github.com/Liza-Tennant/moral_choice_dyadic}}} 
\end{abstract}

\section{Introduction}

An open question in AI research and development is how to represent and specify ethical choices and constraints for this class of technologies in computational terms \cite{ajmeri2020elessar,amodei2016concrete,AWAD2022388,dignum2017responsible,yu2018building,wallach2010robot}. In particular, there is an increasing interest in understanding how certain types of behavior and outcomes might emerge from the interactions of learning agents in artificial societies \cite{de2006learning,foerster2018lola,hughes2018inequity,jaques2019social,leib02017multiagent,mckee2020social,peysakhovich2018consequentialist,peysakhovich2017prosocial,rodriguezsoto2021multi,sandholm1996multiagent} and in interactive systems where humans are in the loop \cite{carroll2019ontheutility,rahwan2019}. We believe that a promising and insightful starting point is the analysis of emergent behavior of Reinforcement Learning (RL) agents that act according to a predefined set of \textit{moral rewards} in situations were there is tension between individual interest and collective social outcomes, namely in social dilemmas \cite{axelrod1981evolution,rapoport1974prisoner,sigmund2010}.

In this work, we present for the first time a systematic analysis of the learning dynamics and choices of RL agents whose rewards are based on moral theories. The goal is to define reward structures that are simplified yet representative of a set of key ethical frameworks. In other words, the (moral) agents play the social dilemmas not trying to maximize their cumulative return considering the payoff matrix of the game, but using their intrinsic reward according to a given ethical system. 
Classic moral theories include Utilitarianism\footnote{In this paper, the term Utilitarian indicates an agent that shows prosocial behavior according to several exponents of Utilitarianism \cite{Bentham1996}, i.e., one that tries to maximize the global reward given by the sum of the individual rewards, in contrast with a Selfish agent, whose goal is the maximization of their own utility.} \cite{Bentham1996}, Deontological morality \cite{kant1981grounding} and Virtue Ethics \cite{aristotle} - each of these explains different aspects of human moral preferences, and has specific implications for social learning agents \cite{anderson2005towards}. We map choices to an intrinsic reward system \cite{chentanez2004intrinsically} according to these ethical frameworks. The majority of existing modeling work has focused on single types of social preference \cite{hughes2018inequity,jaques2019social,KLEIMANWEINER2017107,peysakhovich2017prosocial,peysakhovich2018consequentialist}. However, as suggested by the continued debate between the three moral frameworks \cite{Mabille2021}, and by evidence from human moral psychology \cite{bentahila2021universality,graham2009liberals}, a broad range of moral preferences are likely to exist within and across societies (especially in preferences regarding AI morality \cite{awad2018moral}). Thus, implementing any one theory \textit{top-down} without consideration of how it might interact with other types of moral reasoning risks creating societies in which exploitative and/or defenseless behavior emerges \cite{Wallach2009}. The present study instead is based on a \textit{bottom-up} modeling of interactions between different types of moral learning agents as a way of gathering insights for understanding and improving human, artificial and hybrid societies.

Social dilemma games, which illustrate the trade-off between self-interested and mutually beneficial choices \cite{rapoport1974prisoner}, create a useful testing ground for moral agents \cite{cavagnetto2014game,cunningham1967,hedge2020ethics}. In particular, in this work we consider \textit{iterated} games. The use of learning agents in these games allows us to study how certain behaviors might evolve in a society given a set of initial agent preferences and environmental characteristics such as payoff structure \cite{harper2017reinforcement,sandholm1996multiagent}. 
Evolutionary modeling can effectively be applied to agents' moral choices to study the emergence of specific social outcomes \cite{binmore2005natural}. For example, a growing body of research has explored emergent cooperation within this framework \cite{jaques2019social,leib02017multiagent}, with some studies specifically implementing certain aspects of morality to elicit cooperative behaviors \cite{hughes2018inequity,peysakhovich2018consequentialist}. One study \cite{mckee2020social} has moved towards exploring populations heterogeneous in terms of social preferences. However, to our knowledge no work has yet modeled the behavior of different moral agents learning against one another in such social dilemma settings \cite{tolmejer2021implementations}, so little is known about the potential risks that may evolve from implementing any one type of morality in AI agents who act in diverse social environments. 

The contributions of this work can be summarized as follows:
\begin{itemize}
    \item We introduce a methodological framework for developing moral artificial agents and define intrinsic moral rewards
    inspired by philosophical theories.
    \item We systematically evaluate emergent behaviors and outcomes in pairwise interactions, including between different types of moral agents, in three social dilemma games, namely Iterated Prisoner's Dilemma, Iterated Volunteer's Dilemma and Iterated Stag Hunt.
\end{itemize}
We believe that our approach can be generalized to other types of moral agents or games, and can be used in the future to model agent learning against human opponents. Also for this reason, the code used for this study is available as open source software 
to encourage further work in this area. 

\section{Background and Related Work}
\subsection{Social Dilemma Games}

Social dilemma games simulate social situations in which agents obtain different payoffs from choosing one action or another, and due to the structure of the game, each agent faces a trade-off between individual interest and societal benefit. The most widely studied type is a symmetric matrix game with two players, each choosing one of two possible abstract actions - Cooperate or Defect. 
Players must decide on their respective actions simultaneously, without communicating. 



Three classic games from economics and philosophy which are relevant to moral choice are the Prisoner's Dilemma \cite{rapoport1974prisoner}, Volunteer's Dilemma (or `Chicken') \cite{poundstone1993}, and Stag Hunt \cite{skyrms2001stag} (see payoffs for row vs. column player in Table \ref{tab:three_games}). We implement \textit{iterated} games, in which players repeatedly face one of these dilemmas, and over numerous interactions learn to maximize their cumulative payoff.

In the Prisoner's Dilemma, mutual cooperation would achieve a Pareto-optimal outcome (in which one player cannot be made better off without disadvantaging the other
) - but each individual player's best response is to defect due to \textit{greed} (facing a cooperator, they benefit from defecting) and \textit{fear} (facing a defector, they suffer by cooperating). In the Volunteer's Dilemma, a selfish or rational player may choose to defect due to greed, and if both do so, both obtain the lowest possible payoffs (i.e. no one volunteers, and the society suffers). Finally, in the Stag Hunt game two players can cooperate in hunting a stag and thus obtain the greatest possible payoff each; however, given a lack of trust between the players (i.e., fear of a non-cooperative partner), either may be tempted to defect and hunt a hare on their own instead, decreasing both players' payoffs as a result.

\begin{table}[t]
\begin{tabular}{l|cc}
\textbf{IPD} & C    & D    \\ \hline
C         & 3,3 & 1,4 \\ 
D         & 4,1 & 2,2 
\end{tabular}
\hfill
\begin{tabular}{l|cc}
\textbf{IVD} & C    & D    \\ \hline
C         & 4,4 & 2,5 \\ 
D         & 5,2 & 1,1 
\end{tabular}
\hfill
\begin{tabular}{l|cc}
\textbf{ISH} & C    & D    \\ \hline
C         & 5,5 & 1,4 \\ 
D         & 4,1 & 2,2 
\end{tabular}
\caption{Payoff matrices for each step of the Iterated Prisoner's Dilemma (IPD), Iterated Volunteer's Dilemma (IVD) \& Iterated Stag Hunt (ISH) games, in which players are motivated to defect by either \textit{greed} (IVD), \textit{fear} (ISH), or both (IPD).}
\label{tab:three_games}
\end{table}

\subsection{Repeated Games and Learning}

Iterated versions of the matrix games provide a complex set of strategies, since players can take actions to punish their opponents for past wrongdoing or to influence their future behaviors, causing instability in the environment and dynamic behaviors. Thus, calculating predicted equilibria in these situations is not always computationally plausible, and simulation methods are required in order to study potential emergent behaviors and outcomes. 


Reinforcement Learning (RL) is a well-suited technique for modeling agents that learn by interacting with others in an environment \cite{Sutton2018RLSE}. It can be applied in conjunction with Evolutionary Game Theory \cite{hofbauer_sigmund_1998} to iterated social dilemma games \cite{abel2016reinforcement,littman1994markov,sandholm1996multiagent}, in which payoffs constitute extrinsic rewards, and traits such as moral or social preferences \cite{fehr2002social} can be encoded in the agent's intrinsic reward \cite{chentanez2004intrinsically}.

\subsection{Moral Choices}

The tension between individual and social benefit presents an interesting test-bed for ethics \cite{cavagnetto2014game}: how would different moral theories manifest themselves in these three games? Certain embedded social preferences have been shown to promote cooperation between RL agents in social dilemmas - such as learning via opponent-learning awareness \cite{foerster2018lola}, social influence \cite{jaques2019social}, or prosocial reward functions \cite{peysakhovich2018consequentialist}, including inequity-aversion \cite{hughes2018inequity} and social value orientation \cite{mckee2020social}. 

We anchor our definitions of moral choices in traditional moral philosophy. Of the numerous moral theories which have been proposed over the millennia, we will focus on three established and influential moral frameworks: consequentialist, norm-based and virtue ethics. The methodological approach presented in this work can potentially be applied to a variety of other moral systems. \textit{Consequentialist} morality focuses on the consequences of an action, and includes Utilitarianism \cite{Bentham1996}, which defines actions as moral if they maximize total utility for all agents in a society. \textit{Norm-} or \textit{Duty}-based morality, including Deontological ethics \cite{kant1981grounding}, considers an act moral if it does not contradict the society's external norms. 
Finally, in \textit{virtue ethics} \cite{aristotle}, moral agents must act in line with their certain internal virtues, such as fairness or care \cite{graham2013moral}. Different virtues can matter more or less to different agents \cite{aristotle,graham2009liberals} and can themselves have consequentialist or norm-based foundations. Furthermore, a single agent may rely on more than one type of virtue, so a more expressive way of modeling virtue ethics might be through a multi-objective paradigm. We will investigate this by implementing a mixed virtue agent alongside single-virtue agents. 

In this work we study the implications of each of these theories in social dilemma environments with learning agents. We model moral preferences inspired by these theories though intrinsic reward functions \cite{chentanez2004intrinsically}. 
Intrinsic rewards have been used to model consequentialist traits, including preferences against inequity \cite{hughes2018inequity}, towards prosocial \cite{peysakhovich2018consequentialist,peysakhovich2017prosocial} and/or competitive outcomes \cite{abel2016reinforcement,mckee2020social}.
Work on norm-based moral agents is limited \cite{tolmejer2021implementations}, mainly focused on computational modeling of norm emergence among agents with static policies (i.e., without learning) \cite{salah2022interaction}, or using learning with reputation mechanisms \cite{anastassacos2021cooperation}. However, no studies investigated the impact of pre-defined moral norms in social dilemmas, which is the focus of the present work.

It is worth noting that no work has focused on the learning interactions between consequentialist and norm-based moral agents. In the broader social preferences literature, \cite{mckee2020social} has shown that RL agents trained in populations with diverse social preferences learn more robust and general strategies, while prosocial agents in \cite{foerster2018lola,hughes2018inequity} got exploited when facing agent types other than themselves. Heterogeneous moralities are observed in human societies \cite{graham2013moral}, so as a starting point we explicitly investigate the implications of different moral agent types learning against one another in dyadic interactions. This also gives us an opportunity to test potential criticisms of each moral framework, as well as their interactions, through simulation. In particular, philosophers have criticized consequentialism because it risks exploitation or rights violation of a small number of agents for the good of the majority (one extreme example includes the justification of slavery \cite{kilbridge1993}). 
At the same time, blind compliance with norms may also risk in developing negative consequences if the norm is defined too narrowly. 



\section{Modeling Moral Choice with Reinforcement Learning}
\subsection{Reinforcement Learning for Decision-Making in Social Dilemmas}

\begin{table*}[t]
  \small
  \centering
  \begin{tabular}{lll}\toprule
    \textit{Agent Moral Type} & \textit{Moral Reward Function} & \textit{Source of Morality} \\ \midrule
    Utilitarian  & $R_{M_{intr}}^t=R_{M_{extr}}^t + R_{O_{extr}}^t$ & External/Internal conseq.\\ 
    Deontological  & $R_{M_{intr}}^t= 
\begin{cases}
    $--$\xi,& \text{if } a_M^t=D ,  a_O^{t-1}=C \\ 
    0,              & \text{otherwise}
\end{cases}\ $ & External norm \\ 
    Virtue-equality & $R_{M_{intr}}^t=1-\frac{|R_{M_{extr}}^t-R_{O_{extr}}^t|}{R_{M_{extr}}^t+R_{O_{extr}}^t}$ & Internal consequentialist \\ 
    Virtue-kindness  & $R_{M_{intr}}^t= 
\begin{cases}
    \xi,& \text{if } a_M^t=C  \\
    0,              & \text{otherwise}
\end{cases}\ $ & Internal norm  \\ 
    Virtue-mixed  &
    $R_{M_{intr}}^t= 
\begin{cases}
        \beta*(1-\frac{|R_{M_{extr}}^t-R_{O_{extr}}^t|}{R_{M_{extr}}^t+R_{O_{extr}}^t}) + (1-\beta)*\hat{\xi}
    ,& \text{if } a_M^t=C  \\
        \beta*(1-\frac{|R_{M_{extr}}^t-R_{O_{extr}}^t|}{R_{M_{extr}^t}+R_{O_{extr}}^t})
    ,              & \text{otherwise}
\end{cases}$
& Internal norm \& conseq. \\

    \bottomrule
  \end{tabular}
  \caption{Definitions of the types of intrinsic moral rewards, from the point of view of the moral agent $M$ playing versus an opponent $O$.}
  \label{tab:moral_definitions}
\end{table*}

Our environment is a two-player iterated dilemma game, played over 10000 iterations, which represent a single episode. At each iteration, a moral player $M$ and an opponent $O$ play the one-shot matrix game corresponding to one of three classic social dilemmas - Prisoner's Dilemma, Volunteer's Dilemma and Stag Hunt (see Table \ref{tab:three_games}). In this Markov game \cite{littman1994markov}, both players learn to choose actions from interacting with their opponent. At time $t$, the learning agent $M$ observes a state, which is the pair of actions played by $M$ and $O$ at $t-1$: $s_M^{t}=(a_O^{t-1},a_M^{t-1})$, and chooses an action simultaneously with the opponent $O$: $a^t_M, a^t_O \in \{C, D\}$. The player M then receives a reward $R_M^{t+1}$ and observes a new state $s_M^{t+1}$. 
Both agents use tabular Q-Learning~\cite{watkins1992q} to update the $Q$-value for a given state and action for both agents as follows:
\begin{eqnarray}
\resizebox{.96\linewidth}{!}{$
            \displaystyle
Q(s^t,a^t)\leftarrow Q(s^t,a^t)+\alpha \bigg [R^{t+1}+\gamma\max_a Q(s^{t+1},a)-Q(s^t,a^t) \bigg]
$}
\end{eqnarray}

\noindent where $\alpha$ is the learning rate and $\gamma$ is a discount factor. 

We simulate moral choice using an $\epsilon$-greedy policy, where agents act randomly $\epsilon$\% of the time and otherwise play greedily according to the Q-values learned so far. Player $M$ can learn according to an \textit{extrinsic} game reward $R_{M_{extr}}^{t+1}$, which depends directly on the joint actions $a_M^t,a_O^t$ (as defined in Table \ref{tab:three_games}) or according to an \textit{intrinsic} reward $R_{M_{intr}}^{t+1}$, which we define subsequently. 


\subsection{Modeling Morality as Intrinsic Reward}

\begin{figure}[t]
  \includegraphics[width=1\linewidth]{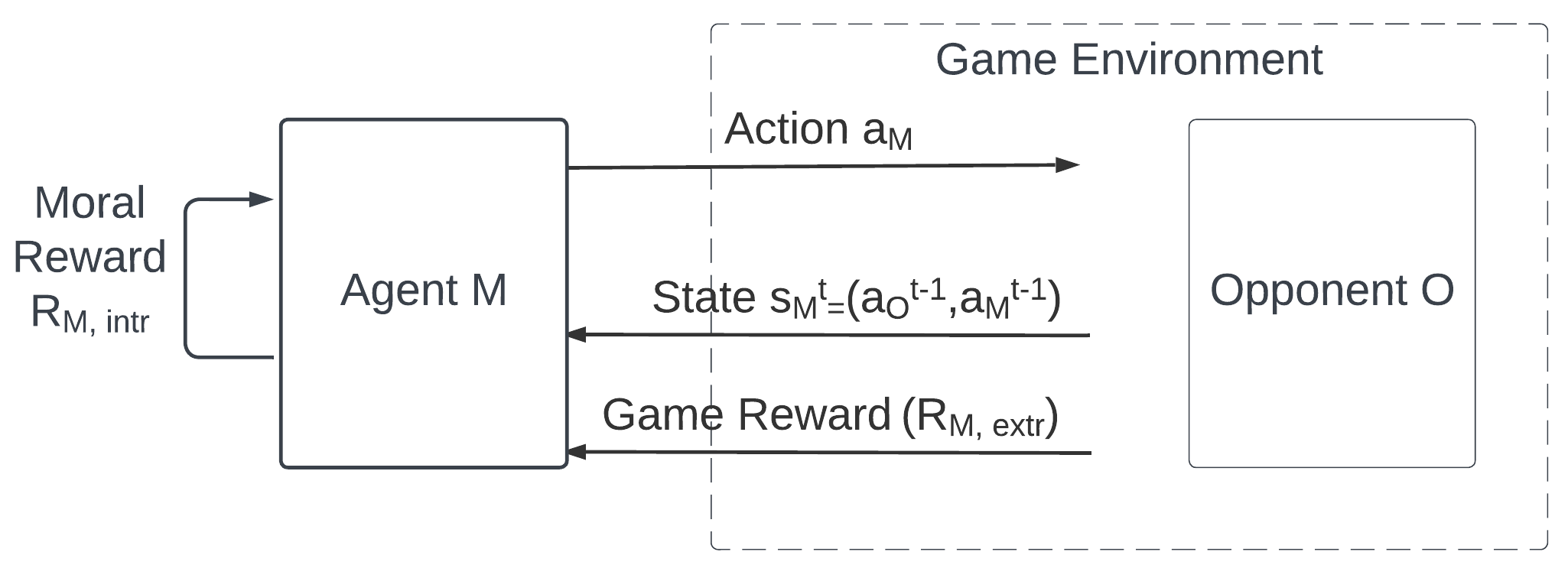}
  \caption{A step in the RL process, from the point of view of a learning moral agent $M$ playing against a learning opponent $O$. Agent $M$ calculates their moral (intrinsic) reward according to their own ethical framework, which may or may not consider the game (extrinsic) reward coming from playing the dilemma.}
  \label{fig:diagram}
\end{figure}

As discussed, intrinsic rewards have been used to nudge RL agents towards learning more cooperatively in social dilemmas via certain social preferences \cite{hughes2018inequity,jaques2019social,peysakhovich2018consequentialist}. We extend this work by defining four Q-learning moral agents $M$ that learn according to various intrinsic rewards $R_{M_{intr}}$ (see Figure \ref{fig:diagram}), and contrasting these against a traditional Q-learning \textit{Selfish} agent\footnote{We use the term moral agents for agents that are not selfish. However, selfishness itself can be considered as a moral choice, expressed as rational egotism. In the case of social dilemmas, selfishness maps with the concept of rationality. For these agents $R_{M_{intr}} = R_{M_{extr}}$.} $M$ who learns to maximize its extrinsic (game) reward $R_{M_{extr}}$ (e.g., \cite{leib02017multiagent}). 

We define four different moral learners: \textit{Utilitarian}, \textit{Deontological}, \textit{Virtue-equality} and \textit{Virtue-kindness}:
\begin{itemize}
\item the \textit{Utilitarian} agent tries to maximize the total payoff for both players \cite{Bentham1996}, and receives an intrinsic reward $R_{intr}^t$ based on collective payoffs from this iteration, equal to $R_{M_{extr}}^t+R_{O_{extr}}^t$;
\item the \textit{Deontological} agent tries to follow the conditional cooperation norm \cite{fehr2004social} and gets punished by $-\xi$ for defecting against an opponent who previously cooperated (i.e., receive a negative reward $R_{M_{intr}}^t=-\xi$ if $a_O^{t-1}=C$ and $a_M^{t}=D$); 
\item the \textit{Virtue-equality} agent tries to maximize equality between the two players' payoffs $R_{M_{extr}}^t$ and $R_{O_{extr}}^t$, measured using a two-agent variation of the Gini coefficient \cite{gini1912variabilita};
\item the \textit{Virtue-kindness} agent gets a reward $\xi$ for cooperating against any opponent on this iteration. 
\end{itemize}
Precise definitions of the reward functions are presented in Table \ref{tab:moral_definitions}. We use the same moral rewards across the three games to investigate the impact of payoff structure on learning. Given the payoff matrices of the three games used (Table \ref{tab:three_games}), we set the parameter $\xi$ in the two norm-based rewards to be $\xi=5$, so it sends a strong signal of a value similar to the maximum payoff available in any of the three games.
\textit{Utilitarian} agents can be defined as consequentialist since their reward depends on consequences in the environment (and, in particular, on the reward of the other agent), but the source of this morality can be external (social norms) or internal (the agent's values). \textit{Deontological} agents depend on an external norm. \textit{Virtue-equality} agents can be considered internal consequentialist, since they follow a consequence-based internal virtue. \textit{Virtue-kindness} agents depend on an internal norm. 

Finally, we consider the fact that the true nature of virtue ethics implies a combination of virtues within a single agent (e.g., \cite{graham2013moral}). Multi-objective models may be needed to represent the complexity of human values - moral or otherwise \cite{peschl2022moral,rodriguez2022instilling,vamplew2018human}. Thus, we implement a \textit{Virtue-mixed} agent as well, which uses a linear combination of the \textit{equality} and \textit{kindness} virtue rewards defined earlier. We use a linear combination because the two types of reward are independent - either value of the \textit{equality} reward is possible given either value of the \textit{kindness} reward. The formal definition of intrinsic reward for the \textit{Virtue-mixed} agent $M$ can be found in Table \ref{tab:moral_definitions}. We normalize the values of the \textit{kindness} reward $\xi$ as $\hat{\xi}$ in order to bound it to $[0,1]$ (to match the range of the \textit{equality} reward). The parameters $\beta$ and ($1-\beta$), with $\beta \in [0,1]$, define the relative weights on the two types of virtues. Here we present results for $\beta = 0.5$; an analysis of the impact of using different values for the weights can be found in Appendix F.
%
%


\subsection{Measuring Social Outcomes}

In order to investigate the impact of the presence of different intrinsically-motivated moral learners, we define three social outcome metrics, calculated as cumulative return values after the 10000 iterations - $G_{collective}$, $G_{gini}$, $G_{min}$:

\begin{align} 
G_{collective}&=\sum_{t=0}^{10000} {(R_{M_{extr}}^t+R_{O_{extr}}^t)} \\
G_{Gini}&=\sum_{t=0}^{10000}{(1-\frac{|R_{M_{extr}}^t-R_{O_{extr}}^t|}{R_{M_{extr}}^t+R_{O_{extr}}^t})} \\
G_{min}&=\sum_{t=0}^{10000}{\min(R_{M_{extr}}^t,R_{O_{extr}}^t)} . \end{align} 
The collective return $G_{collective}$ measures social welfare, identical to the \textit{Utilitarian} reward accumulated over time; $G_{Gini}$ measures the equality between rewards using the Gini coefficient \cite{gini1912variabilita}, identical to the \textit{Virtue-equality} reward accumulated over time; and $G_{min}$ measures the minimum reward obtained by either of the players, also accumulated over time to reflect long-term impacts. 

In summary, we model pairs of agents learning against one another, and consider pairs that are characterized by the same or different moral reward functions. We explore all possible combinations of the six agents (one \textit{Selfish} and five non-selfish intrinsically-motivated ones - \textit{Utilitarian}, \textit{Deontological}, \textit{Virtue-equality}, \textit{Virtue-kindness} and \textit{Virtue-mixed}) in three social dilemma games (Iterated Prisoner's Dilemma, Iterated Volunteer's Dilemma, and Iterated Stag Hunt). As an additional benchmark and for potential intellectual curiosity, we also study the learning processes of these six agents against static agents from traditional Game Theory tournaments \cite{axelrod1981evolution}, namely \textit{Always Cooperate}, \textit{Always Defect}, \textit{Tit for Tat}, and a \textit{Random} agent. The results of these experiments are available in Appendix B.


\section{Evaluation}

\newcommand{\subf}[2]{%
  {\small\begin{tabular}[h]{@{}c@{}}
  #1\\#2
  \end{tabular}}%
}

\newcommand{\subt}[1]{%
  {\small\begin{tabular}[h]{c@{}}
  #1
  \end{tabular}}%
}

\newcommand{\subfh}[2]{%
  {\small\begin{tabular}[h]{@{}c@{}}
  #1#2
  \end{tabular}}%
}

\newcommand{\subfthree}[3]{%
  {\small\begin{tabular}[h]{@{}c@{}c@{}}
  #1#2#3
  \end{tabular}}%
}

\renewcommand\theadfont{}
\renewcommand\theadalign{cc}

\subsection{Experimental Setup}
We use a linearly decaying exploration rate $\epsilon$ (from 1.0 to 0; see Appendix E for a discussion of the effects of a smaller exploration rate), a steady learning rate $\alpha=0.01$ (a parameter search on \textit{Selfish} versus \textit{Selfish} experiments demonstrated that the results were insensitive to the choice of alpha, hence we chose the smallest value), and discount factor $\gamma=0.90$. At the start of the game, all $Q$-values are initialized to 0. Each pair of agents interact in one episode for 10000 iterations of a given social dilemma game. We repeat each episode 100 times, randomizing seeds and initial states at every run. All pairwise comparisons are independent. 

\subsection{Systematic Comparison of Dyadic Interactions}
Across the three games, we investigate behaviors and outcomes emerging between each pair of agents $M$ and $O$ at the end of an episode using two metrics: simultaneous actions chosen (Figures \ref{fig:actions_IPD}-\ref{fig:actions_STH}) and social outcomes obtained (as defined in Equations 2-4; see Figures \ref{fig:outcomes_IPD}-\ref{fig:outcomes_STH}). An analysis of reward is presented in Appendix C.

\subsubsection{\textbf{Simultaneous Actions}}

\begin{figure*}[t]
\centering
\includegraphics[width=28.2mm]{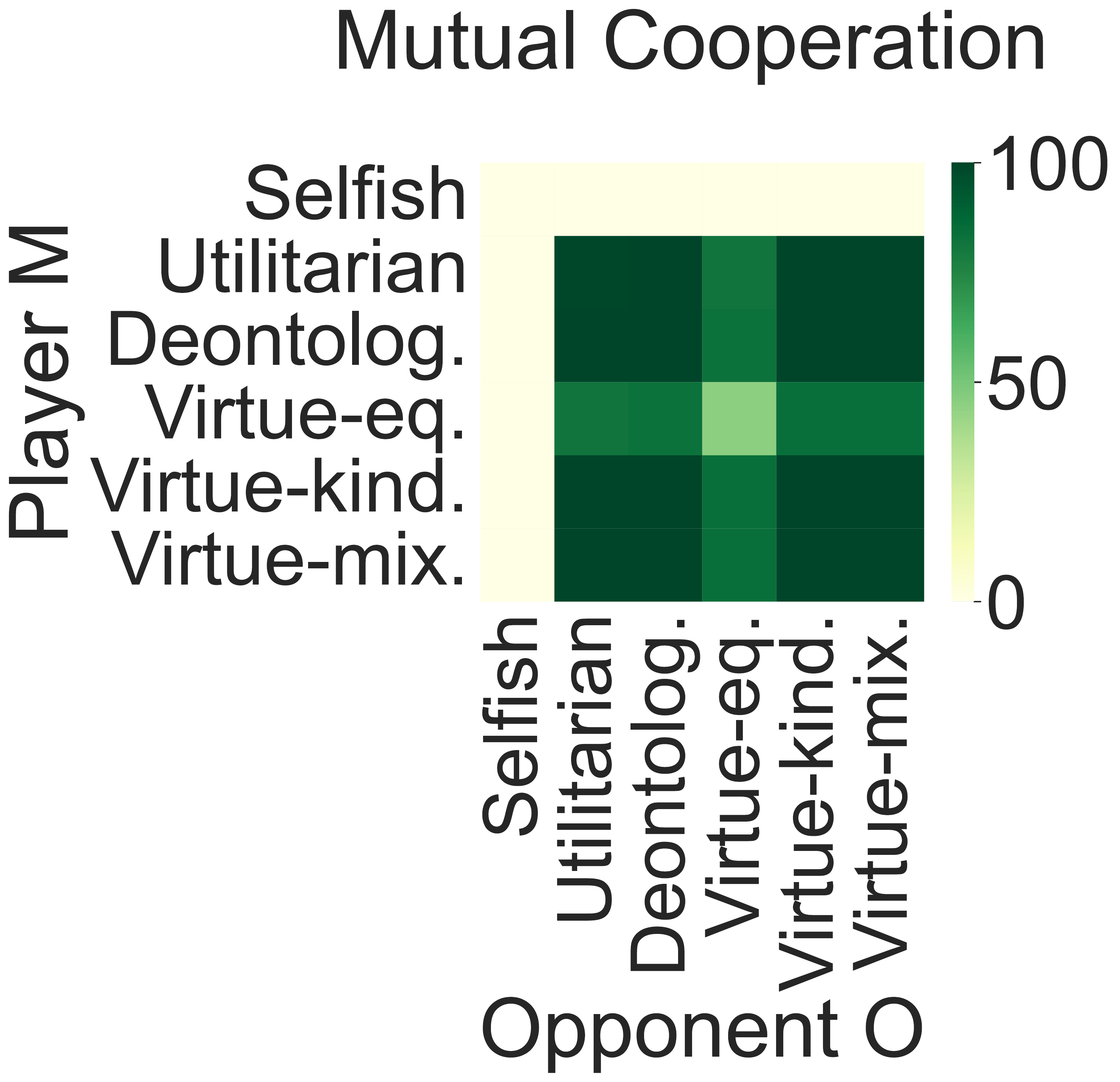}
\includegraphics[width=28.2mm]{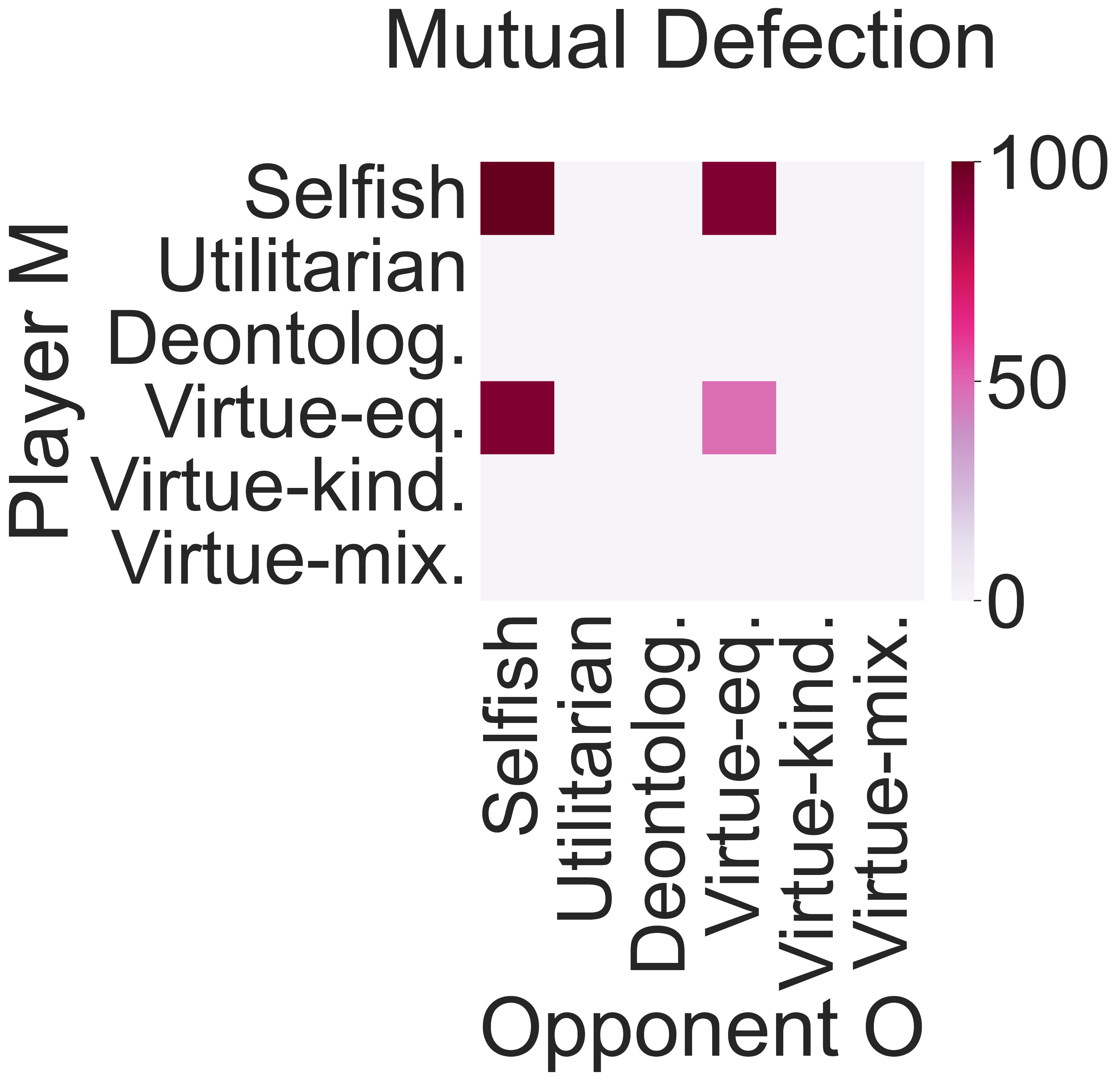}
\includegraphics[width=28.2mm]{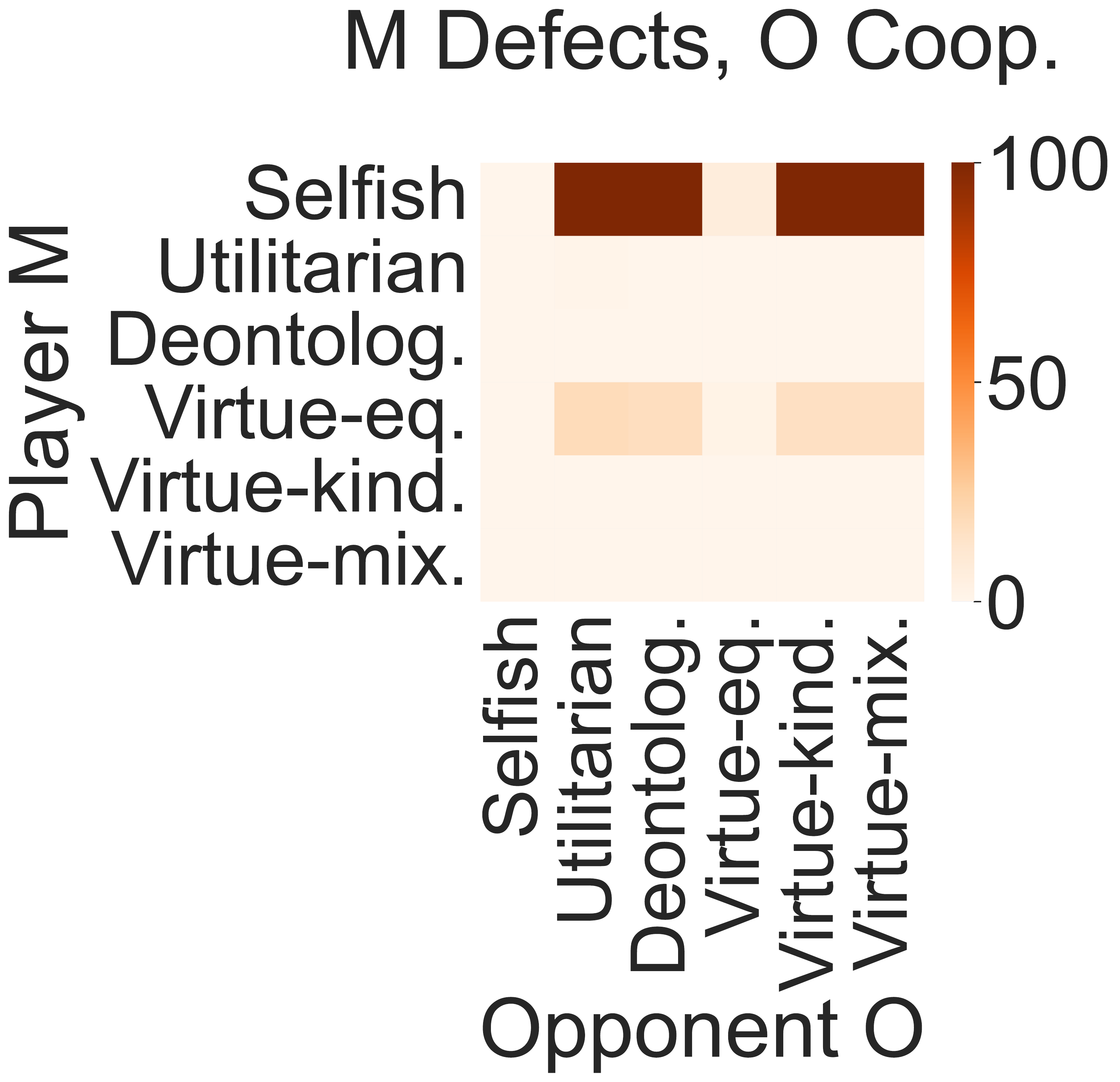}
\includegraphics[width=28.2mm]{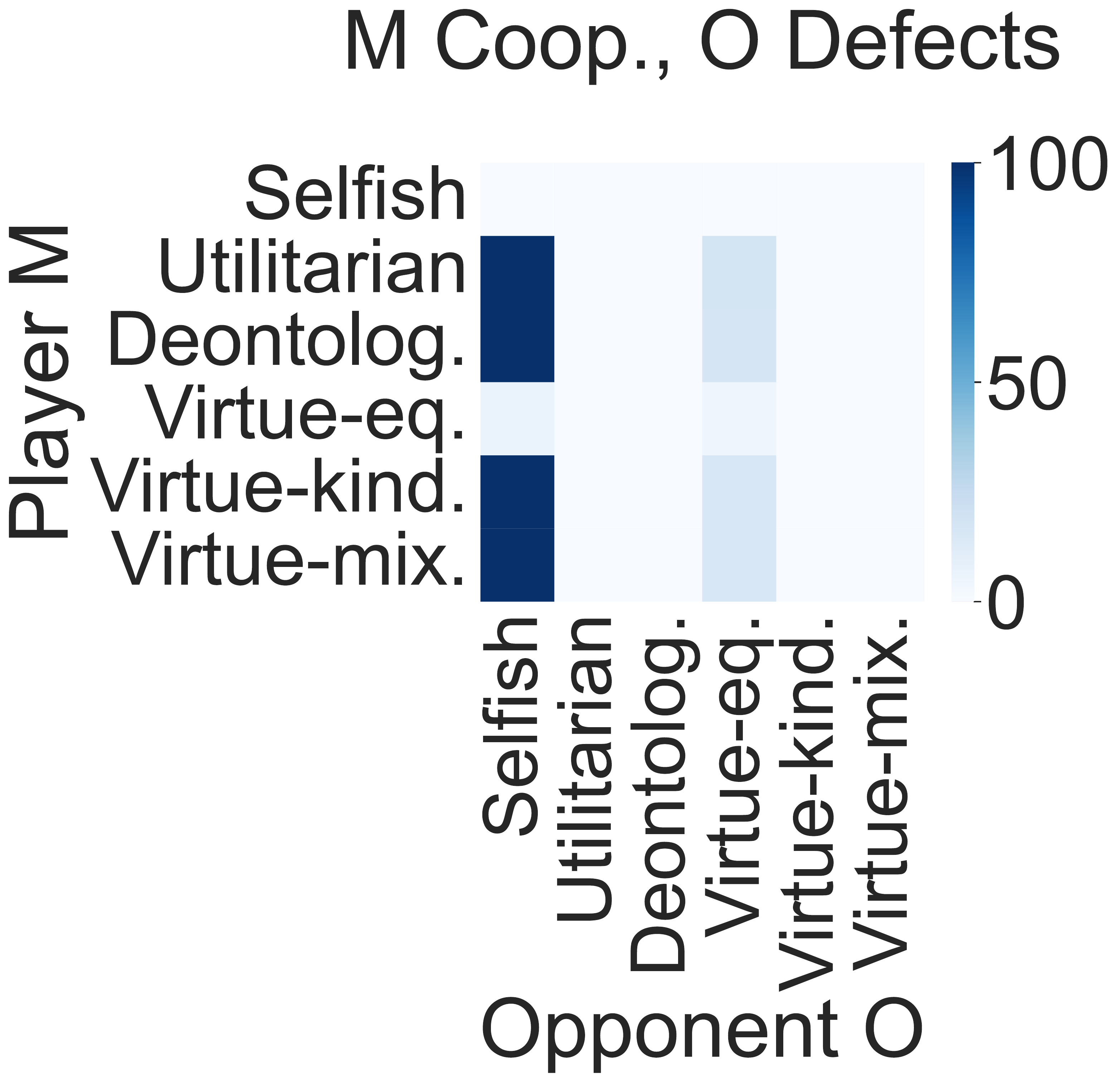}  
\caption{Iterated Prisoner's Dilemma game. Simultaneous actions played by each player $M$ type (row) and the opponent $O$ type (column) at the end of the learning period (10000 iterations). Action pairs are displayed as a percentage over 100 runs.}
\label{fig:actions_IPD}
\end{figure*}
\begin{figure*}[h]
\centering
\includegraphics[width=28.2mm]{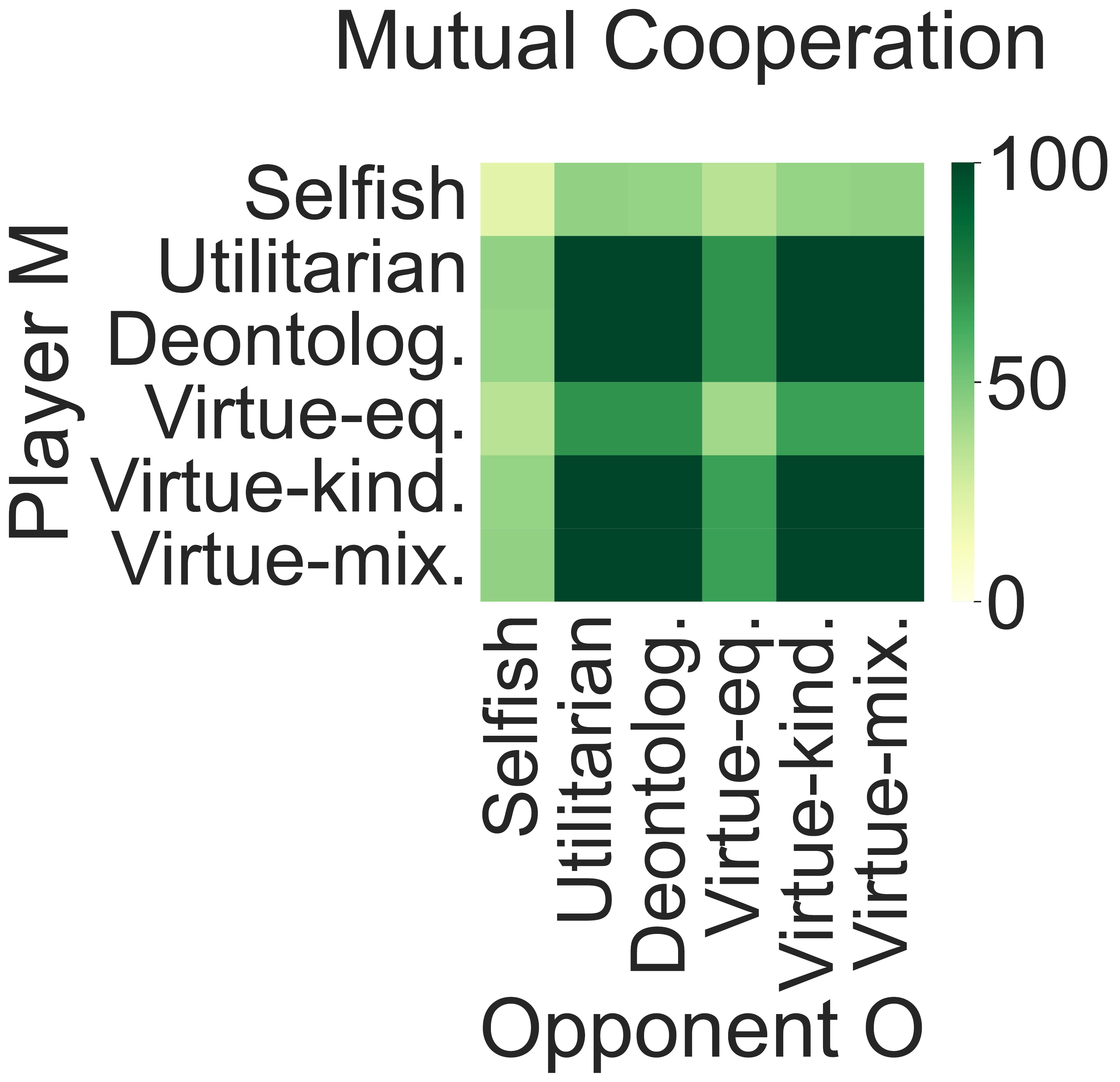}
\includegraphics[width=28.2mm]{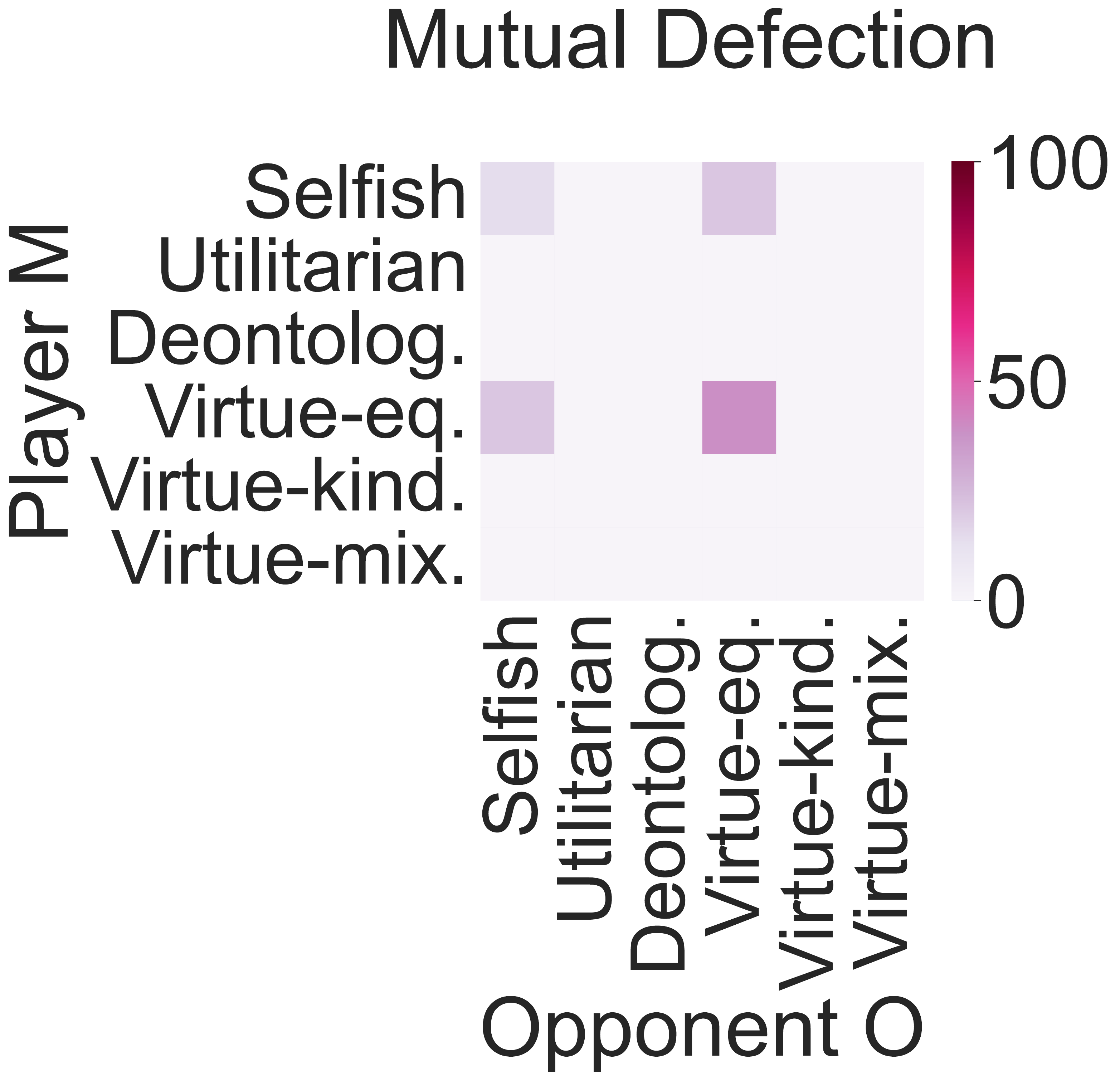}
\includegraphics[width=28.2mm]{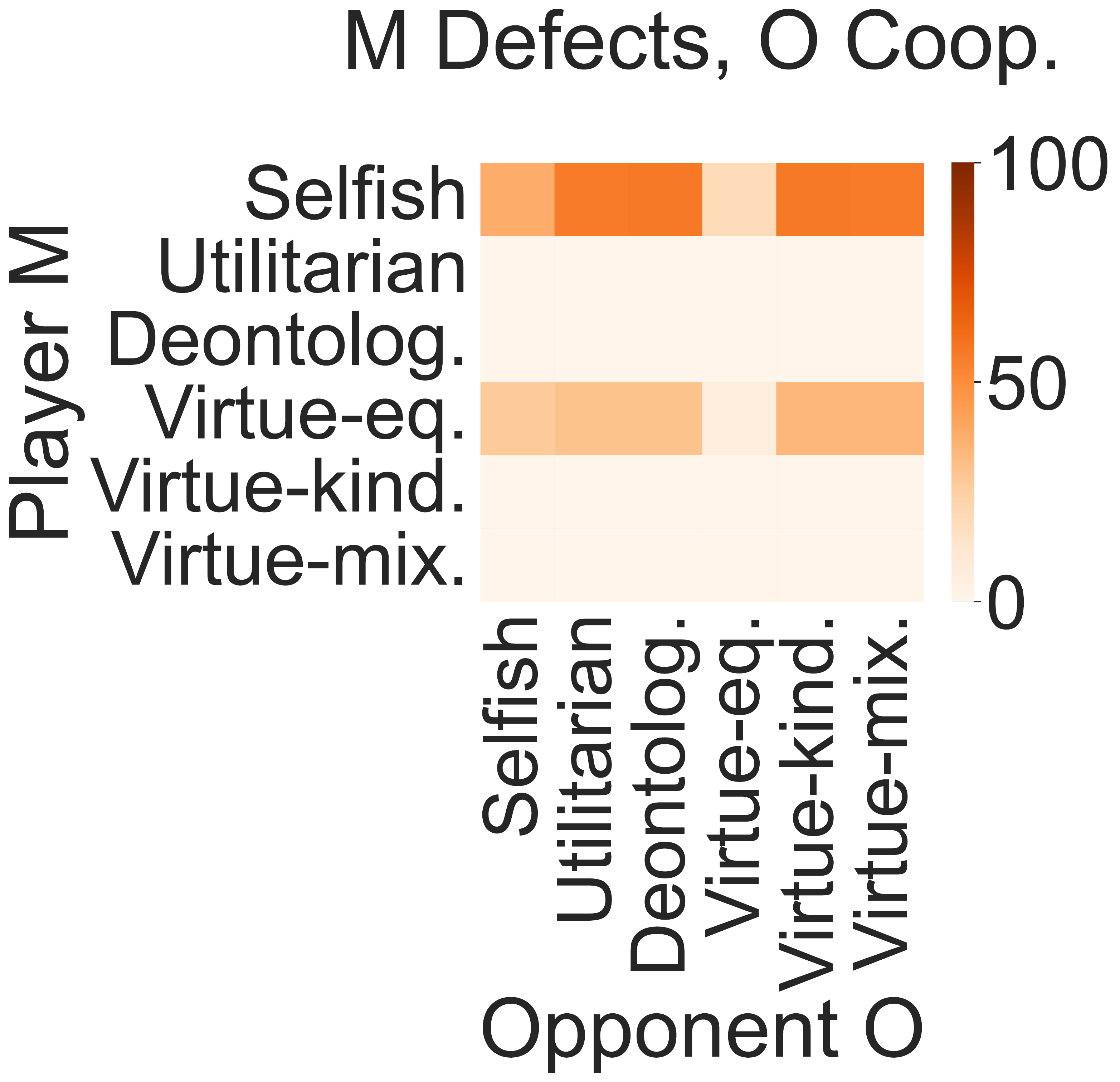}
\includegraphics[width=28.2mm]{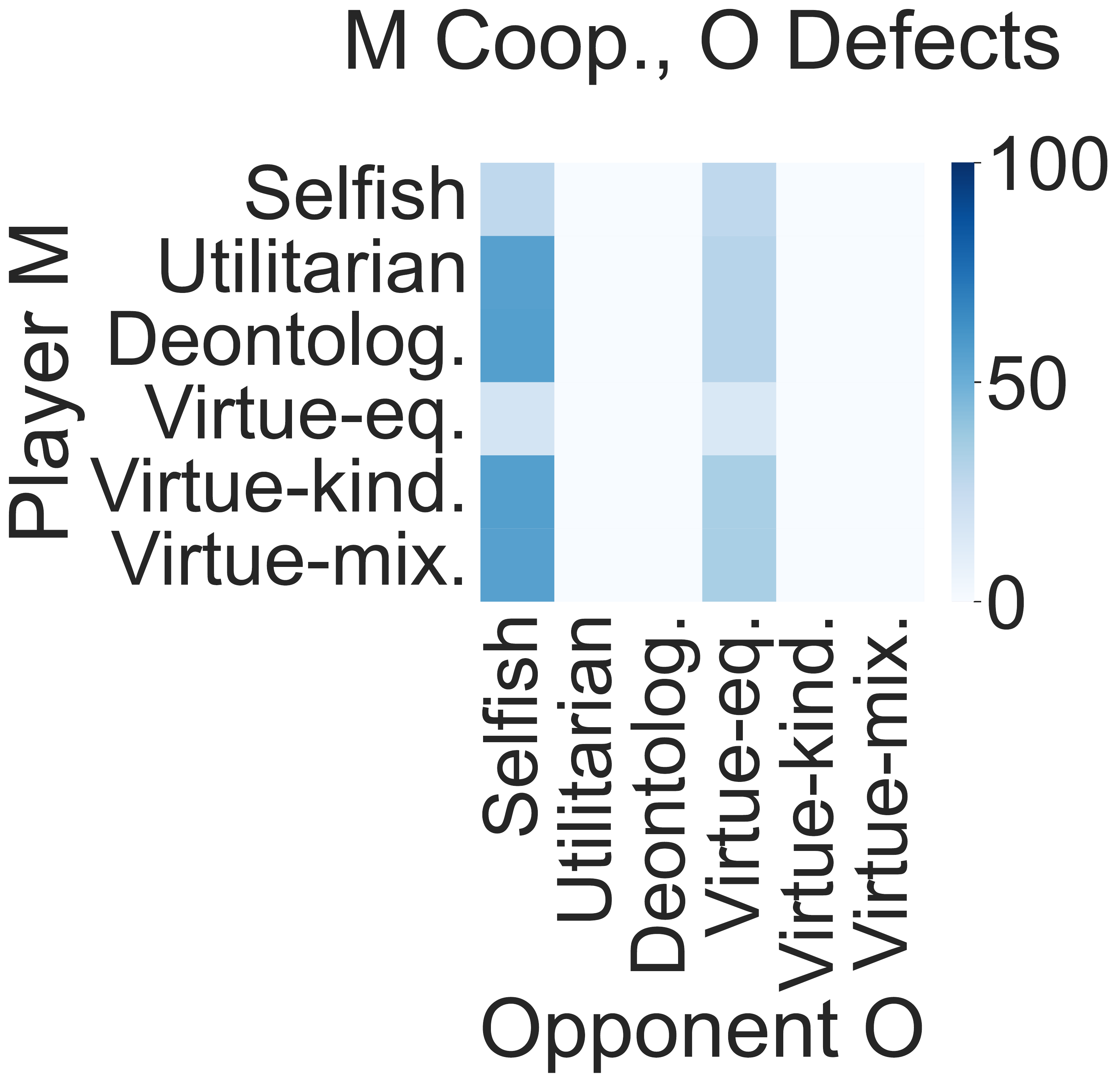}  
\caption{Iterated Volunteer's Dilemma game. Simultaneous actions played by player $M$ type (row) and the opponent $O$ type (column) at the end of the learning period (10000 iterations). Action pairs are displayed as a percentage over the 100 runs.}
\label{fig:actions_VOL}
\end{figure*}
\begin{figure*}[h!]
\centering
\includegraphics[width=28.2mm]{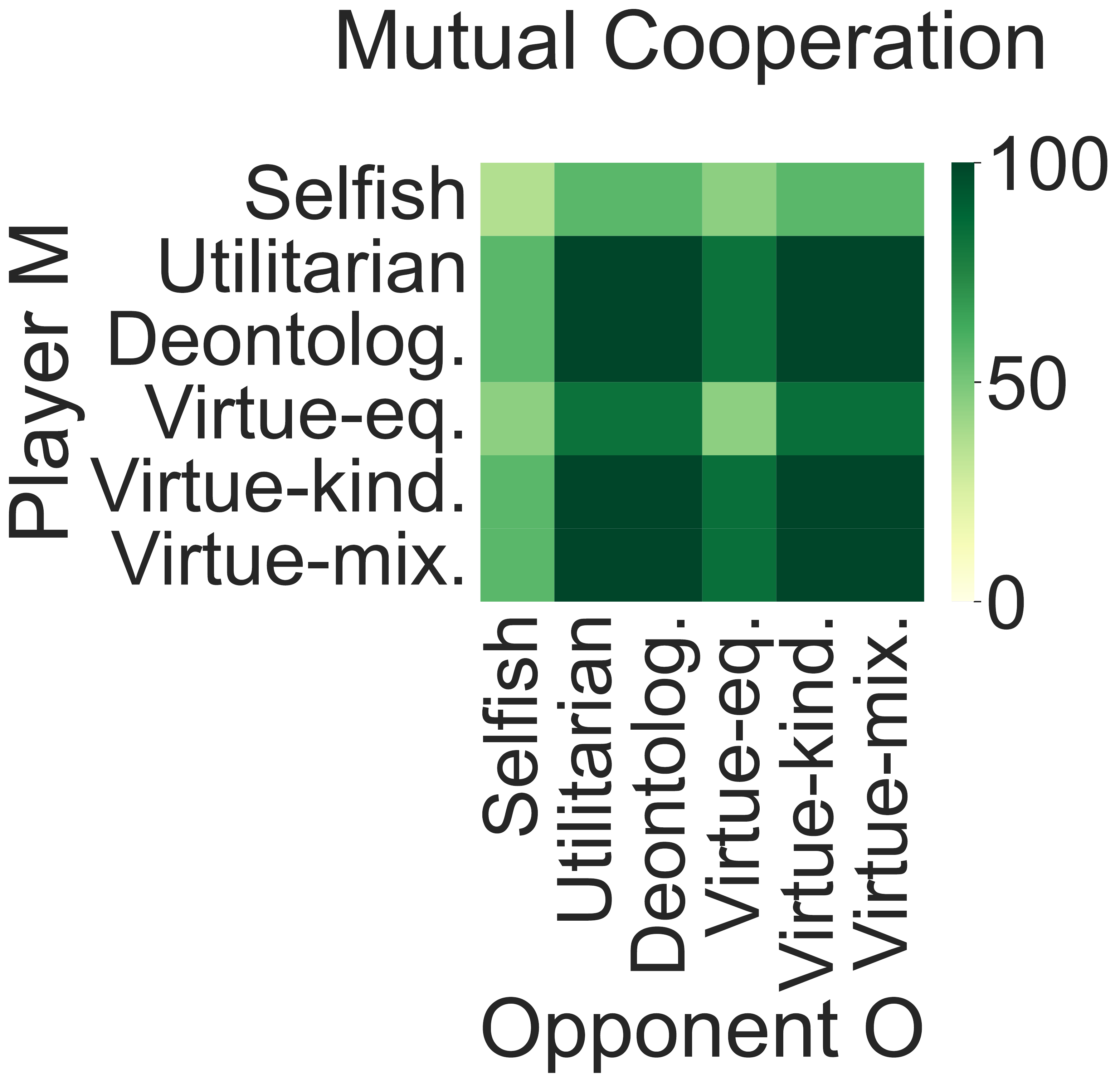}
\includegraphics[width=28.2mm]{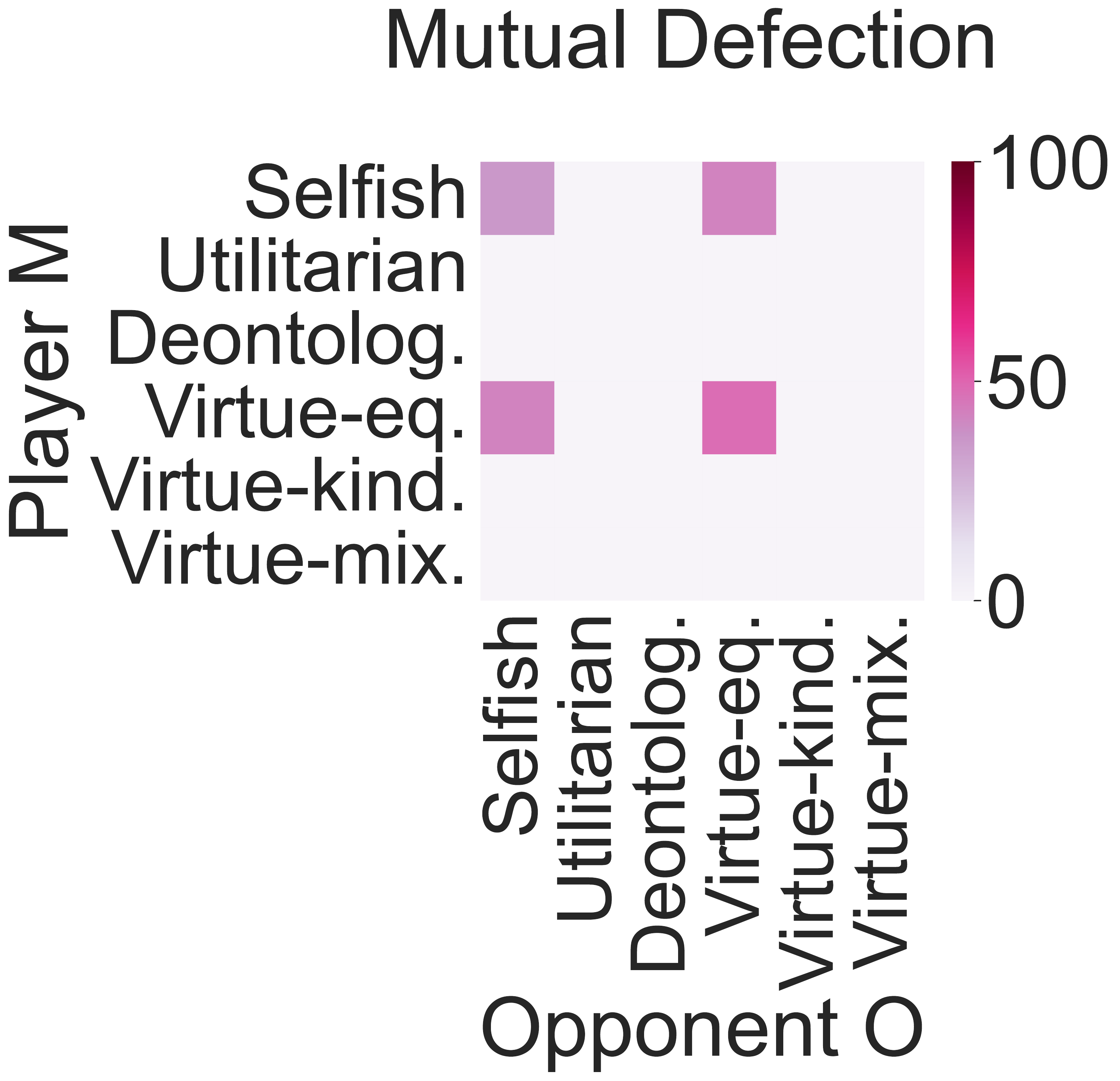}
\includegraphics[width=28.2mm]{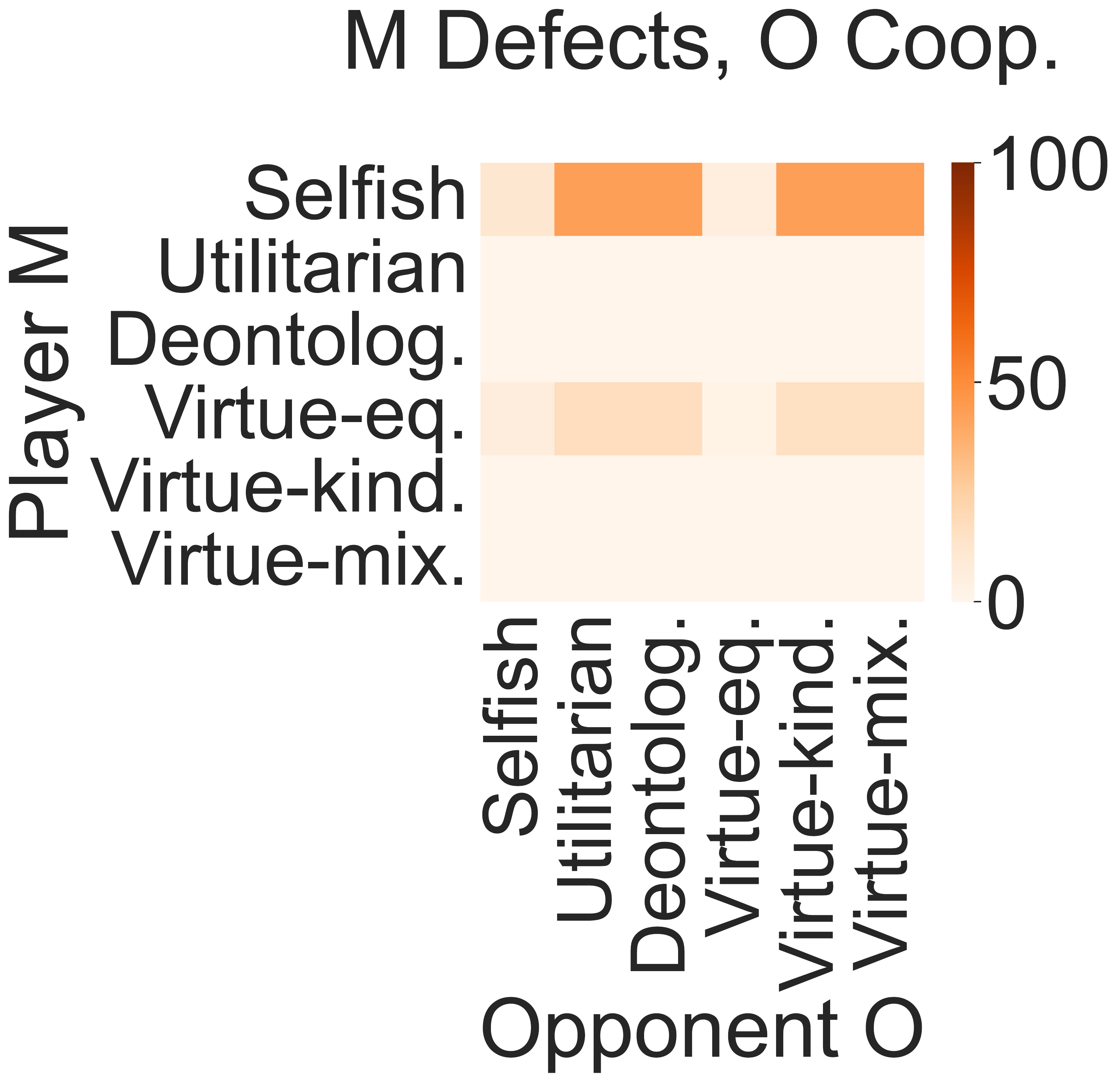}
\includegraphics[width=28.2mm]{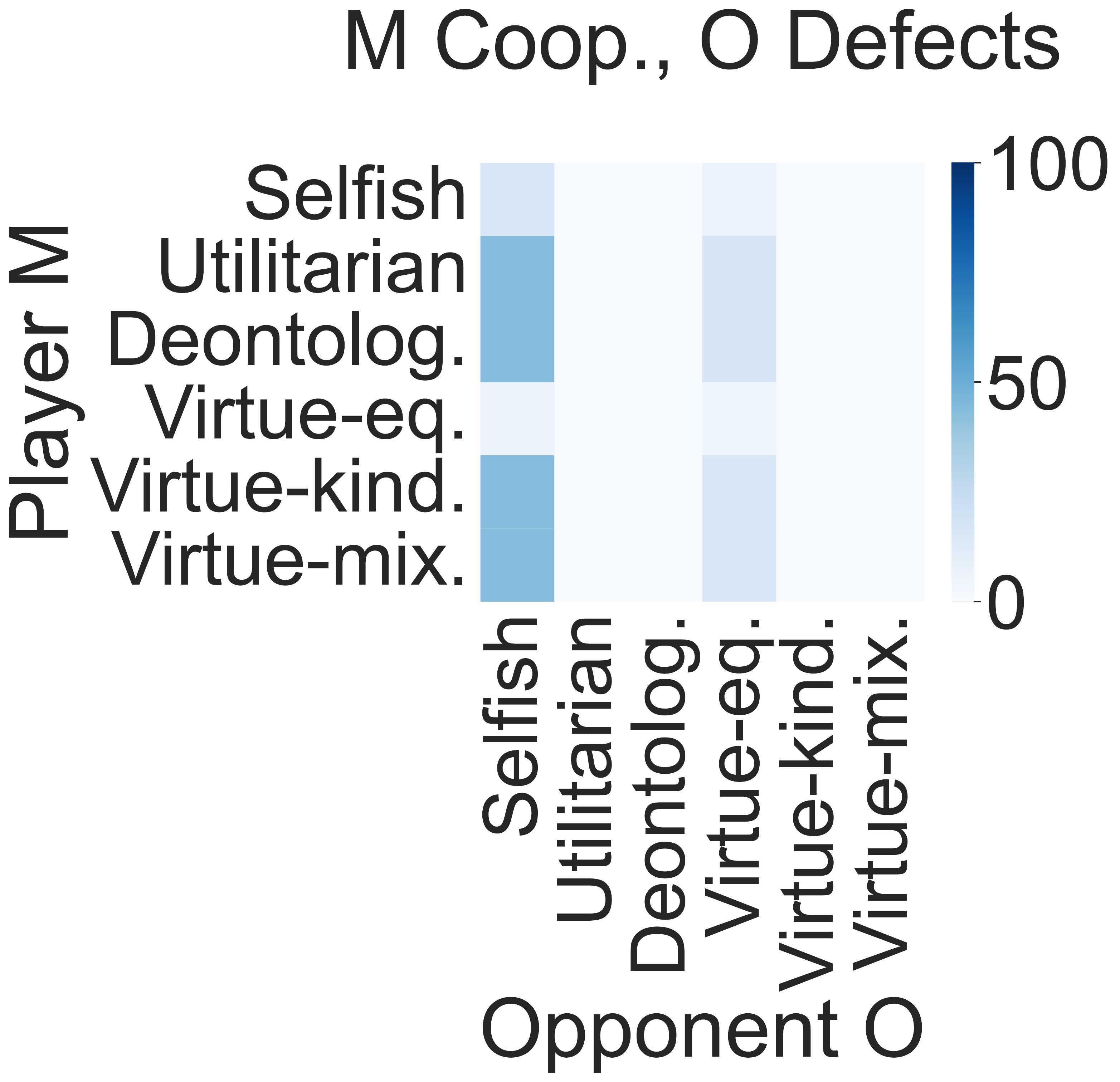}  
\caption{Iterated Stag Hunt game. Simultaneous actions played by player $M$ type (row) and the opponent $O$ type (column) at the end of the learning period (10000 iterations). Action pairs are displayed as a percentage over the 100 runs.}
\label{fig:actions_STH}
\end{figure*}

For every pair of agents, we measure the proportion of action pairs (as a percentage across the 100 runs) that corresponds to mutual cooperation, one player \textit{exploiting} the other (i.e., defecting when the other cooperates), and mutual defection. Results are presented in Figures \ref{fig:actions_IPD}-\ref{fig:actions_STH}. Visualizations of temporal dynamics are provided in Appendix A.

In the Iterated Prisoner's Dilemma (Figure \ref{fig:actions_IPD}), \textit{Selfish} players are motivated by greed and fear. As expected \cite{leib02017multiagent,sandholm1996multiagent}, the traditional \textit{Selfish} agent learns a `safe' Always Defect strategy against all agents. This results in it mutually defecting against itself and \textit{Virtue-equality} (shown in pink), and exploiting all other agents (shown in orange). For the non-selfish players, the \textit{Utilitarian}, \textit{Deontological}, \textit{Virtue-kindness} and \textit{Virtue-mixed} agents (i.e., all but \textit{Virtue-equality}) never defect against any other agent, and as a result achieve mutual cooperation when facing one another (shown in dark green), or face 100\% exploitation when facing a \textit{Selfish} agent (shown in dark blue). The \textit{Virtue-equality} agent, on the other hand, learns a more defensive but less cooperative strategy - it mutually defects on 100\% of the runs against a \textit{Selfish} opponent, or 50\% against itself, and learns to exploit all other non-selfish agents on 15-20\% of the runs (shown in light orange). 
Further analyses (see Appendix G) show that this is due to non-convergence of the agent's learning over the duration used in this set of experiments. Running the experiments for a higher number of steps results in slow convergence to fully cooperative strategies by the \textit{Virtue-equality} agent, similar to other non-selfish agents, which is an interesting finding per se.

In the Iterated Volunteer's Dilemma game (Figure \ref{fig:actions_VOL}), in which players no longer fear an uncooperative partner, even the \textit{Selfish} agent is able to avoid the least desirable outcome of mutual defection more than 75\% of the time (the values in pink do not go above 25\%). Furthermore, given this payoff structure, even the \textit{Selfish} agent can achieve mutual cooperation, 21\% of the time against itself, 34\% against \textit{Virtue-equality}, and over 40\% against other non-selfish agents. This provides a stark contrast to the Prisoner's Dilemma, in which a \textit{Selfish} agent could not achieve mutual cooperation at all. Two \textit{Virtue-equality} agents playing against each other, on the other hand, mutually defect in 40\% of the runs - more than against the \textit{Selfish} agent. In contrast, the \textit{Utilitarian}, \textit{Deontological}, \textit{Virtue-kindness} or \textit{Virtue-mixed} agents never exploit an opponent (on this or the other two games - see panel in orange). However, these agents themselves might get exploited by a \textit{Selfish} or \textit{Virtue-equality} agent 56-57\% of the time (shown in blue), though not as much as on the Prisoner's Dilemma. 

Removing the greed motivation in the Iterated Stag Hunt game (Figure \ref{fig:actions_STH}) leads to an even greater likelihood of a \textit{Selfish} agent achieving mutual cooperation against all other agents (boosting it to 45\% against \textit{Virtue-equality}, and over 55\% against other moral learners). The \textit{Selfish} agent is also better at avoiding exploiting its opponents than in the other two games (43\% maximum, shown in orange), but it is more likely to end up mutually defecting against itself (36\% of the runs) or a \textit{Virtue-equality} agent (42\% of the runs, shown in pink) than in the Volunteer's Dilemma. The \textit{Virtue-equality} agent here also converges to mutual defection against itself 48\% of the time. Against other non-selfish agents, it mutually cooperates on 83\% of the runs, yet still exploits them on 13\% of the runs (see discussion of non-convergence in Appendix G). The \textit{Utilitarian}, \textit{Deontological}, \textit{Virtue-kindness} and \textit{Virtue-mixed} agents achieve mutual cooperation 100\% of the time in the Iterated Stag Hunt, as in the other two games. 

\subsubsection{\textbf{Social Outcomes}}
\begin{figure}[h]
\centering
\includegraphics[width=28mm]{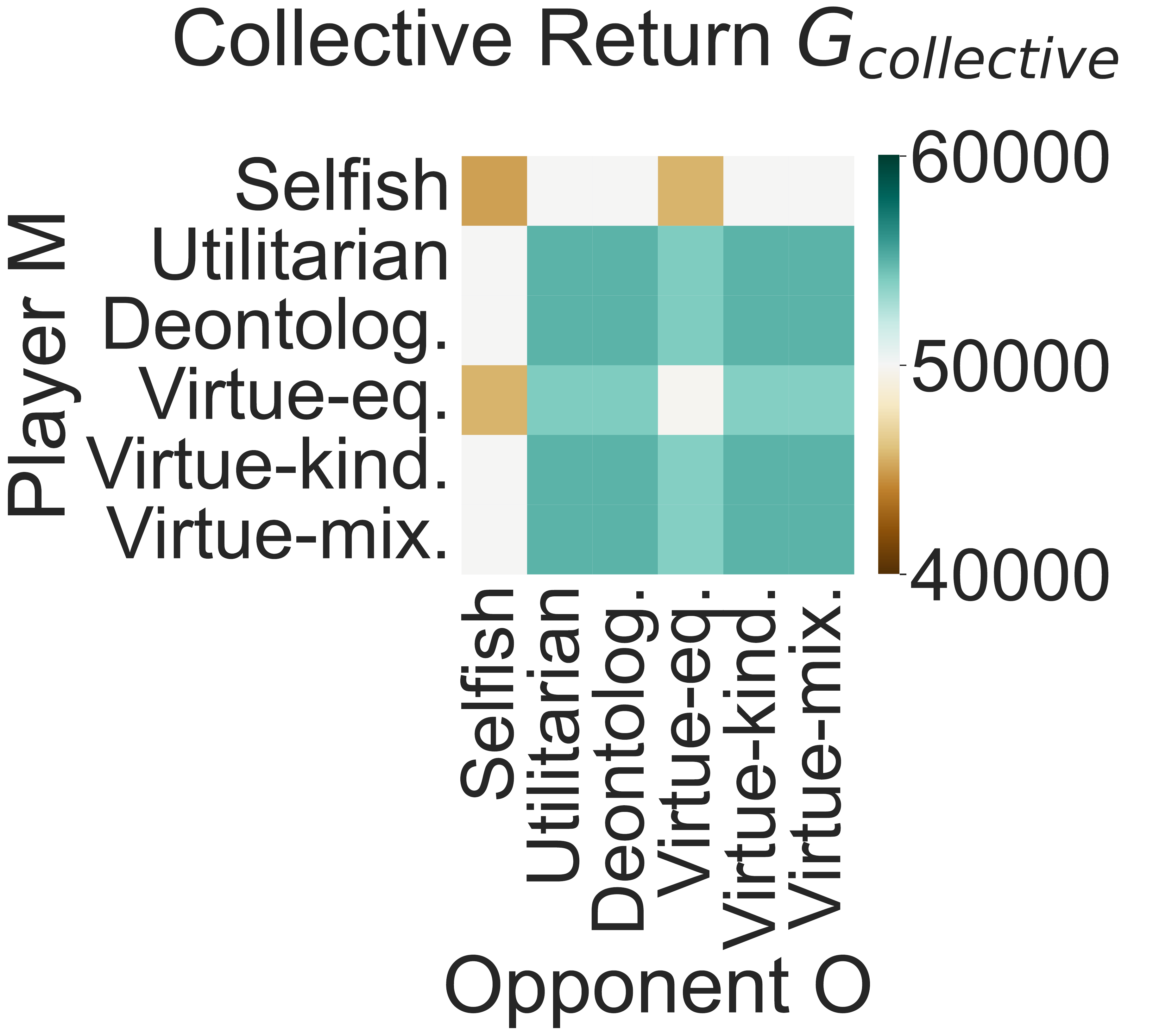}
\includegraphics[width=28mm]{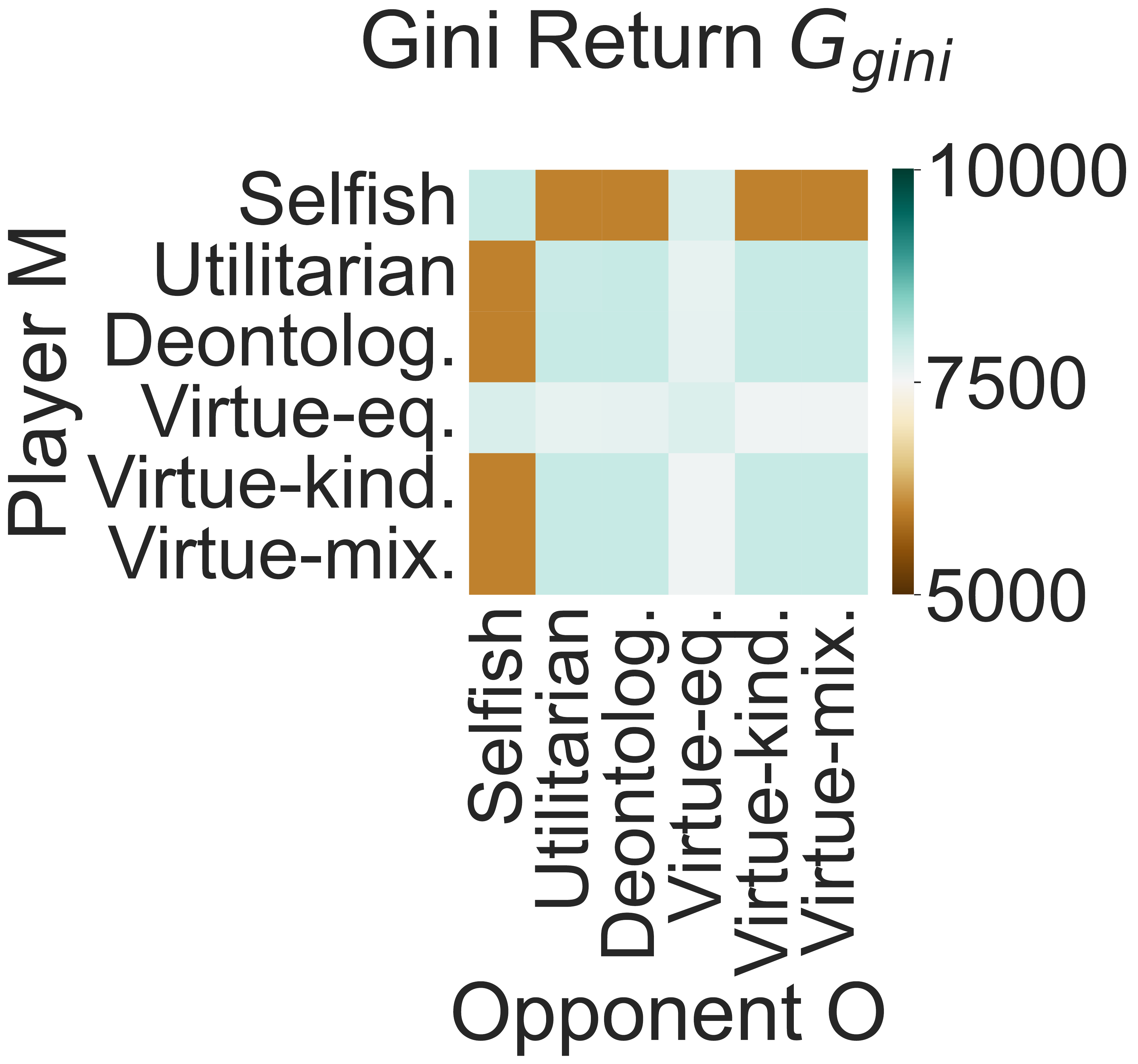}
\includegraphics[width=28mm]{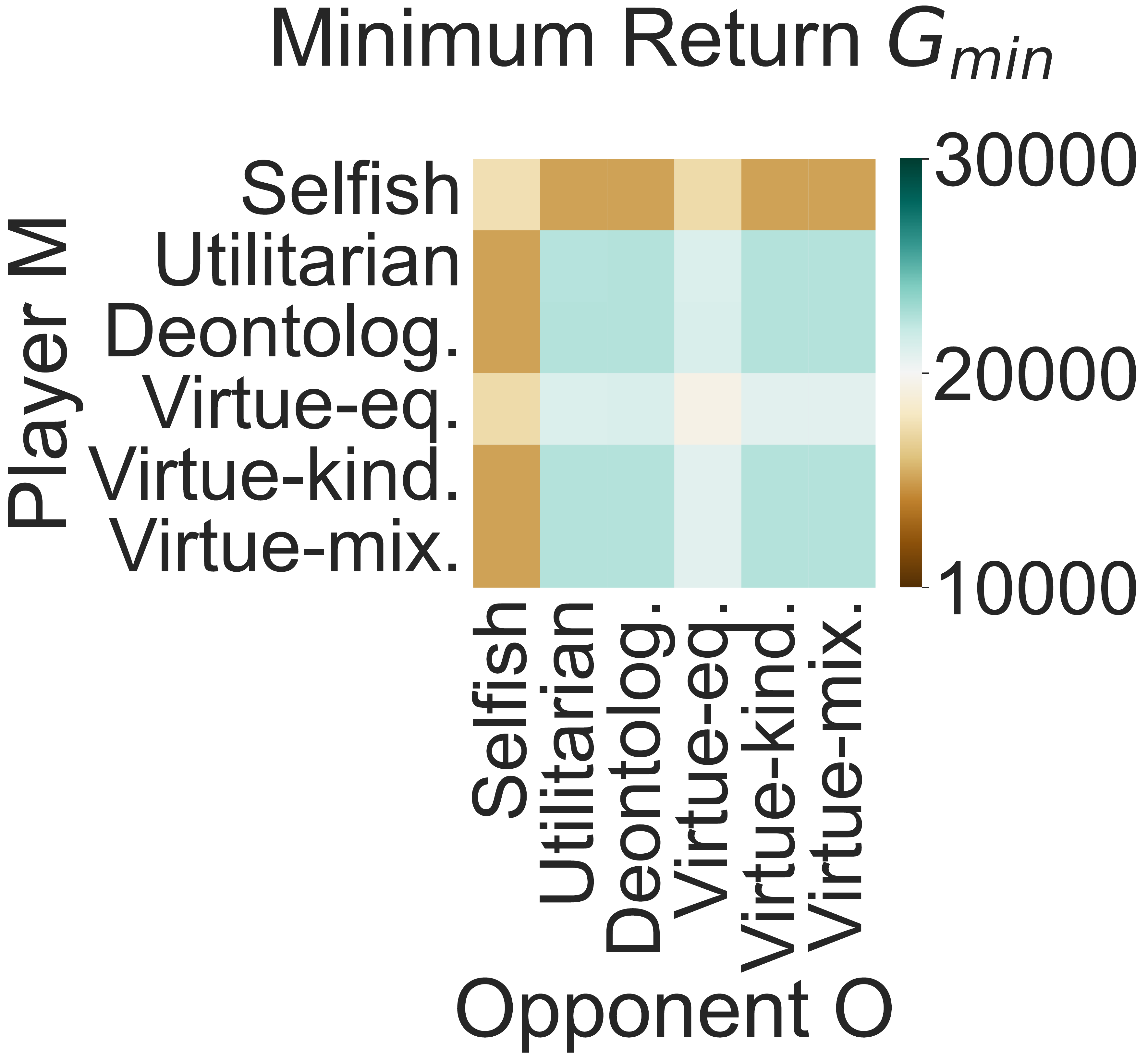}
\caption{Iterated Prisoner's Dilemma game. Relative social outcomes observed for player type $M$ vs. all possible opponents $O$. The plots display averages across the 100 runs.}
\label{fig:outcomes_IPD}
\end{figure}

\begin{figure}[h!]
\centering
\includegraphics[width=28mm]{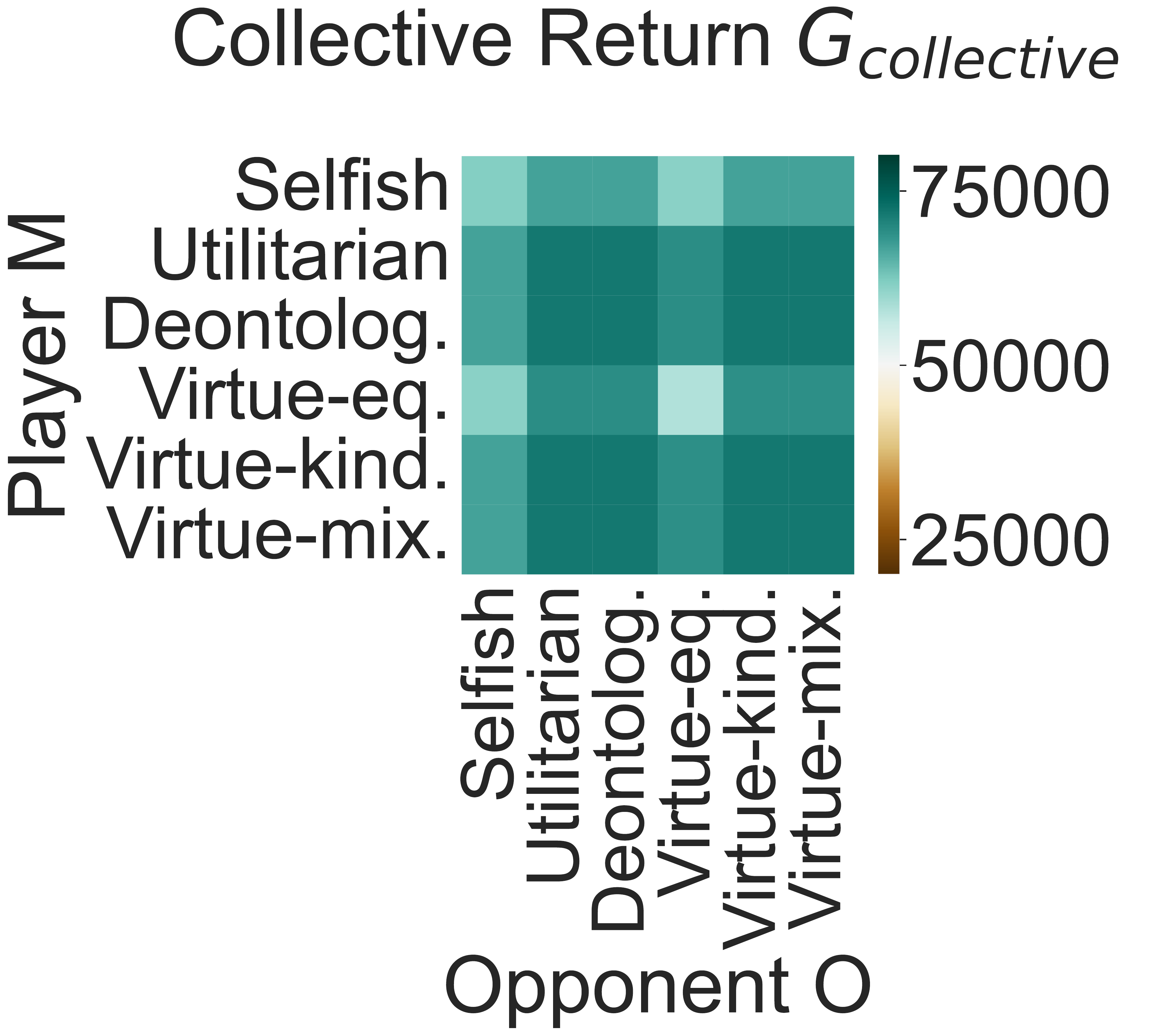}
\includegraphics[width=28mm]{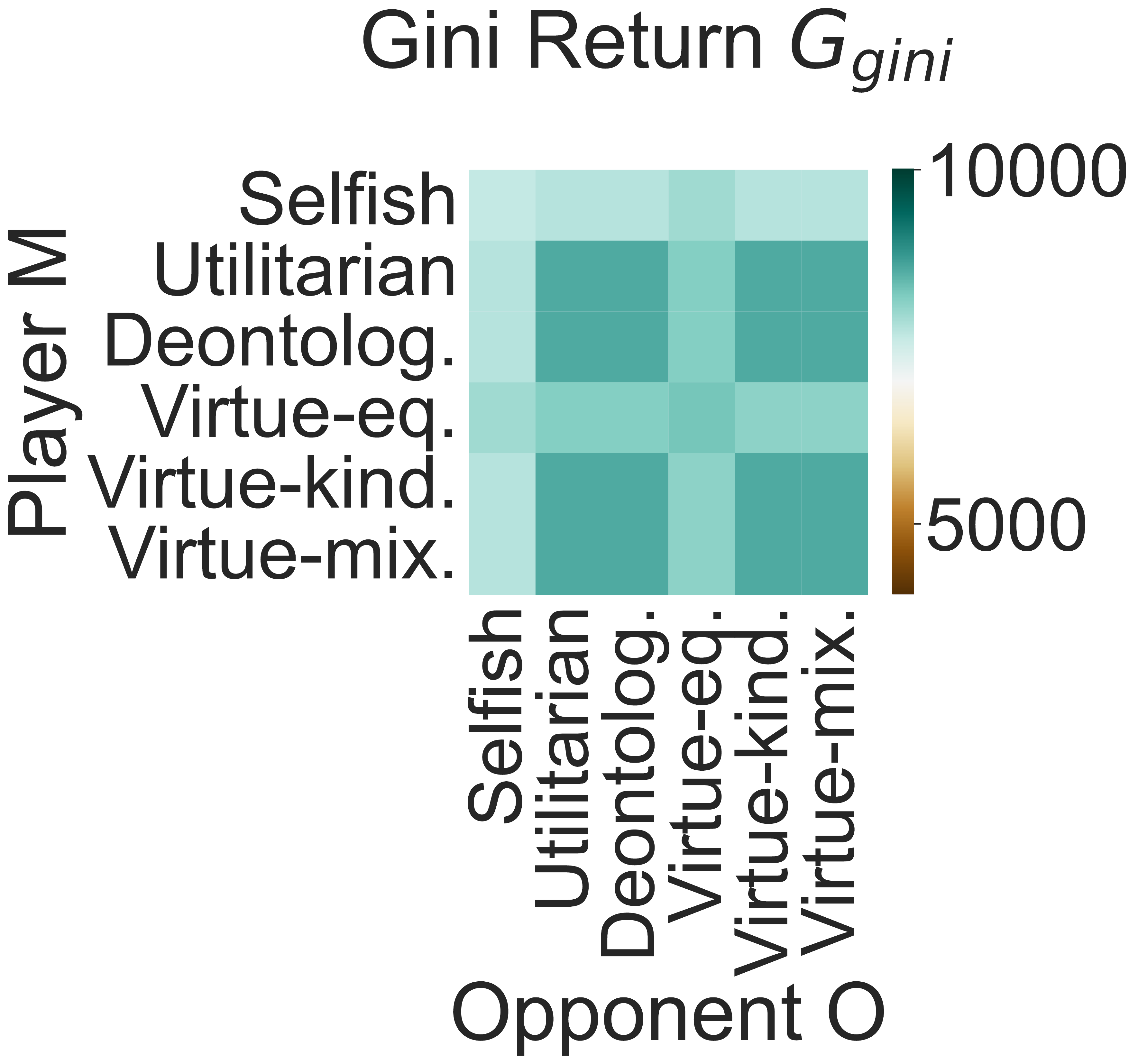}
\includegraphics[width=28mm]{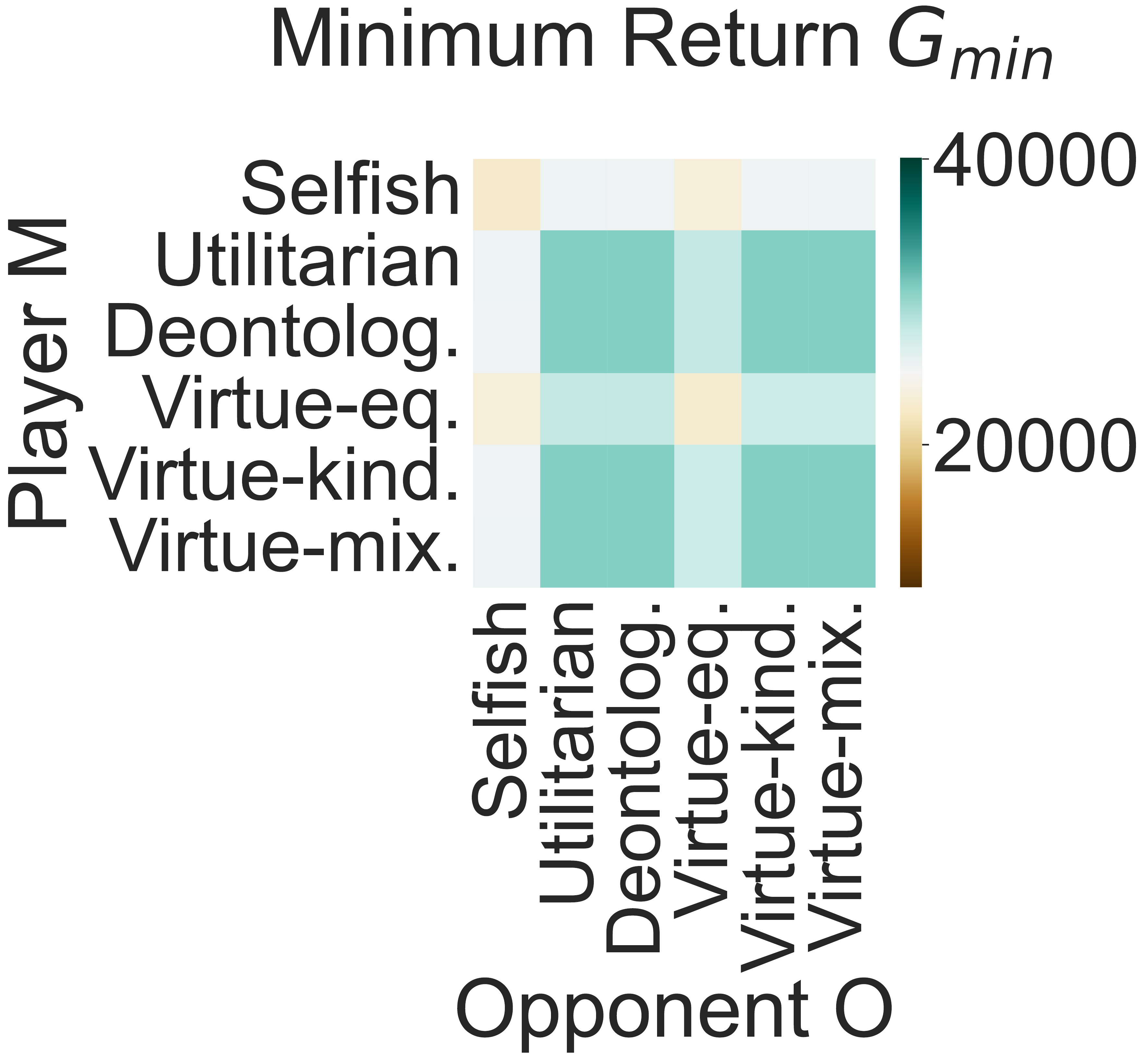}
\caption{Iterated Volunteer's Dilemma game. Relative social outcomes observed for player type $M$ vs. all possible opponents $O$. The plots display averages across the 100 runs.}
\label{fig:outcomes_VOL}
\end{figure}

\begin{figure}[h!]
\centering
\includegraphics[width=28mm]{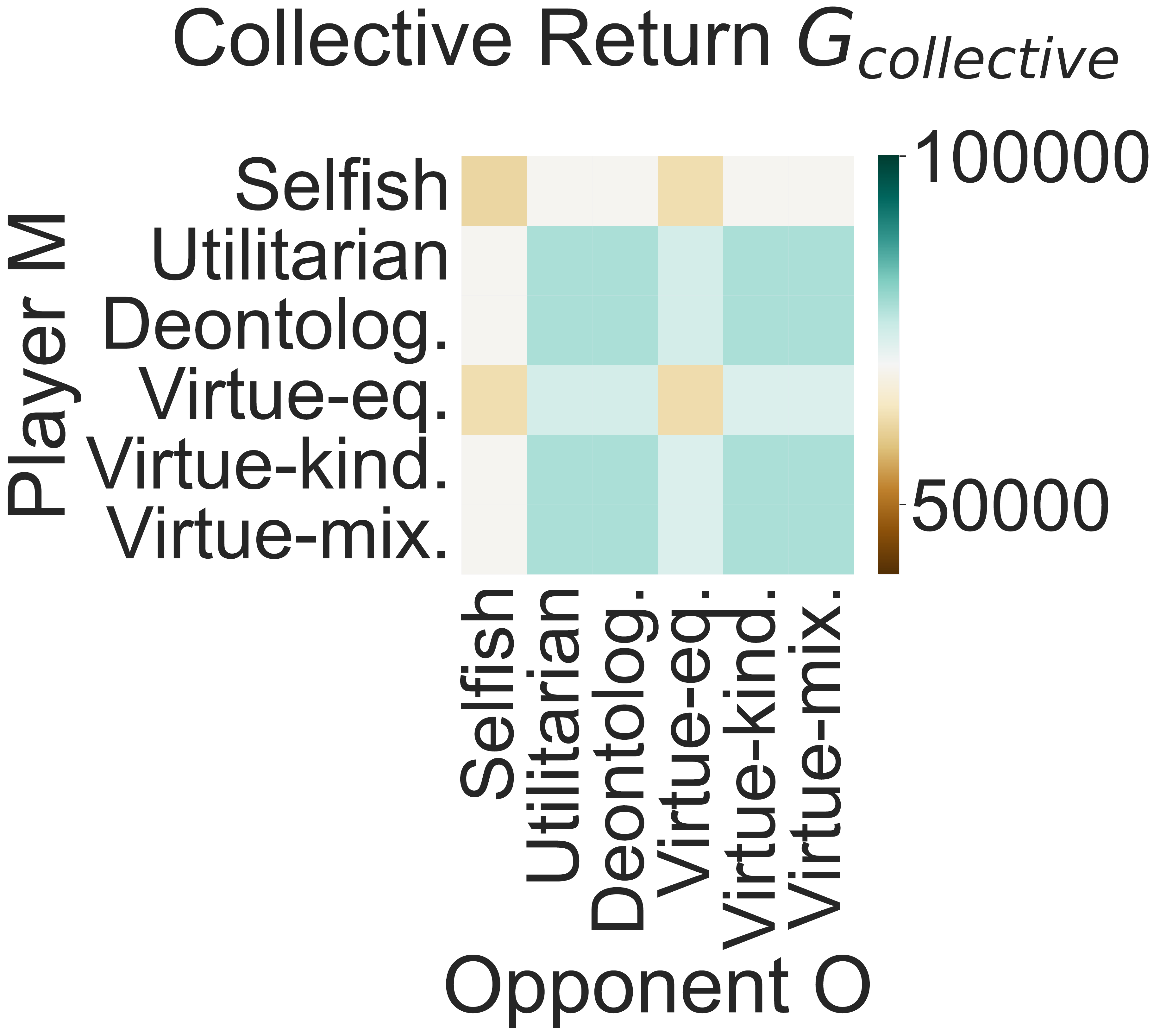}
\includegraphics[width=28mm]{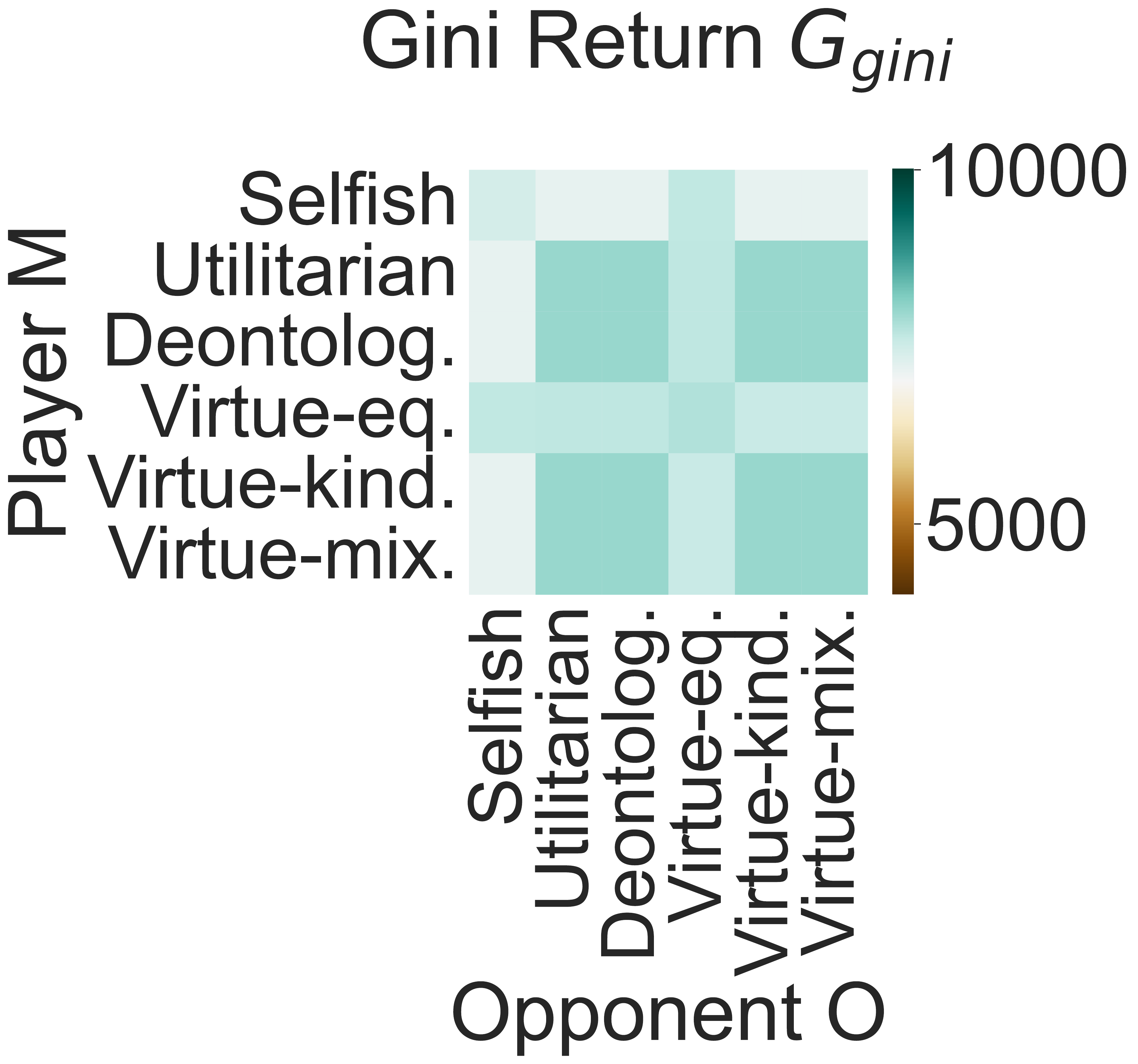}
\includegraphics[width=28mm]{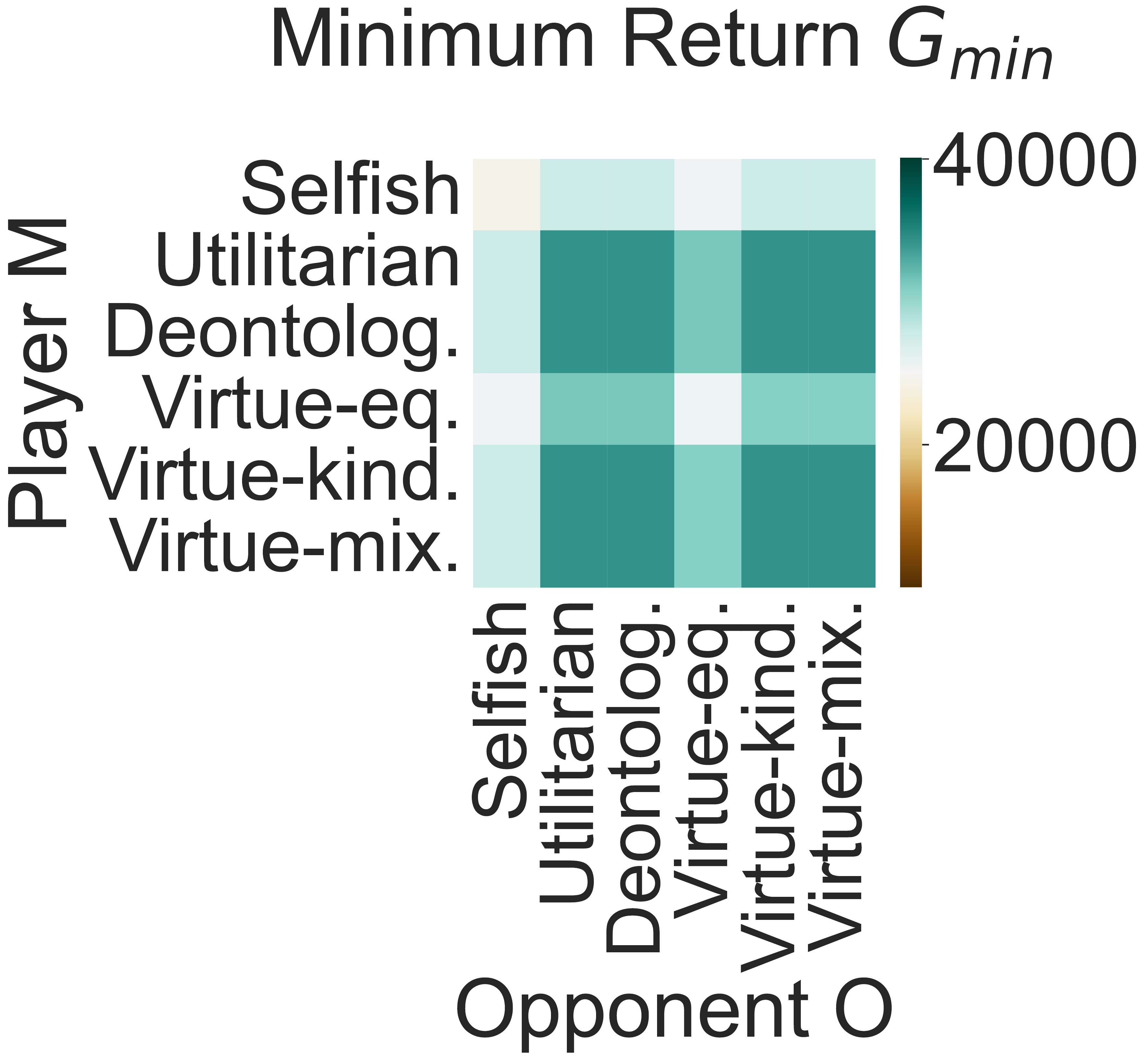}
\caption{Iterated Stag Hunt game. Relative social outcomes observed for player type $M$ vs. all possible opponents $O$. The plots display averages across the 100 runs.}
\label{fig:outcomes_STH}
\end{figure}

We calculate three social outcome values, as defined in Equations 2-4. Figures \ref{fig:outcomes_IPD}-\ref{fig:outcomes_STH} present average values across 100 runs. To aid comparison between games, the color saturation varies from the possible minimum to the possible maximum cumulative value for each respective outcome in each game. Confidence Intervals around these means are presented in Appendix D - they do not change the interpretation of the results.

Broadly, we observe that the best social outcomes on all three games are achieved by any combination of the \textit{Utilitarian}, \textit{Deontological}, \textit{Virtue-kindness} or \textit{Virtue-mixed} agents. The \textit{Virtue-equality} agent's strategy leads to a negative social outcome by either occasionally exploiting these four agents (this can be rectified via longer training episodes), or mutually defecting against itself or a \textit{Selfish} opponent (as seen in Figures \ref{fig:actions_IPD}-\ref{fig:actions_STH}). The \textit{Selfish} agent also creates negative outcomes by learning a defective policy.

We observe different outcomes depending on the characteristics of the game. In the fear-only game Stag Hunt (Figure \ref{fig:outcomes_STH}), the agents achieve lower collective return (left panel) than in the other games (Figures \ref{fig:outcomes_IPD}-\ref{fig:outcomes_VOL}), but greater benefit for the lowest-achieving agent (i.e., minimum return, right panel). The greatest collective return is instead obtained on the game where traditional rational agents are motivated by greed - the Volunteer's Dilemma (Figure \ref{fig:outcomes_VOL}, left panel). Interestingly, among non-selfish agents, the gain in collective return in this game is not achieved through exploitative actions (shown by the low values in the orange panel in Figure \ref{fig:actions_VOL}), and as a result minimum reward (right panel) stays rather high on average too. Finally, the greatest equality between agents over the 10000 iterations, as measured by the Gini return (middle panel), is also achieved on the Volunteer's Dilemma game (Figure \ref{fig:outcomes_VOL}), and Figure \ref{fig:actions_VOL} shows that this mostly involves mutual cooperation (green) rather than defection (pink).

Lastly, we can consider how each agent's goals were achieved by their reward function versus others, given the parallels between the outcome metrics defined in Equations 2-4, and the moral rewards defined in Table \ref{tab:moral_definitions}. The \textit{Utilitarian} agent, in practice, aimed to maximize collective return (left panel, Figures \ref{fig:outcomes_IPD}-\ref{fig:outcomes_STH}), and it did so better than the \textit{Selfish} or \textit{Virtue-equality} agent on all three games, but \textit{Deontological}, \textit{Virtue-kindness} and \textit{Virtue-mixed} agents maximized collective return just as well in these environments. The \textit{Virtue-equality} agent, which in practice aimed at maximizing Gini return (middle panel, Figures \ref{fig:outcomes_IPD}-\ref{fig:outcomes_STH}), did worse than other non-selfish agents on all three games due to non-convergence on some training runs (see Appendix G). Thus, the simple equality-based reward function was less effective at achieving equality as an outcome than other moral rewards. Finally, no agent focused directly on maximizing the least-earning agent's payoff (i.e., minimum return, right panel in Figures \ref{fig:outcomes_IPD}-\ref{fig:outcomes_STH}), but we observe that the \textit{Selfish} and \textit{Virtue-equality} agents achieved their goals by means of bringing another agent down, whereas all other agents were able to play morally without disadvantaging their opponent. No outcome metric directly maps to the goals of the two norm-based agents (\textit{Deontological} and \textit{Virtue-kindness}), but we observe that they achieve their moral goals without negative externalities to others, since they perform as well as the \textit{Utilitarian} and \textit{Virtue-kindness} agents on all three metrics across all games. 


\subsubsection{\textbf{Learning by a Mixed Moral Agent}}
\label{subsection:mixed_agents}
On all three games, the \textit{Virtue-mixed} agent is essentially indistinguishable from \textit{Virtue-kindness} in terms of actions (Figures \ref{fig:actions_IPD}-\ref{fig:actions_STH}) and social outcomes obtained (Figures \ref{fig:outcomes_IPD}-\ref{fig:outcomes_STH}). Thus, an equally-weighted combination of the \textit{equality} and \textit{kindness} elements in a \textit{Virtue-mixed} agent did not protect it from exploitation, but kept it just as likely as a pure-\textit{kindness} agent to achieve the Pareto-optimal outcome of mutual cooperation. Further experiments with different weights on each virtue showed that only very high values of $\beta$ (i.e., a high weight on \textit{equality} versus \textit{kindness}) make the mixed agent switch to more defensive and less cooperative behavior (see Appendix F for detail). 

\section{Discussion}

First of all, we are well aware that the agents implement very simplified forms of ethical systems. Nonetheless, this work can be considered as a starting point for the implementation of more complex decision-making mechanisms with agents grounded on moral views rather than purely selfish (i.e., rational) principles. This is especially important for the design of next-generation AI systems, which might have to implement constraints derived from AI regulations or codes of practice.

In comparison with past studies of Reinforcement Learning agents acting according to certain preferences in social dilemmas, this work provides a systematic analysis of behaviors and outcomes emerging from interactions between a variety of moral agent types. This investigation has allowed us to derive insights into behavioral patterns and strategies that emerge in social dilemmas. In particular, we find that simple norm-based (\textit{Deontological} and \textit{Virtue-kindness}) and \textit{Utilitarian} definitions of moral reward steer agents towards learning cooperative but easy-to-exploit policies, whereas a \textit{Virtue-equality}-based definition is slower to converge so results in more exploitative behavior.
The norm-based agents were defined in a way that promoted prosocial behavior. It remains to be seen what behavior emerges from agents who follow very different moral norms, such as punishing defecting opponents by defection. 

Additionally, we find that a multi-objective reward function is able to steer a mixed virtue agent away from uncooperative behavior, but makes them more likely to get exploited. A potential development could be to study whether reward functions that mix selfish and moral objectives or implement continual curriculum learning \cite{bengio2009curriculum} create more robust moral agents. Another interesting next step may be to model the behavior of these moral agents in more complex societies, for example with a partner selection mechanism \cite{Anastassacos2020partner}. Partner selection by \textit{Utilitarian} agents may largely result in cooperative outcomes, since these agents obtain greater reward when playing against other cooperative moral agents (i.e., \textit{Utilitarian}, \textit{Deontological}, \textit{Virtue-kindness} or \textit{Virtue-mixed}). It is harder to predict the behavior of \textit{Virtue-equality} agents with partner selection - would they drive other non-selfish agents away due to the small amount of exploitation that they perform? Finally, an extreme case would be to assume a society where norms are followed universally by all agents (an idealized situation considered, among the others, by Kant \cite{kant1981grounding}).

\section{Conclusion}


In this work, we have presented for the first time a systematic investigation of the learning of RL agents whose rewards are based on moral theories using classic social dilemmas. In order to do so, we have defined reward structures that are simplified yet representative of three key ethical frameworks. 

In our study, the Utilitarian, Deontological, Virtue-kindness and Virtue-mixed agents learn cooperative policies across every game. At the same time, they are exploited by the Selfish opponent. For the Virtue-equality agent, exploitative behavior emerges during the learning process before convergence. For the Virtue-mixed agent, the `kindness’ signal is stronger than `equality’.

Our main contribution is of methodological nature: this work provides a platform on which future research can build. For example, our approach can be applied to the study of more complex societies with additional or different mechanisms, possibly characterized by a large population of agents. Given its generality, we believe it can also be used as starting point to explore interactions between human and AI agents for real-world applications.

\bibliographystyle{named}
\bibliography{modeling_moral_choice}

\onecolumn
\appendix

\date{}




\begin{center}
\huge{Modeling Moral Choices in Social Dilemmas \\with Multi-Agent Reinforcement Learning: \\ Appendix}
\\
\end{center}


\section{Simultaneous Pairs of Actions over Time - Learning Player vs Learning Opponent}

\begin{figure}[!h]
\centering
\begin{tabular}{|c|cccccc}
\toprule
 & Selfish & Utilitarian & Deontological & Virtue-equality & Virtue-kindness & Virtue-mixed\\
\midrule
\makecell[cc]{\rotatebox[origin=c]{90}{Selfish}} & 
\subt{\includegraphics[width=22mm]{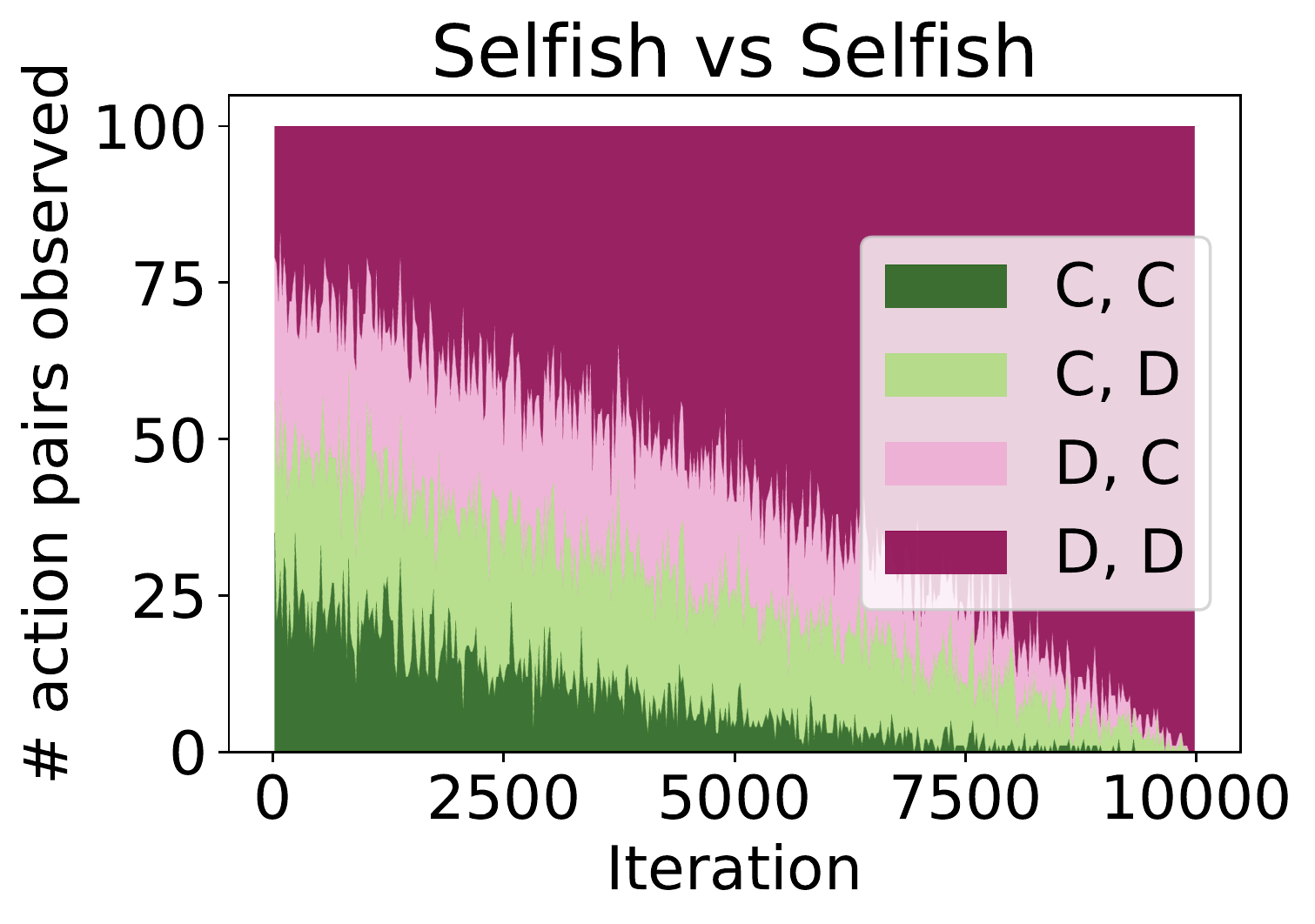}}
&
&
&
&
\\
\makecell[cc]{\rotatebox[origin=c]{90}{ Utilitarian }} & 
\subt{\includegraphics[width=22mm]{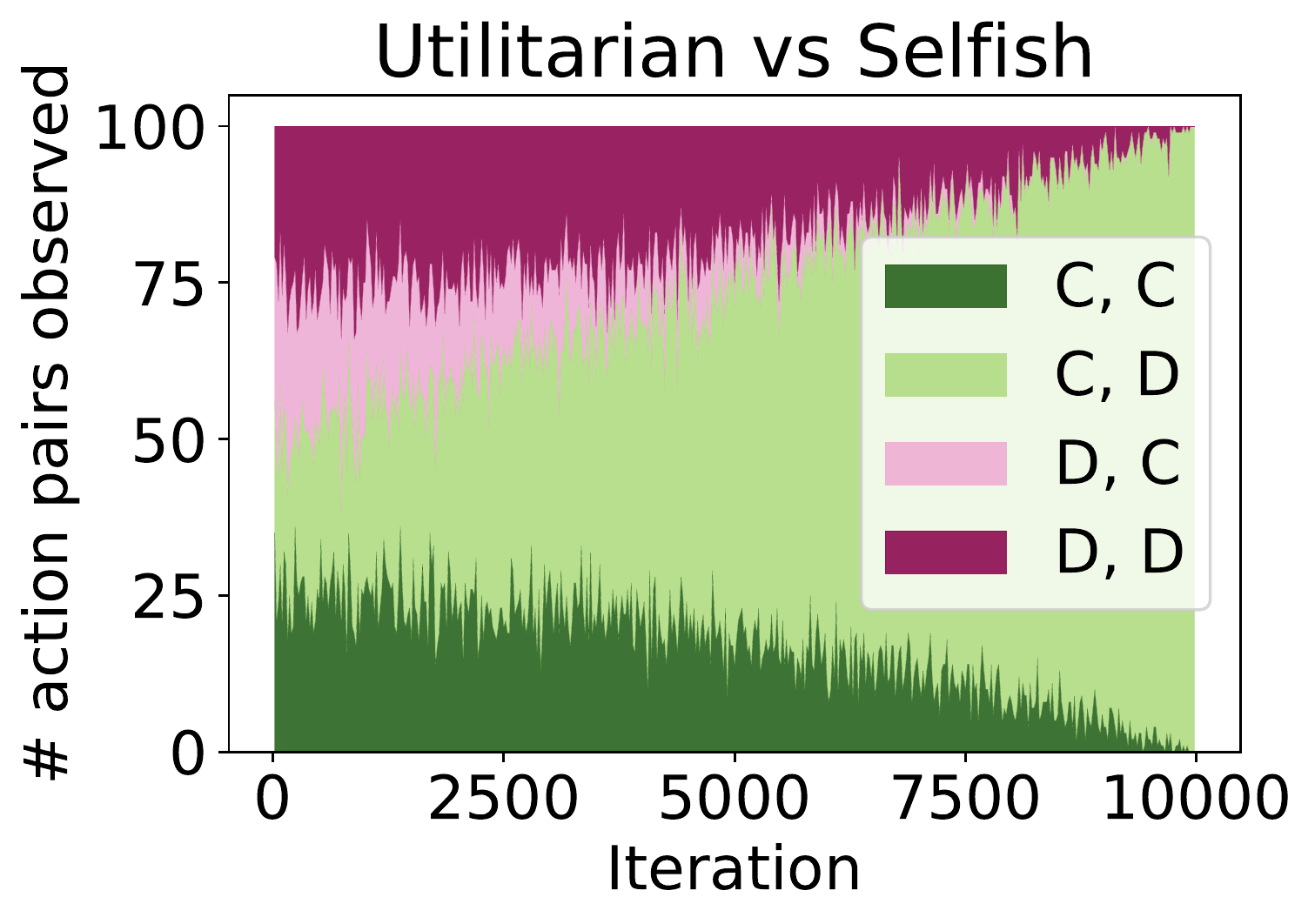}}
&\subt{\includegraphics[width=22mm]{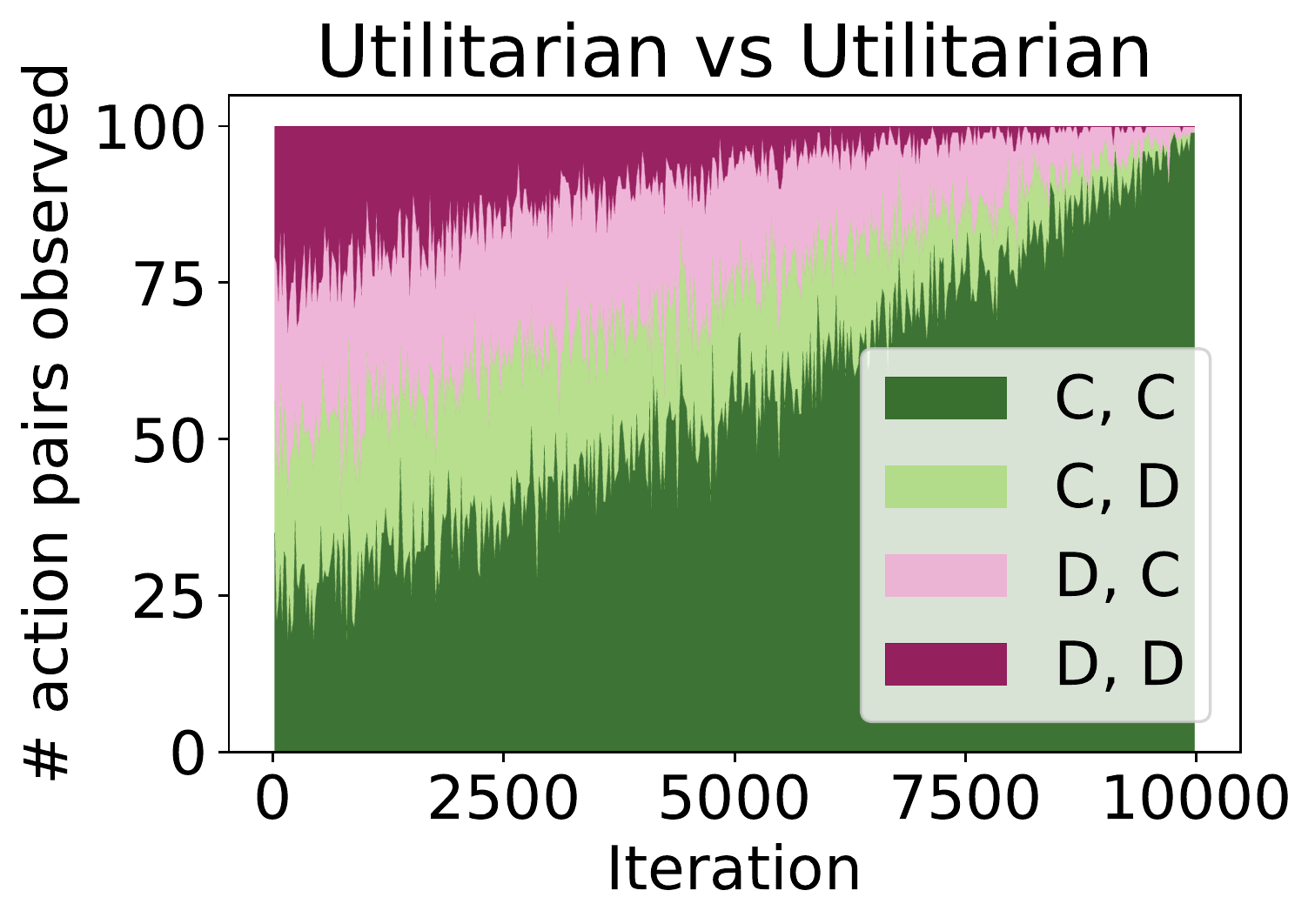}}
&
&
&
\\
\makecell[cc]{\rotatebox[origin=c]{90}{ Deontological }} &
\subt{\includegraphics[width=22mm]{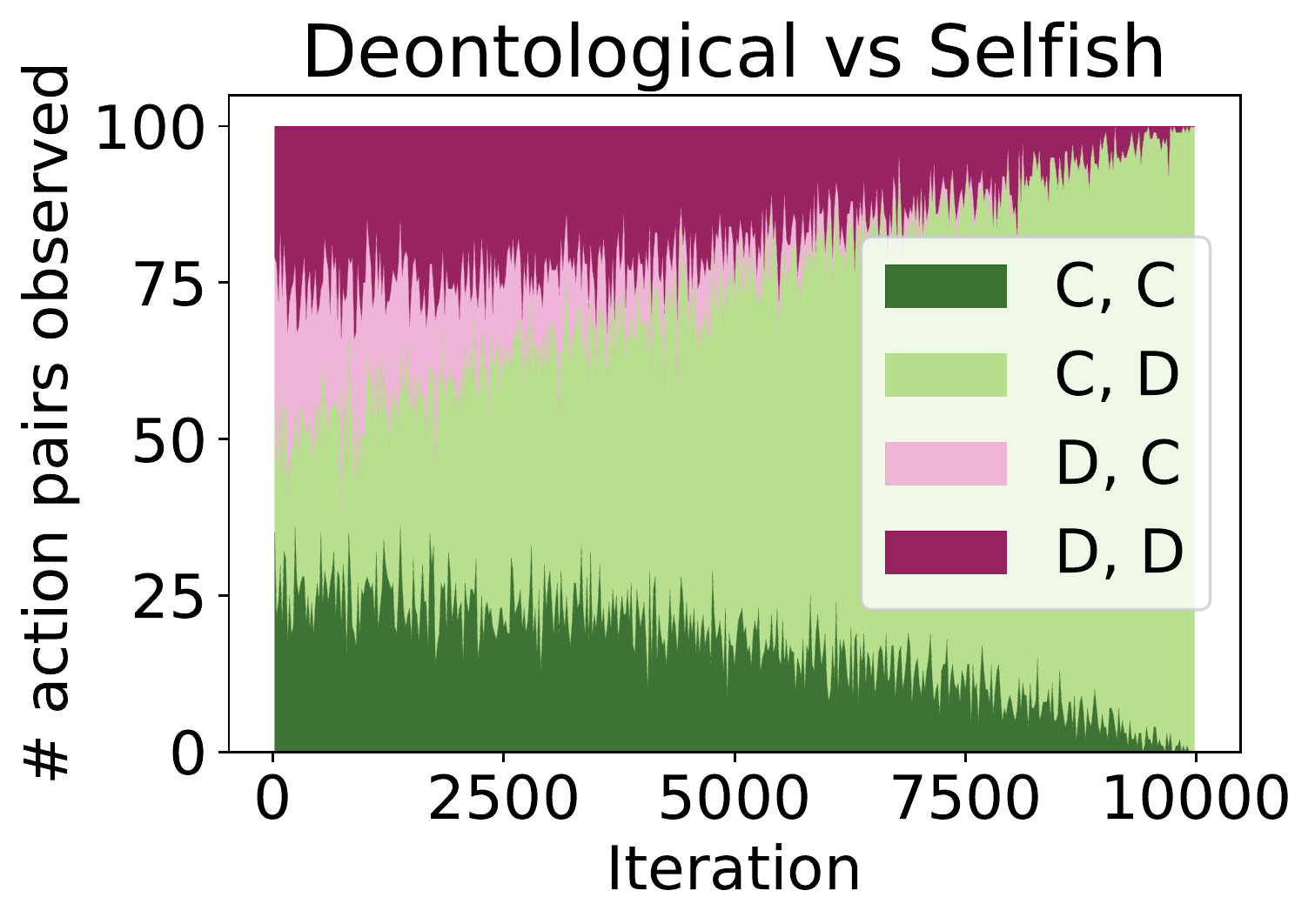}}
&\subt{\includegraphics[width=22mm]{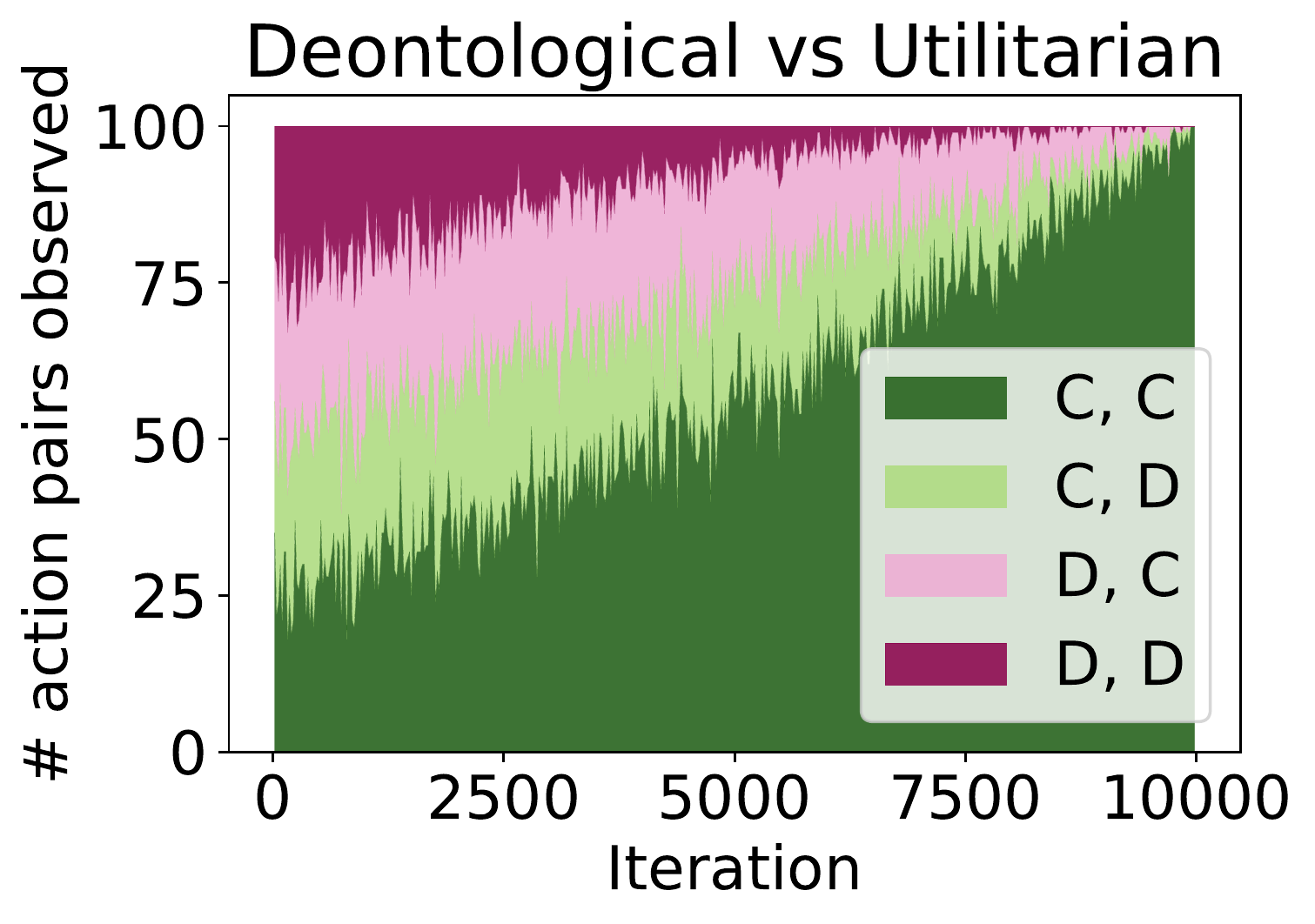}}
&\subt{\includegraphics[width=22mm]{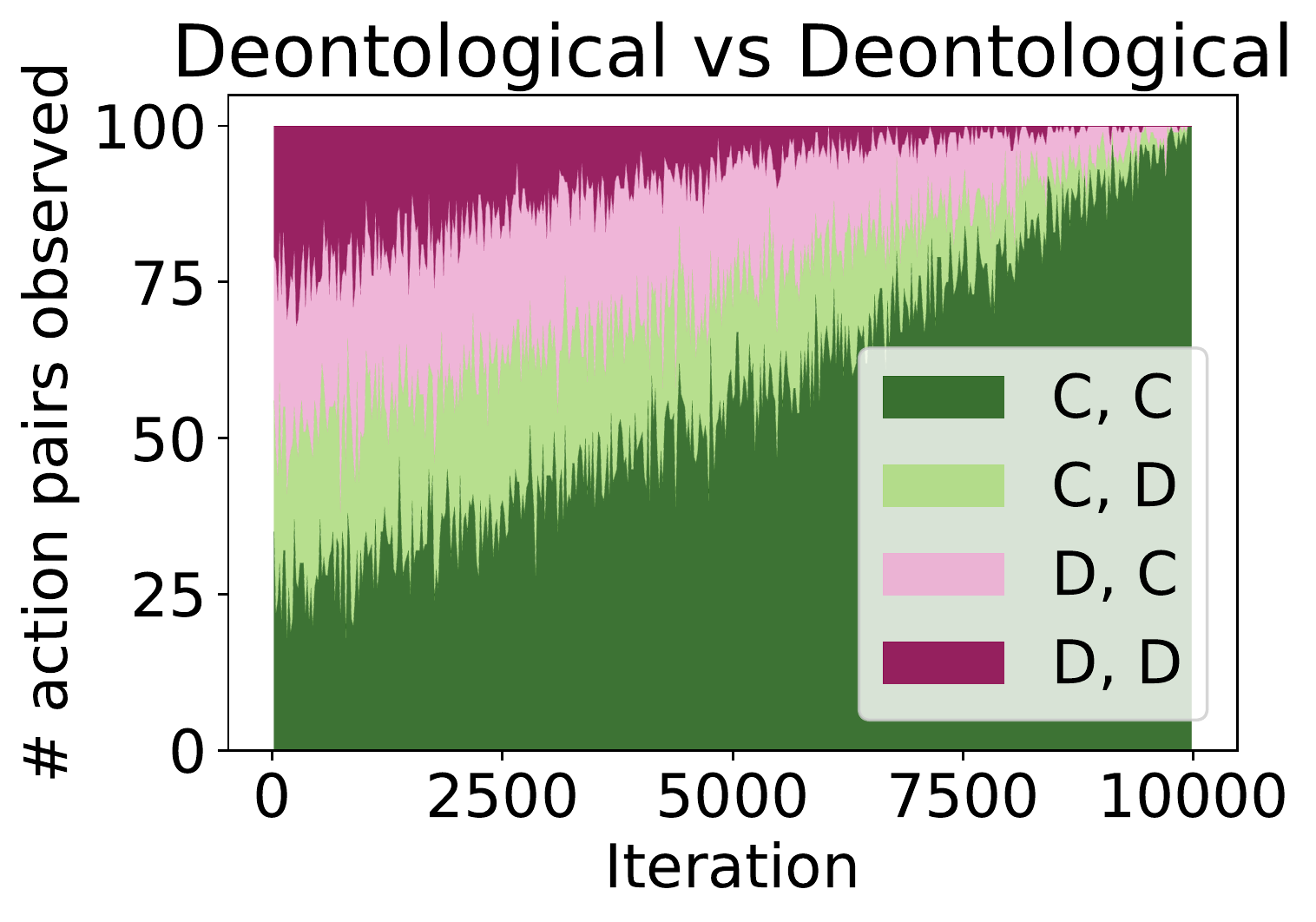}}
&
&
&
\\
\makecell[cc]{\rotatebox[origin=c]{90}{ Virtue-eq. }} &
\subt{\includegraphics[width=22mm]{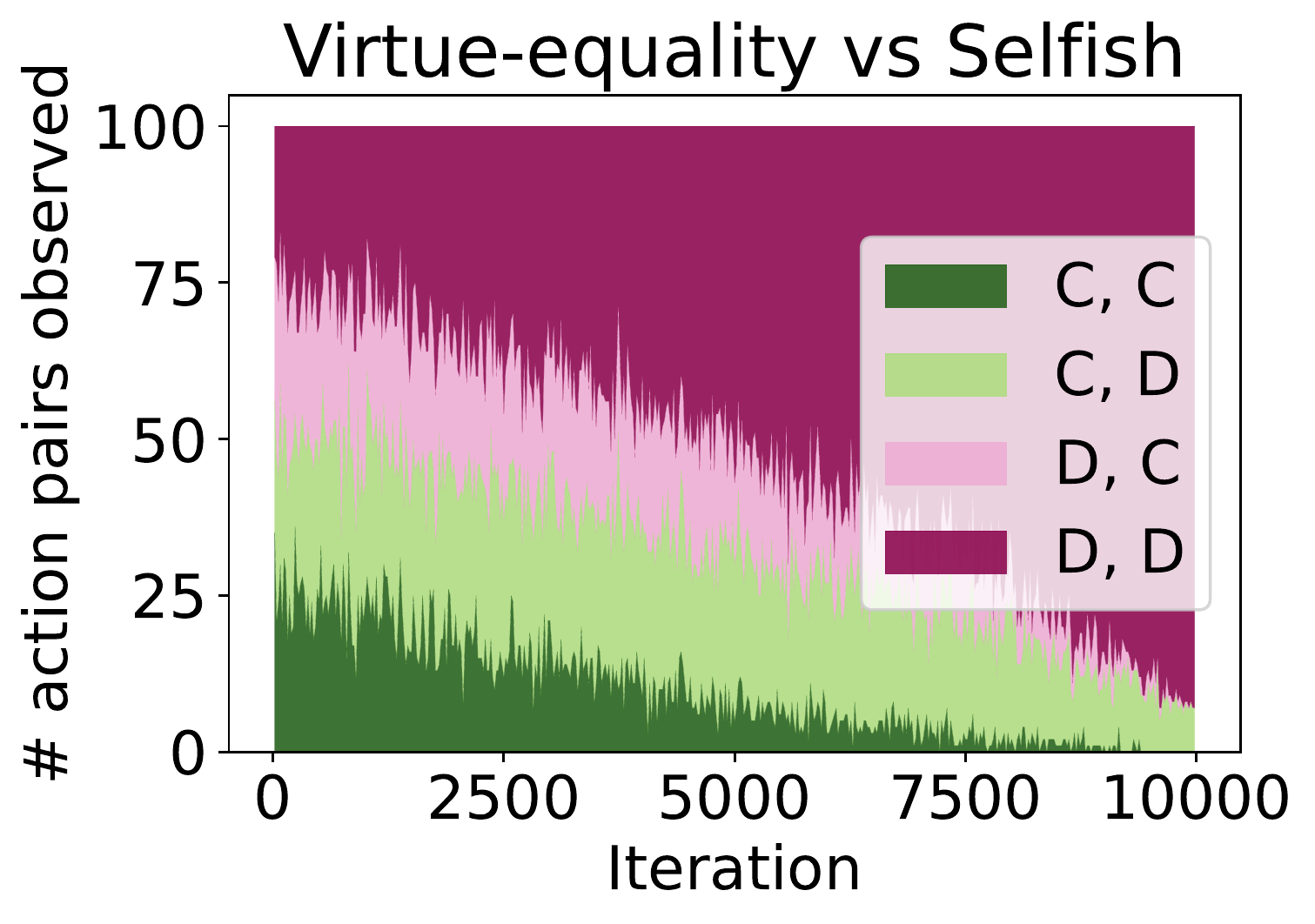}}
&\subt{\includegraphics[width=22mm]{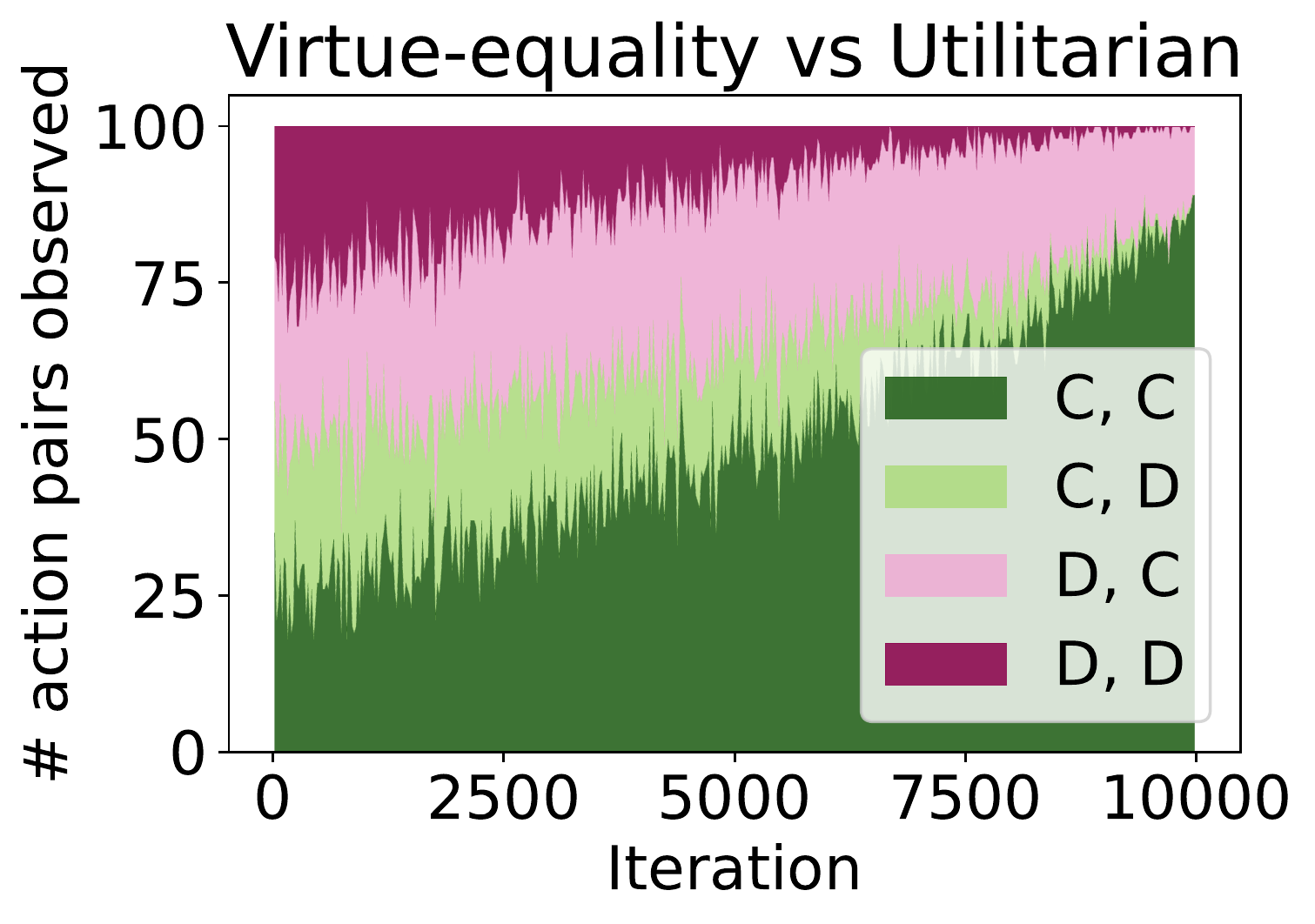}}
&\subt{\includegraphics[width=22mm]{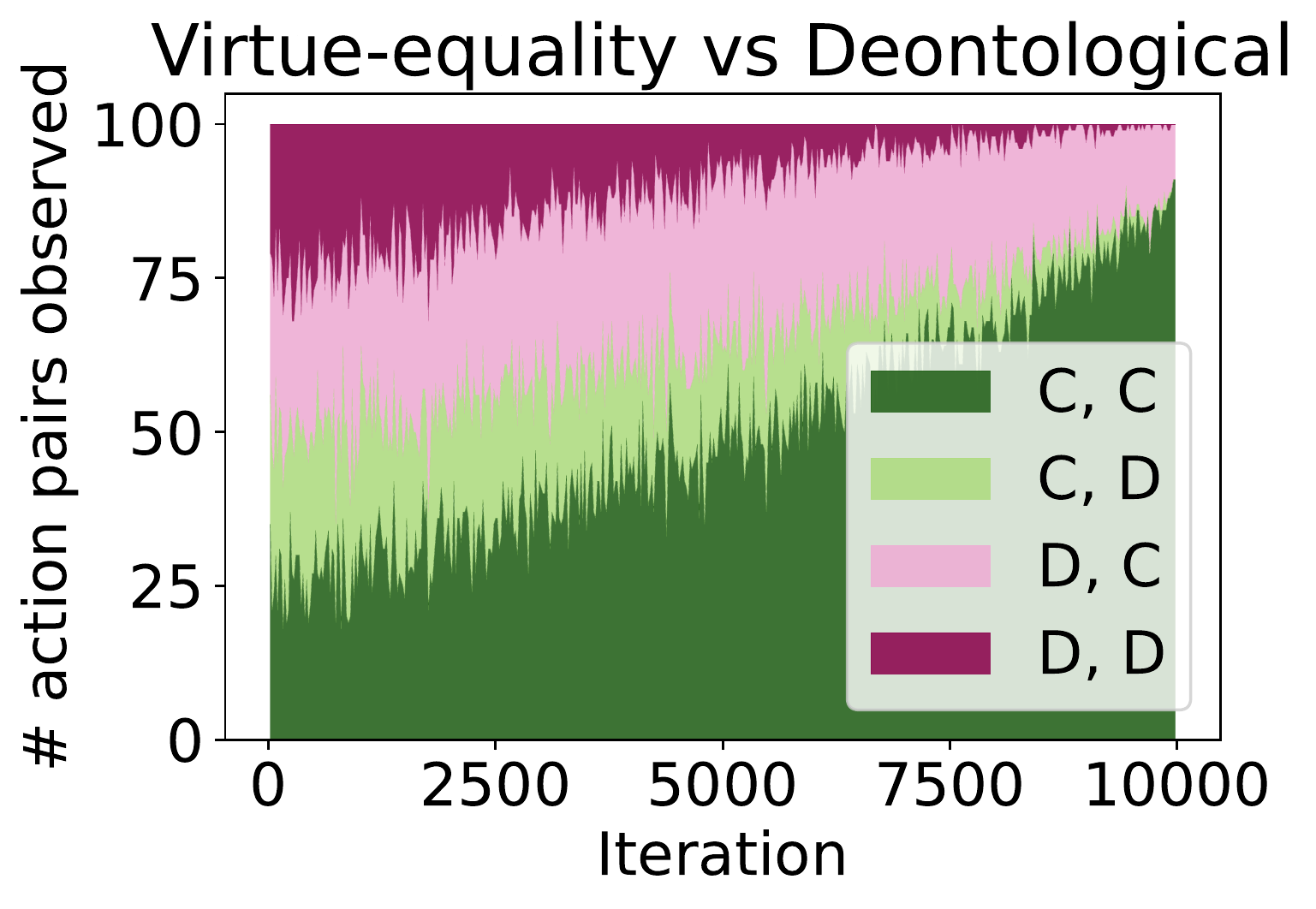}}
&\subt{\includegraphics[width=22mm]{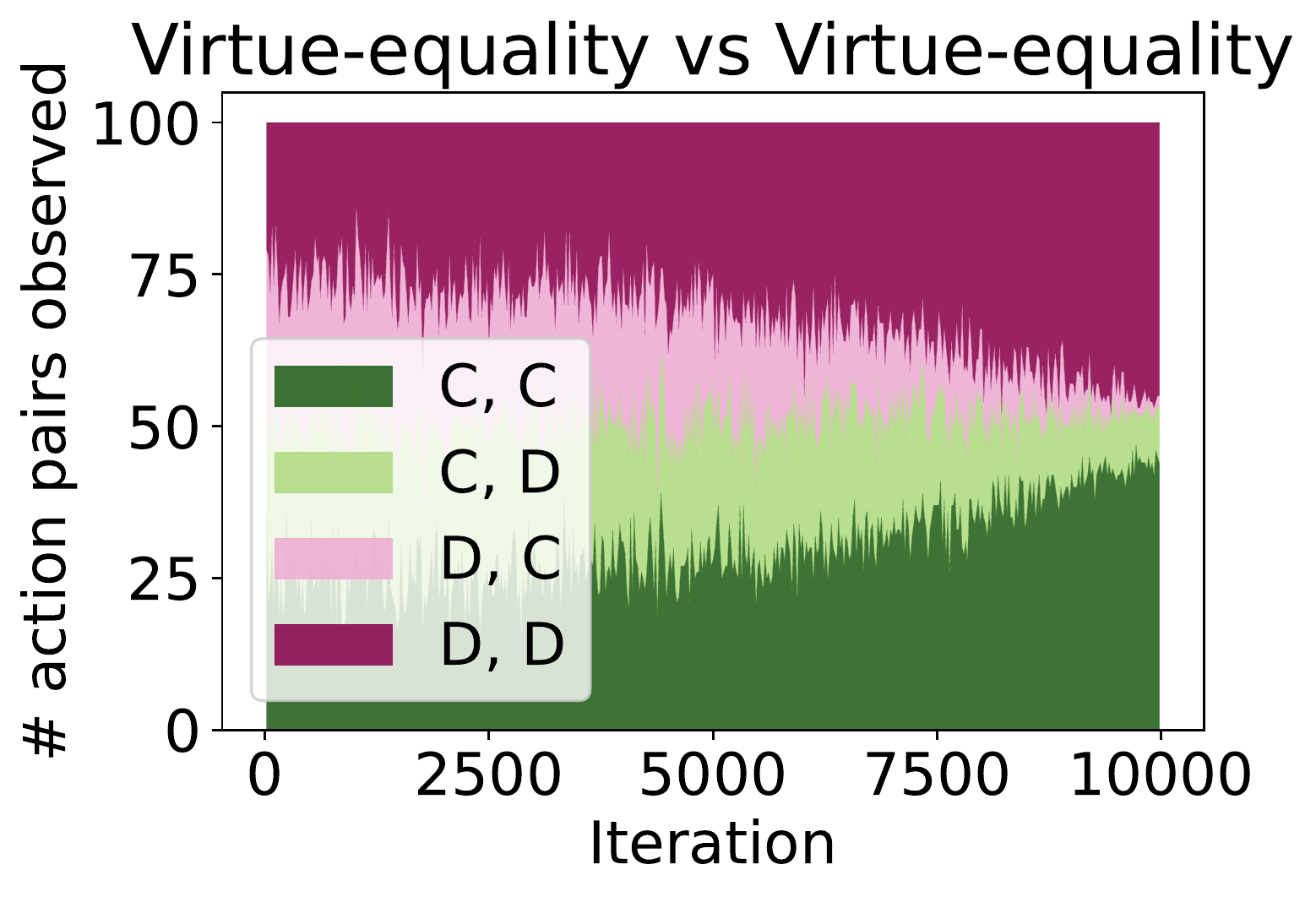}}
&
&
\\
\makecell[cc]{\rotatebox[origin=c]{90}{ Virtue-kind. }} &
\subt{\includegraphics[width=22mm]{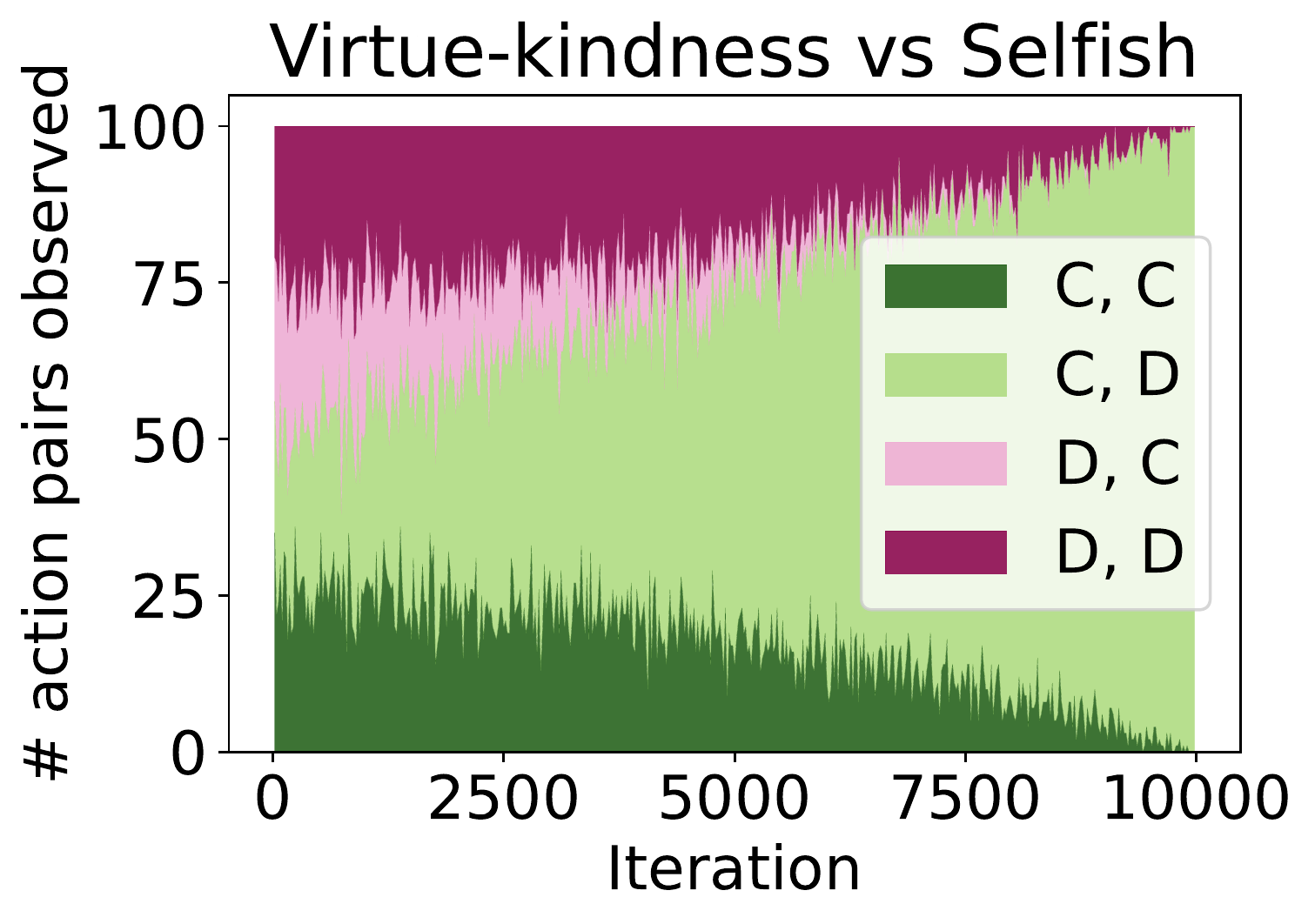}}
&\subt{\includegraphics[width=22mm]{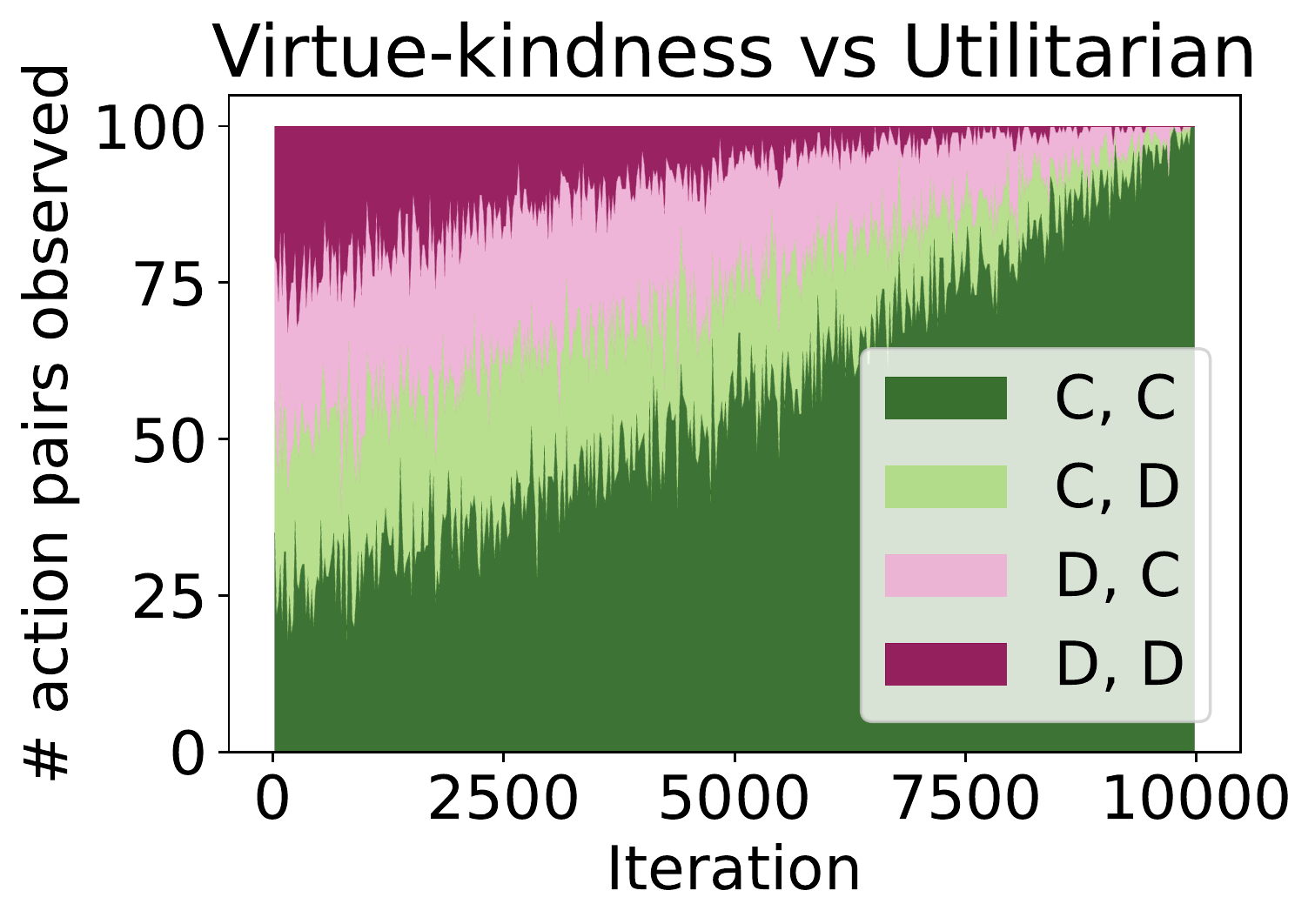}}
&\subt{\includegraphics[width=22mm]{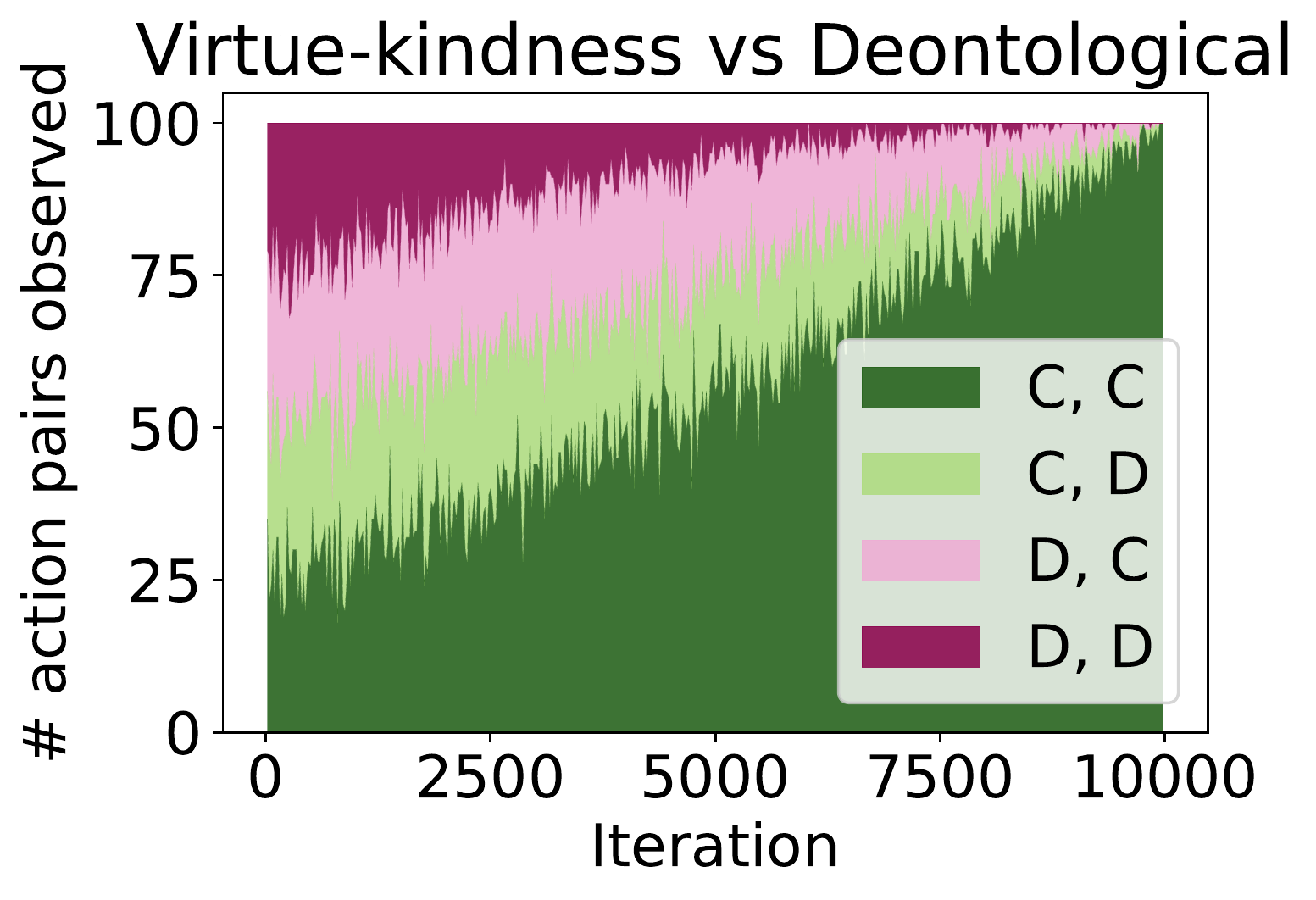}}
&\subt{\includegraphics[width=22mm]{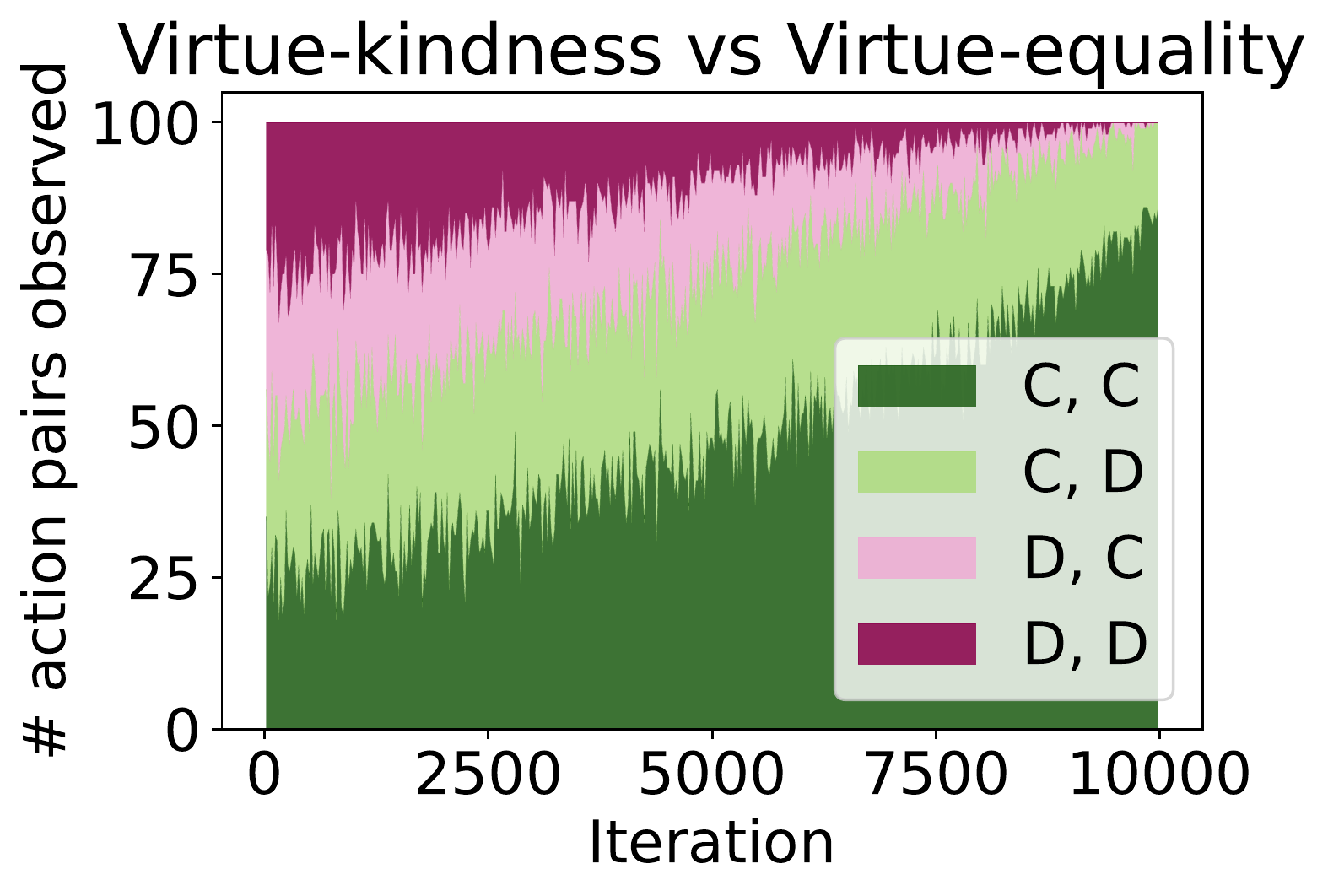}}
&\subt{\includegraphics[width=22mm]{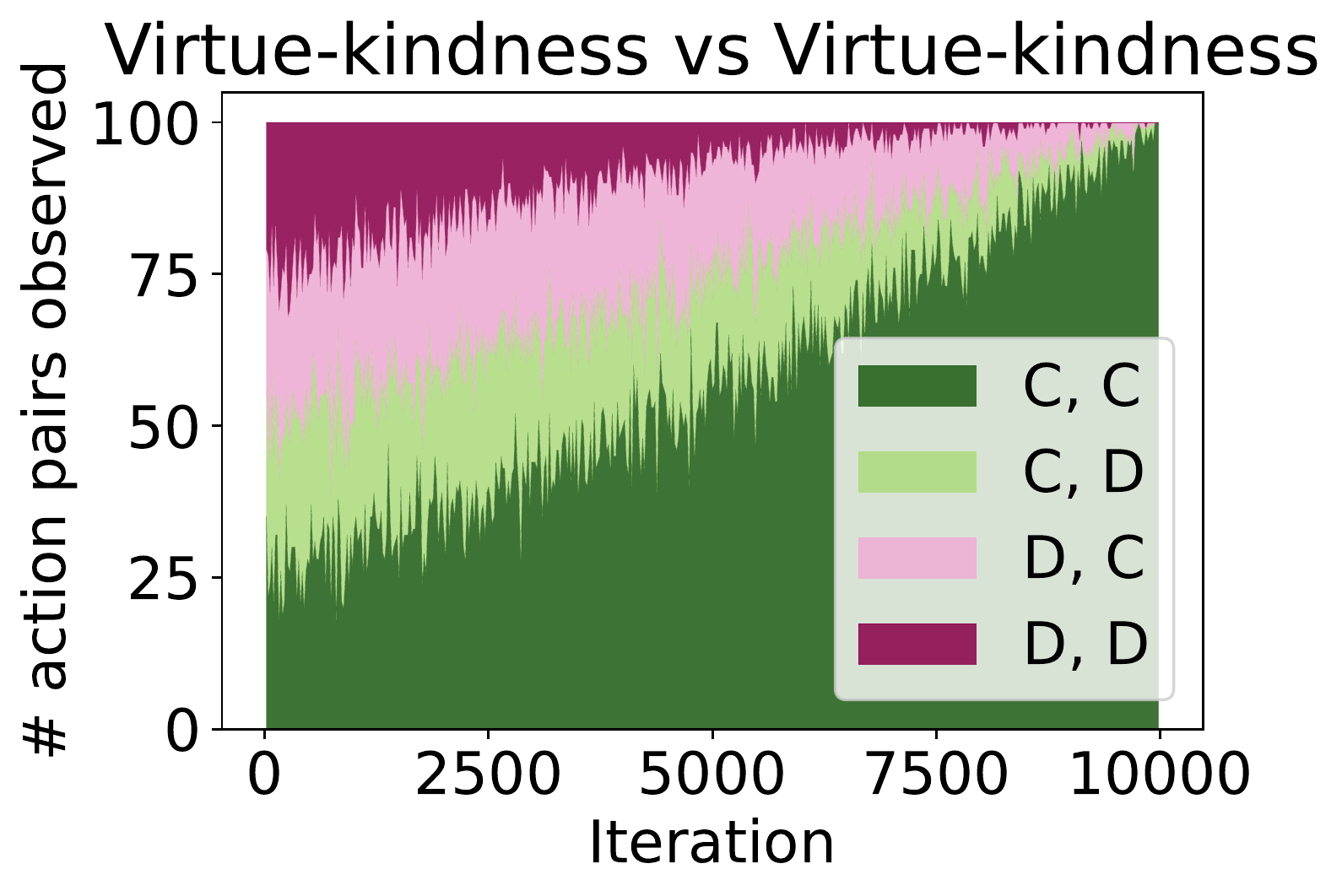}}
&
\\
\makecell[cc]{\rotatebox[origin=c]{90}{ Virtue-mix. }} &
\subt{\includegraphics[width=22mm]{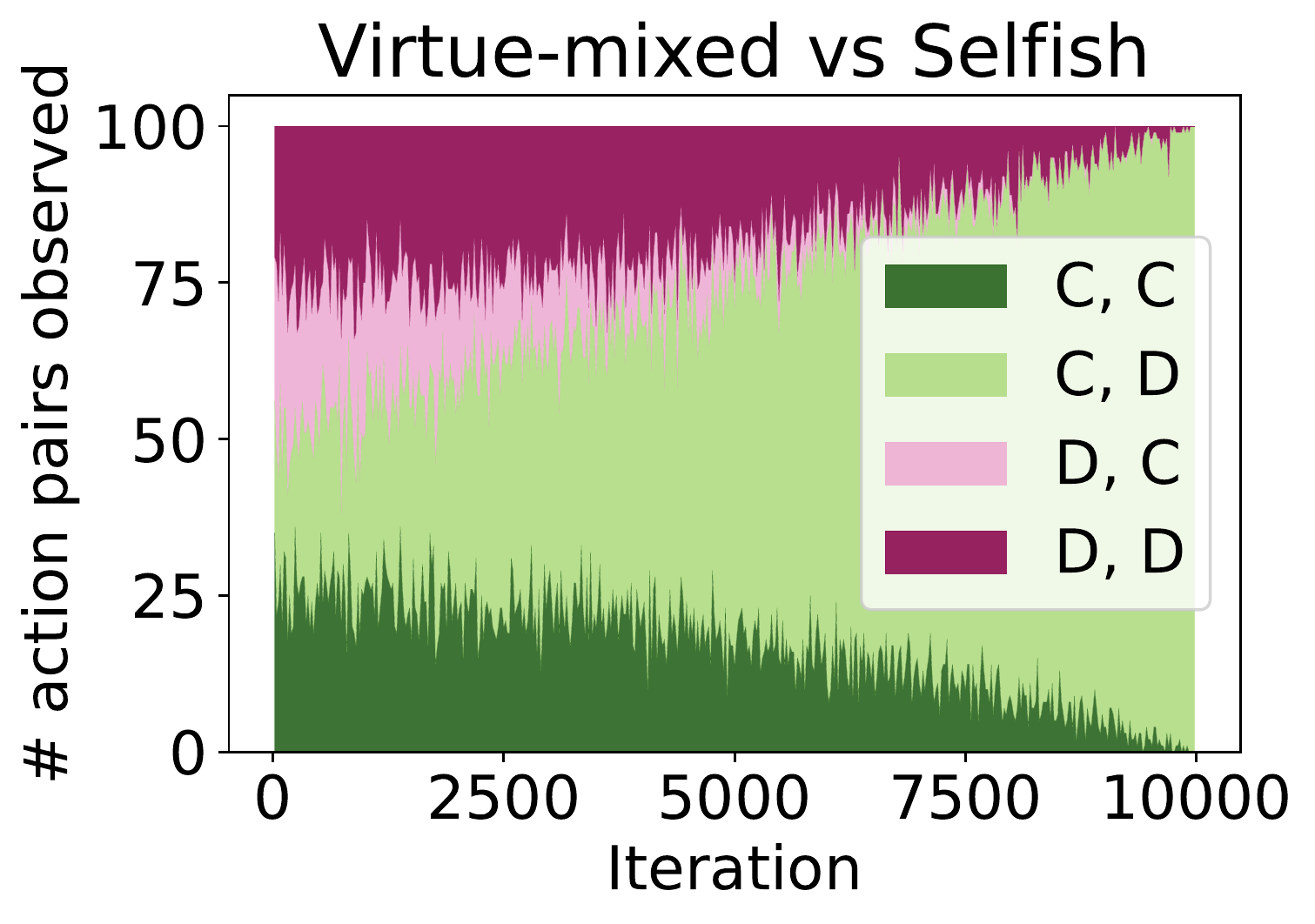}}
&\subt{\includegraphics[width=22mm]{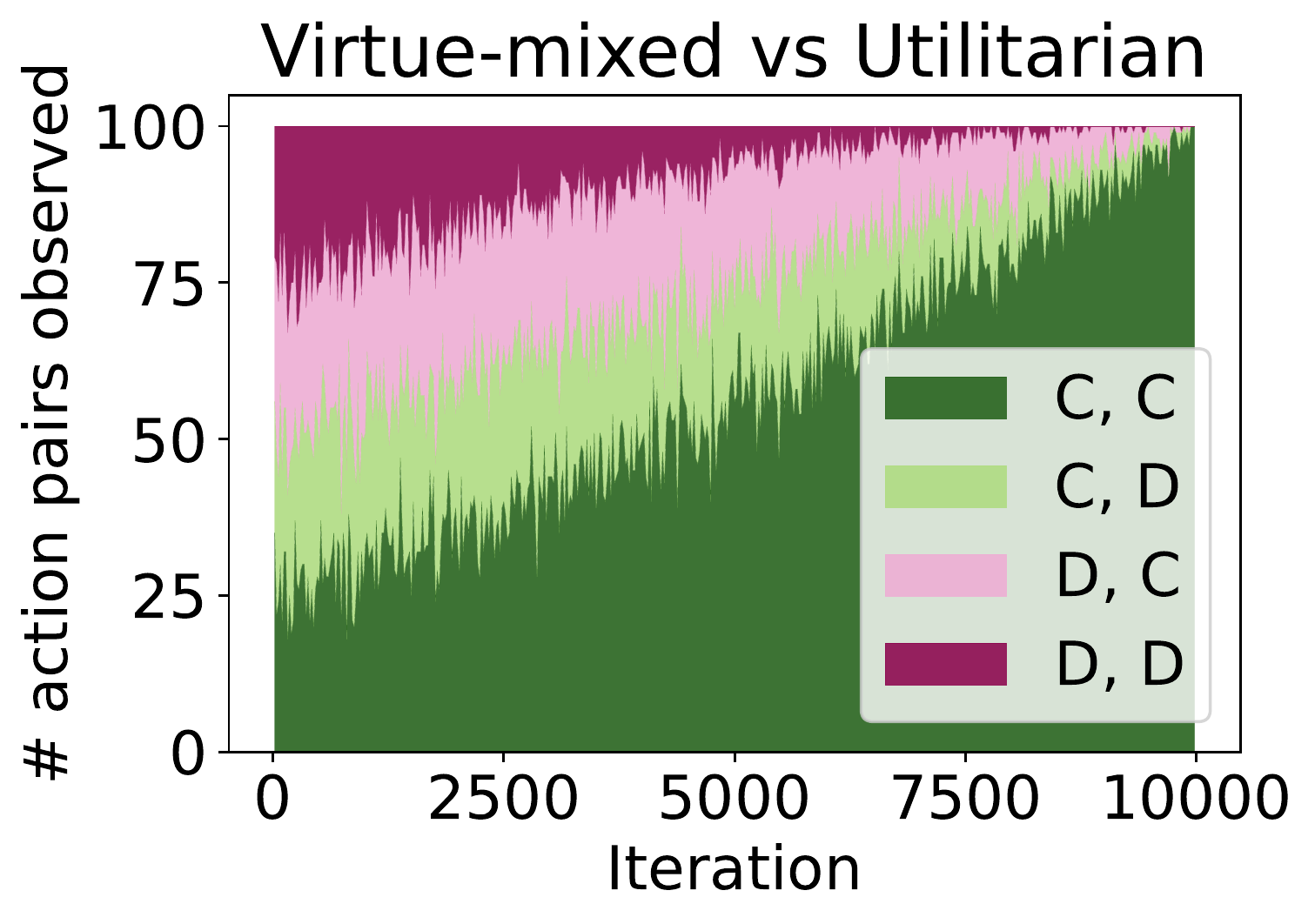}}
&\subt{\includegraphics[width=22mm]{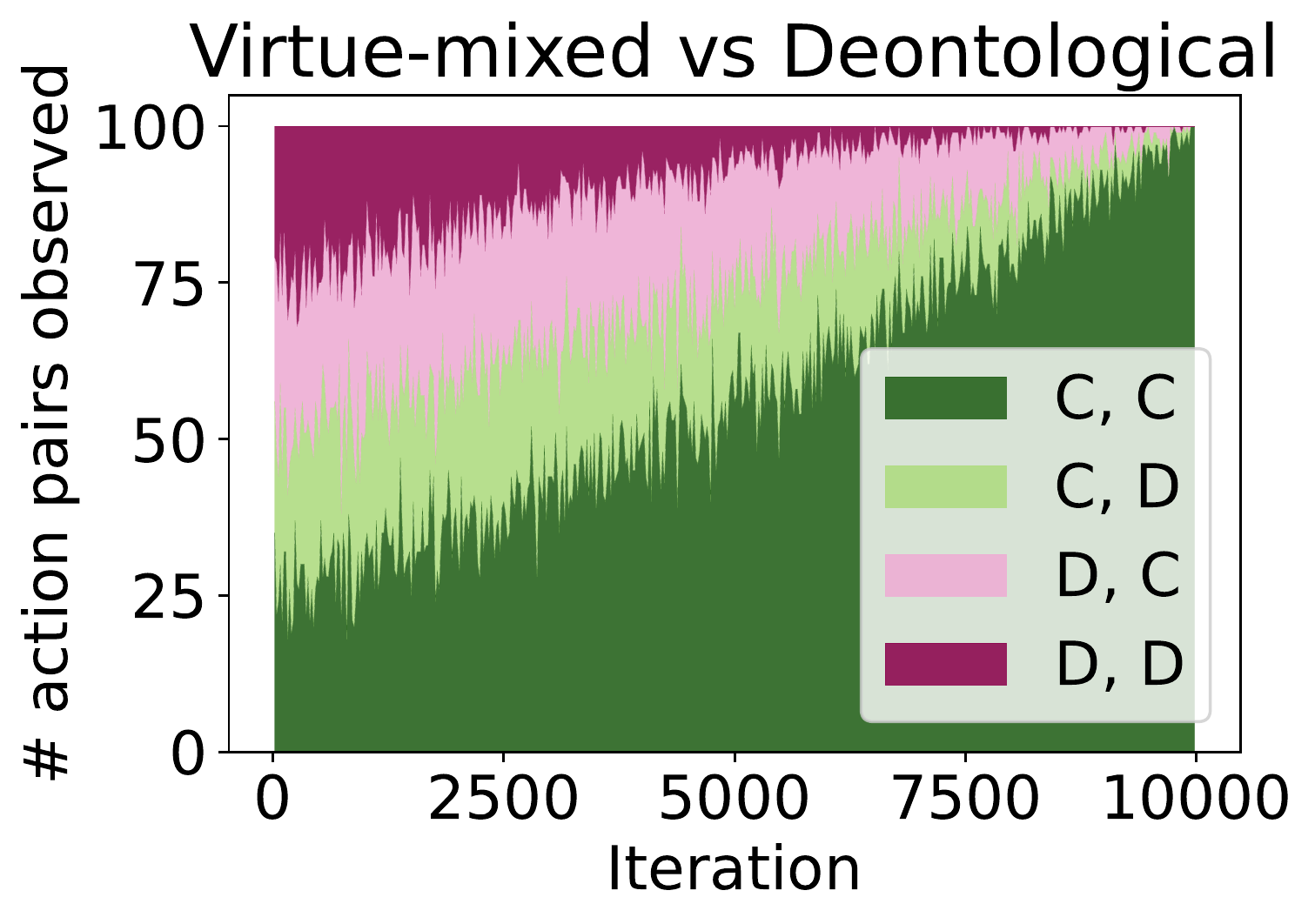}}
&\subt{\includegraphics[width=22mm]{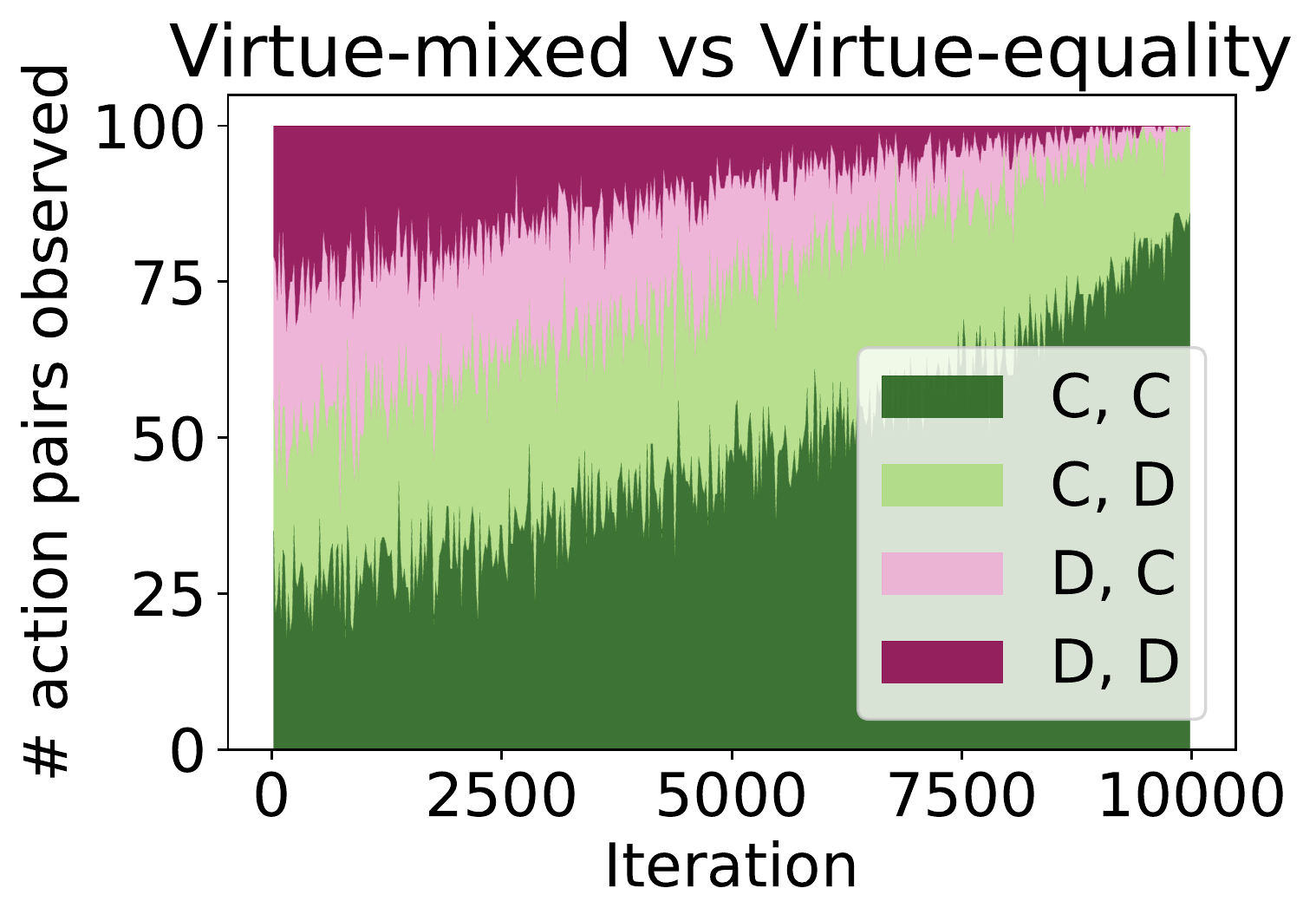}}
&\subt{\includegraphics[width=22mm]{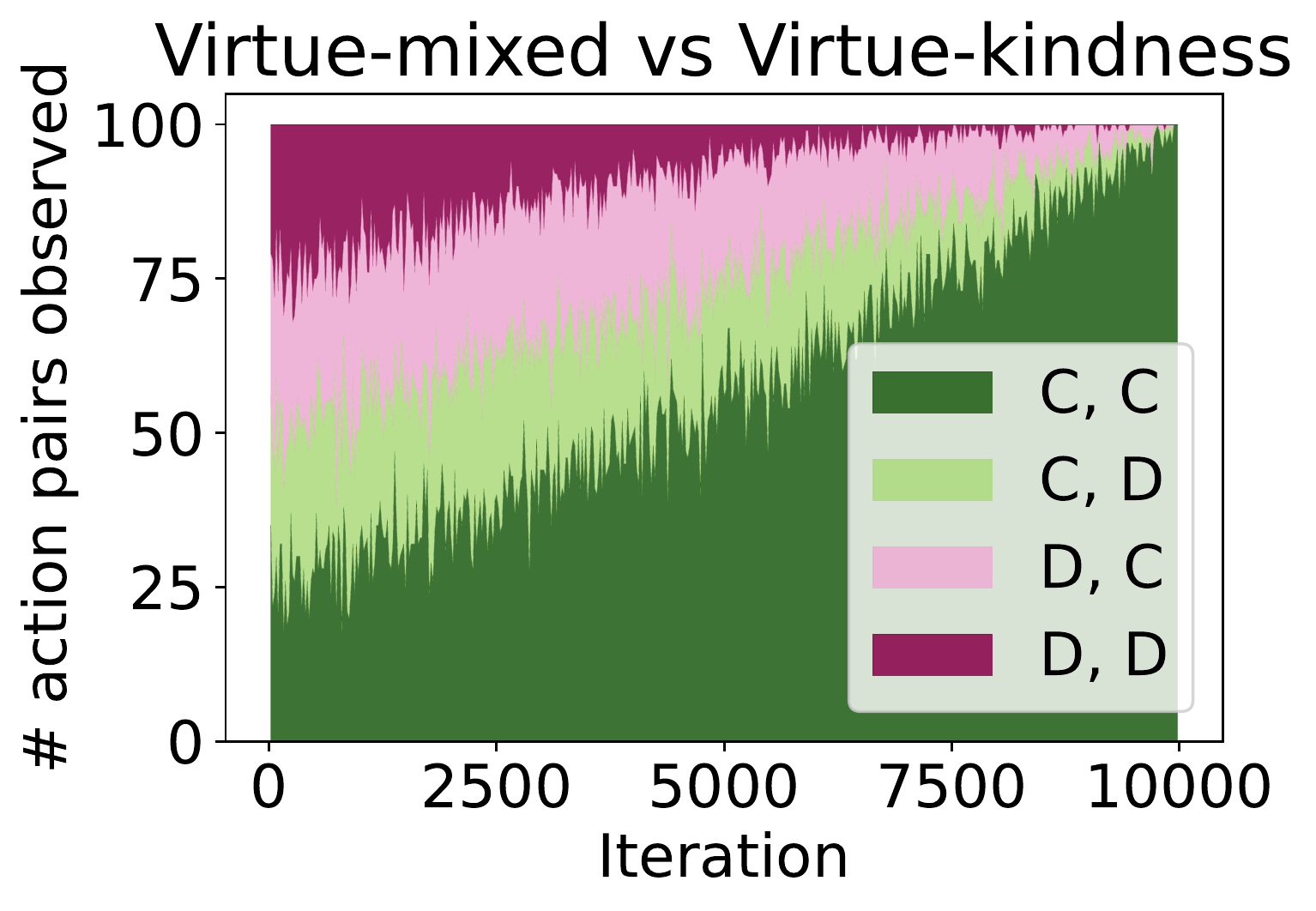}}
&\subt{\includegraphics[width=22mm]{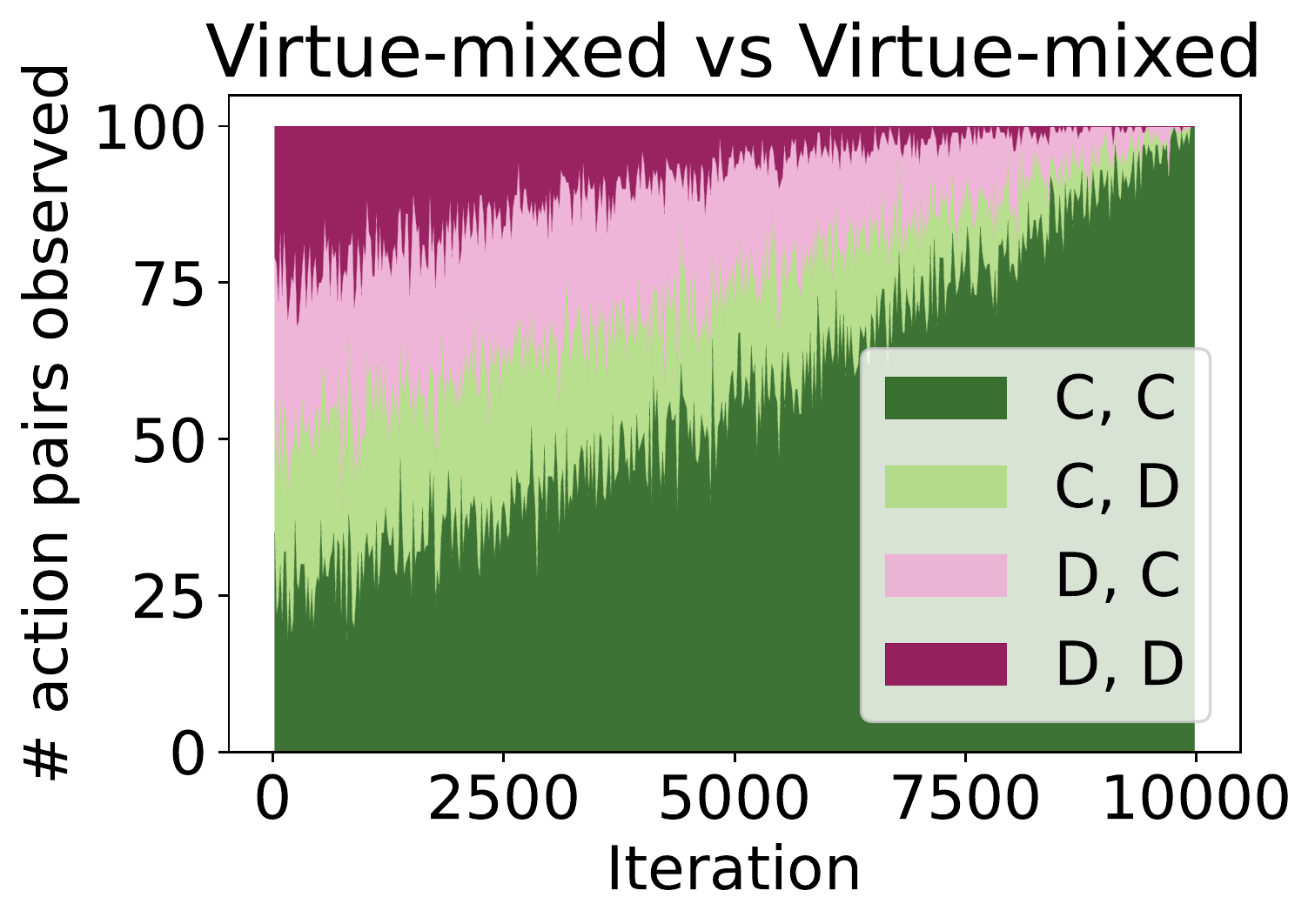}}
\\
\bottomrule
\end{tabular}
\caption{Iterated Prisoner's Dilemma game. Simultaneous pairs of actions observed over time. Learning player $M$ (row) vs. learning opponent $O$ (column).}
\label{fig:action_pairs_learning_IPD}
\end{figure}

\begin{figure*}[!h]
\centering
\begin{tabular}{|c|cccccc}
\toprule
 & Selfish & Utilitarian & Deontological & Virtue-equality & Virtue-kindness & Virtue-mixed\\
\midrule
\makecell[cc]{\rotatebox[origin=c]{90}{ Selfish }} & 
\subt{\includegraphics[width=22mm]{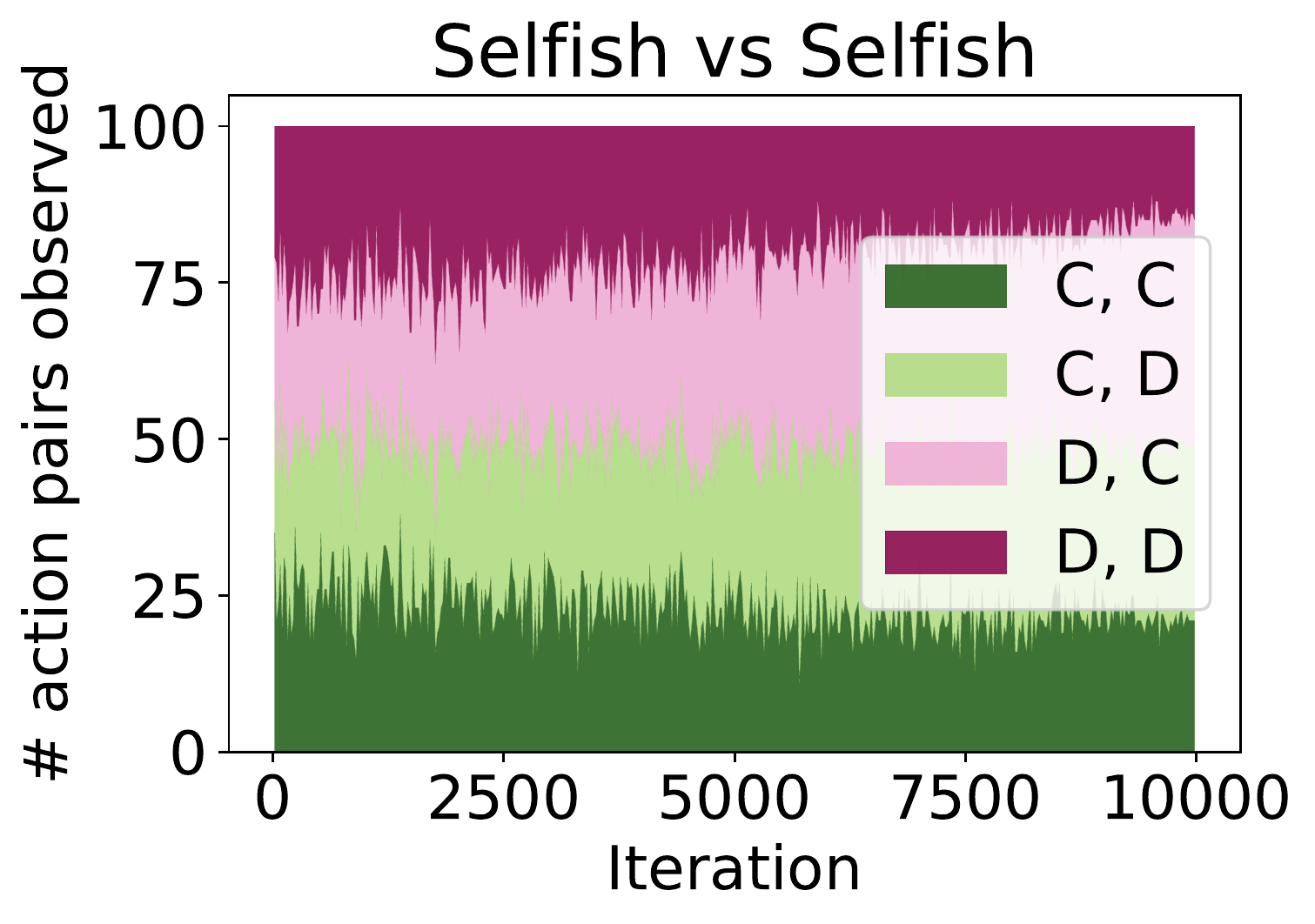}}
&
&
&
&
\\
\makecell[cc]{\rotatebox[origin=c]{90}{ Utilitarian }} & 
\subt{\includegraphics[width=22mm]{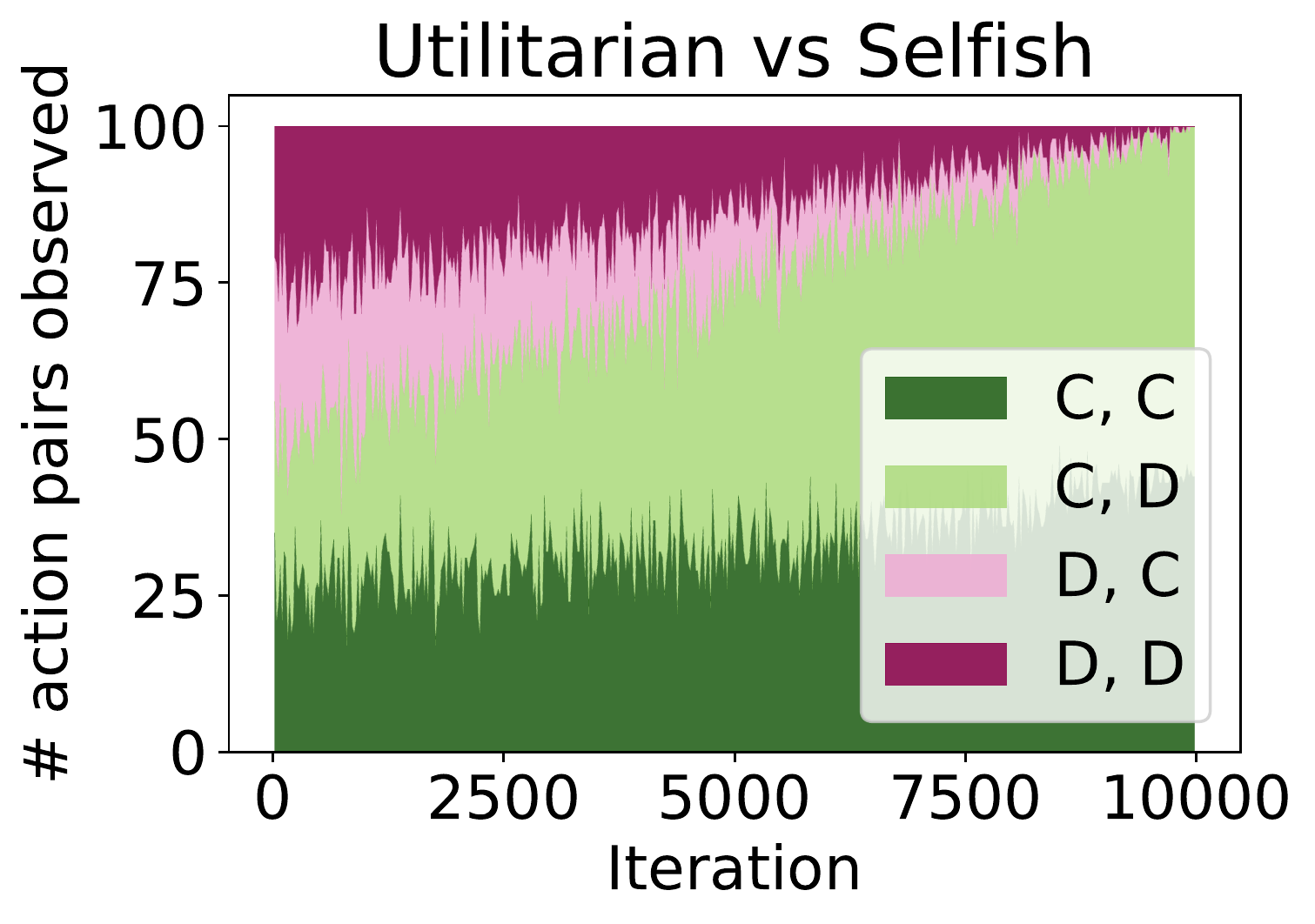}}
&\subt{\includegraphics[width=22mm]{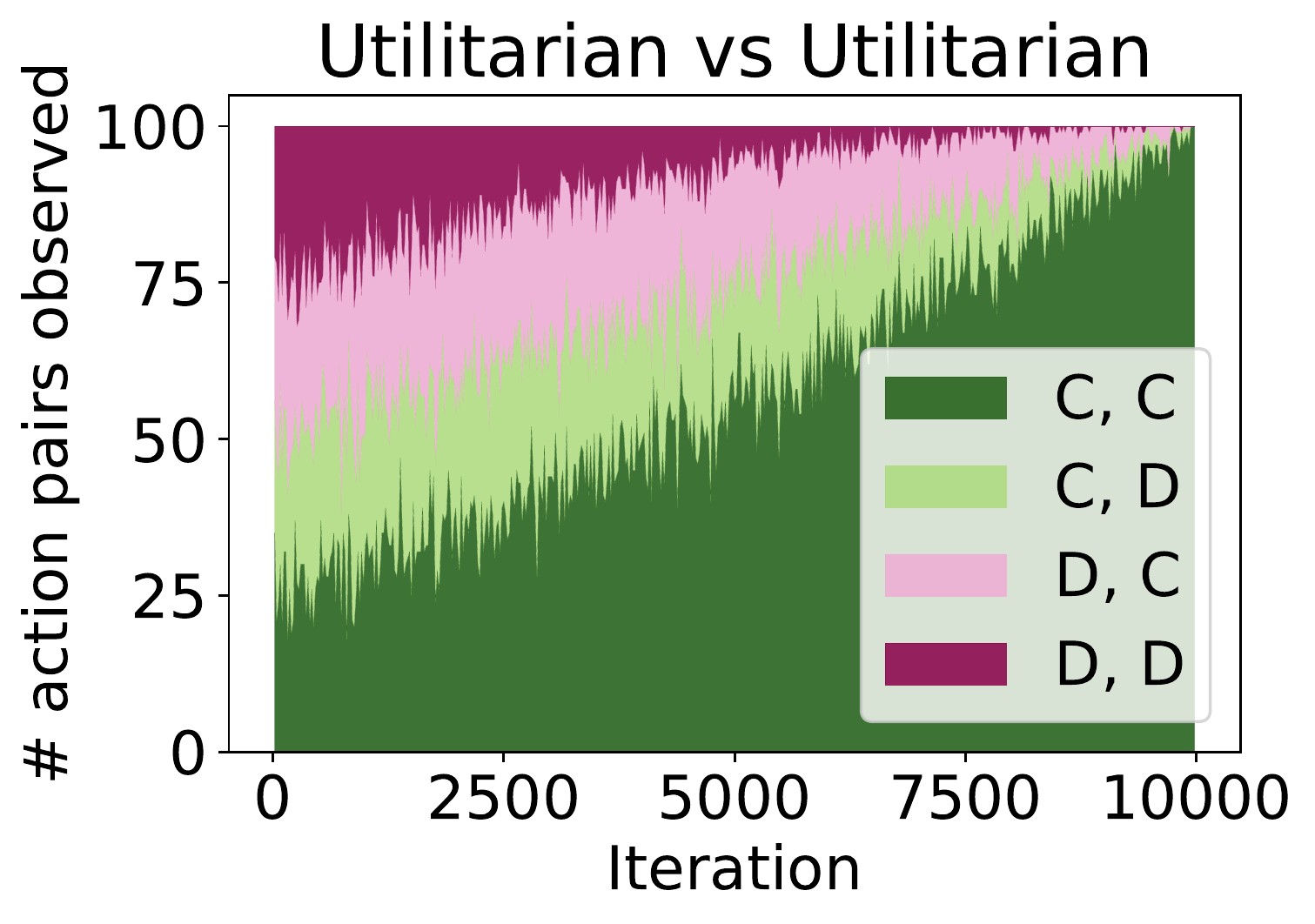}}
&
&
&
\\
\makecell[cc]{\rotatebox[origin=c]{90}{ Deontological }} &
\subt{\includegraphics[width=22mm]{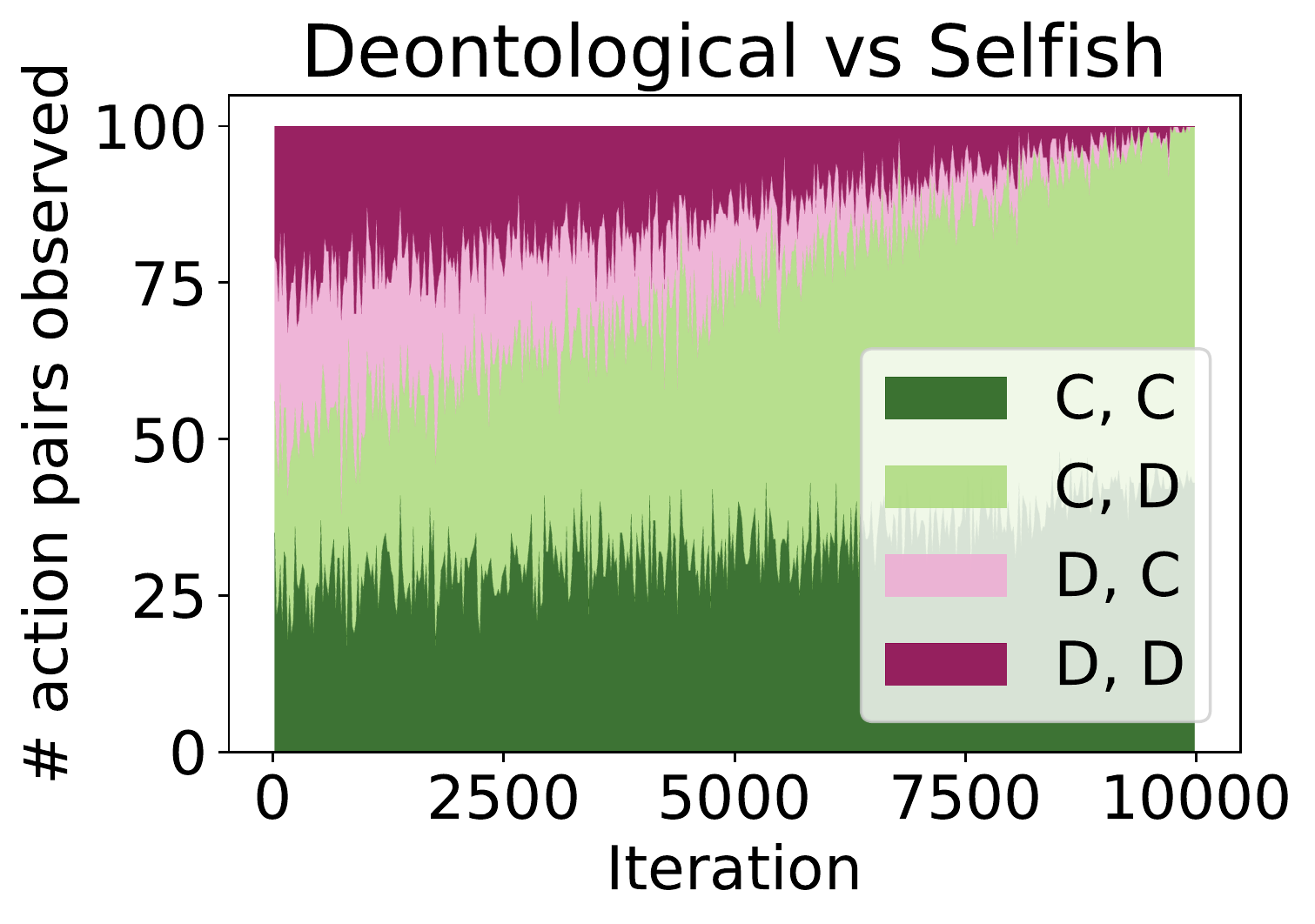}}
&\subt{\includegraphics[width=22mm]{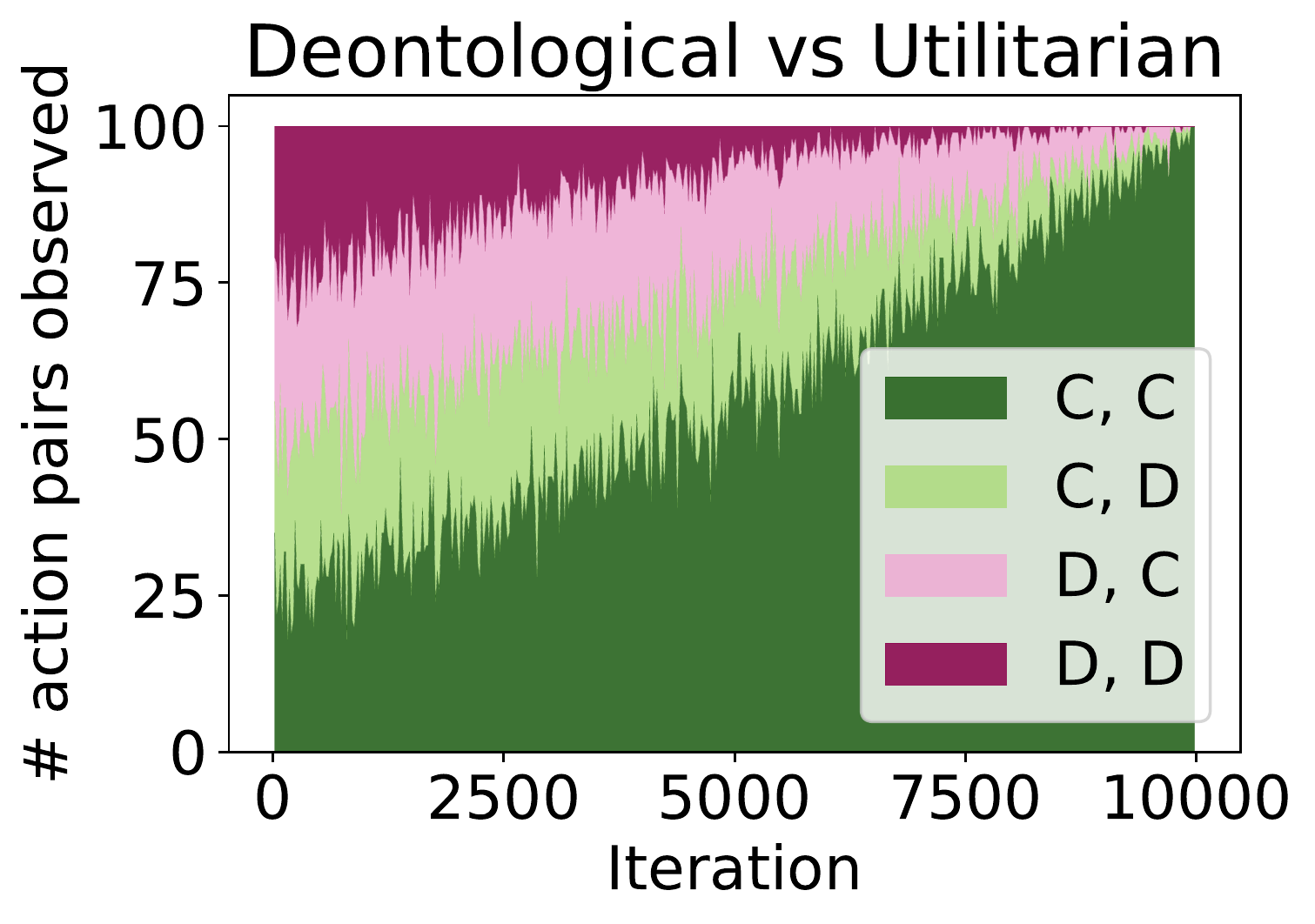}}
&\subt{\includegraphics[width=22mm]{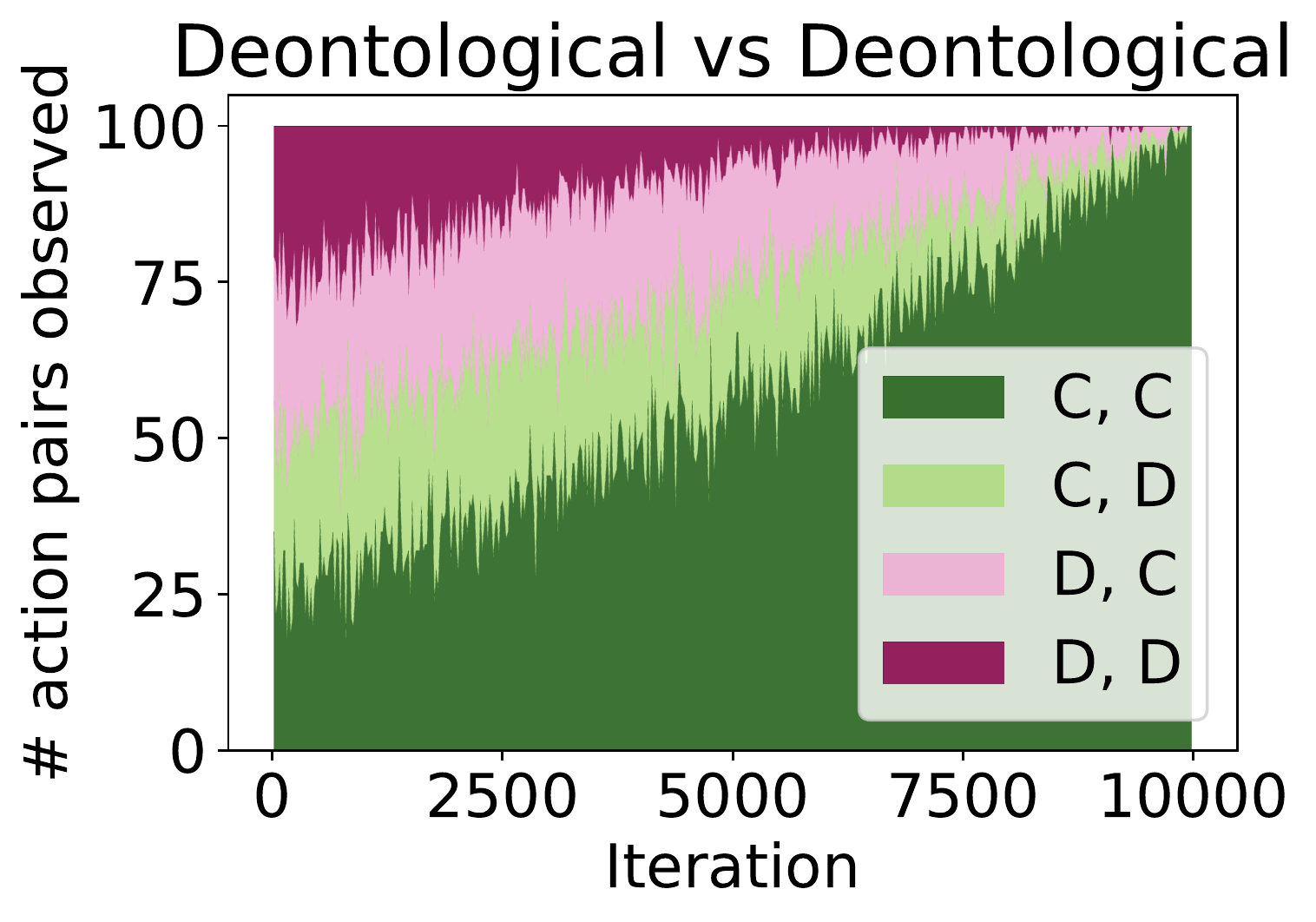}}
&
&
&
\\
\makecell[cc]{\rotatebox[origin=c]{90}{ Virtue-eq. }} &
\subt{\includegraphics[width=22mm]{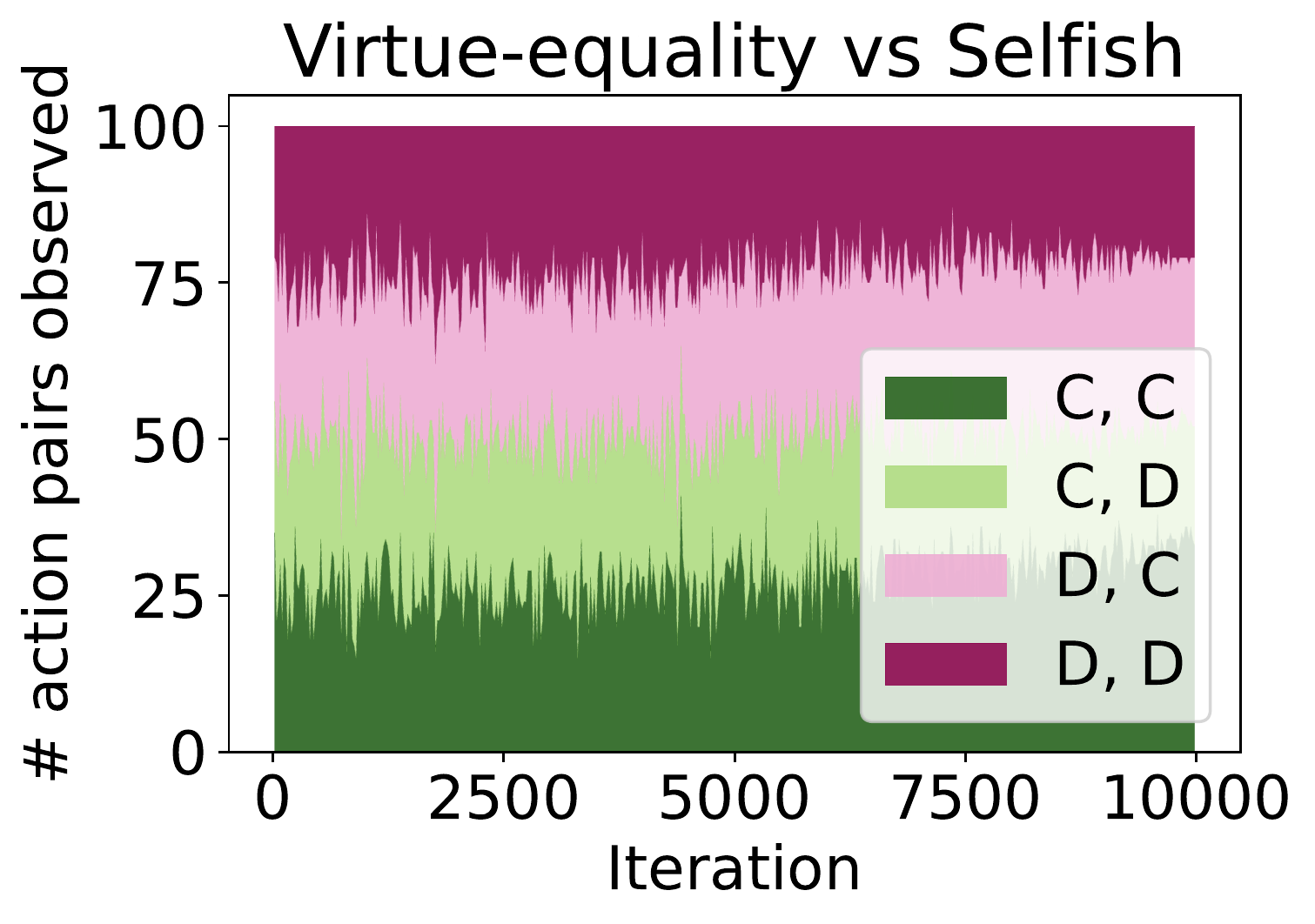}}
&\subt{\includegraphics[width=22mm]{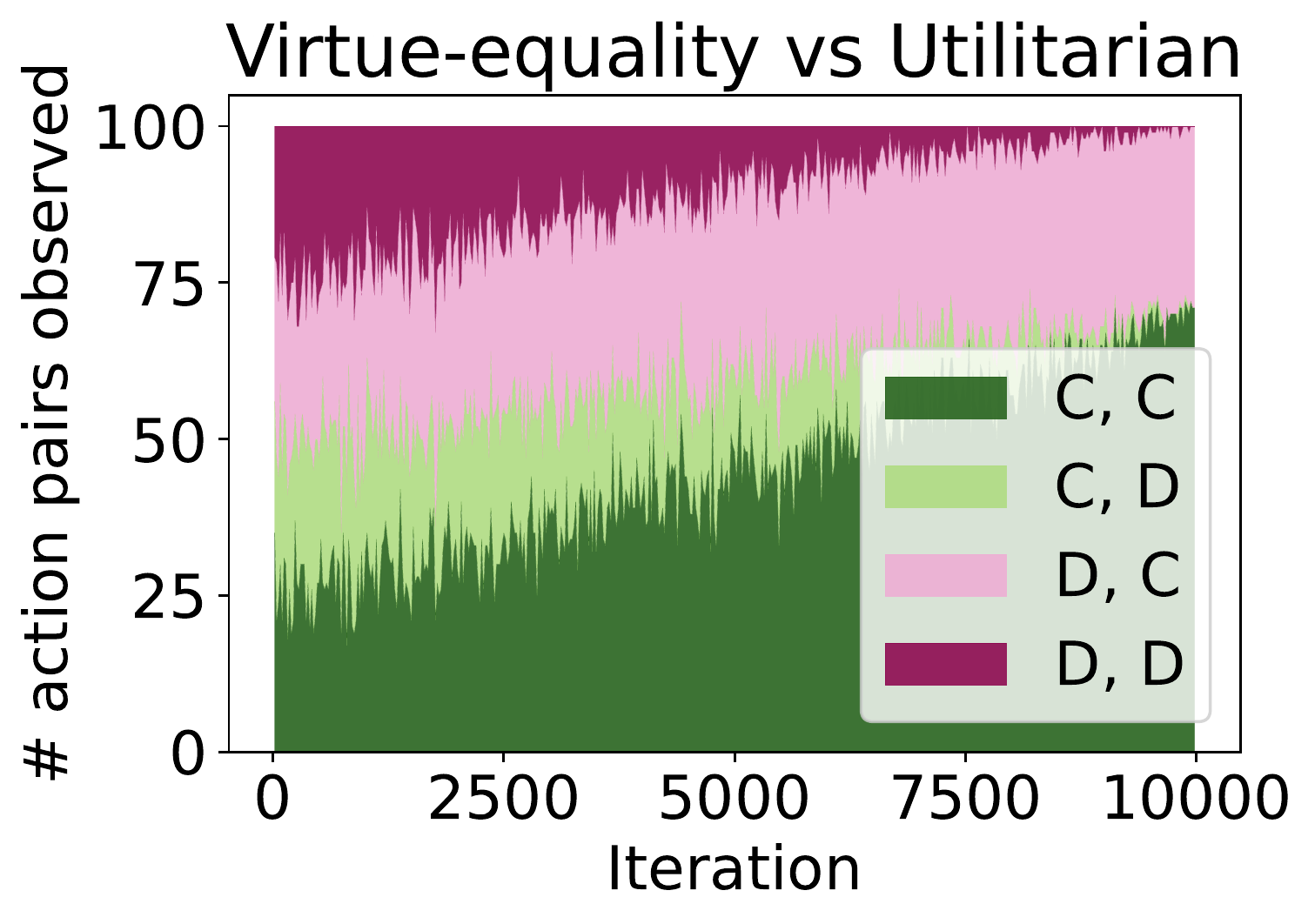}}
&\subt{\includegraphics[width=22mm]{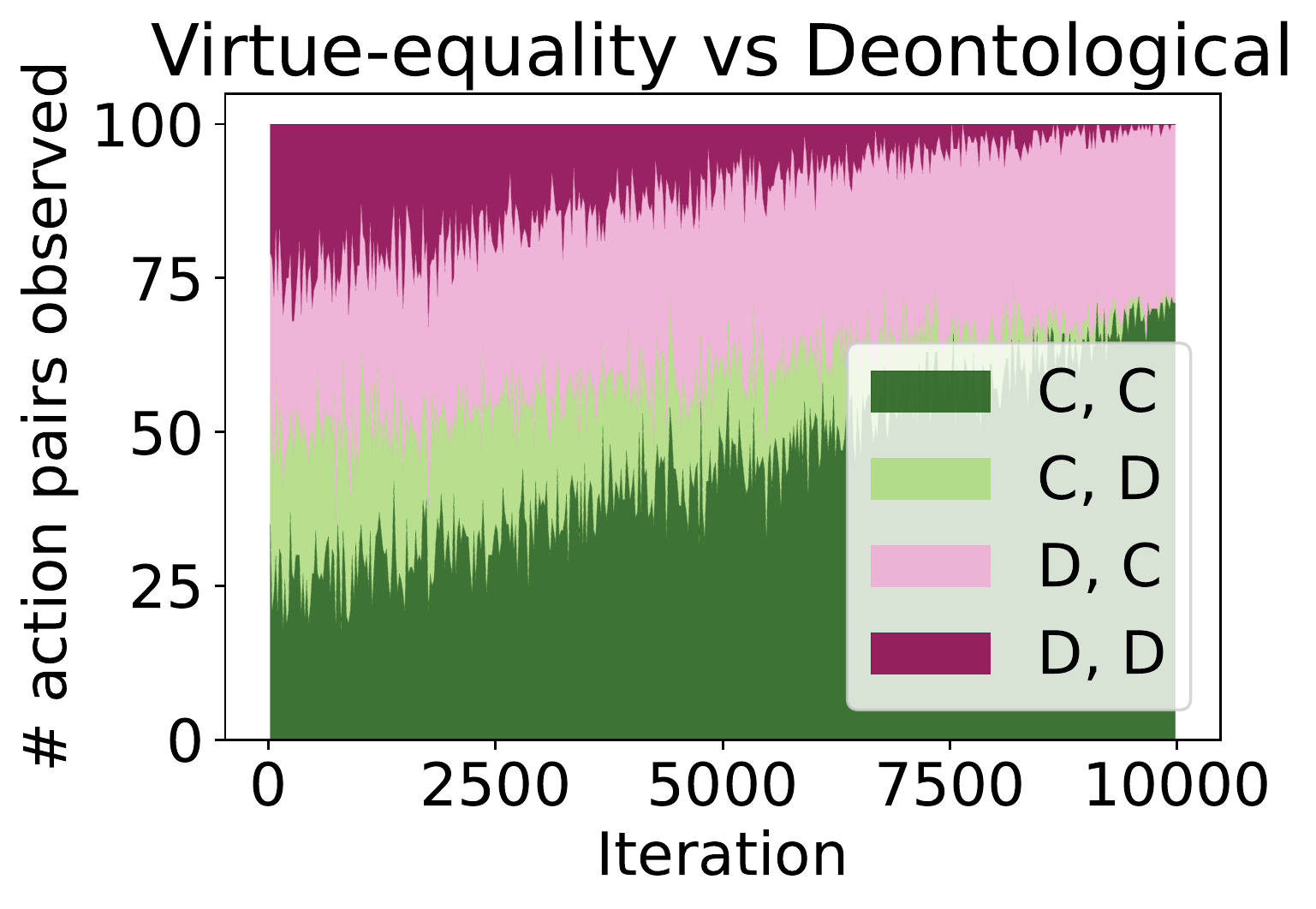}}
&\subt{\includegraphics[width=22mm]{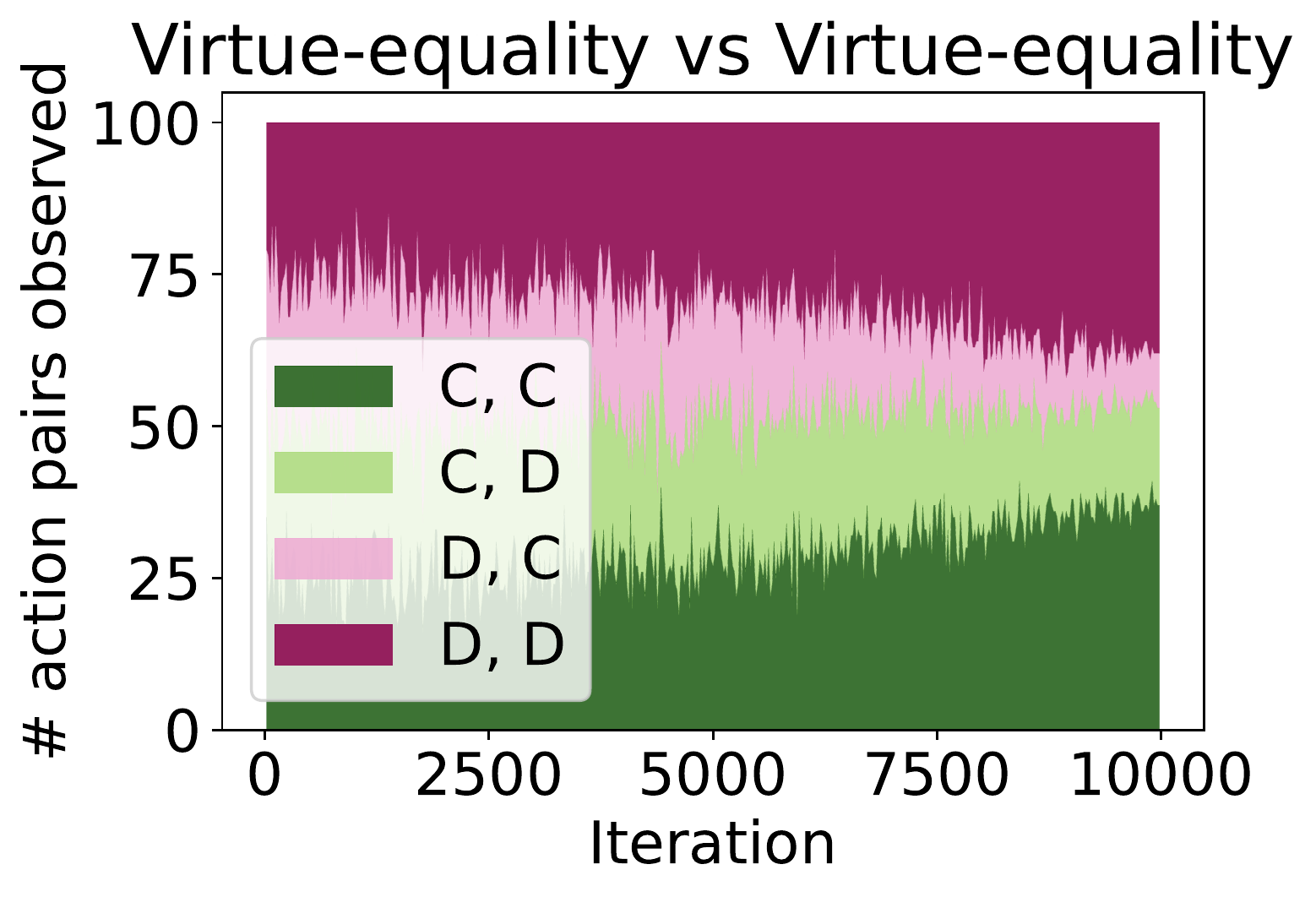}}
&
&
\\
\makecell[cc]{\rotatebox[origin=c]{90}{ Virtue-kind. }} &
\subt{\includegraphics[width=22mm]{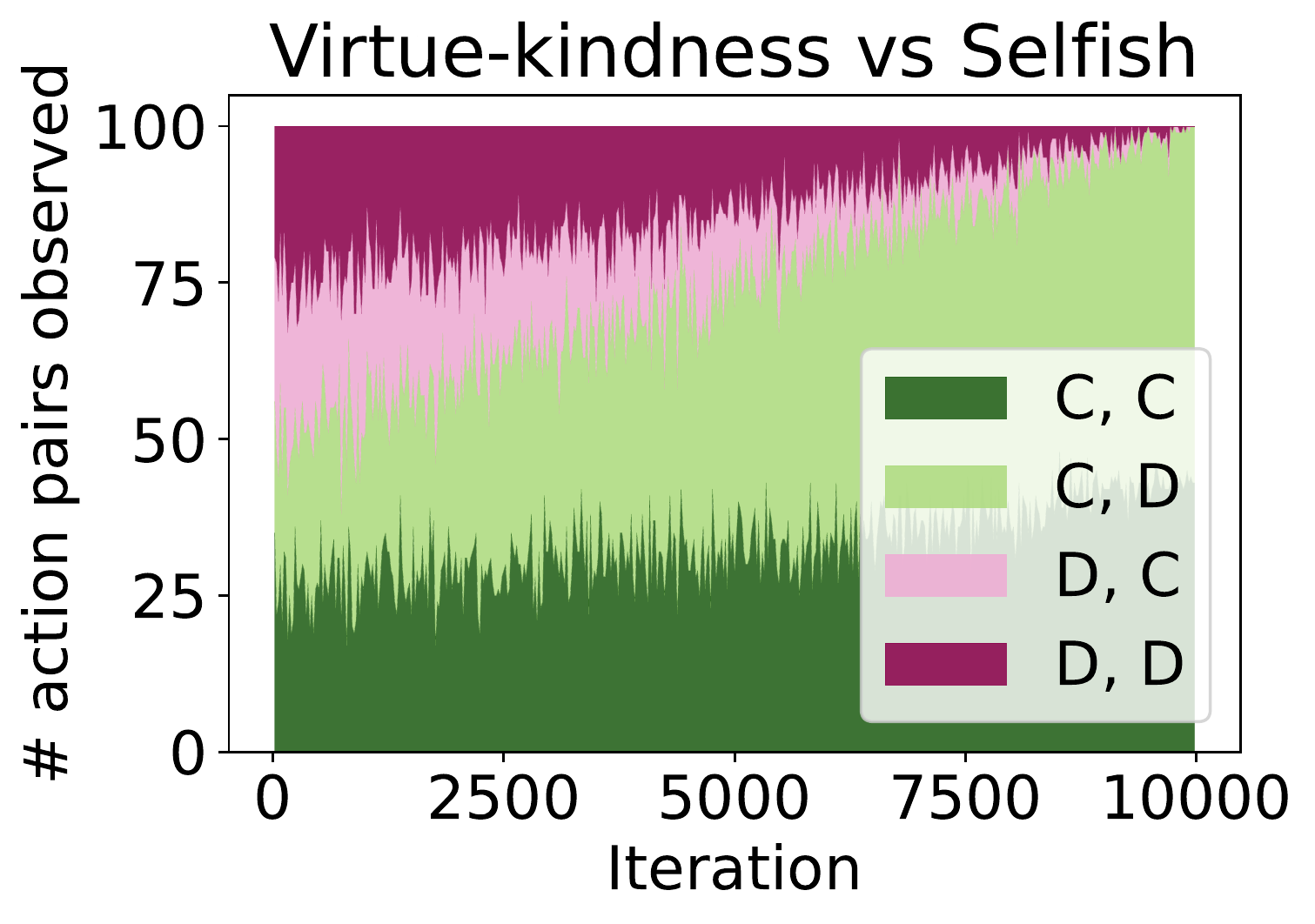}}
&\subt{\includegraphics[width=22mm]{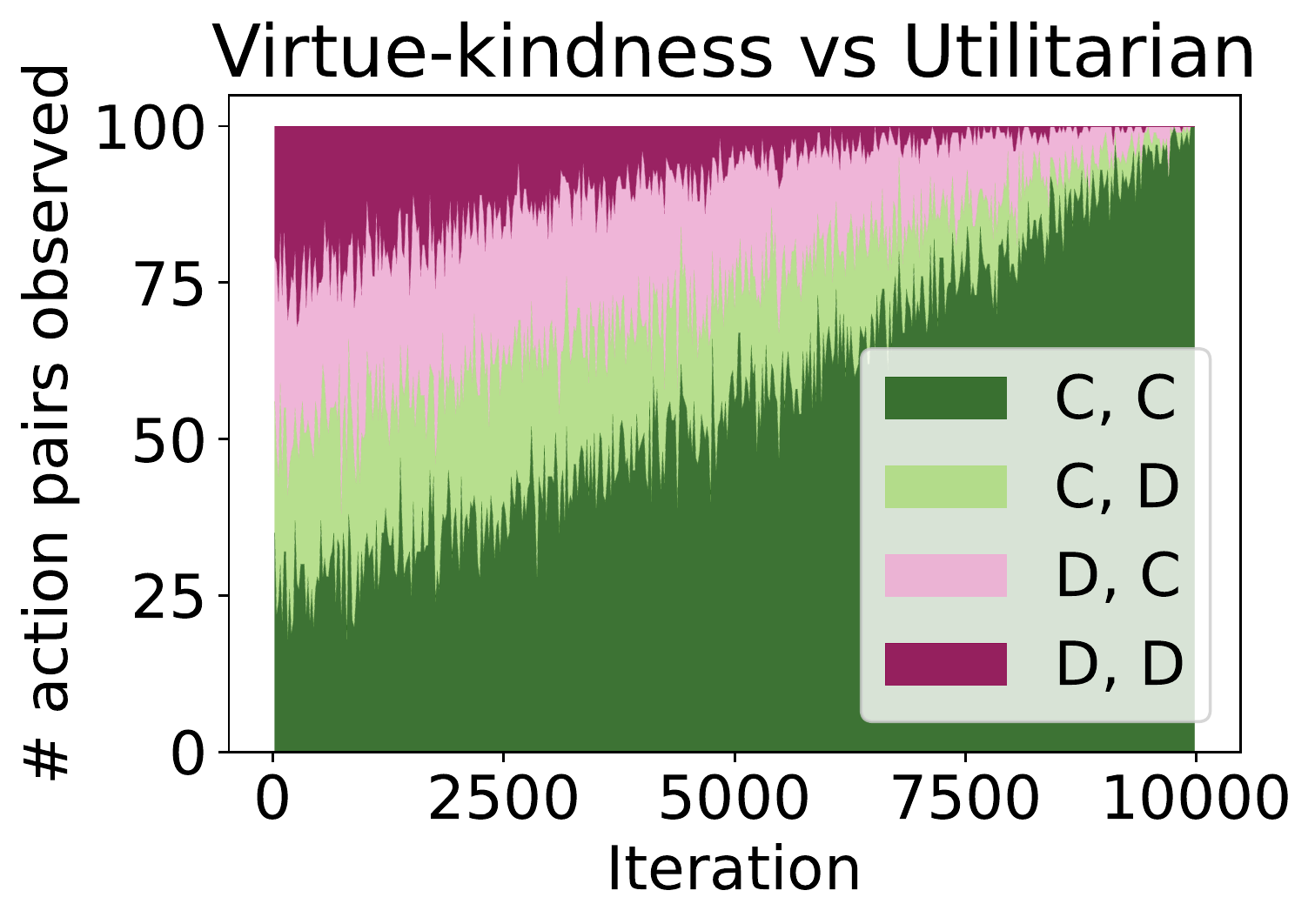}}
&\subt{\includegraphics[width=22mm]{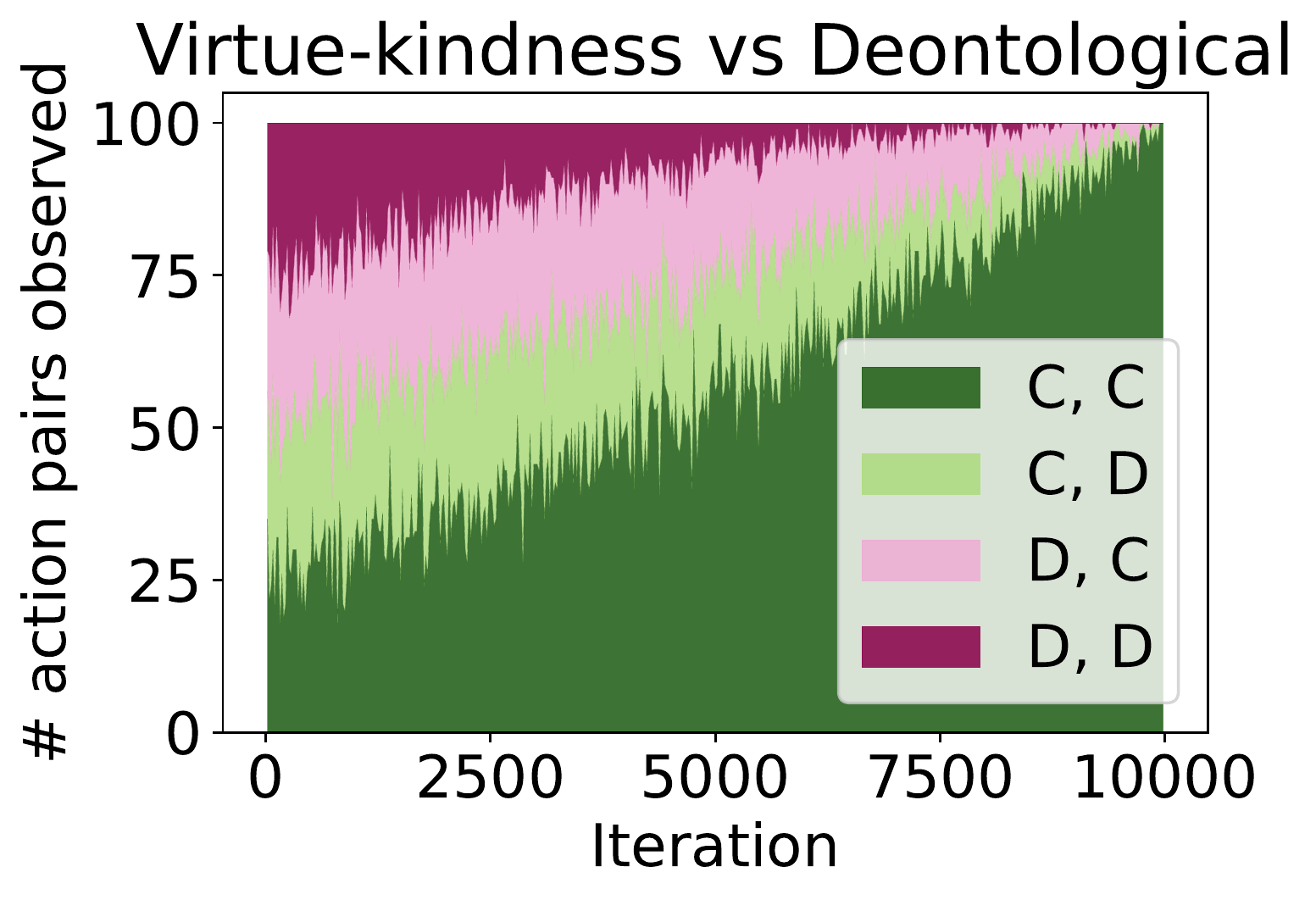}}
&\subt{\includegraphics[width=22mm]{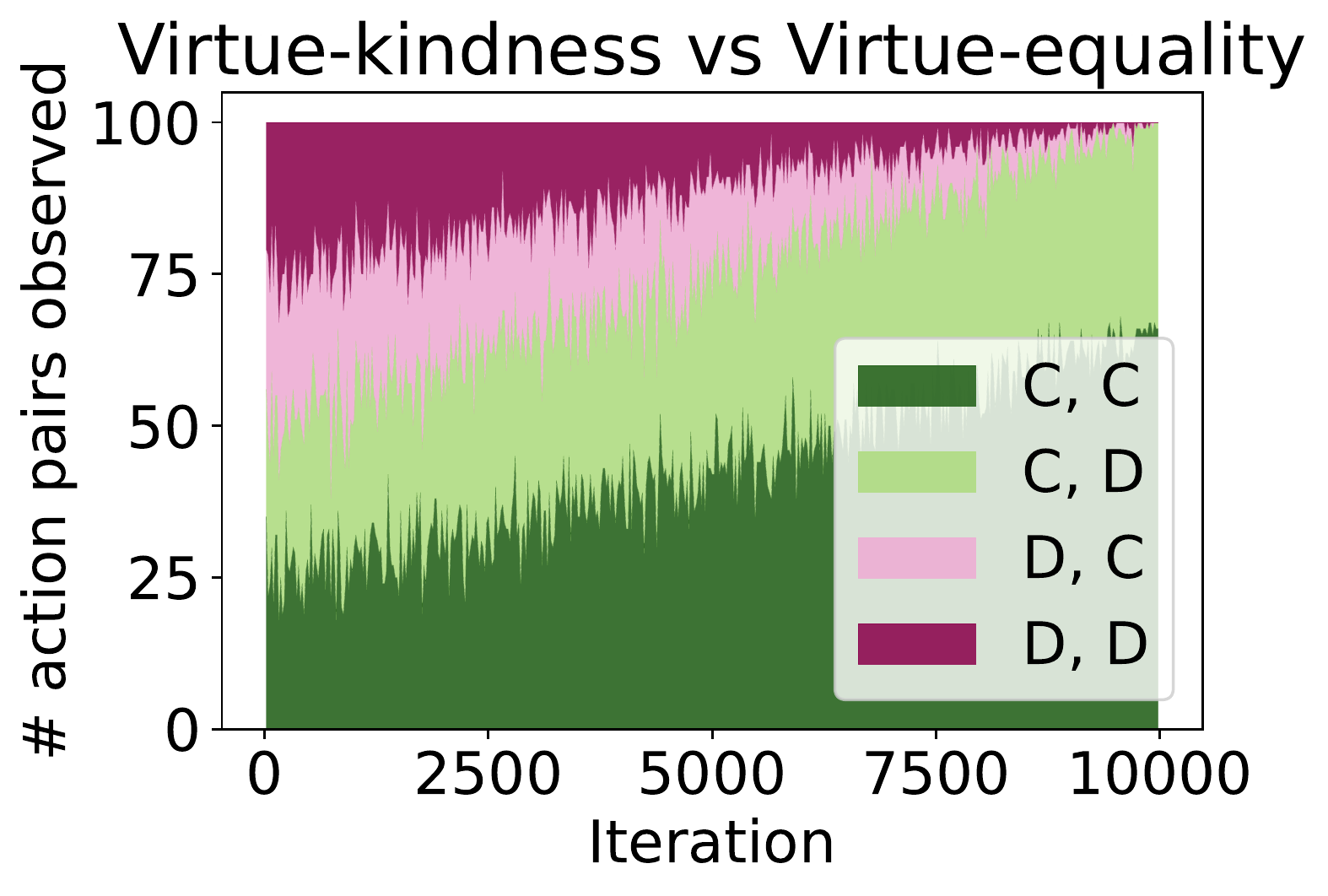}}
&\subt{\includegraphics[width=22mm]{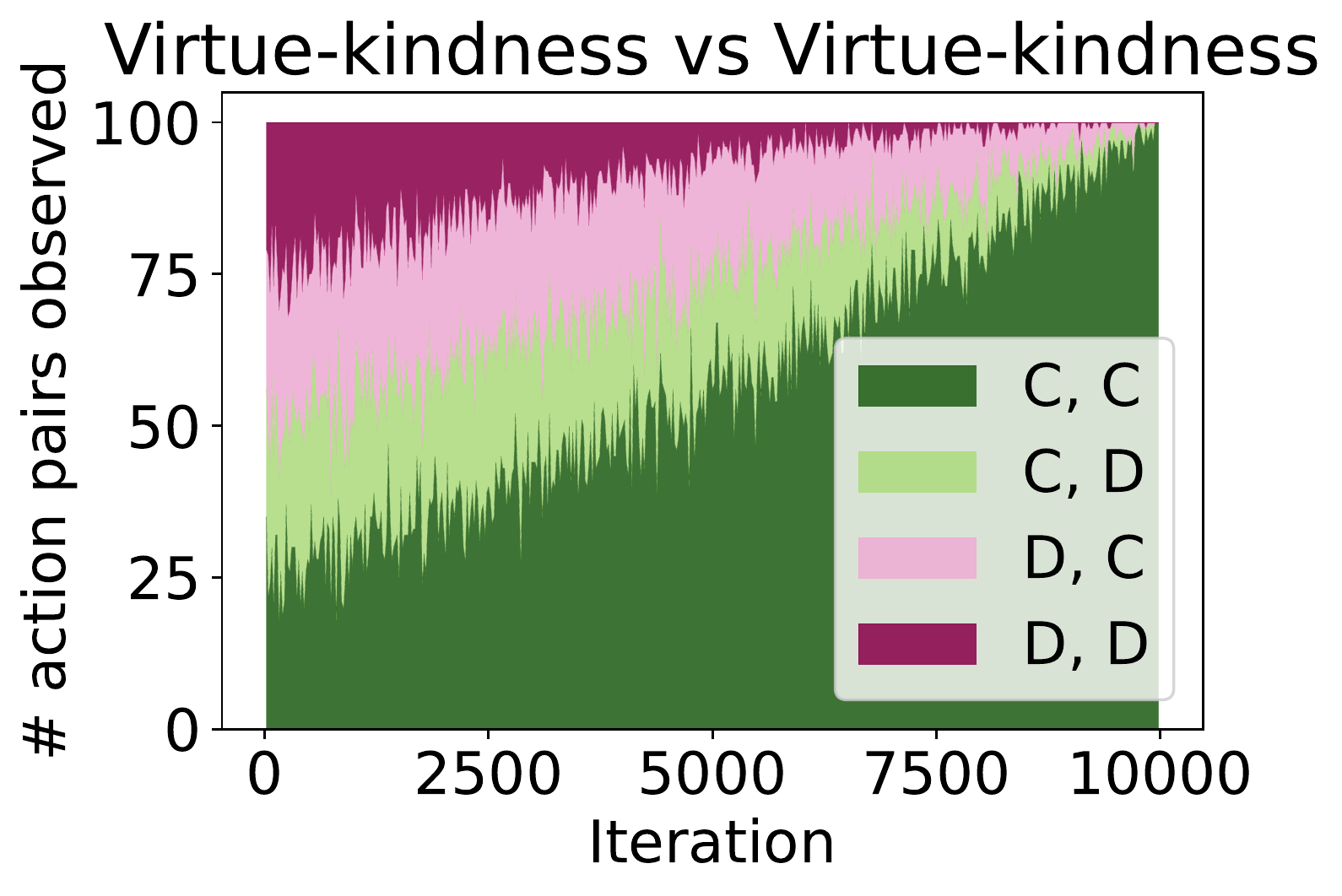}}
&
\\
\makecell[cc]{\rotatebox[origin=c]{90}{ Virtue-mix. }} &
\subt{\includegraphics[width=22mm]{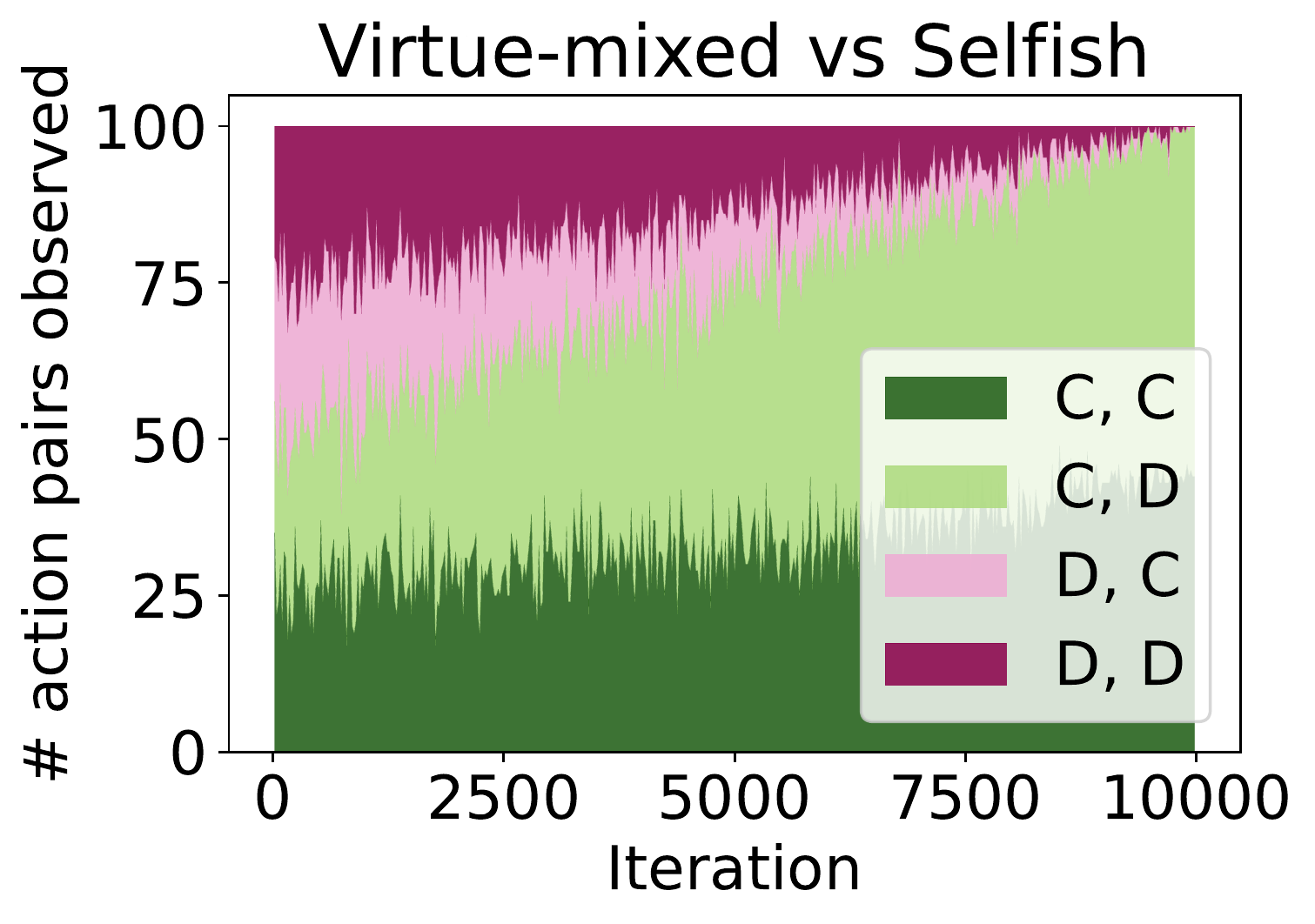}}
&\subt{\includegraphics[width=22mm]{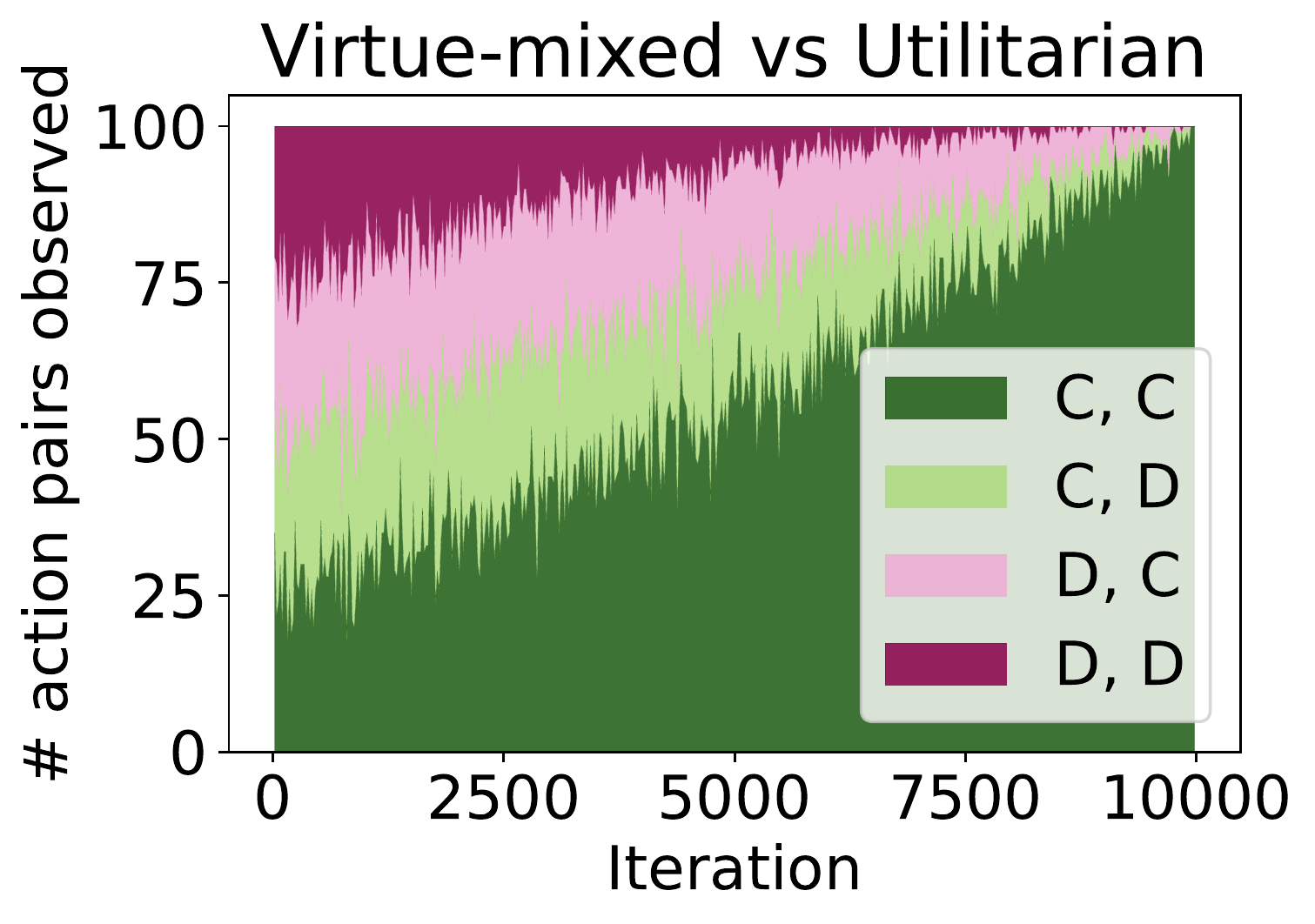}}
&\subt{\includegraphics[width=22mm]{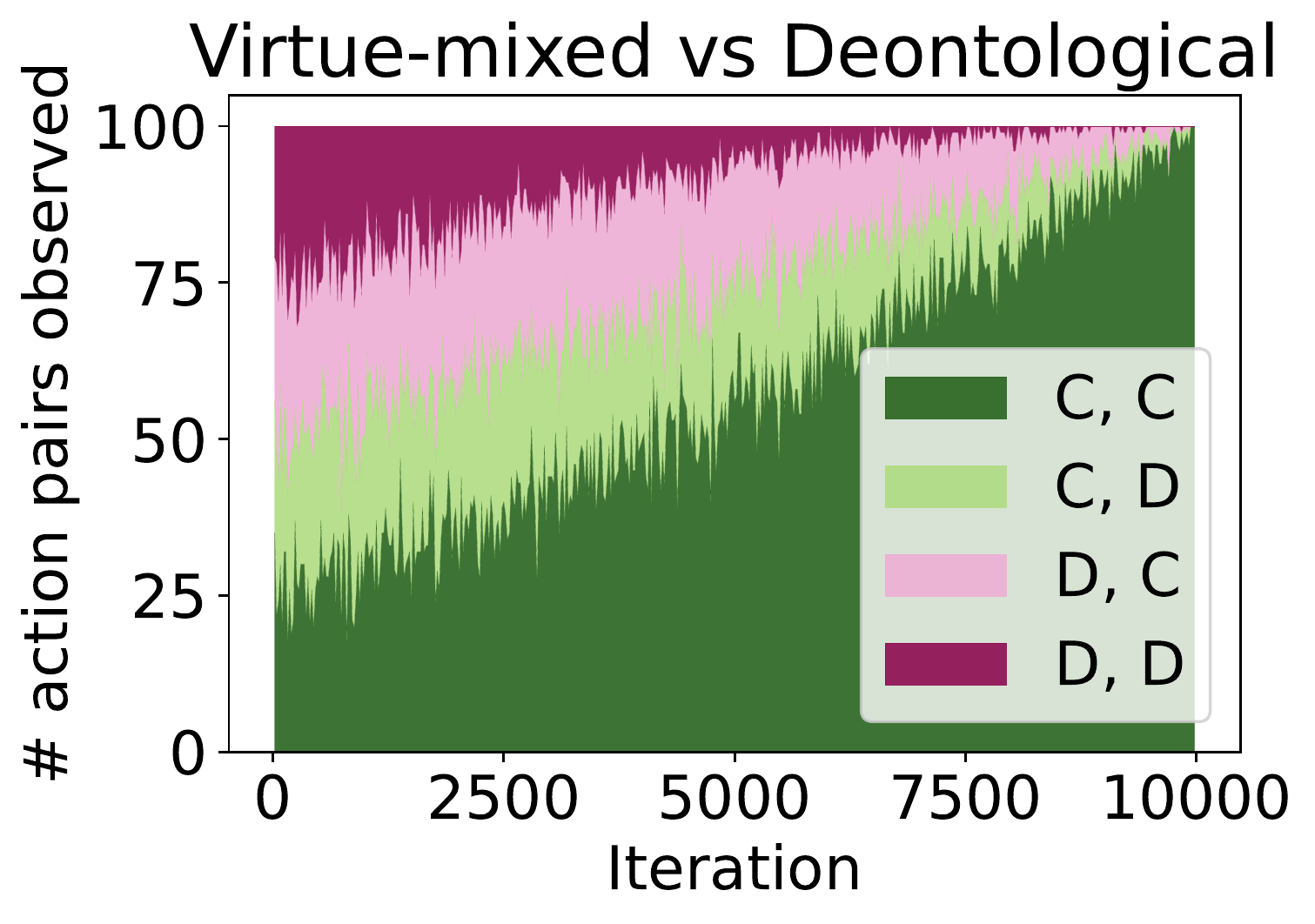}}
&\subt{\includegraphics[width=22mm]{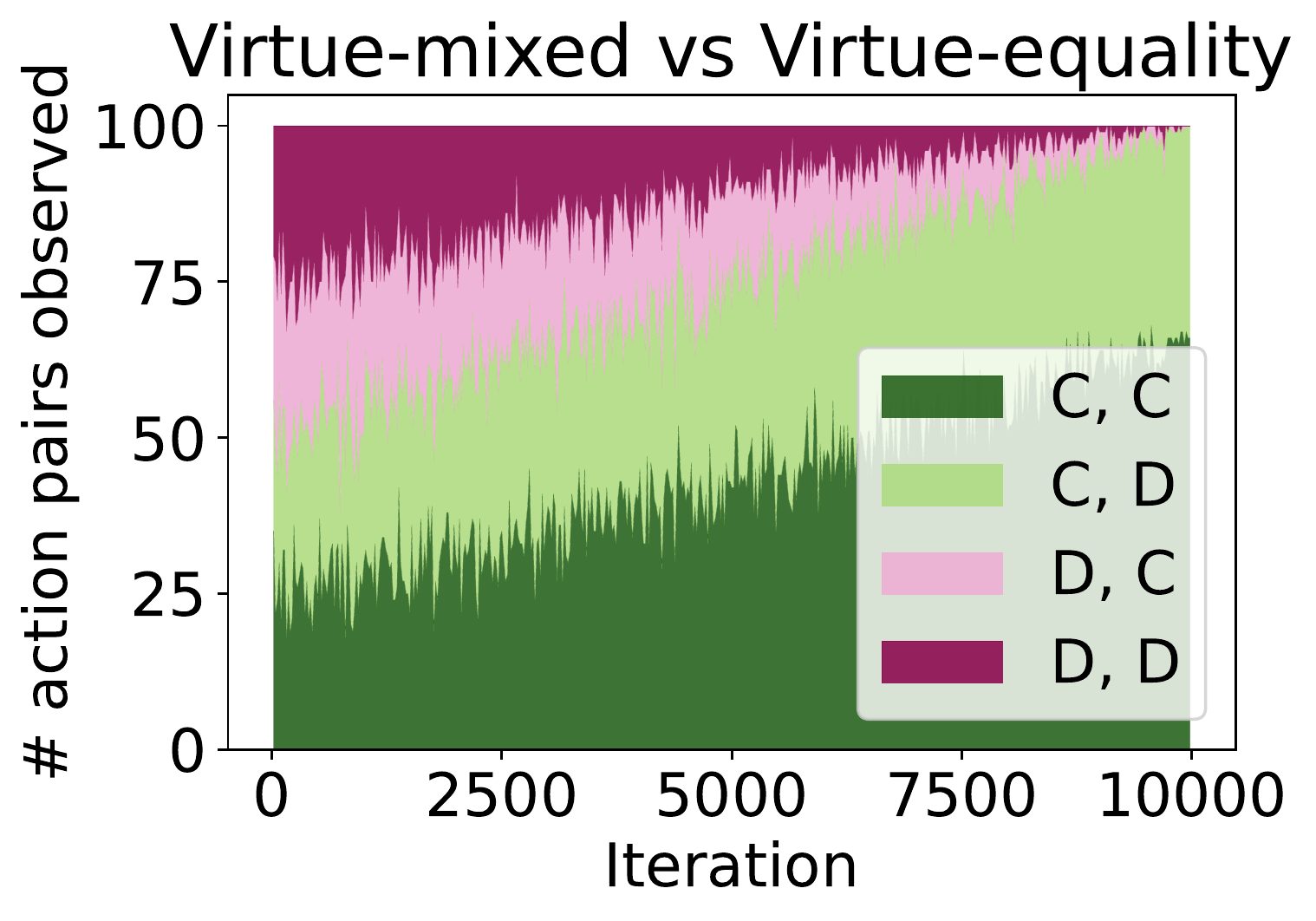}}
&\subt{\includegraphics[width=22mm]{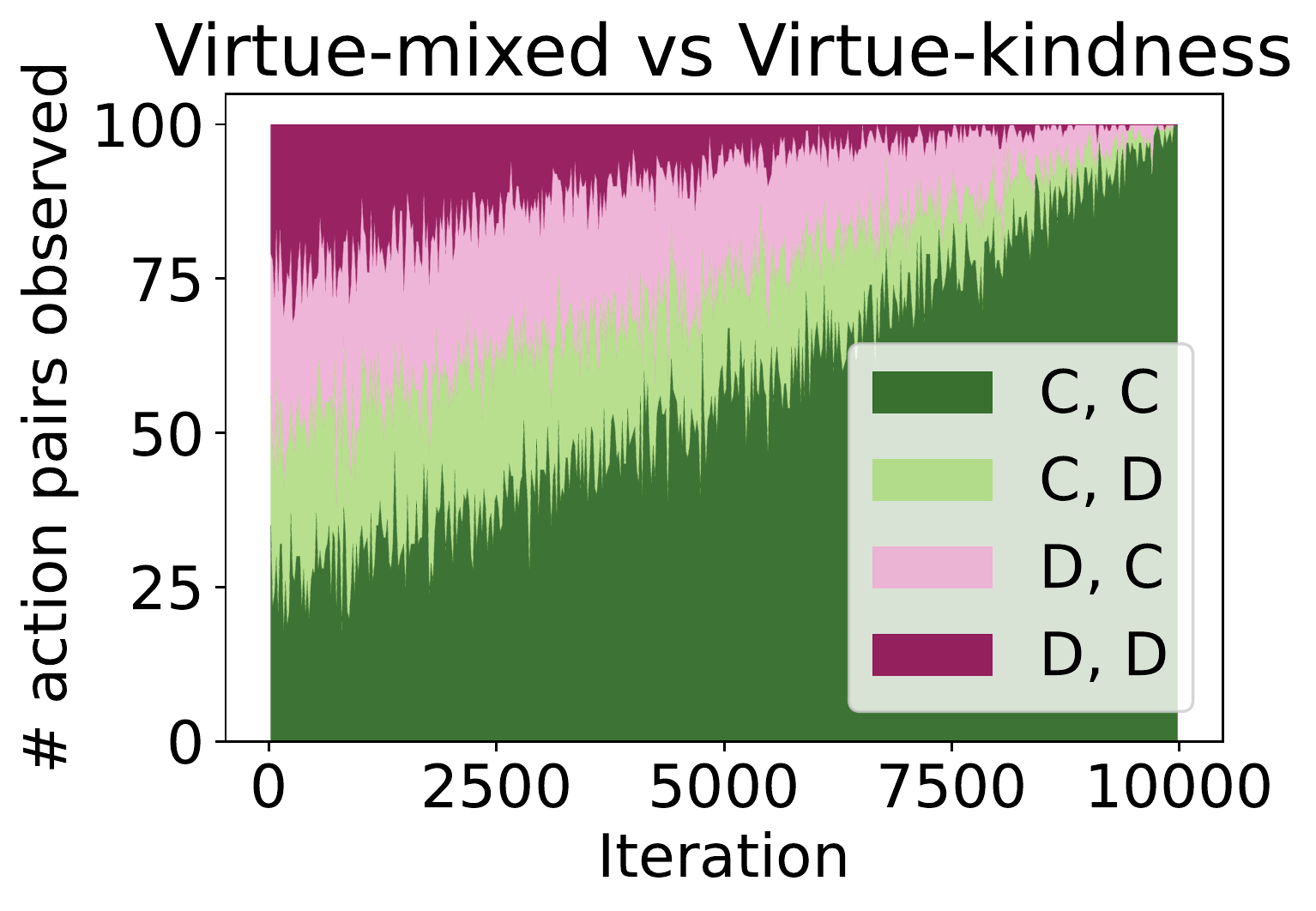}}
& \subt{\includegraphics[width=22mm]{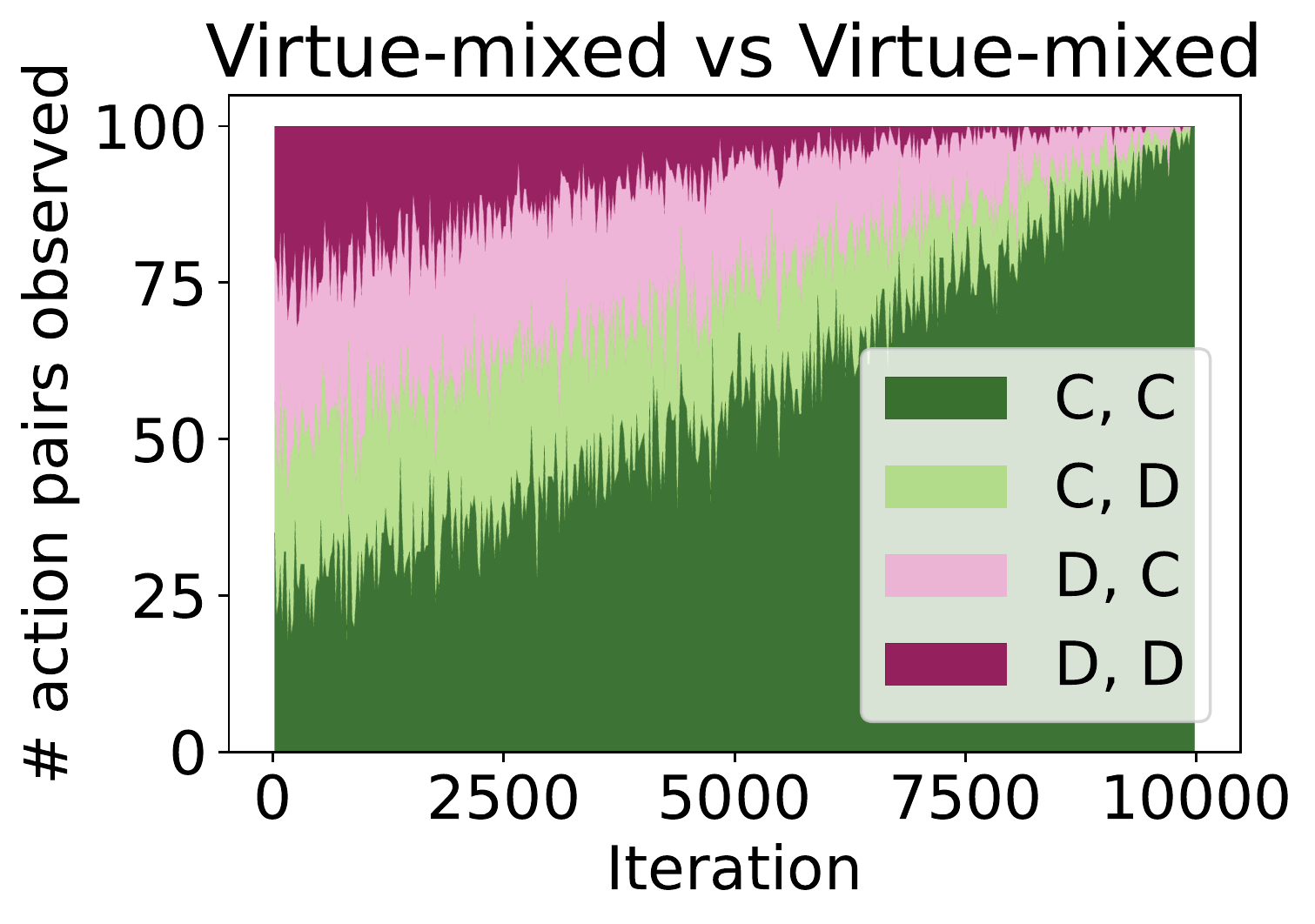}}
\\
\bottomrule
\end{tabular}
\caption{Iterated Volunteer's Dilemma game. Simultaneous pairs of actions observed over time. Learning player $M$ (row) vs. learning opponent $O$ (column).}
\label{fig:action_pairs_learning_VOL}
\end{figure*}

\begin{figure*}[!h]
\centering
\begin{tabular}{|c|cccccc}
\toprule
 & Selfish & Utilitarian & Deontological & Virtue-equality & Virtue-kindness & Virtue-mixed\\
\midrule
{\rotatebox[origin=c]{90}{ Selfish }} & 
\subt{\includegraphics[width=22mm]{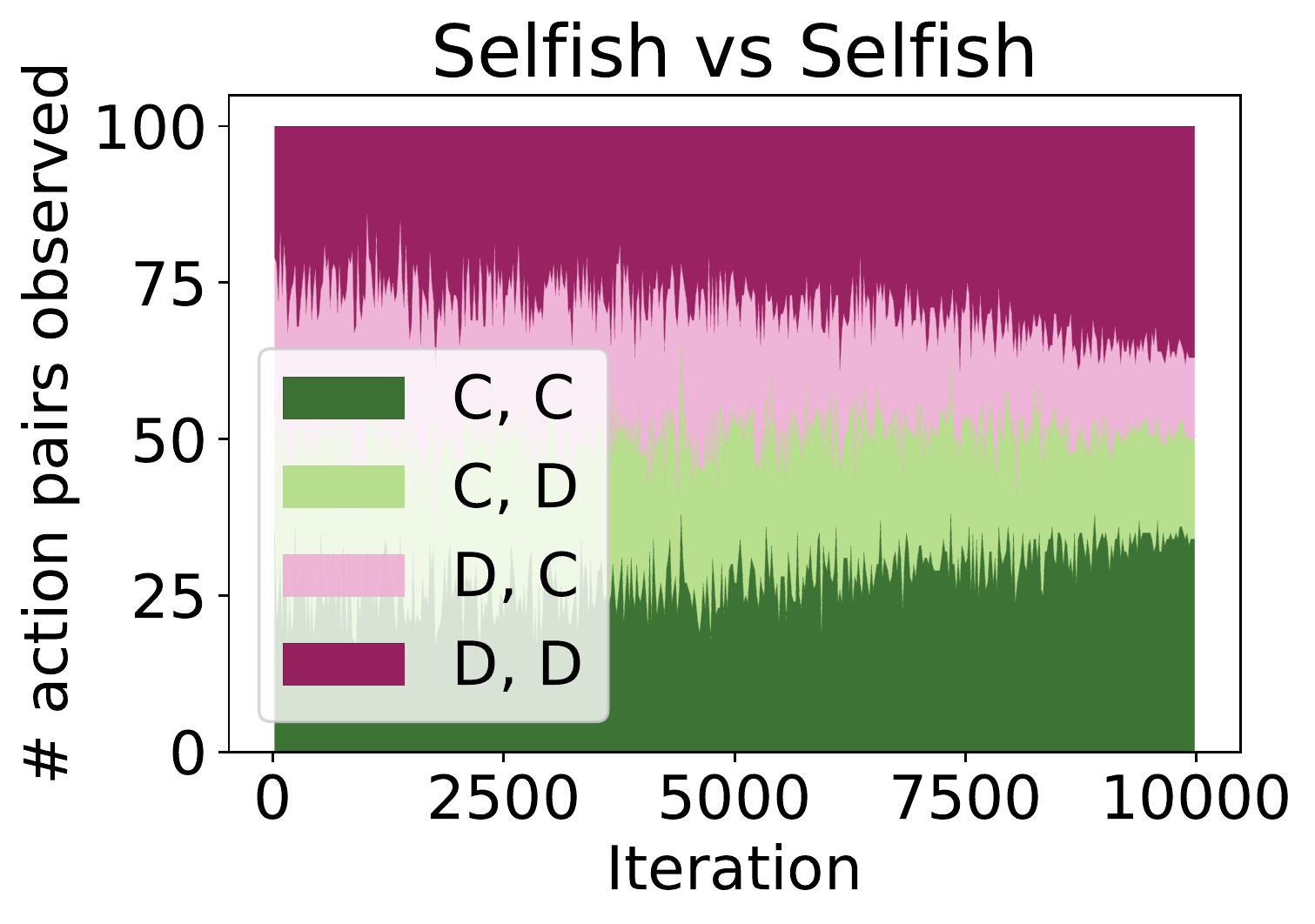}}
&
&
&
&
\\
\makecell[cc]{\rotatebox[origin=c]{90}{ Utilitarian }} & 
\subt{\includegraphics[width=22mm]{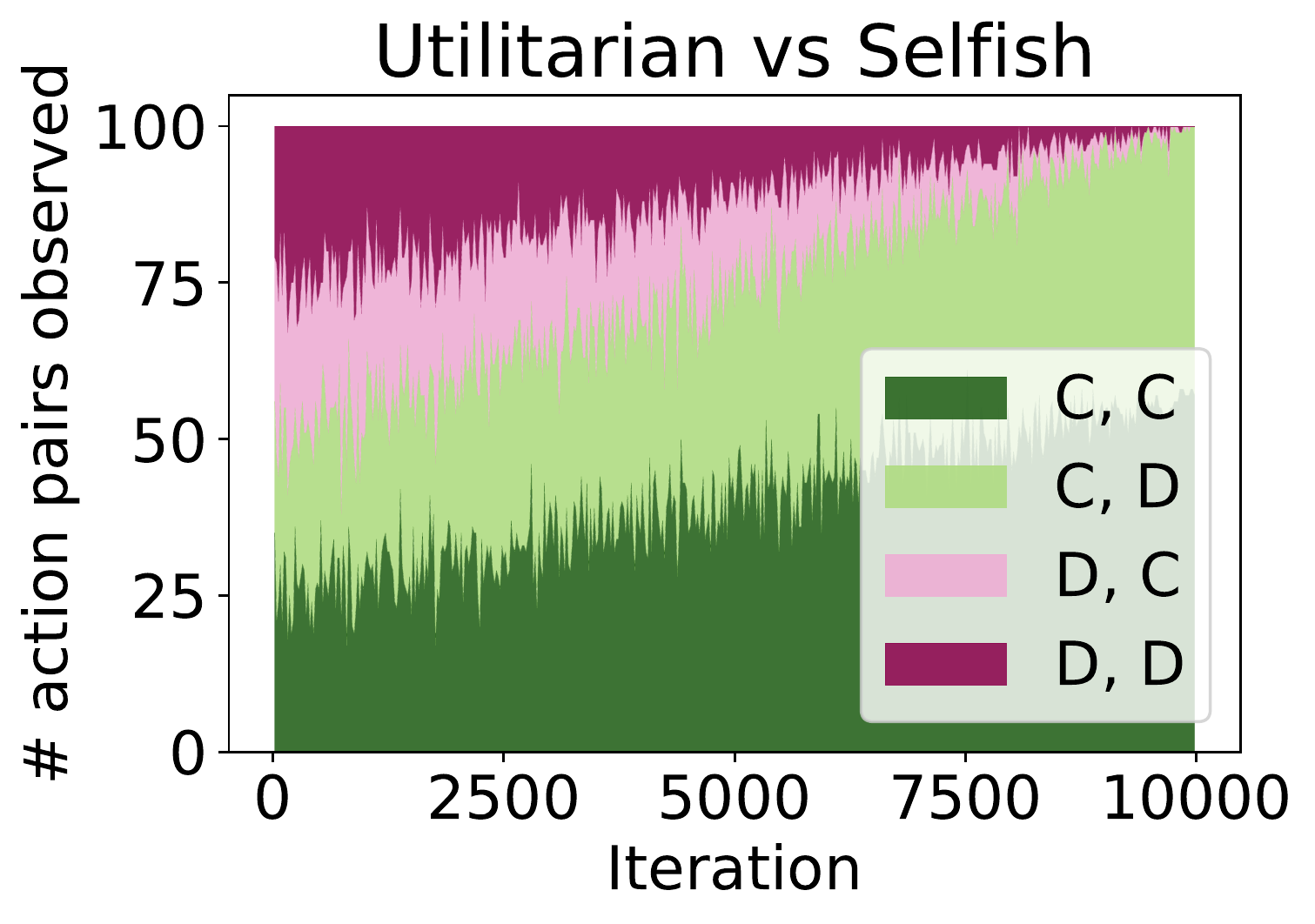}}
&\subt{\includegraphics[width=22mm]{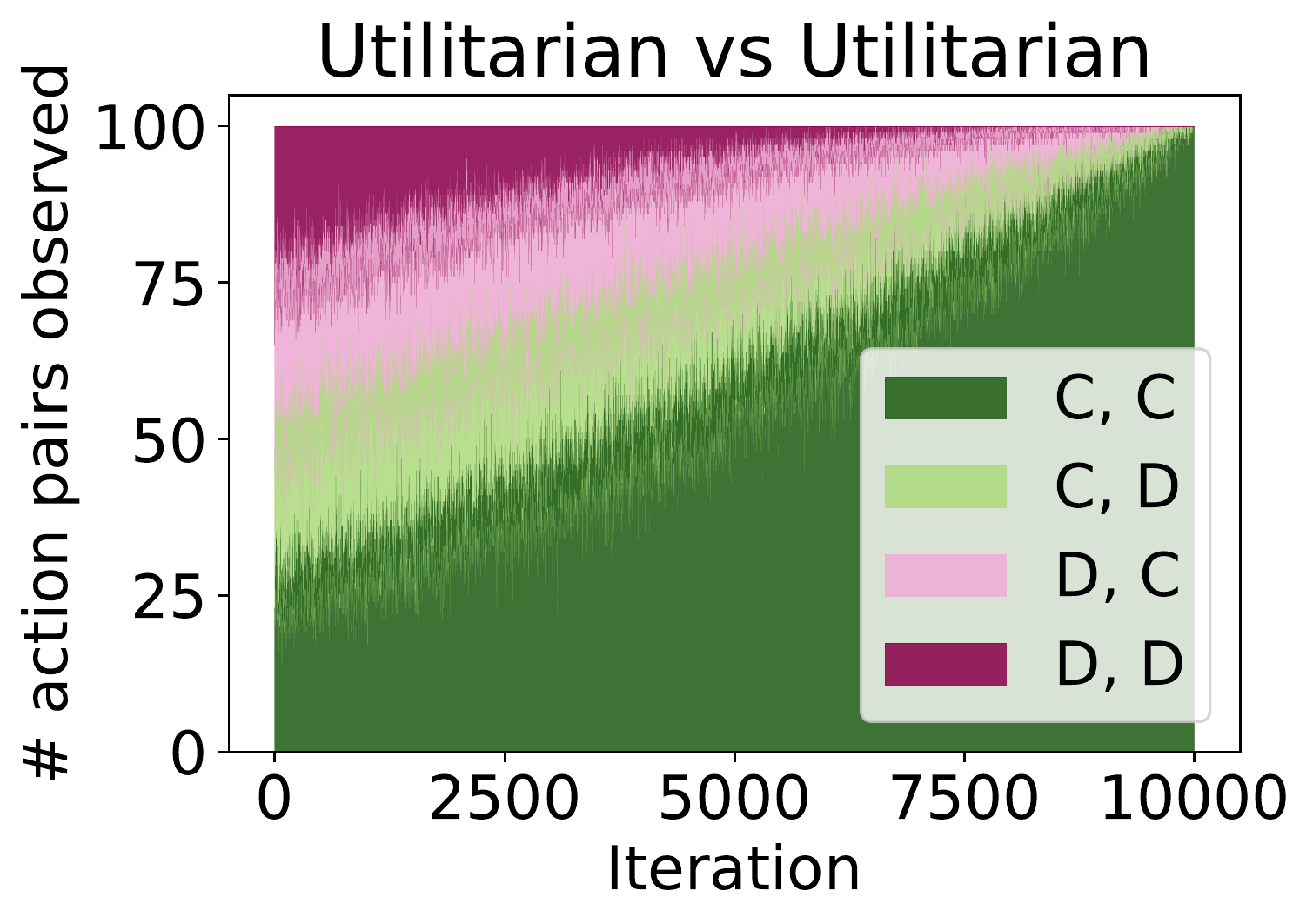}}
&
&
&
\\
\makecell[cc]{\rotatebox[origin=c]{90}{ Deontological }} &
\subt{\includegraphics[width=22mm]{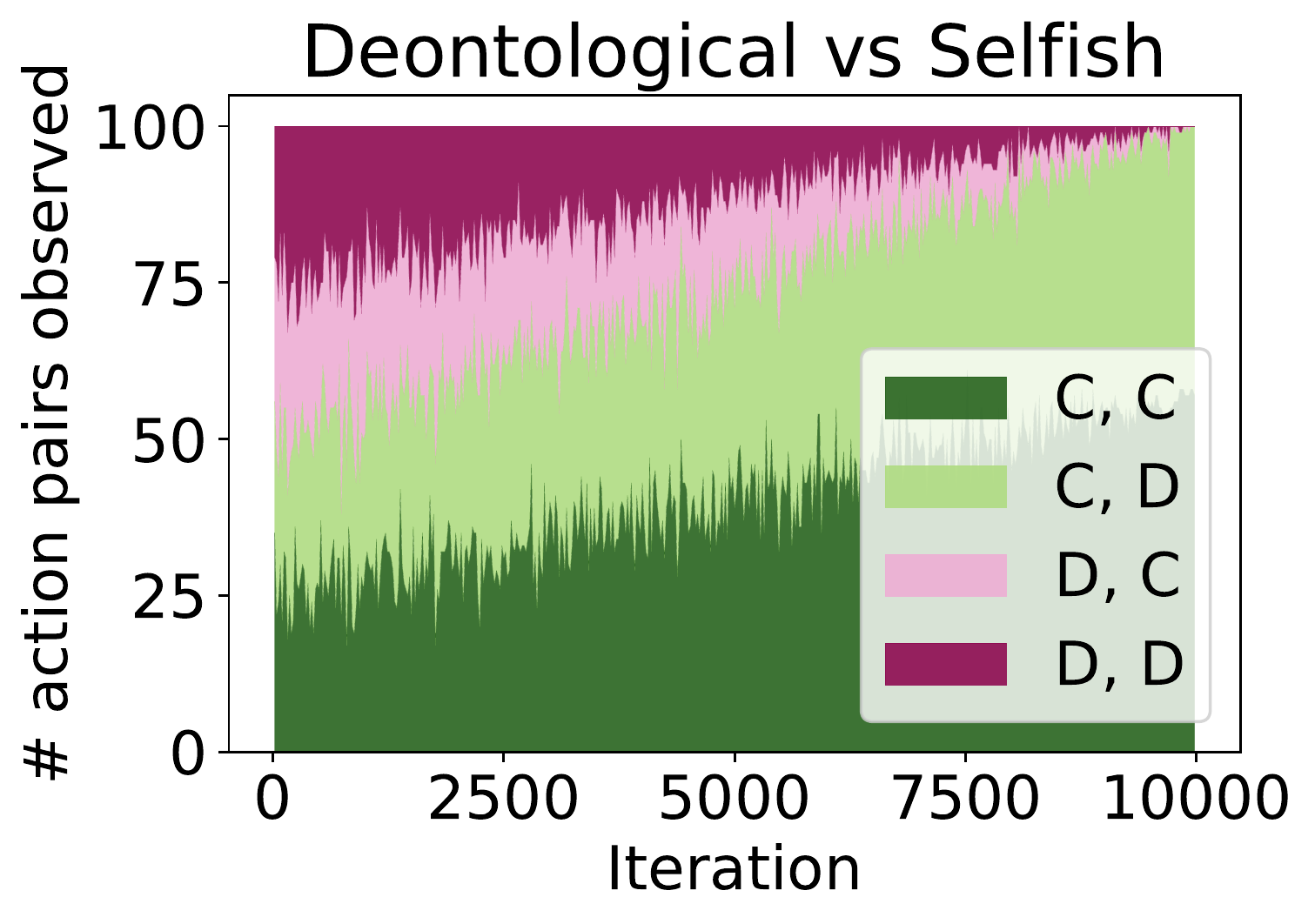}}
&\subt{\includegraphics[width=22mm]{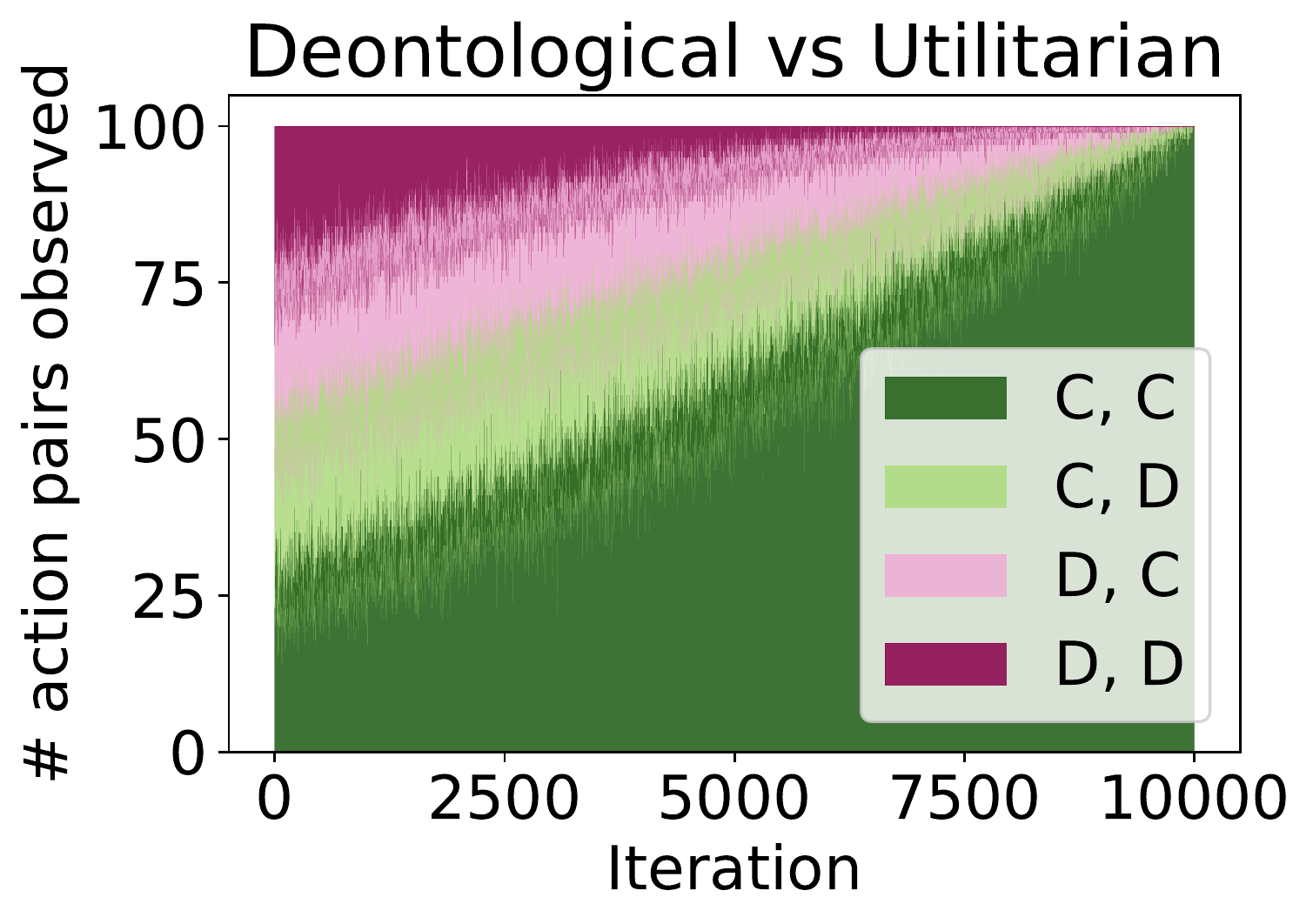}}
&\subt{\includegraphics[width=22mm]{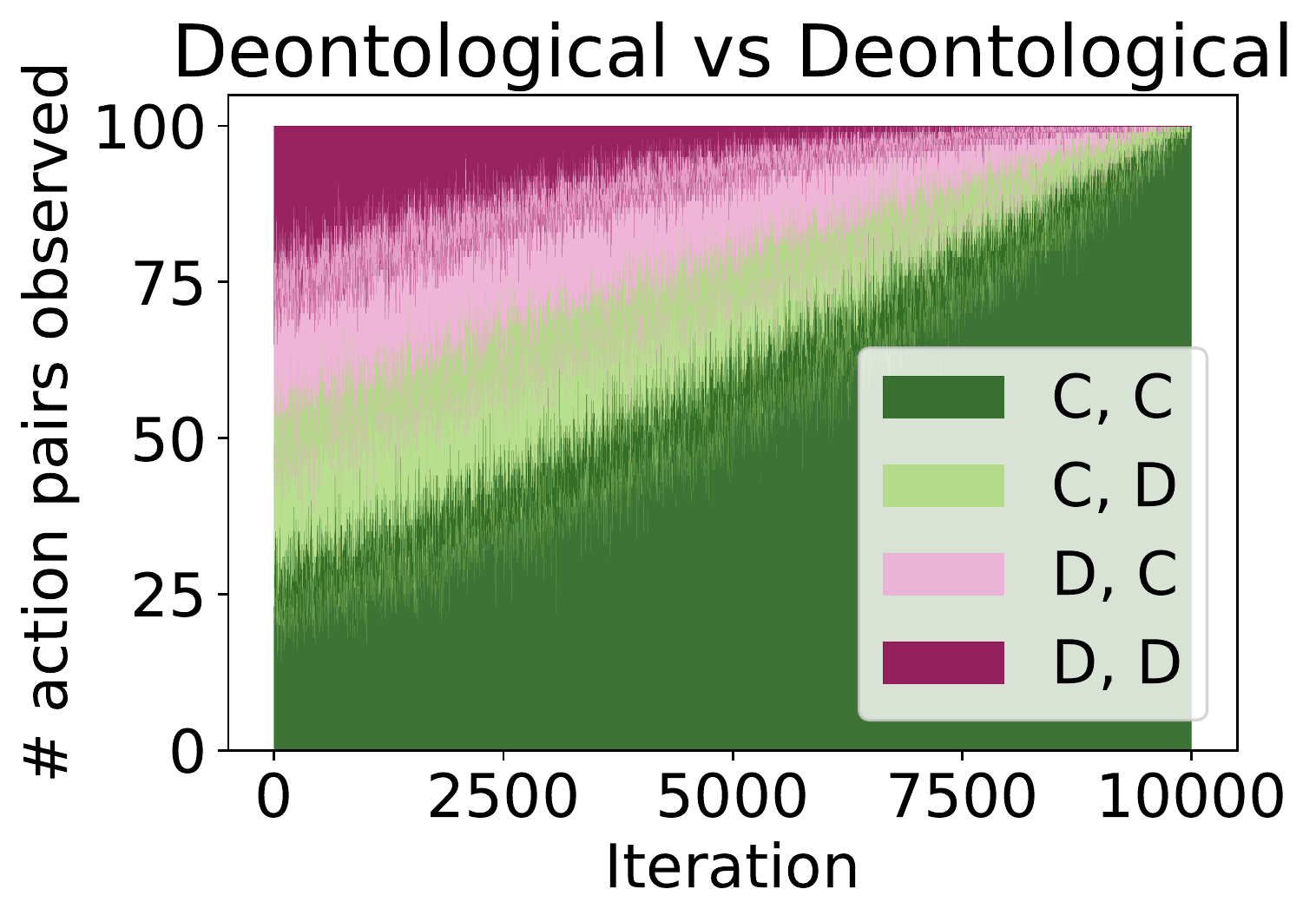}}
&
&
&
\\
\makecell[cc]{\rotatebox[origin=c]{90}{ Virtue-eq. }} &
\subt{\includegraphics[width=22mm]{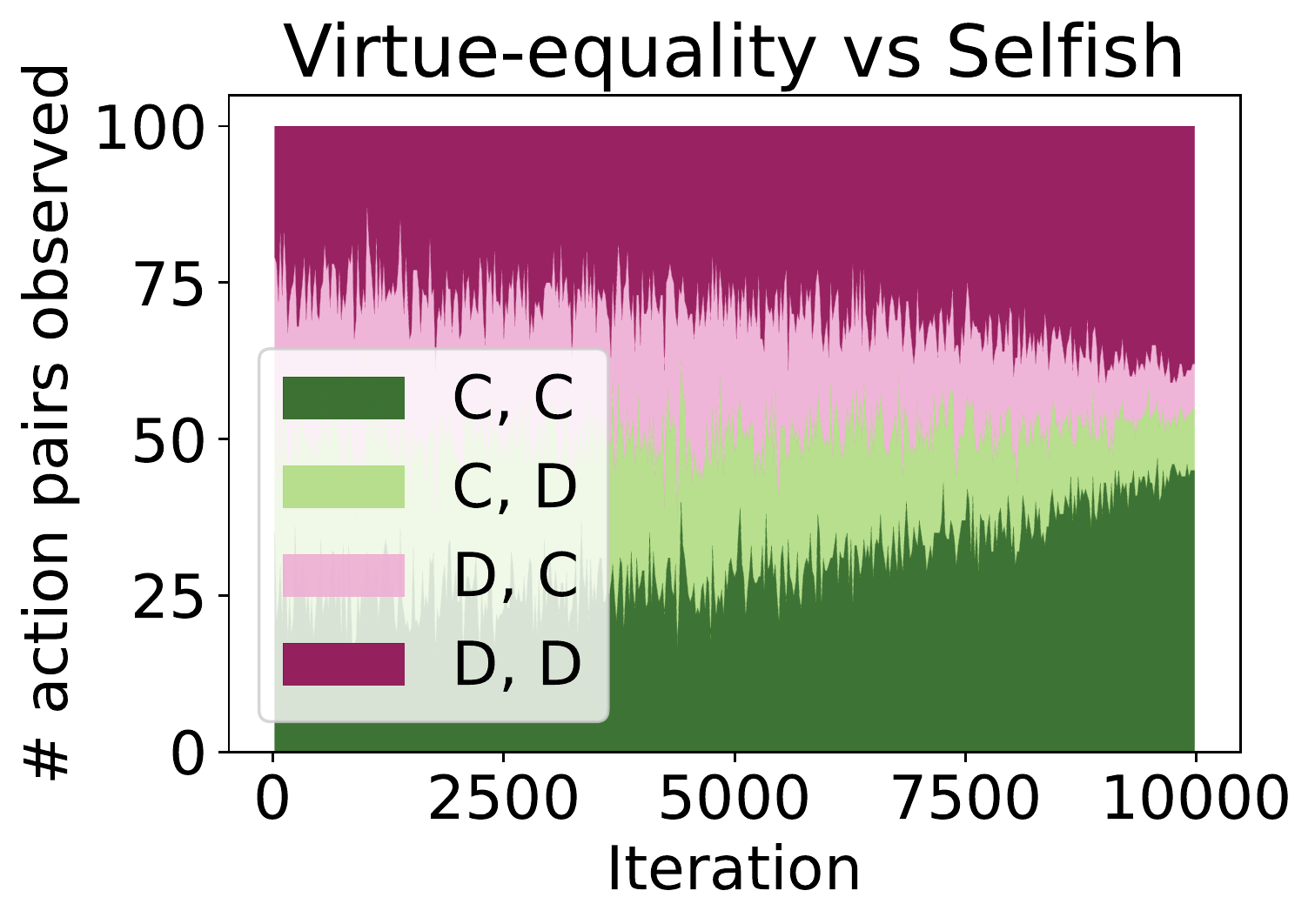}}
&\subt{\includegraphics[width=22mm]{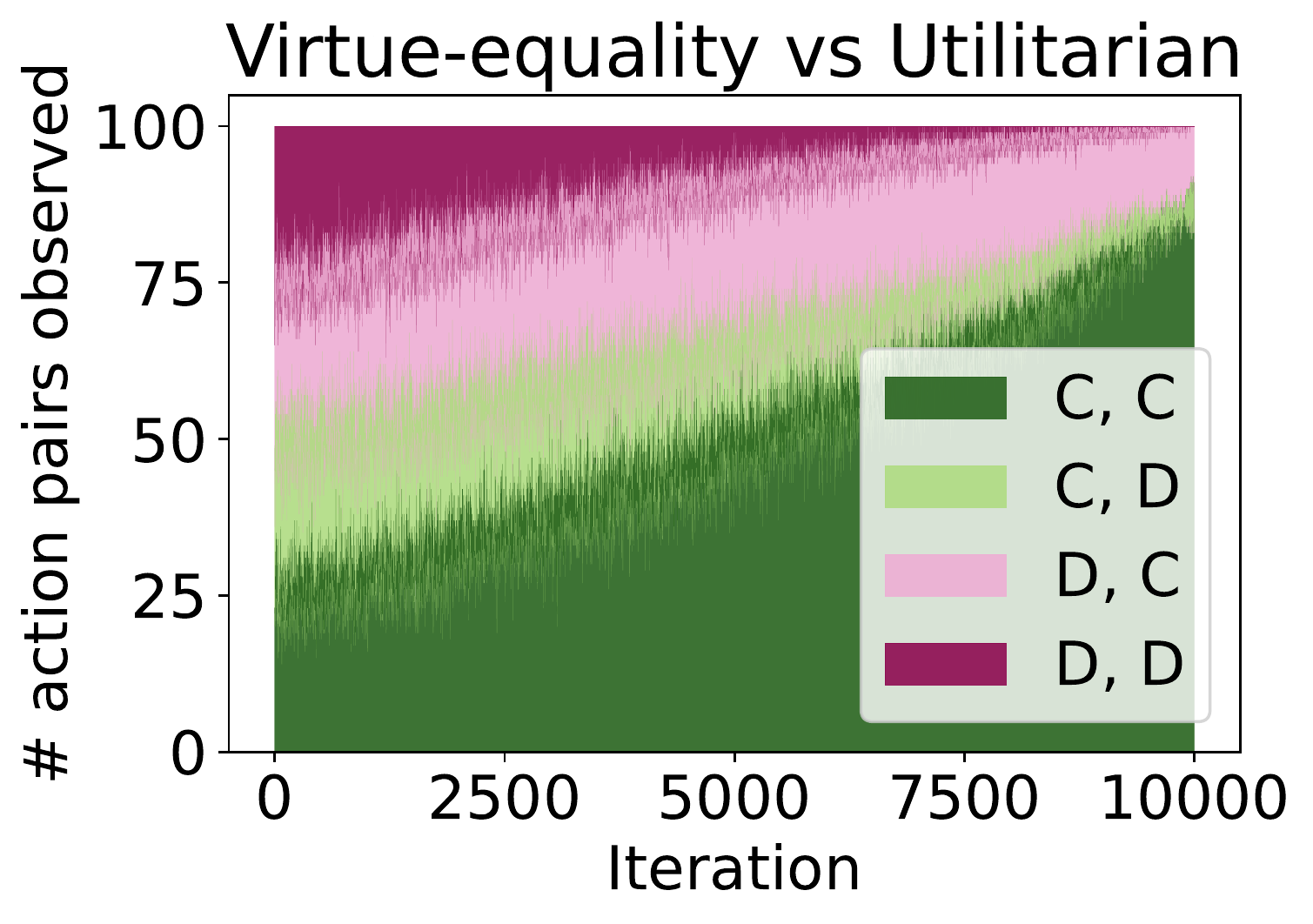}}
&\subt{\includegraphics[width=22mm]{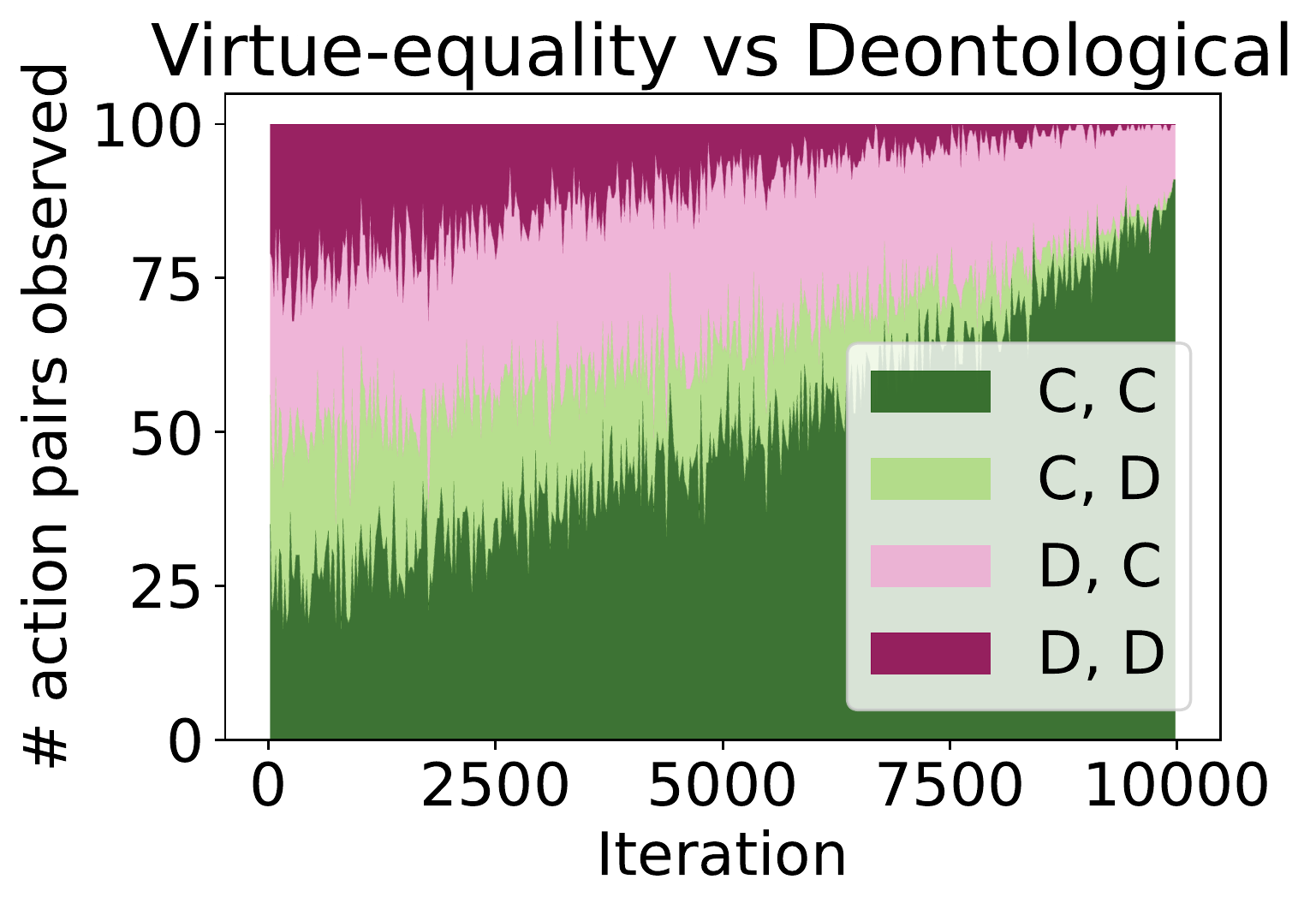}}
&\subt{\includegraphics[width=22mm]{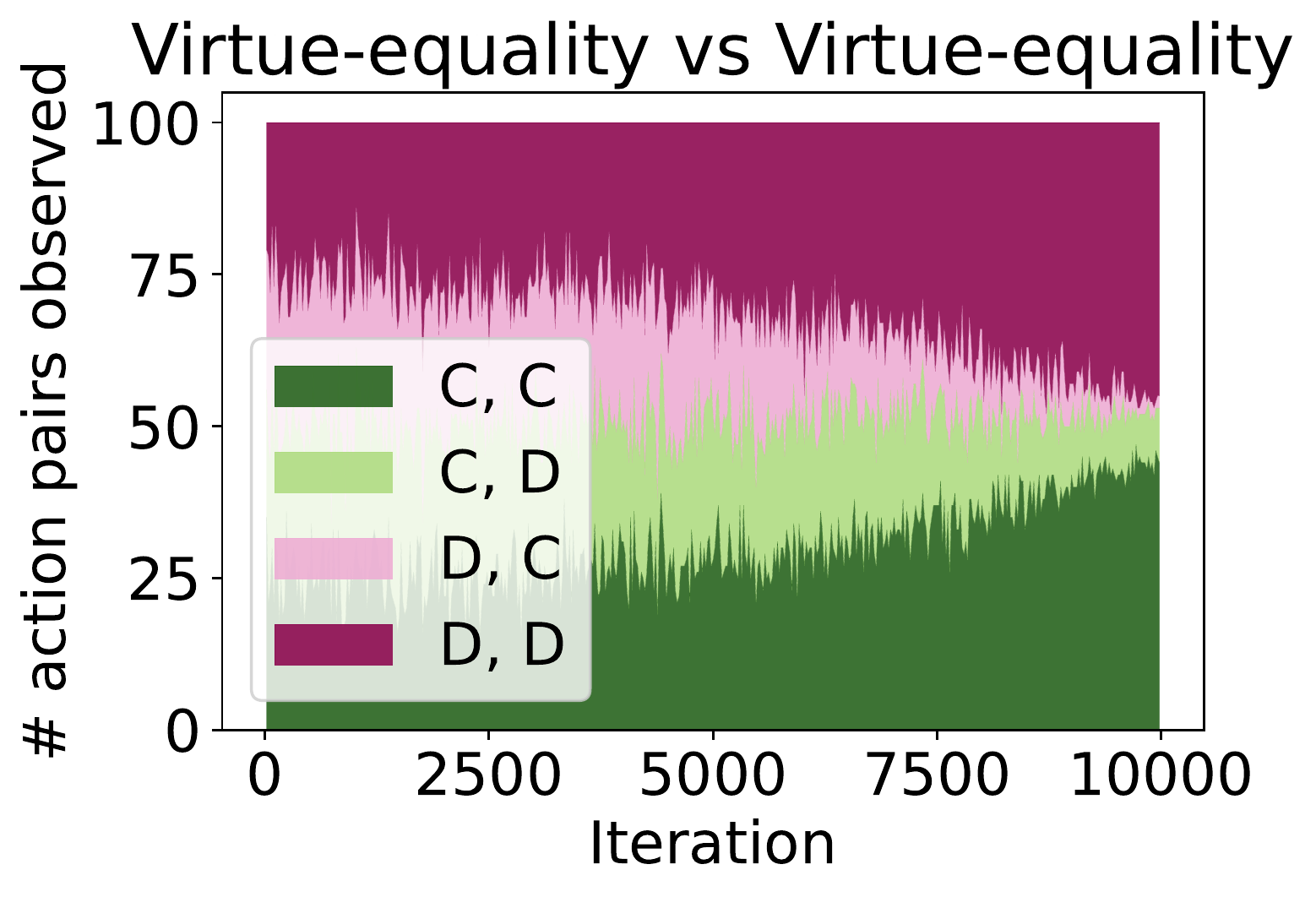}}
&
&
\\
\makecell[cc]{\rotatebox[origin=c]{90}{ Virtue-kind. }} &
\subt{\includegraphics[width=22mm]{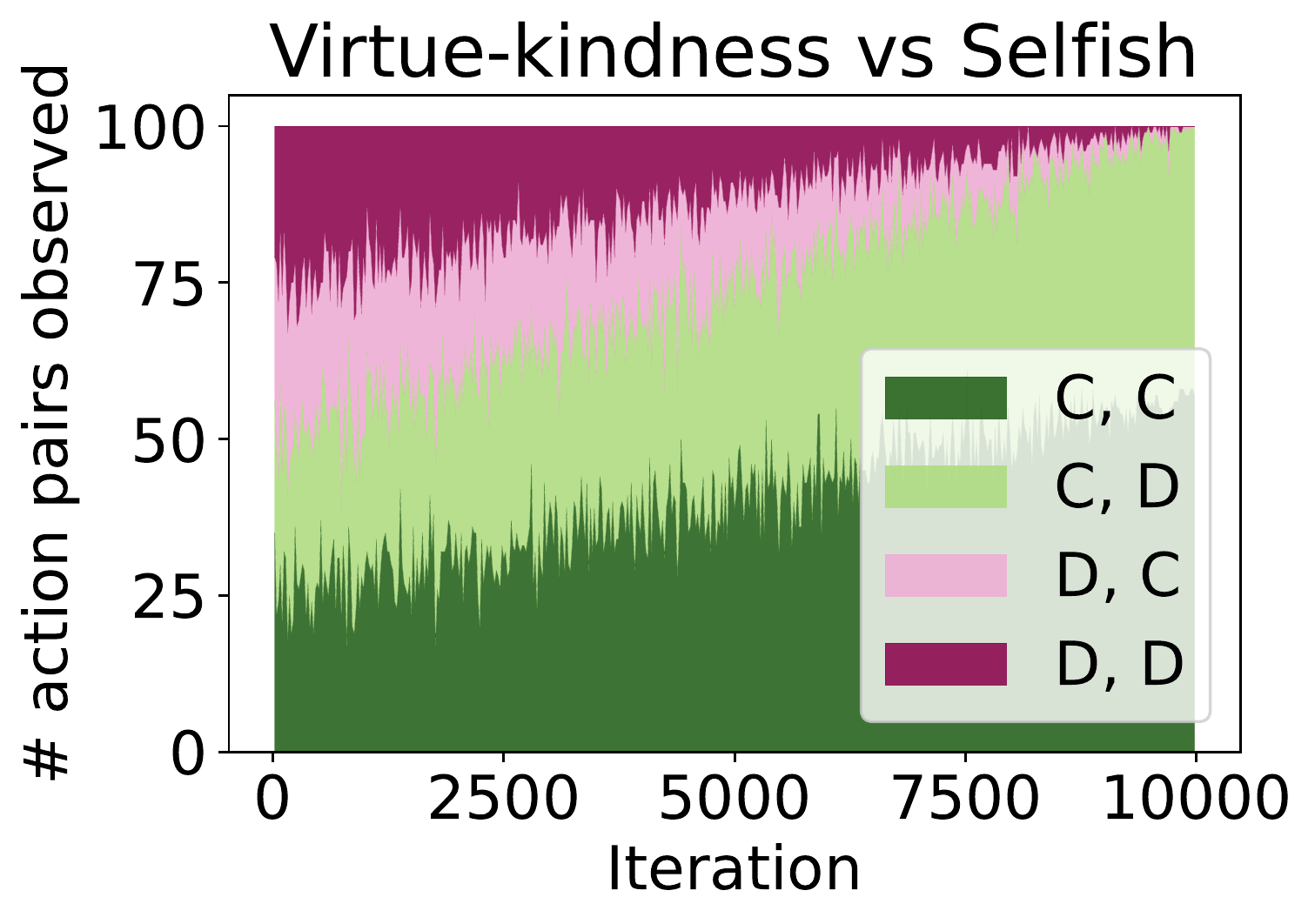}}
&\subt{\includegraphics[width=22mm]{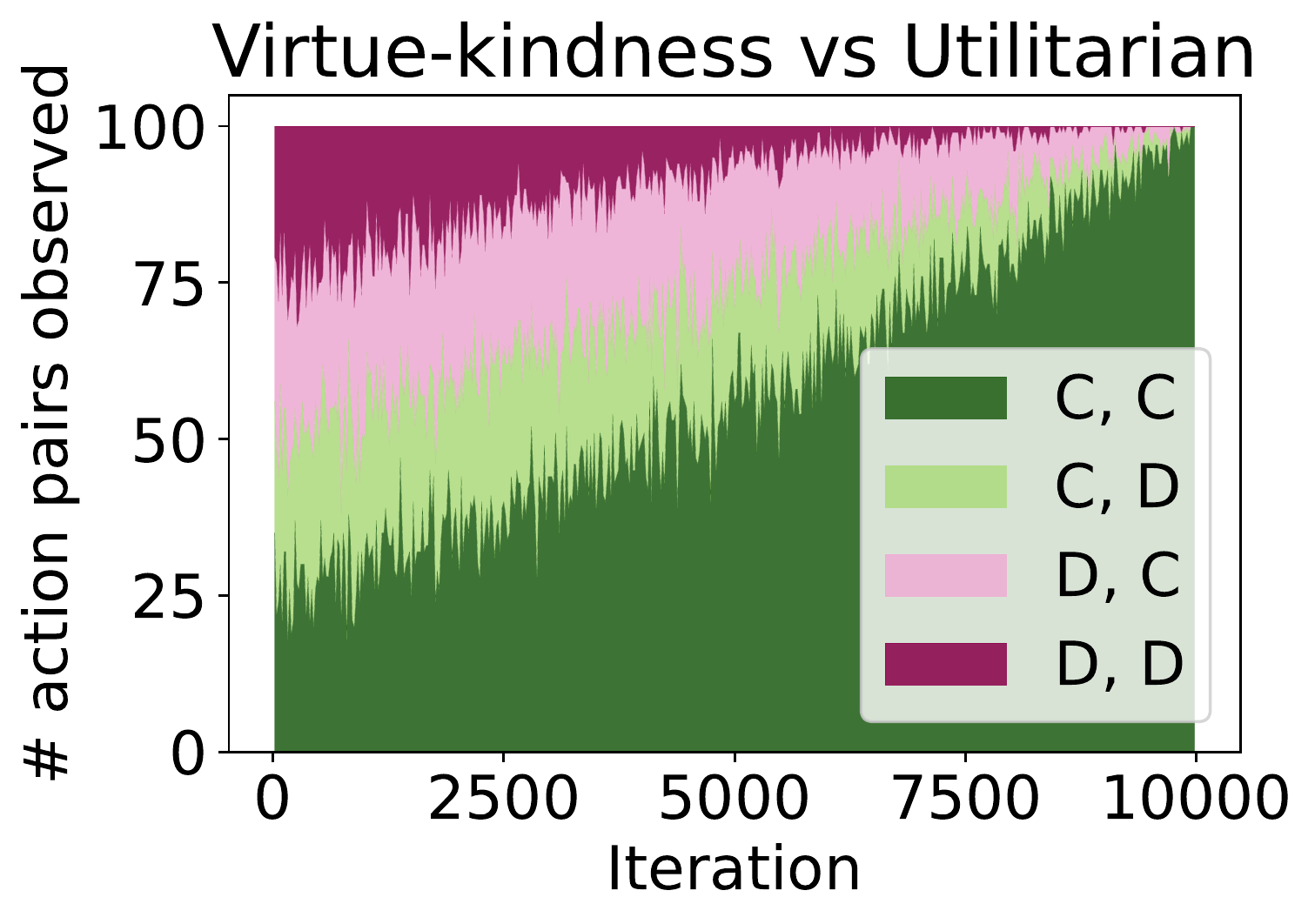}}
&\subt{\includegraphics[width=22mm]{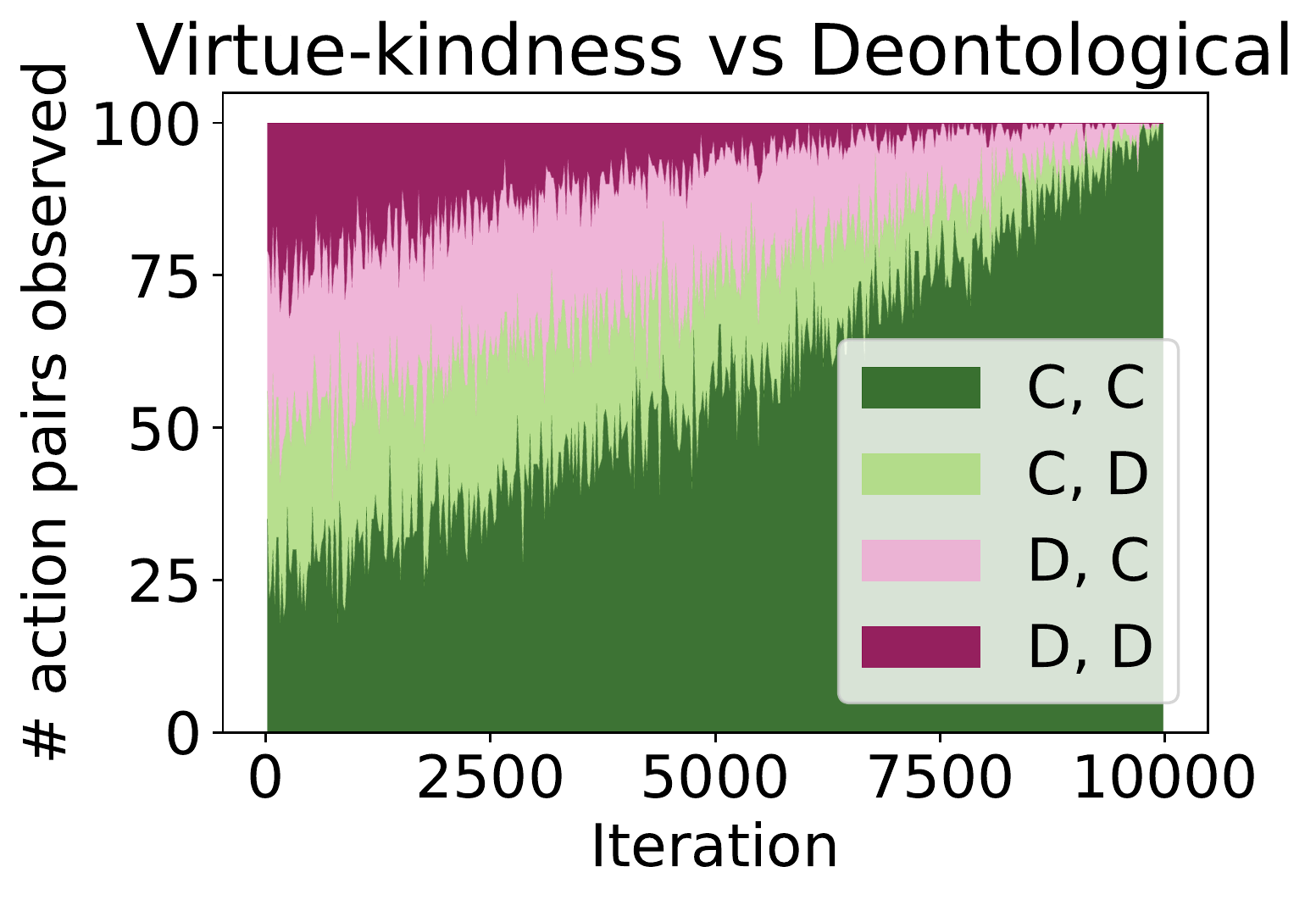}}
&\subt{\includegraphics[width=22mm]{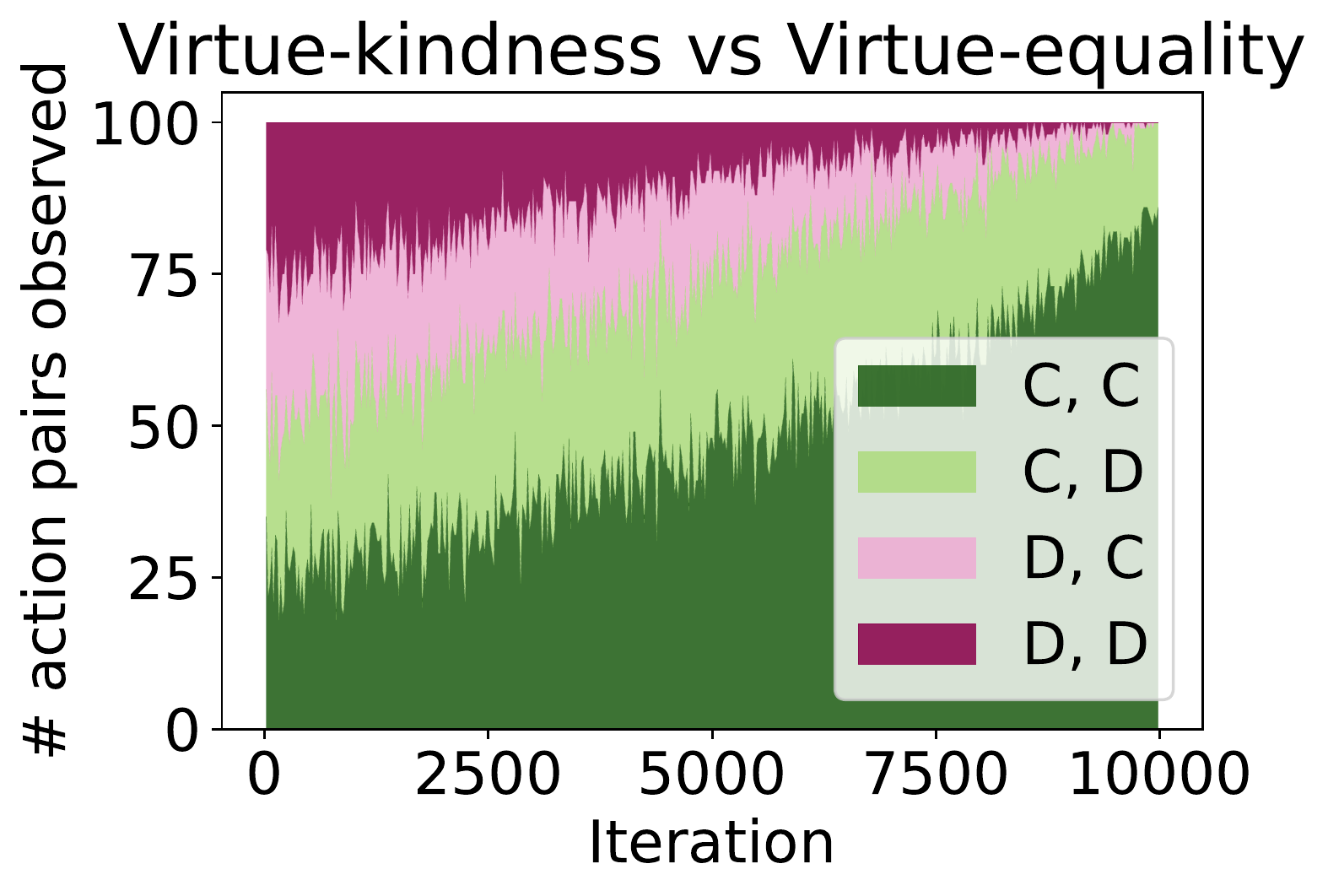}}
&\subt{\includegraphics[width=22mm]{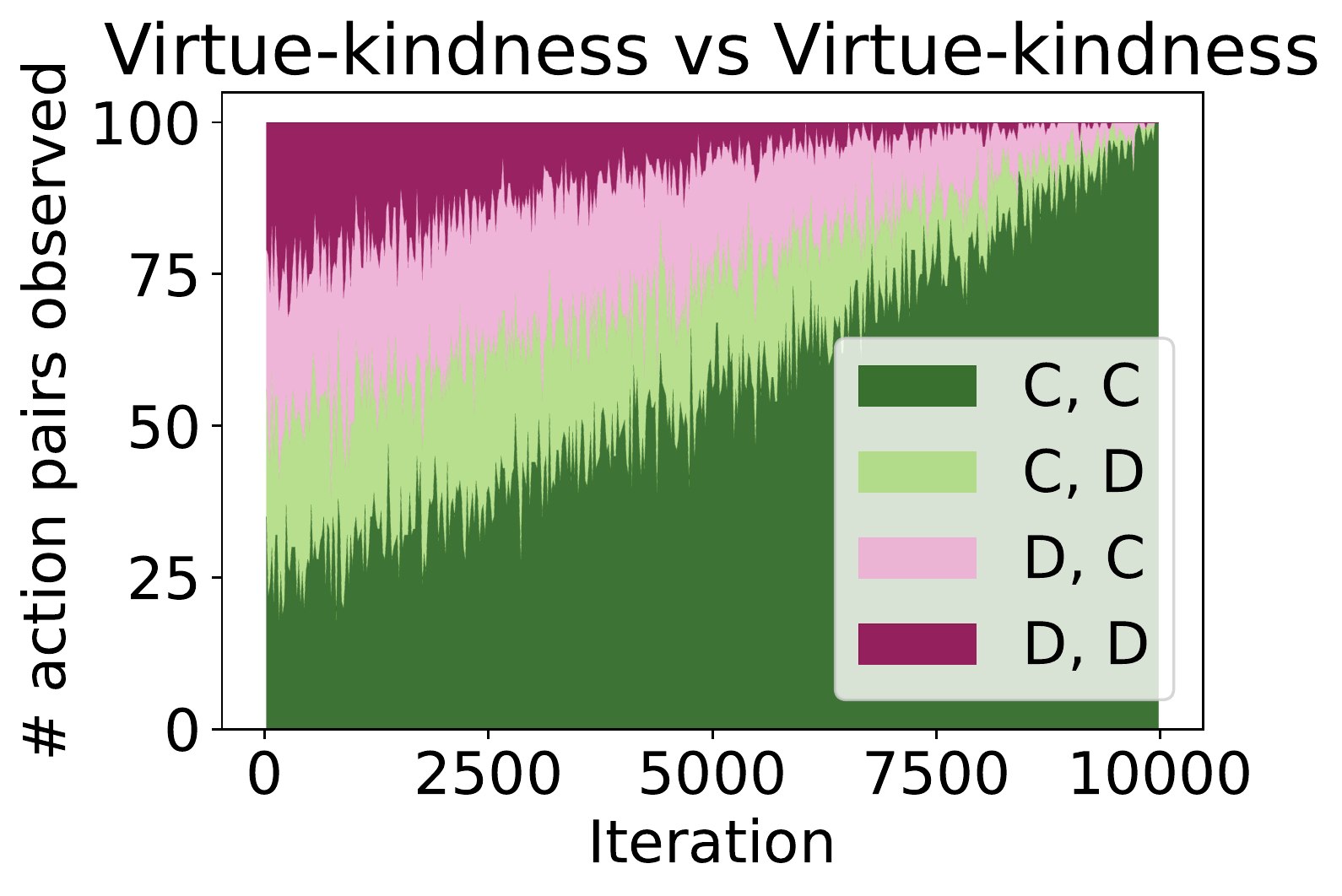}}
&
\\
\makecell[cc]{\rotatebox[origin=c]{90}{ Virtue-mix. }} &
\subt{\includegraphics[width=22mm]{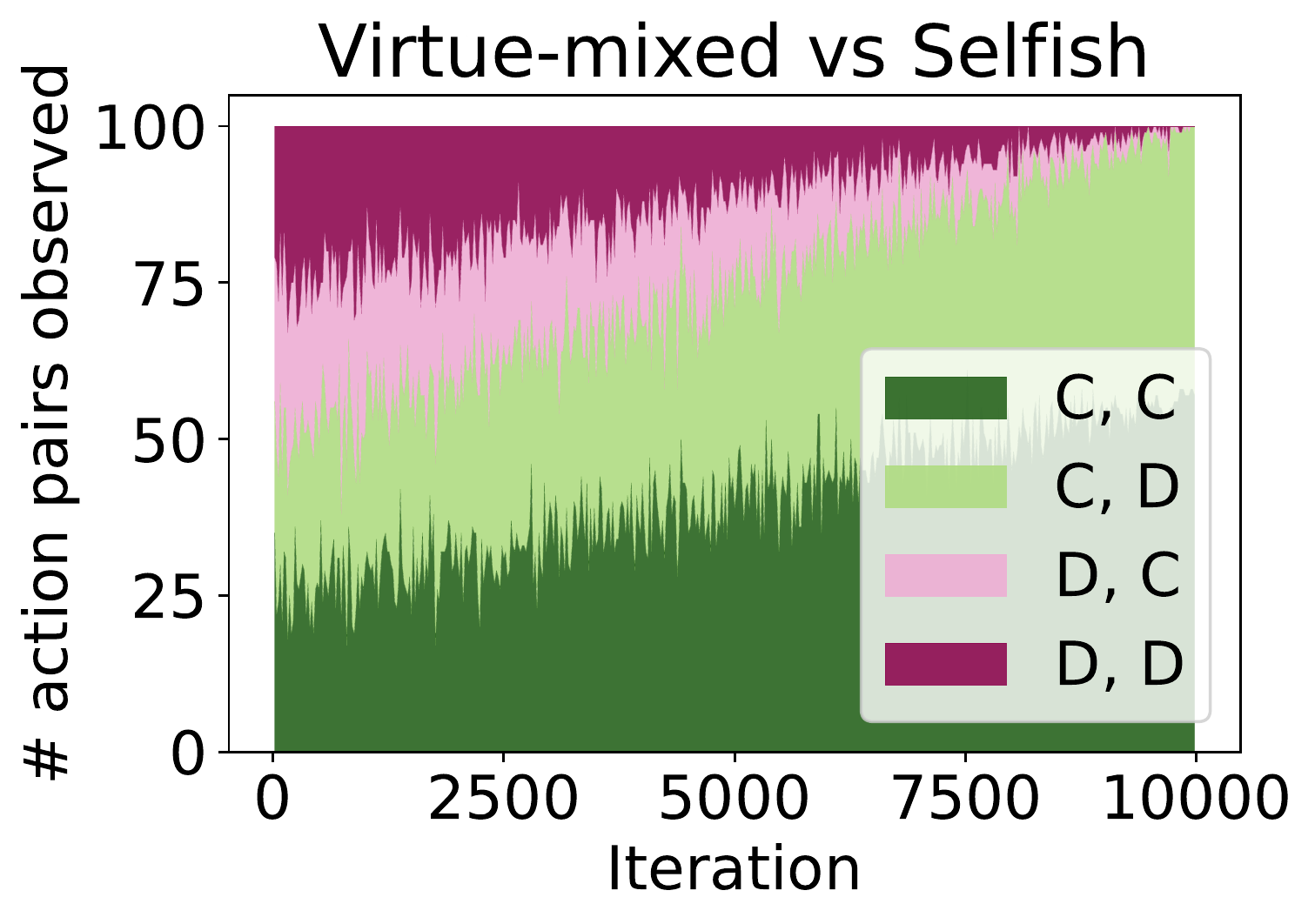}}
&\subt{\includegraphics[width=22mm]{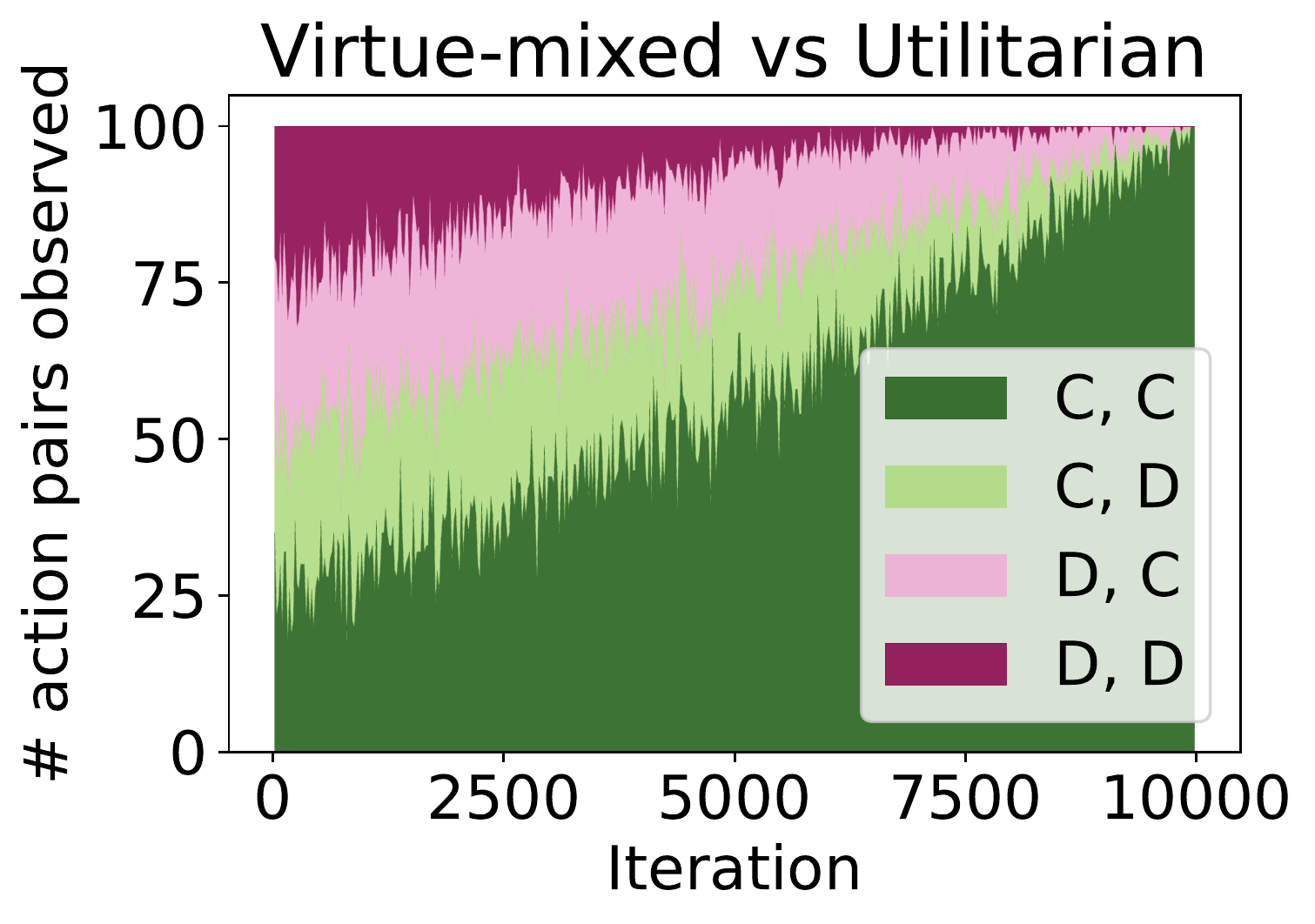}}
&\subt{\includegraphics[width=22mm]{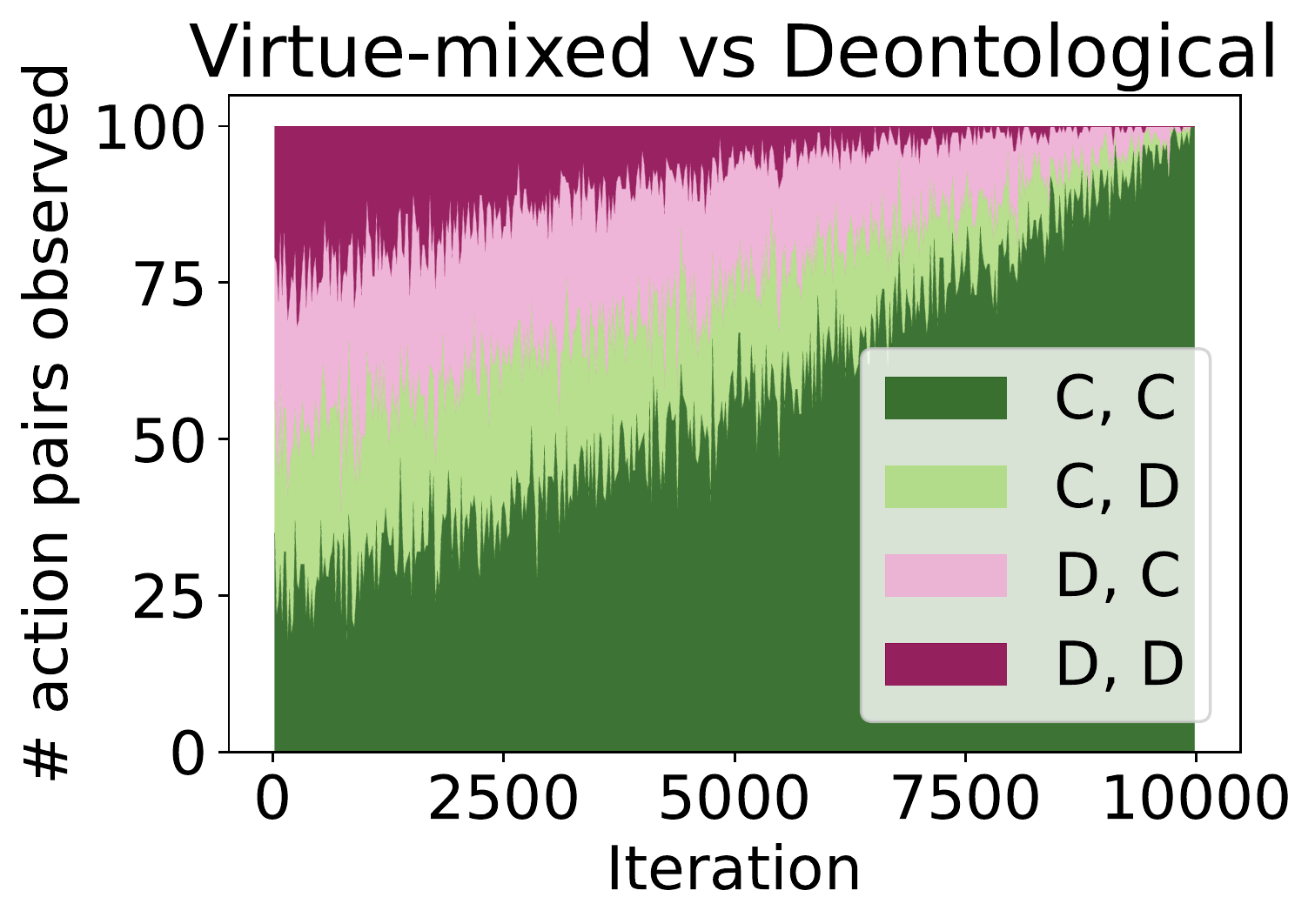}}
&\subt{\includegraphics[width=22mm]{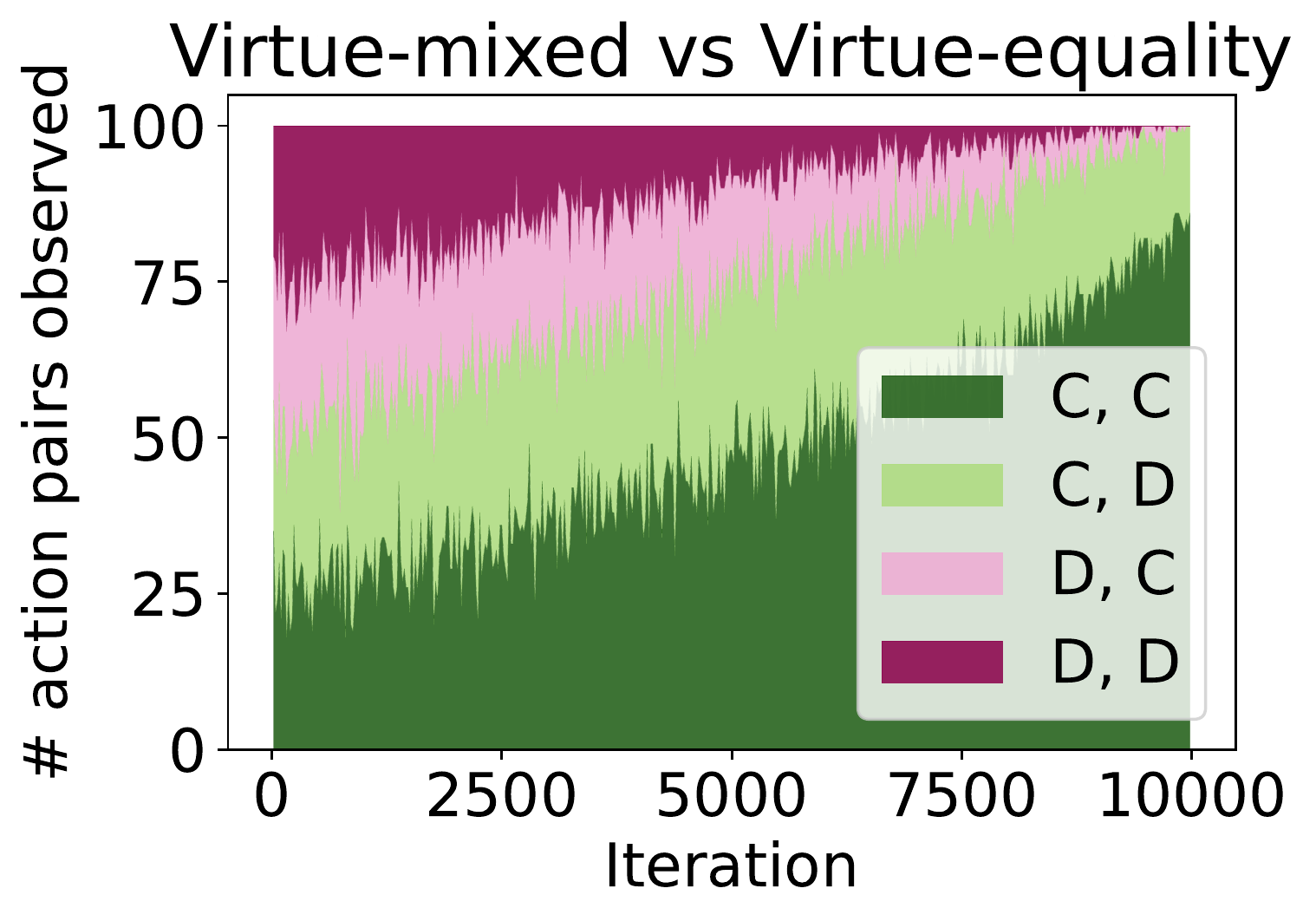}}
&\subt{\includegraphics[width=22mm]{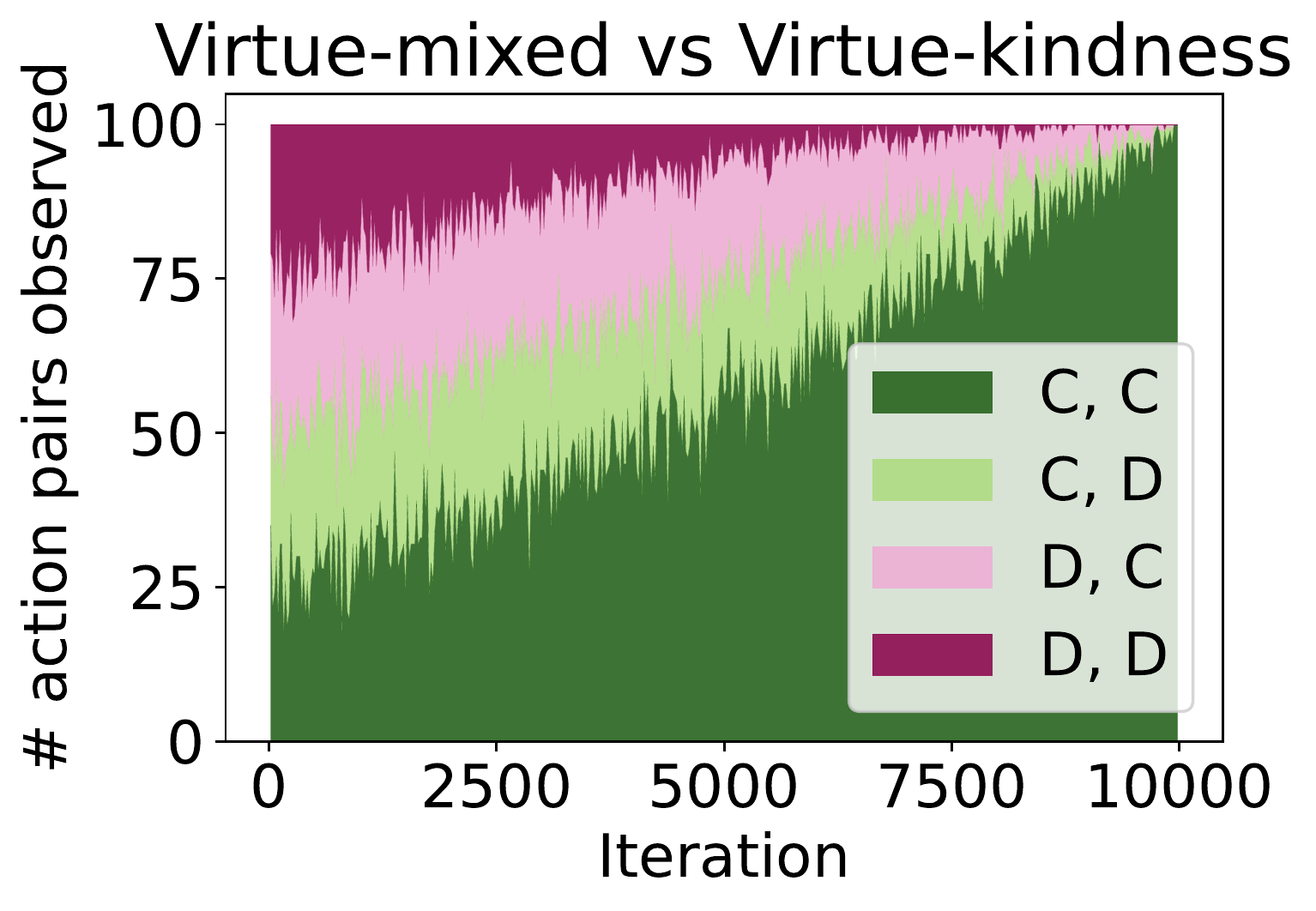}}
&\subt{\includegraphics[width=22mm]{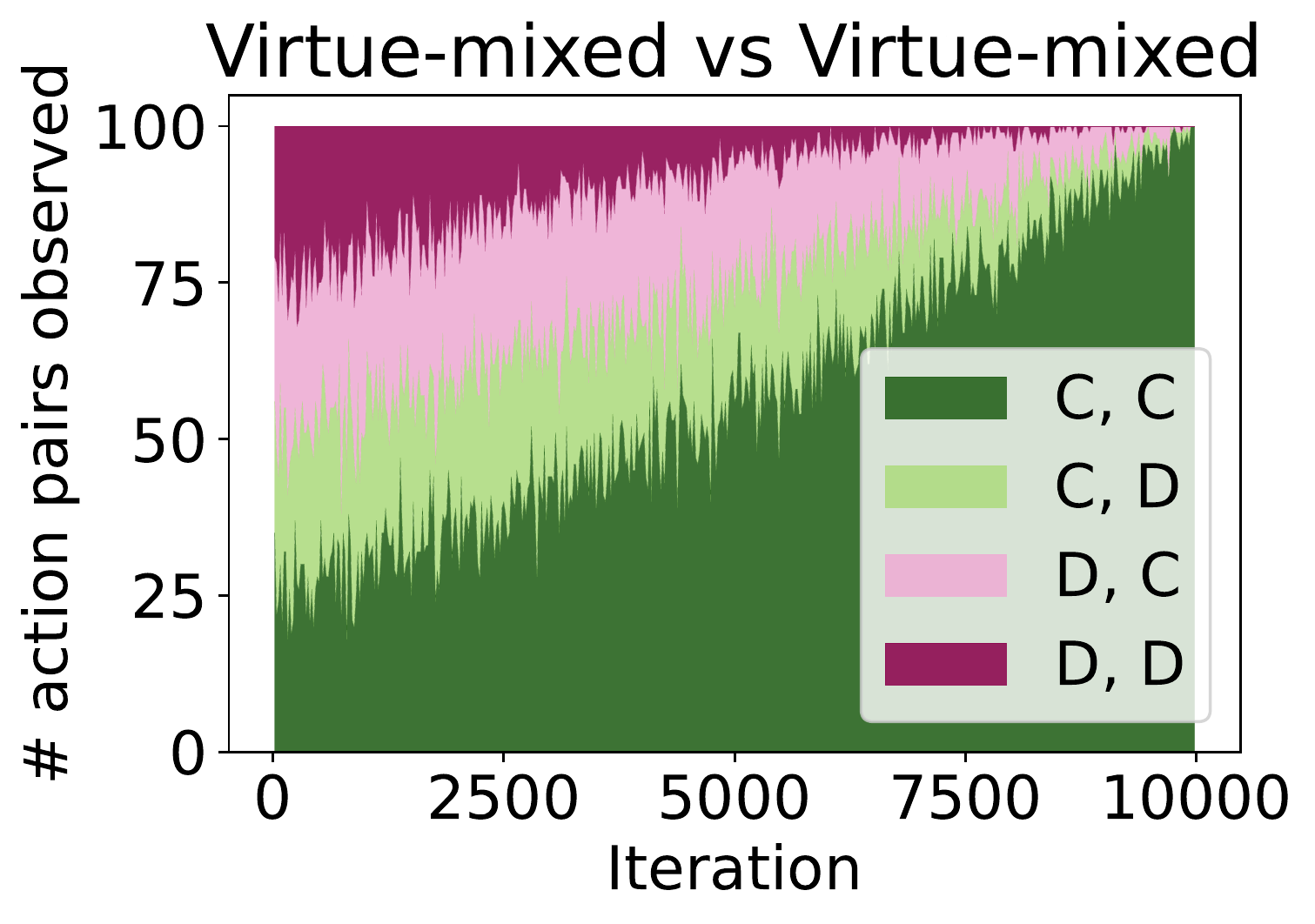}}
\\
\bottomrule
\end{tabular}
\caption{Iterated Stag Hunt game. Simultaneous pairs of actions observed over time. Learning player $M$ (row) vs. learning opponent $O$ (column).}
\label{fig:action_pairs_learning_STH}
\end{figure*}

In the paper we present simultaneous actions played on the final iteration, after the learning period of 10000 iterations is complete, and when is there is no longer any exploration ($\epsilon=0$). In Figures \ref{fig:action_pairs_learning_IPD}-\ref{fig:action_pairs_learning_STH} we present the simultaneous actions played over time, to show the dynamics of learning for every pair of agents. 

\section{Learning Against Static (Baseline) Agents - Results}
\subsection{Simultaneous Pairs of Actions over Time - Learning Player vs Static Opponent}

\begin{figure*}[!h]
\centering
\begin{tabular}{|c|cccc}
\toprule
 & AC & AD & TFT & Random \\
\midrule
\makecell[cc]{\rotatebox[origin=c]{90}{Selfish}} & 
\subt{\includegraphics[width=35mm]{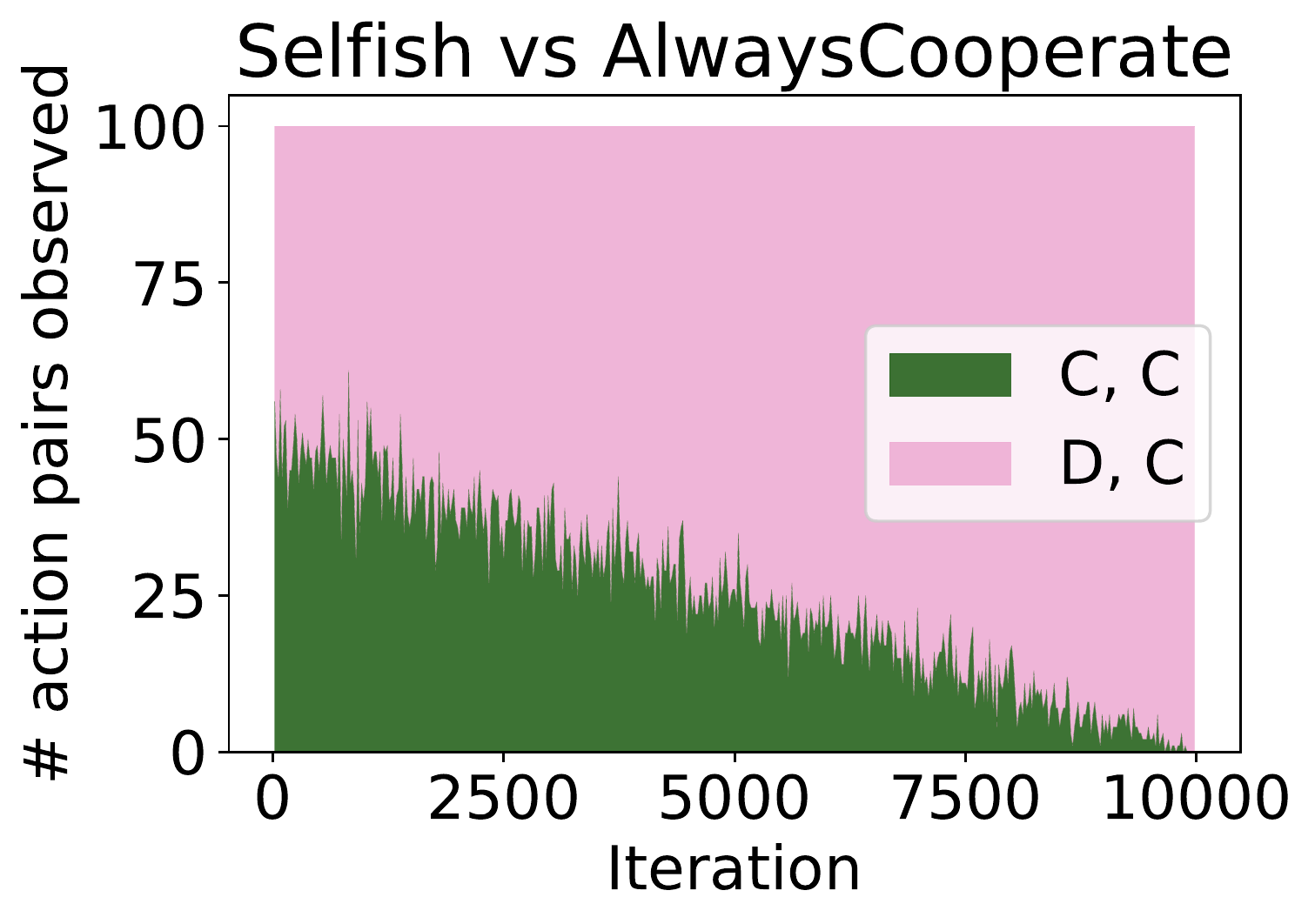}}
&\subt{\includegraphics[width=35mm]{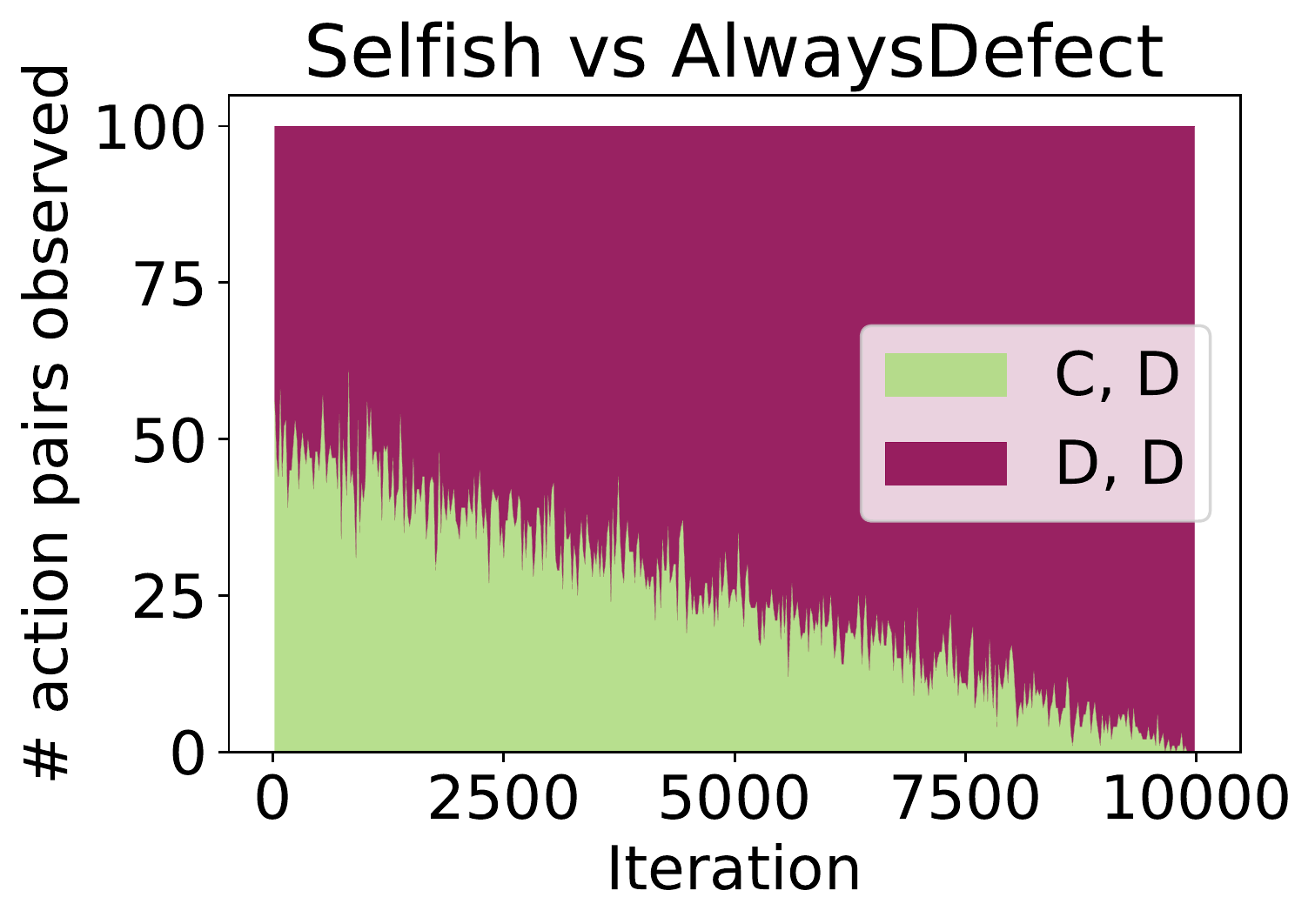}}
&\subt{\includegraphics[width=35mm]{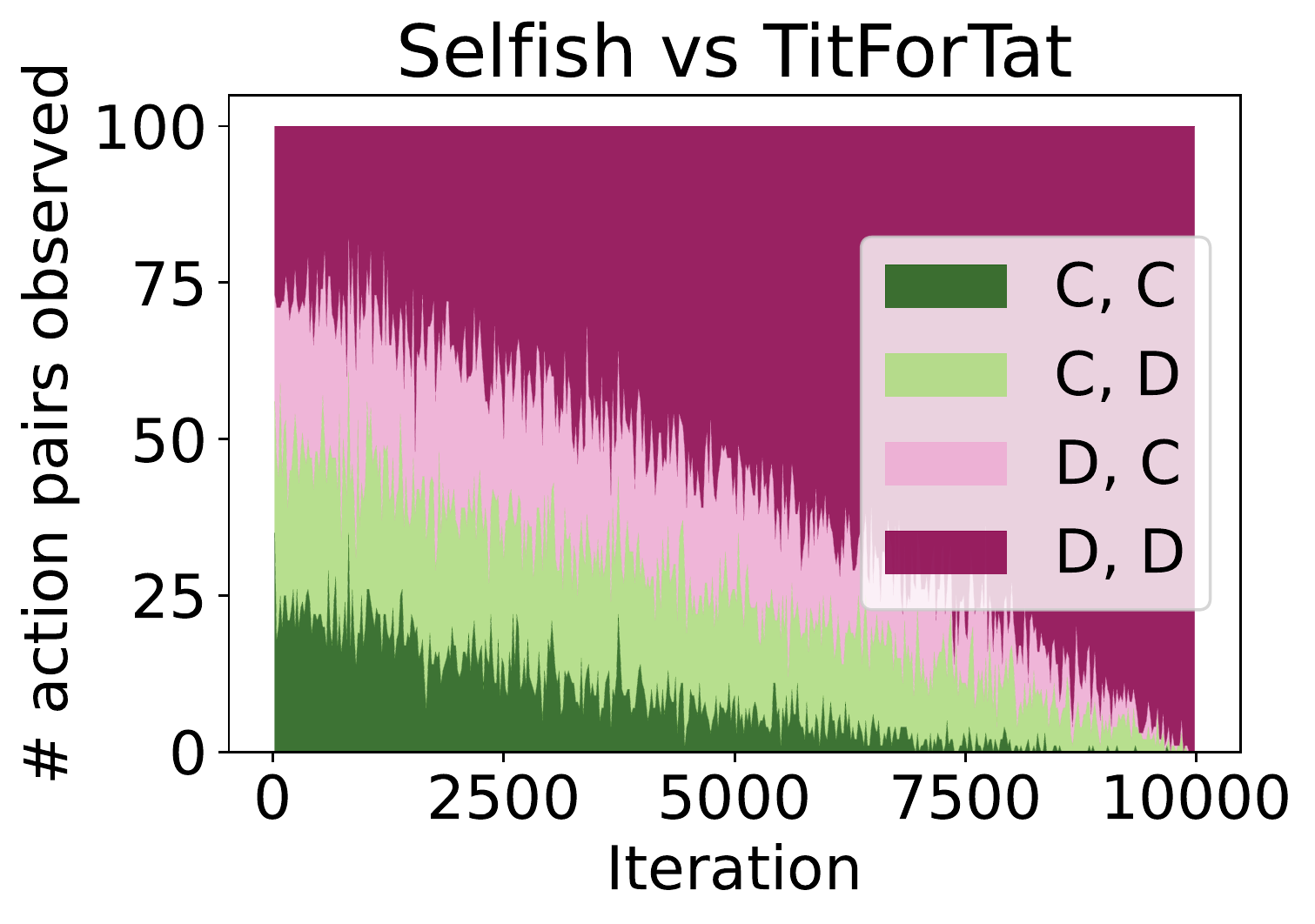}}
&\subt{\includegraphics[width=35mm]{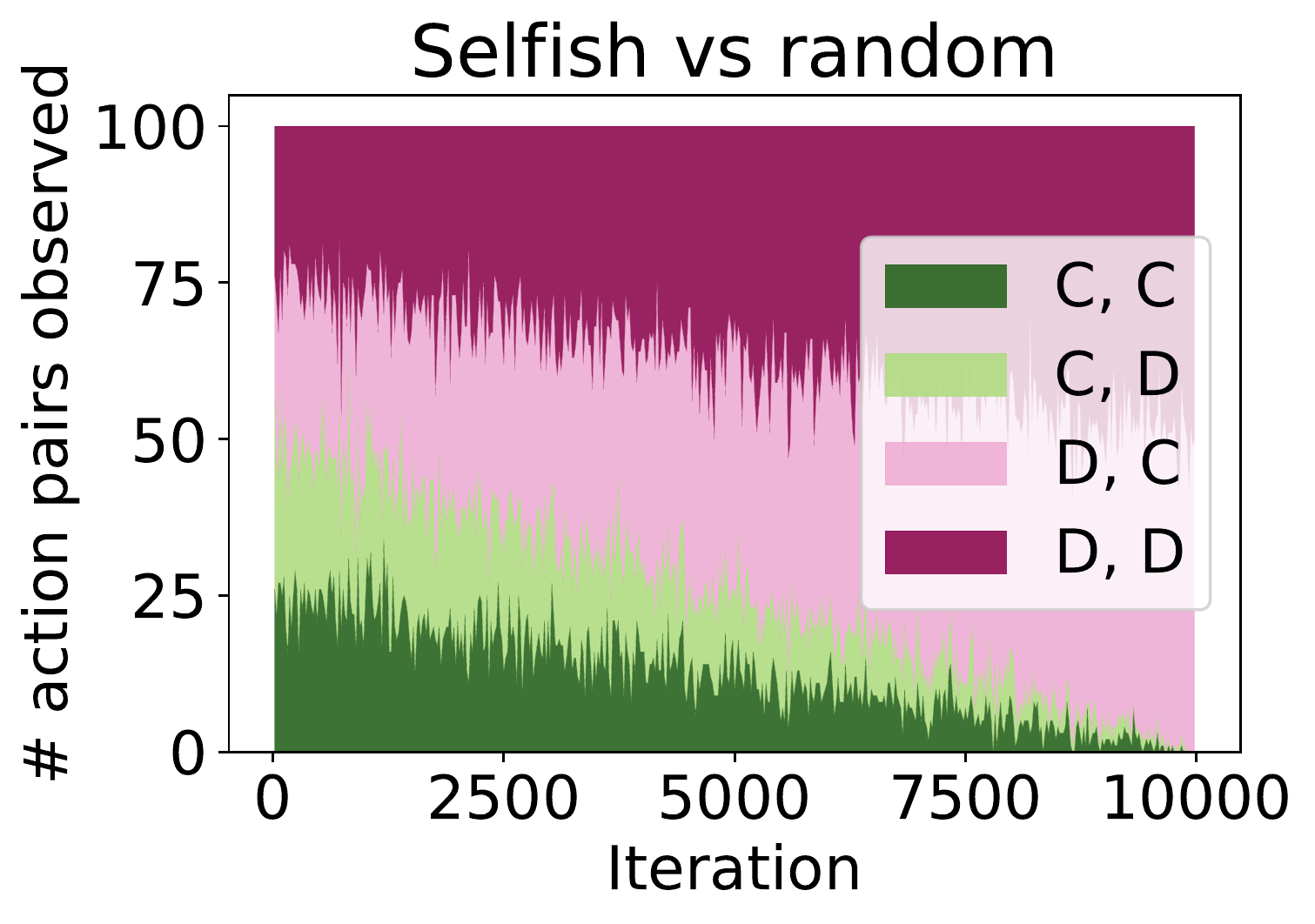}}
\\
\makecell[cc]{\rotatebox[origin=c]{90}{ Utilitarian }} & 
\subt{\includegraphics[width=35mm]{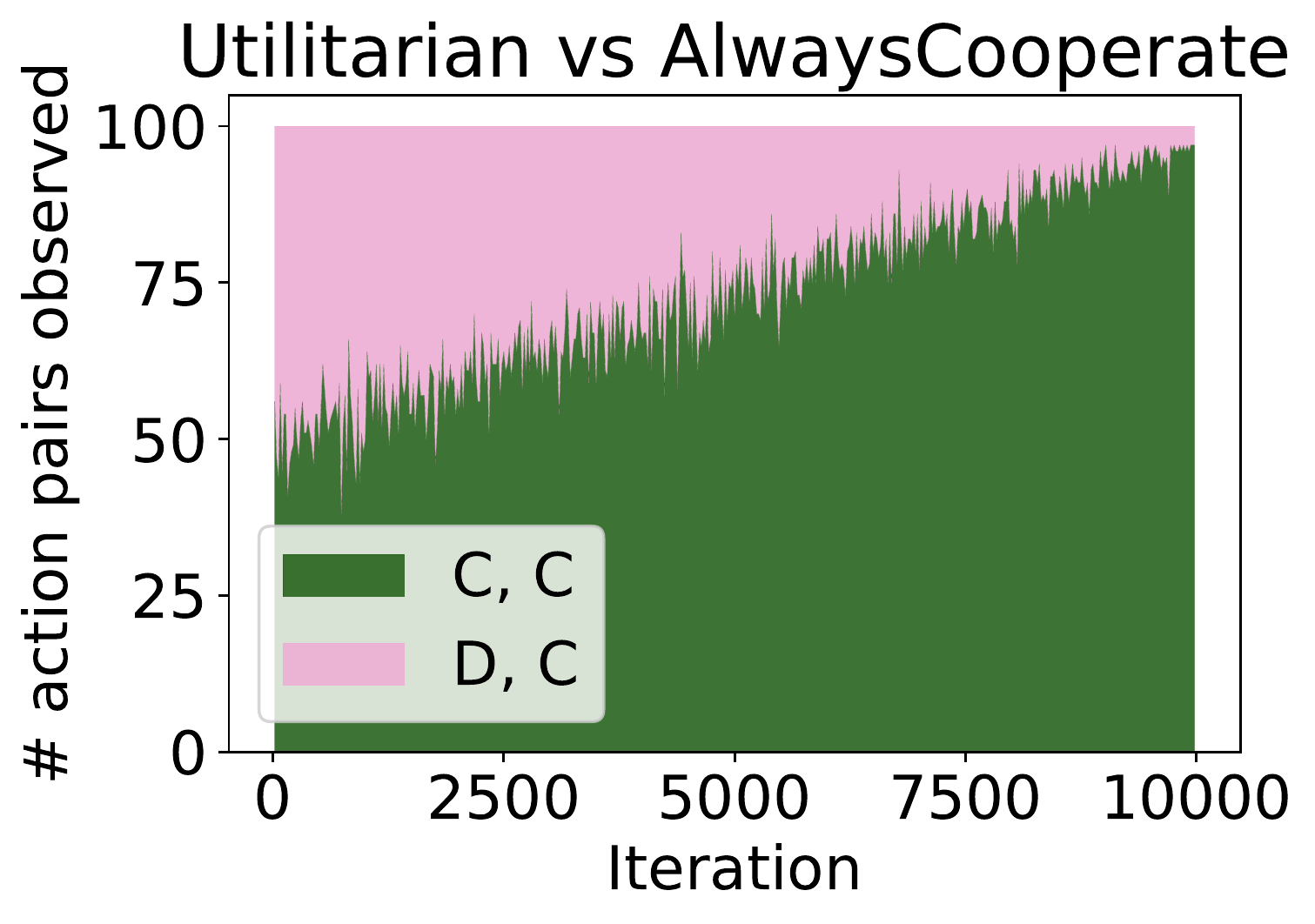}}
&\subt{\includegraphics[width=35mm]{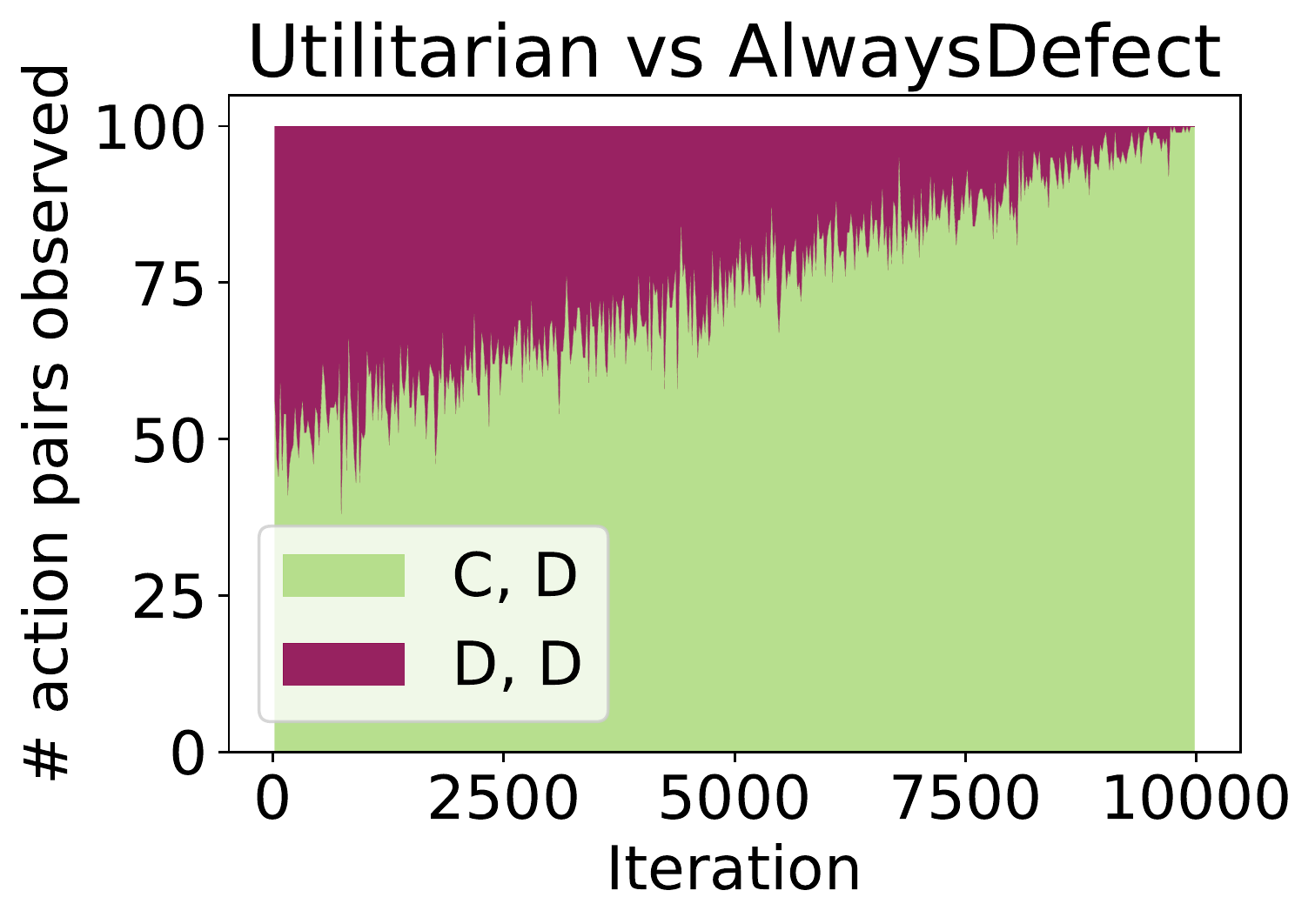}}
&\subt{\includegraphics[width=35mm]{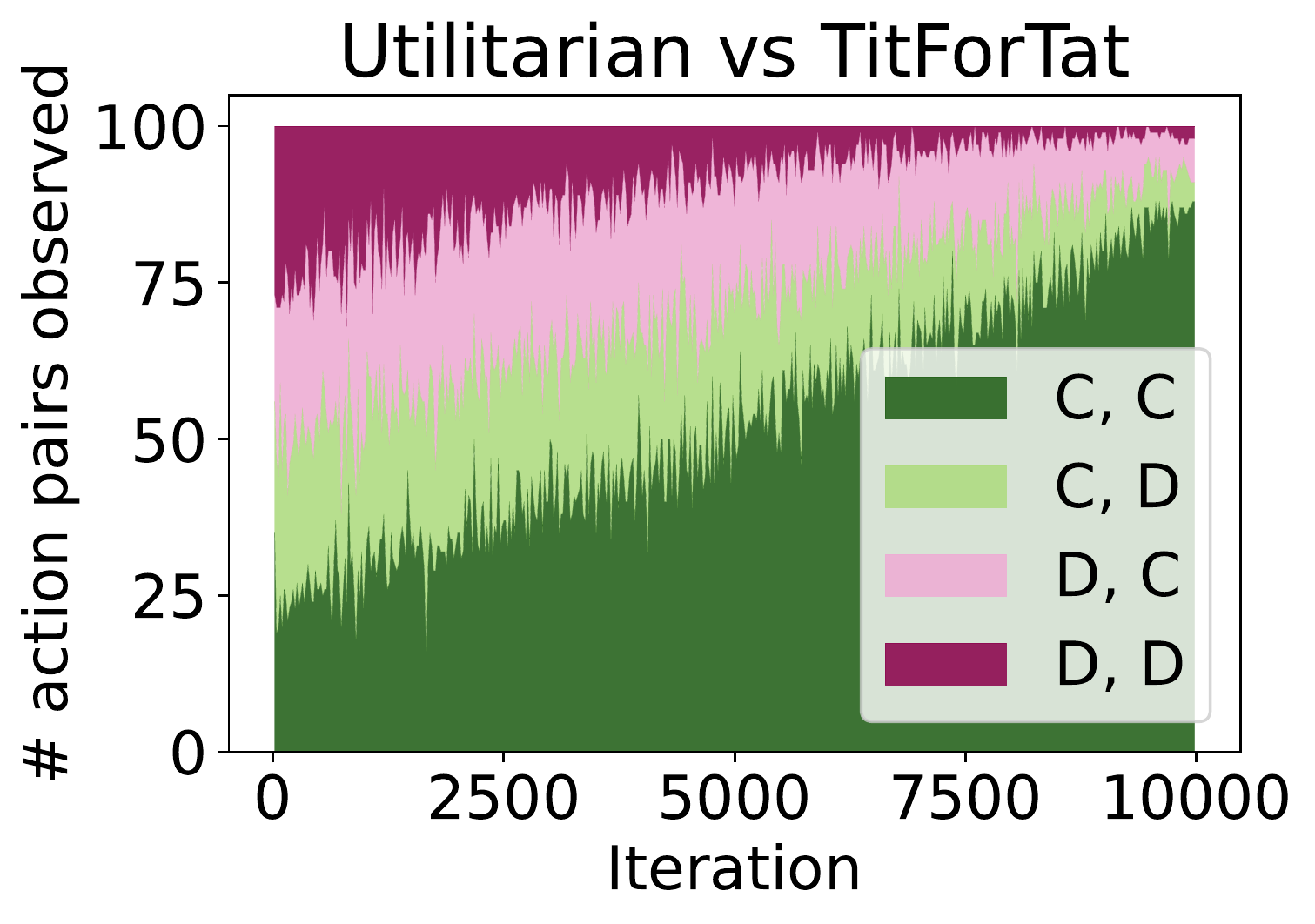}}
&\subt{\includegraphics[width=35mm]{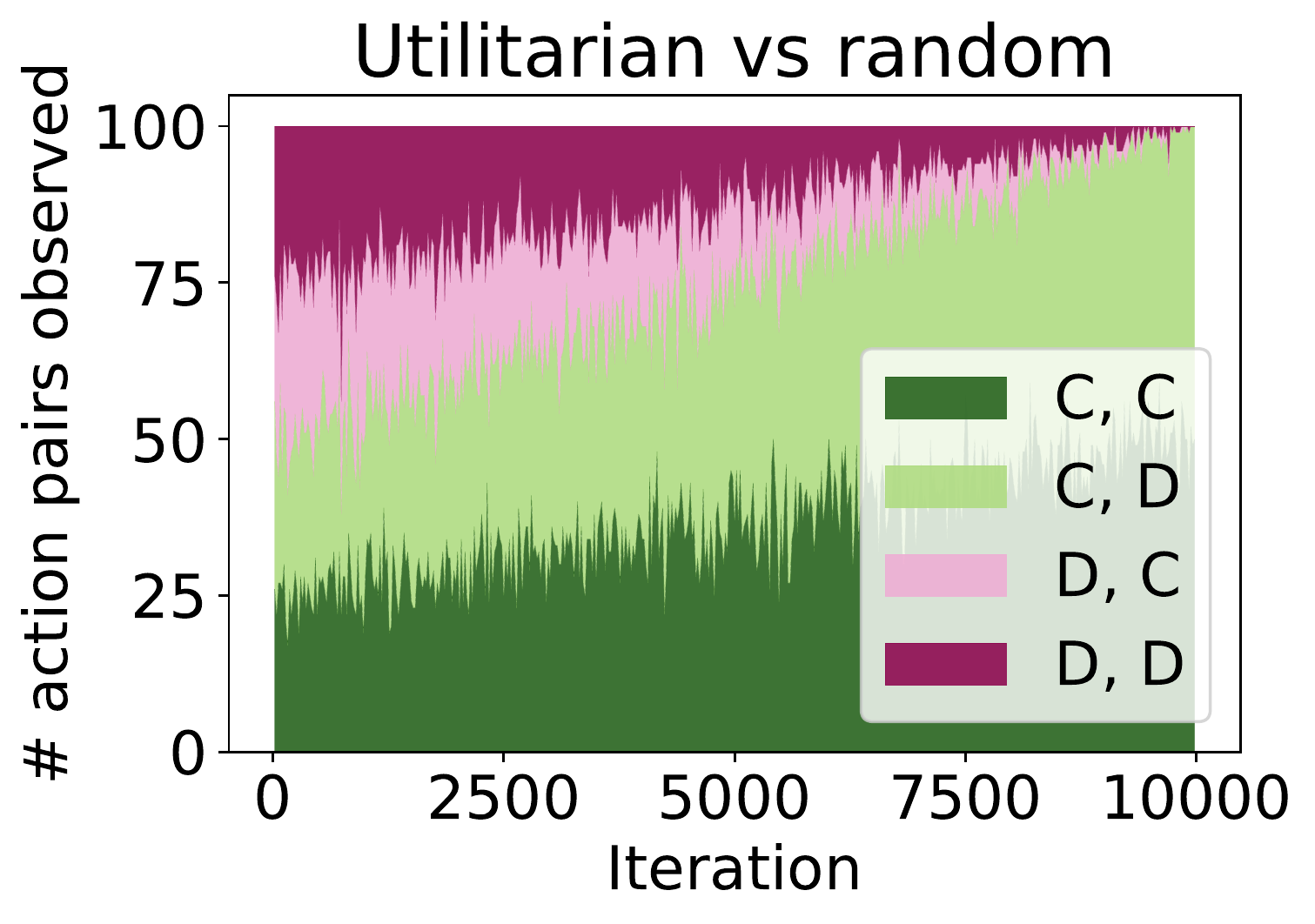}}
\\
\makecell[cc]{\rotatebox[origin=c]{90}{ Deontological }} &
\subt{\includegraphics[width=35mm]{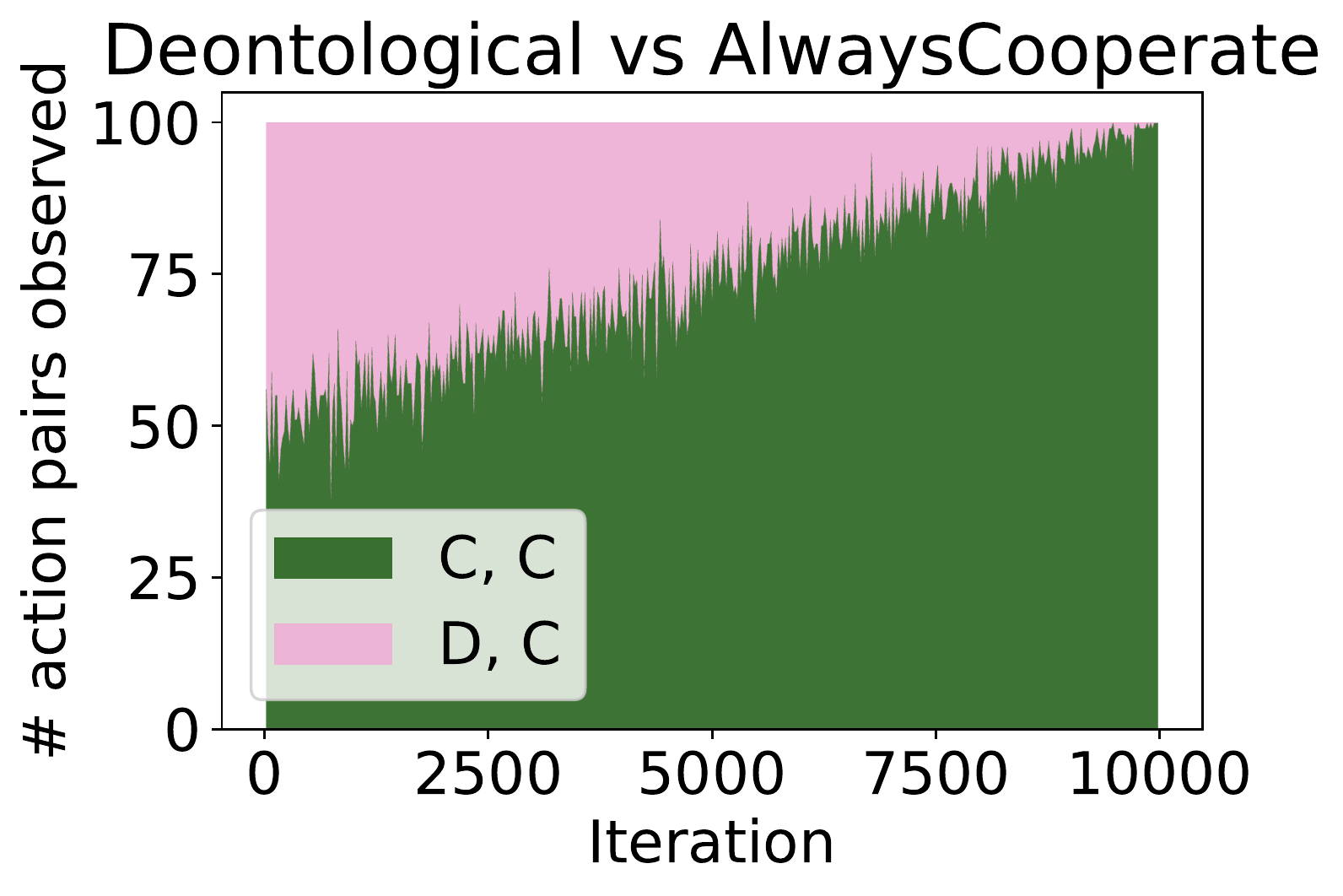}}
&\subt{\includegraphics[width=35mm]{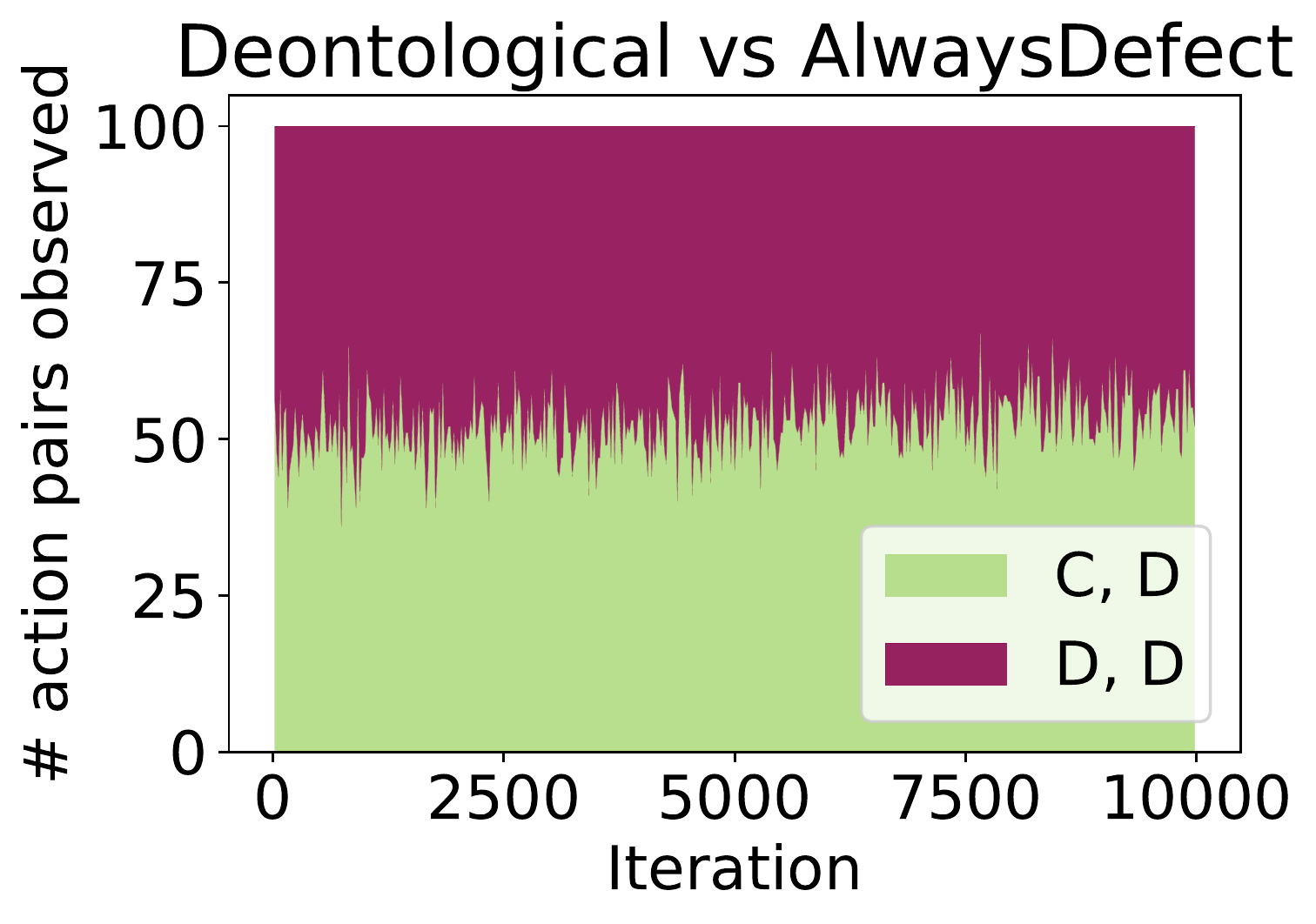}}
&\subt{\includegraphics[width=35mm]{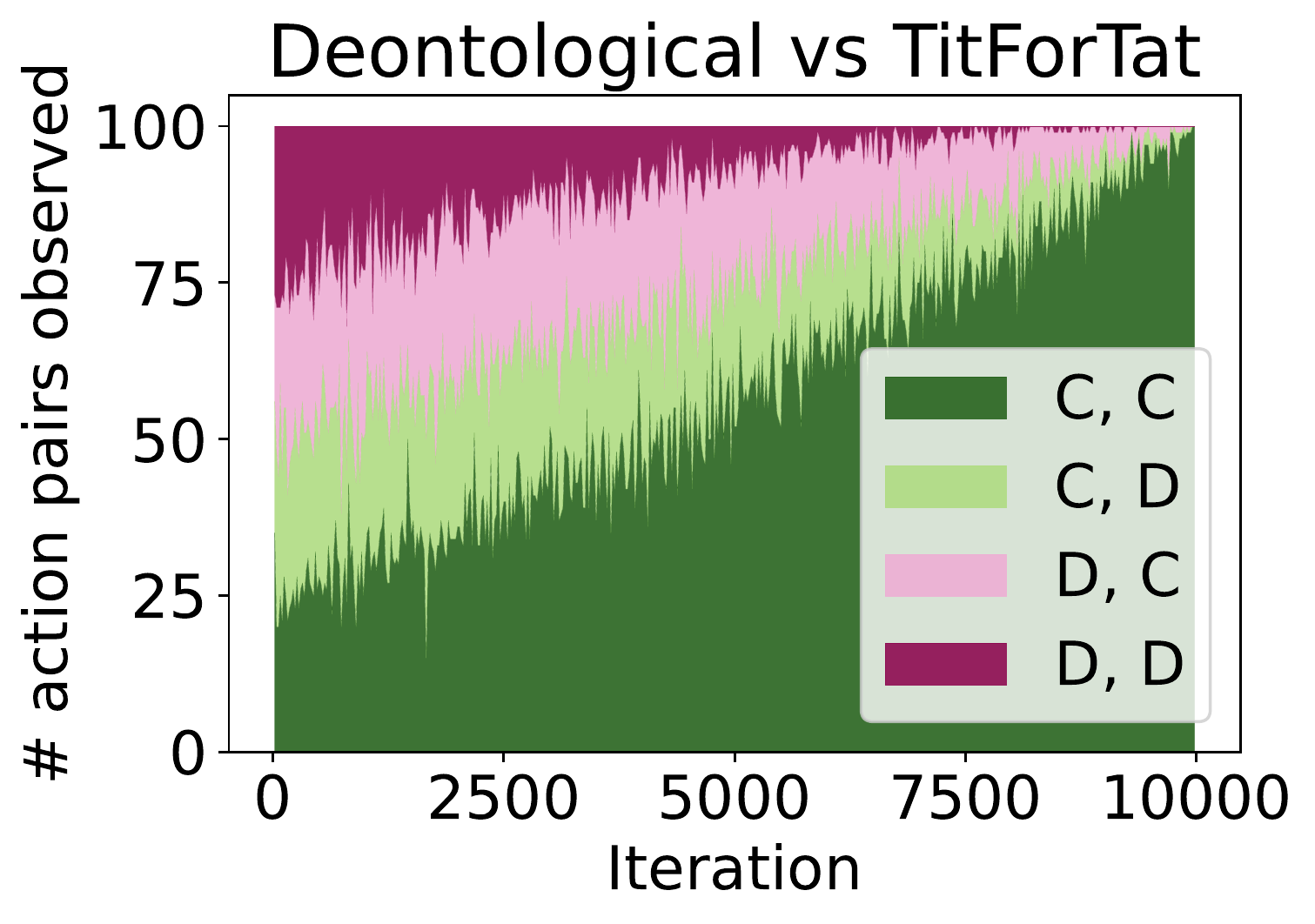}}
&\subt{\includegraphics[width=35mm]{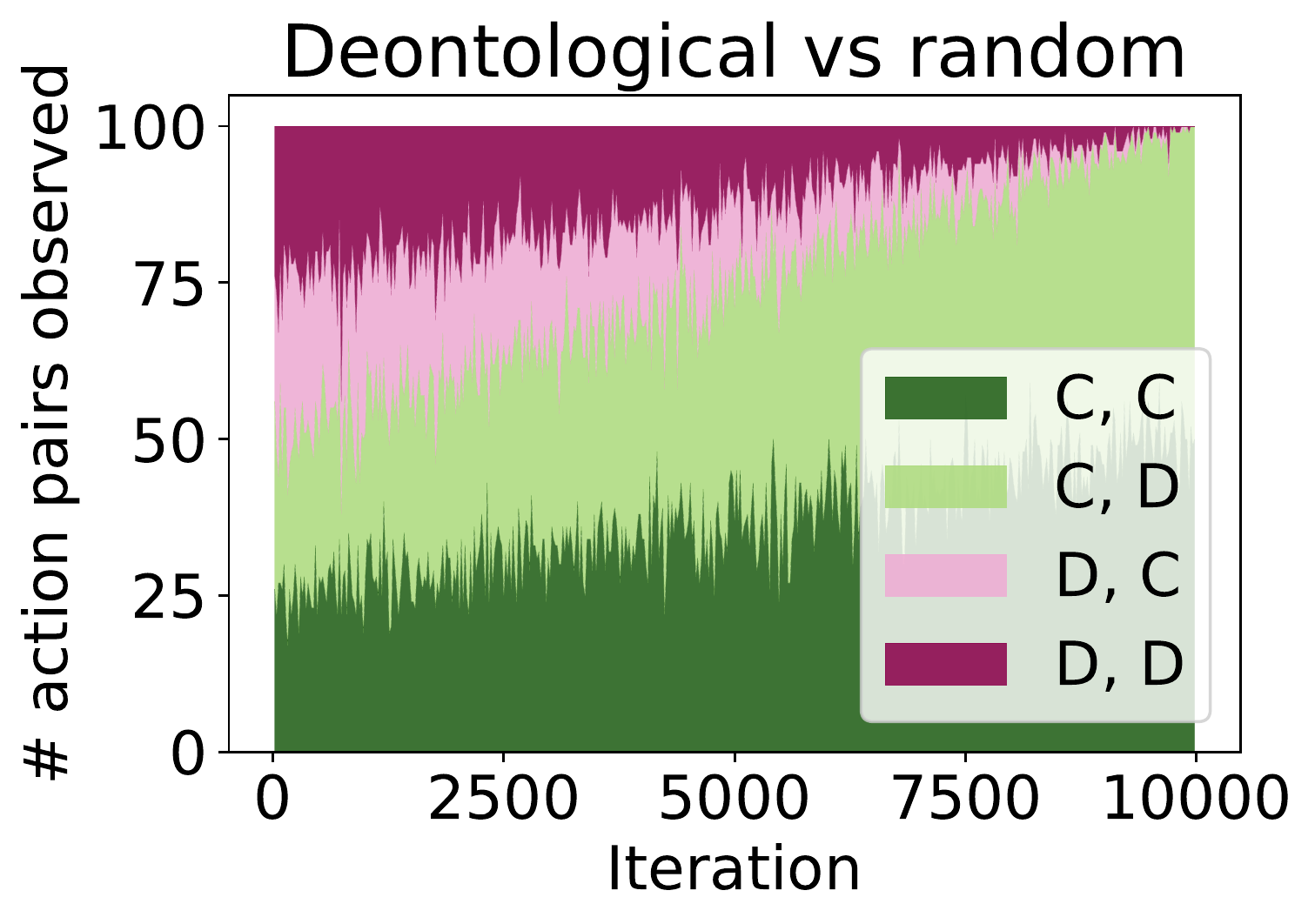}}
\\
\makecell[cc]{\rotatebox[origin=c]{90}{ Virtue-eq. }} &
\subt{\includegraphics[width=35mm]{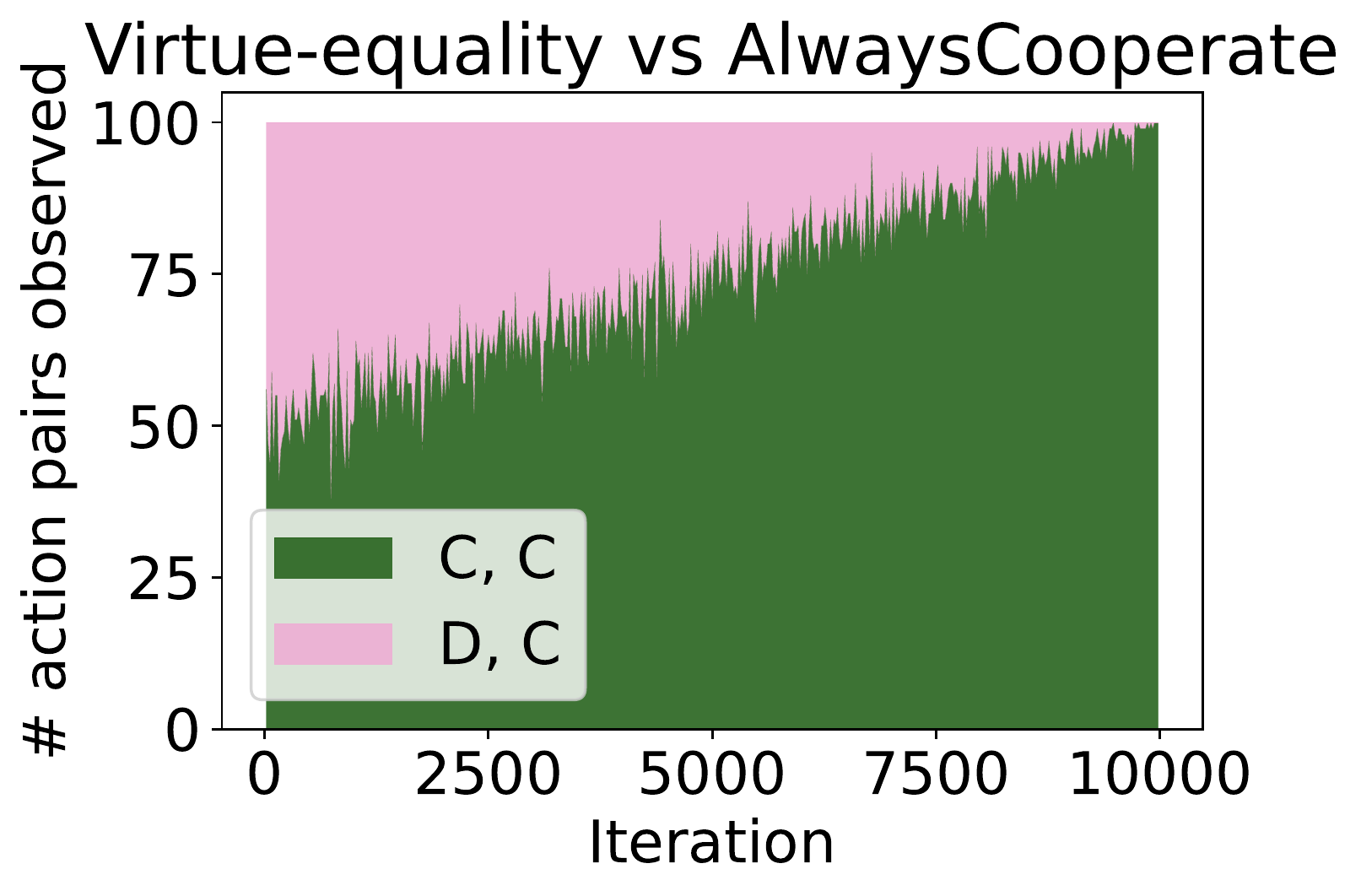}}
&\subt{\includegraphics[width=35mm]{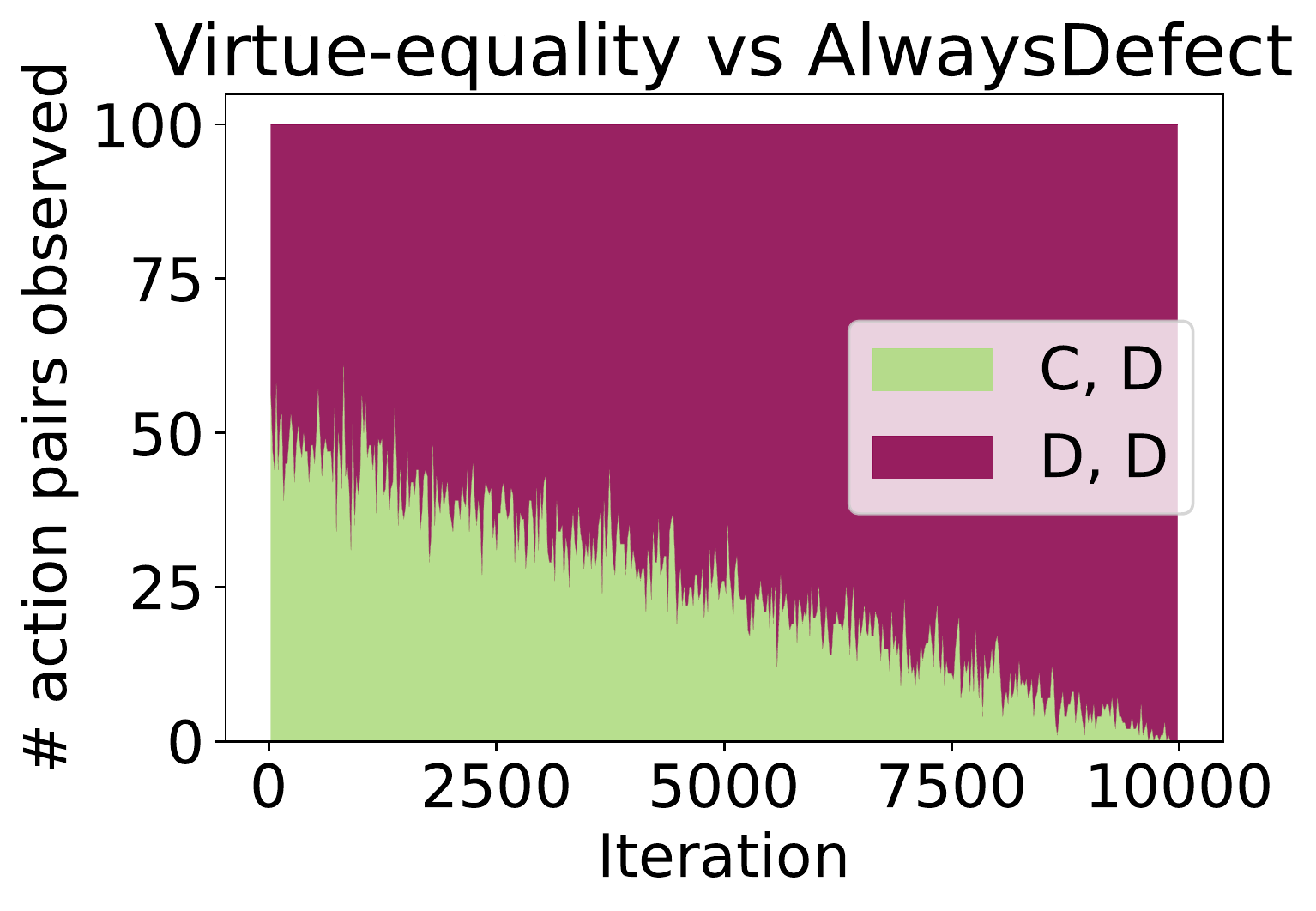}}
&\subt{\includegraphics[width=35mm]{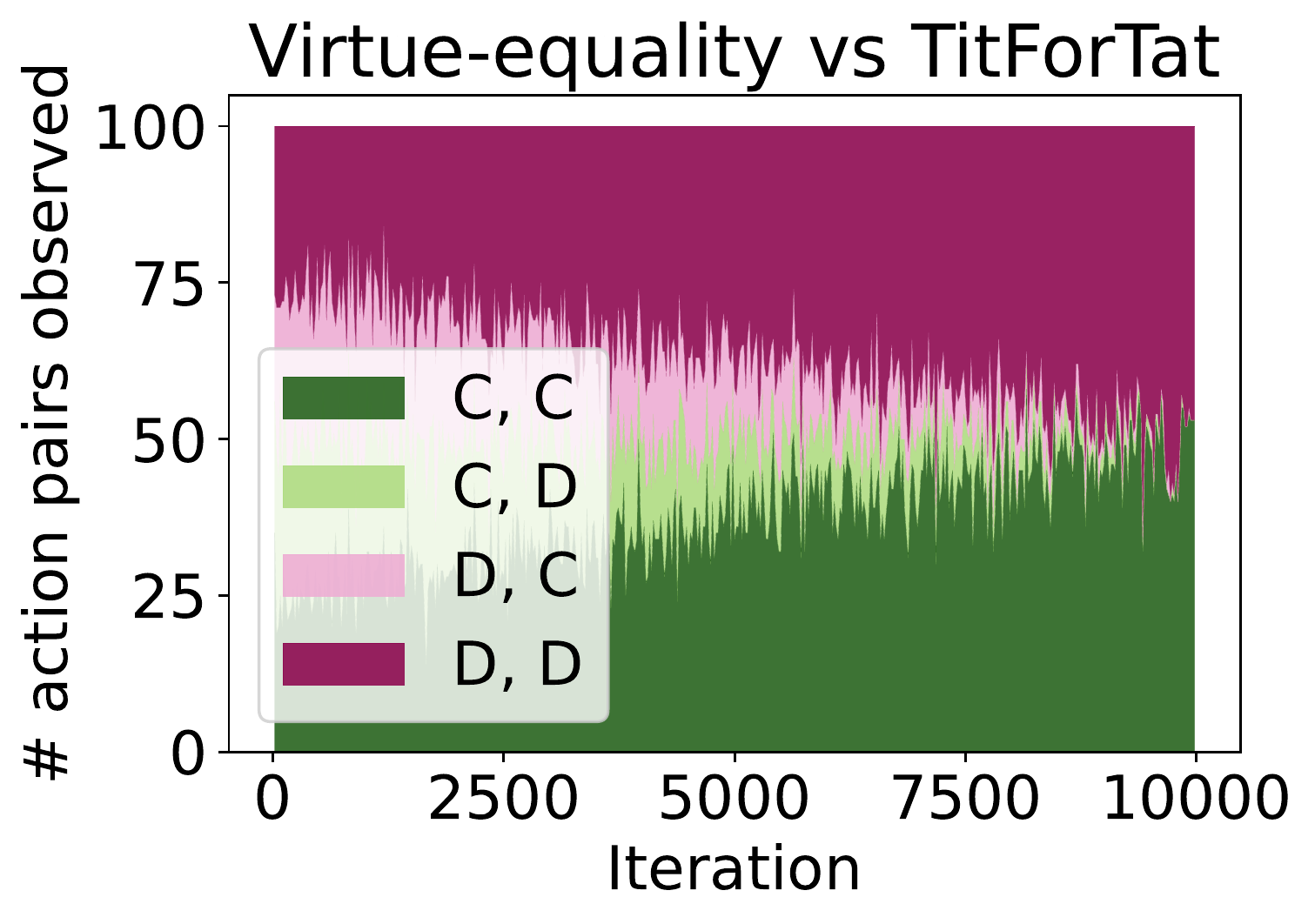}}
&\subt{\includegraphics[width=35mm]{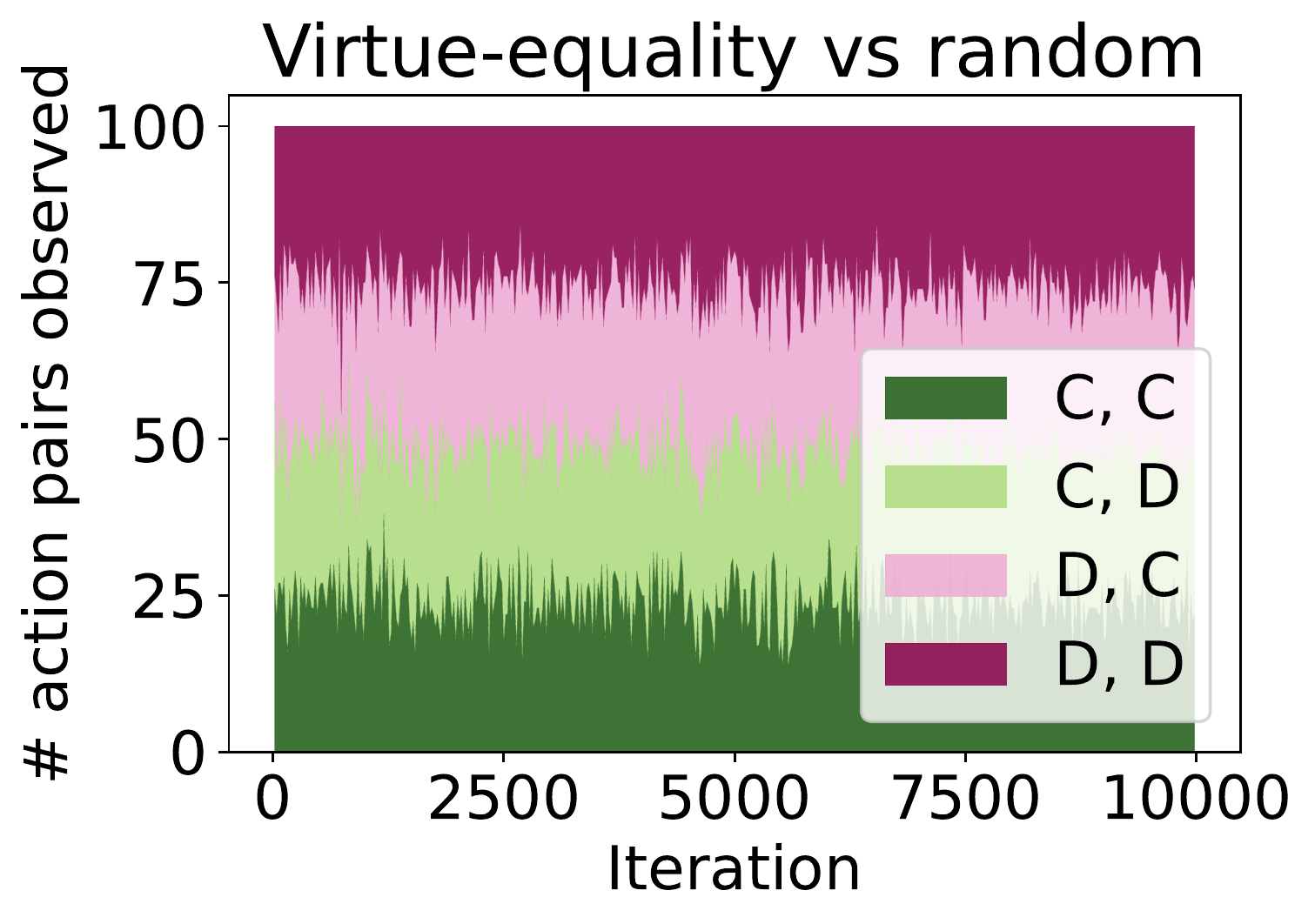}}
\\
\makecell[cc]{\rotatebox[origin=c]{90}{ Virtue-kind. }} &
\subt{\includegraphics[width=35mm]{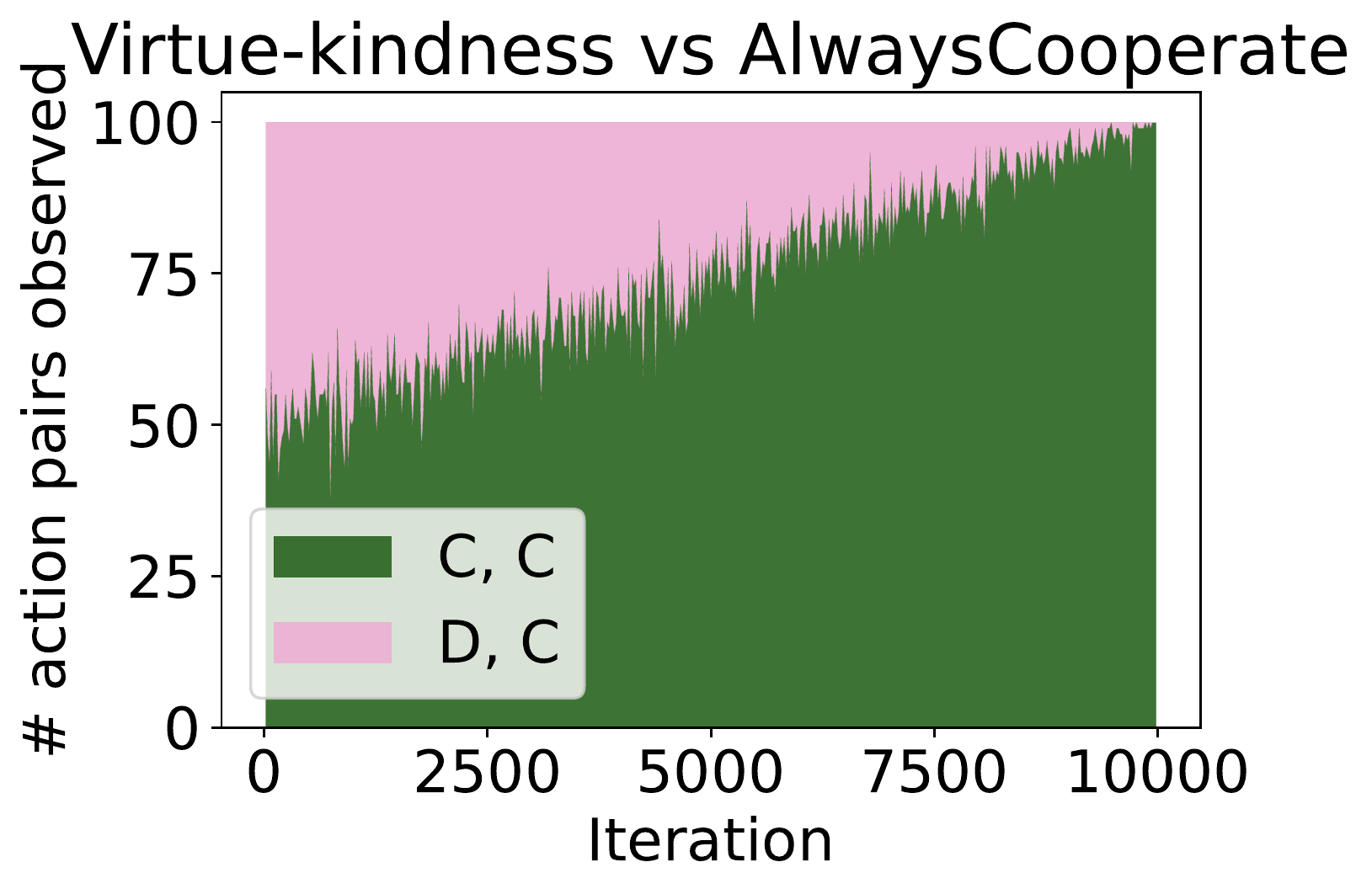}}
&\subt{\includegraphics[width=35mm]{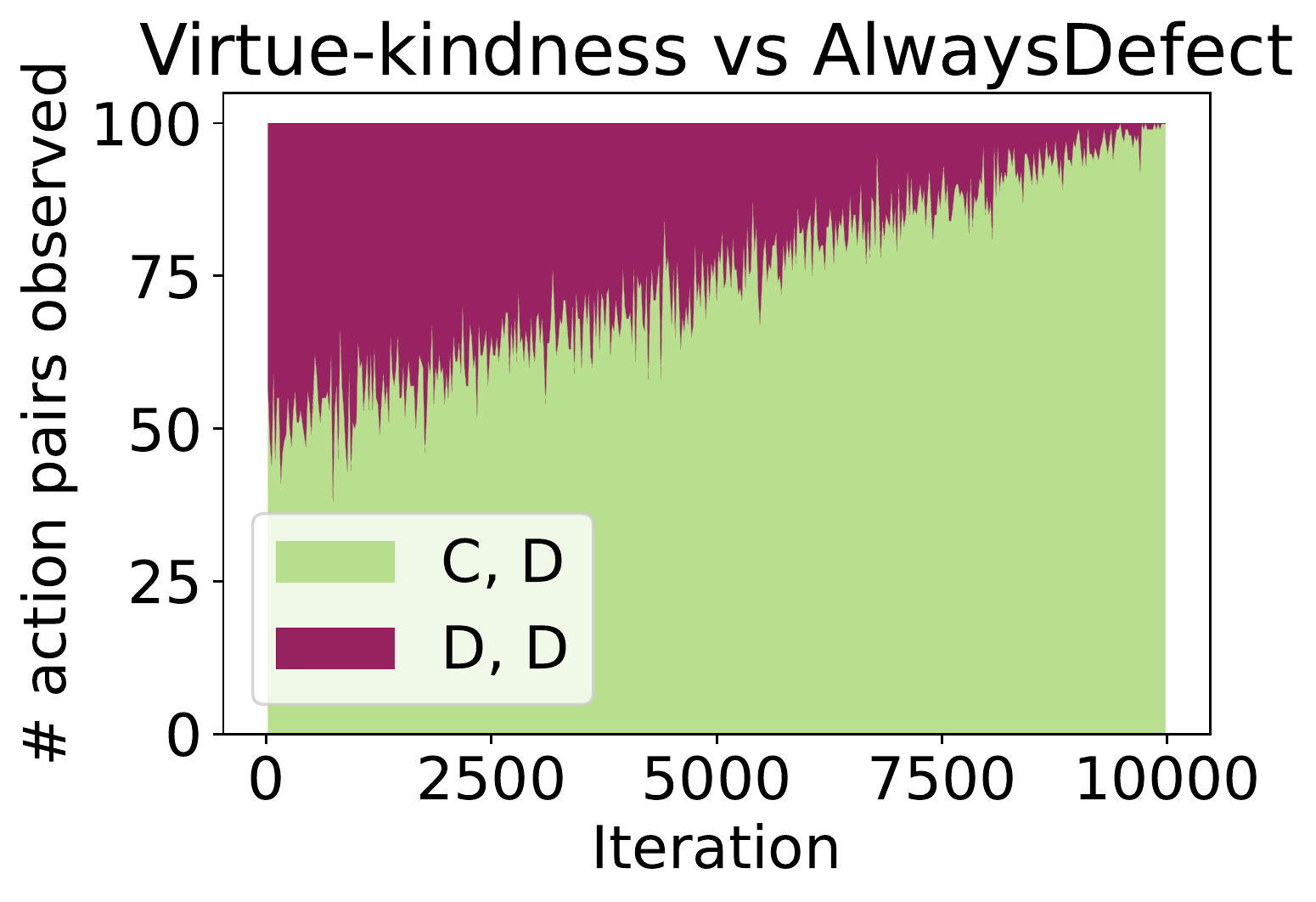}}
&\subt{\includegraphics[width=35mm]{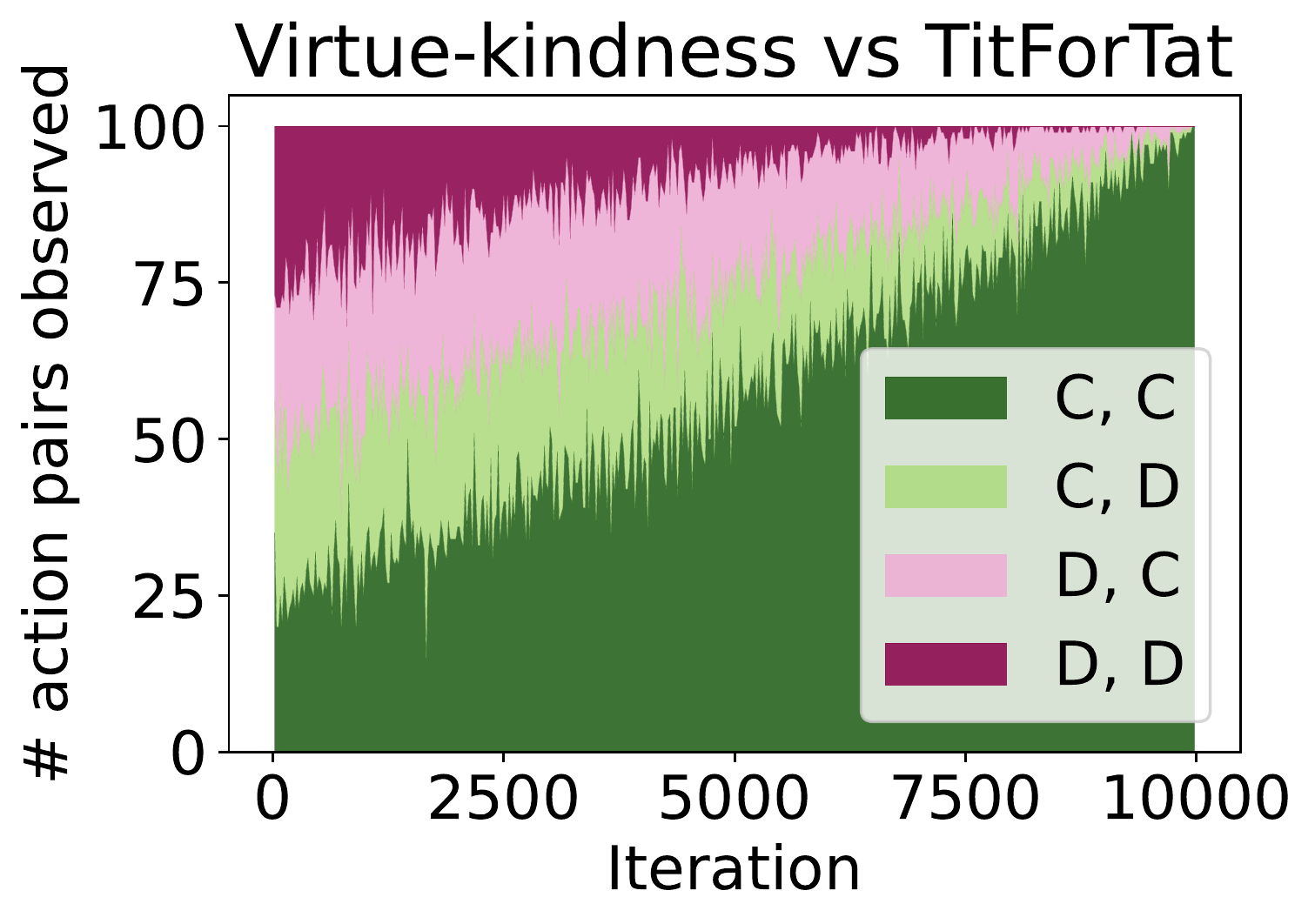}}
&\subt{\includegraphics[width=35mm]{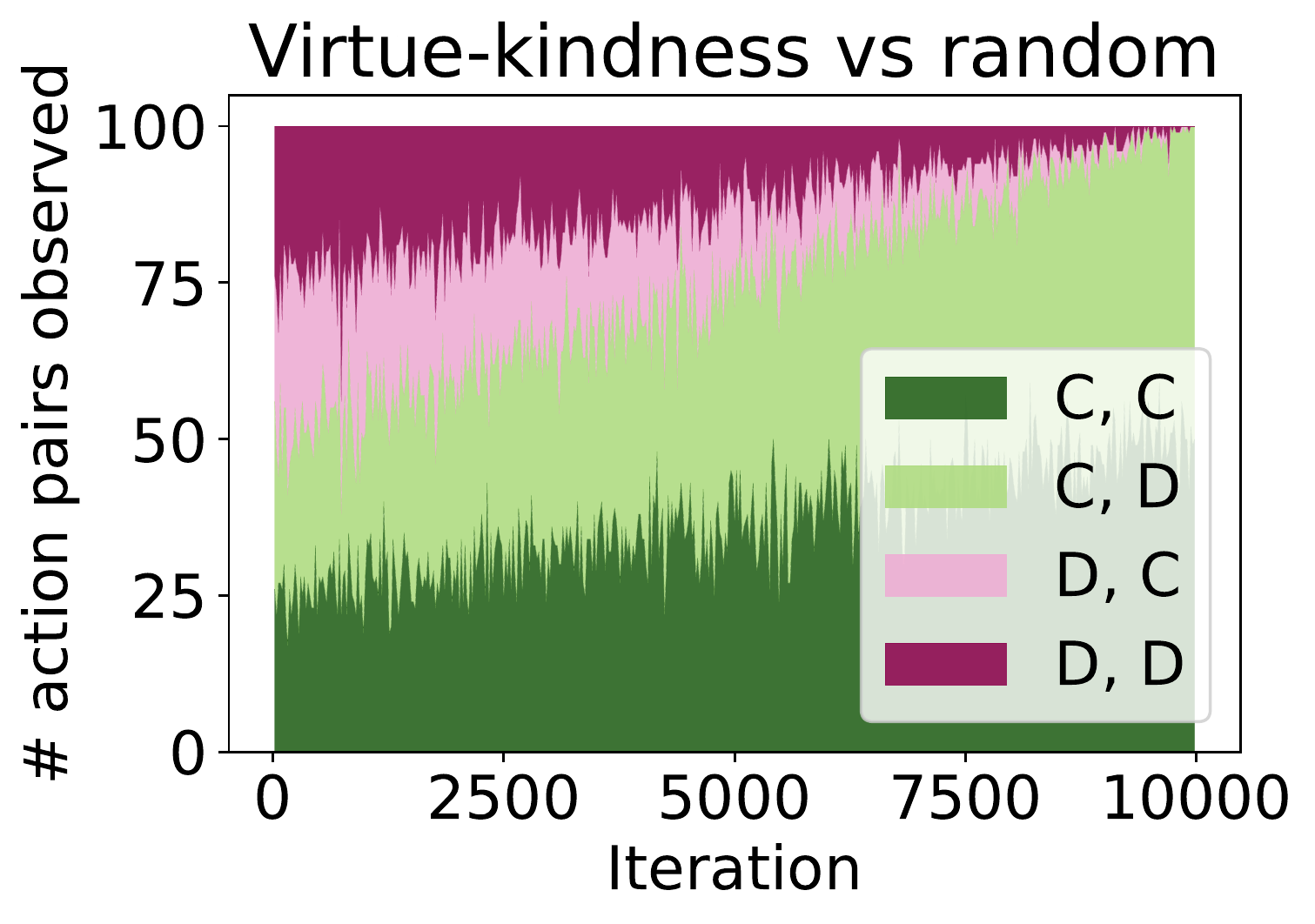}}
\\
\makecell[cc]{\rotatebox[origin=c]{90}{ Virtue-mix. }} &
\subt{\includegraphics[width=35mm]{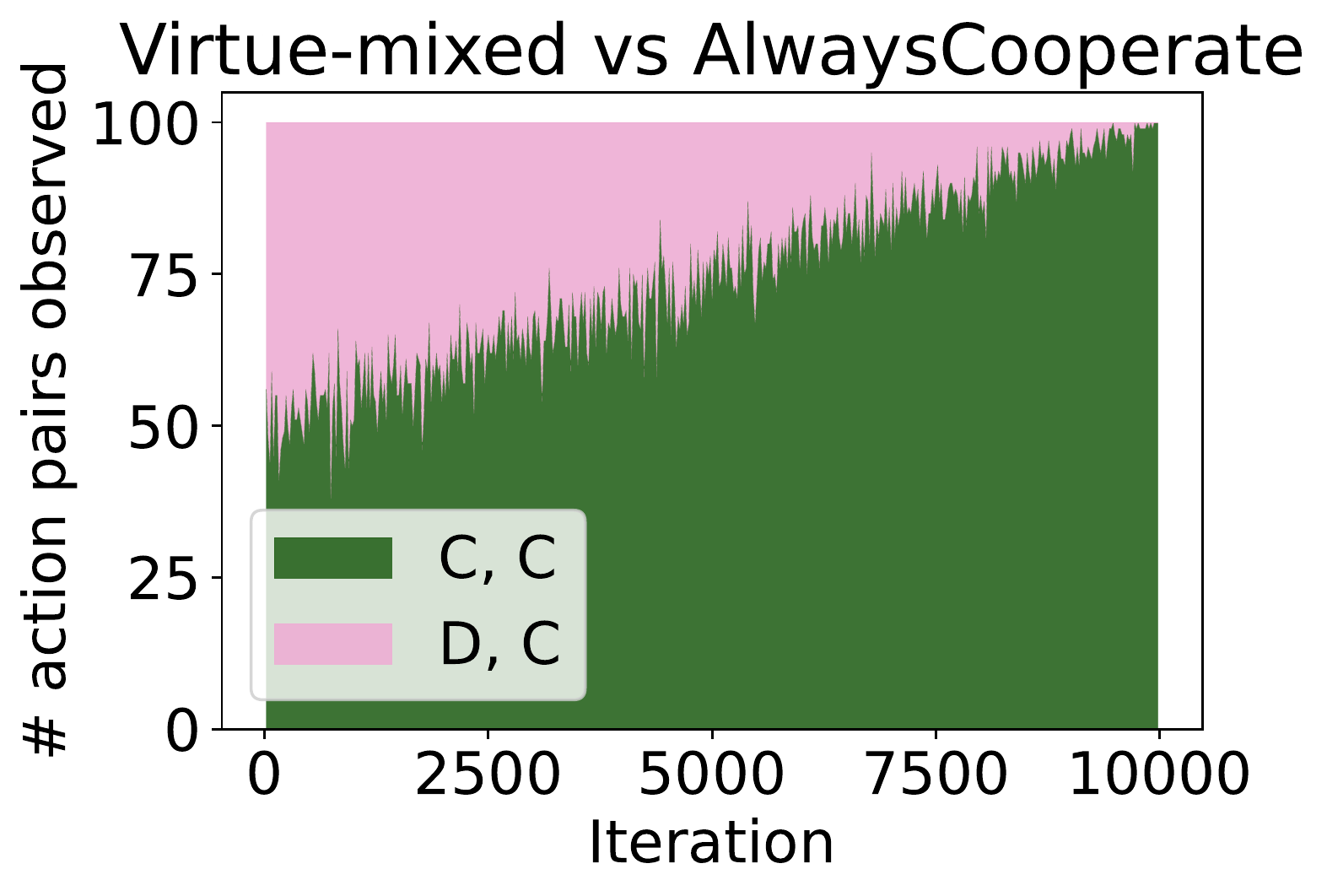}}
&\subt{\includegraphics[width=35mm]{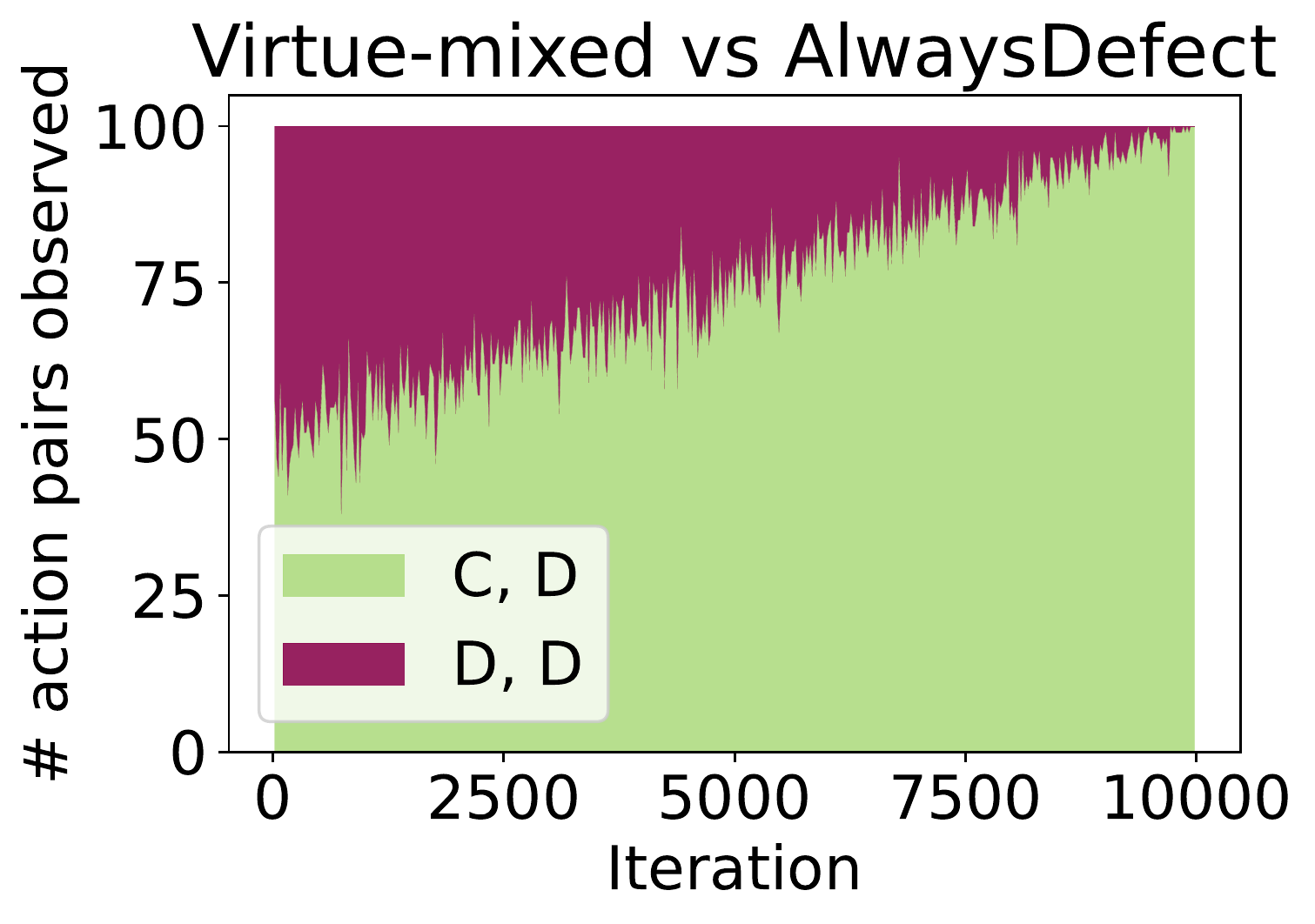}}
&\subt{\includegraphics[width=35mm]{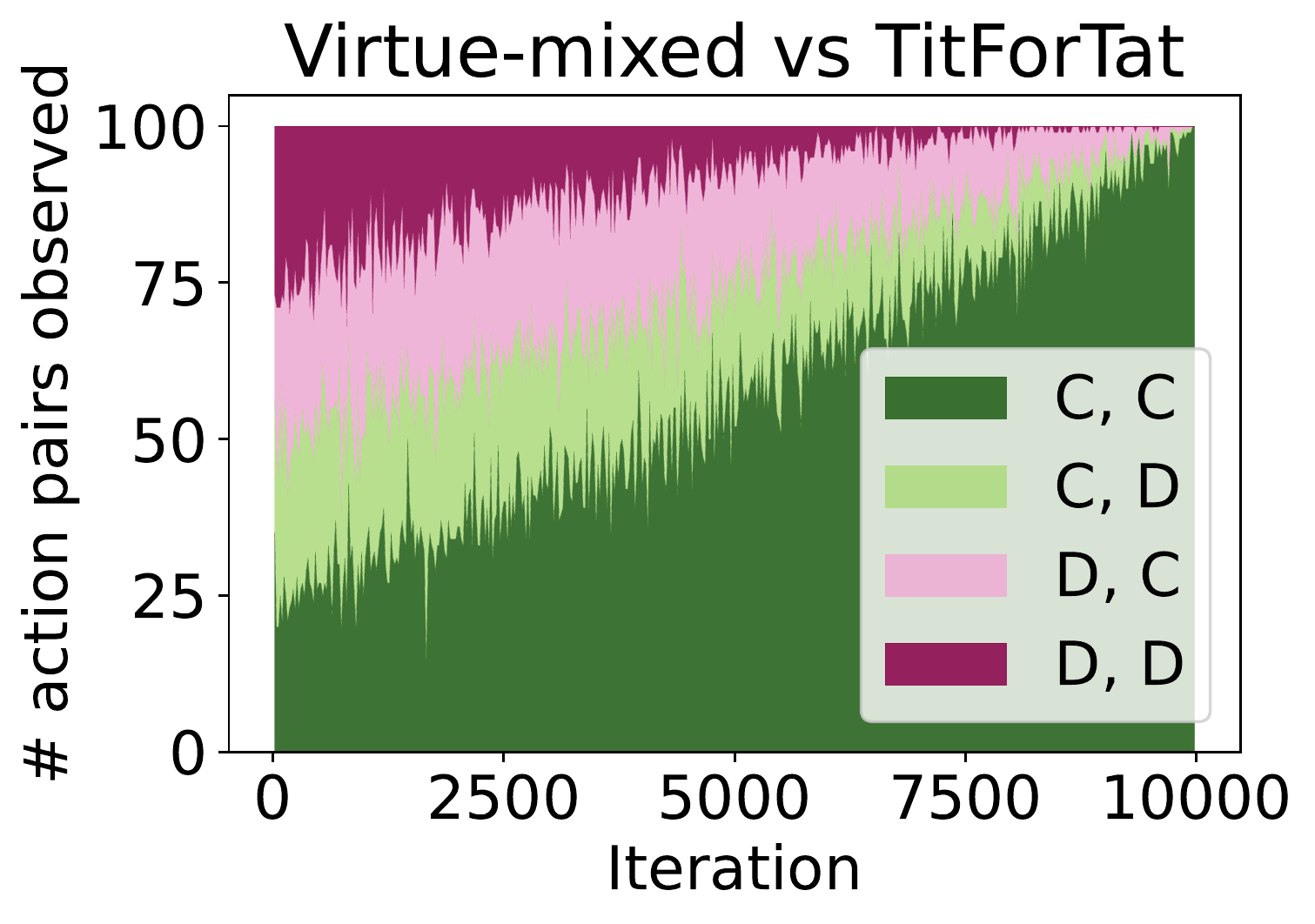}}
&\subt{\includegraphics[width=35mm]{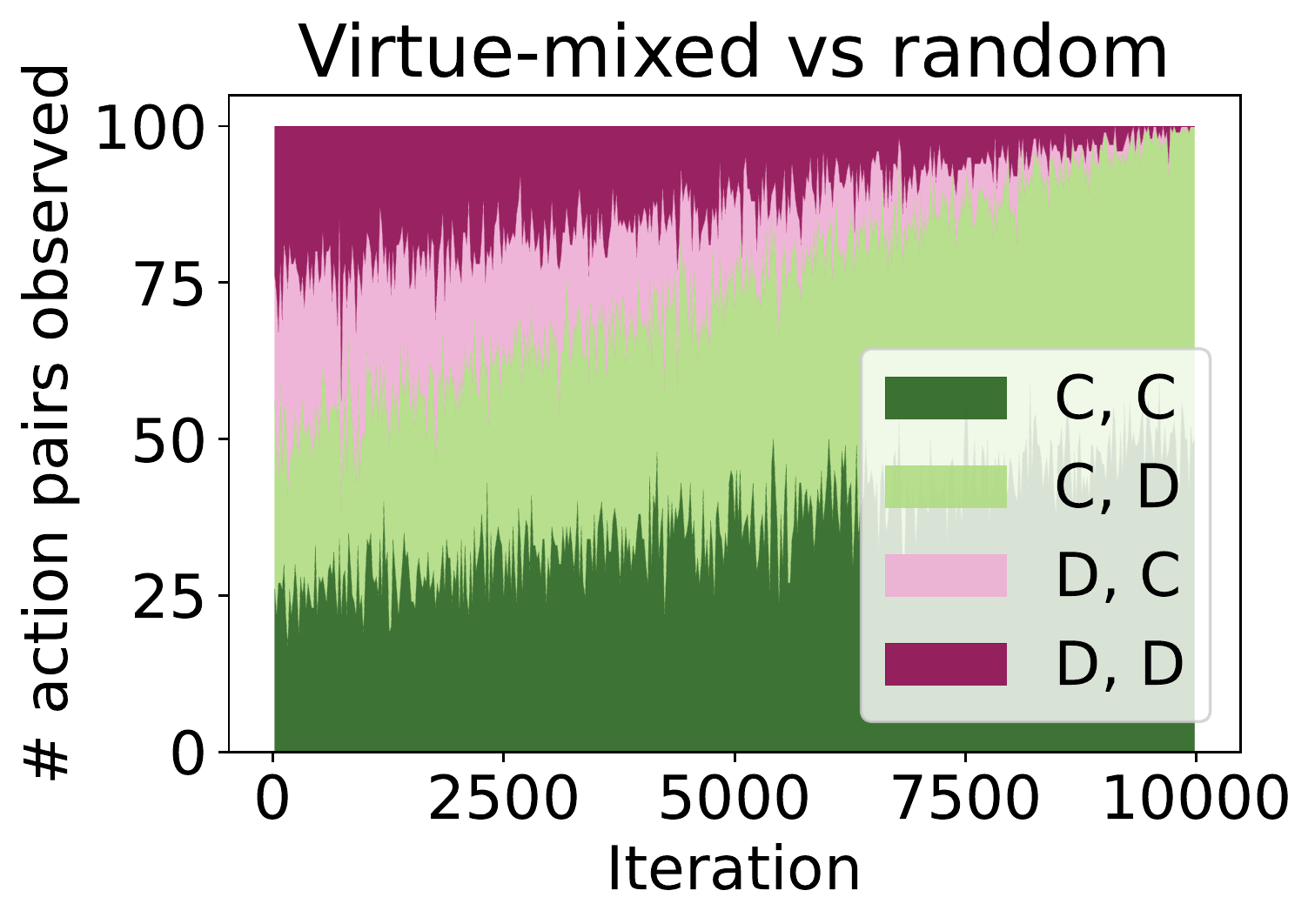}}
\\
\bottomrule
\end{tabular}
\caption{Iterated Prisoner's Dilemma game. Simultaneous pairs of actions observed over time. Learning player $M$ (row) vs. static opponent $O$ (column).}
\label{fig:action_pairs_baseline_IPD}
\end{figure*}

\begin{figure*}[!h]
\centering
\begin{tabular}{|c|cccc}
\toprule
 & AC & AD & TFT & Random \\
\midrule
\makecell[cc]{\rotatebox[origin=c]{90}{ Selfish }} & 
\subt{\includegraphics[width=35mm]{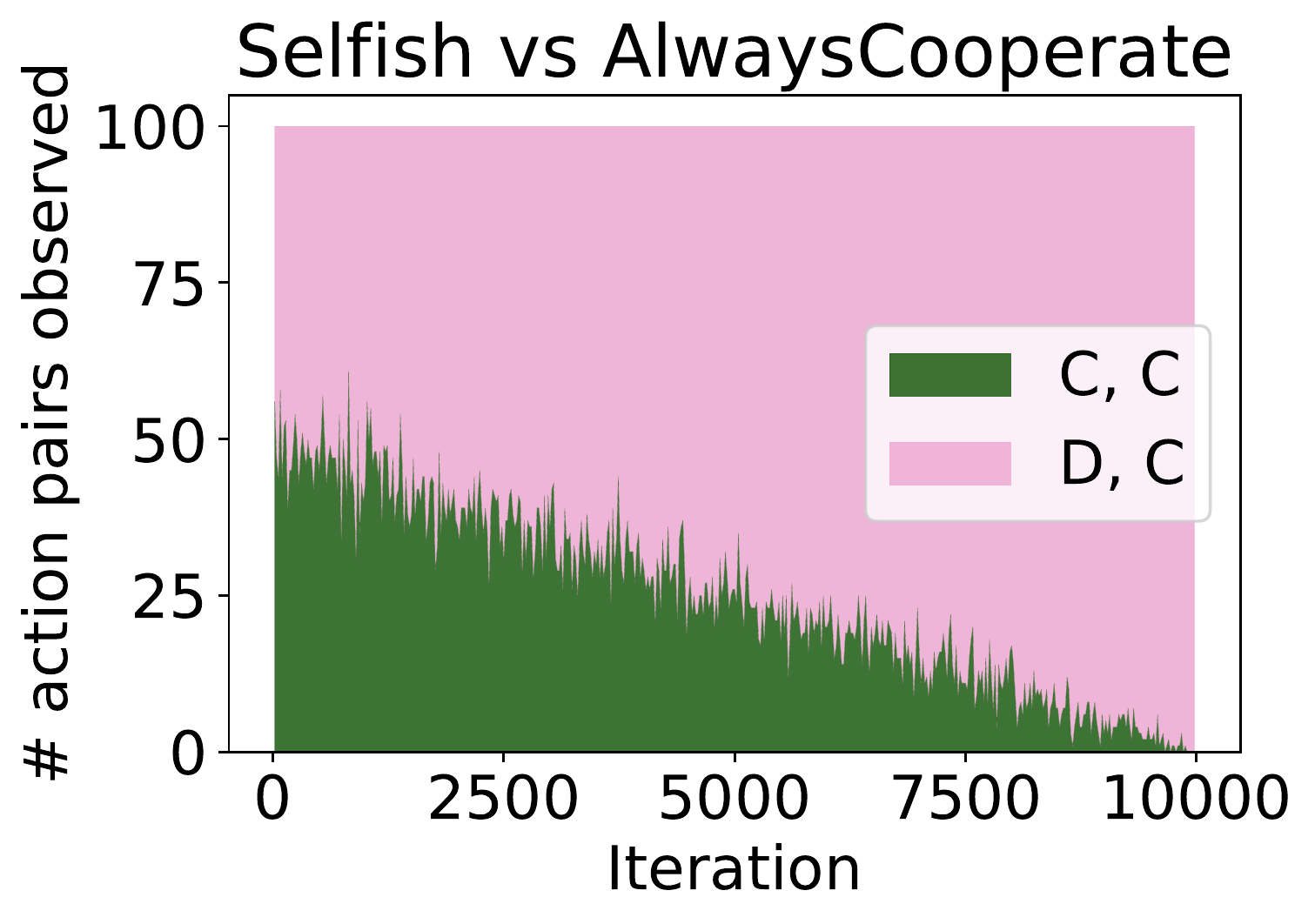}}
&
\subt{\includegraphics[width=35mm]{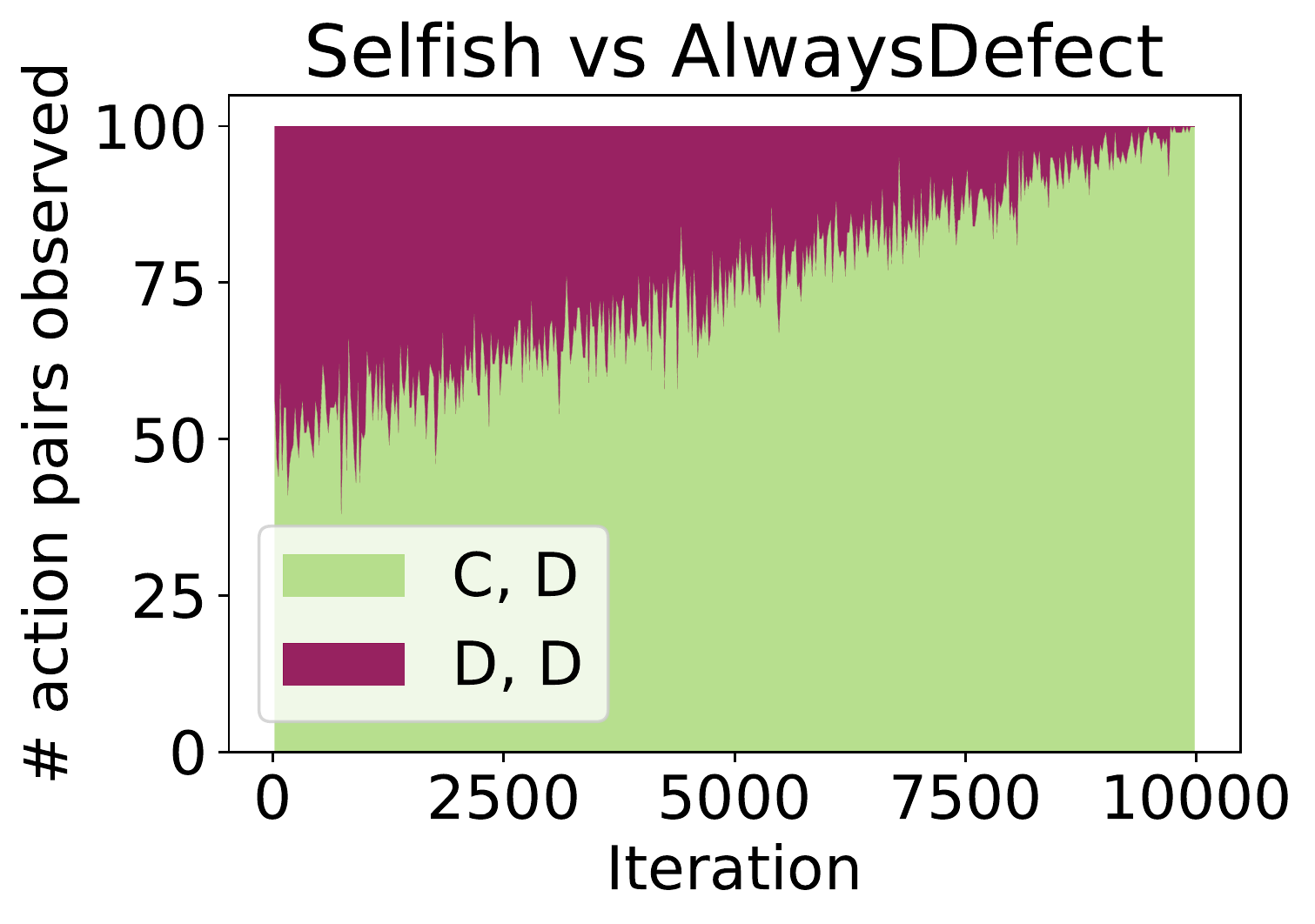}}
&
\subt{\includegraphics[width=35mm]{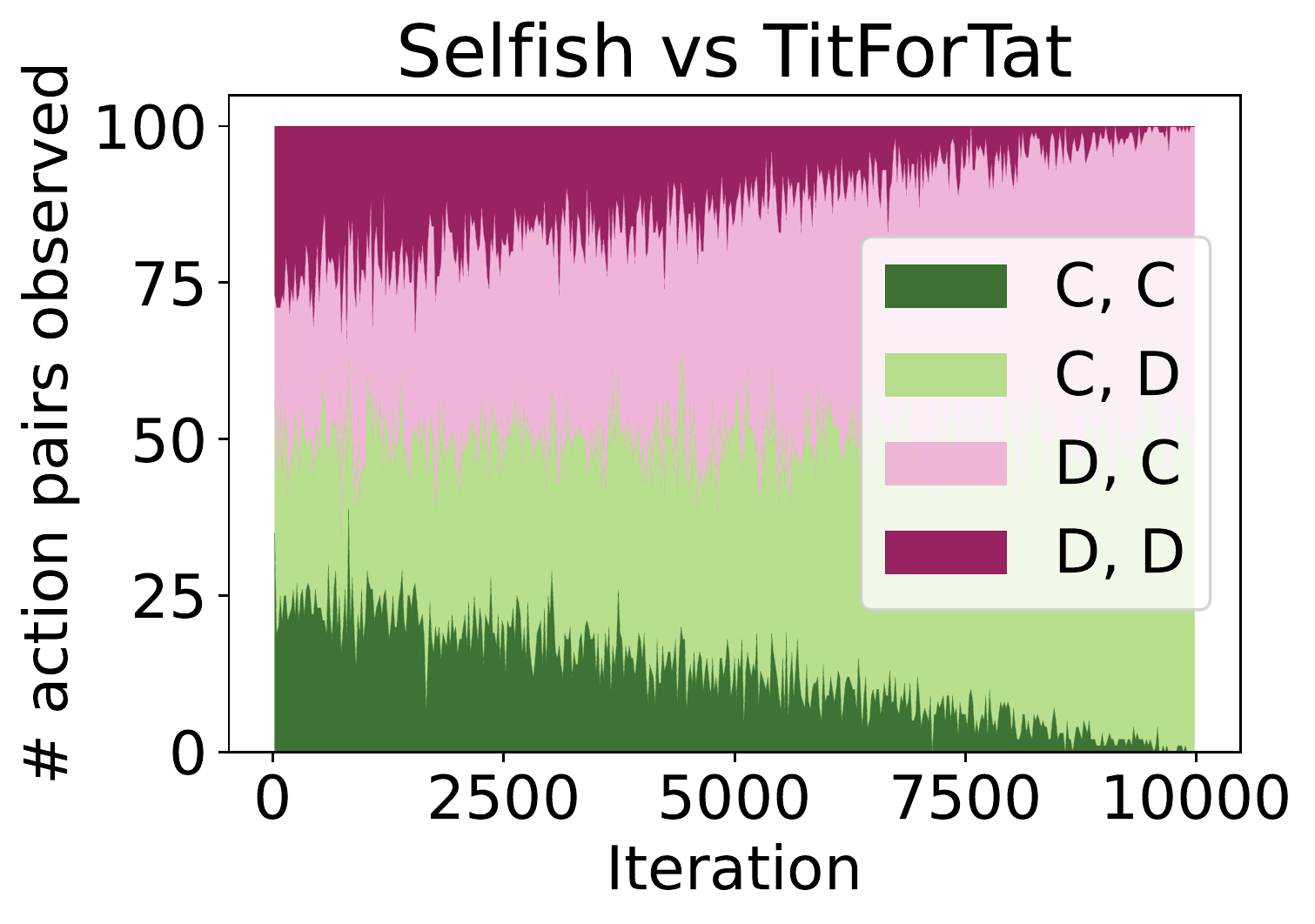}}
&
\subt{\includegraphics[width=35mm]{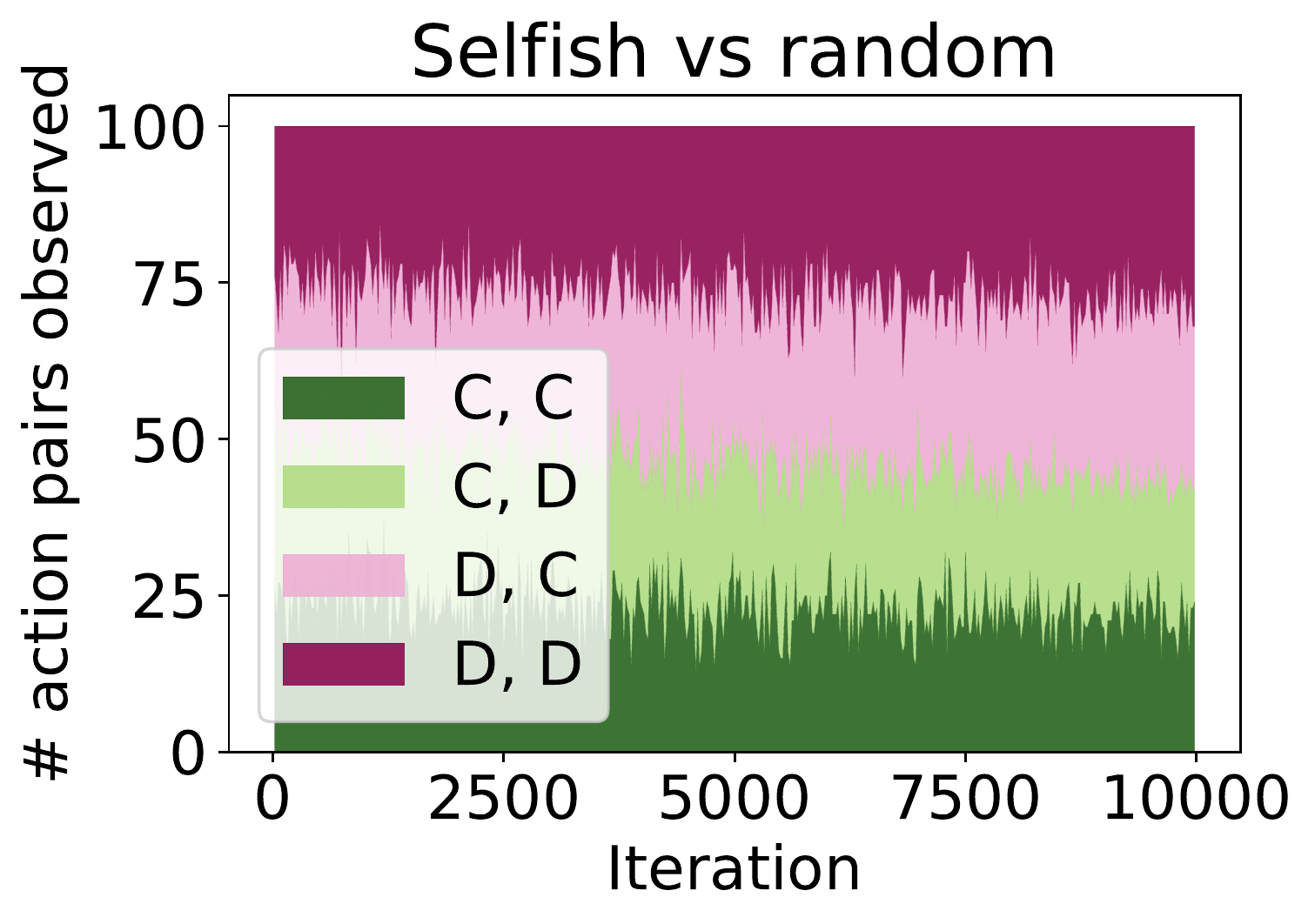}}
\\
\makecell[cc]{\rotatebox[origin=c]{90}{ Utilitarian }} & 
\subt{\includegraphics[width=35mm]{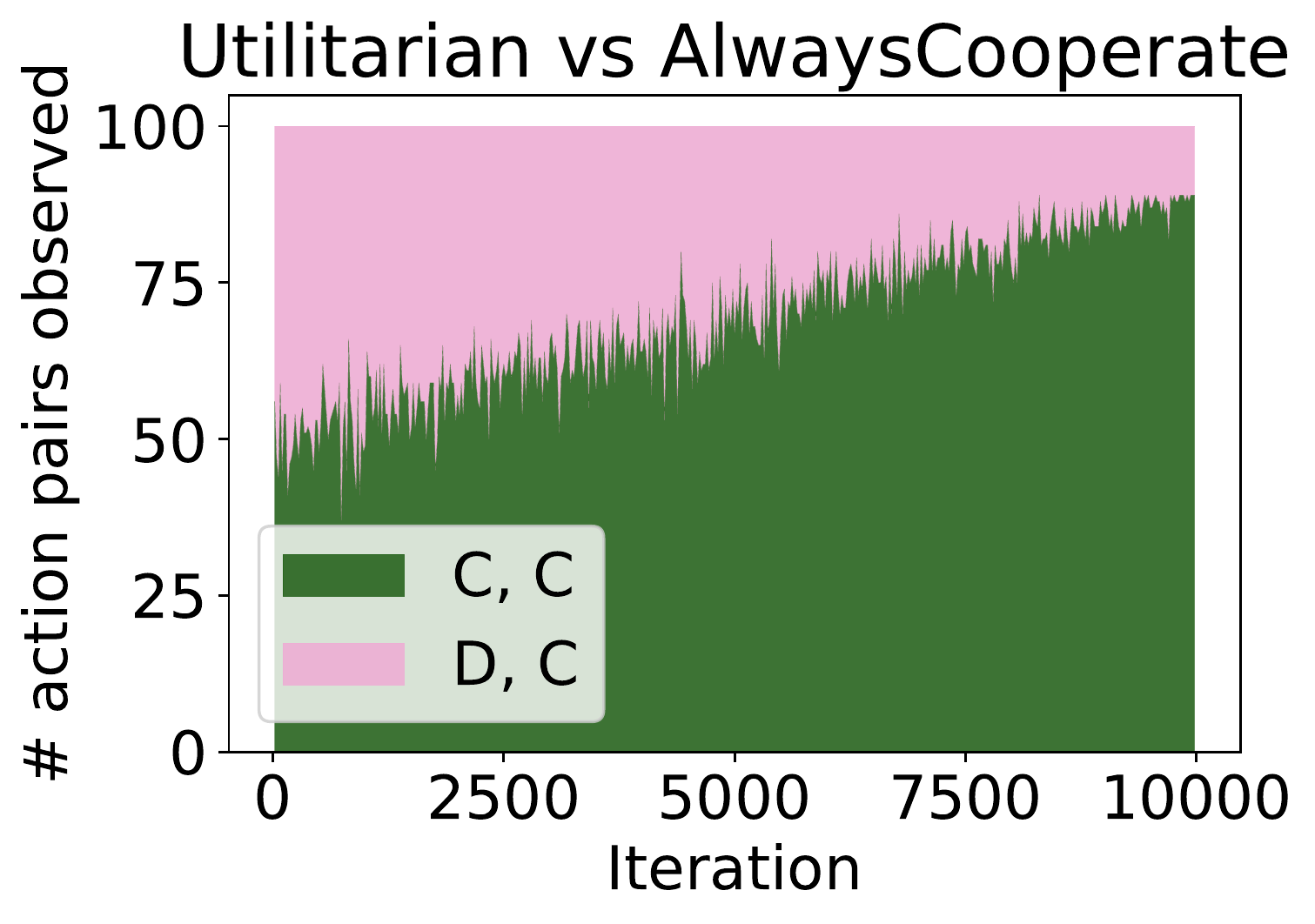}}
&
\subt{\includegraphics[width=35mm]{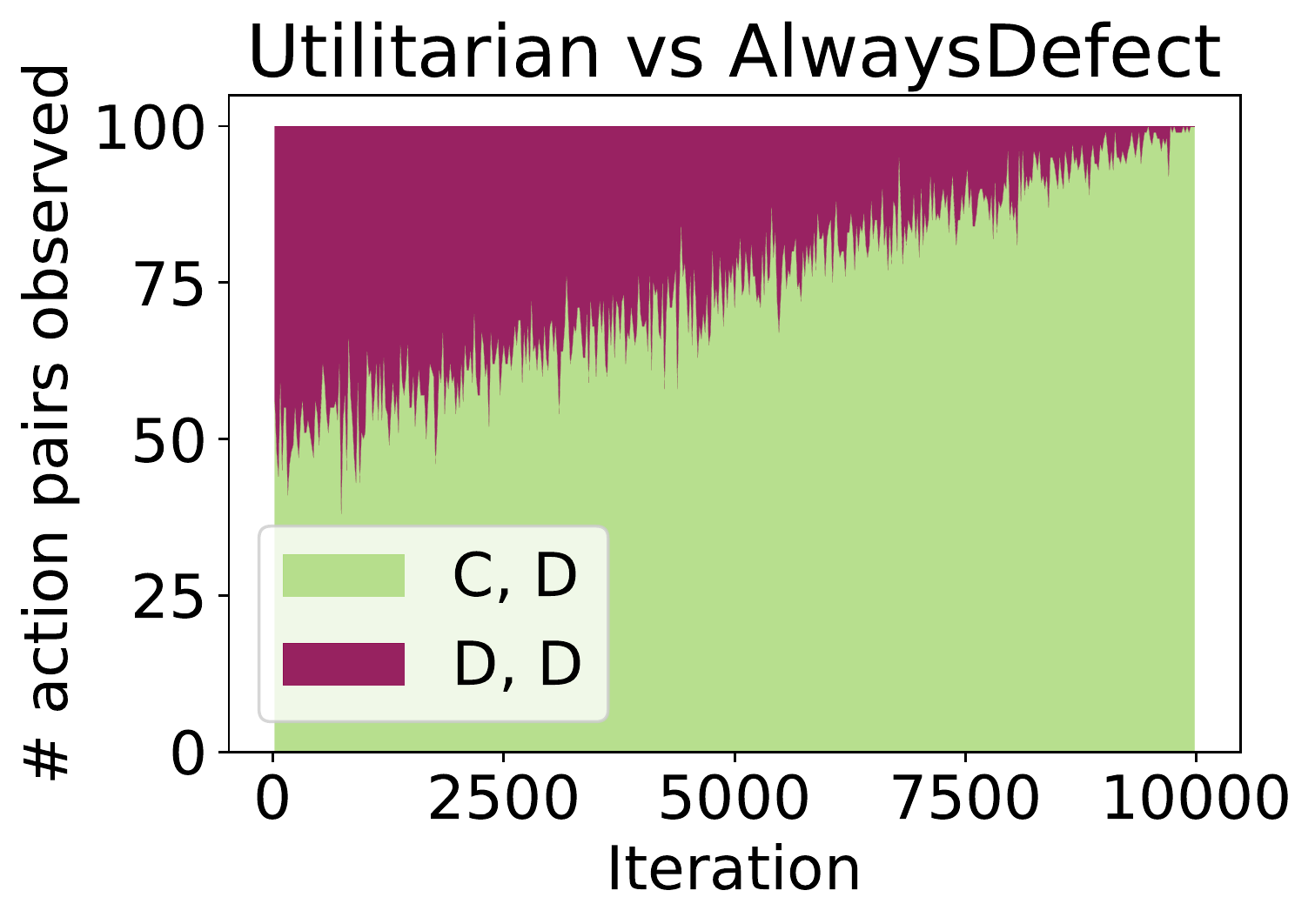}}
&
\subt{\includegraphics[width=35mm]{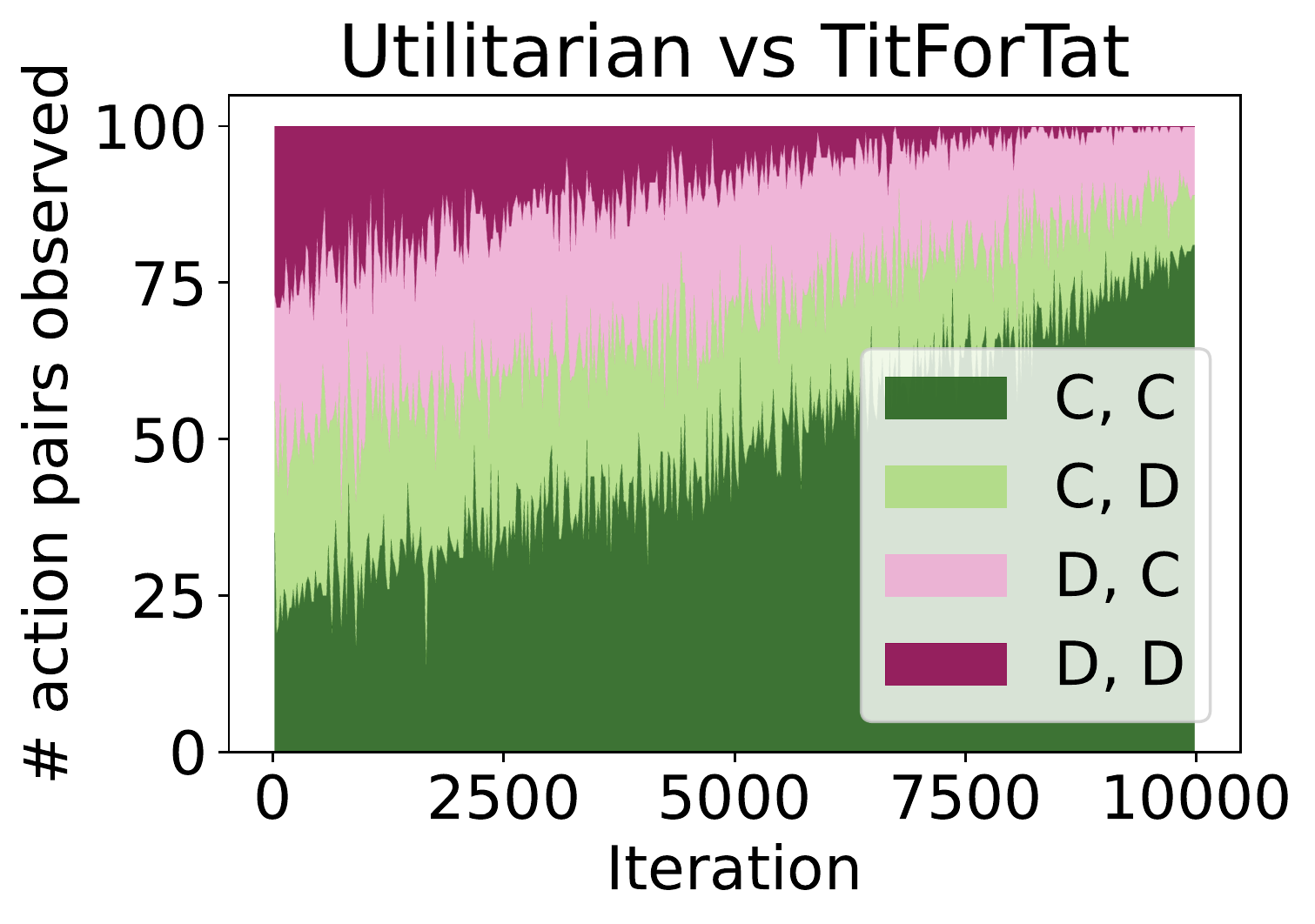}}
&
\subt{\includegraphics[width=35mm]{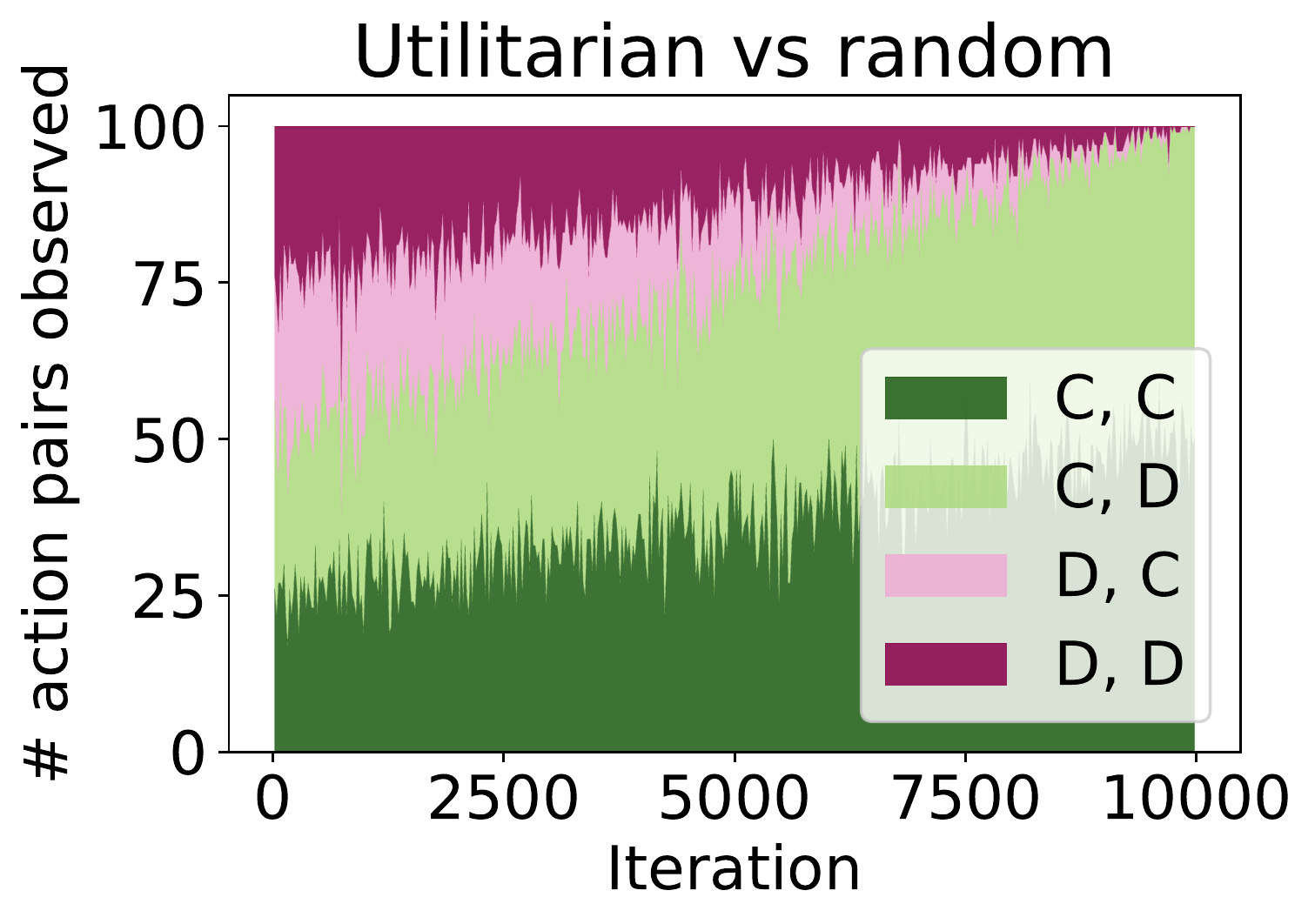}}
\\
\makecell[cc]{\rotatebox[origin=c]{90}{ Deontological }} &
\subt{\includegraphics[width=35mm]{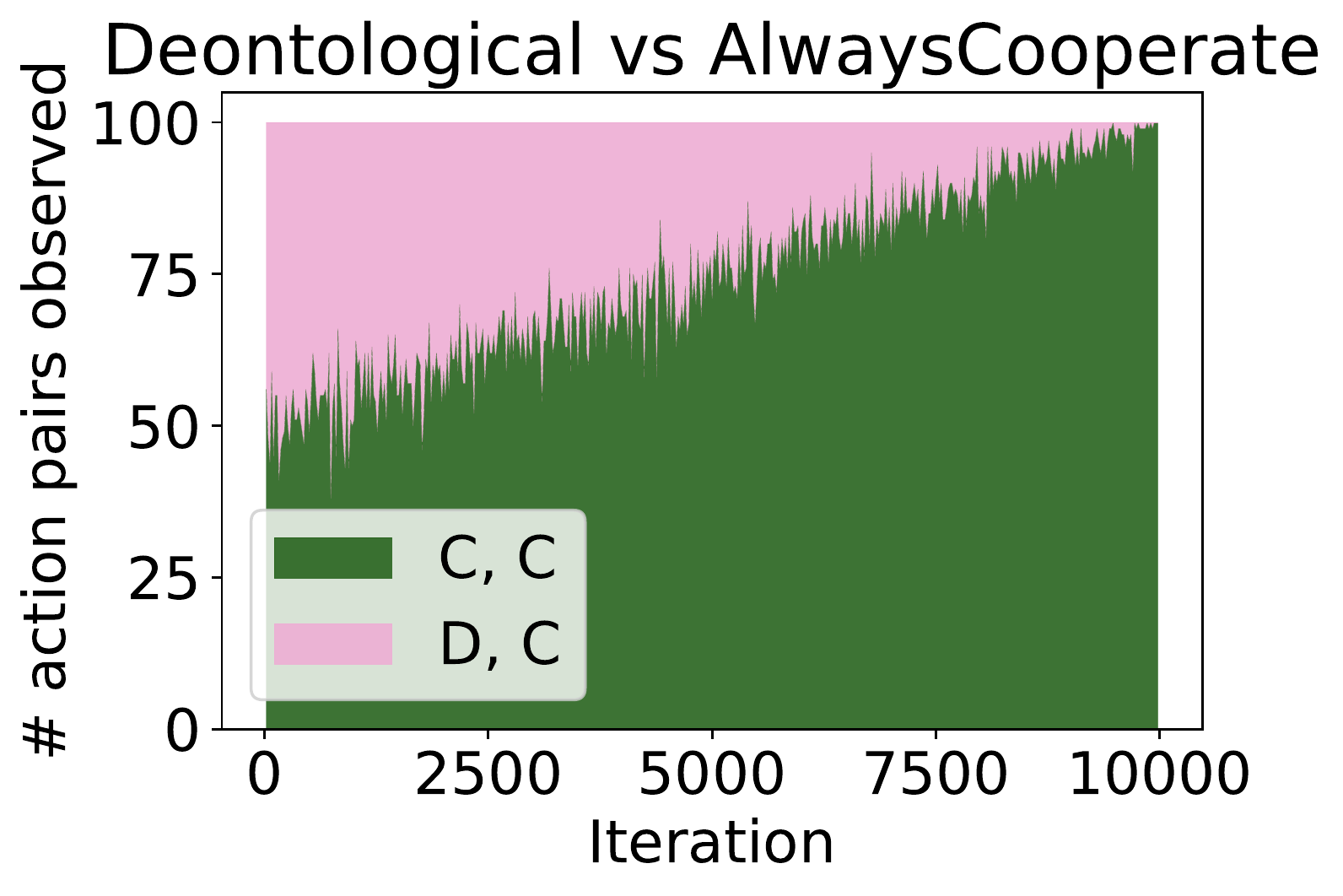}}
&
\subt{\includegraphics[width=35mm]{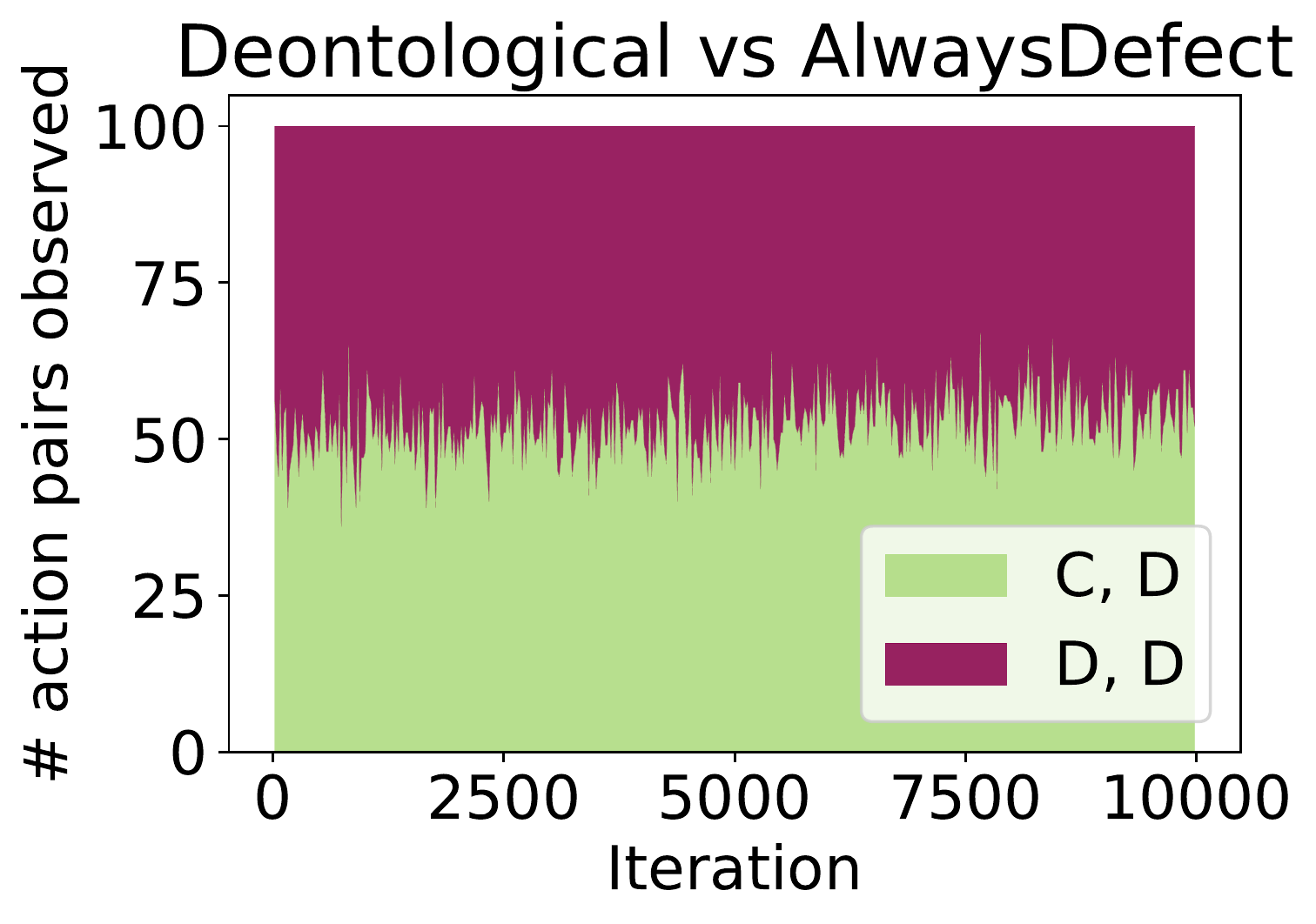}}
&
\subt{\includegraphics[width=35mm]{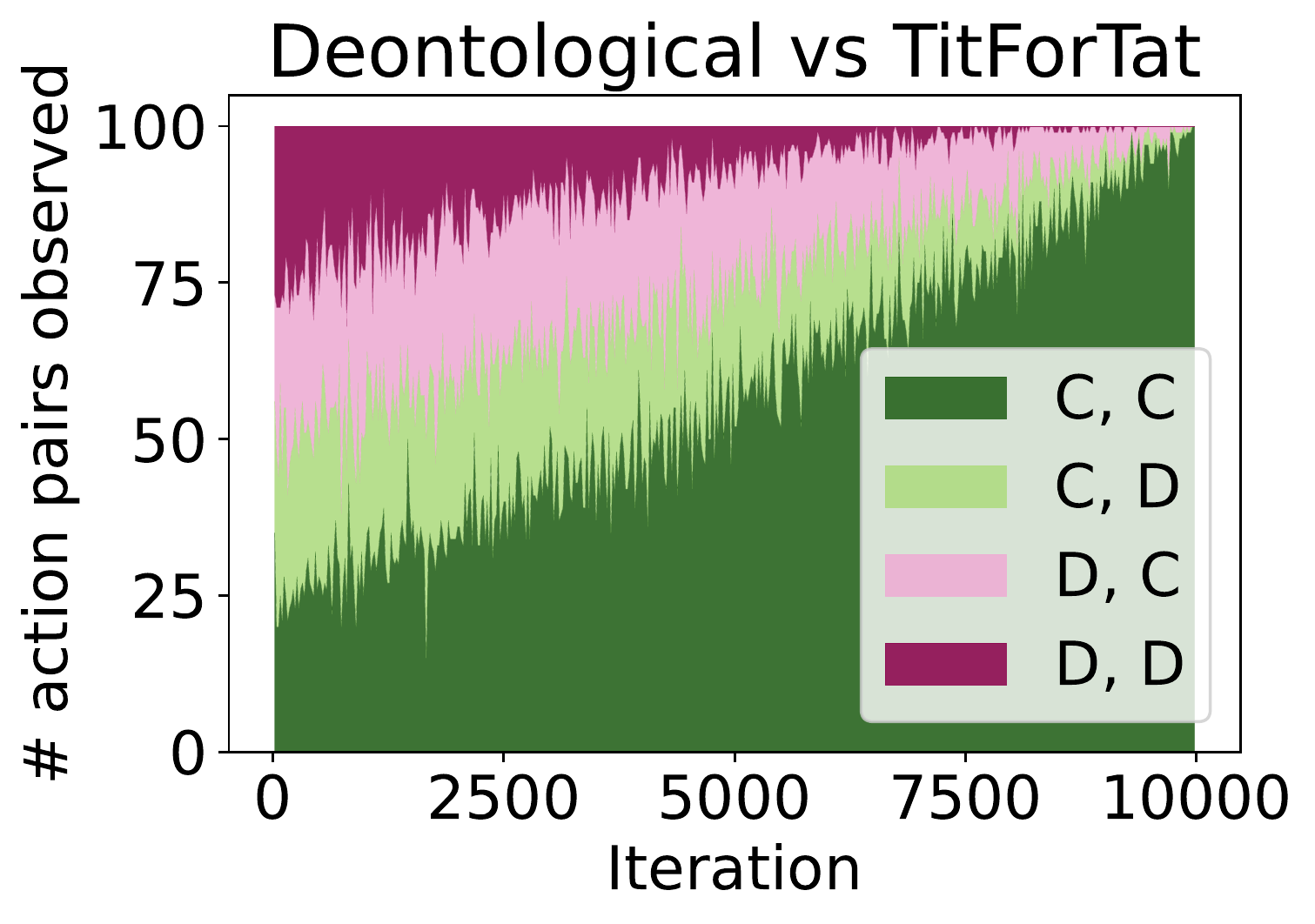}}
&
\subt{\includegraphics[width=35mm]{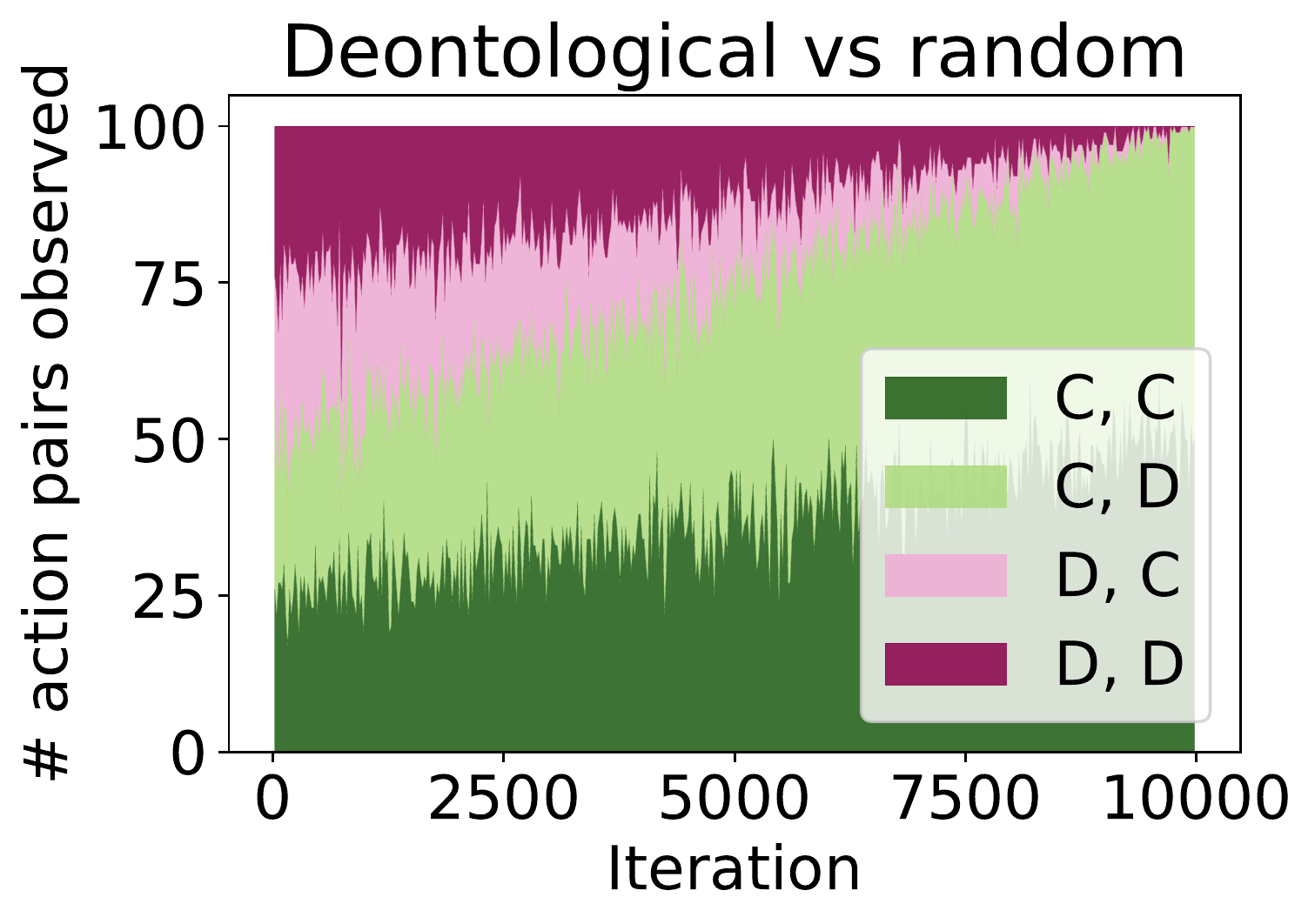}}
\\
\makecell[cc]{\rotatebox[origin=c]{90}{ Virtue-eq. }} &
\subt{\includegraphics[width=35mm]{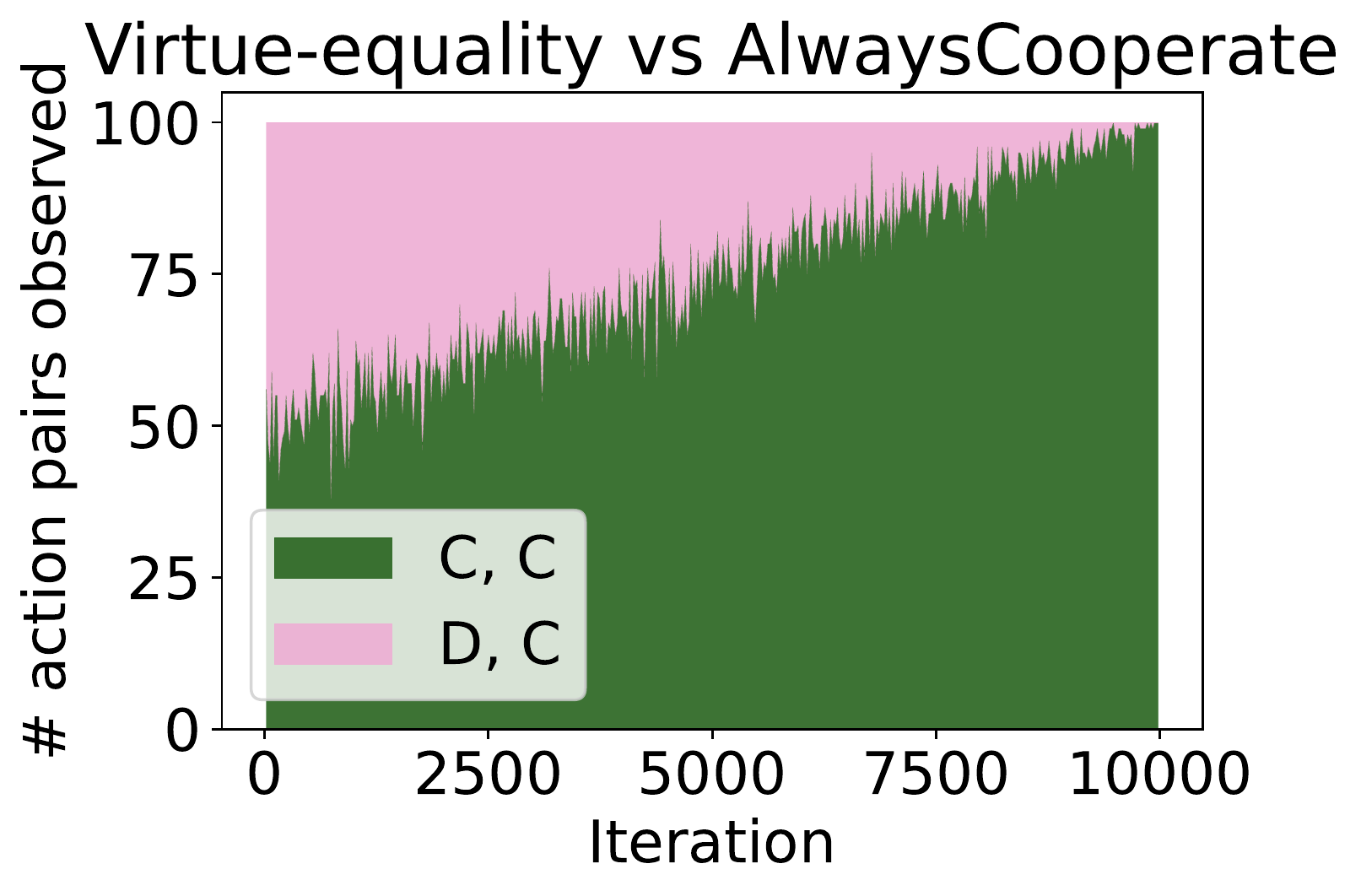}}
&
\subt{\includegraphics[width=35mm]{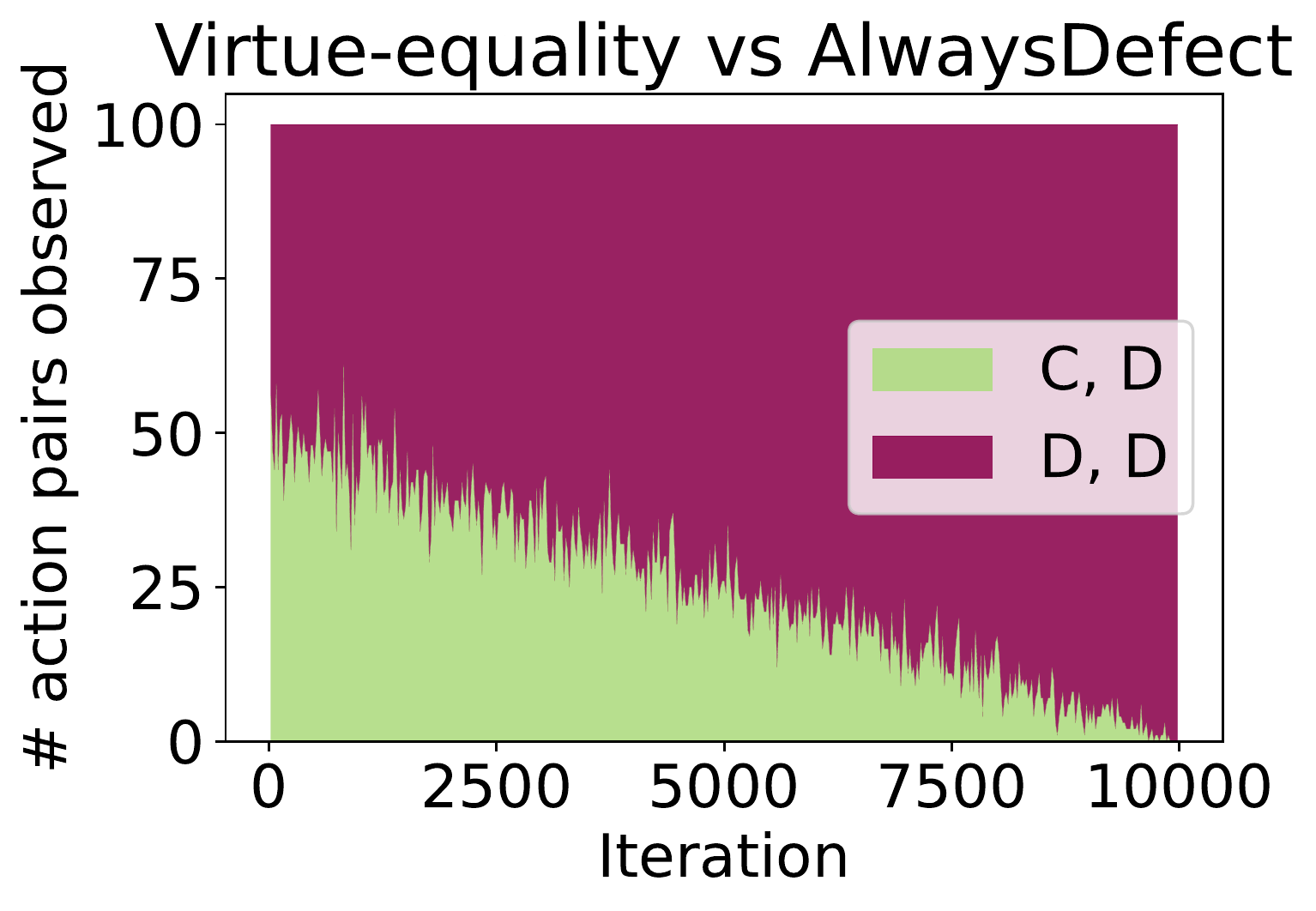}}
&
\subt{\includegraphics[width=35mm]{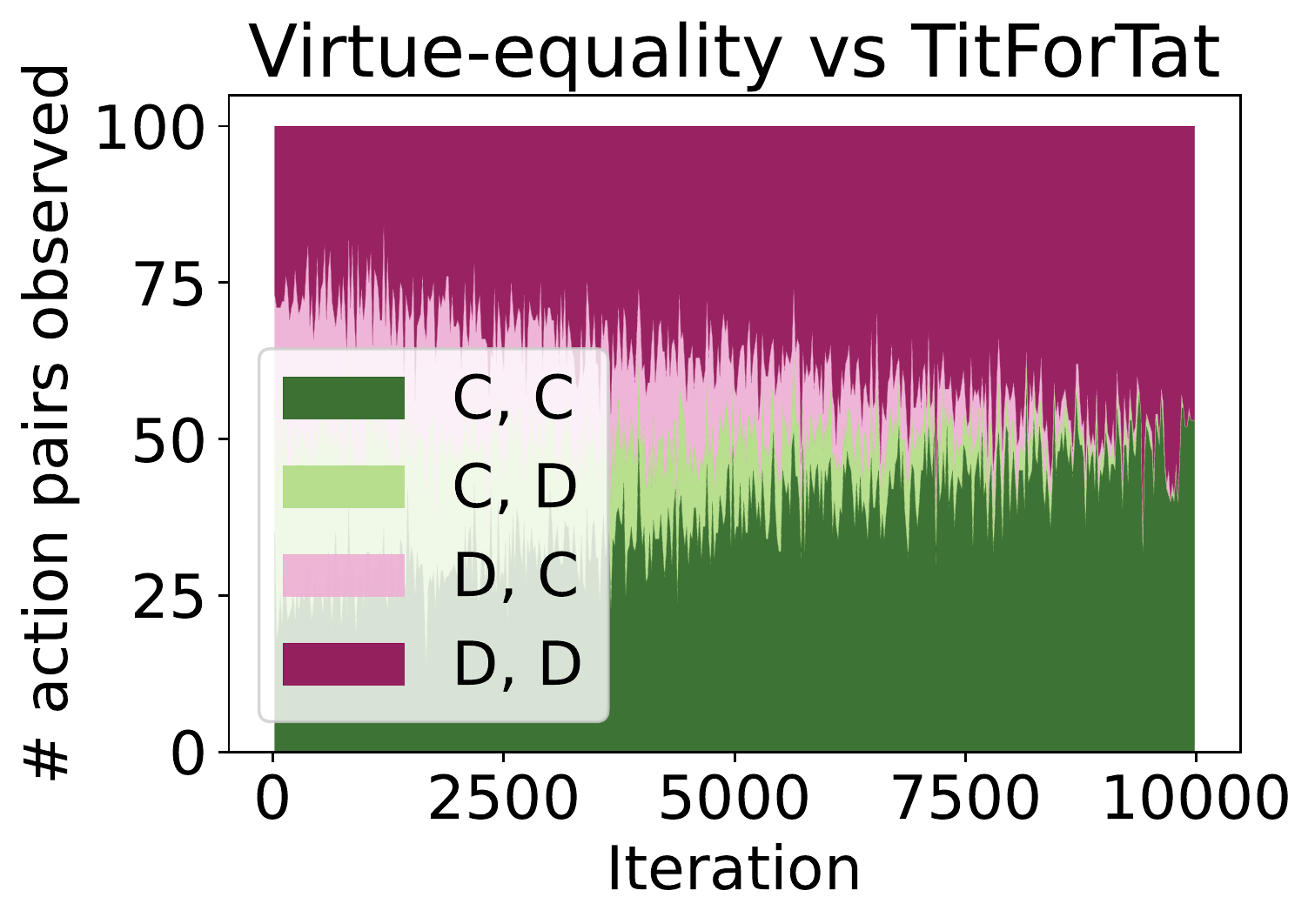}}
&
\subt{\includegraphics[width=35mm]{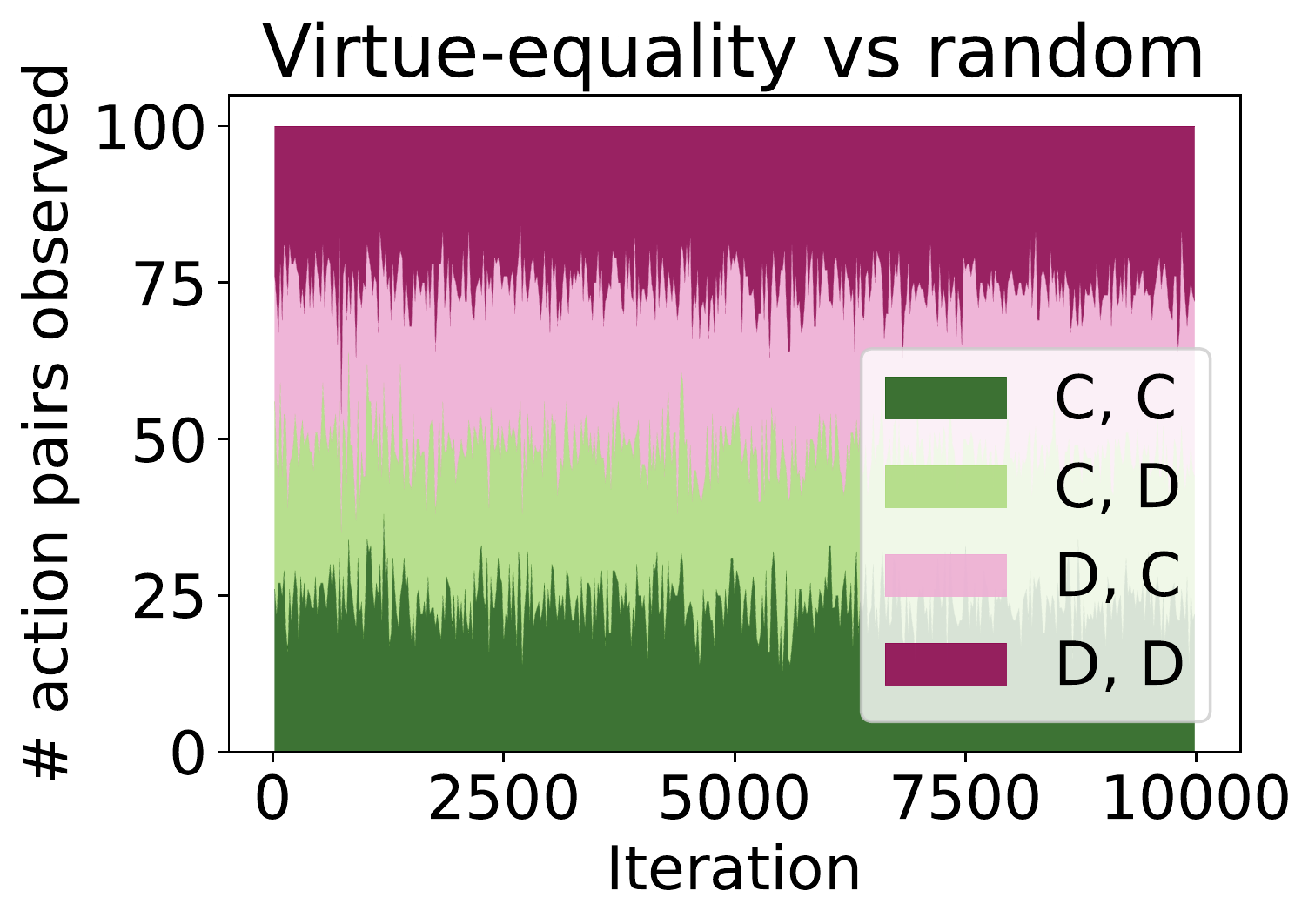}}
\\
\makecell[cc]{\rotatebox[origin=c]{90}{ Virtue-kind. }} &
\subt{\includegraphics[width=35mm]{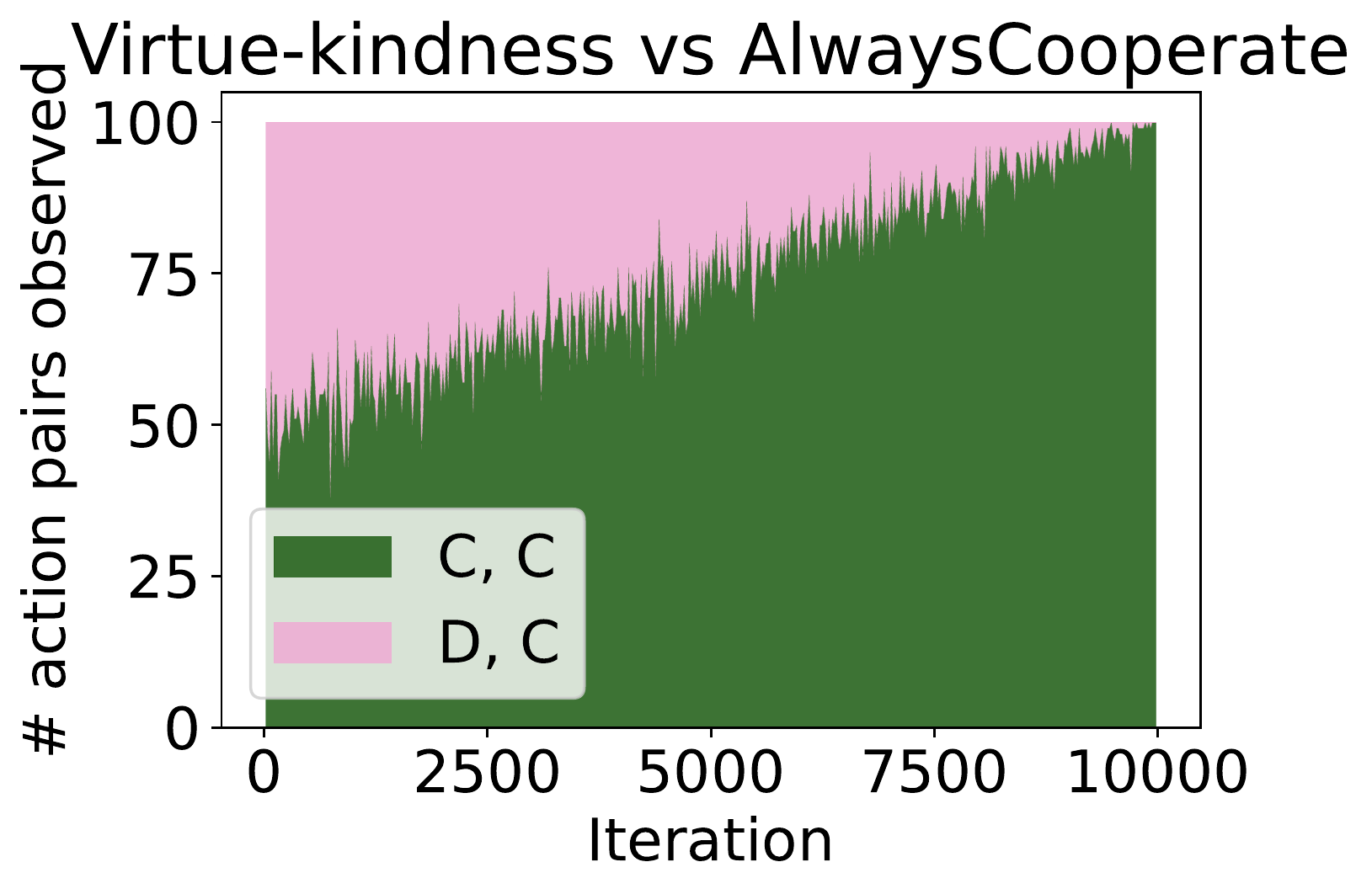}}
&
\subt{\includegraphics[width=35mm]{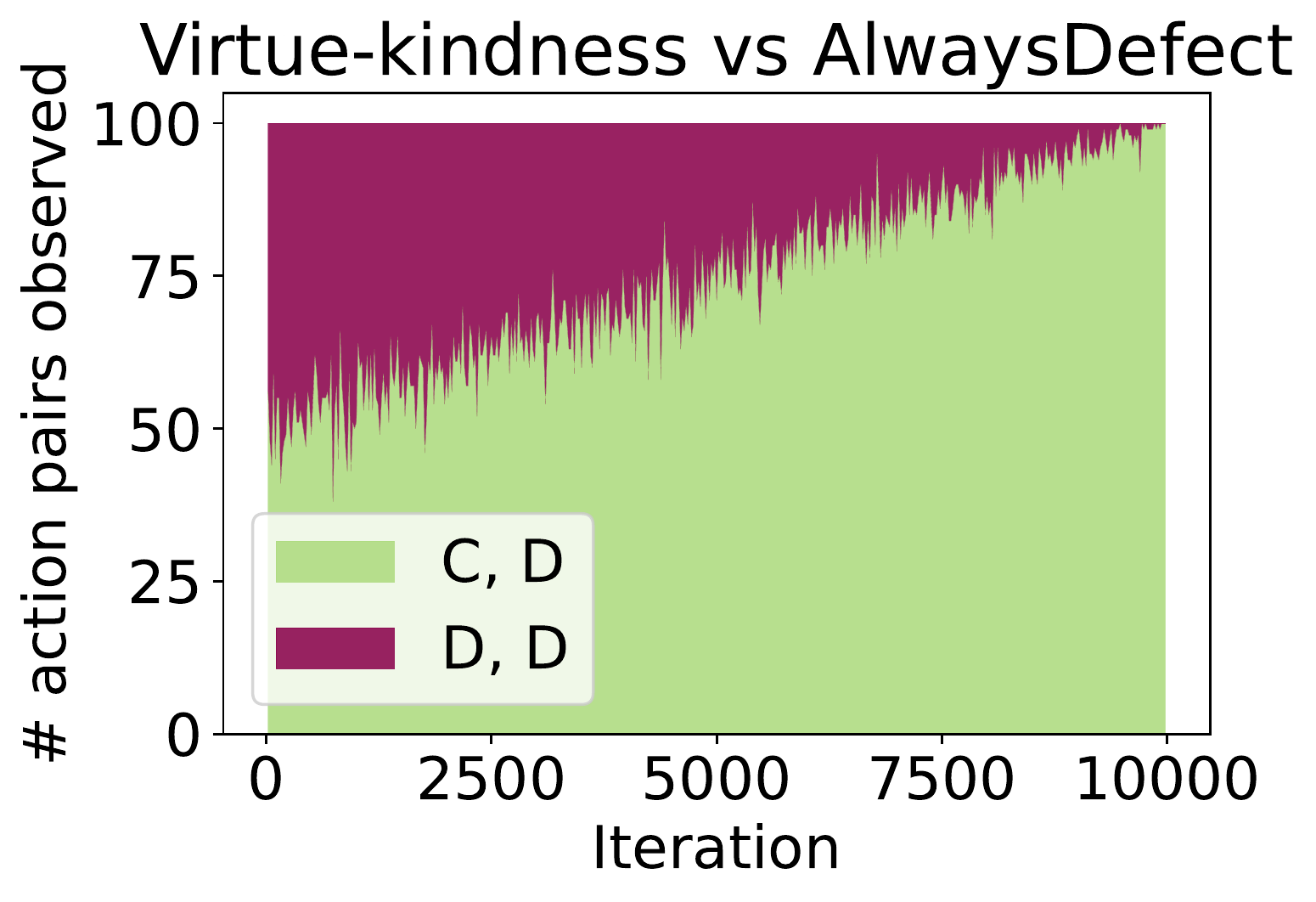}}
&
\subt{\includegraphics[width=35mm]{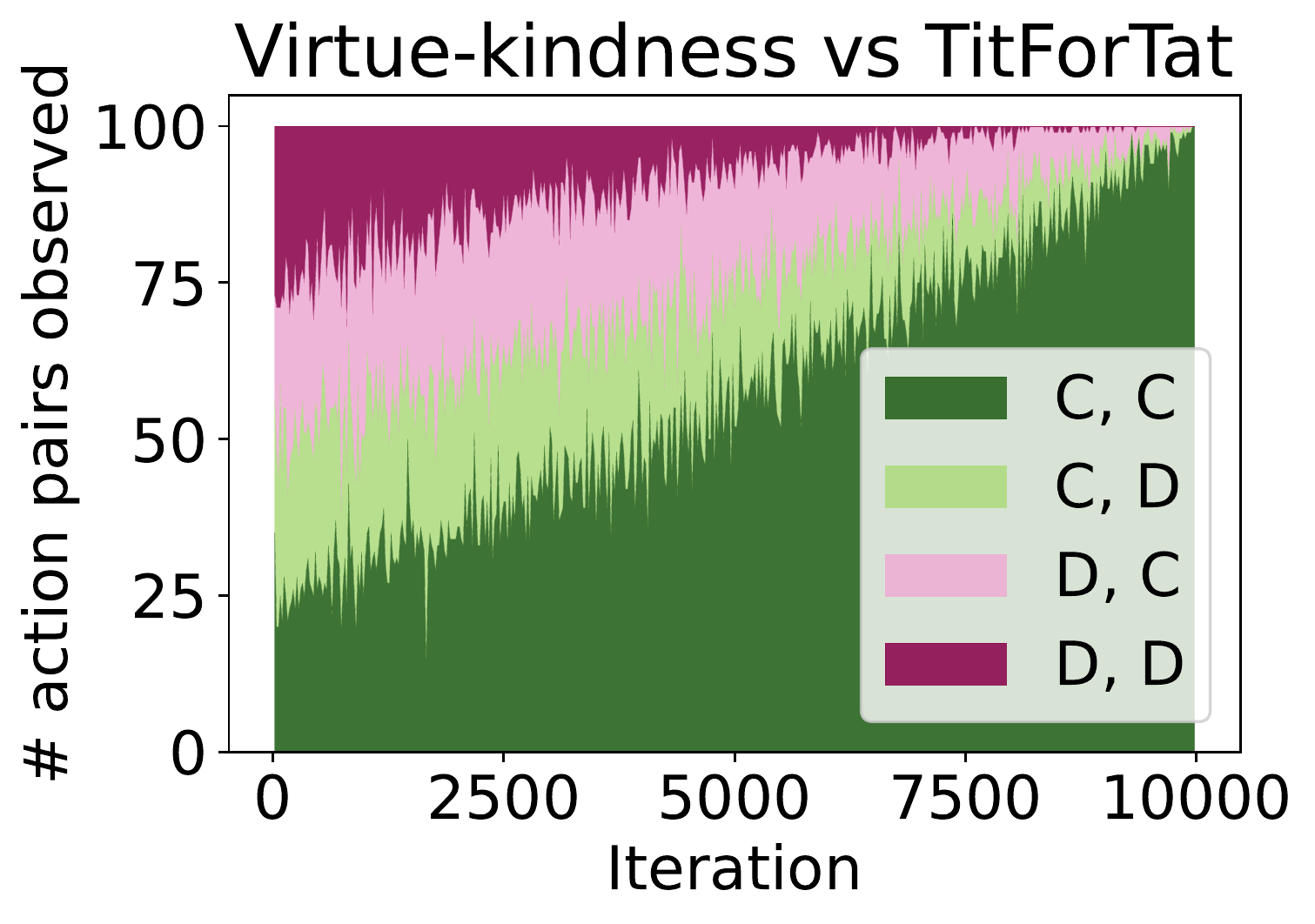}}
&
\subt{\includegraphics[width=35mm]{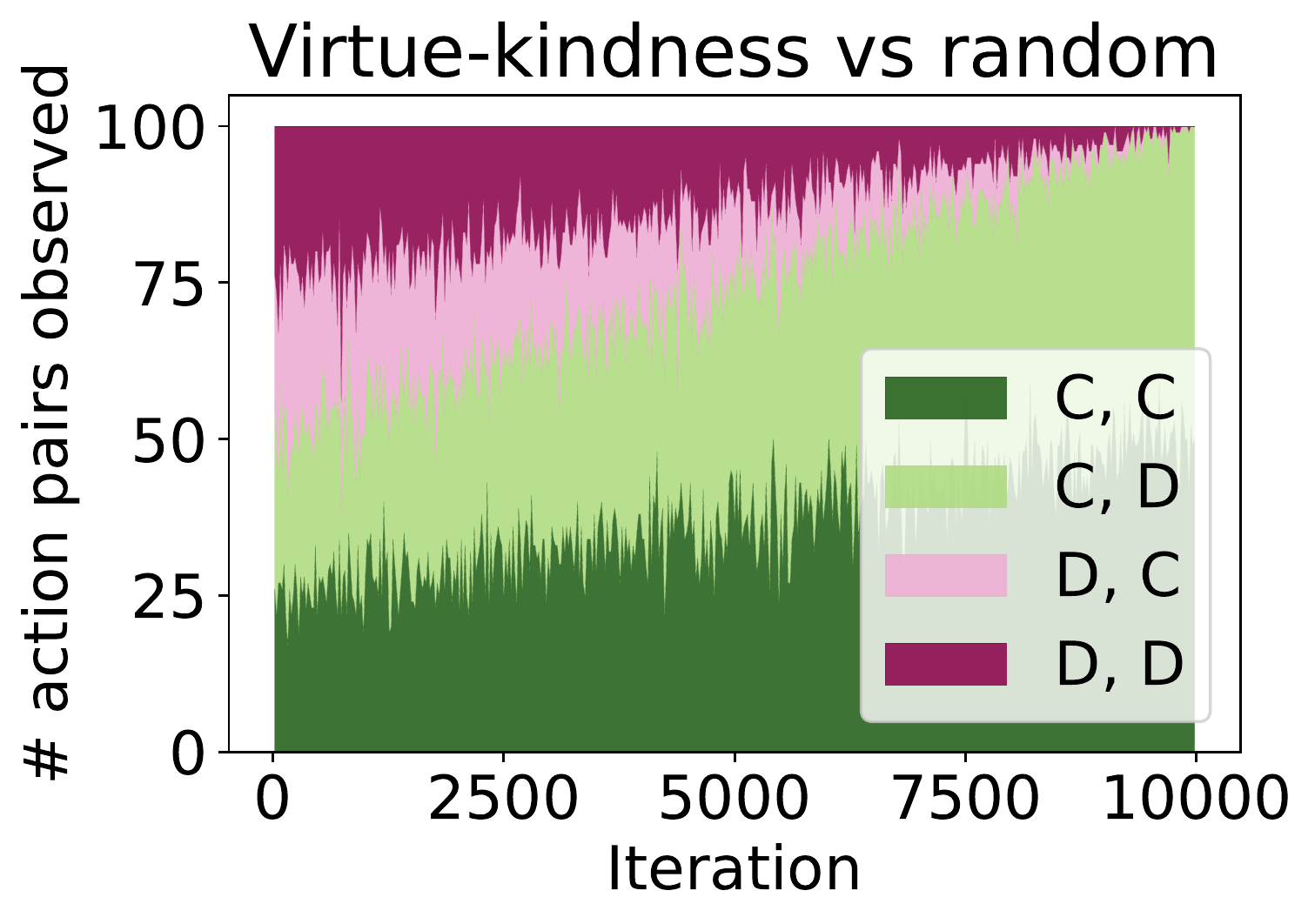}}
\\
\makecell[cc]{\rotatebox[origin=c]{90}{ Virtue-mix. }} &
\subt{\includegraphics[width=35mm]{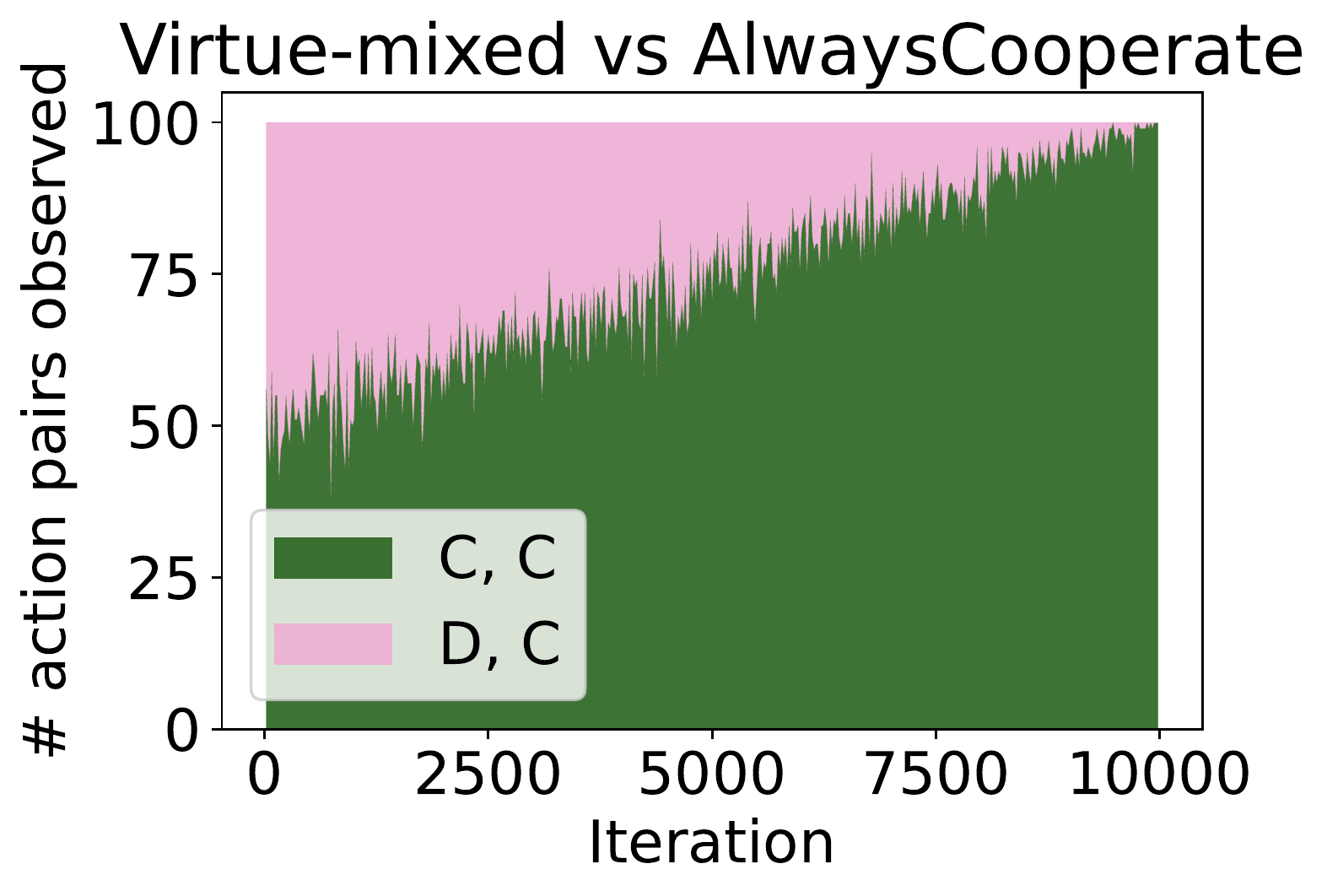}}
&
\subt{\includegraphics[width=35mm]{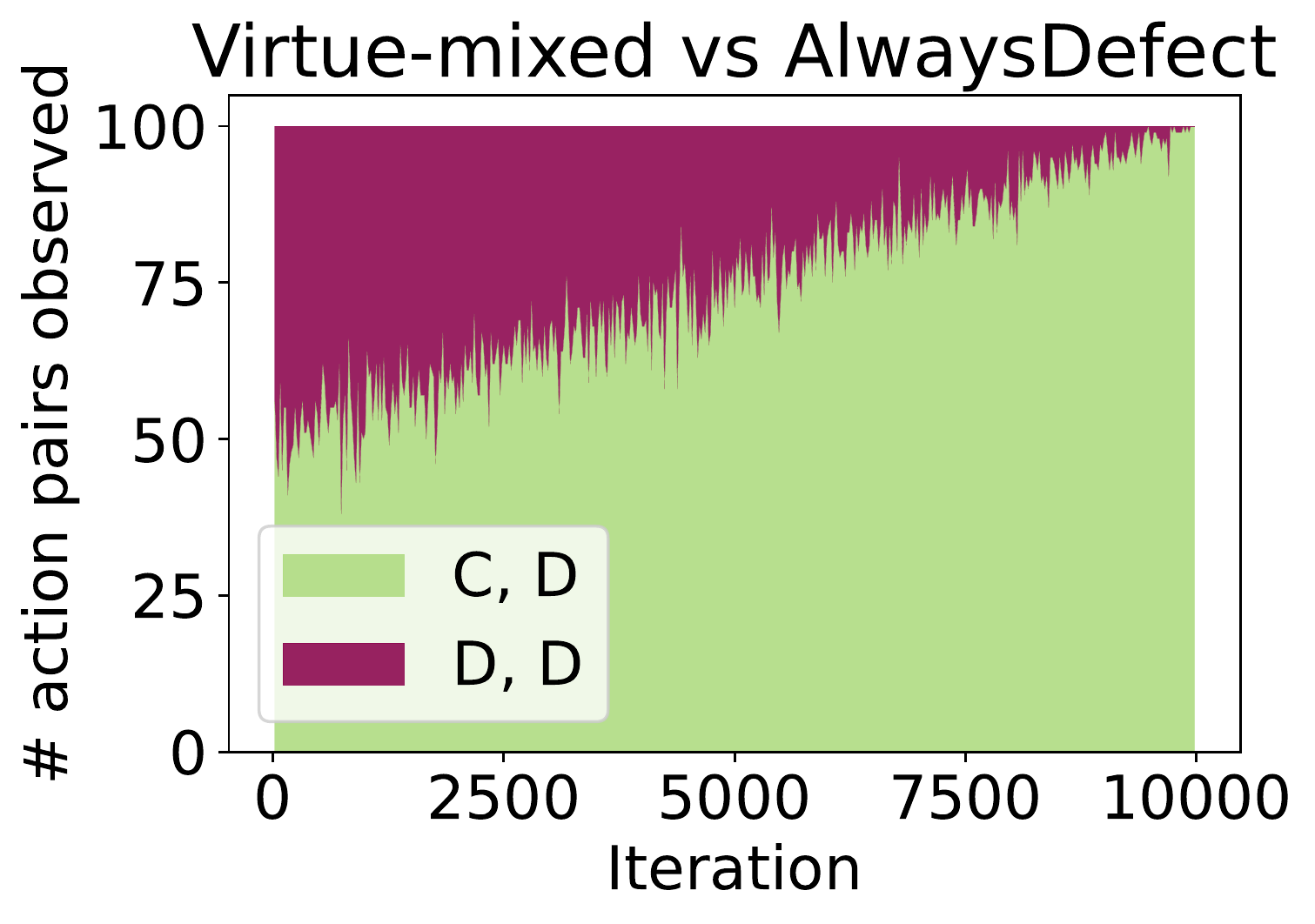}}
&
\subt{\includegraphics[width=35mm]{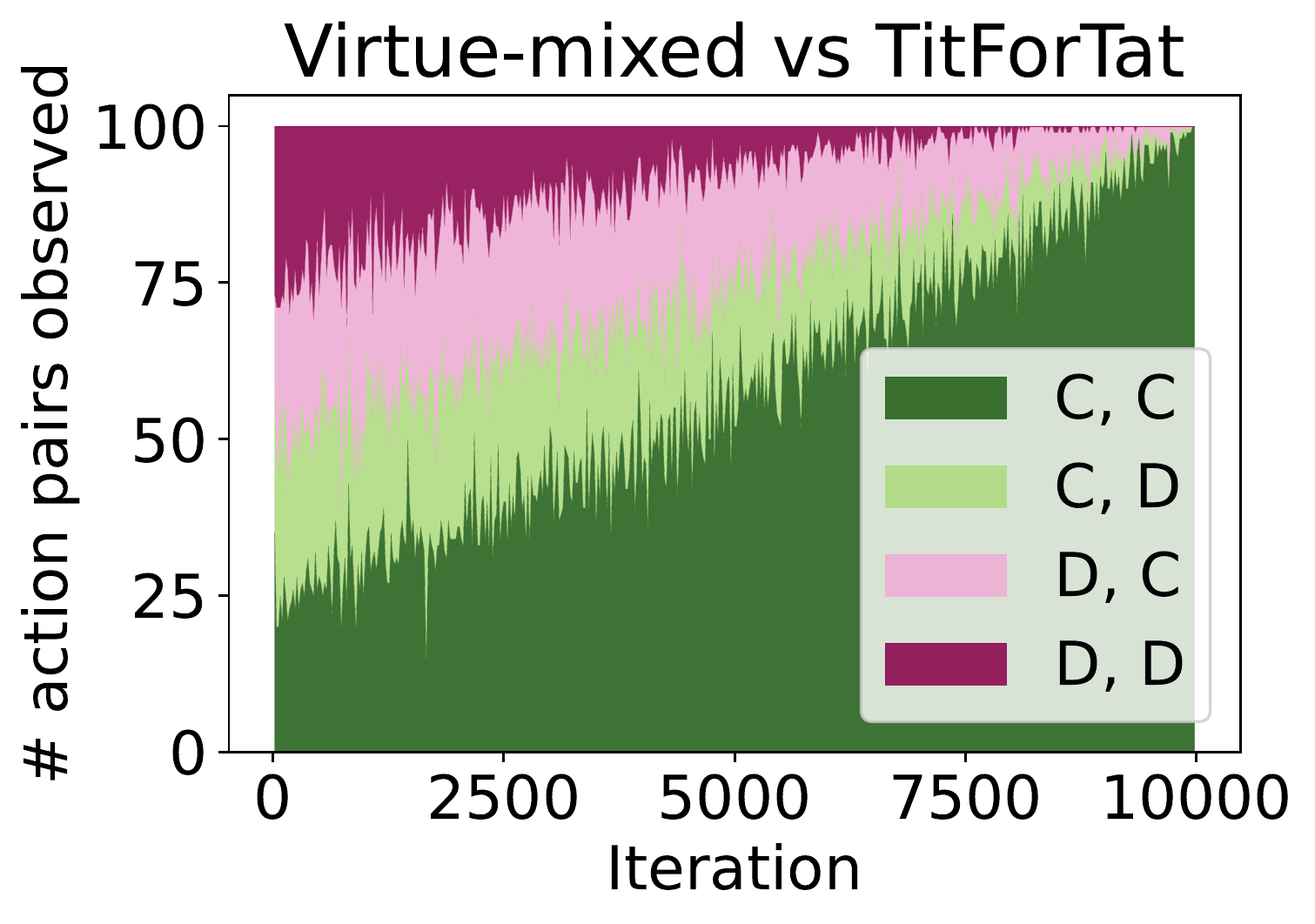}}
&
\subt{\includegraphics[width=35mm]{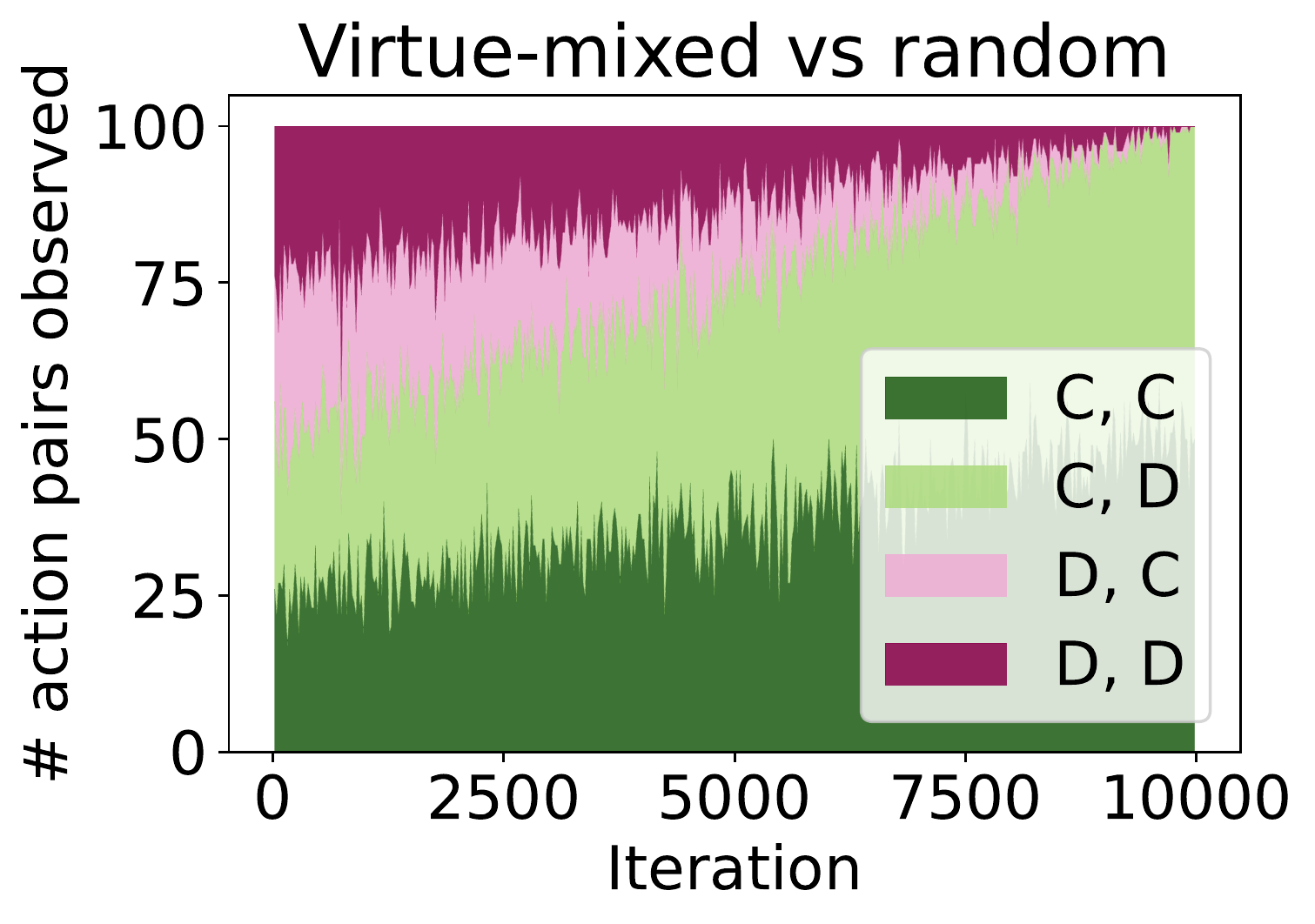}}
\\
\bottomrule
\end{tabular}
\caption{Iterated Volunteer's Dilemma game. Simultaneous pairs of actions observed over time. Learning player $M$ (row) vs. static opponent $O$ (column). }
\label{fig:action_pairs_baseline_VOL}
\end{figure*}

\begin{figure*}[!h]
\centering
\begin{tabular}{|c|cccc}
\toprule
 & AC & AD & TFT & Random \\
\midrule
\makecell[cc]{\rotatebox[origin=c]{90}{ Selfish }} & 
\subt{\includegraphics[width=35mm]{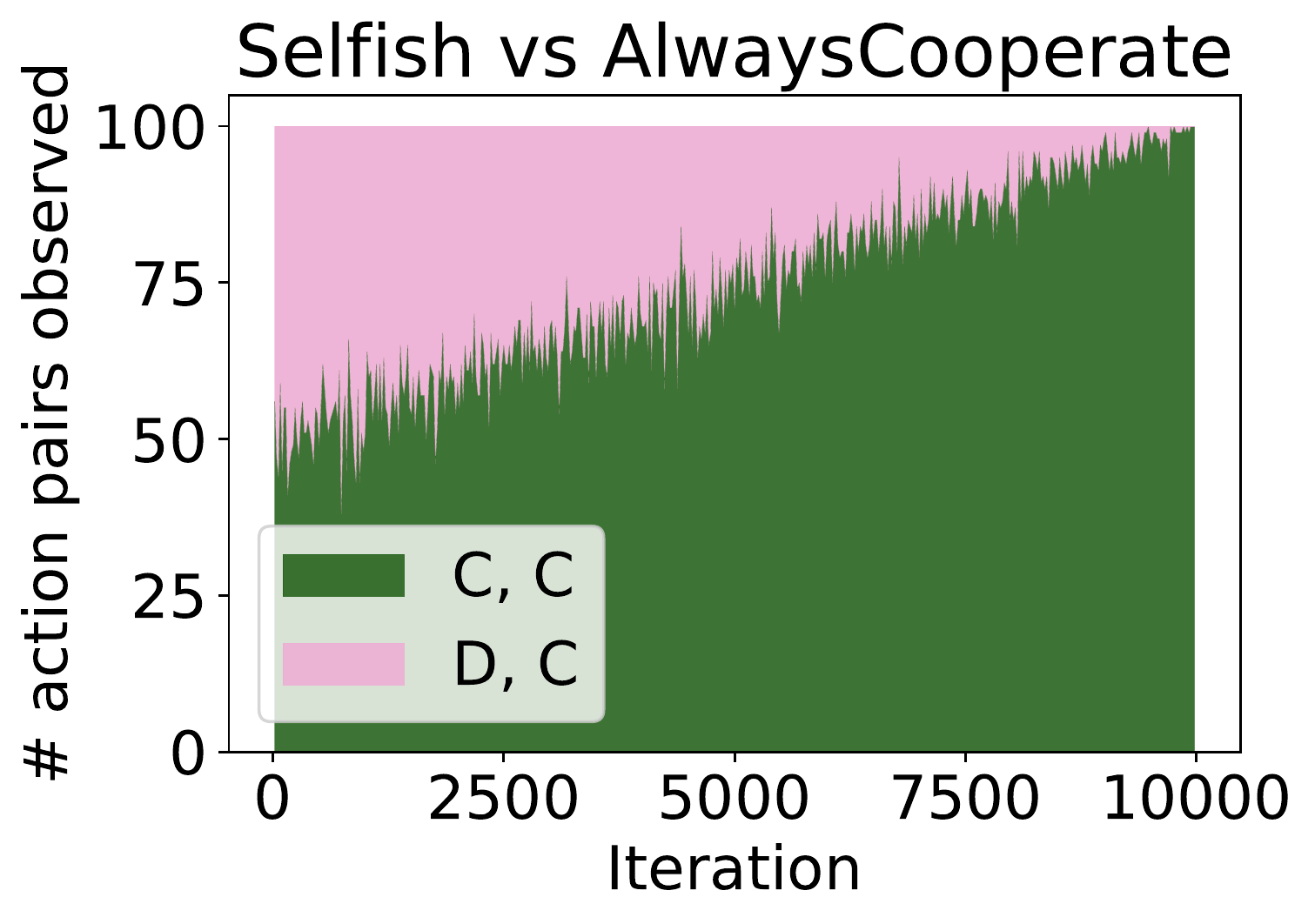}}
&\subt{\includegraphics[width=35mm]{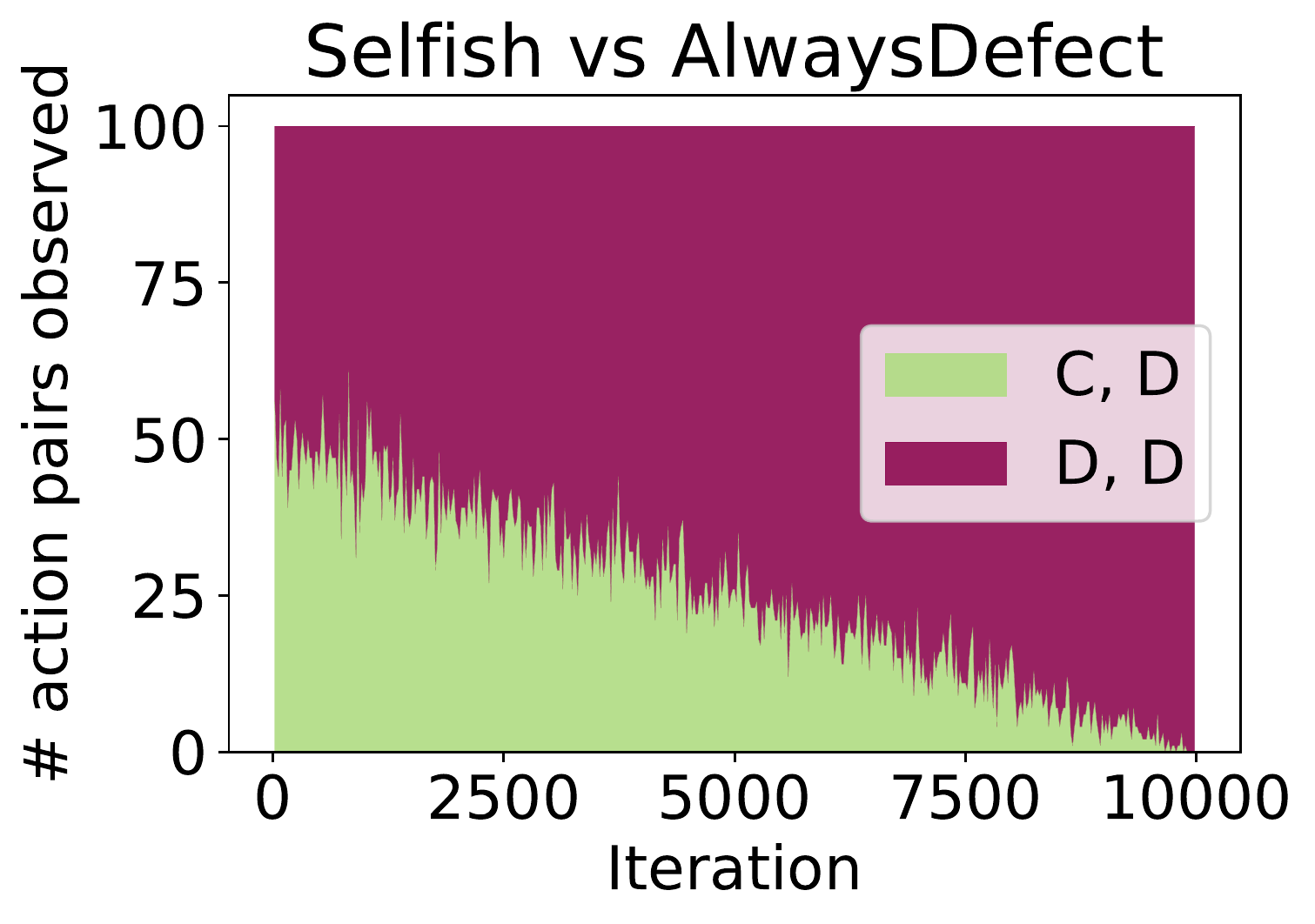}}
&\subt{\includegraphics[width=35mm]{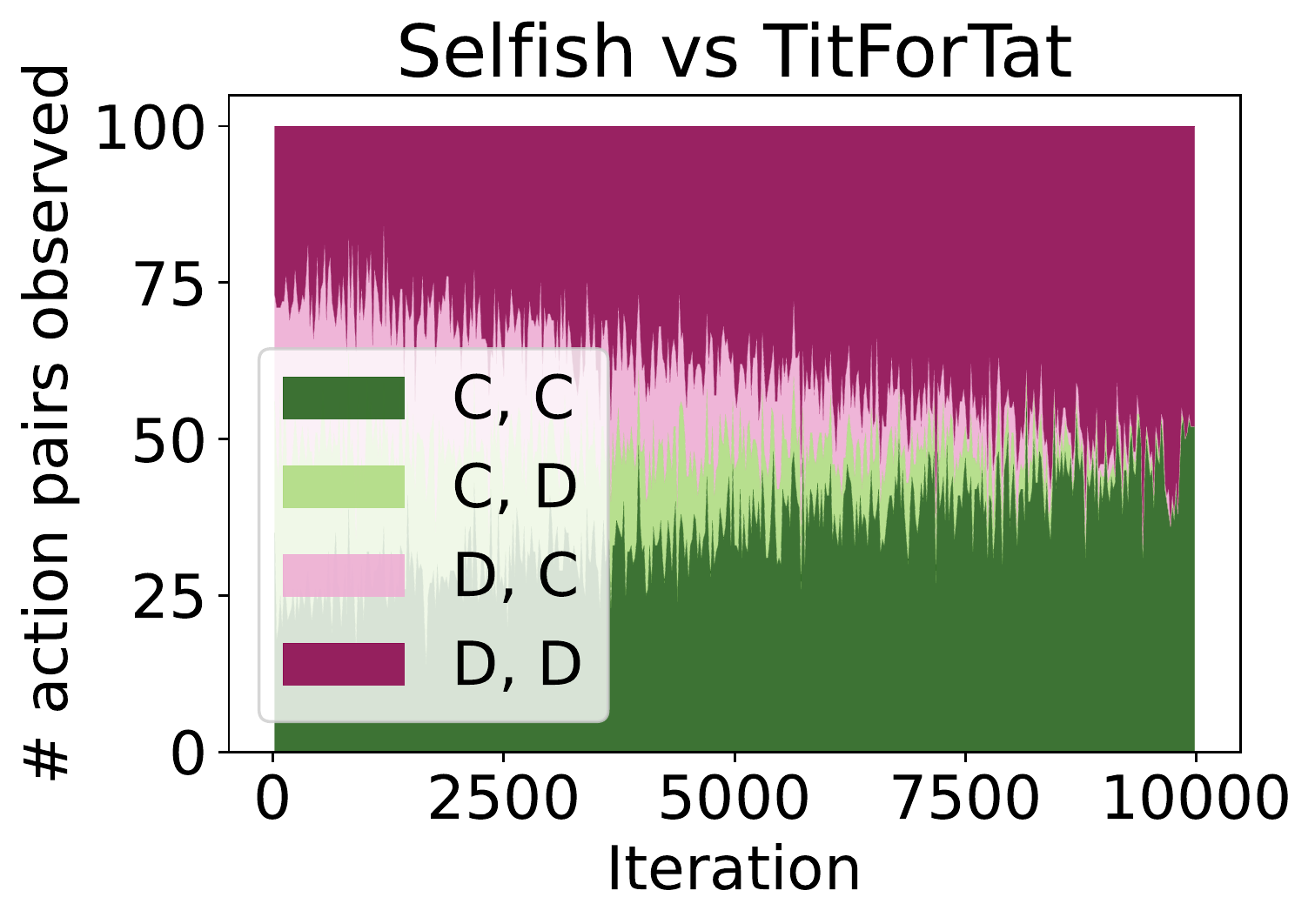}}
&\subt{\includegraphics[width=35mm]{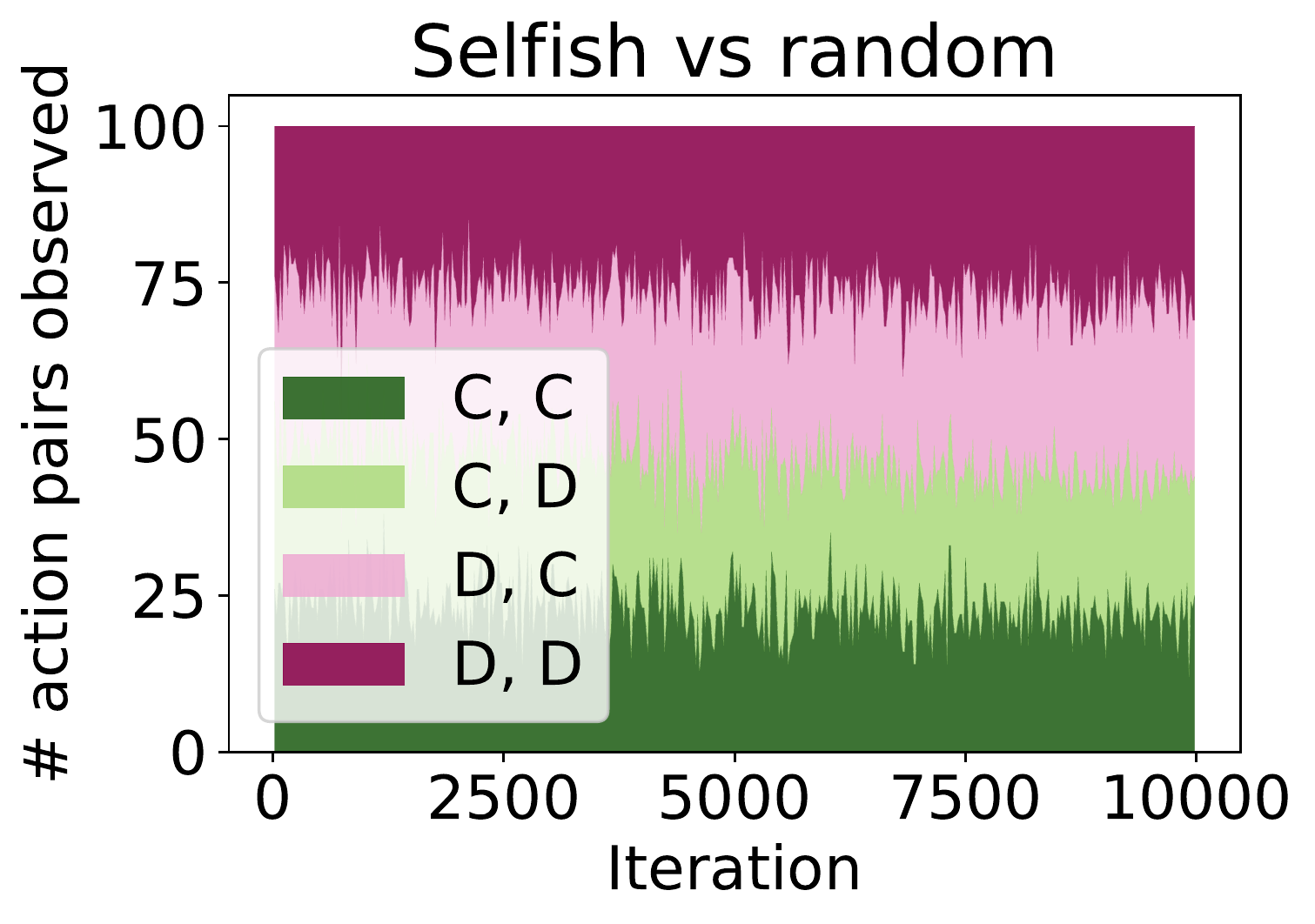}}
\\
\makecell[cc]{\rotatebox[origin=c]{90}{ Utilitarian }} & 
\subt{\includegraphics[width=35mm]{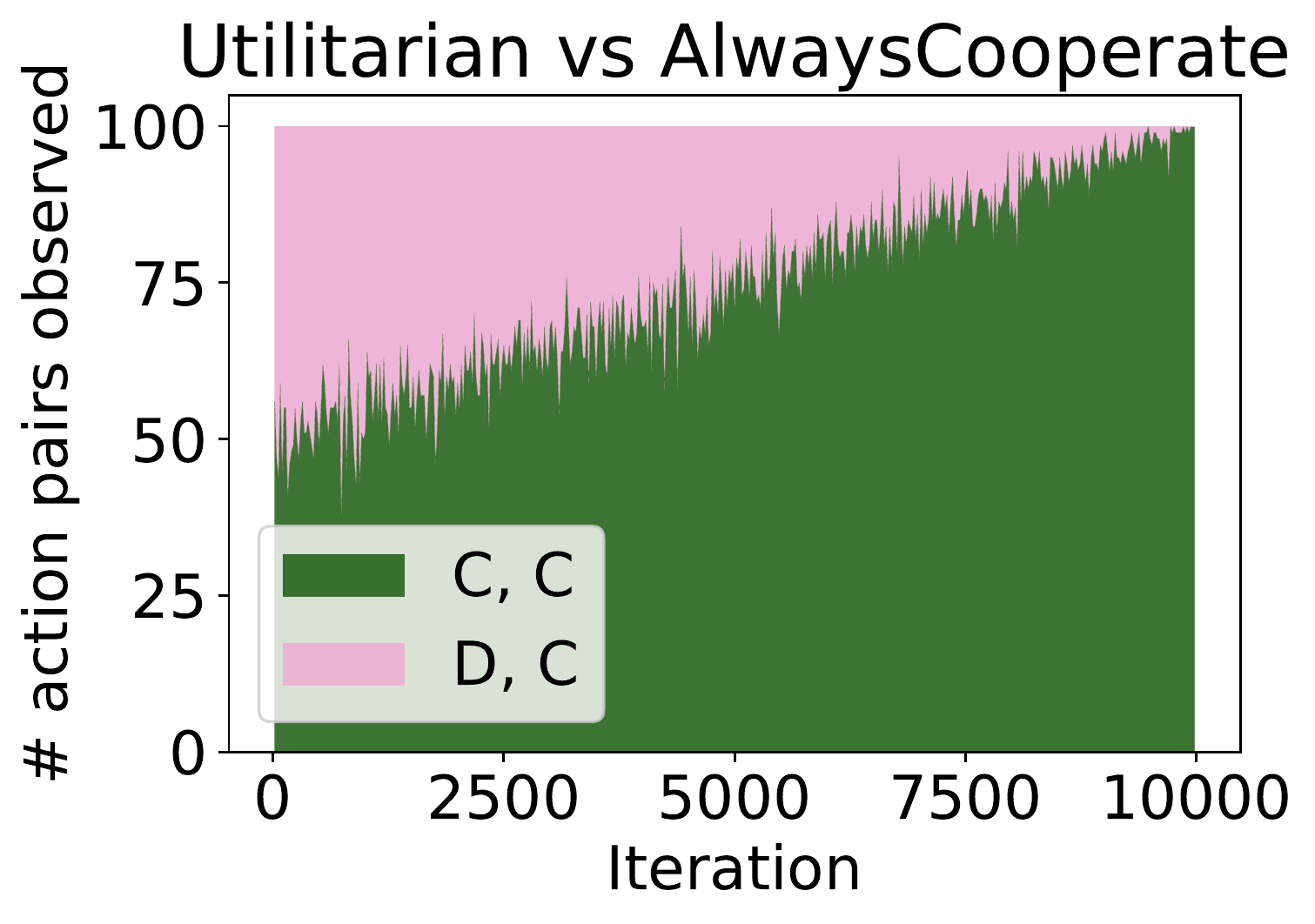}}
&\subt{\includegraphics[width=35mm]{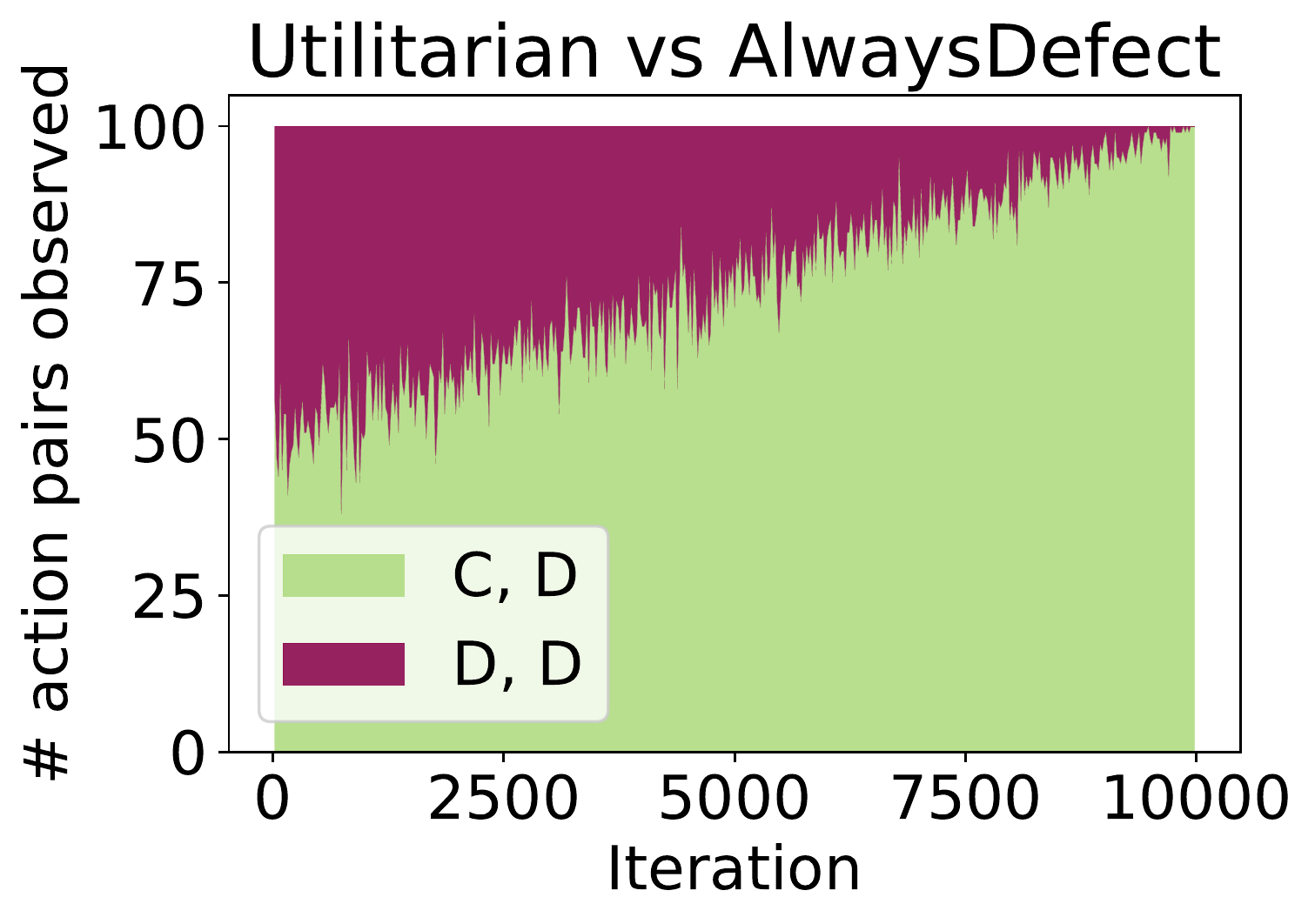}}
&\subt{\includegraphics[width=35mm]{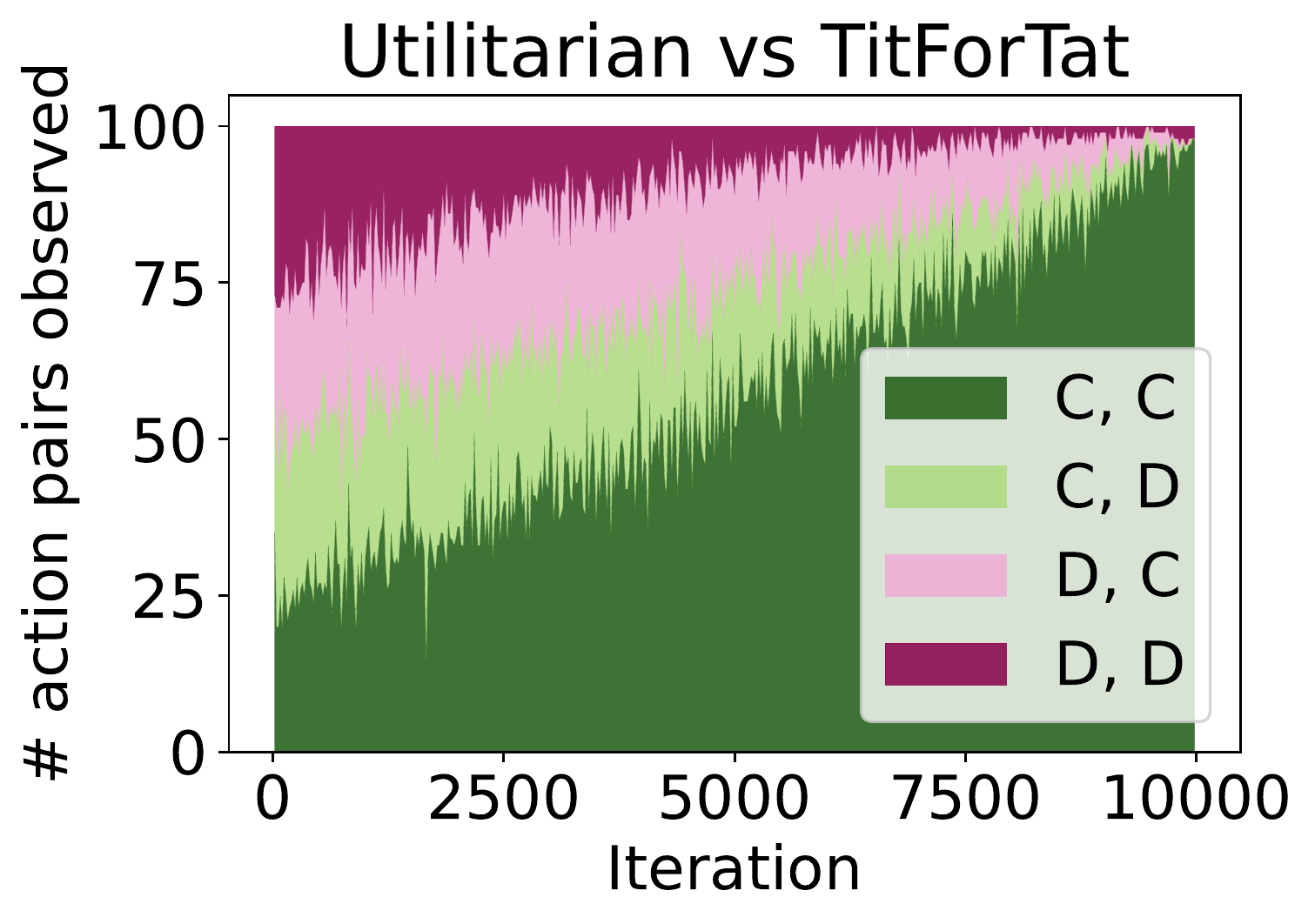}}
&\subt{\includegraphics[width=35mm]{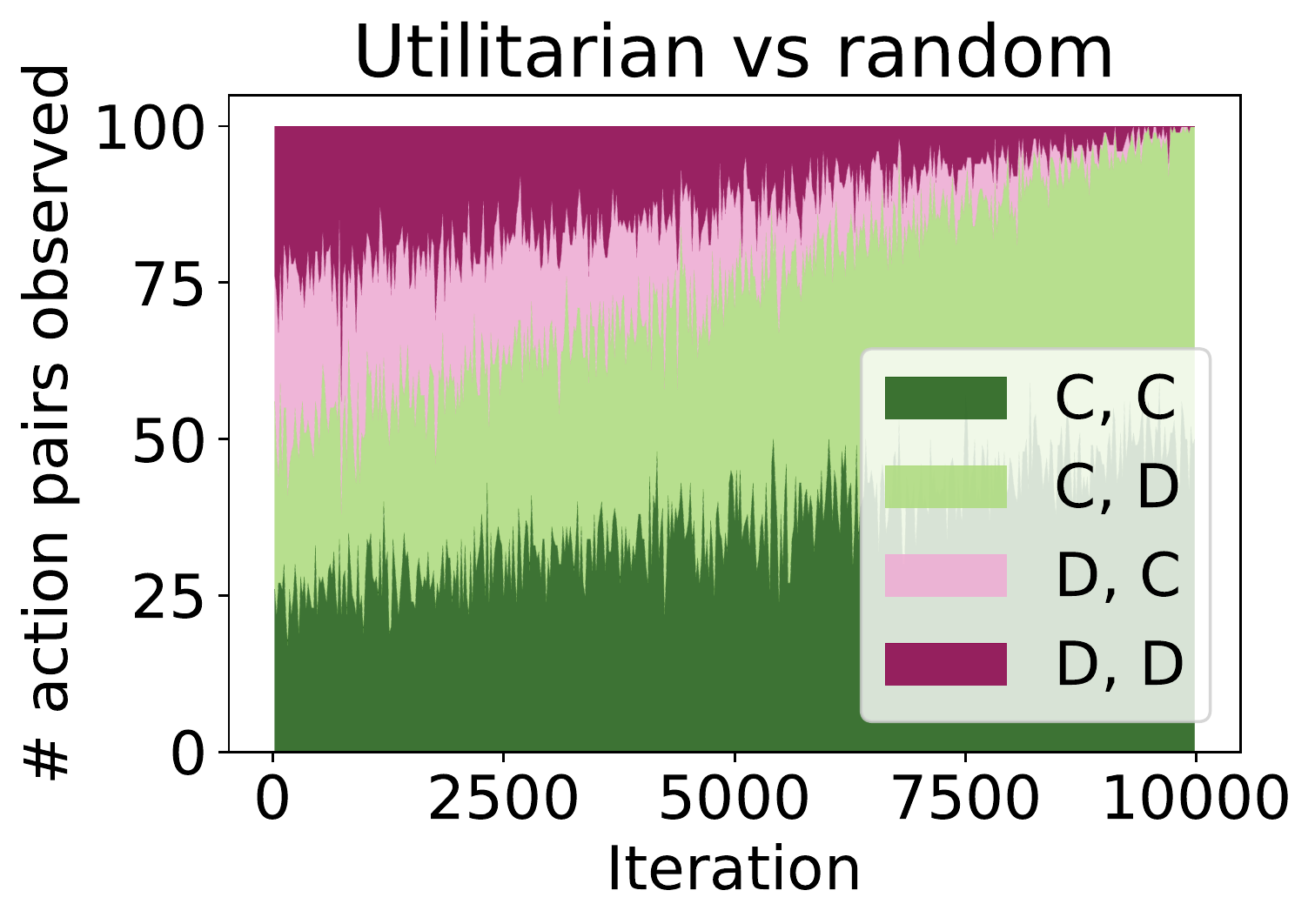}}
\\
\makecell[cc]{\rotatebox[origin=c]{90}{ Deontological }} &
\subt{\includegraphics[width=35mm]{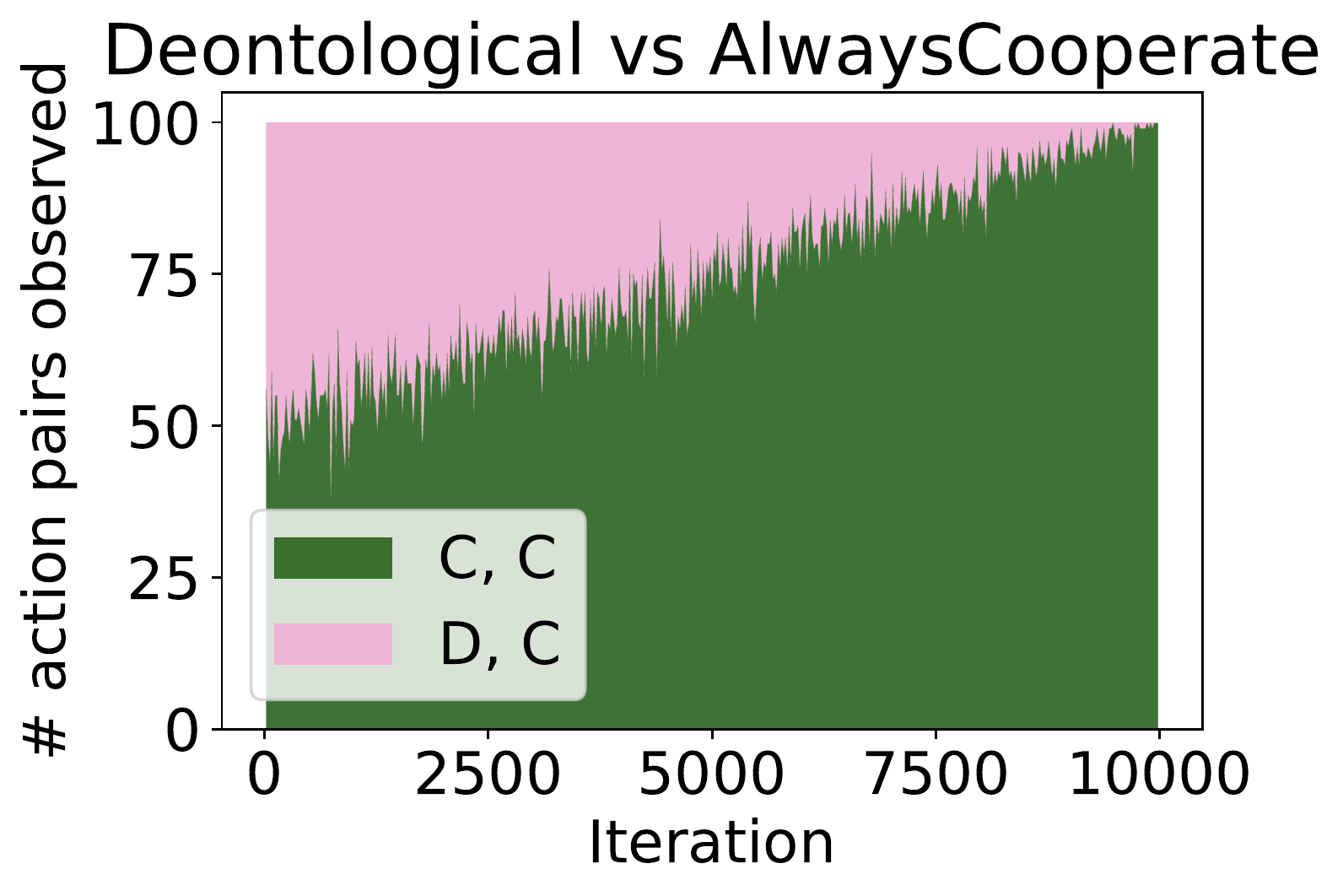}}
&\subt{\includegraphics[width=35mm]{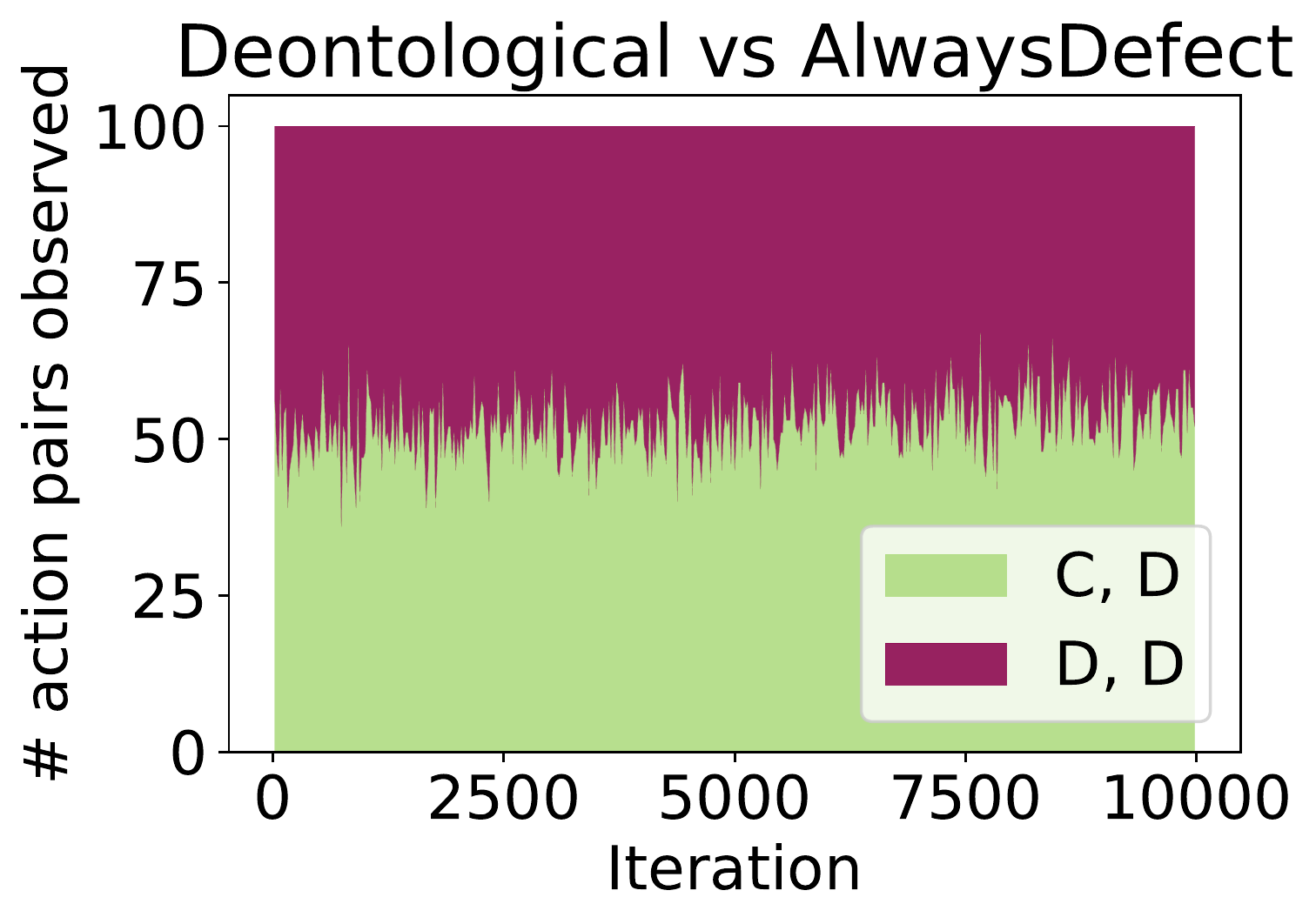}}
&\subt{\includegraphics[width=35mm]{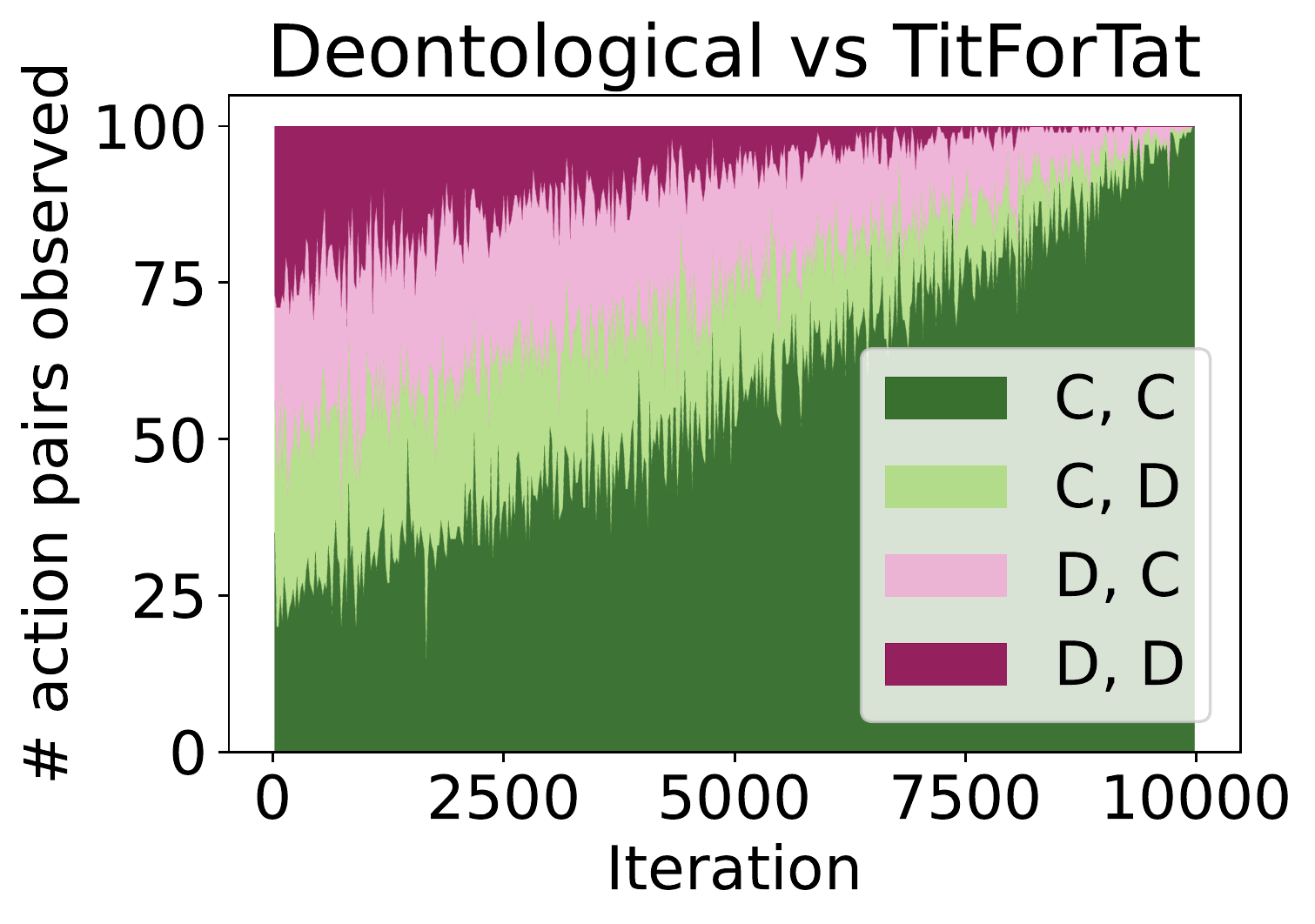}}
&\subt{\includegraphics[width=35mm]{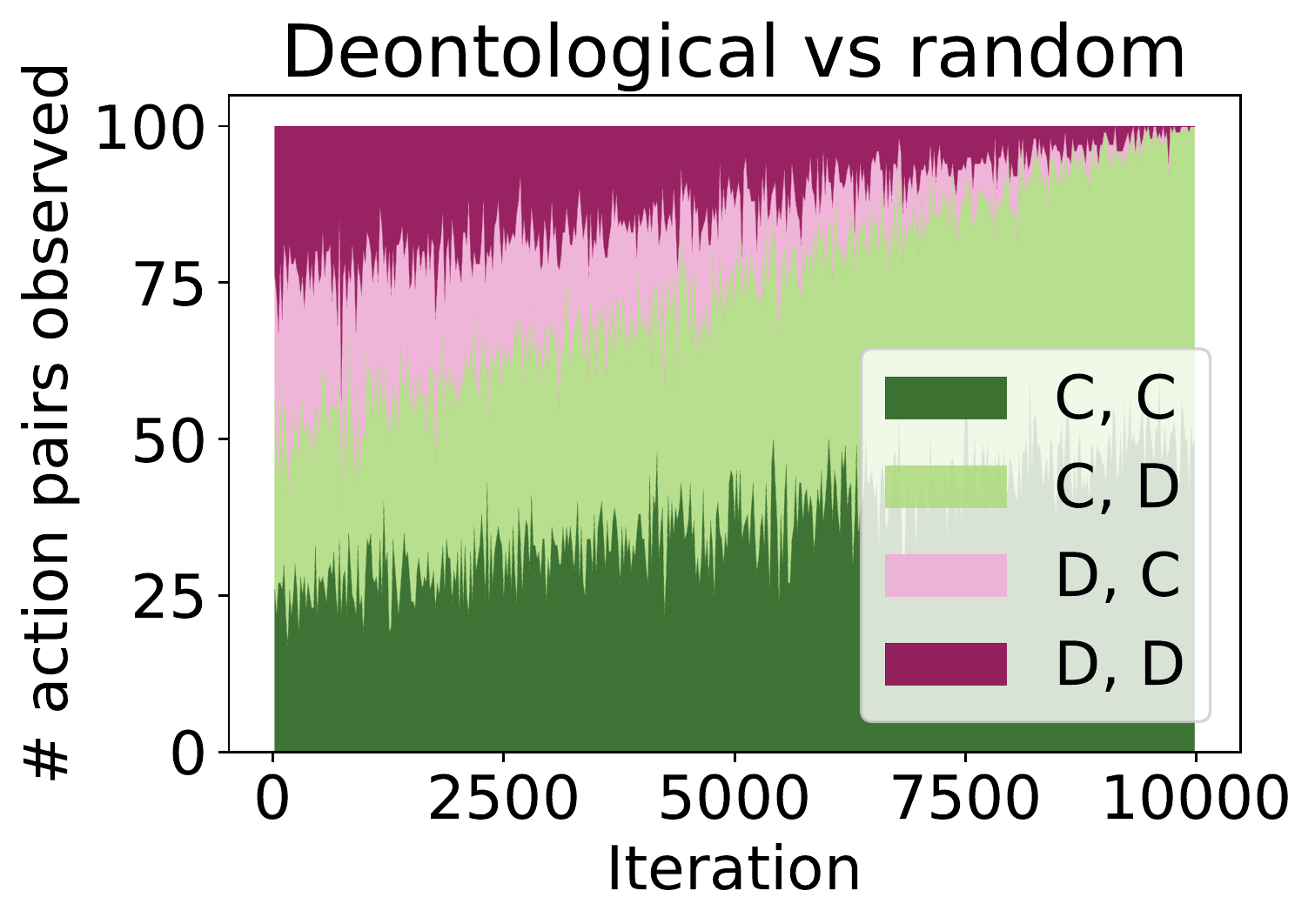}}
\\
\makecell[cc]{\rotatebox[origin=c]{90}{ Virtue-eq. }} &
\subt{\includegraphics[width=35mm]{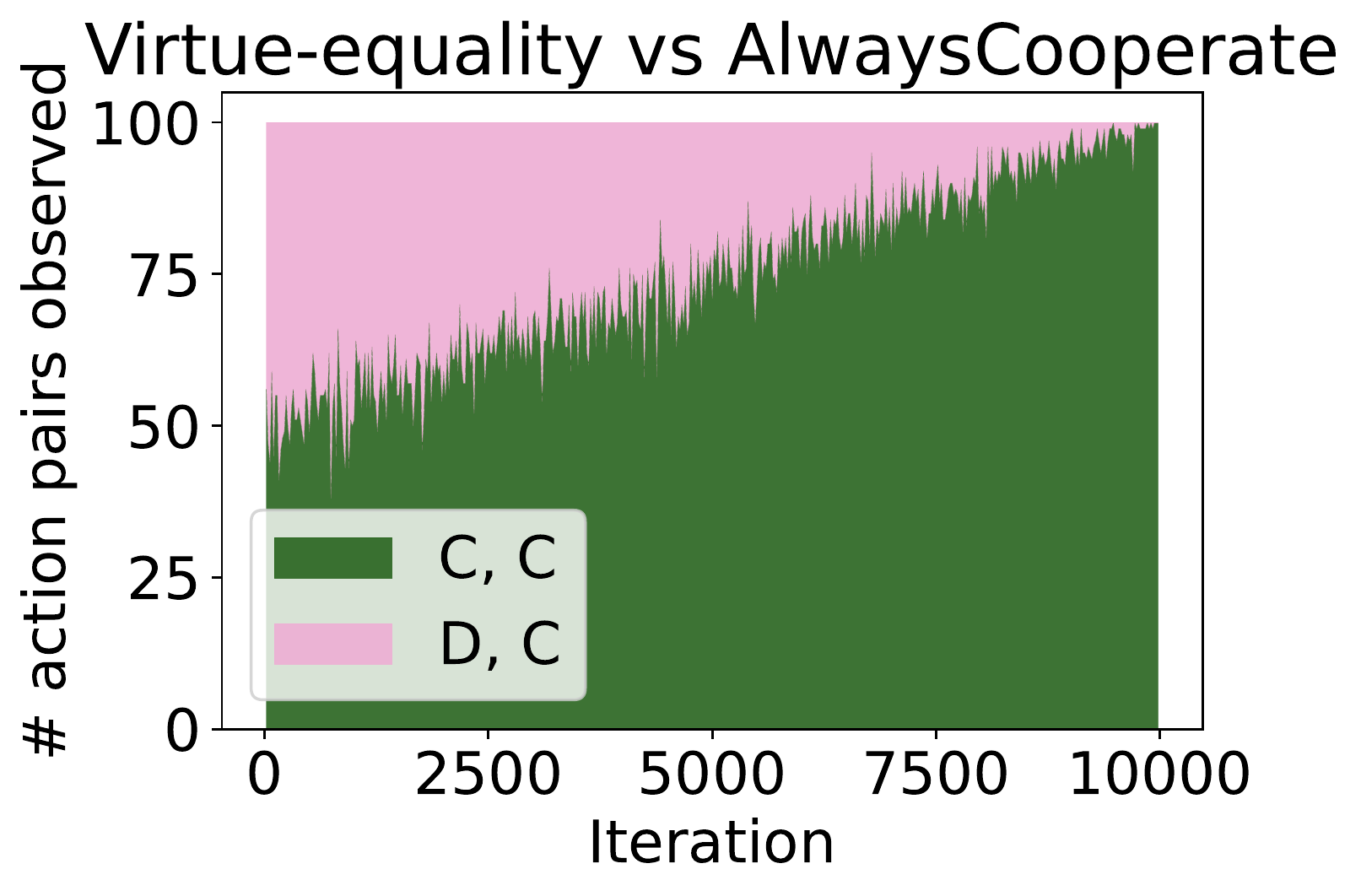}}
&\subt{\includegraphics[width=35mm]{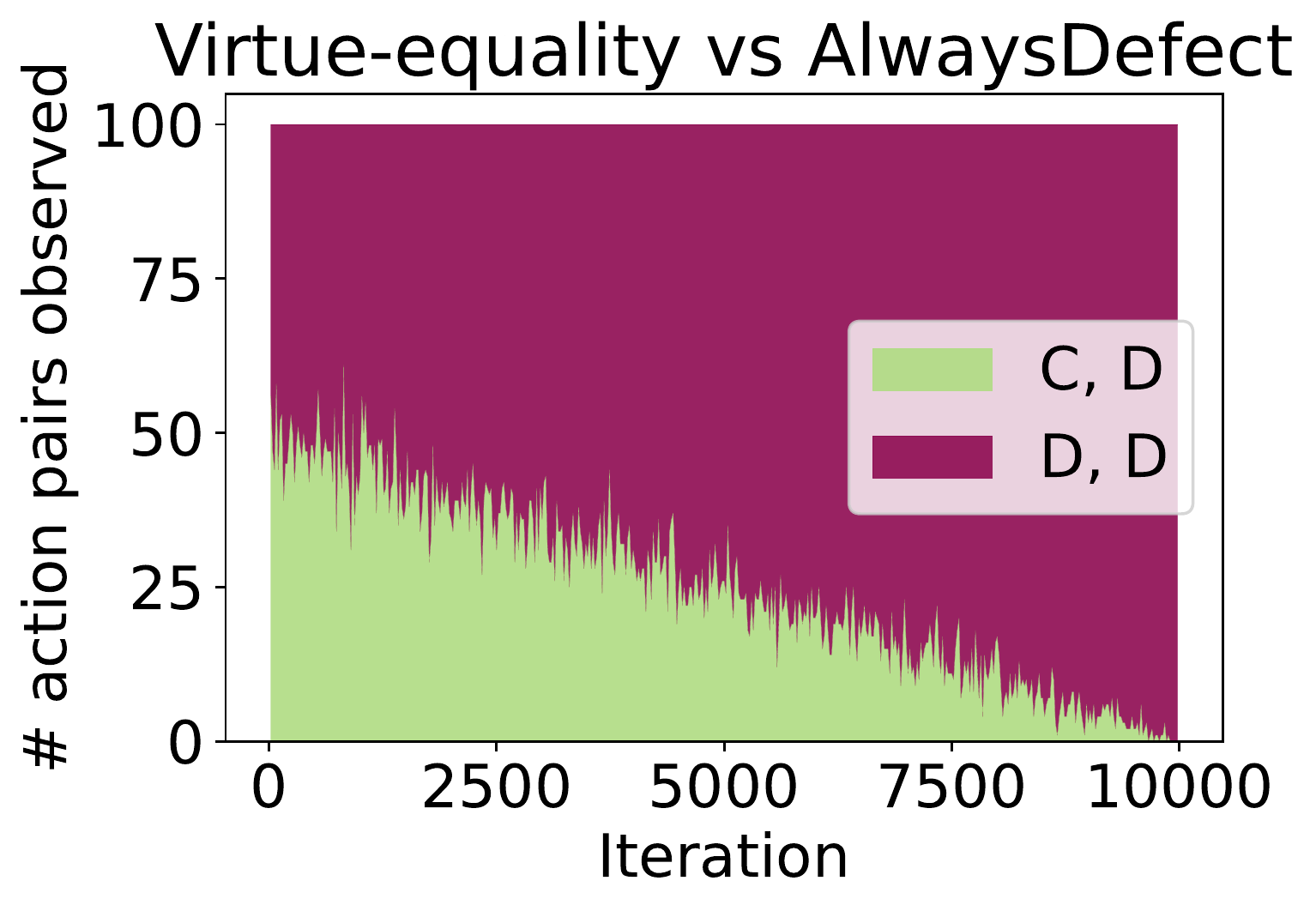}}
&\subt{\includegraphics[width=35mm]{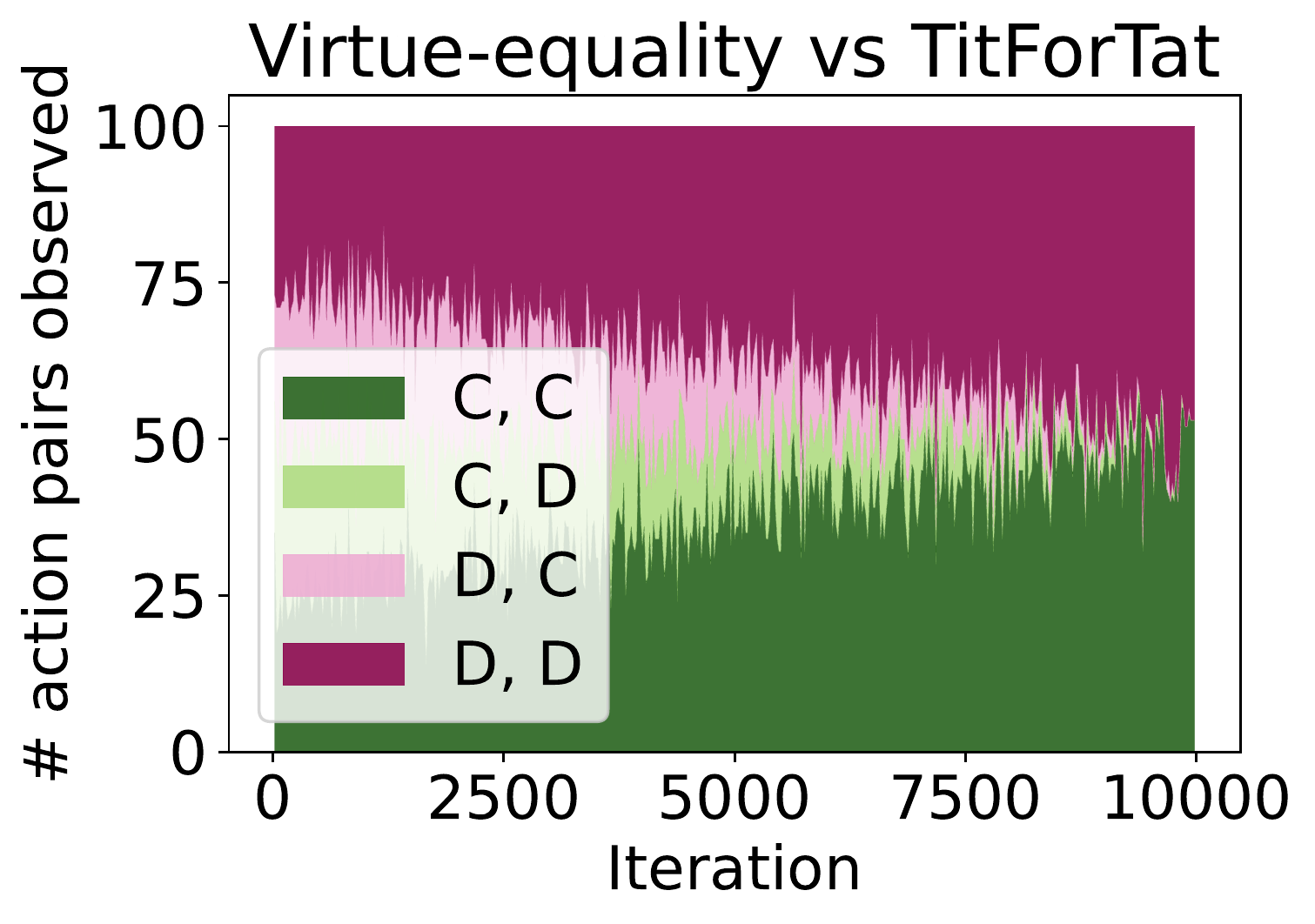}}
&\subt{\includegraphics[width=35mm]{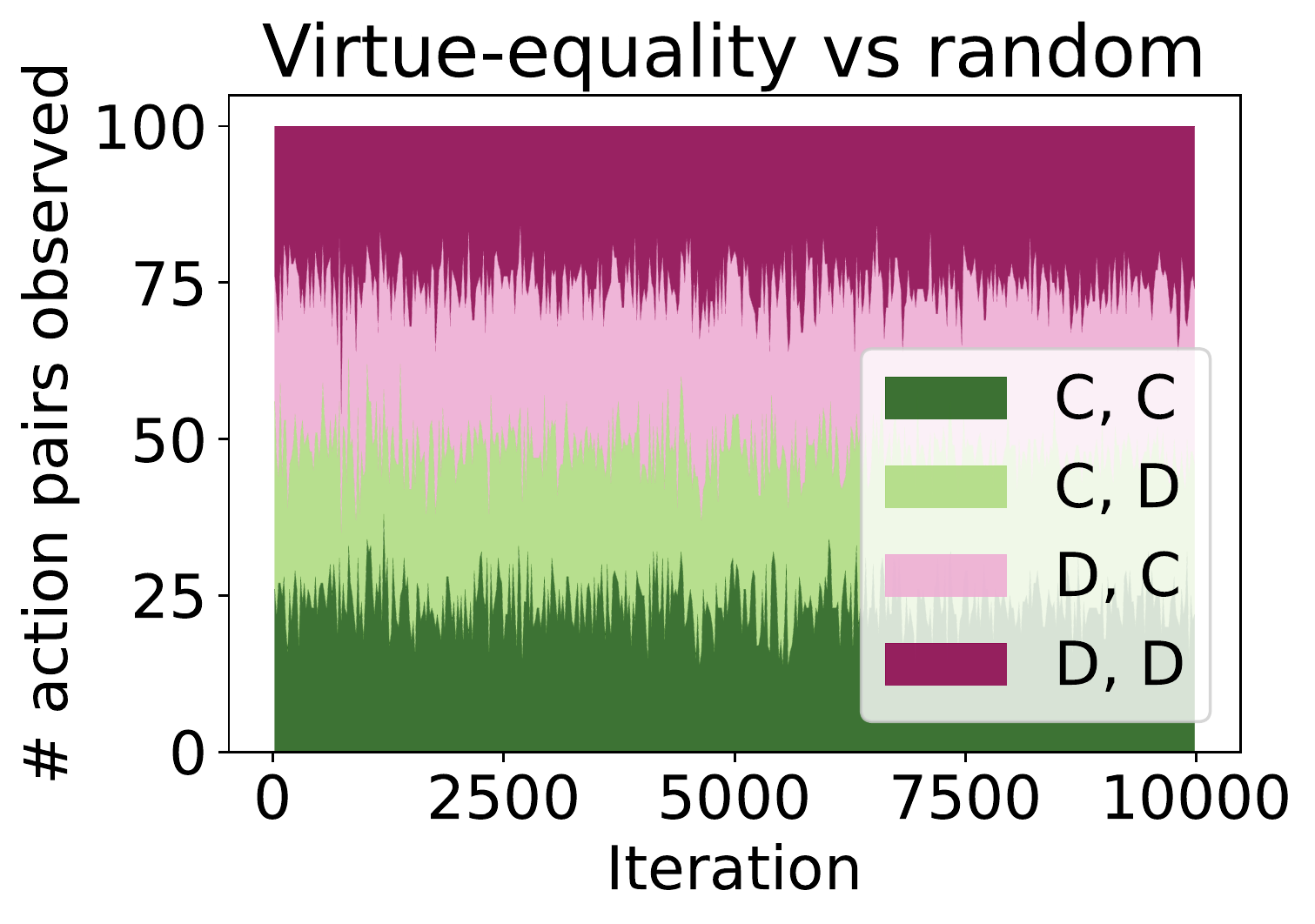}}
\\
\makecell[cc]{\rotatebox[origin=c]{90}{ Virtue-kind. }} &
\subt{\includegraphics[width=35mm]{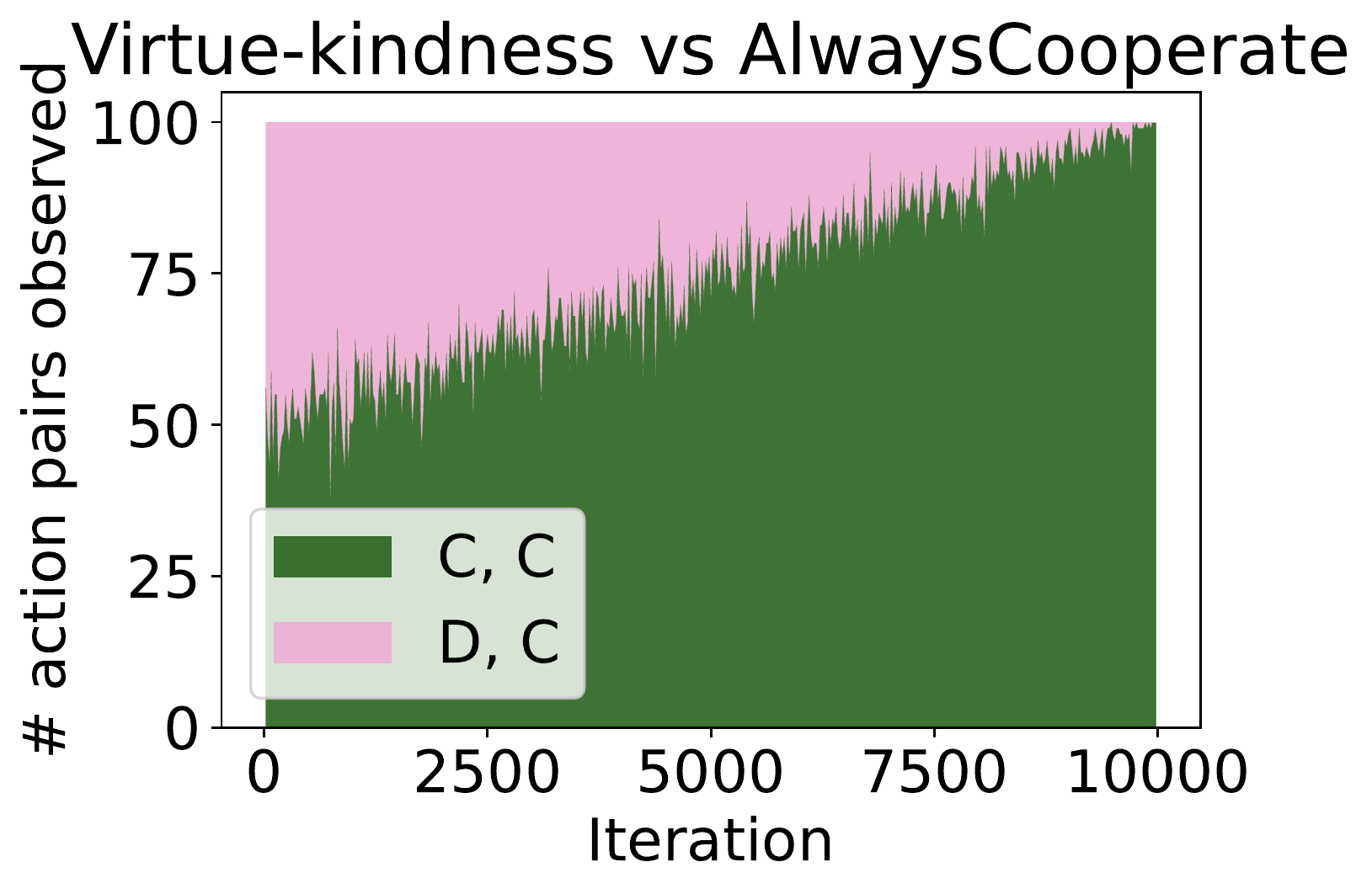}}
&\subt{\includegraphics[width=35mm]{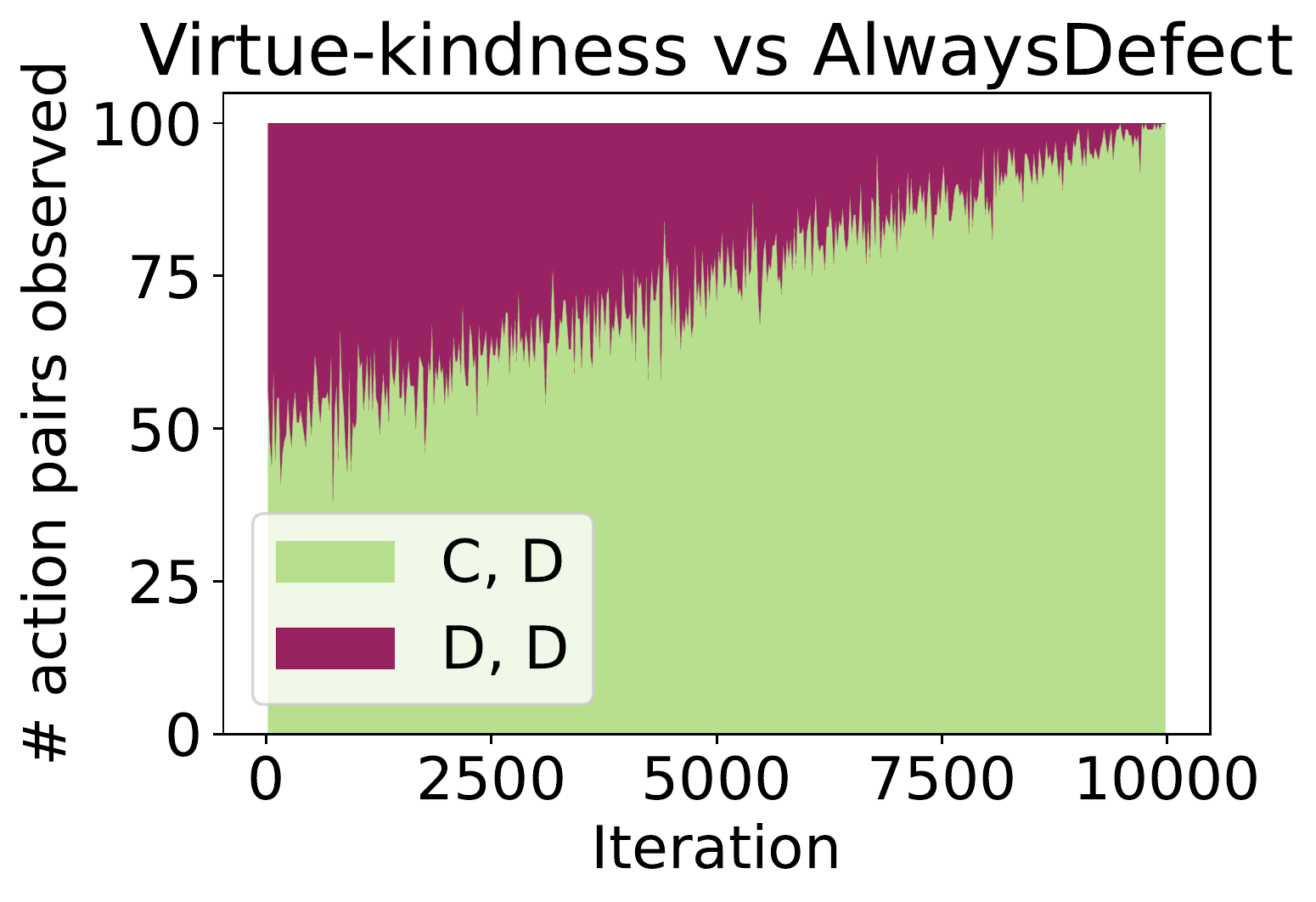}}
&\subt{\includegraphics[width=35mm]{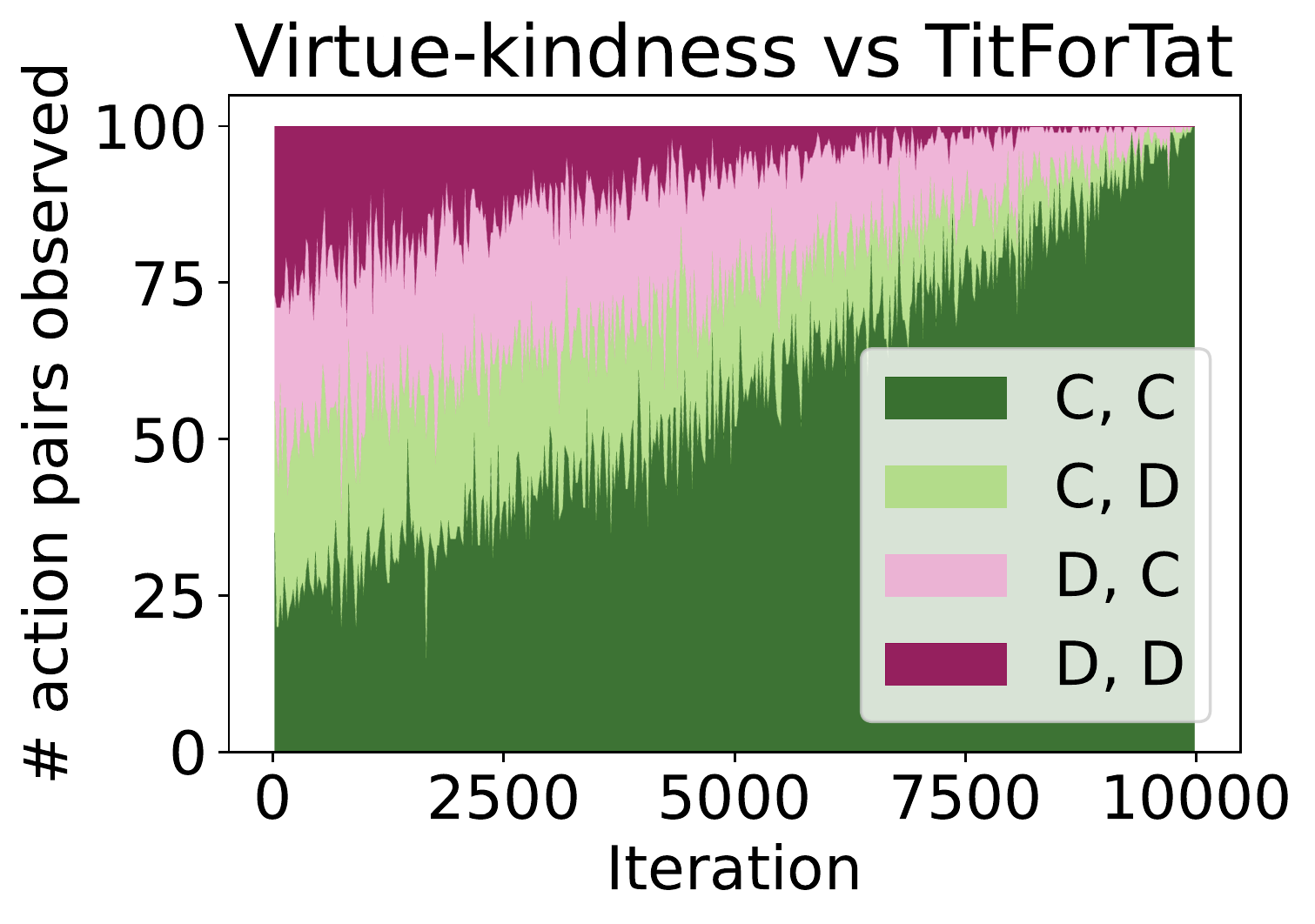}}
&\subt{\includegraphics[width=35mm]{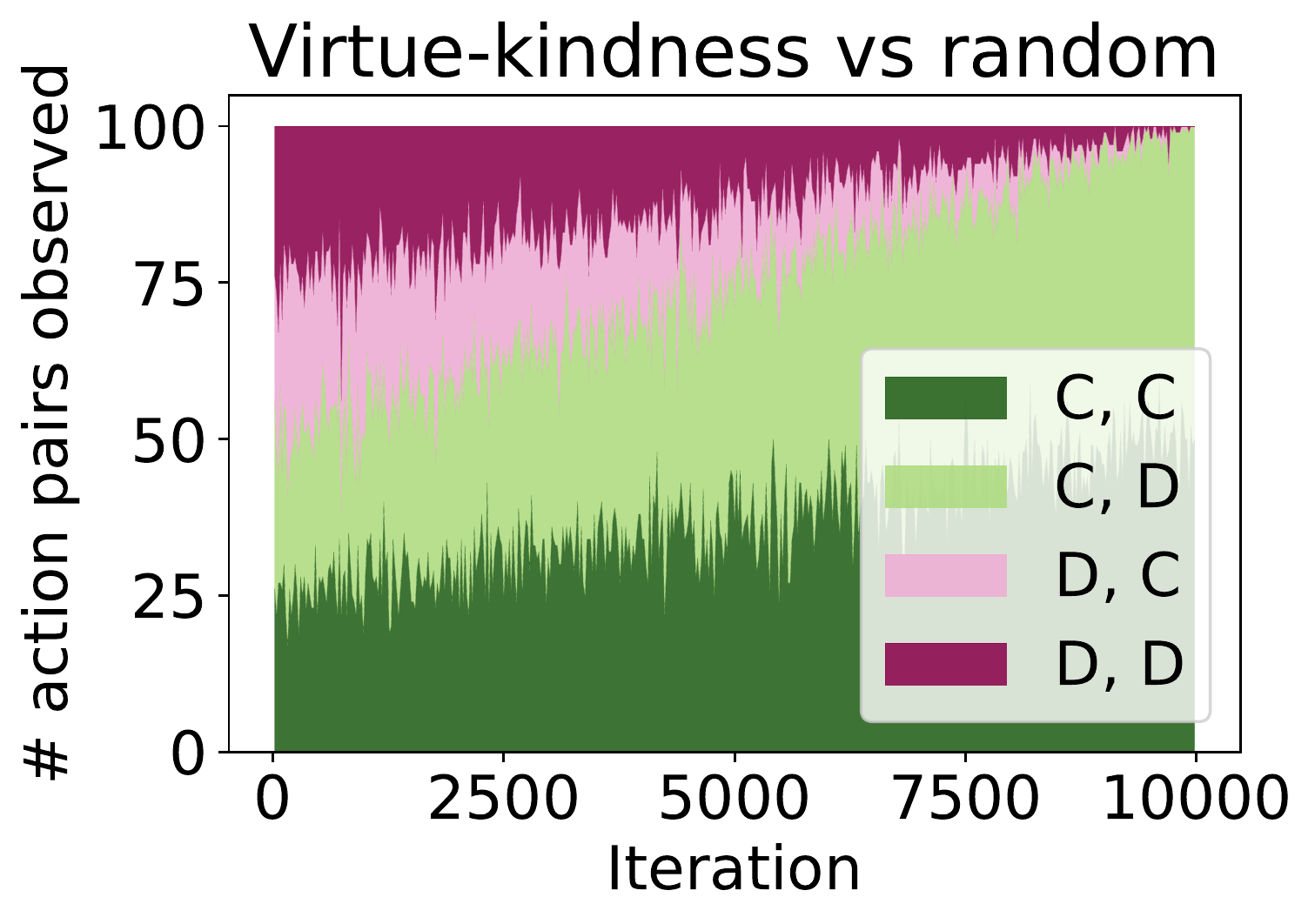}}
\\
\makecell[cc]{\rotatebox[origin=c]{90}{ Virtue-mix. }} &
\subt{\includegraphics[width=35mm]{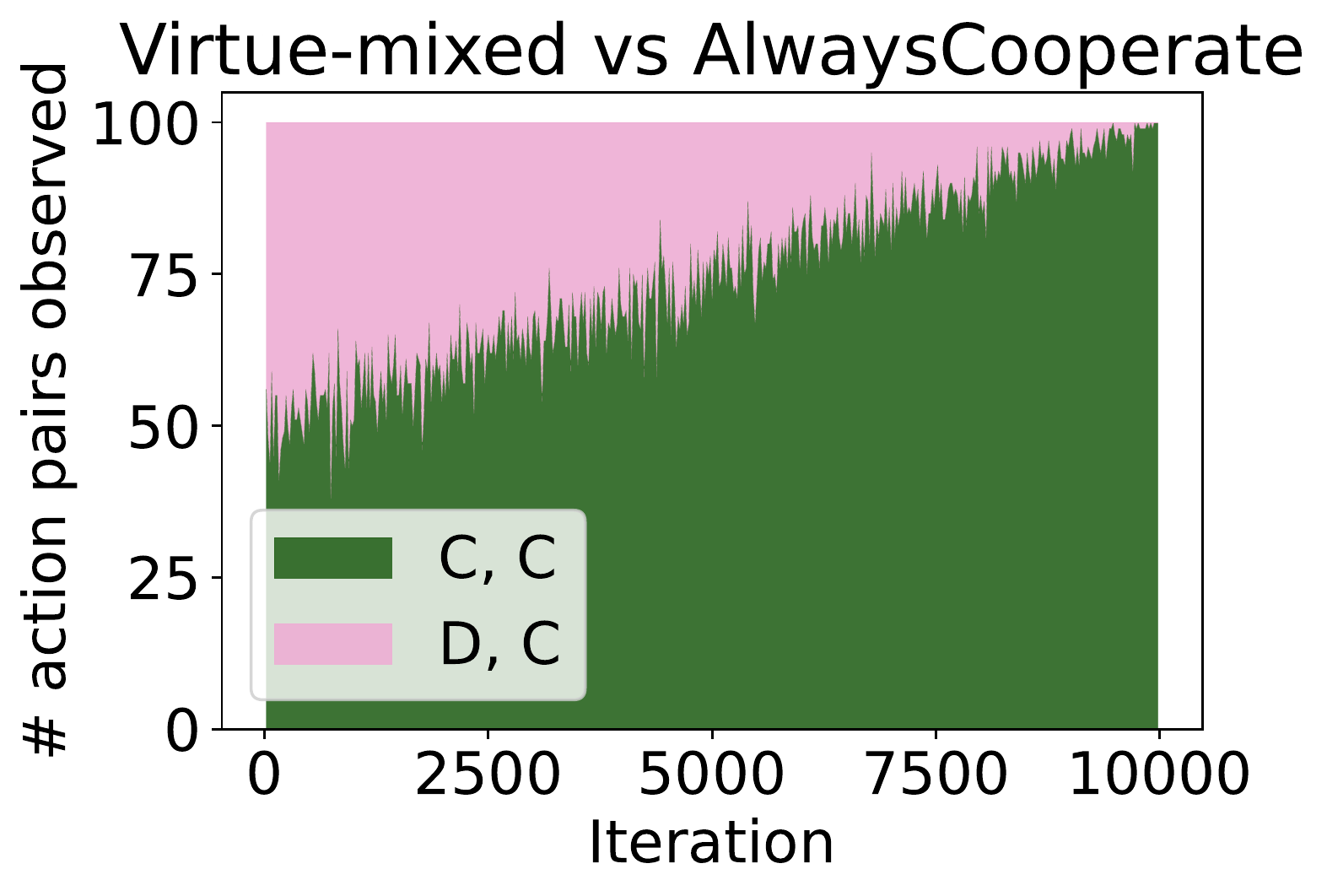}}
&\subt{\includegraphics[width=35mm]{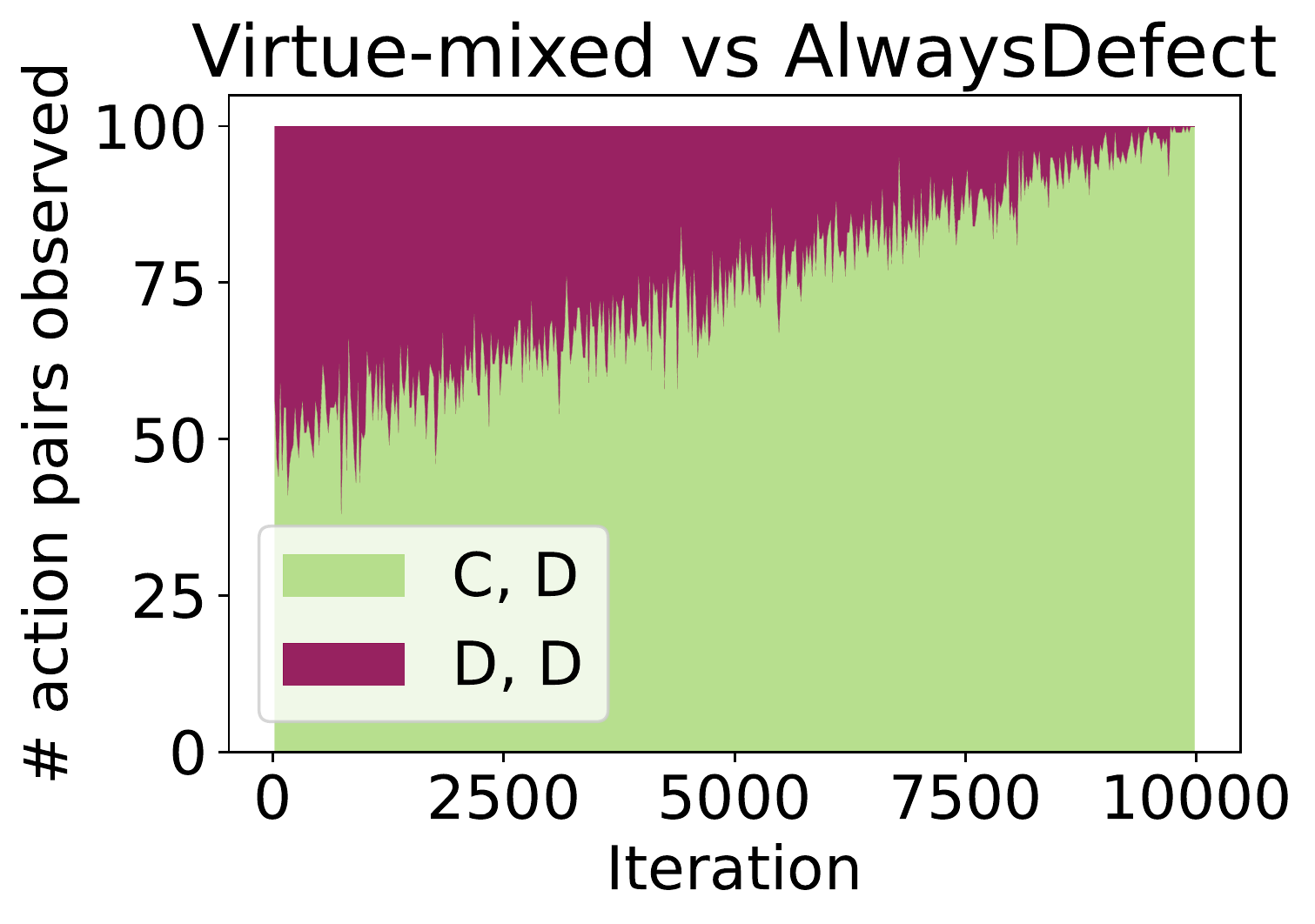}}
&\subt{\includegraphics[width=35mm]{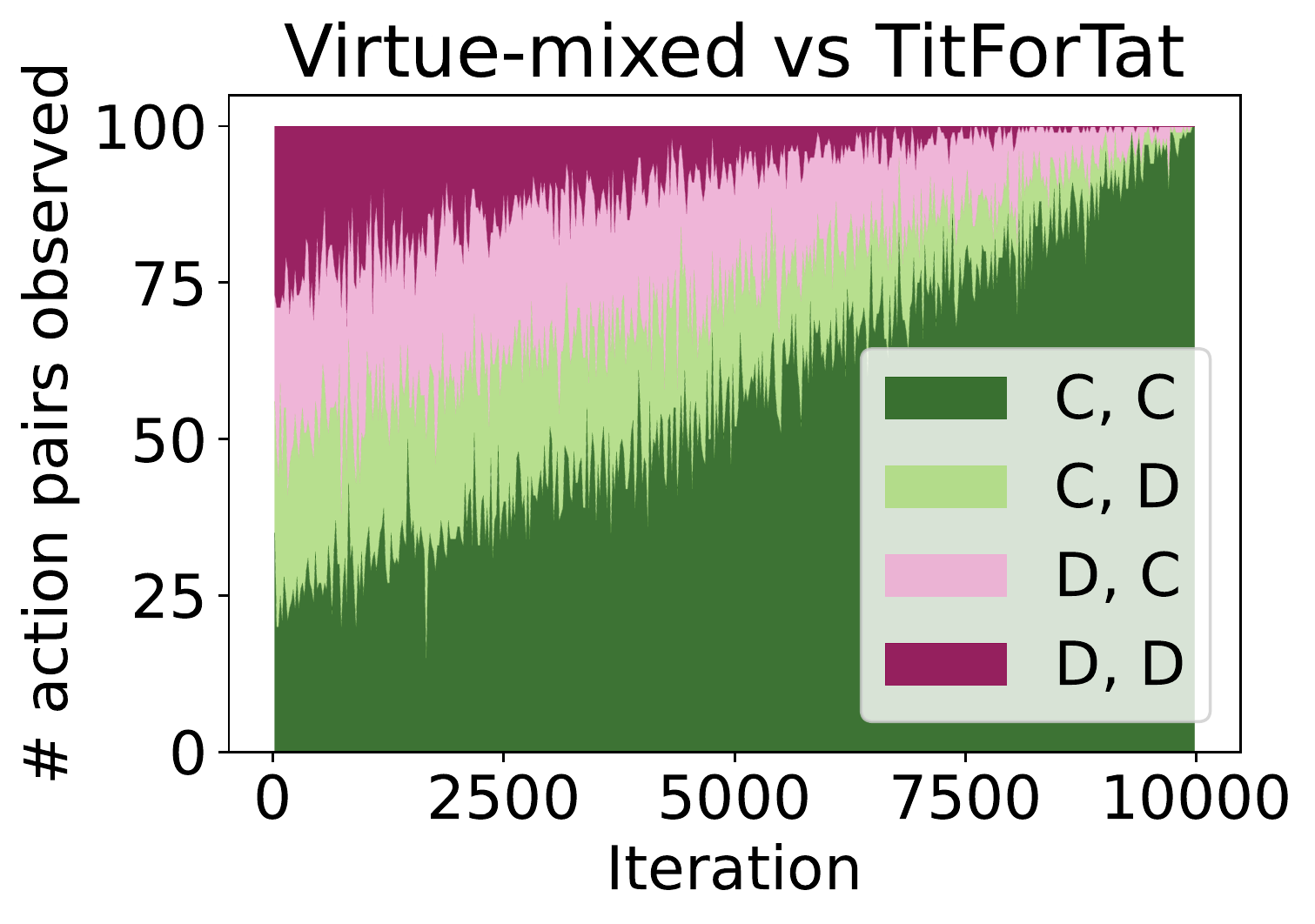}}
&\subt{\includegraphics[width=35mm]{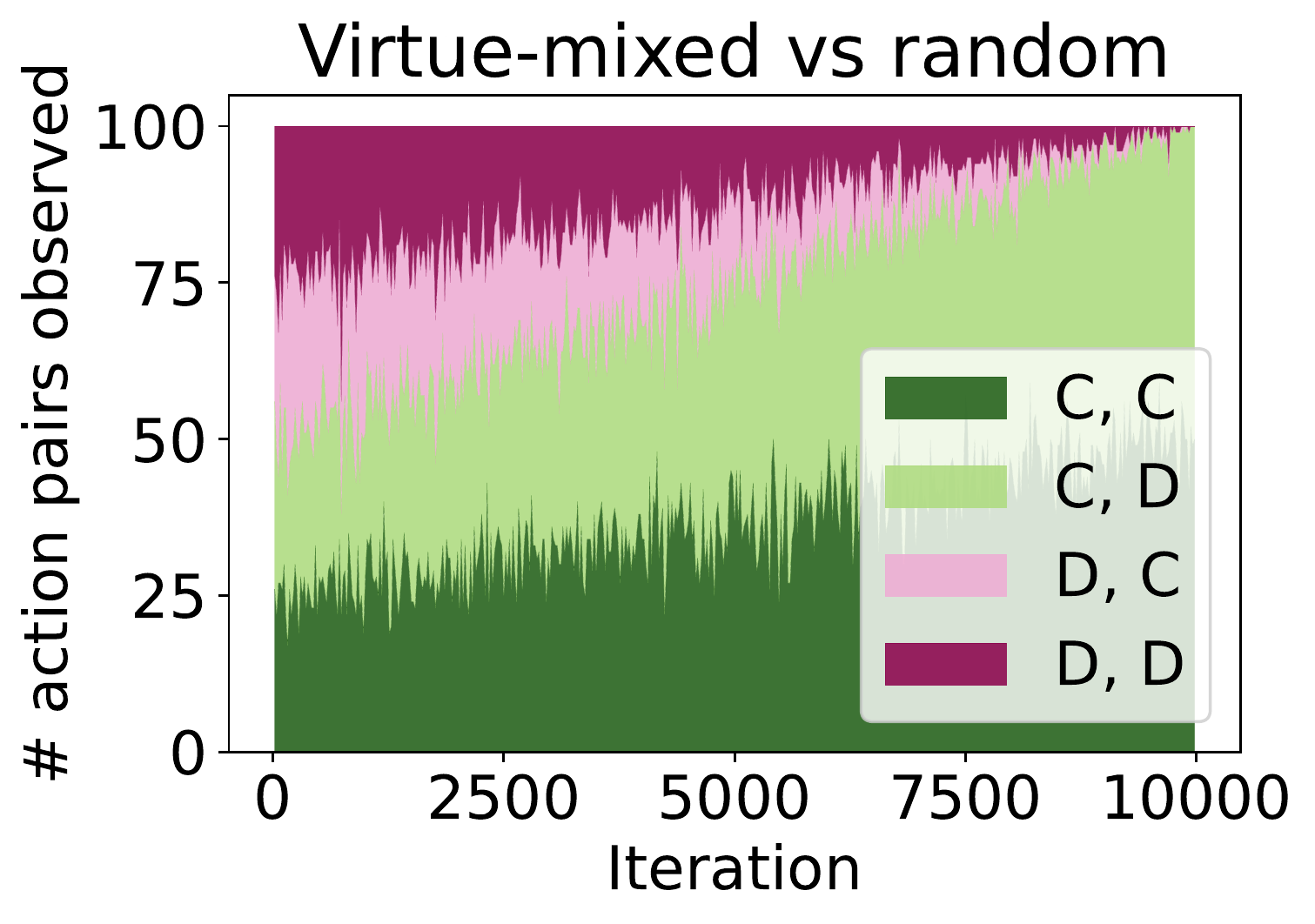}}
\\
\bottomrule
\end{tabular}
\caption{Iterated Stag Hunt game. Simultaneous pairs of actions observed over time. Learning player $M$ (row) vs. static opponent $O$ (column).}
\label{fig:action_pairs_baseline_STH}
\end{figure*}

In Figures \ref{fig:action_pairs_baseline_IPD}-\ref{fig:action_pairs_baseline_STH} we present simultaneous actions over time for learning agents versus static opponents - \textit{Always Cooperate, Always Defect, Tit for Tat and Random}. These were run as a benchmark before implementing learning pairs of agents - but provide clear insights into the behavior of our moral agents against predictable opponents whose behavior is stable. 

\subsection{Summary of Simultaneous Actions - Learning Player vs Static Opponent}

In this section we provide a summary of the simultaneous actions performed on the last iteration against static (predictable) opponents.

\begin{figure*}[!h]
\centering
\includegraphics[width=35mm]{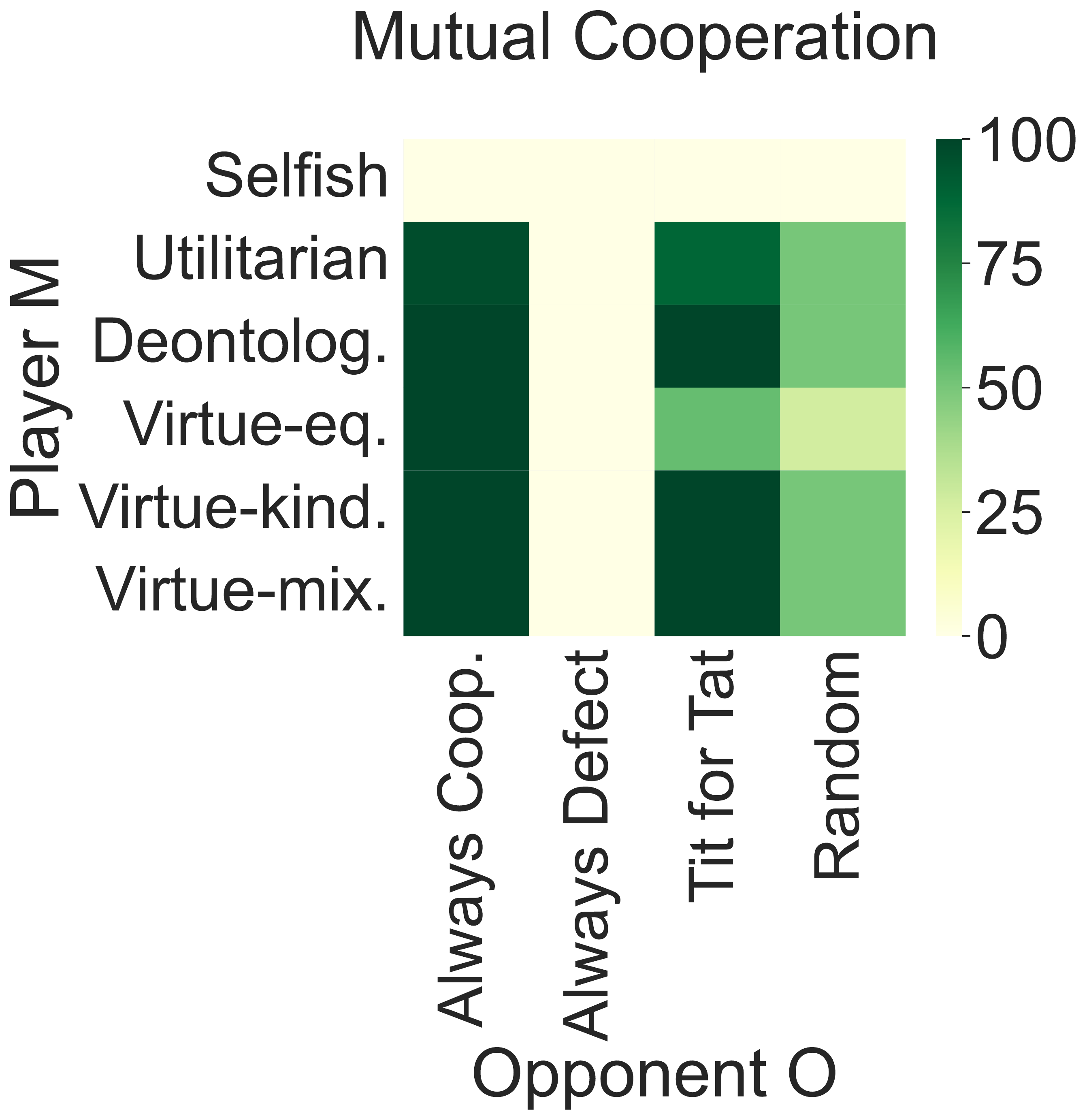}
\includegraphics[width=35mm]{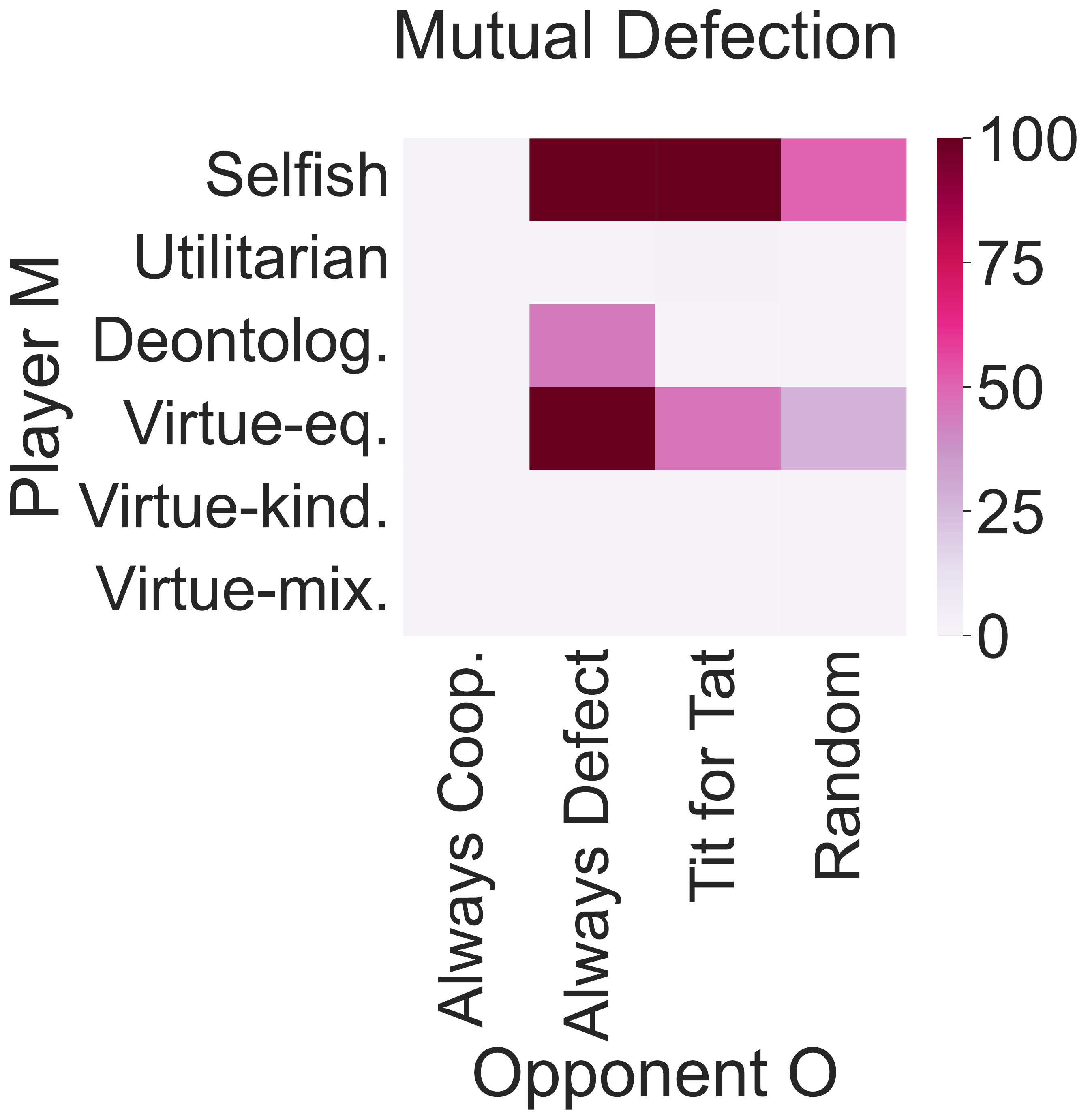}
\includegraphics[width=35mm]{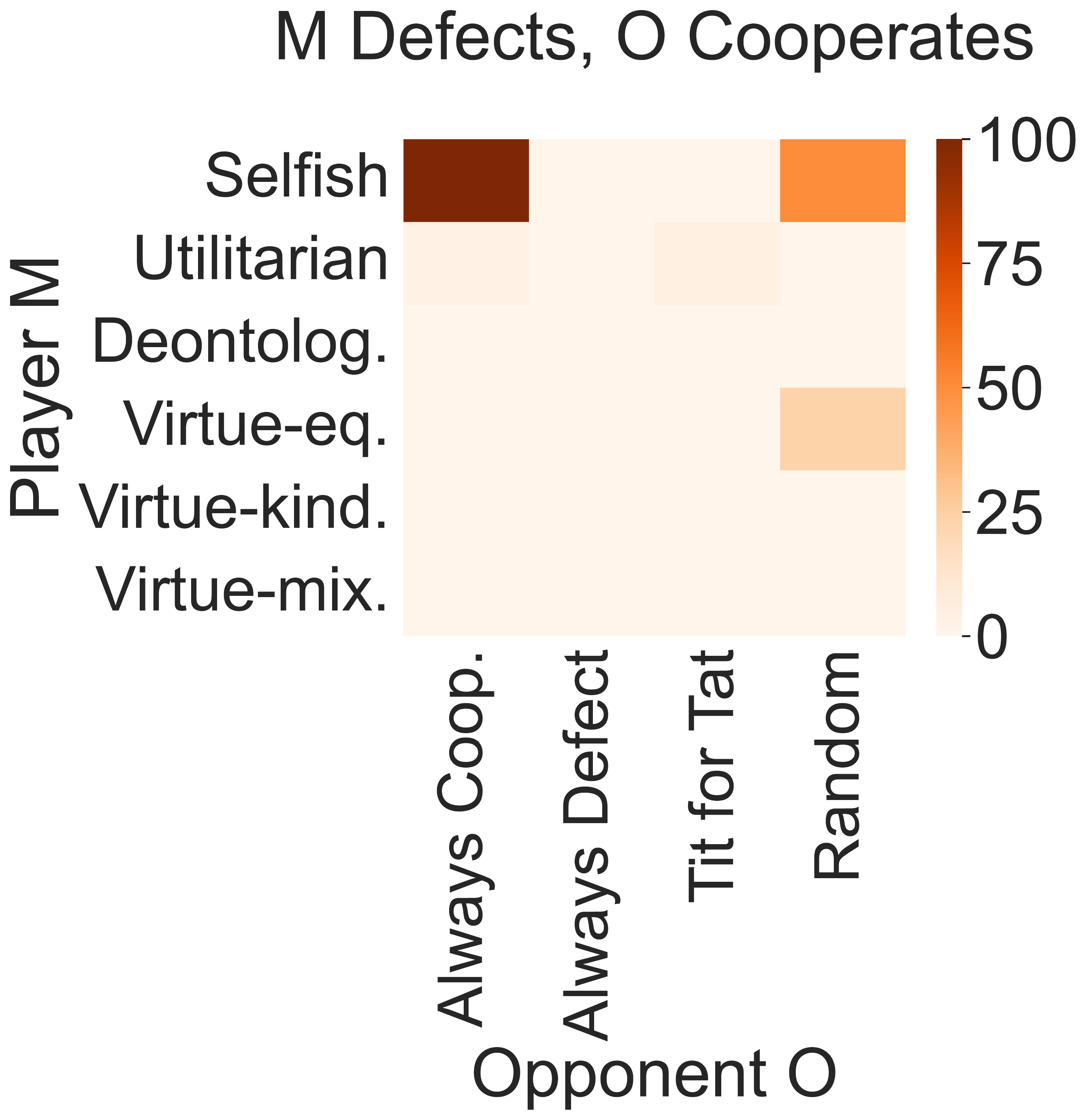}
\includegraphics[width=35mm]{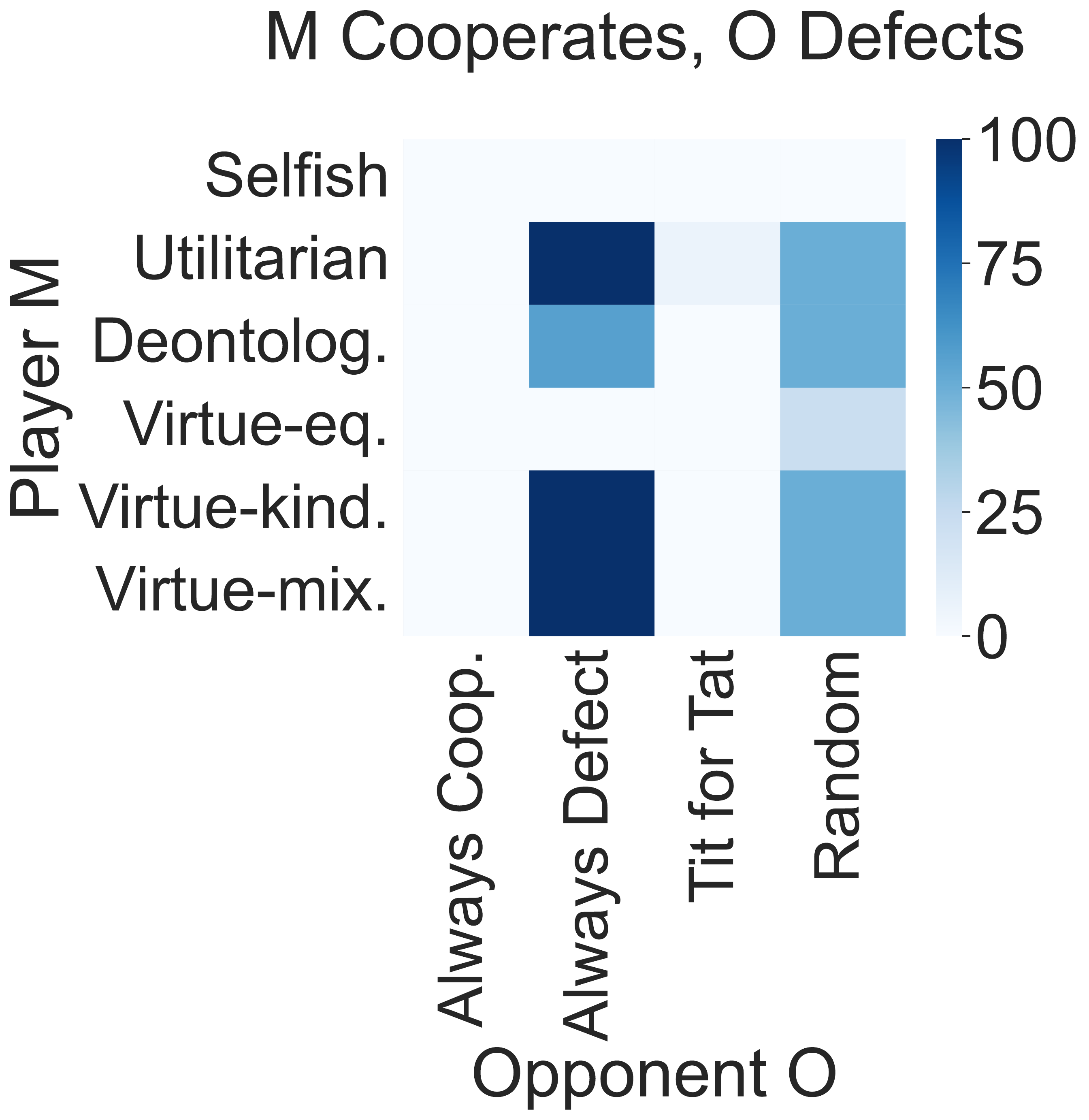}
\caption{Iterated Prisoner's Dilemma game. Simultaneous actions played by player $M$ type and the static opponent $O$ type at the end of the learning period (10000 iterations). Action pairs are displayed as a percentage over the 100 runs.}
\label{fig:actions_static_IPD}
\end{figure*}

Considering learning against static opponents on the Iterated Prisoner's Dilemma (Figure \ref{fig:actions_static_IPD}), the traditional \textit{Selfish} agent learns to Defect on 100\% of the runs against everyone. The \textit{Virtue-equality} agent learns efficiently against \textit{Always Cooperate} and defends itself against exploitation by \textit{Always Defect}. The \textit{Utilitarian}, \textit{Virtue-kindness} and \textit{Virtue-mixed} agents also learn efficiently against \textit{Always Cooperate}, but do not protect themselves from exploitation by an \textit{Always Defect} agent (see exploitation in blue). The \textit{Deontological} agent is able to defend itself somewhat better against an \textit{Always Defect} agent (by achieving mutual defection on half of the runs) because its reward function allows it to play randomly against a defector. Against the reciprocal \textit{Tit for Tat}, most moral agents learn to always cooperate, but the \textit{Virtue-equality} agent converges to 50\% mutual defection. This once again highlights that a dyadic interaction between agents focused on equality (\textit{Virtue-equality}), reciprocity (\textit{Tit for Tat}) or maximizing their own game reward (\textit{Selfish}) can end up in the inefficient equilibrium. Finally, against a \textit{Random} agent most moral agents implement a `safe' \textit{Always Cooperate} strategy (which results in them being exploited half of the time) -  except the equality agent, which plays a random strategy against the Random opponent and thus gets exploited less frequently.

\begin{figure*}[!h]
\centering
\includegraphics[width=35mm]{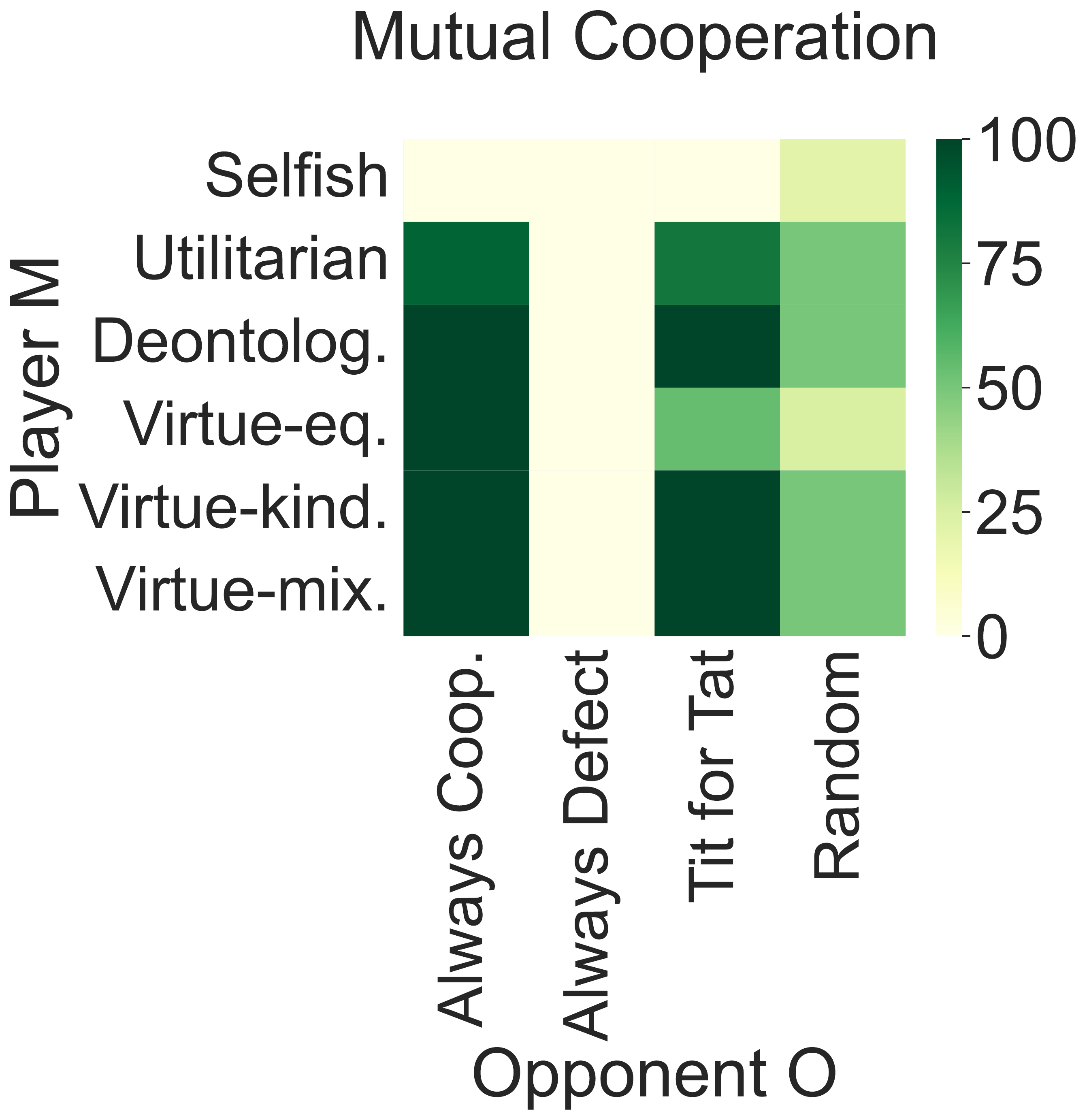}
\includegraphics[width=35mm]{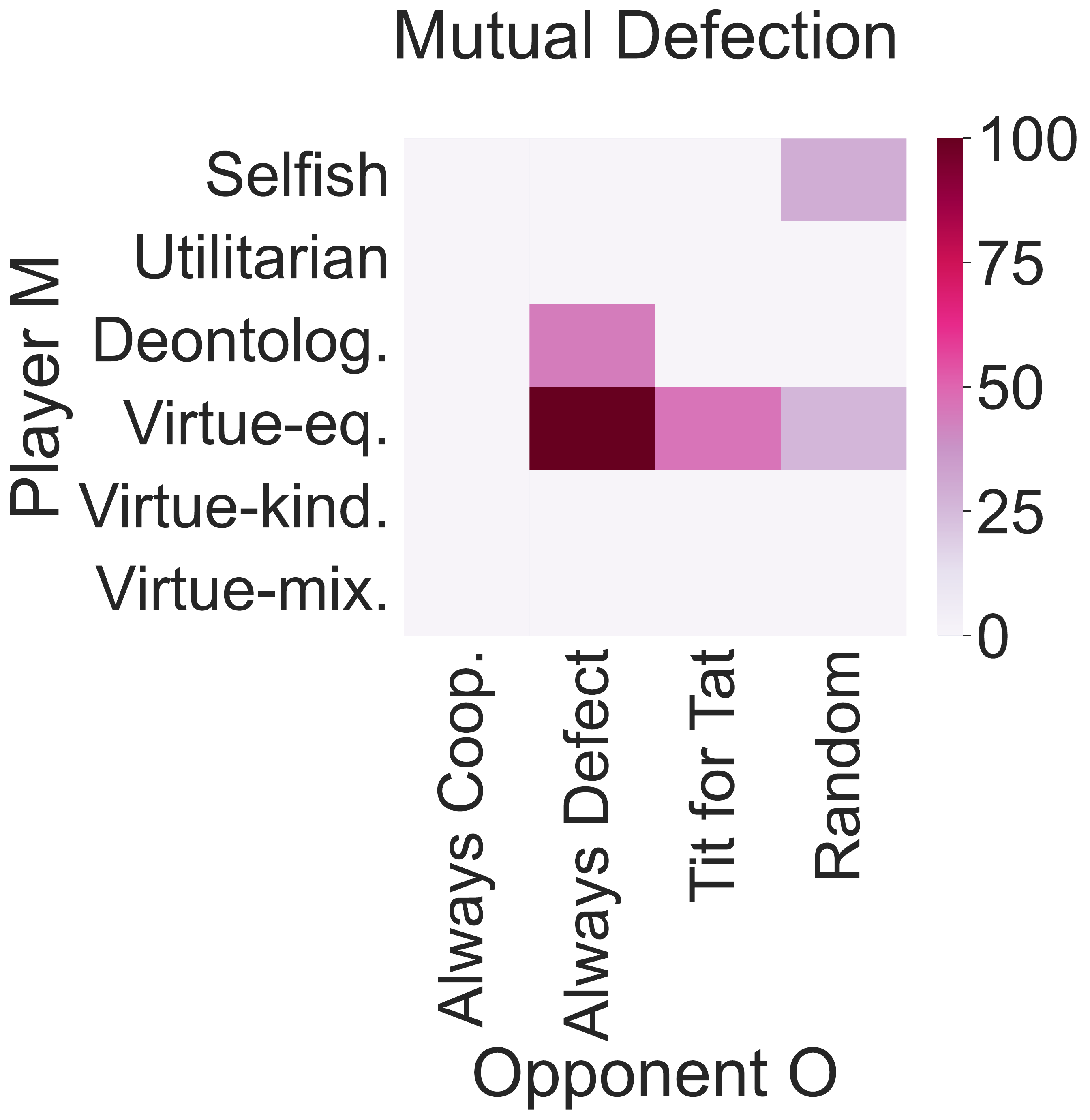}
\includegraphics[width=35mm]{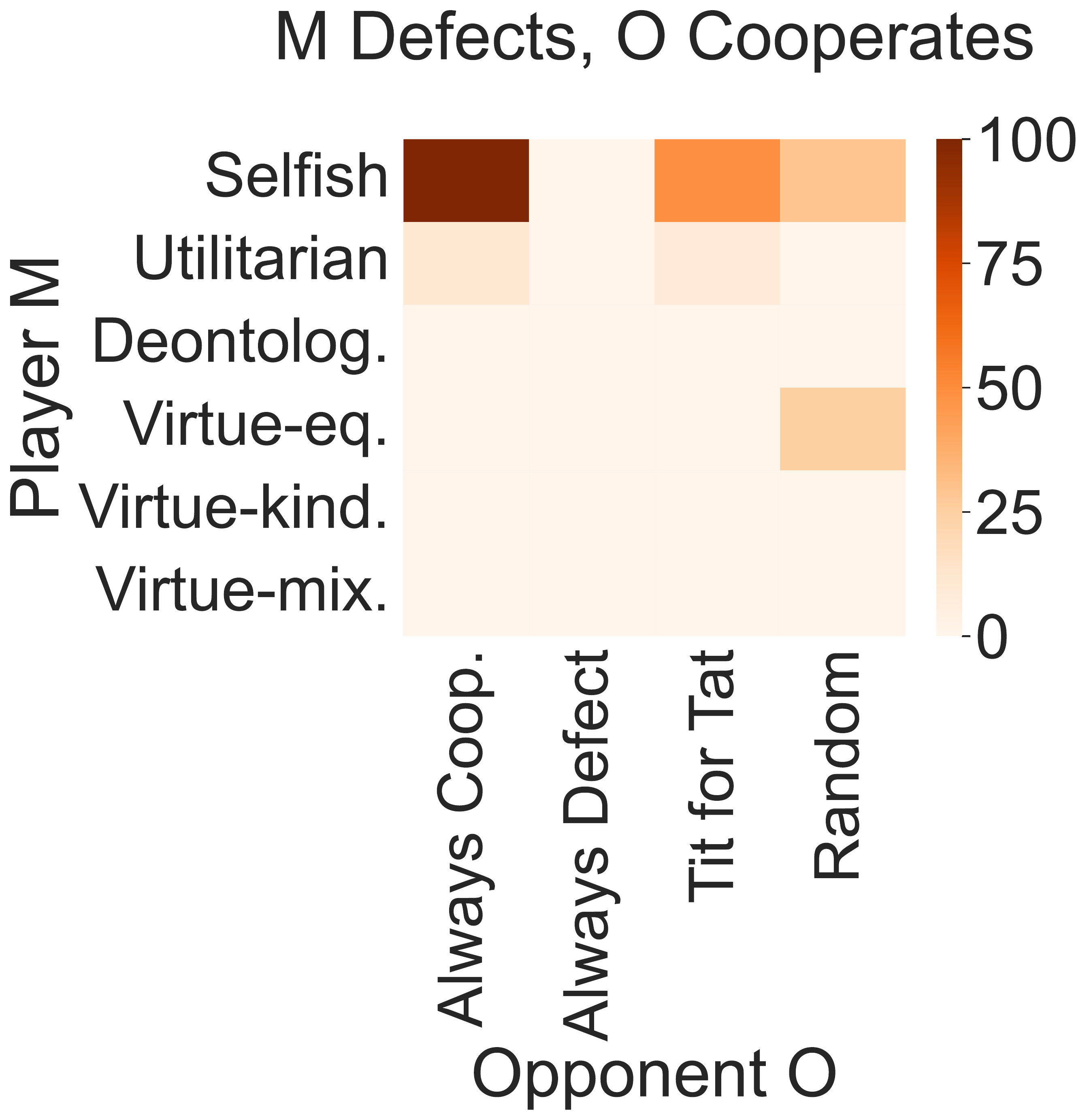}
\includegraphics[width=35mm]{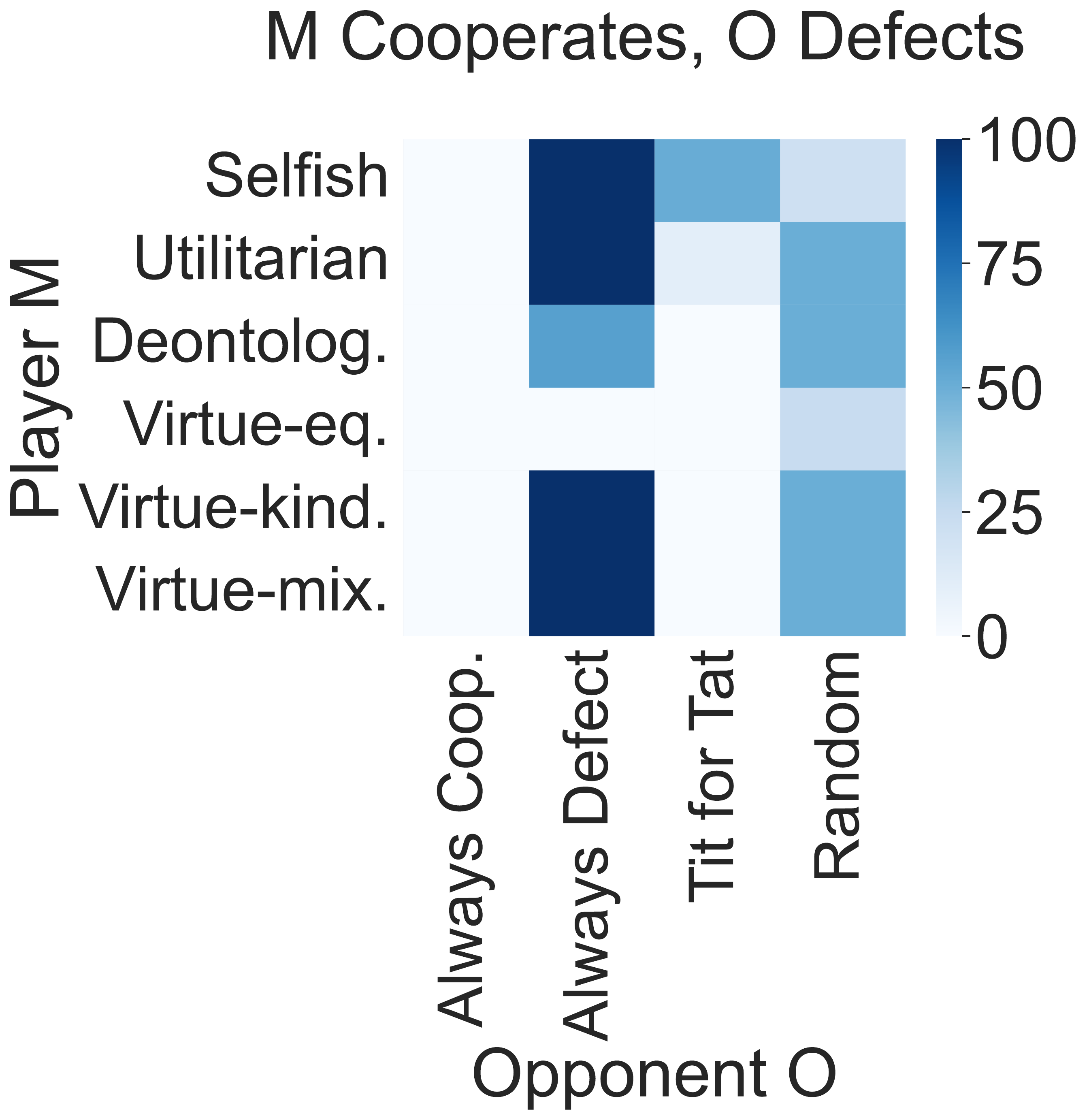}
\caption{Iterated Volunteer's Dilemma game. Simultaneous actions played by player $M$ type and the static opponent $O$ type at the end of the learning period (10000 iterations). Action pairs are displayed as a percentage over the 100 runs.}
\label{fig:actions_static_VOL}
\end{figure*}

Learning with moral rewards against static agents in the Iterated Volunteer's Dilemma (Figure \ref{fig:actions_static_VOL}) results in pairs of actions similar to the Iterated Prisoner's Dilemma. The equality agent is now the only one to end up in the inefficient mutual defection situation - 100\% of the time against \textit{Always Defect}, and half the time and a quarter of the time against \textit{Tit for Tat} or \textit{Random} respectively. The \textit{Selfish} agent here is more likely to end up in an exploitative situation - either it exploits an \textit{Always Cooperate} agent (see orange), or it gets exploited by \textit{Always Defect} (see blue), or half-half against \textit{Tit for Tat}. Against a \textit{Random} agent the \textit{Selfish} learner now learns a \textit{Random} strategy, instead of an \textit{Always Defect} strategy as in the Iterated Prisoner's Dilemma. 

\begin{figure*}[!h]
\centering
\includegraphics[width=35mm]{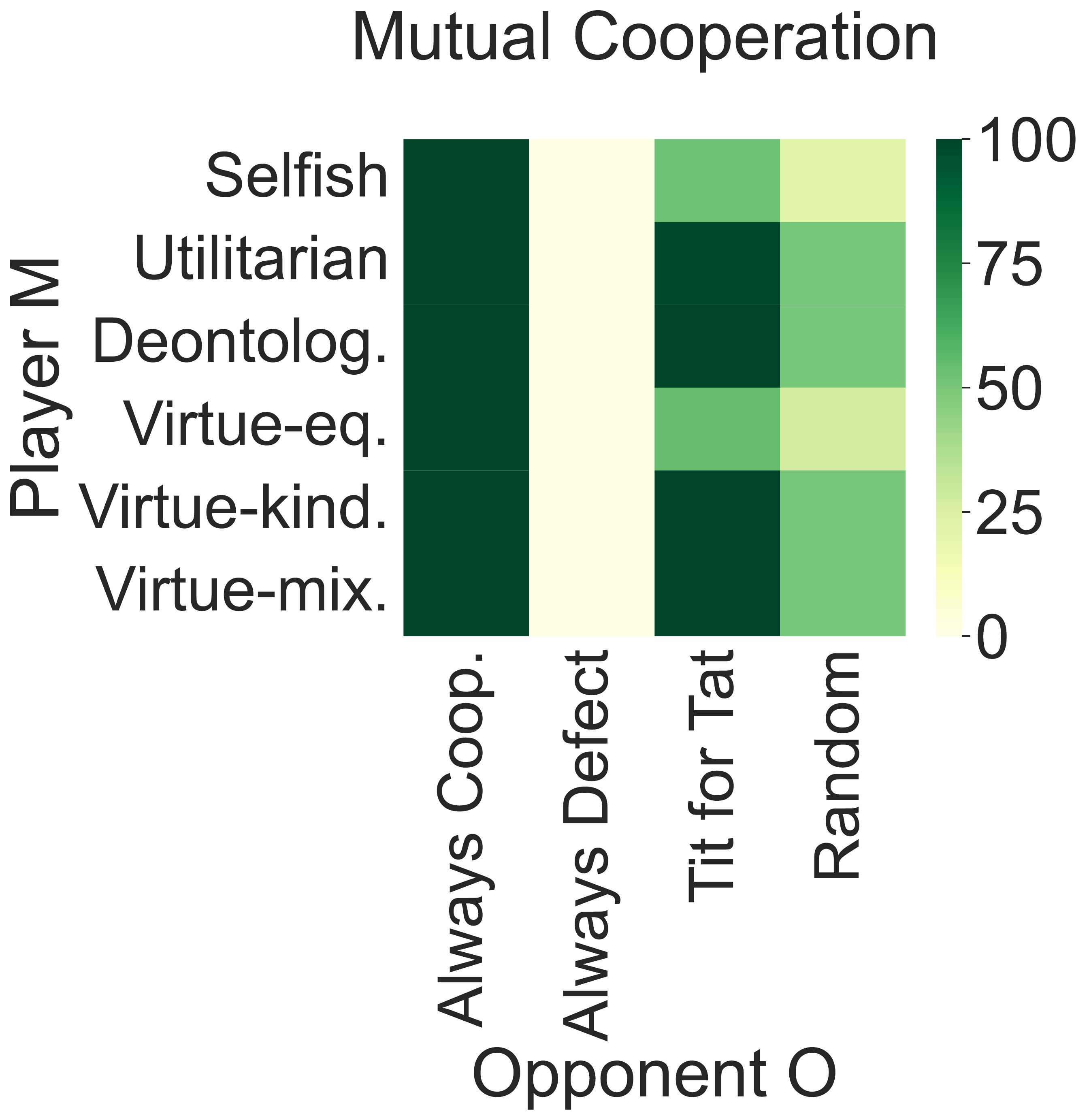}
\includegraphics[width=35mm]{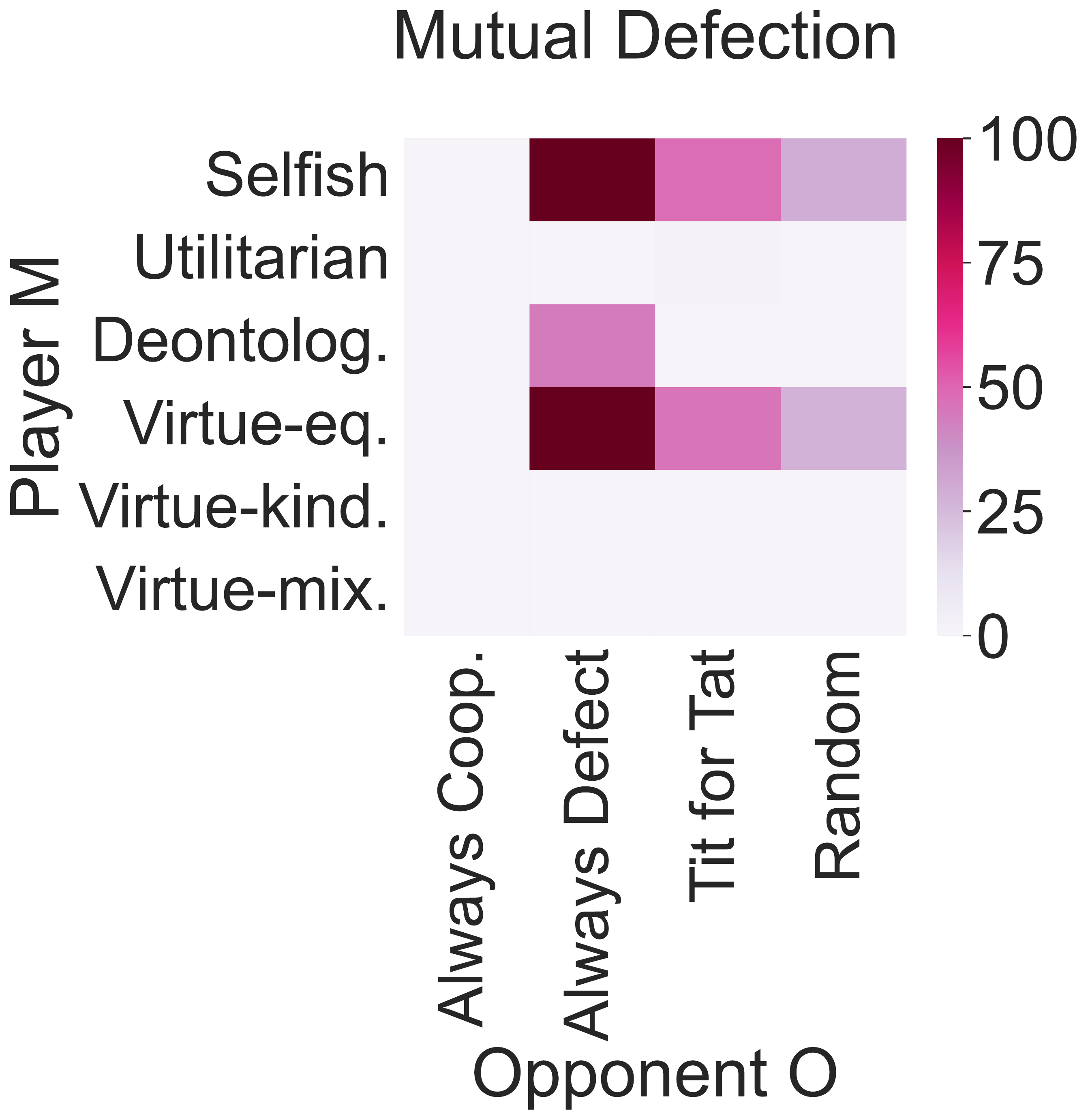}
\includegraphics[width=35mm]{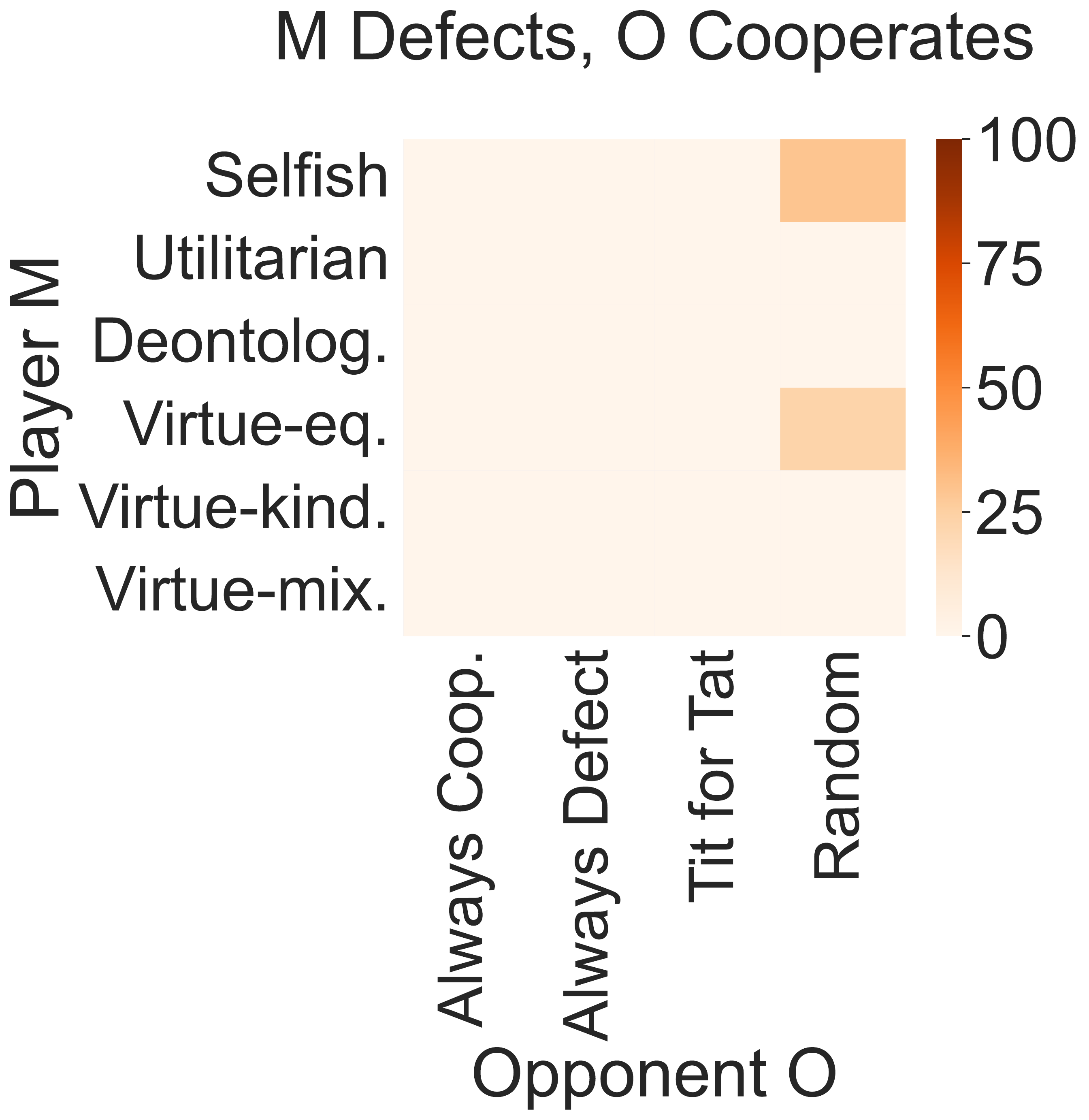}
\includegraphics[width=35mm]{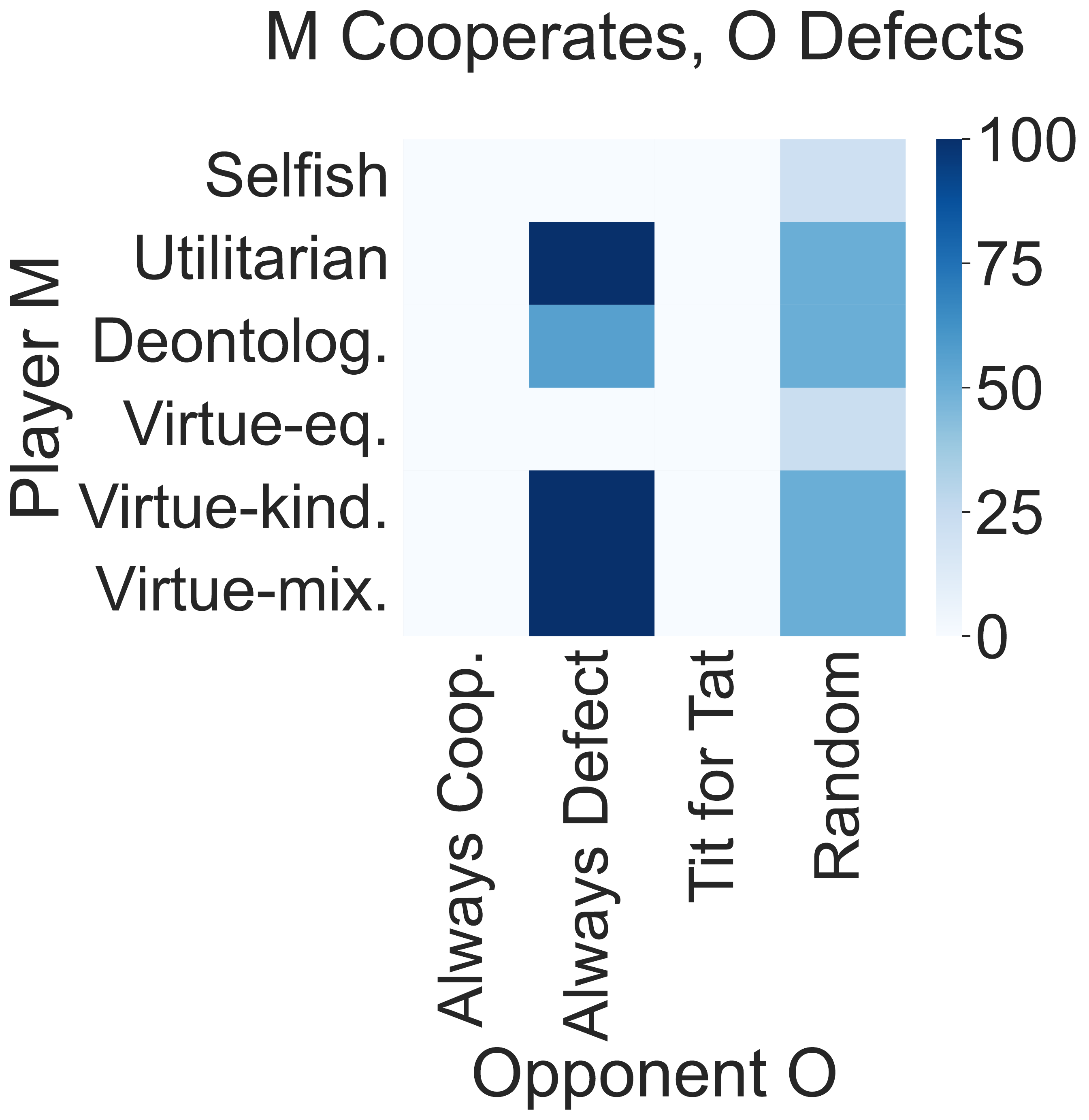}
\caption{Iterated Stag Hunt game. Simultaneous actions played by player $M$ type and the static opponent $O$ type at the end of the learning period (10000 iterations). Action pairs are displayed as a percentage over the 100 runs.}
\label{fig:actions_static_STH}
\end{figure*}

Finally, learning against static opponents in the Iterated Stag Hunt game (Figure \ref{fig:actions_static_STH}), all agents including the \textit{Selfish} one converge to the Pareto-optimal mutual cooperation against \textit{Always Cooperate}. Against \textit{Always Defect}, once again the \textit{Utilitarian}, \textit{Virtue-kindness} and \textit{Virtue-mixed} agents are defenseless (and get exploited 100\% of the time - see plots in blue), the \textit{Deontological} agent is protected from exploitation half of the time because they play randomly against a defector, and the \textit{equality} and \textit{Selfish} agents are able to achieve 100\% mutual defection. \textit{Tit for Tat} elicits mutual defection half of the time from \textit{Virtue-equality} and \textit{Selfish} agents - and the other half the time it elicits mutual cooperation.  

\subsection{Summary of Reward - Learning Player vs Static Opponent}
\begin{figure*}[!h]
\centering
\begin{tabular}[t]{|c|cccccc}
\toprule
& Selfish & Utilitarian & Deontological & Virtue-equality & Virtue-kindness & Virtue-mixed \\
\midrule
\makecell[cc]{\rotatebox[origin=c]{90}{\thead{Game Reward}}} & 
\subt{\includegraphics[height=20mm]{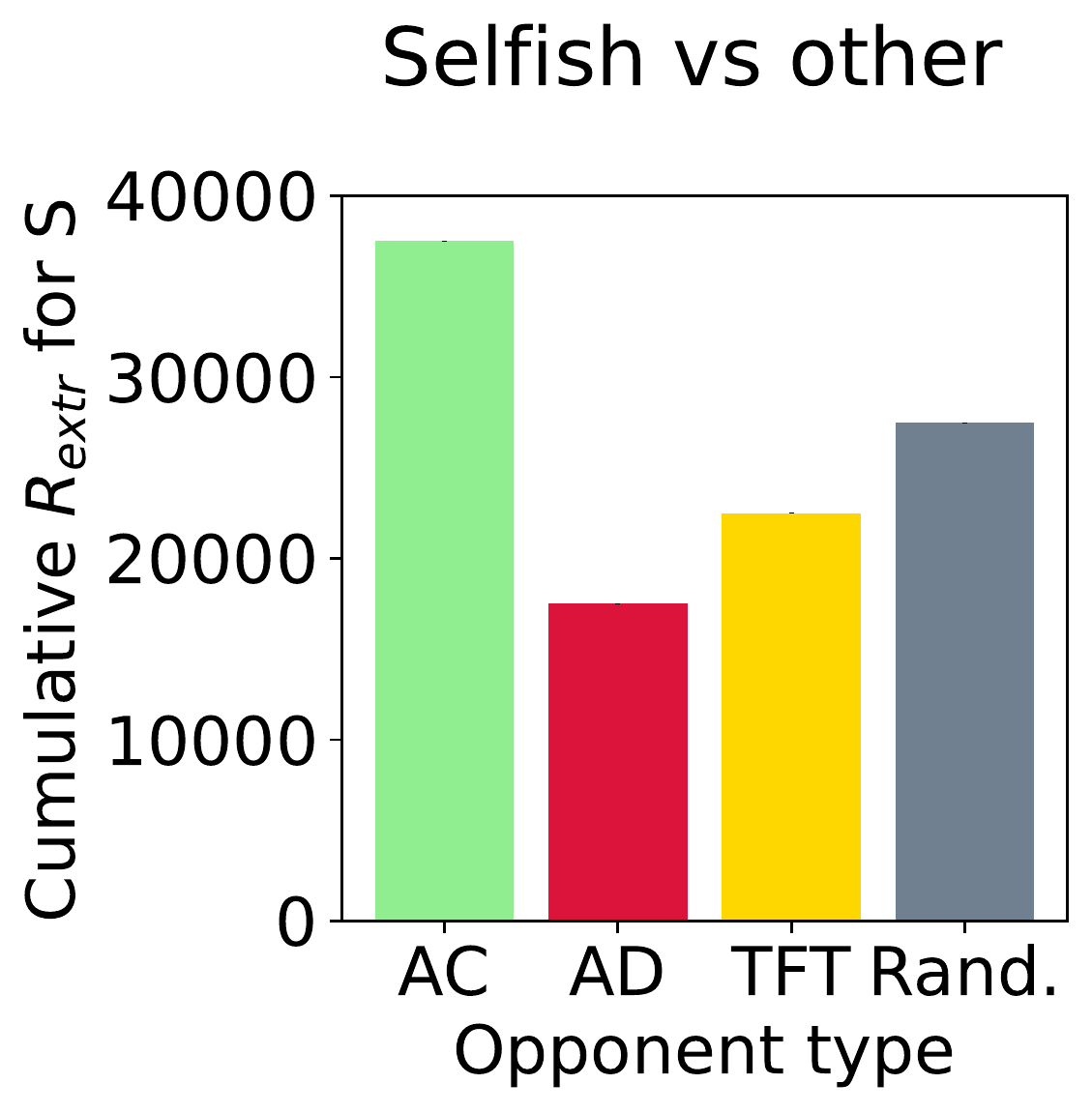}} & \subt{\includegraphics[height=20mm]{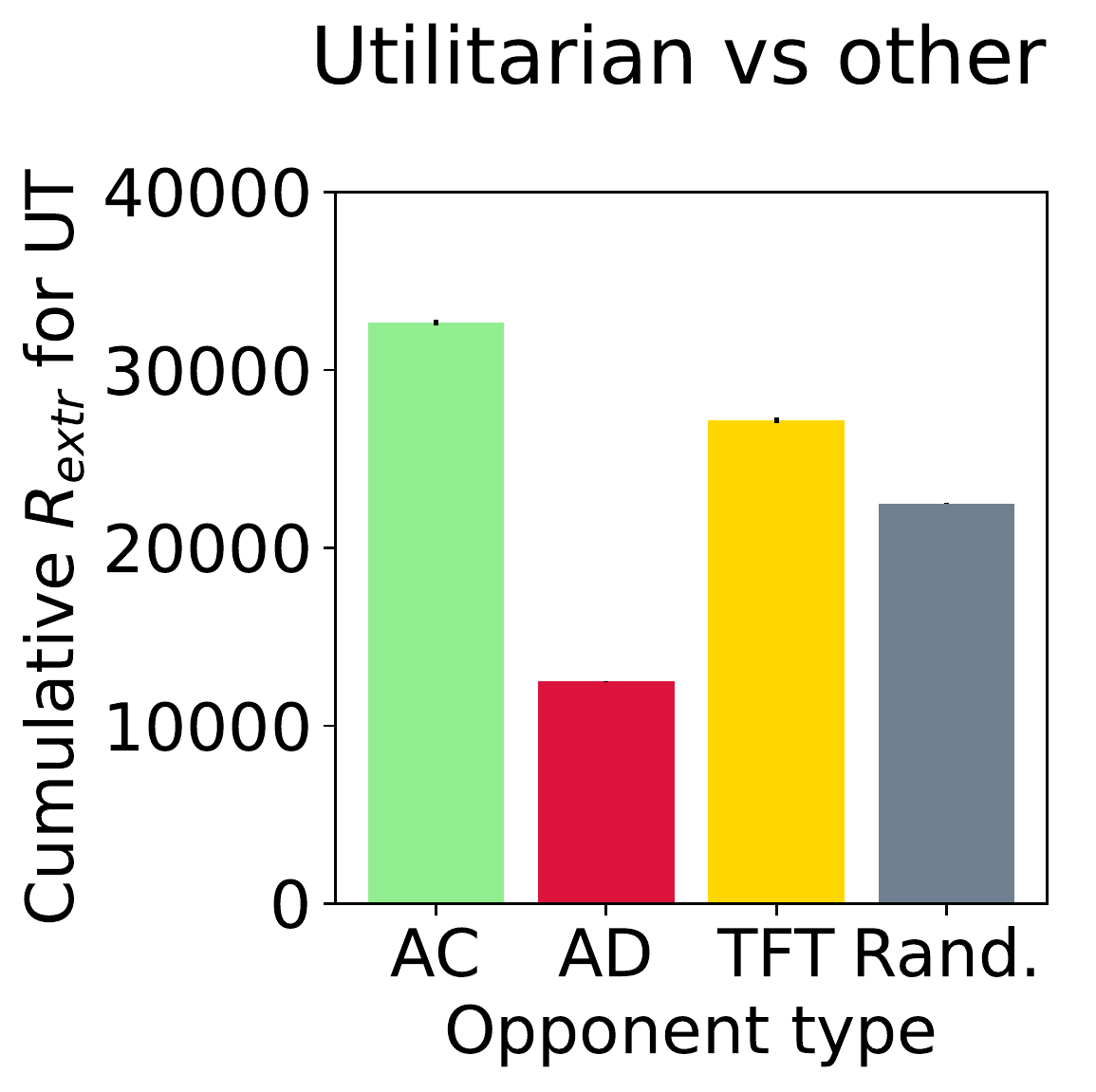}} & \subt{\includegraphics[height=20mm]{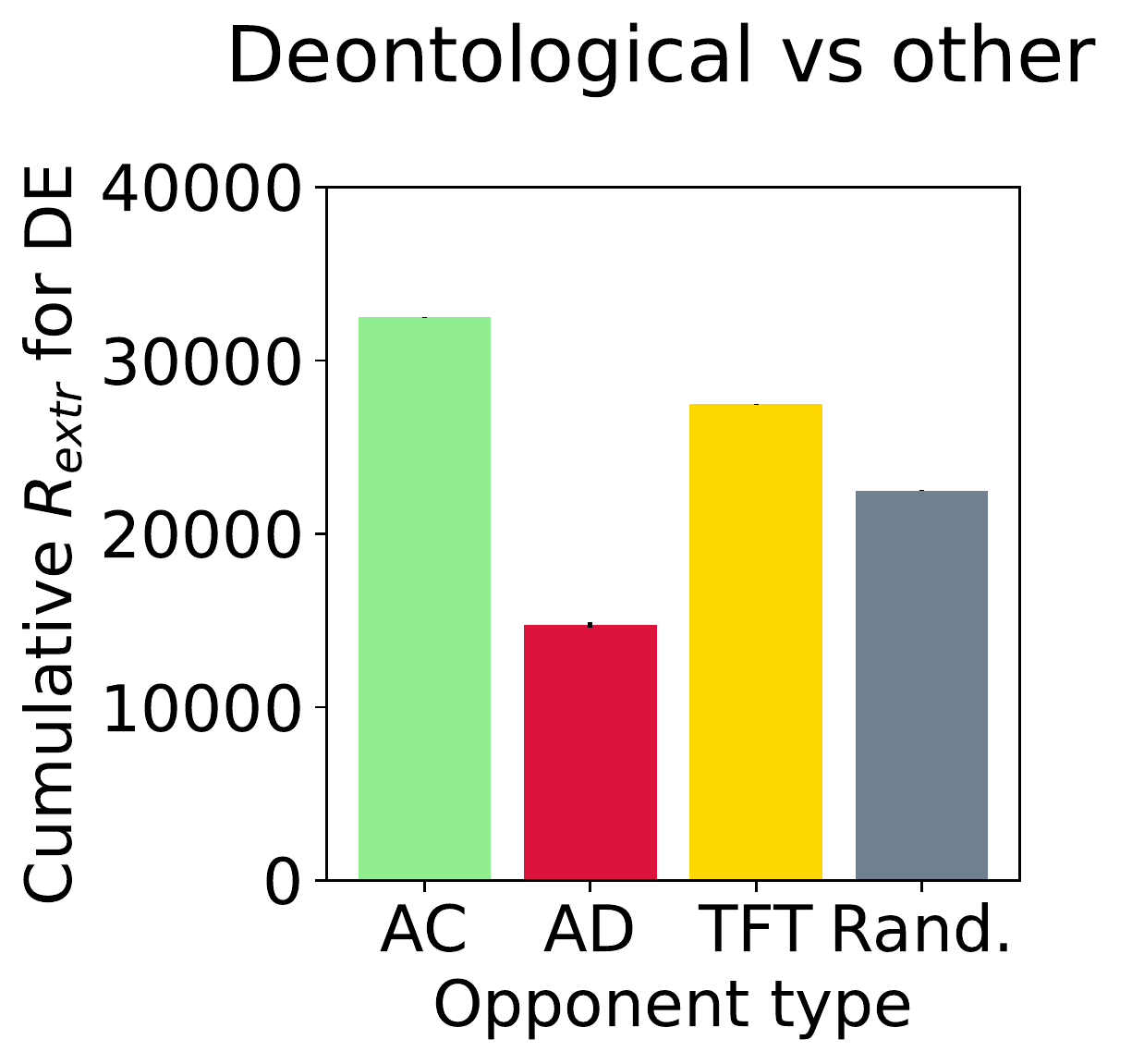}} & \subt{\includegraphics[height=20mm]{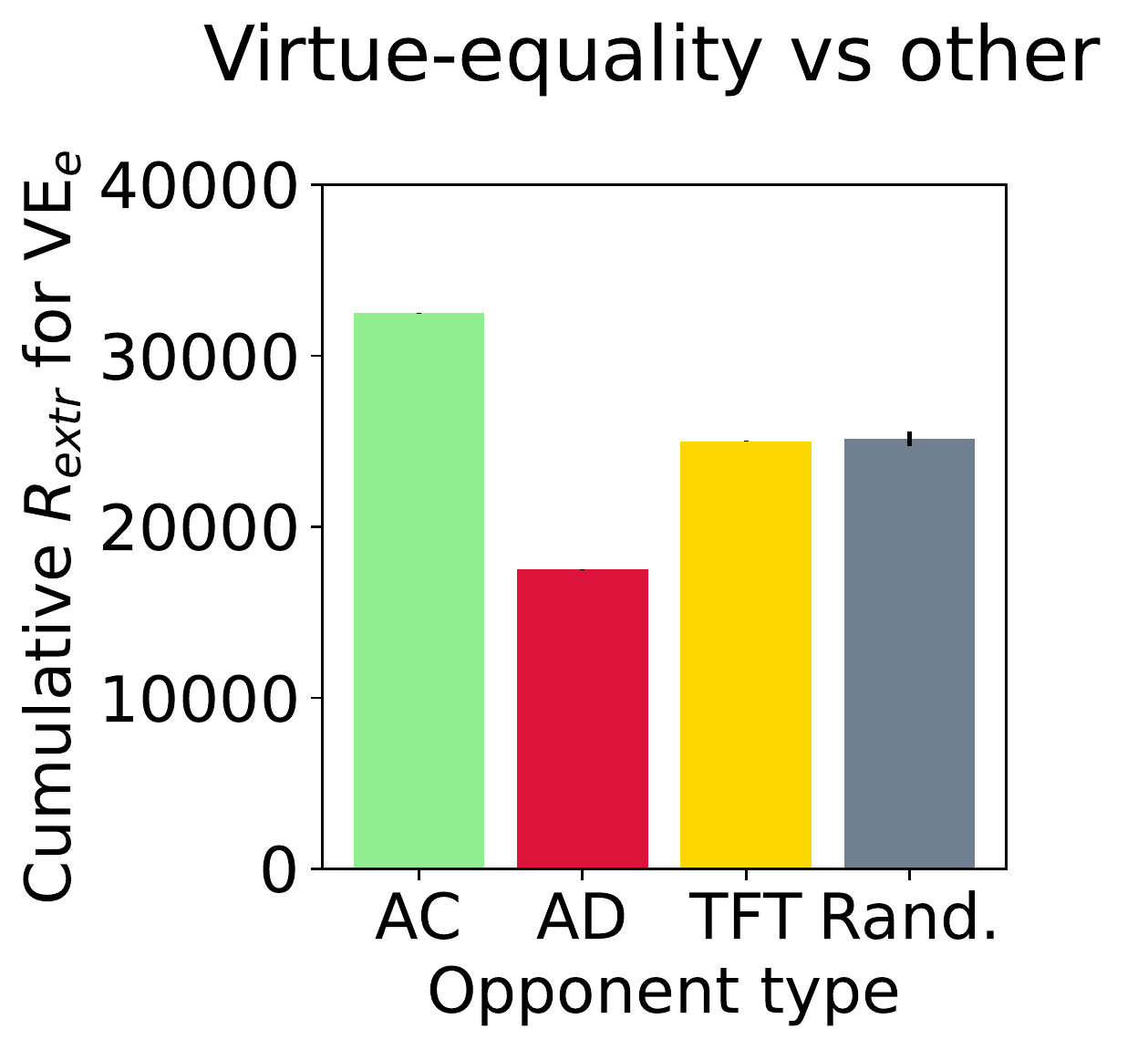}} & \subt{\includegraphics[height=20mm]{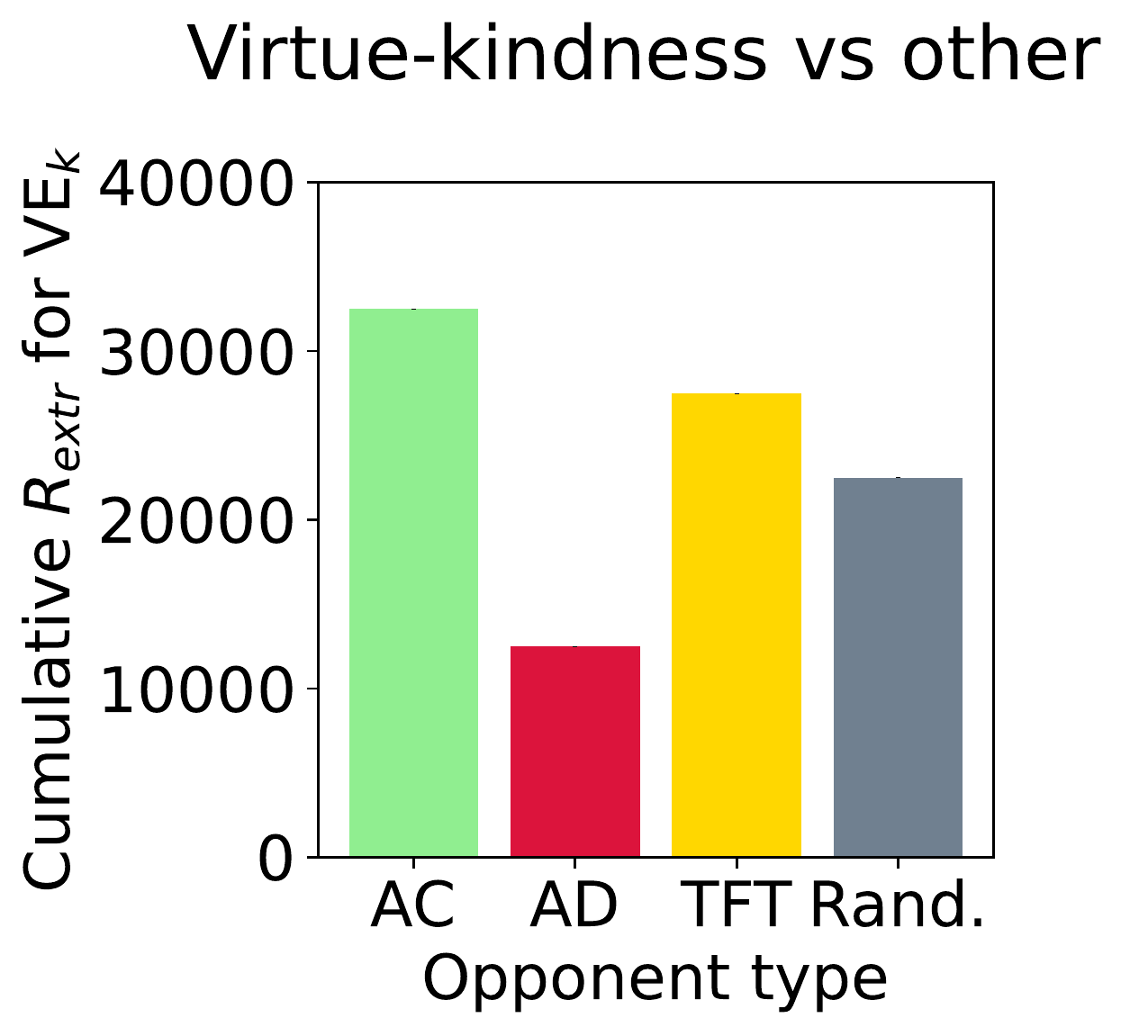}} & \subt{\includegraphics[height=20mm]{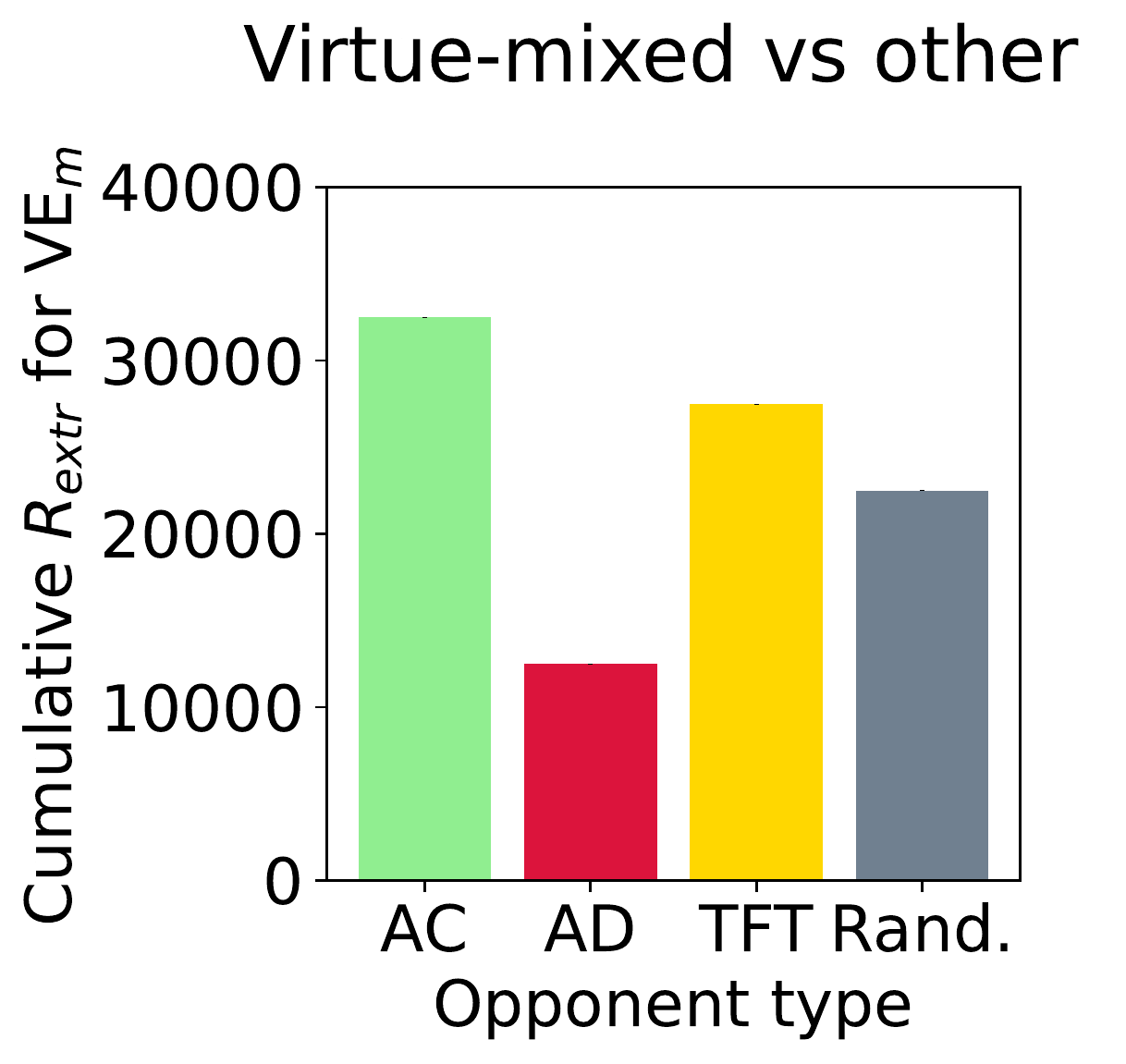}}
\\ 
\makecell[cc]{\rotatebox[origin=c]{90}{\thead{Moral Reward}}} & 
& \subt{\includegraphics[height=20mm]{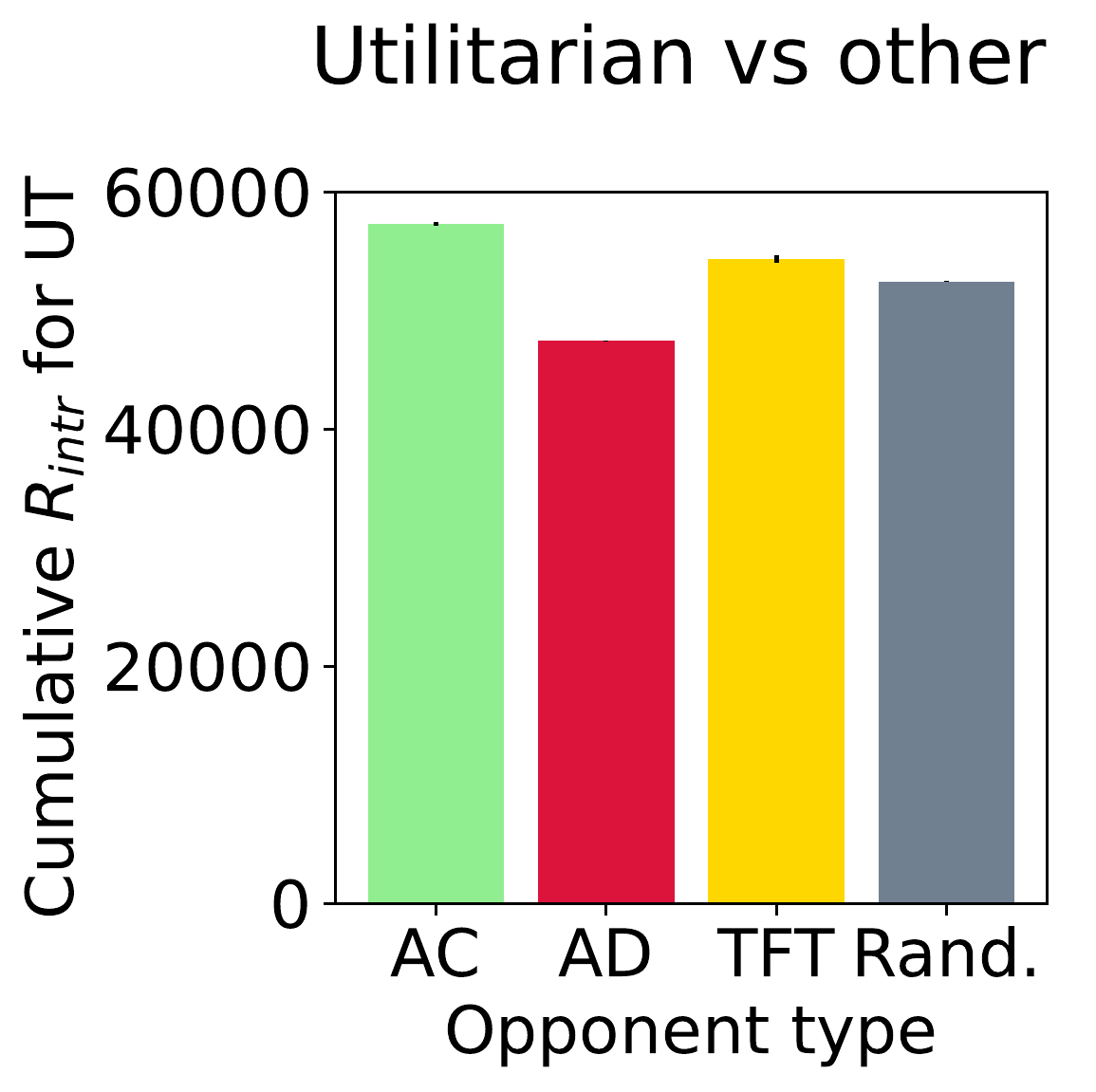}} & \subt{\includegraphics[height=20mm]{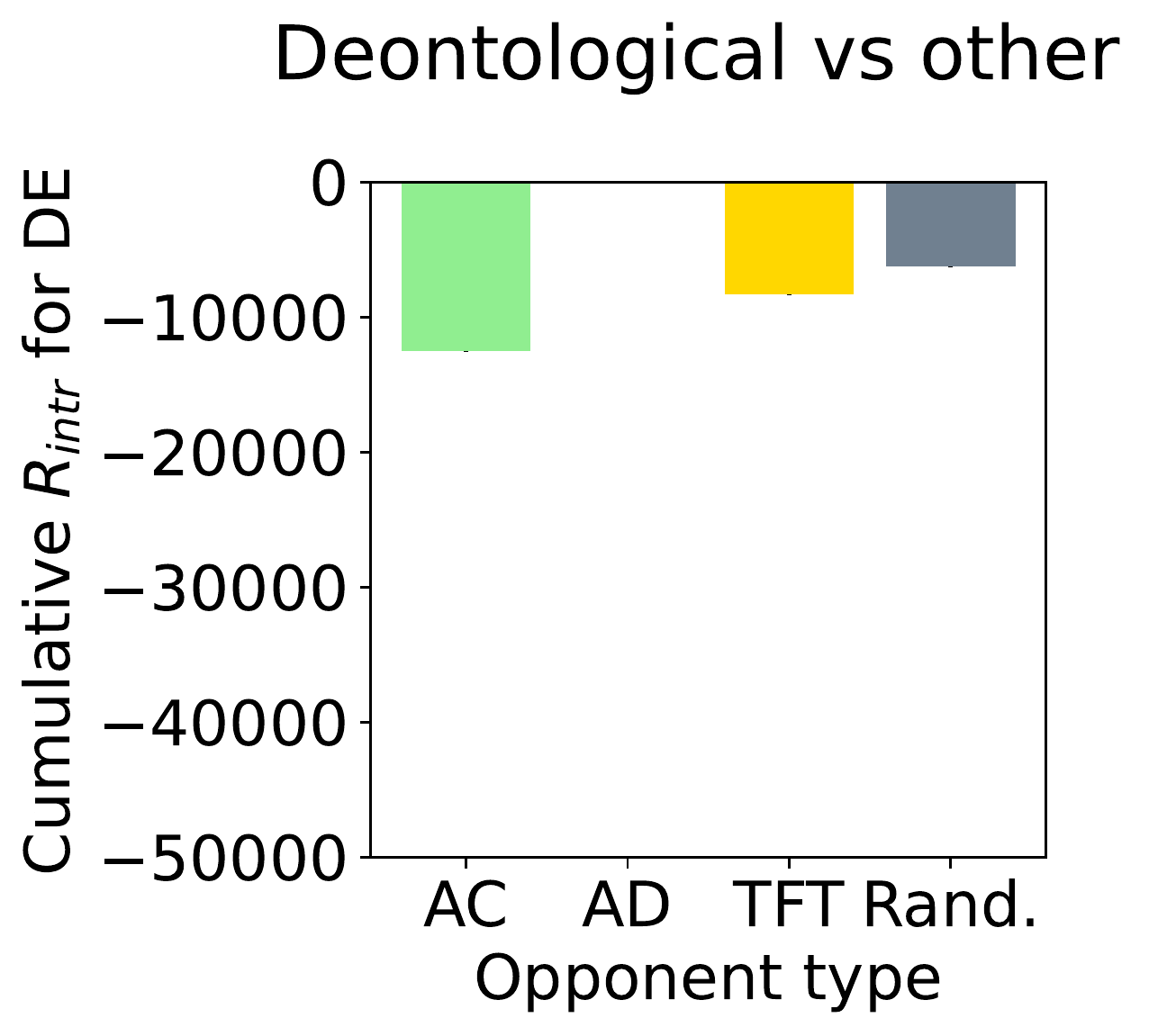}} & \subt{\includegraphics[height=20mm]{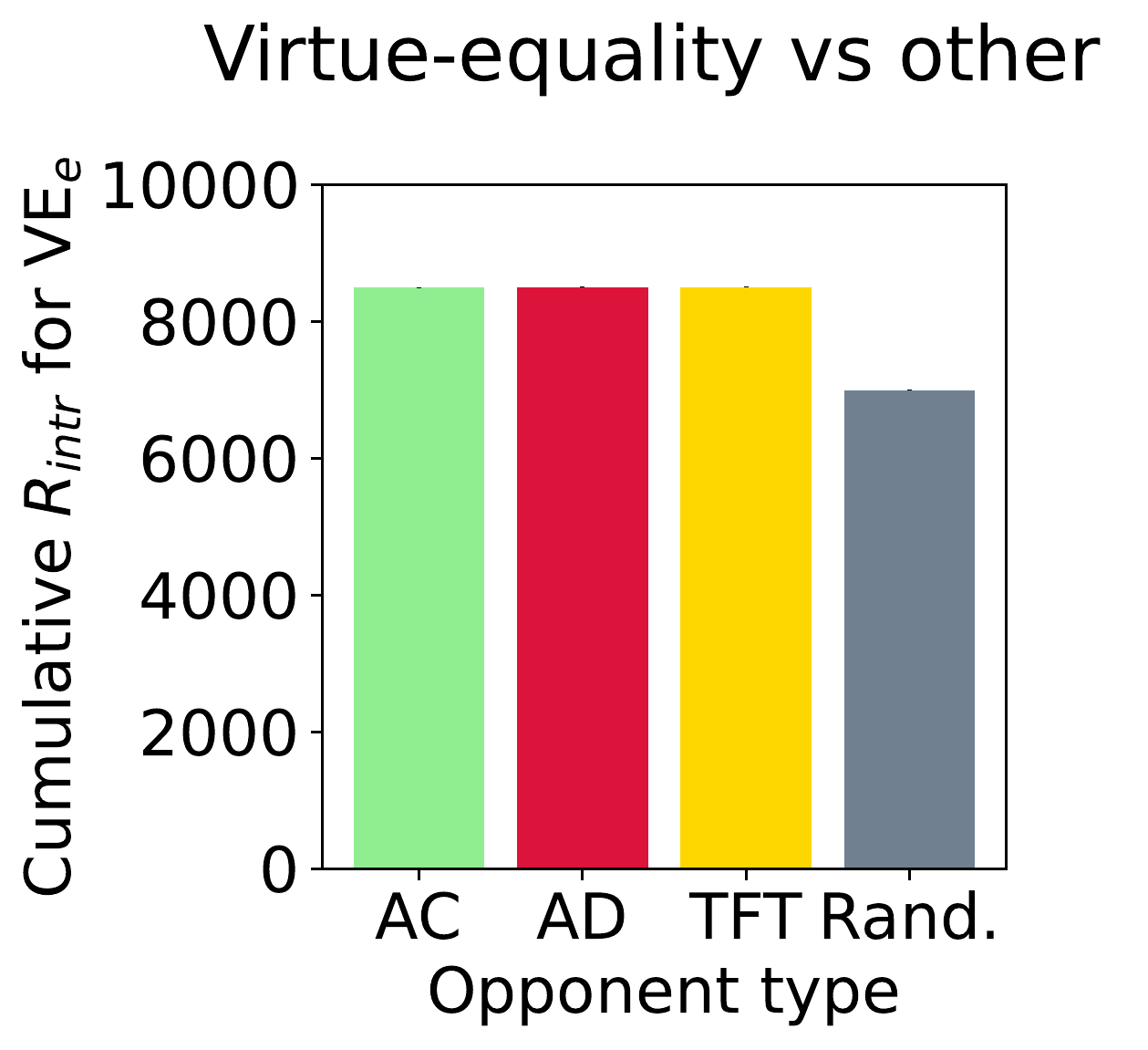}} & \subt{\includegraphics[height=20mm]{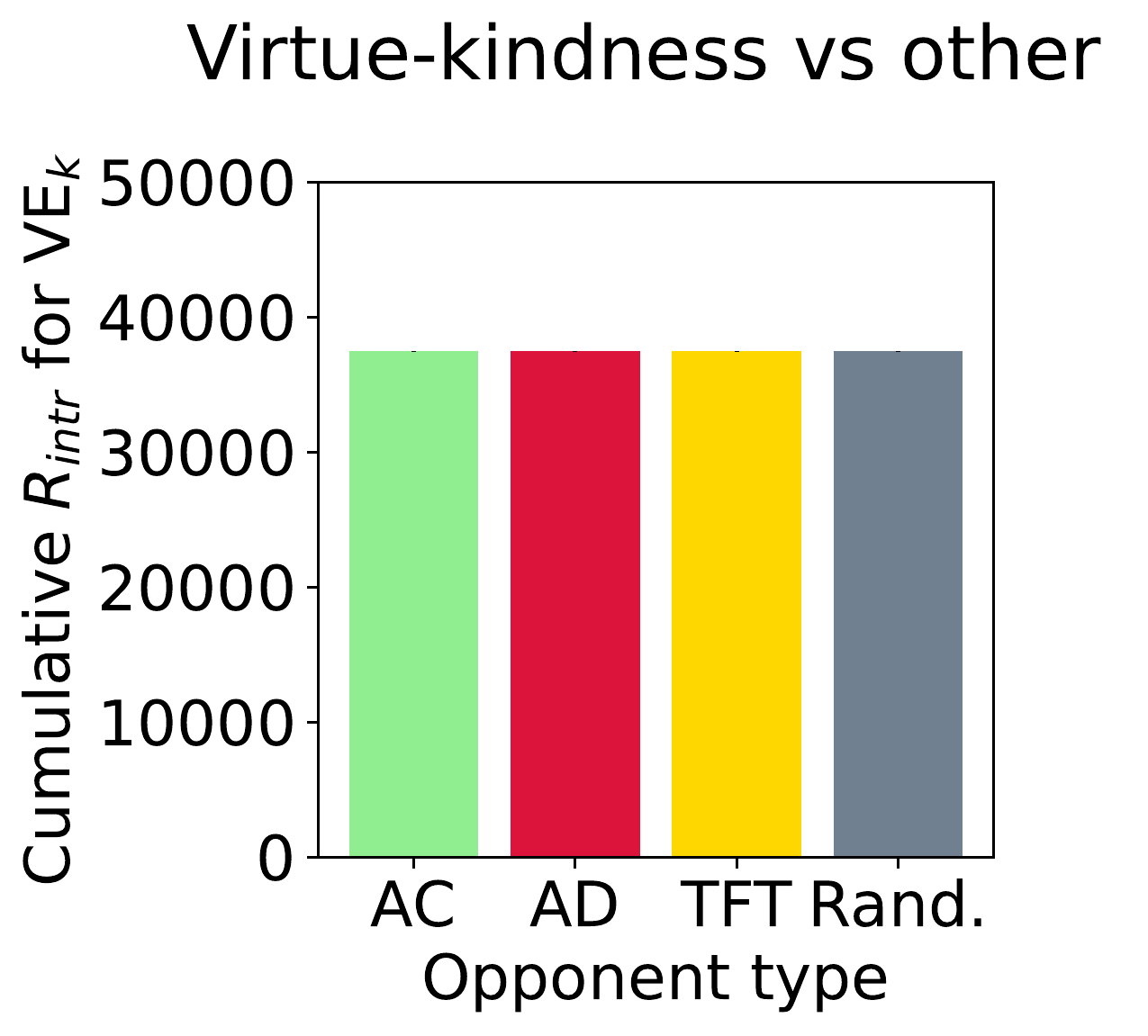}} & \subt{\includegraphics[height=20mm]{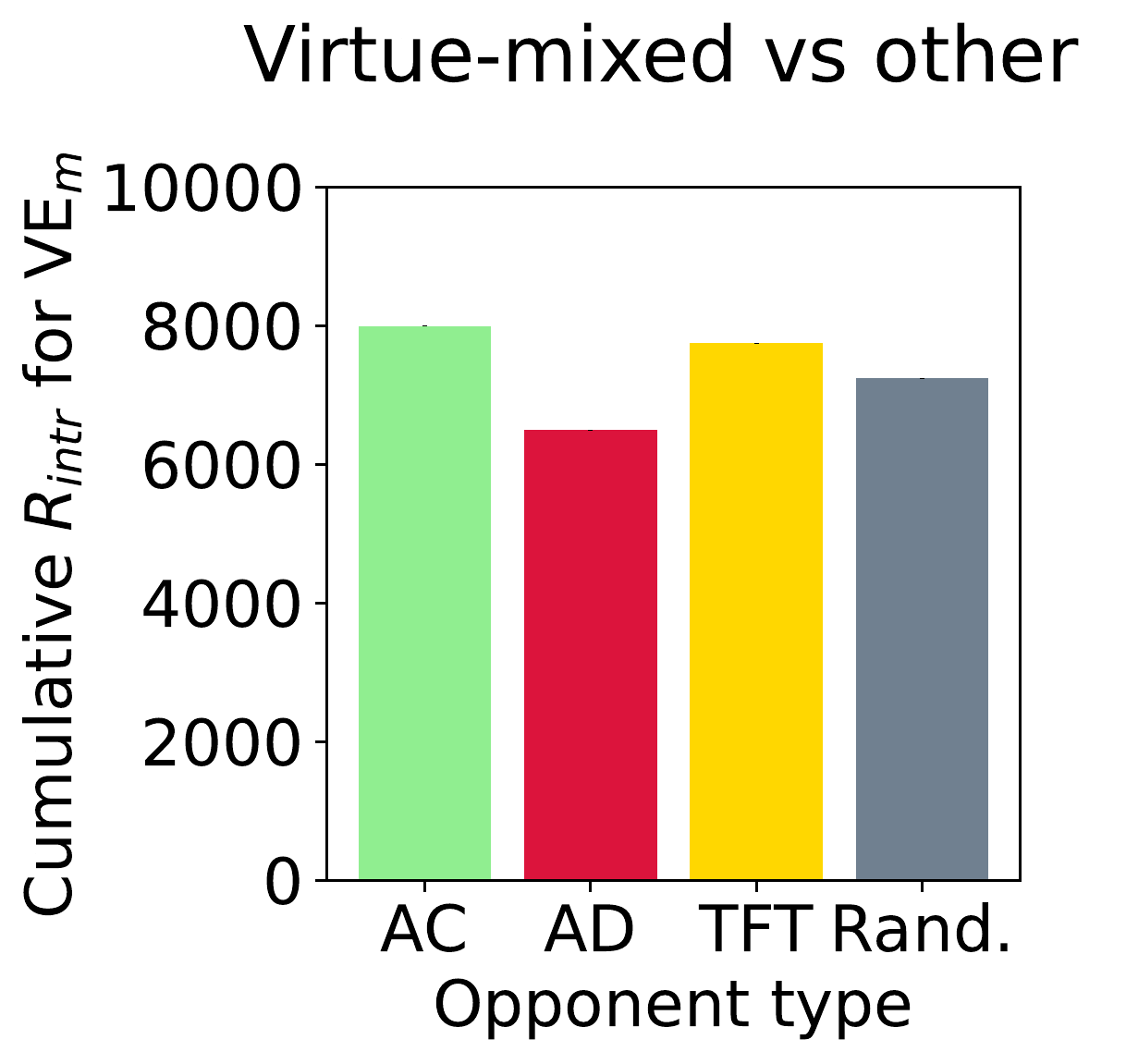}}
\\
\end{tabular}
\\ A. Iterated Prisoner's Dilemma \\

\begin{tabular}[t]{|c|cccccc}
\toprule
\makecell[cc]{\rotatebox[origin=c]{90}{\thead{Game Reward}}} & \subt{\includegraphics[height=20mm]{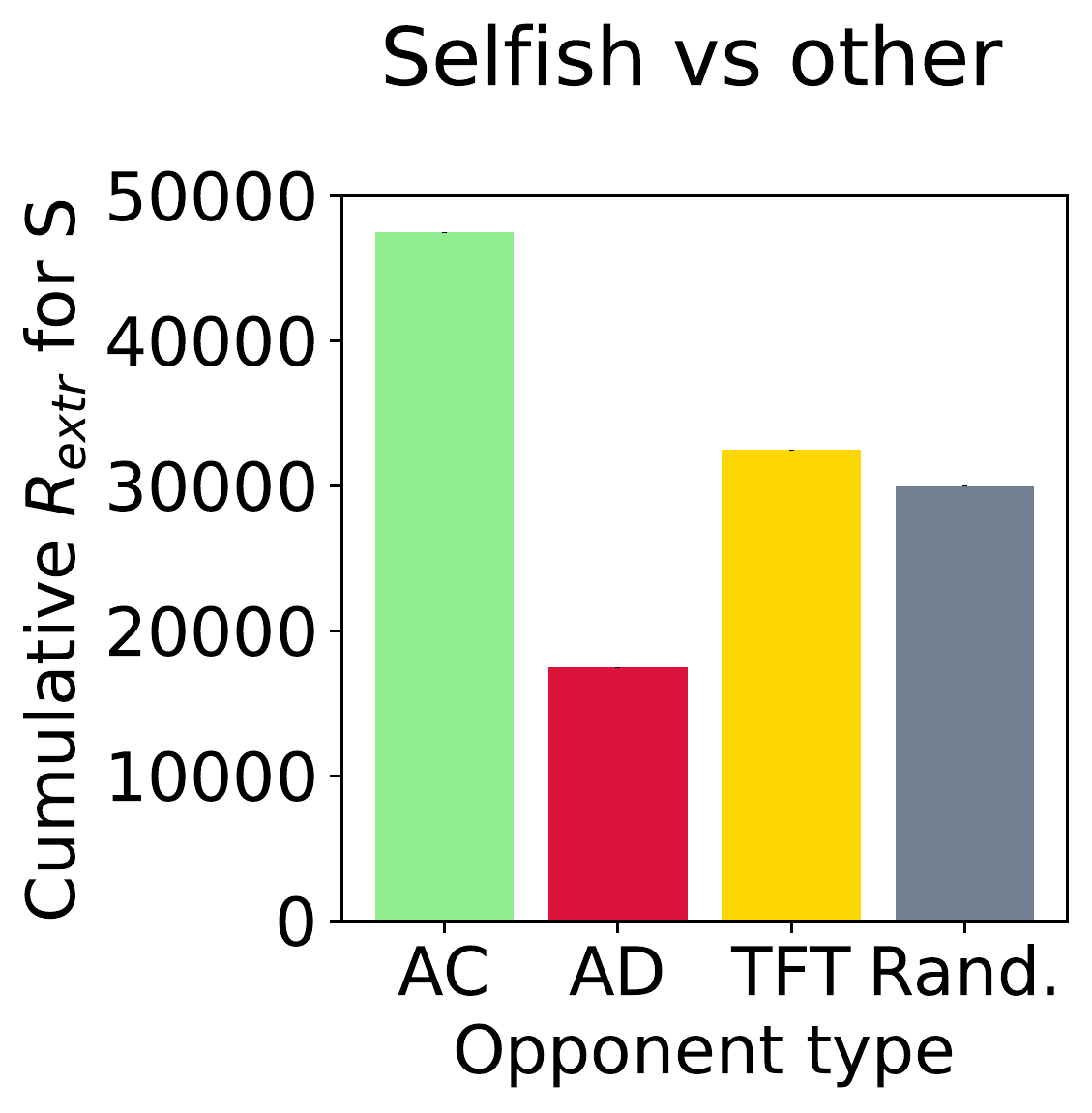}} & \subt{\includegraphics[height=20mm]{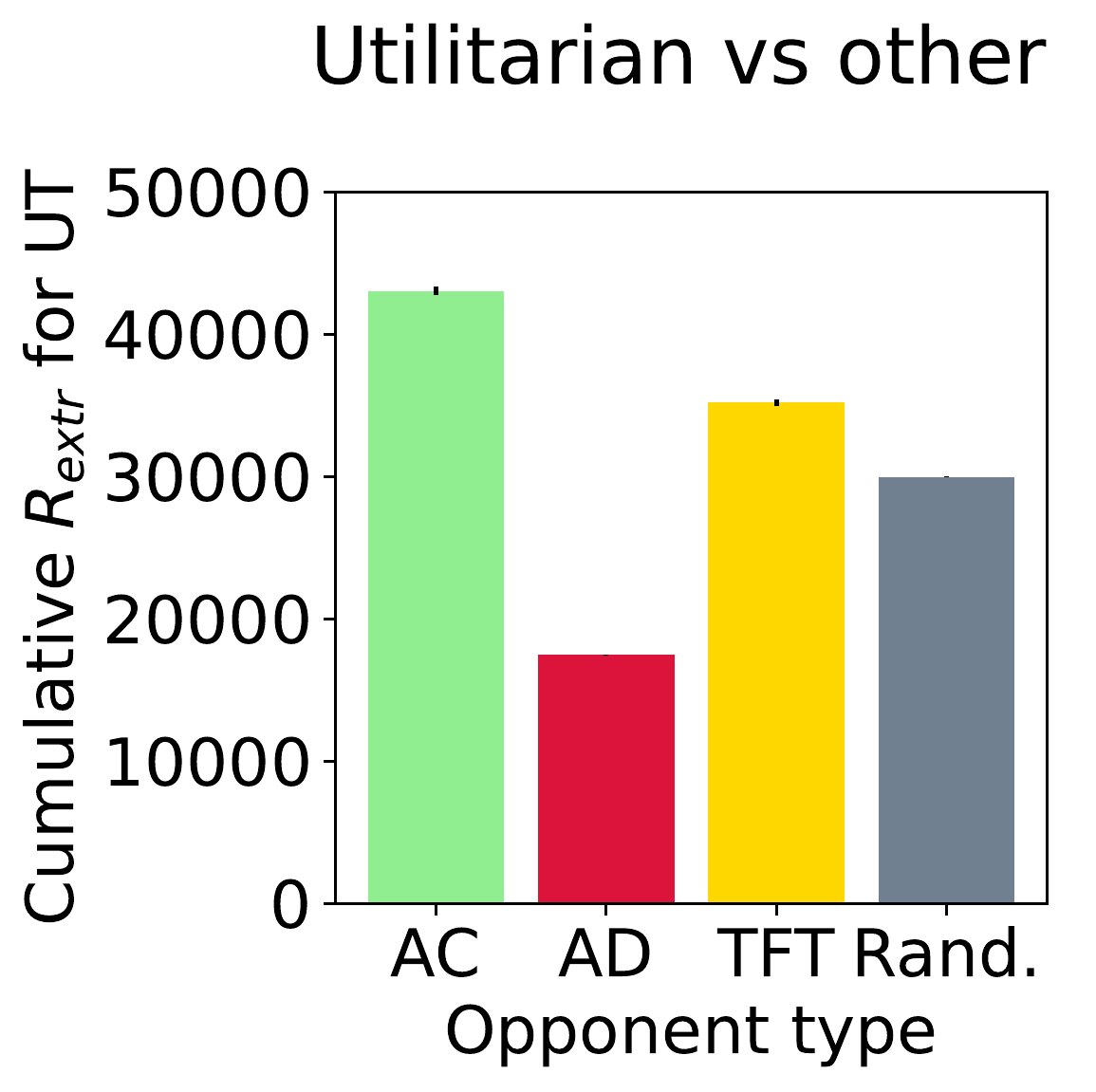}} & \subt{\includegraphics[height=20mm]{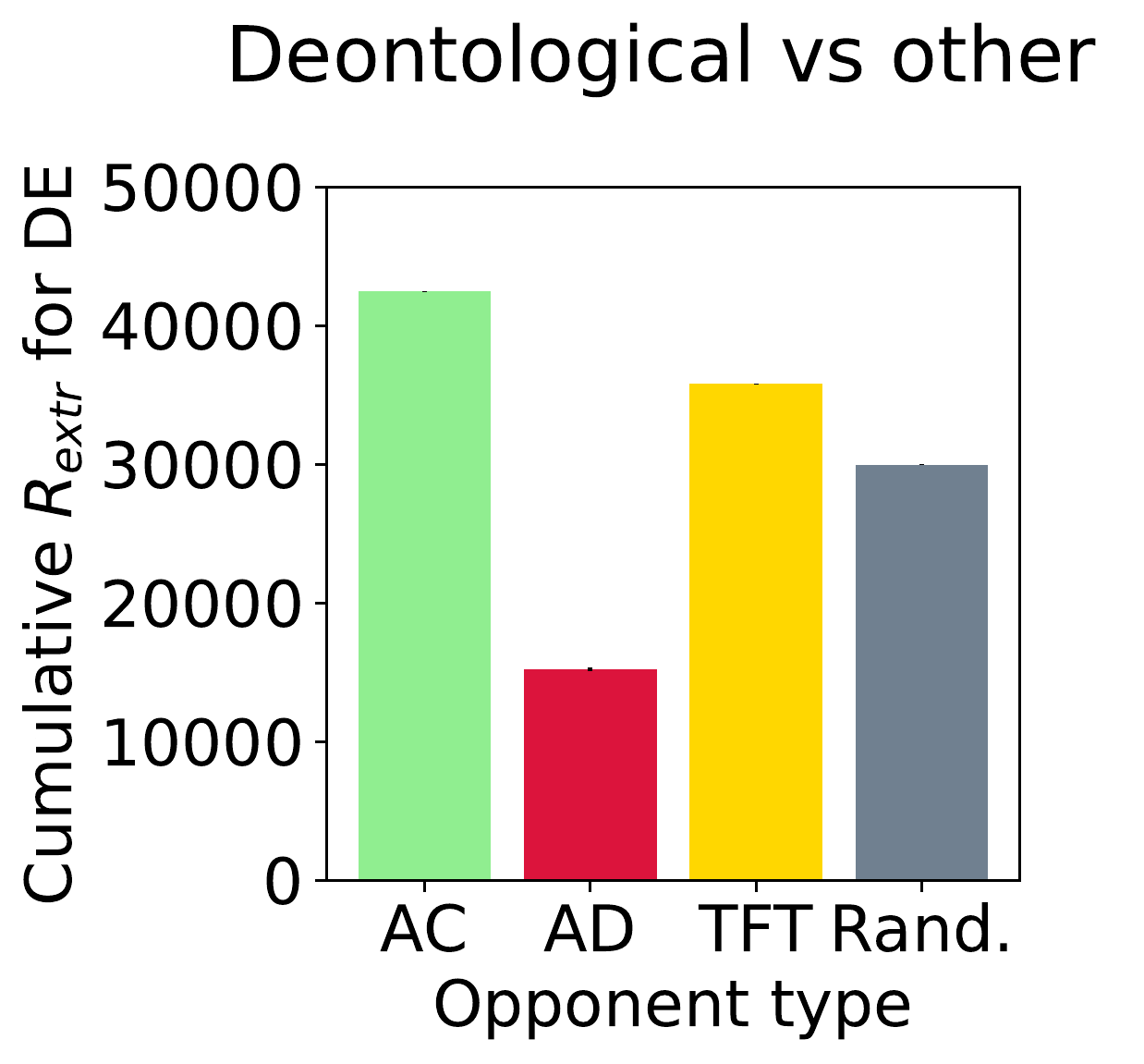}} & \subt{\includegraphics[height=20mm]{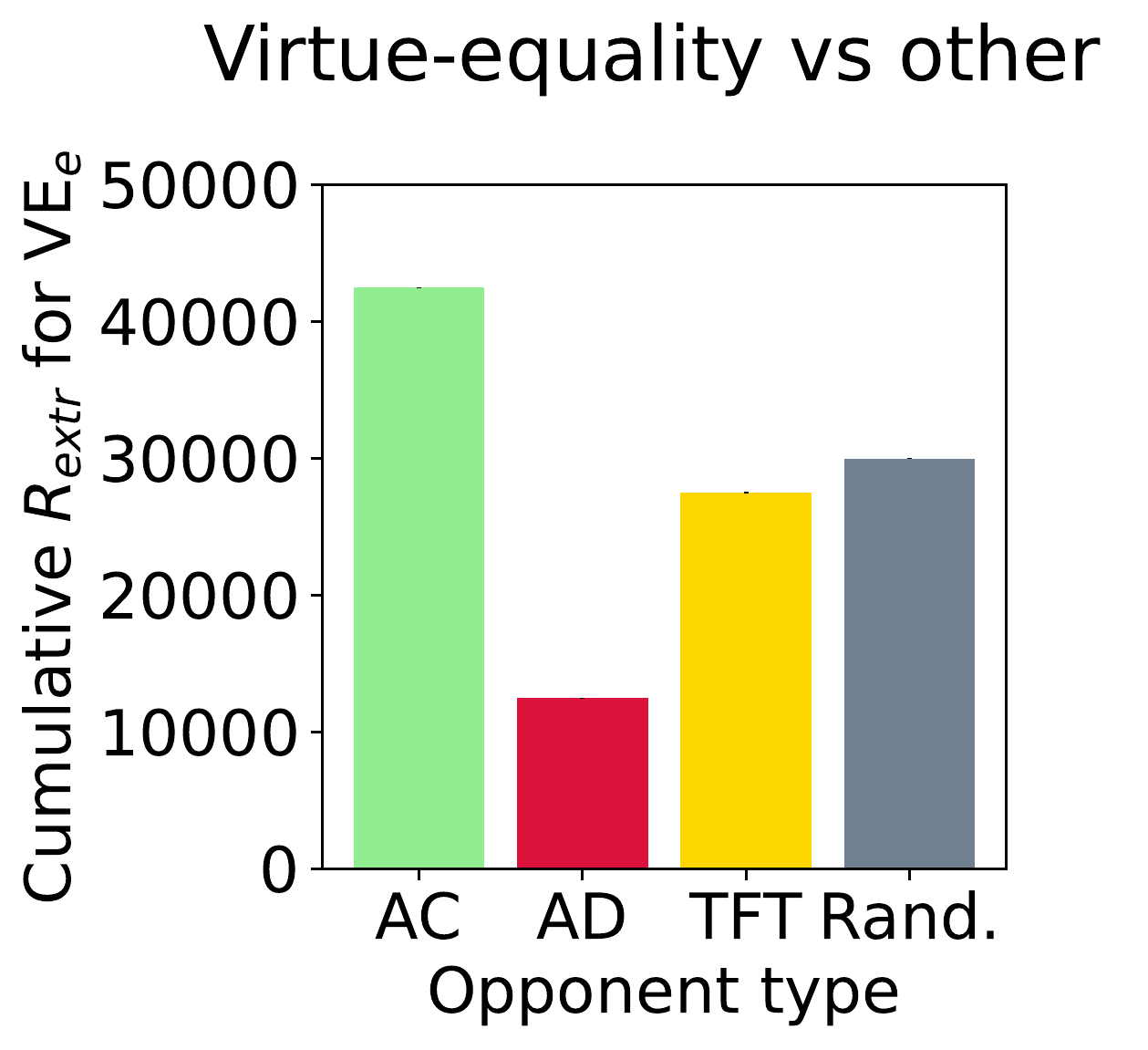}} & \subt{\includegraphics[height=20mm]{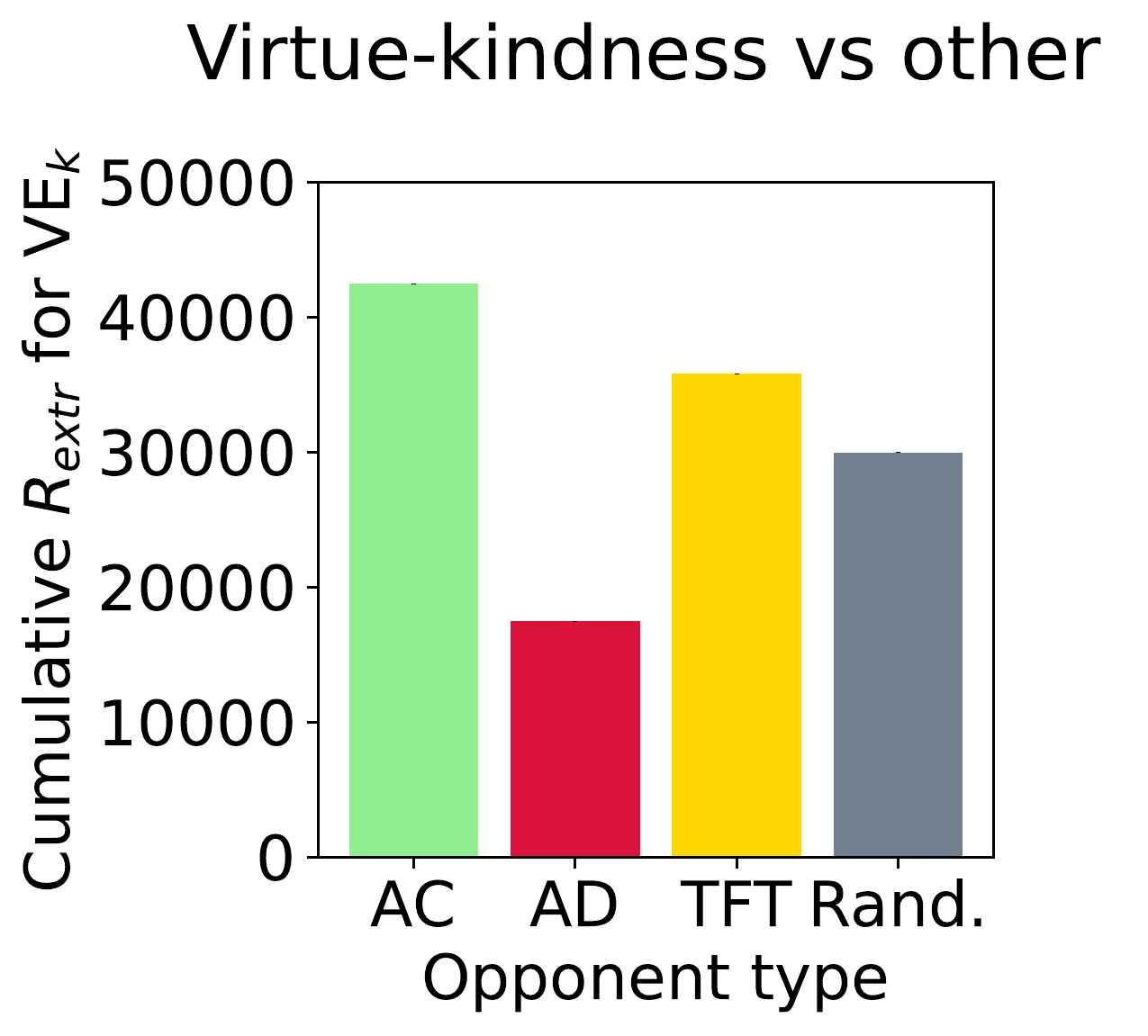}} & \subt{\includegraphics[height=20mm]{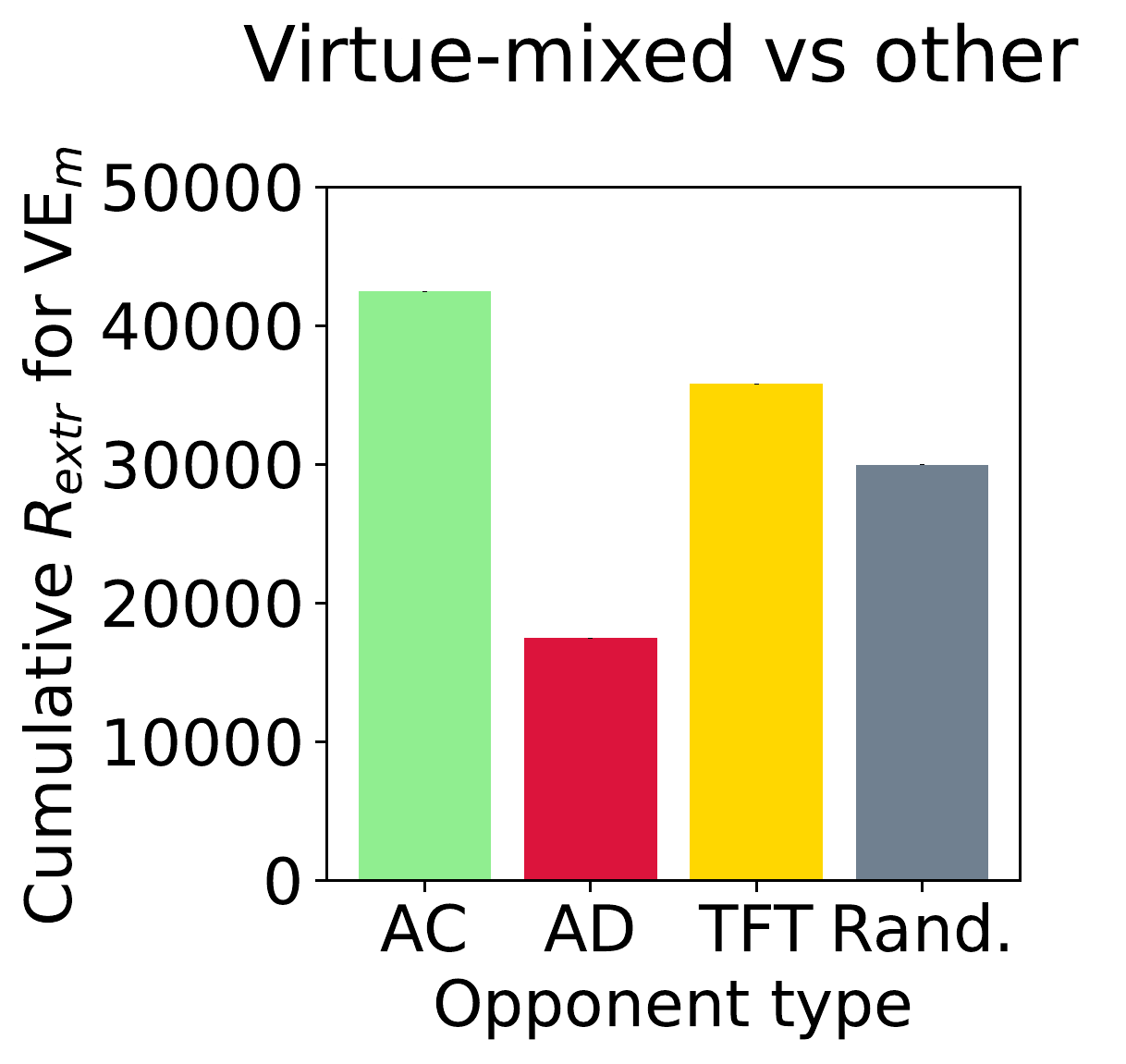}}
\\ 
\makecell[cc]{\rotatebox[origin=c]{90}{\thead{Moral Reward}}} &  & \subt{\includegraphics[height=20mm]{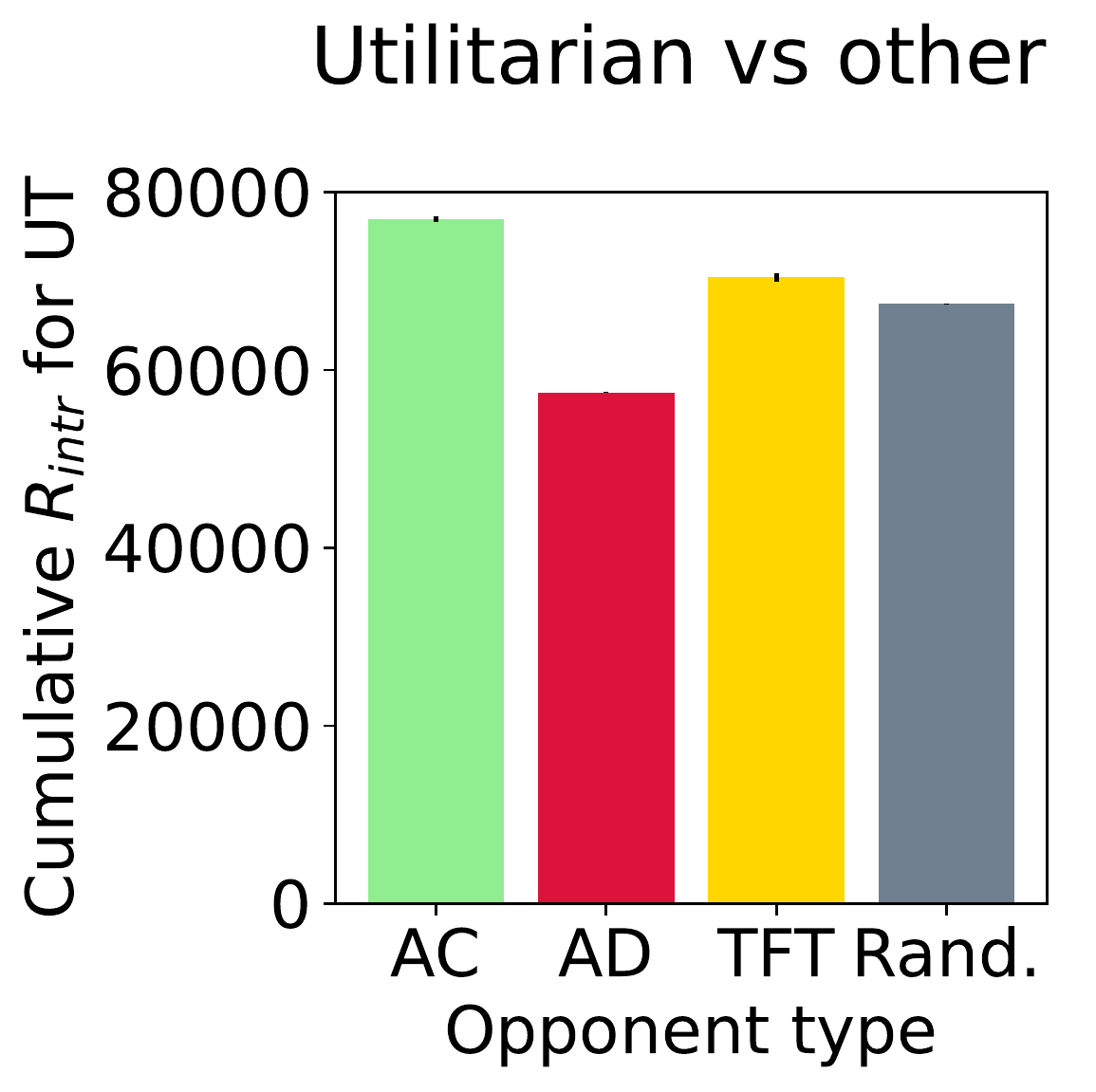}} & \subt{\includegraphics[height=20mm]{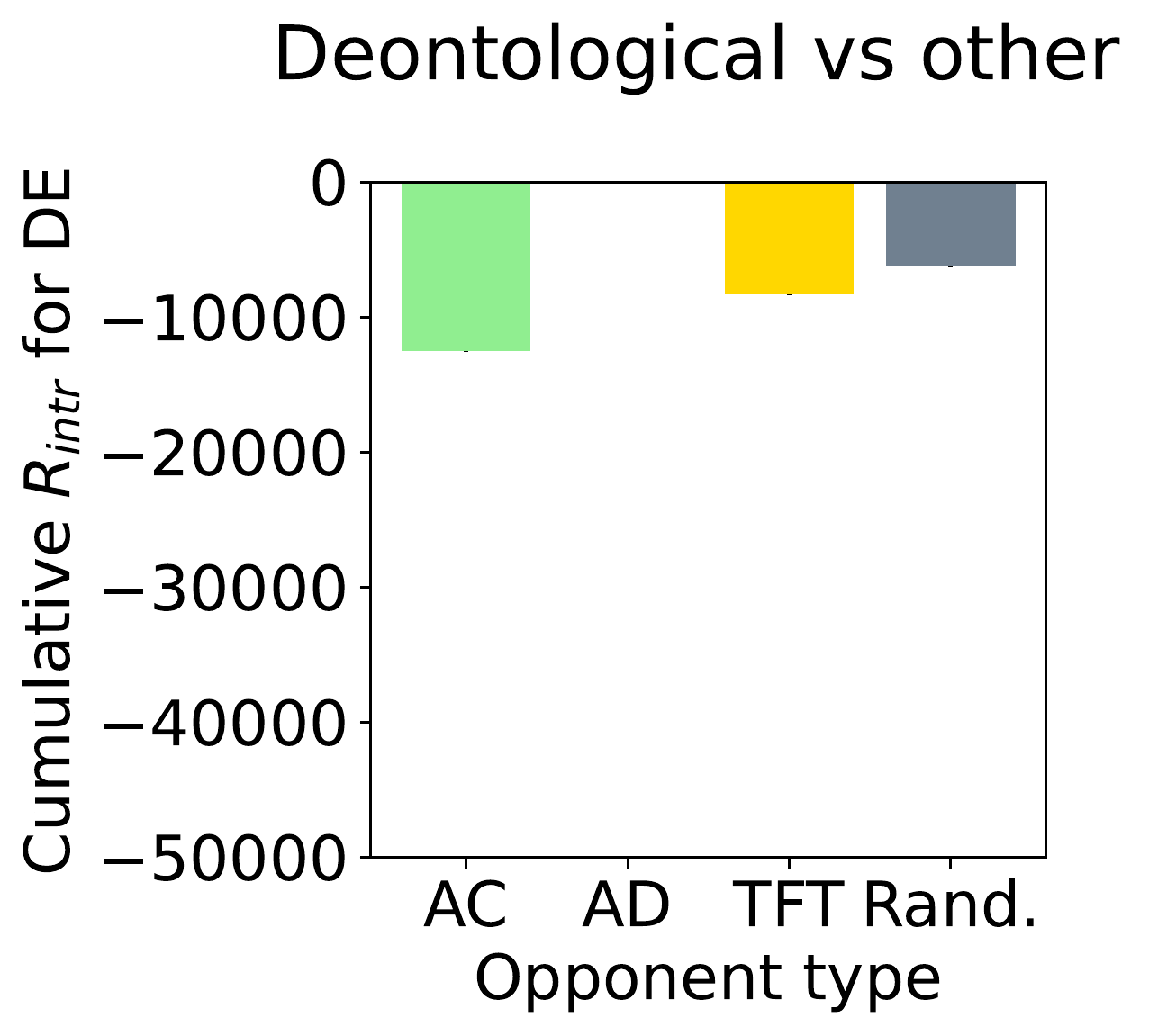}} & \subt{\includegraphics[height=20mm]{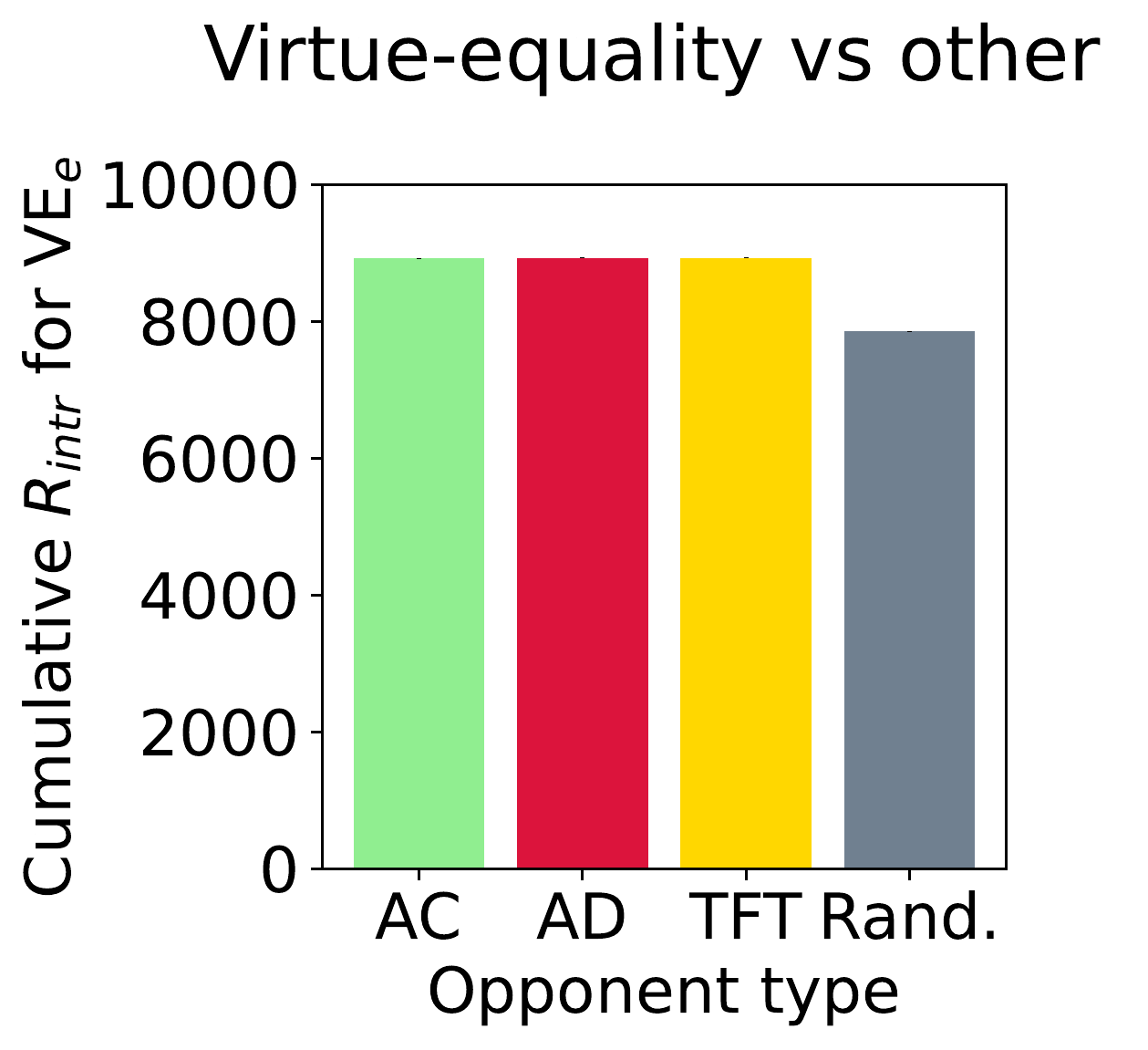}} & \subt{\includegraphics[height=20mm]{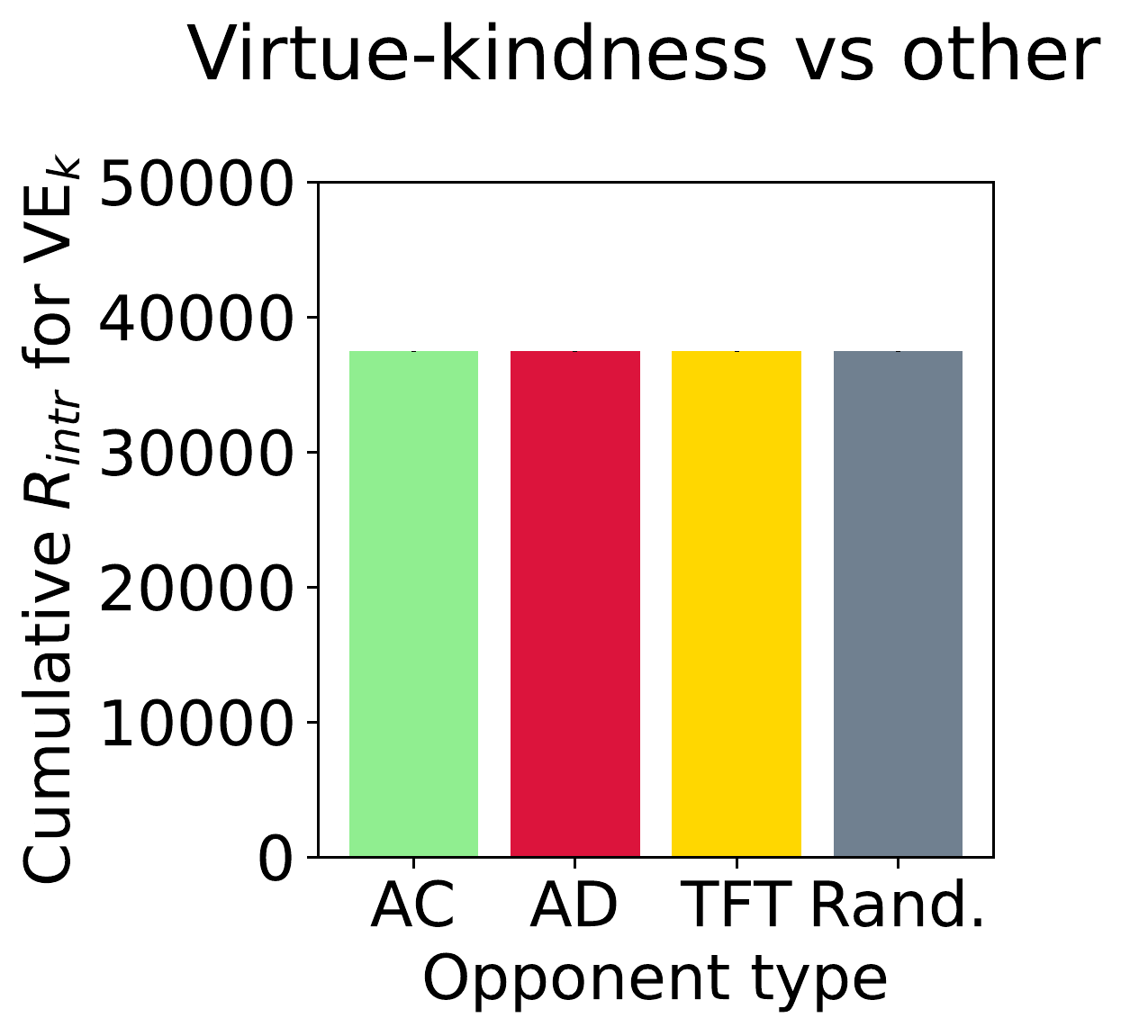}} & \subt{\includegraphics[height=20mm]{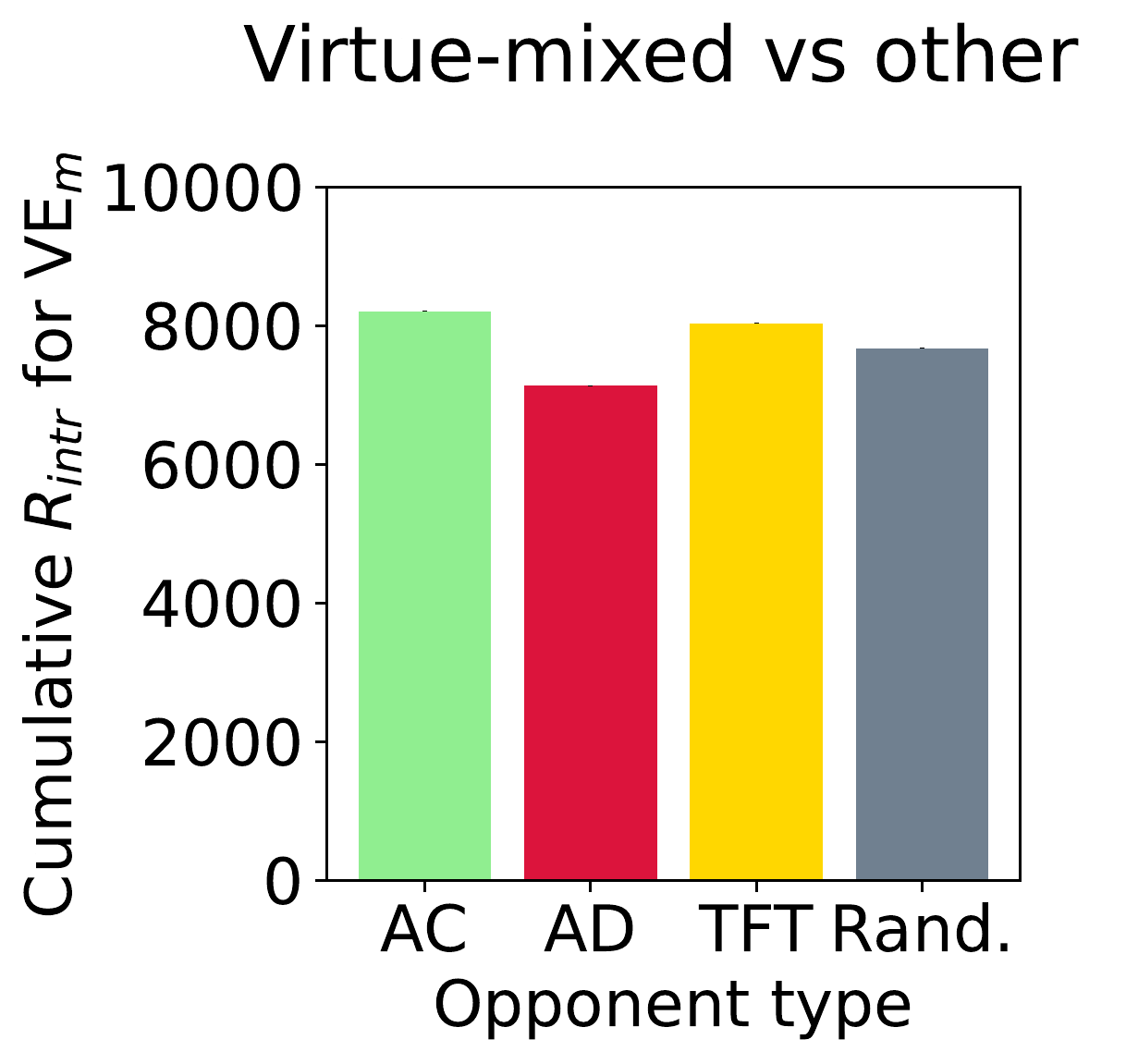}}
\\
\end{tabular}
\\ B. Iterated Volunteer's Dilemma \\

\begin{tabular}[t]{|c|cccccc}
\toprule
\makecell[cc]{\rotatebox[origin=c]{90}{\thead{Game Reward}}} & \subt{\includegraphics[height=20mm]{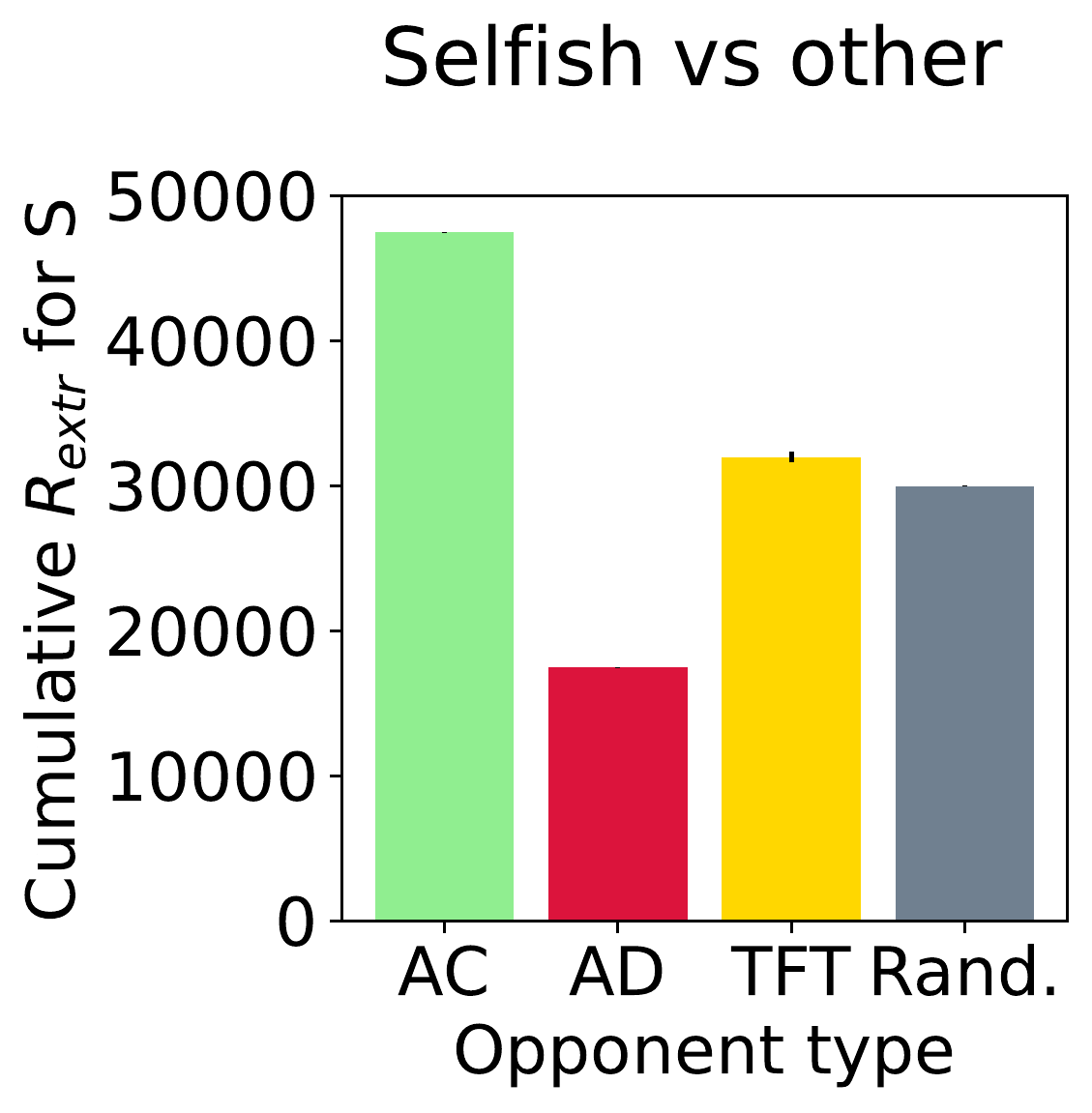}} & \subt{\includegraphics[height=20mm]{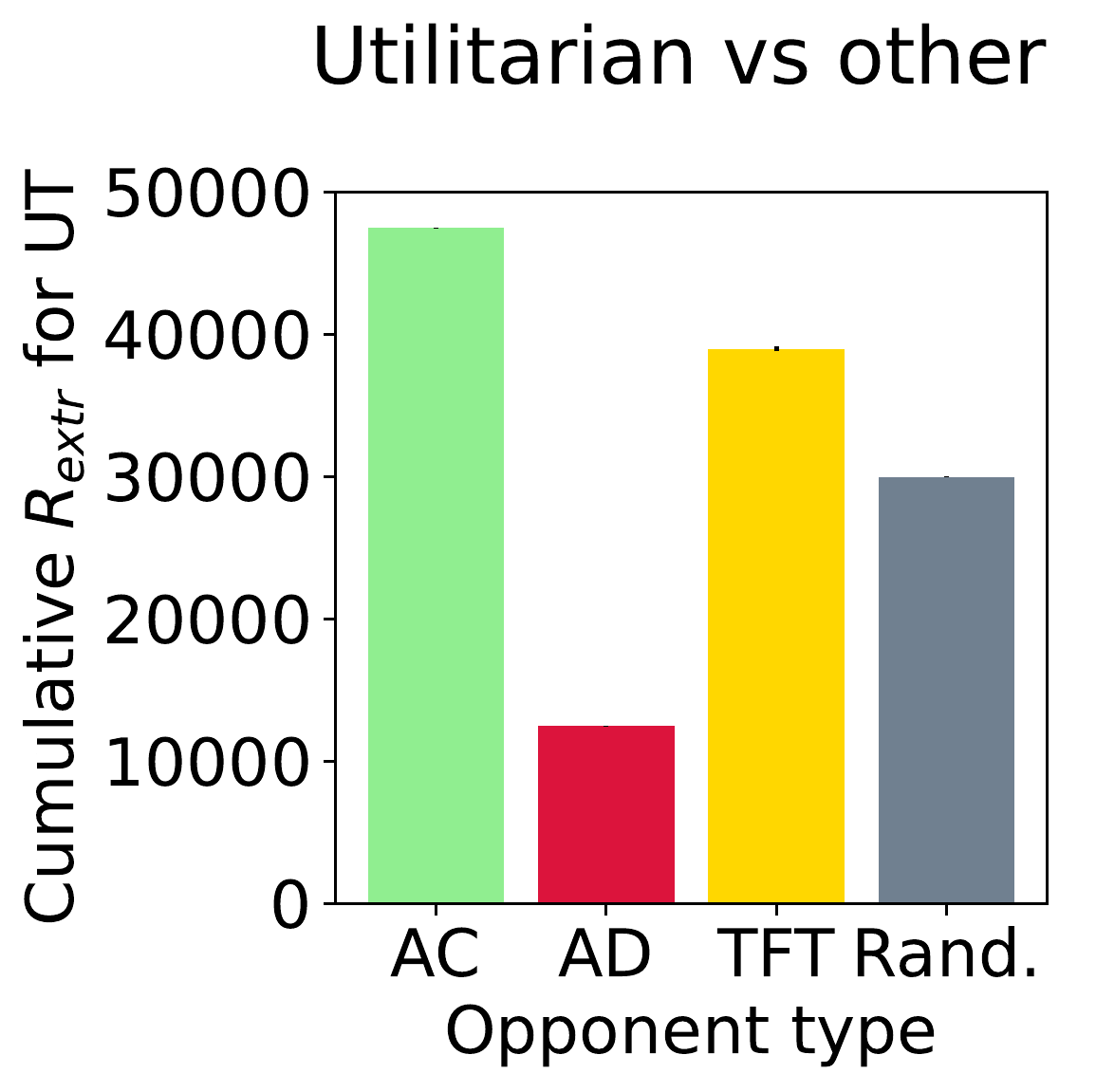}} & \subt{\includegraphics[height=20mm]{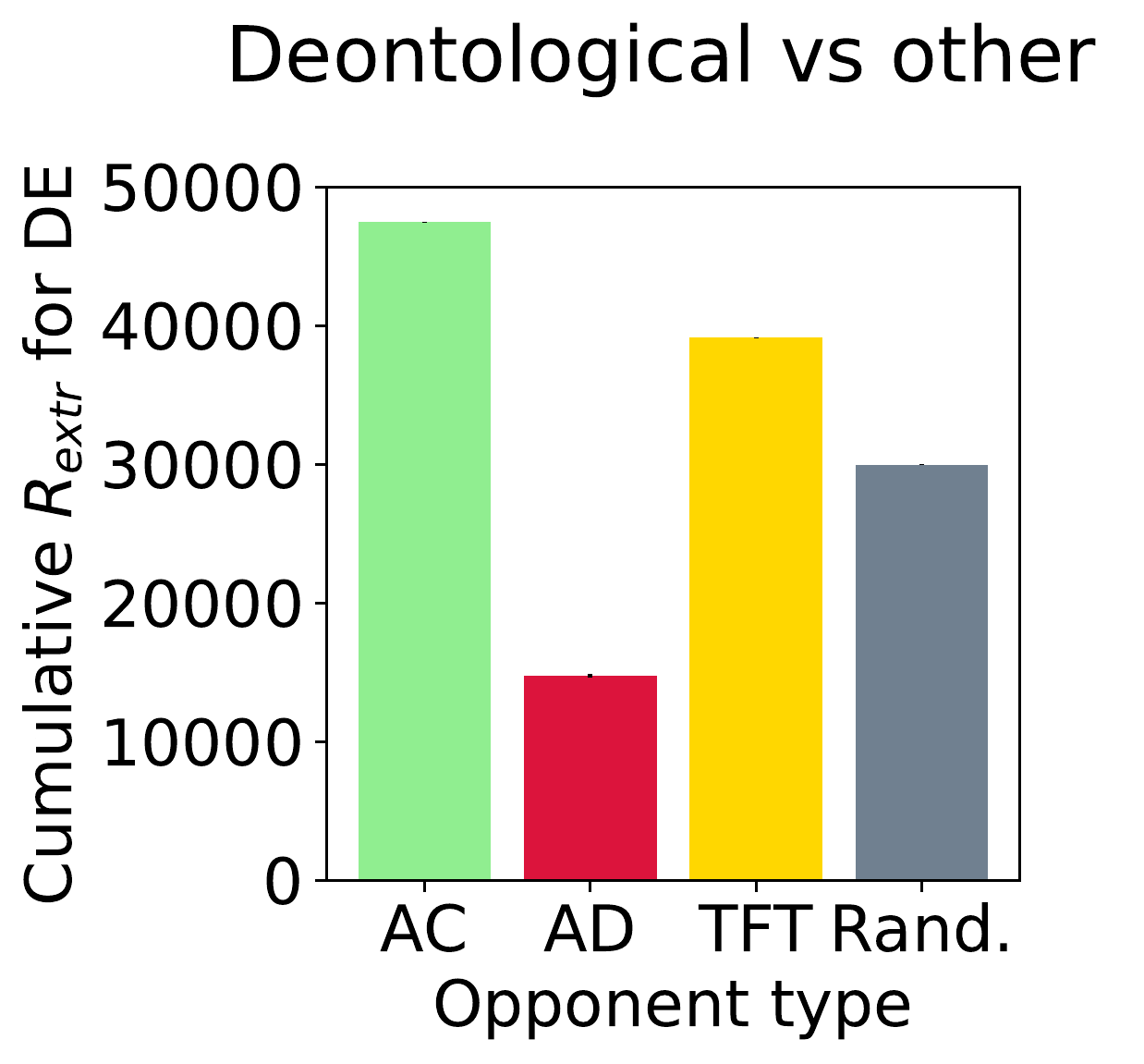}} & \subt{\includegraphics[height=20mm]{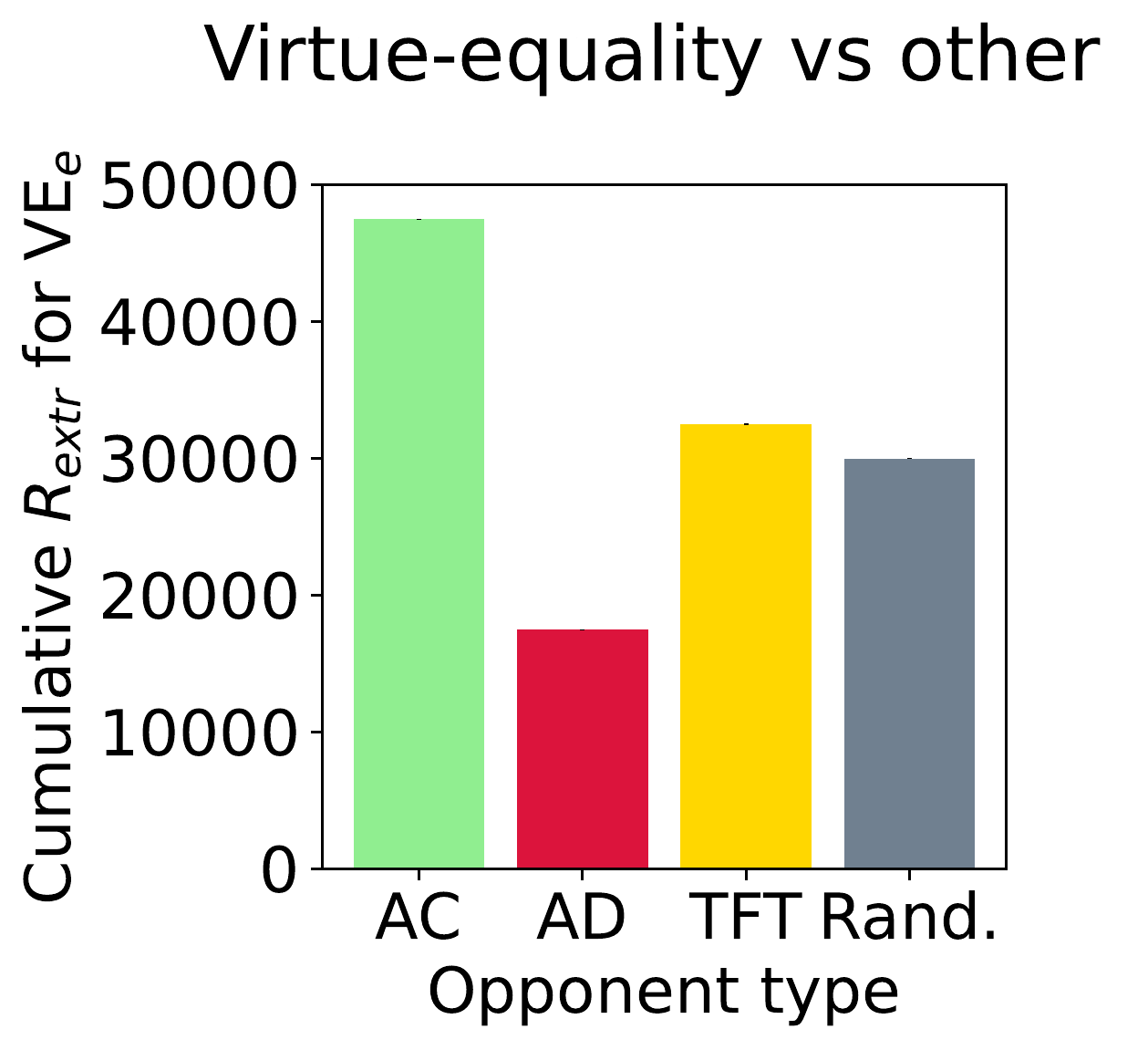}} & \subt{\includegraphics[height=20mm]{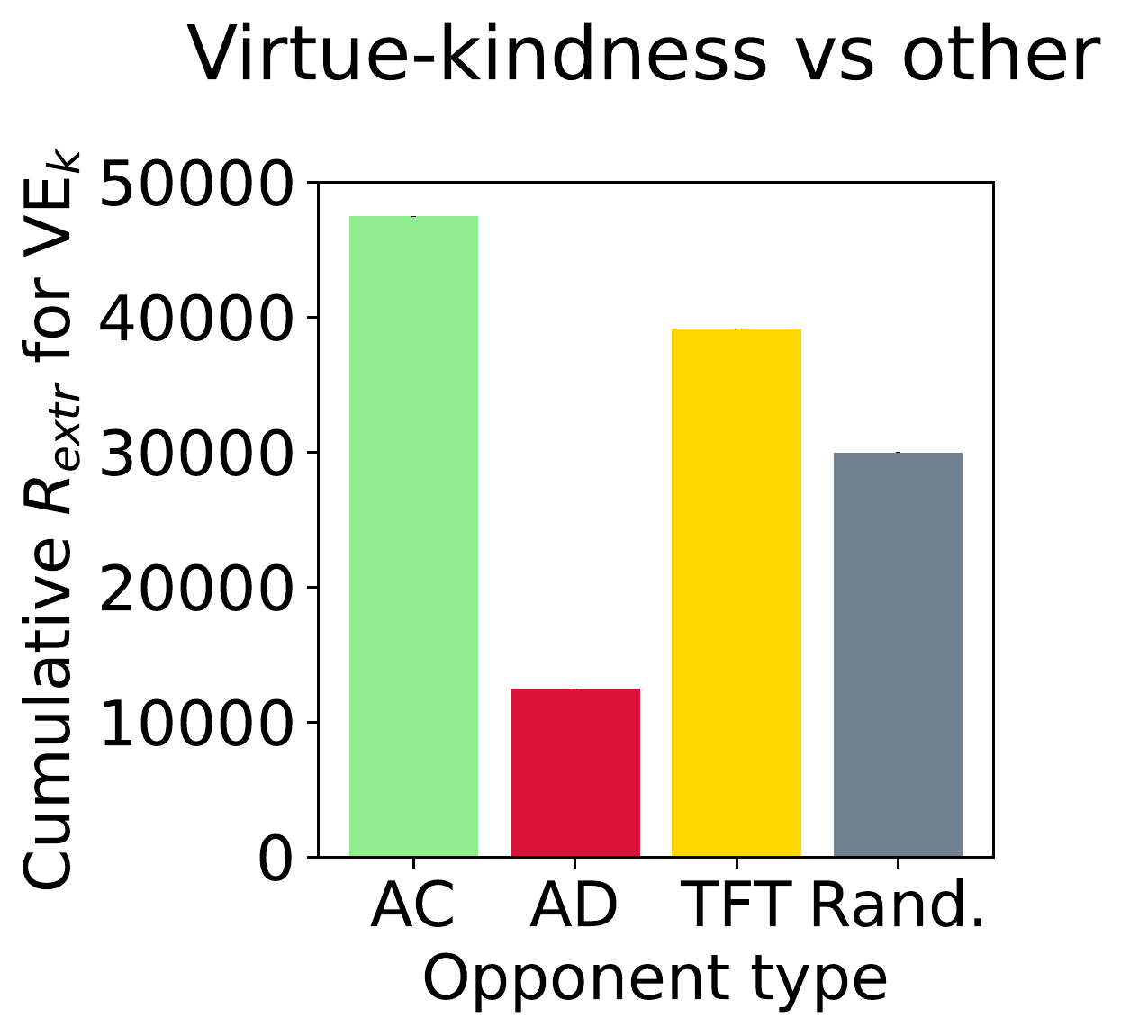}} & \subt{\includegraphics[height=20mm]{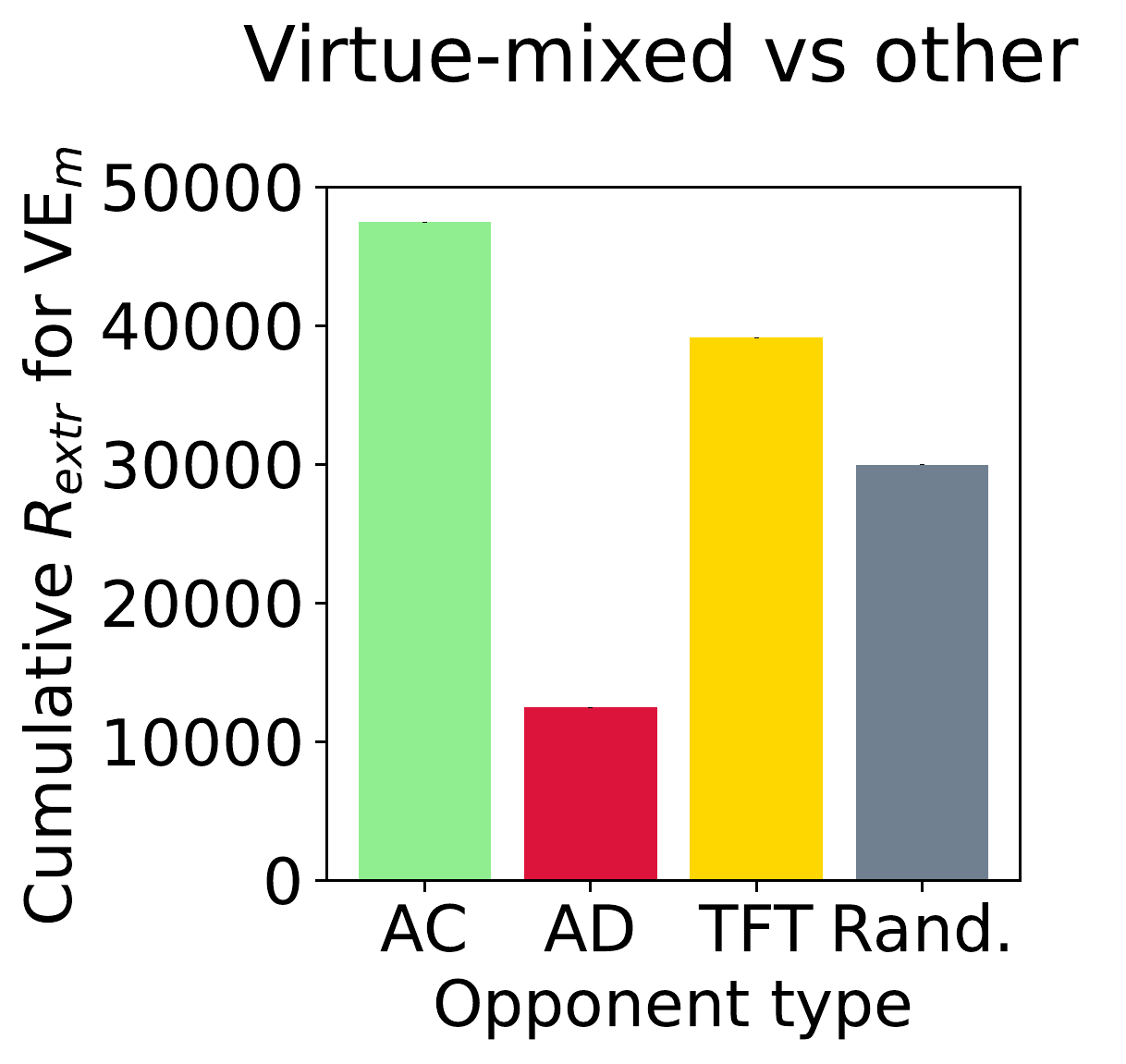}}
\\ 
\makecell[cc]{\rotatebox[origin=c]{90}{\thead{Moral Reward}}} &  & \subt{\includegraphics[height=20mm]{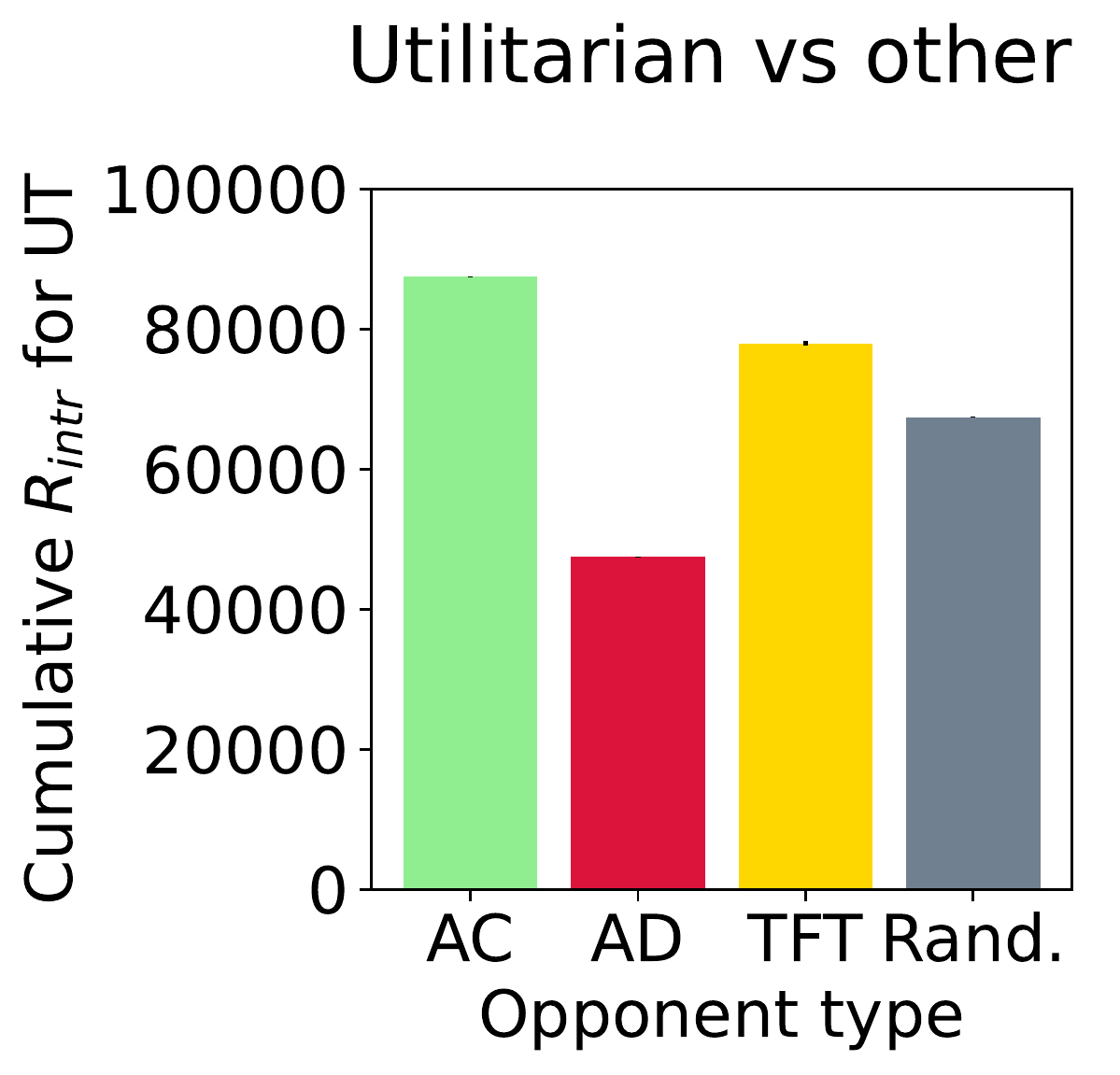}} & \subt{\includegraphics[height=20mm]{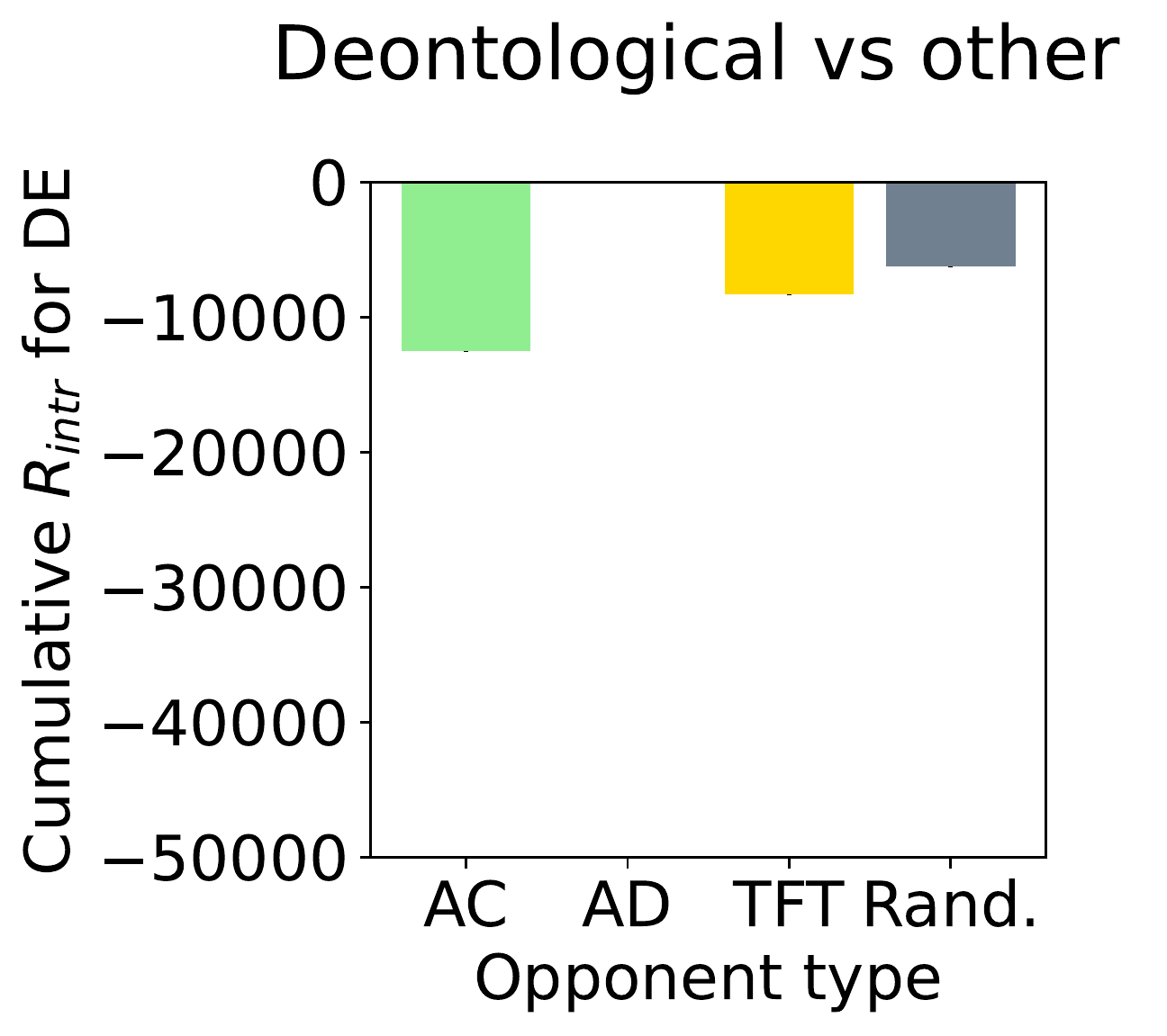}} & \subt{\includegraphics[height=20mm]{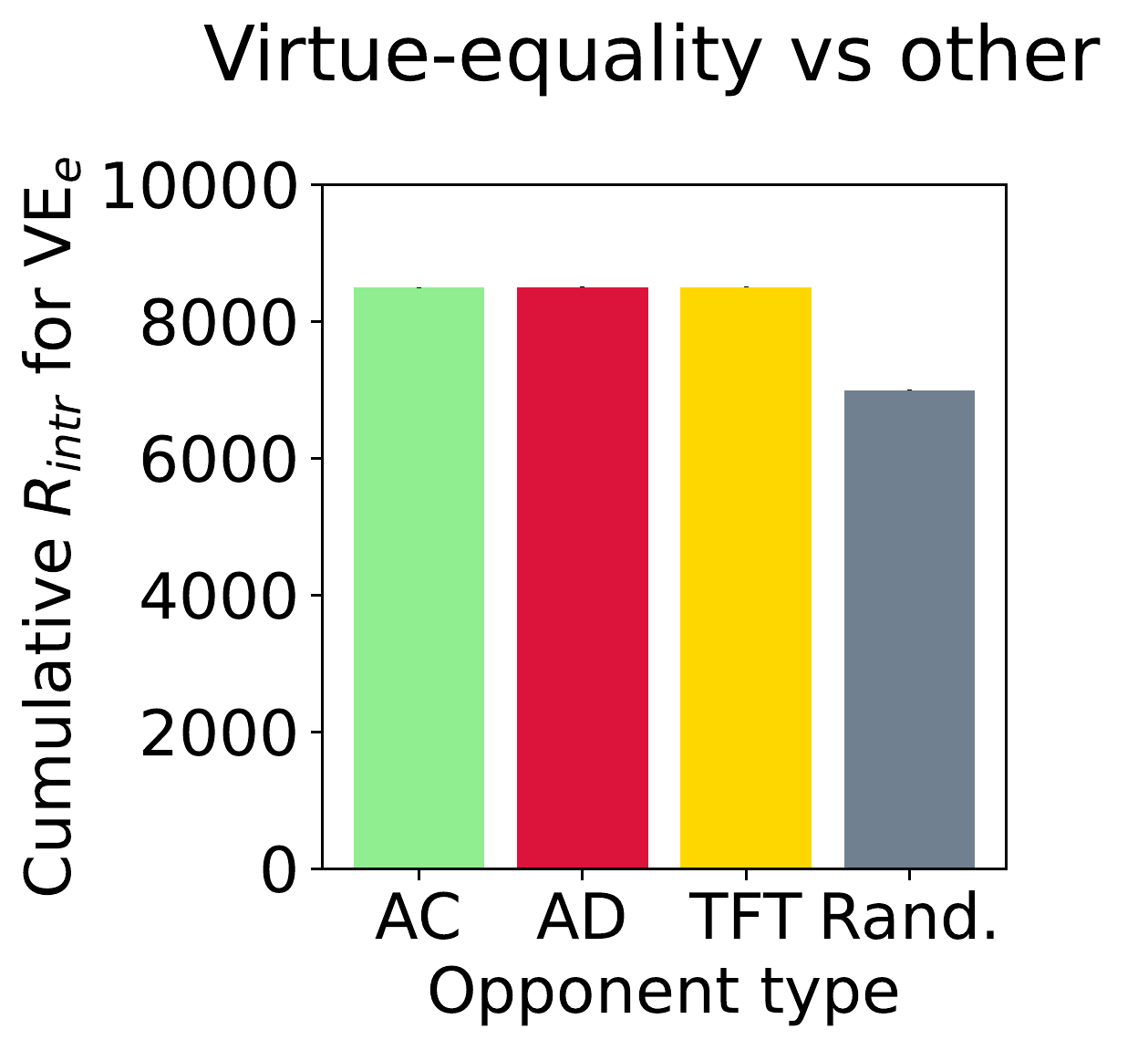}} & \subt{\includegraphics[height=20mm]{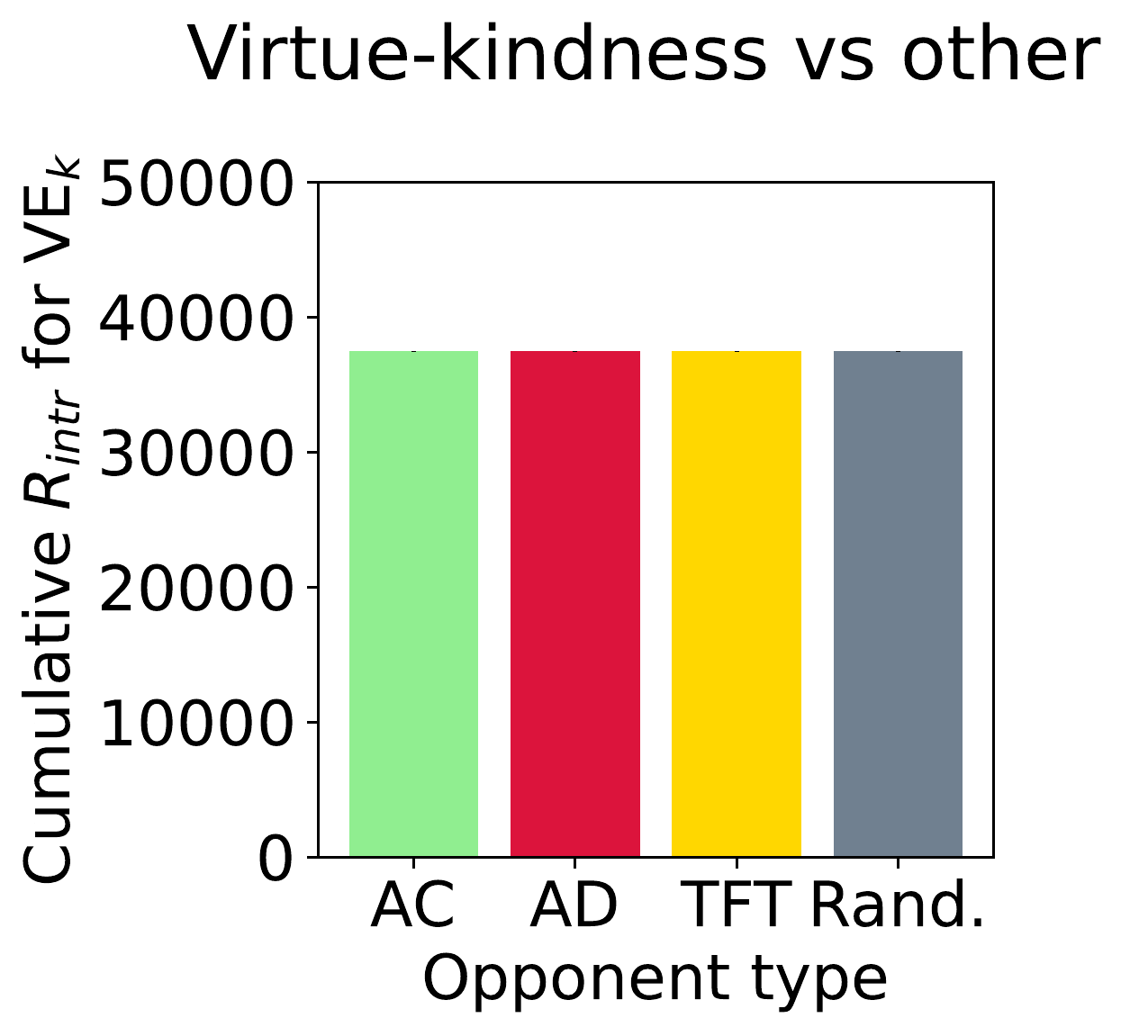}} & \subt{\includegraphics[height=20mm]{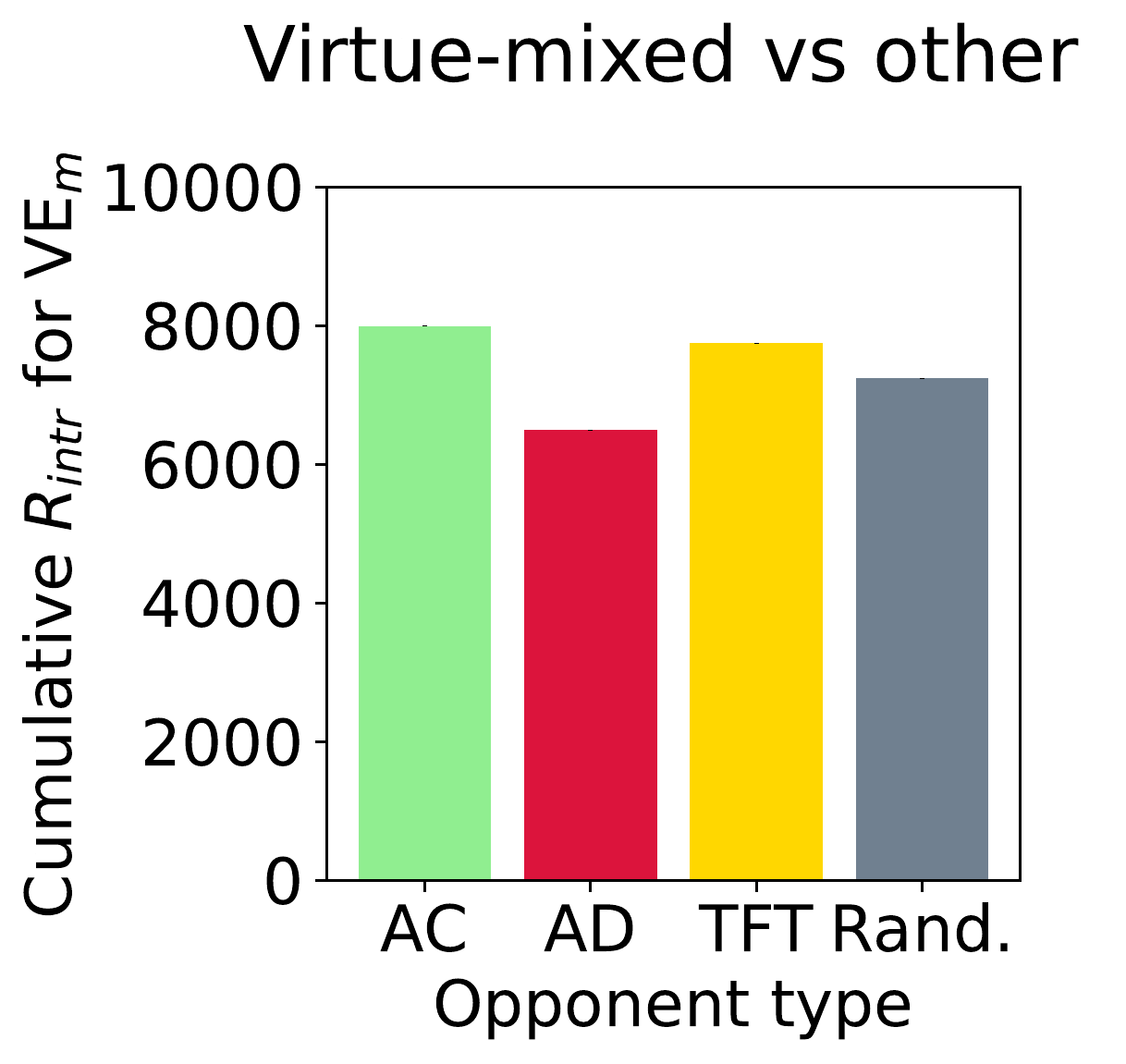}}
\\
\end{tabular}
\\ C. Iterated Stag Hunt \\ 

\caption{Game \& moral reward (cumulative) obtained after 10000 iterations by a given player type $M$ (column) vs. all possible static opponents $O$ - for all three games (panels A-C). The plots display averages across the 100 runs $\pm$ 95\%CI.}
\label{fig:reward_baseline}
\end{figure*}

Figure \ref{fig:reward_baseline} visualizes the game and moral reward obtained by each moral player against all static agent types - \textit{Always Cooperate, Always Defect, Tit for Tat} and \textit{Random}.

\subsection{Summary of Social Outcomes - Learning Player vs Static Opponent}

\begin{figure*}[!h]
\centering
\includegraphics[width=35mm]{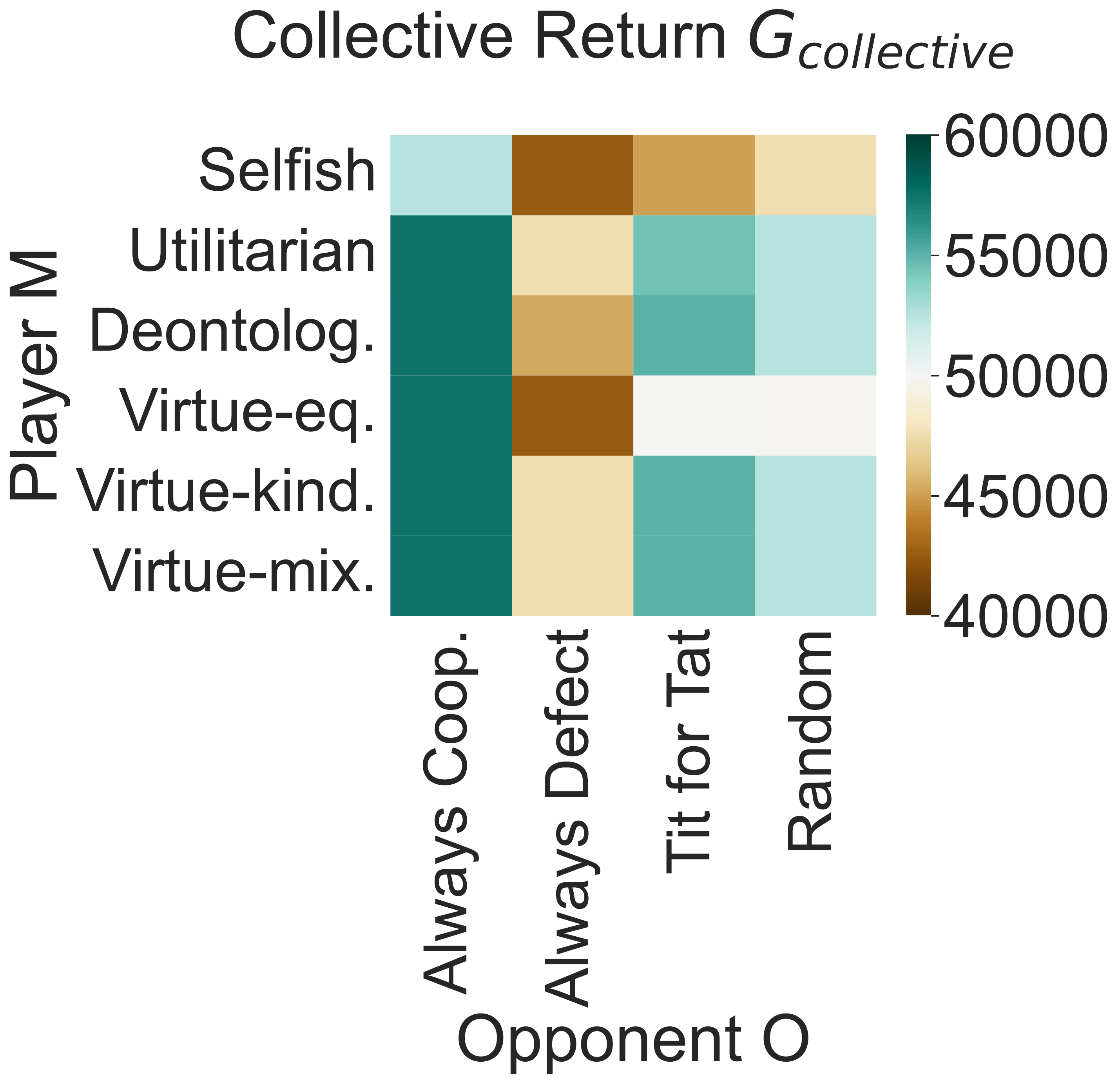}
\includegraphics[width=35mm]{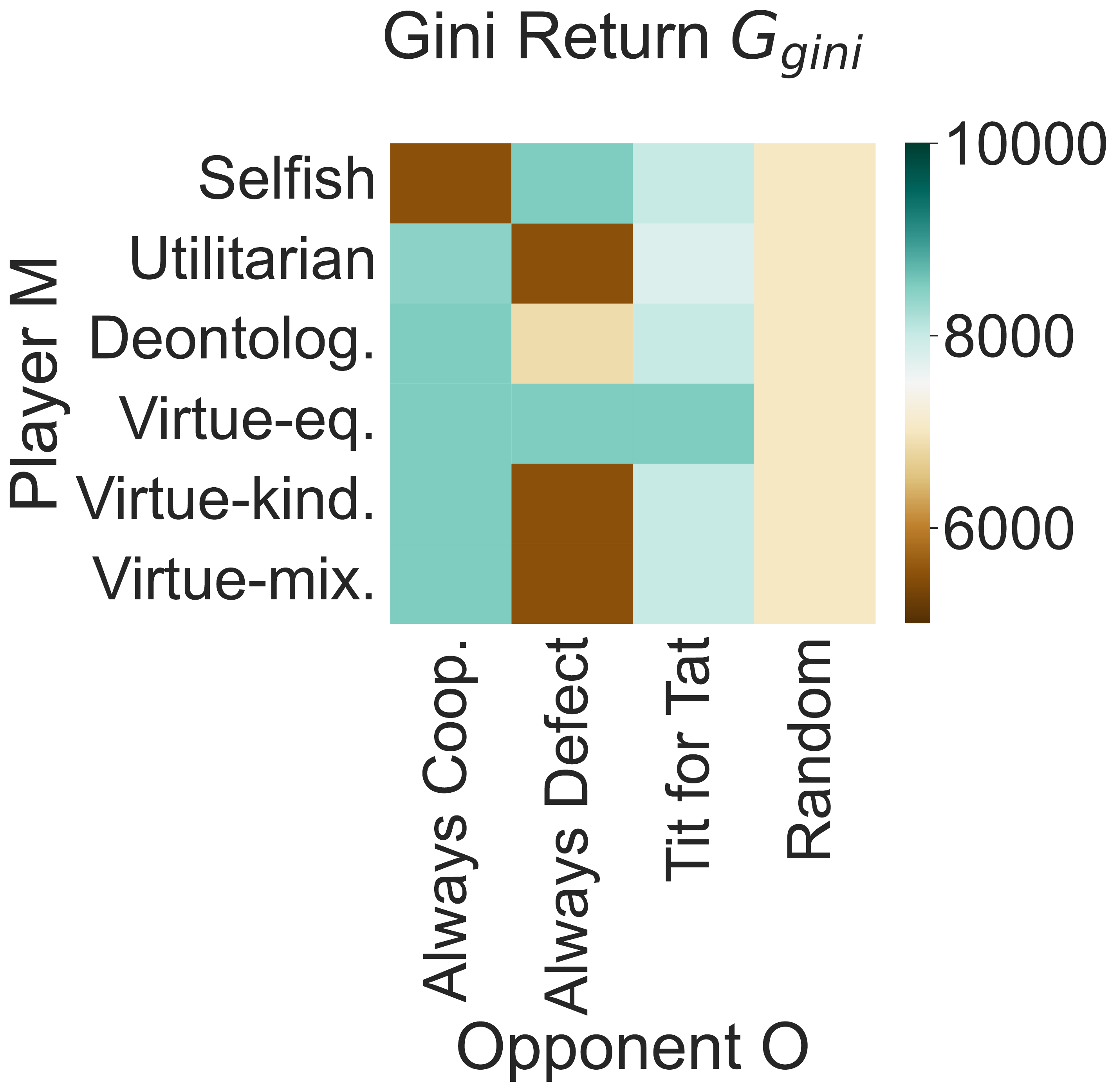}
\includegraphics[width=35mm]{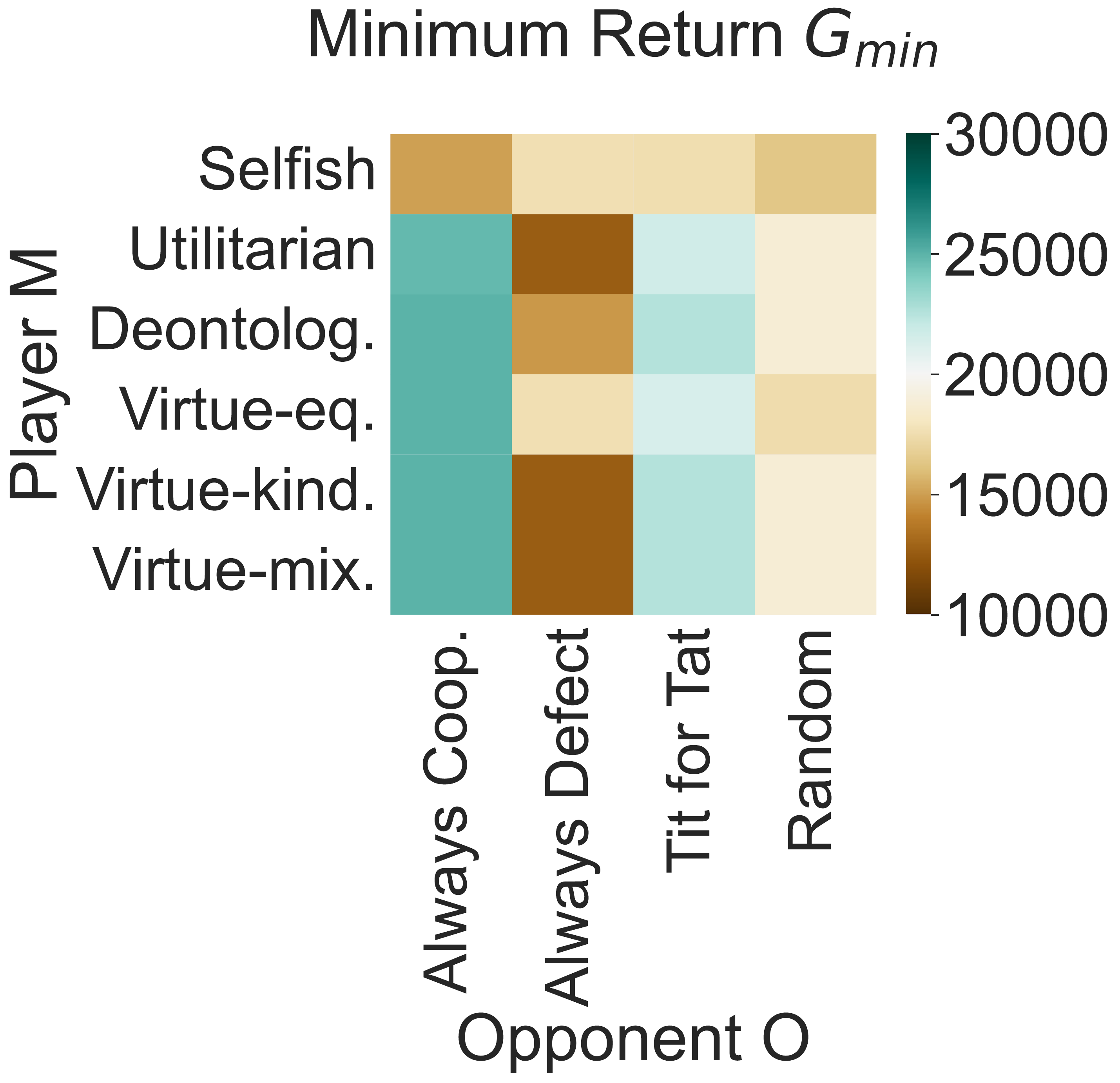}
\caption{Iterated Prisoner's Dilemma game. Relative social outcomes observed after 10000 iterations for learning player type $M$ (row) vs. all possible static opponents $O$. The plots display averages across the 100 runs.}
\label{fig:baseline_outcomes_IPD}
\end{figure*}

\begin{figure*}[!h]
\centering
\includegraphics[width=35mm]{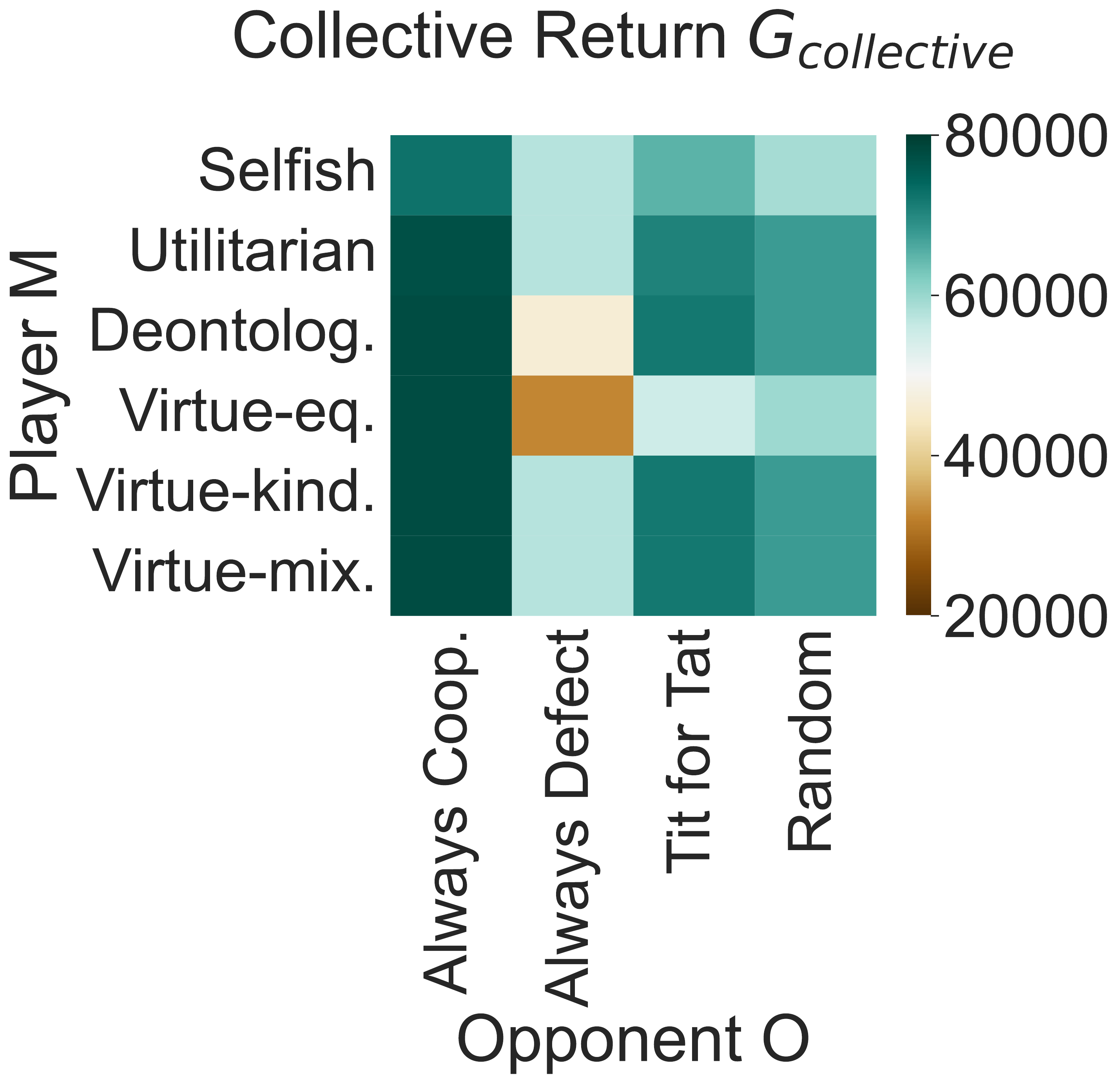}
\includegraphics[width=35mm]{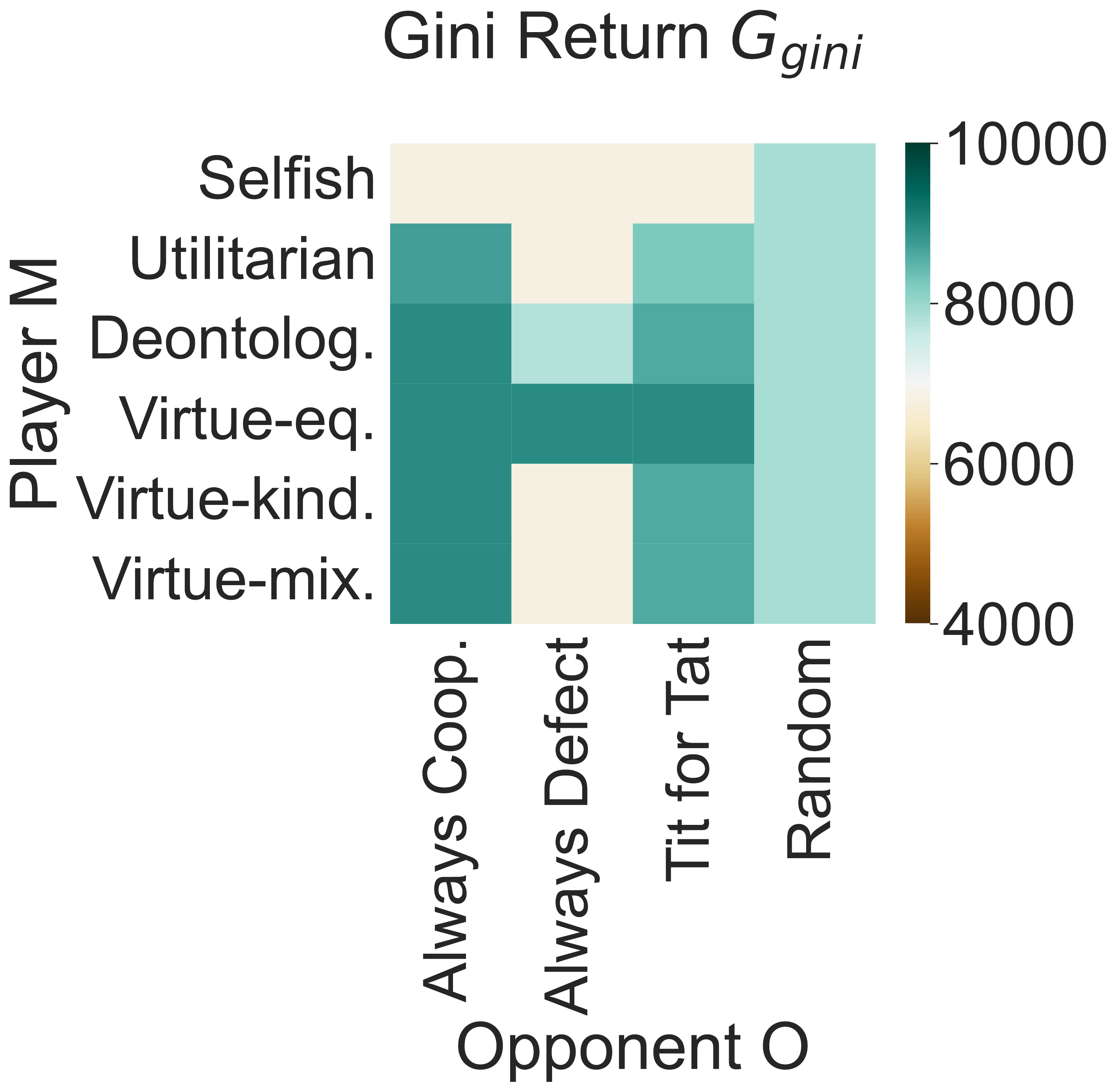}
\includegraphics[width=35mm]{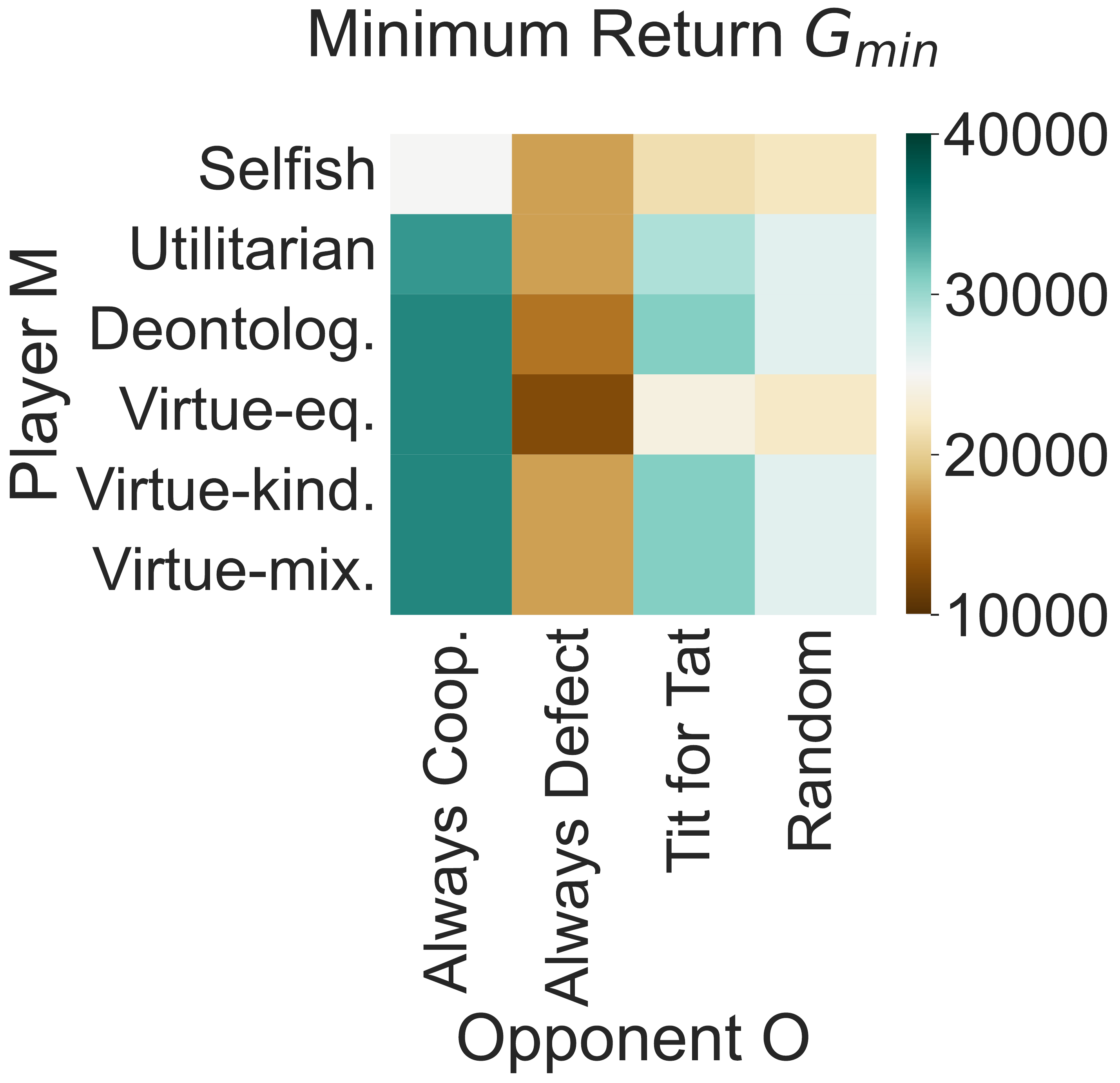}
\caption{Iterated Volunteer's Dilemma game. Relative social outcomes observed after 10000 iterations for learning player type $M$ (row) vs. all possible static opponents $O$. The plots display averages across the 100 runs.}
\label{fig:baseline_outcomes_VOL}
\end{figure*}

\begin{figure*}[!h]
\centering
\includegraphics[width=35mm]{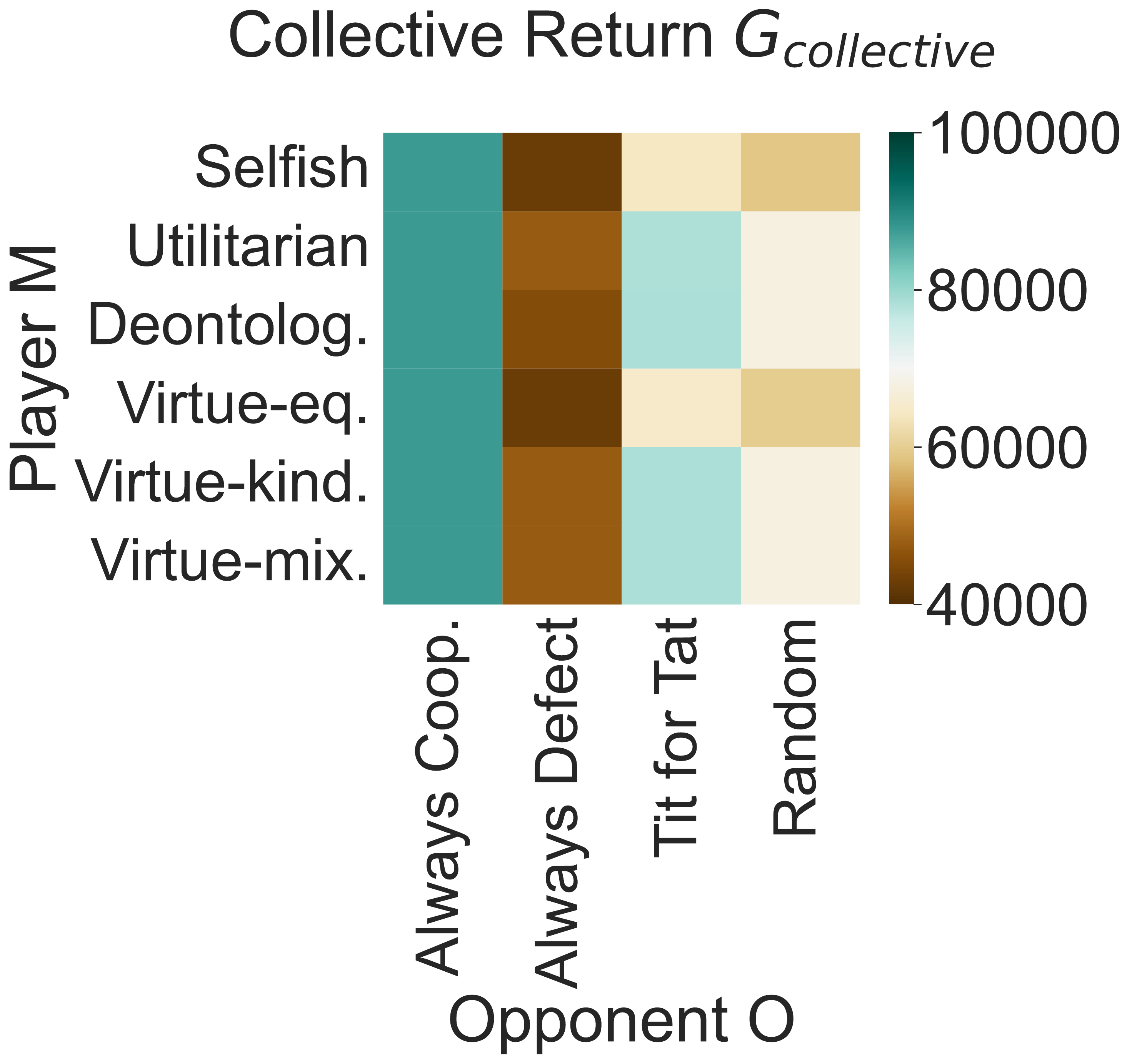}
\includegraphics[width=35mm]{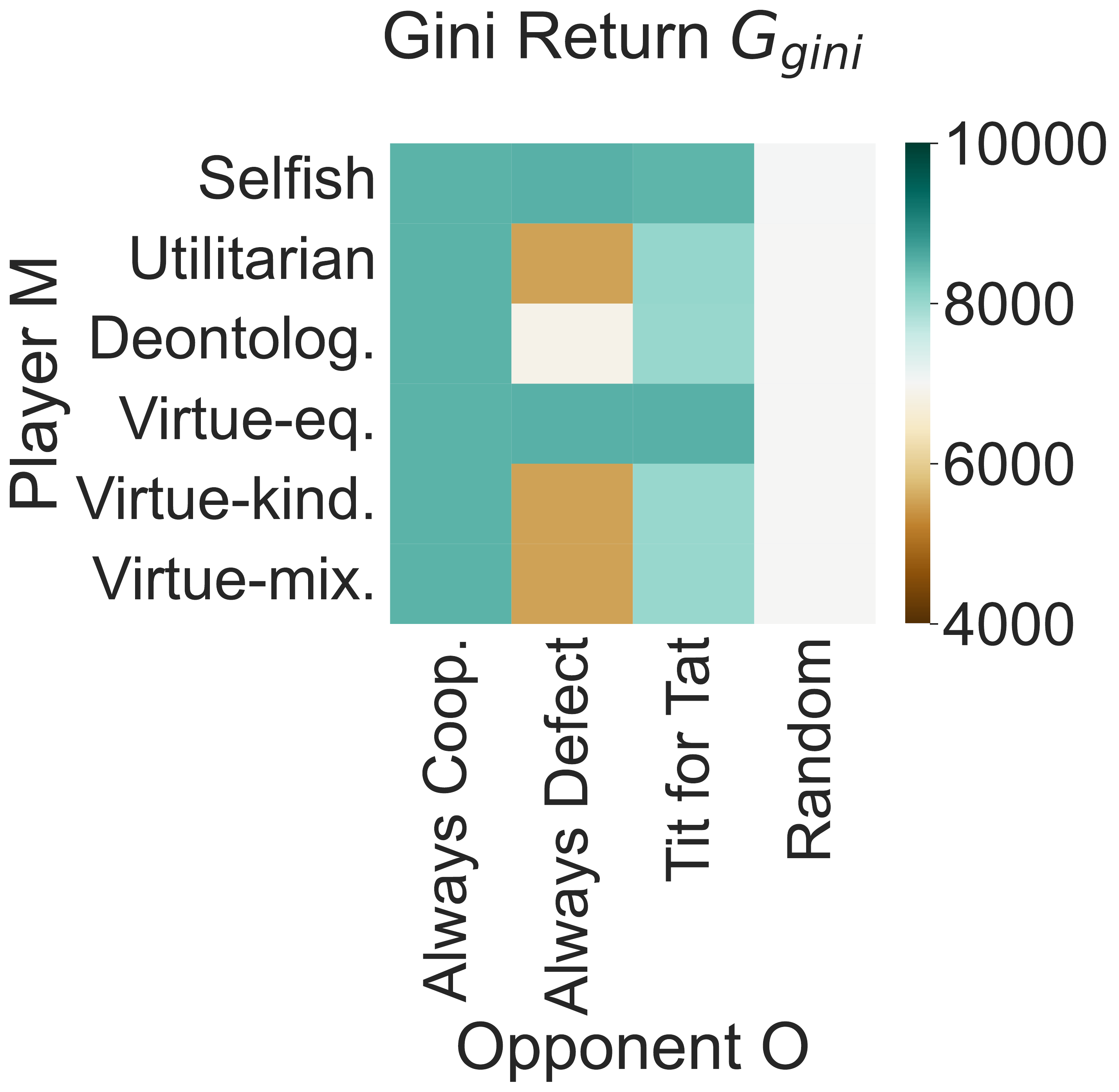}
\includegraphics[width=35mm]{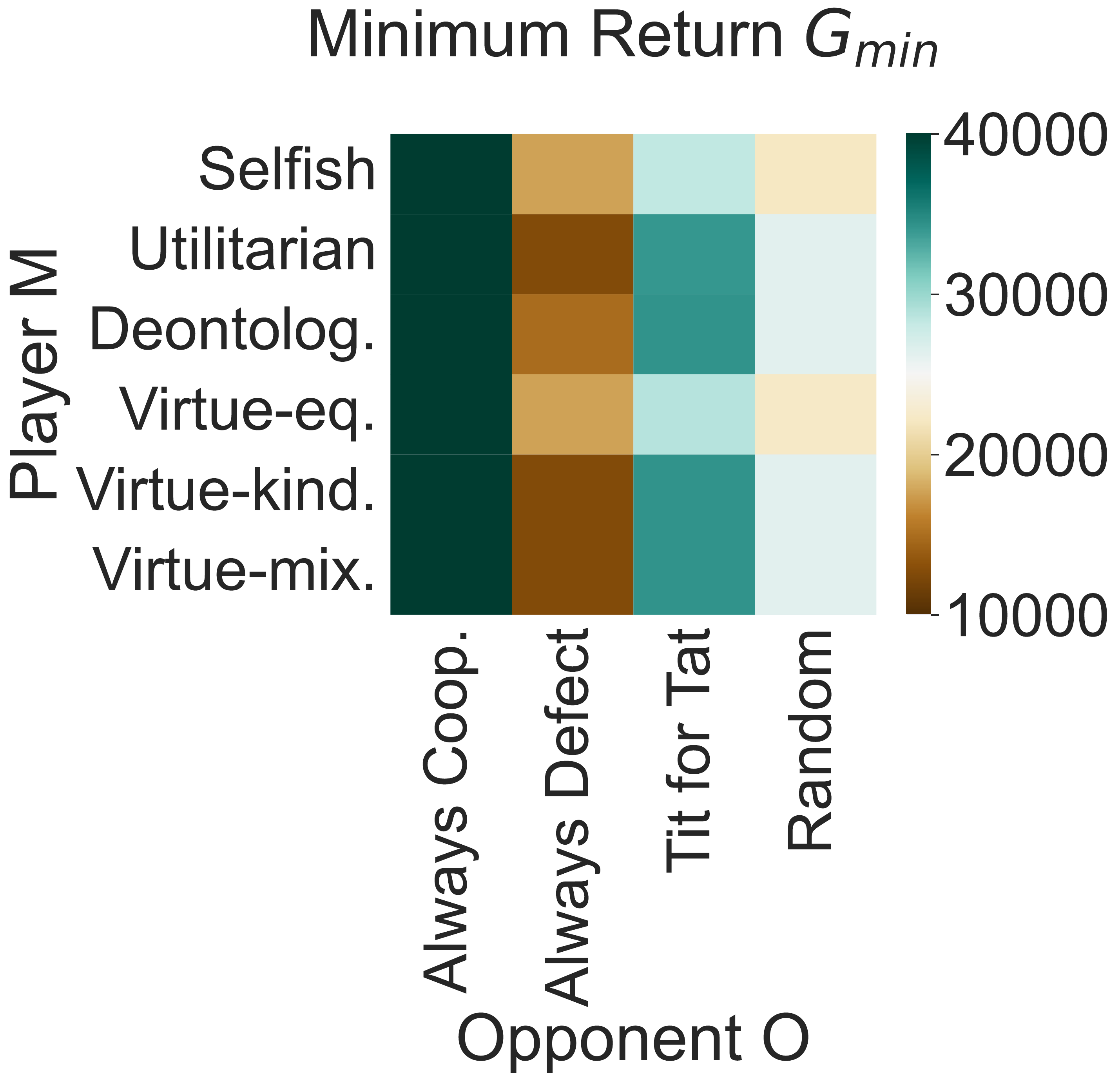}
\caption{Iterated Stag Hunt game. Relative social outcomes observed after 10000 iterations for learning player type $M$ (row) vs. all possible learning static $O$. The plots display averages across the 100 runs.}
\label{fig:baseline_outcomes_STH}
\end{figure*}

In Figures \ref{fig:baseline_outcomes_IPD}-\ref{fig:baseline_outcomes_STH} we provide a summary of the social outcomes obtained when moral players learn against static opponents.

\section{Game and Moral Reward}
\subsection{Reward obtained by the end of the training}
\label{subsection:individual_outcomes}

\begin{figure*}[h!]
\centering
\begin{tabular}[t]{|c|cccccc}
\toprule
& Selfish & Utilitarian & Deontological & Virtue-equality & Virtue-kindness & Virtue-mixed \\
\midrule
\makecell[cc]{\rotatebox[origin=c]{90}{\thead{Game Reward}}} & 
\subt{\includegraphics[height=21mm]{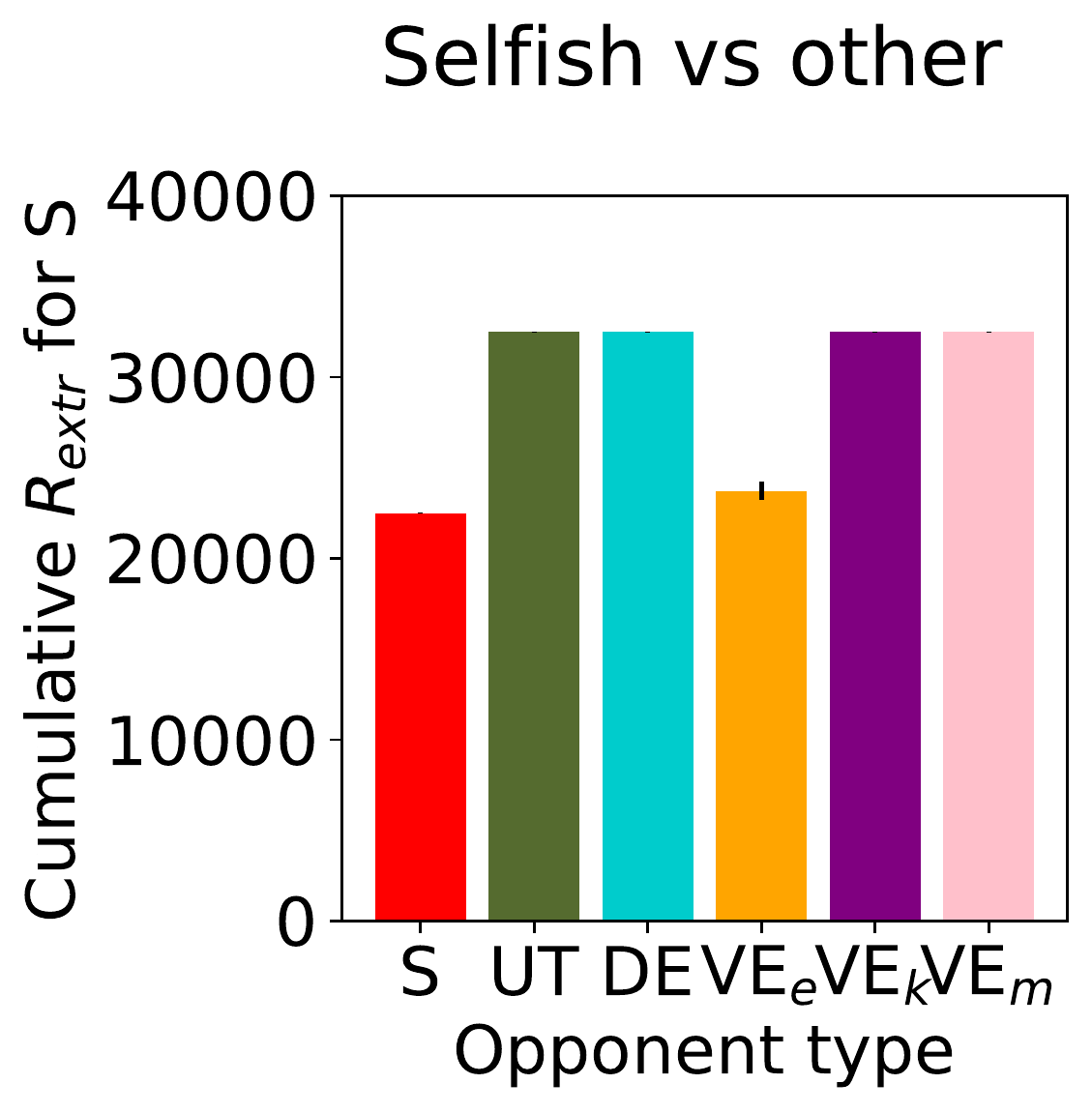}} & \subt{\includegraphics[height=21mm]{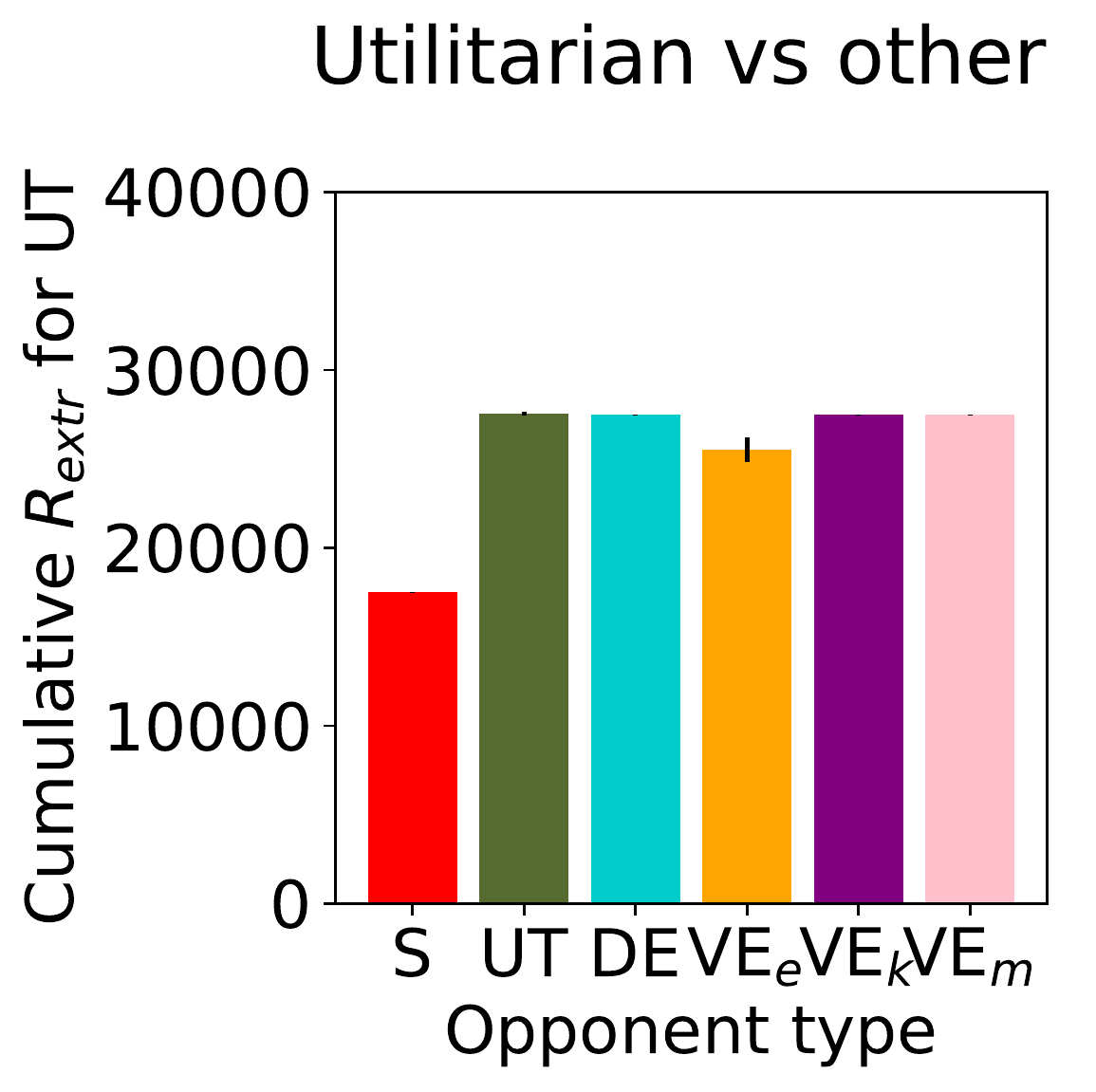}} & \subt{\includegraphics[height=21mm]{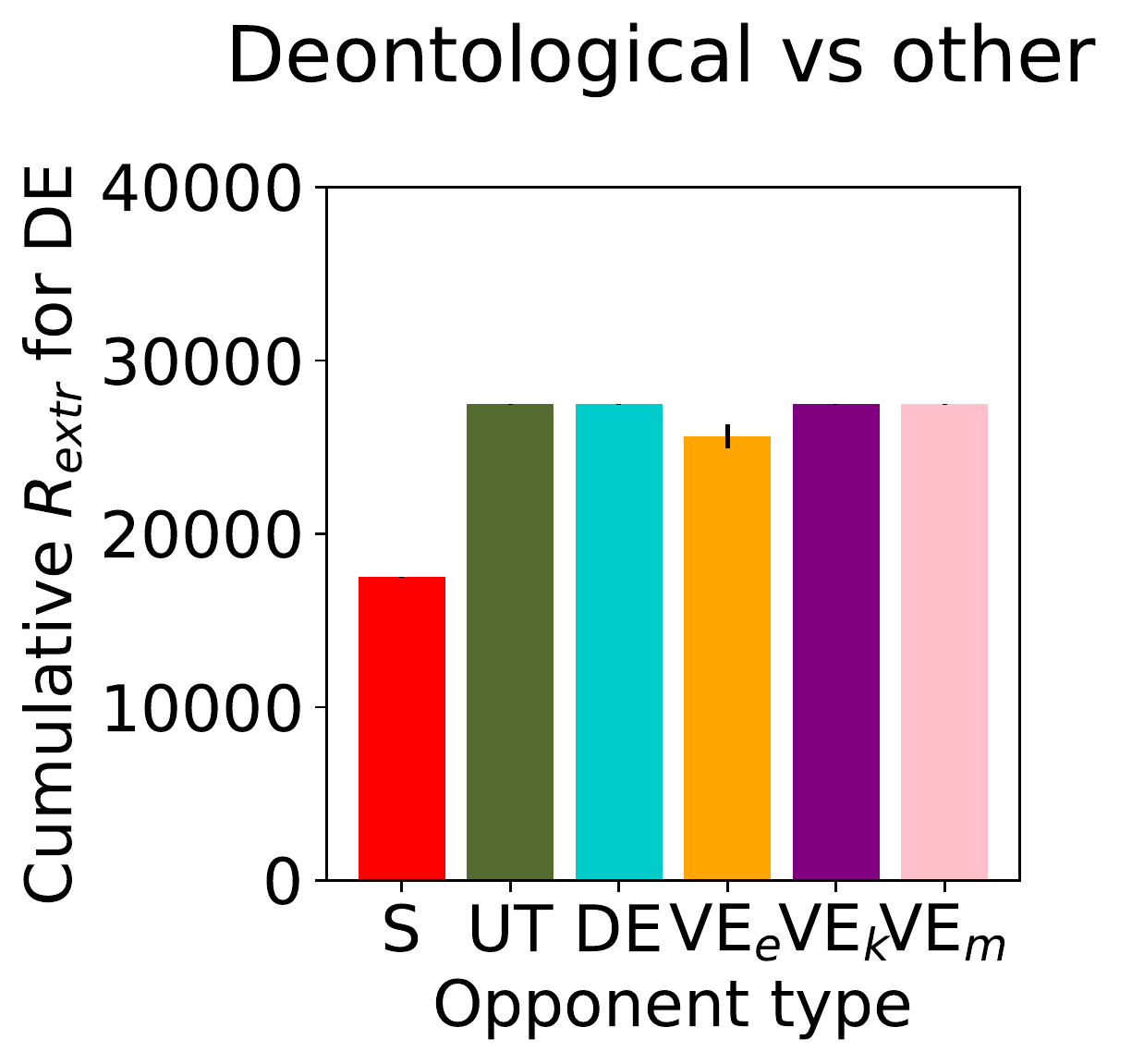}} & \subt{\includegraphics[height=21mm]{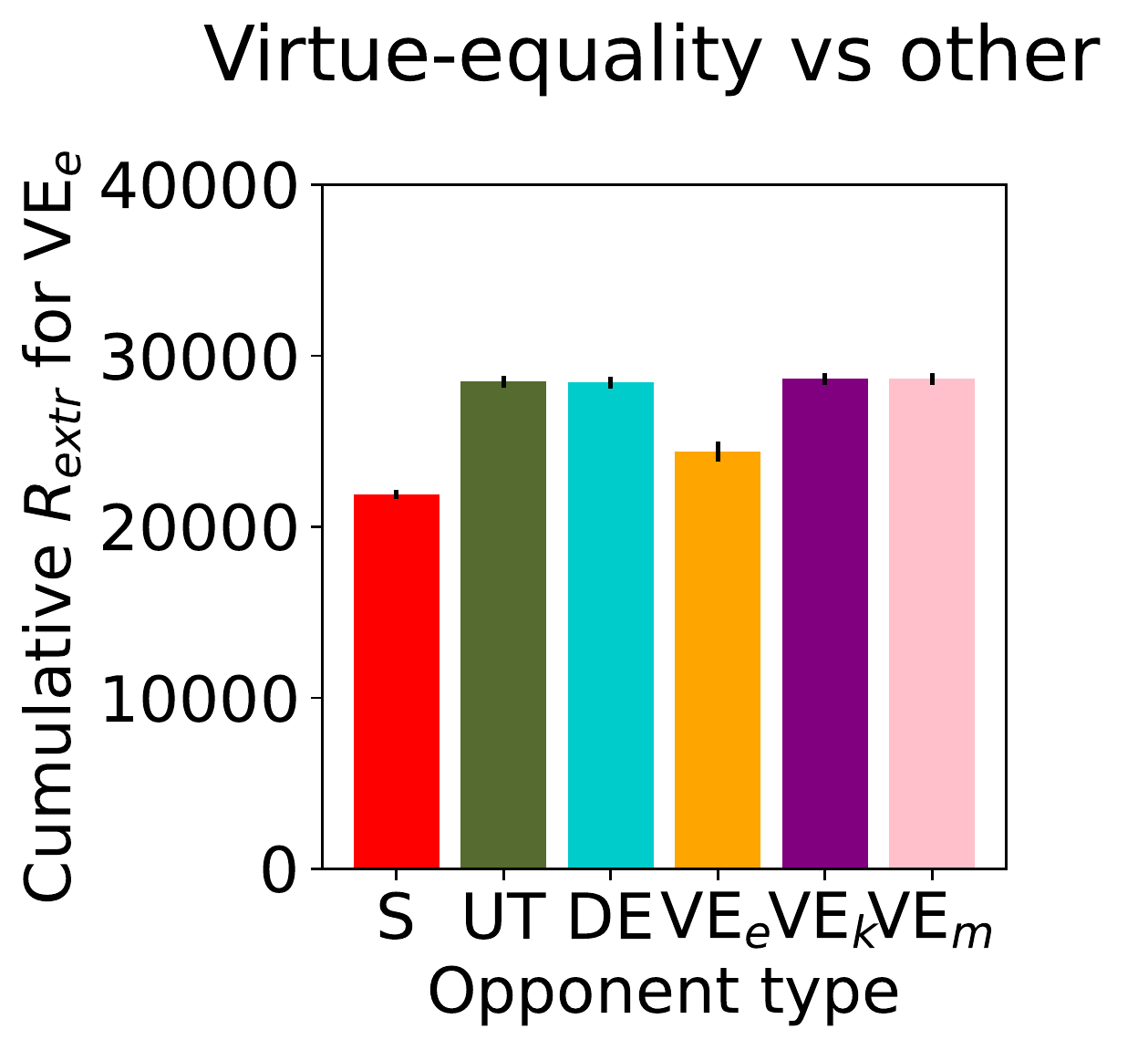}} & \subt{\includegraphics[height=21mm]{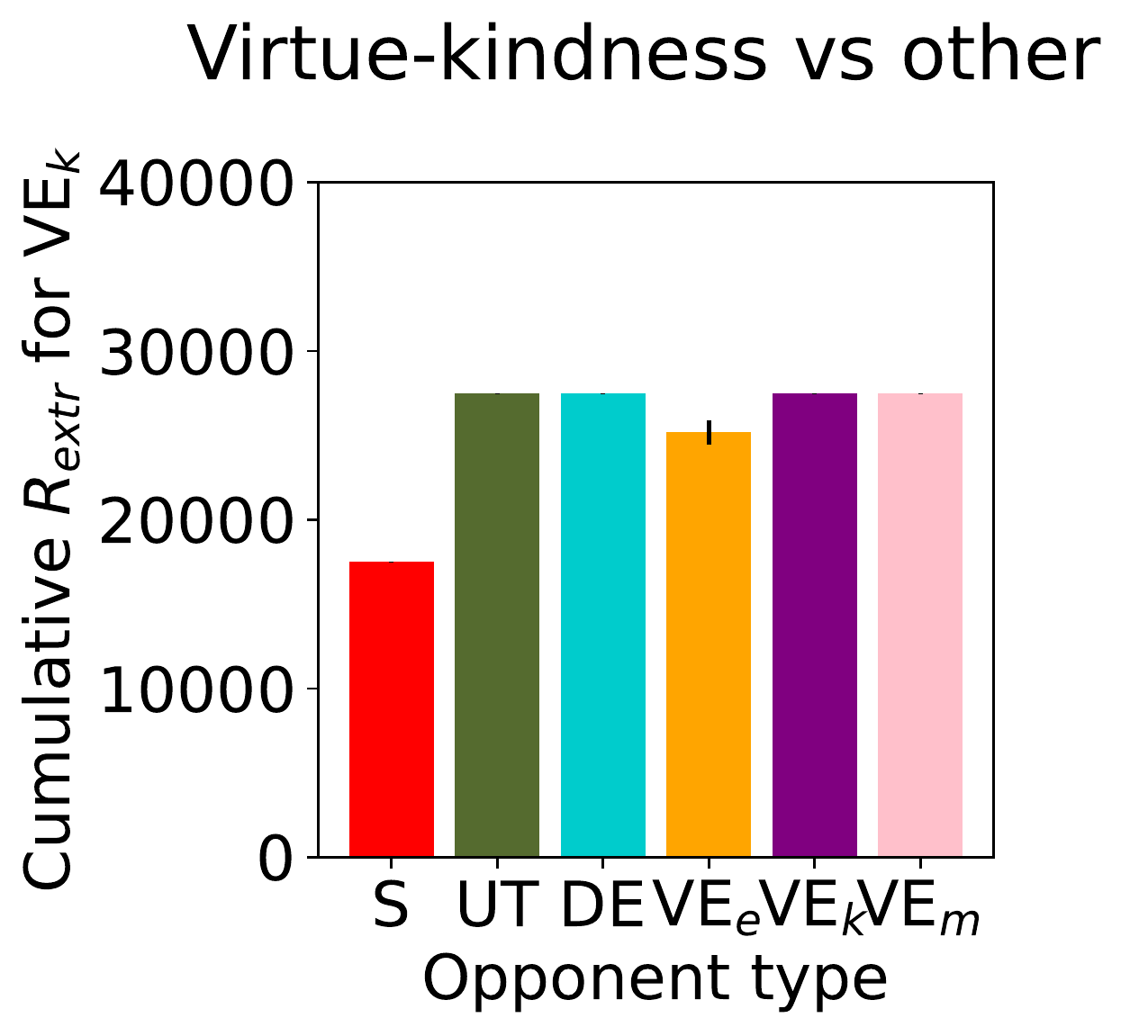}} & \subt{\includegraphics[height=21mm]{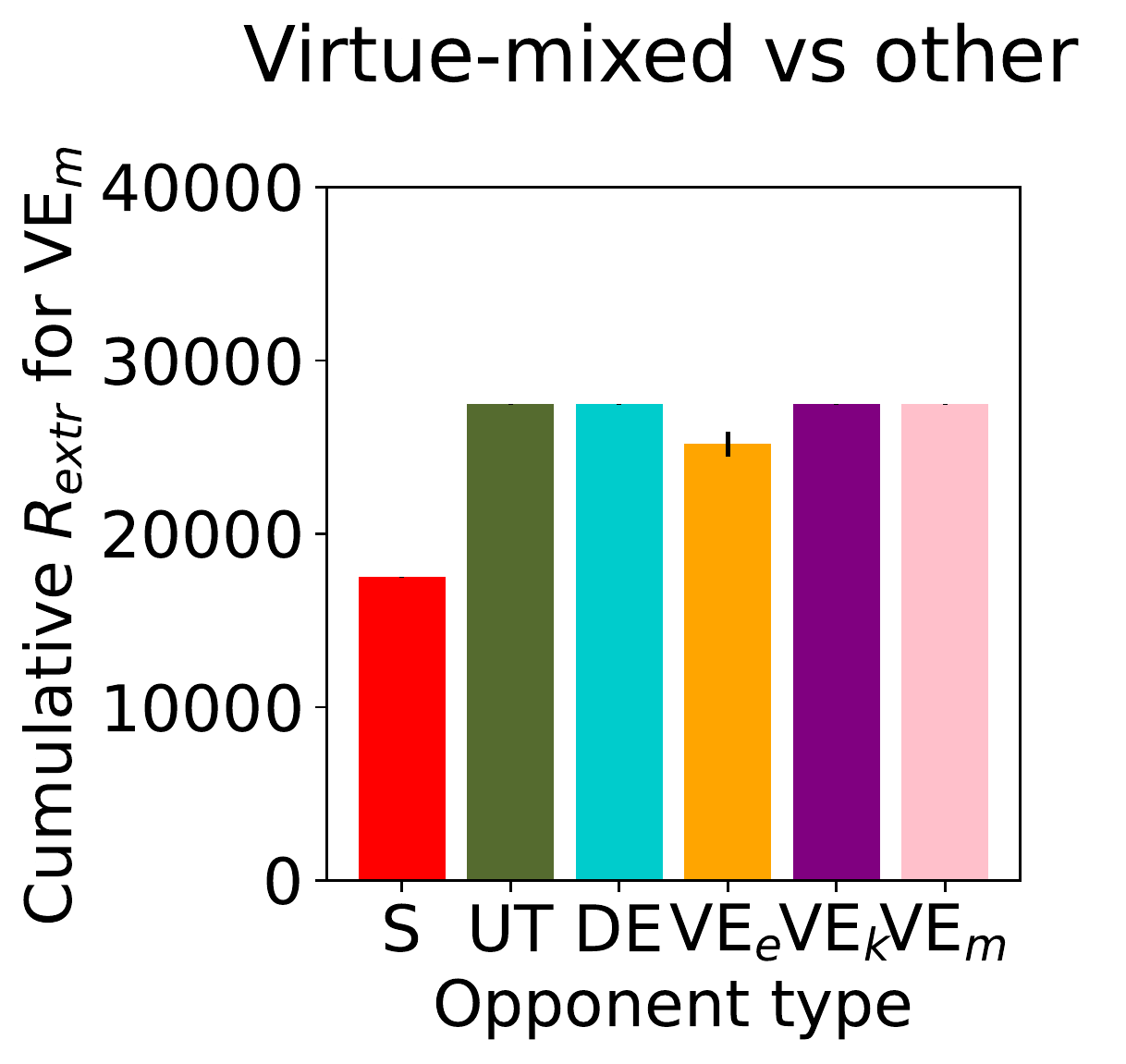}}
\\ 
\makecell[cc]{\rotatebox[origin=c]{90}{\thead{Moral Reward}}} & 
& \subt{\includegraphics[height=21mm]{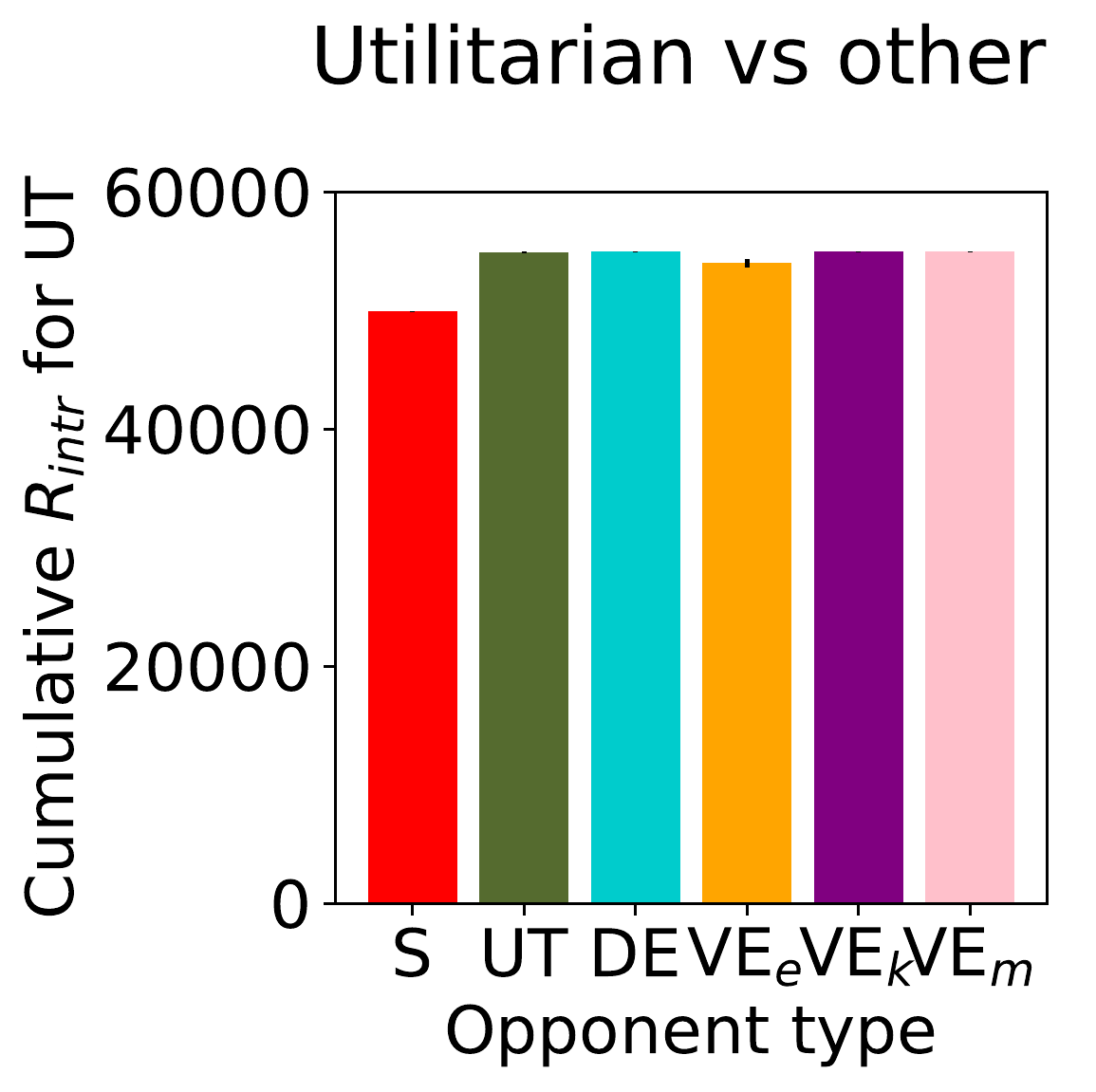}} & \subt{\includegraphics[height=21mm]{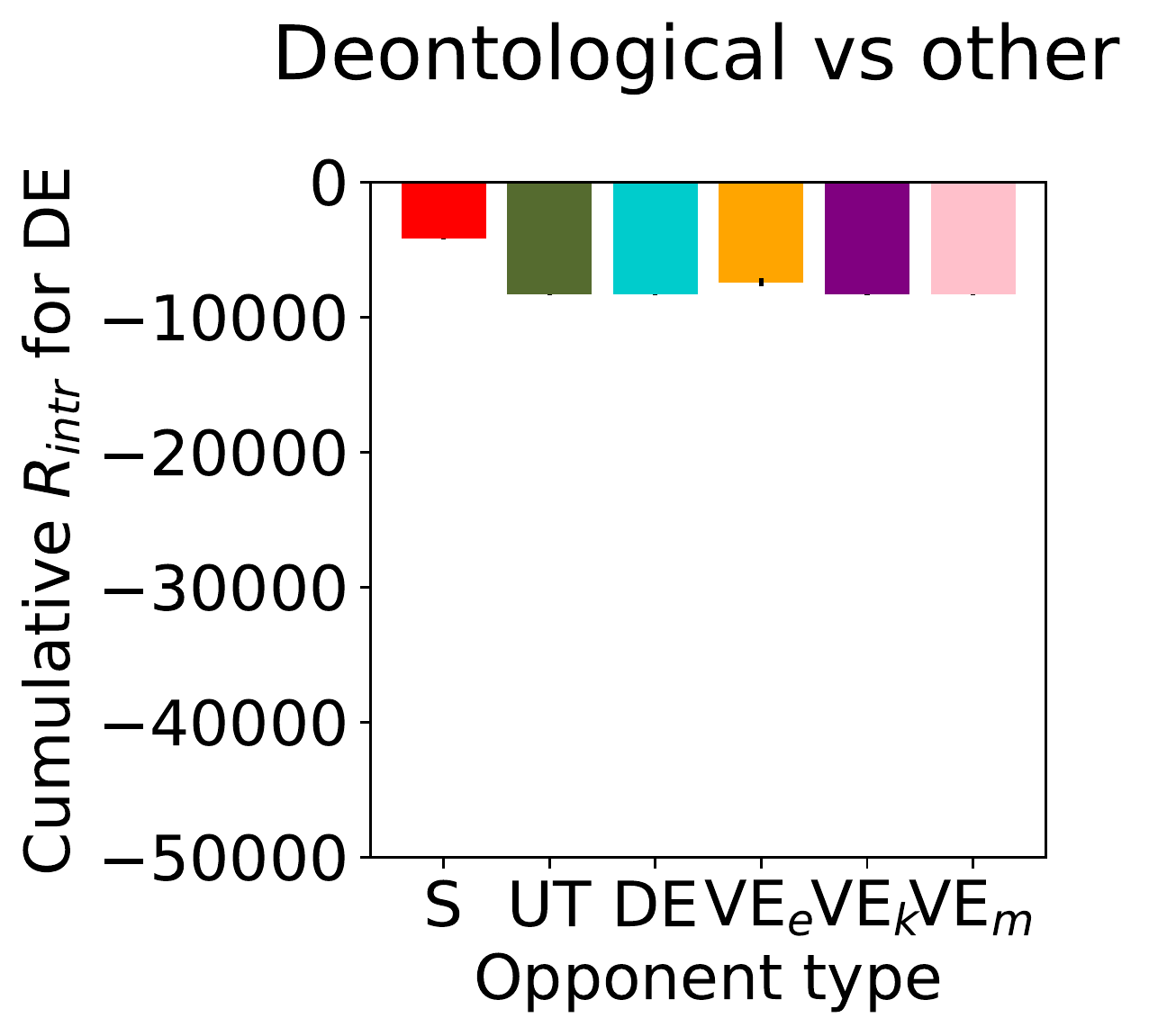}} & \subt{\includegraphics[height=21mm]{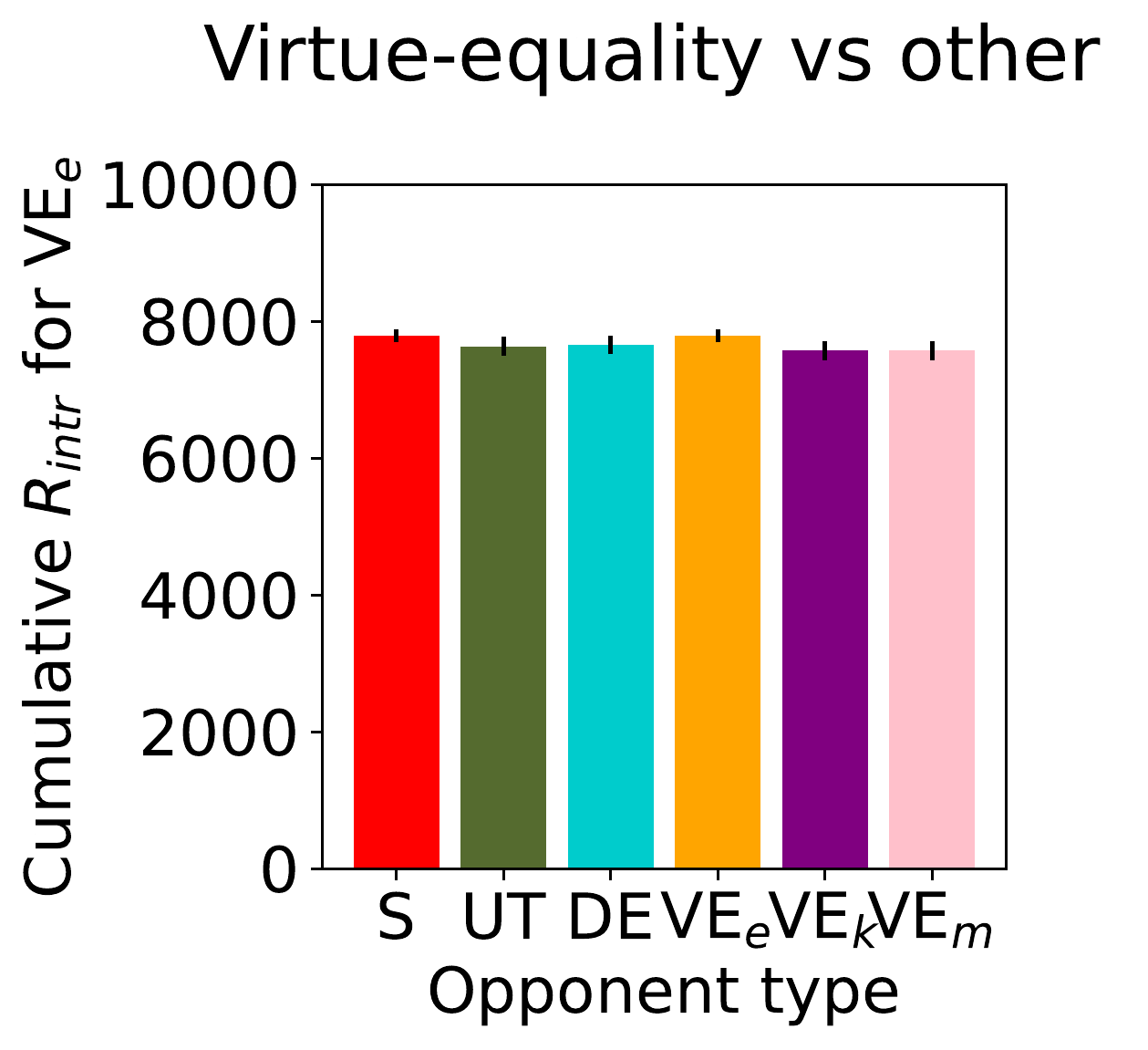}} & \subt{\includegraphics[height=21mm]{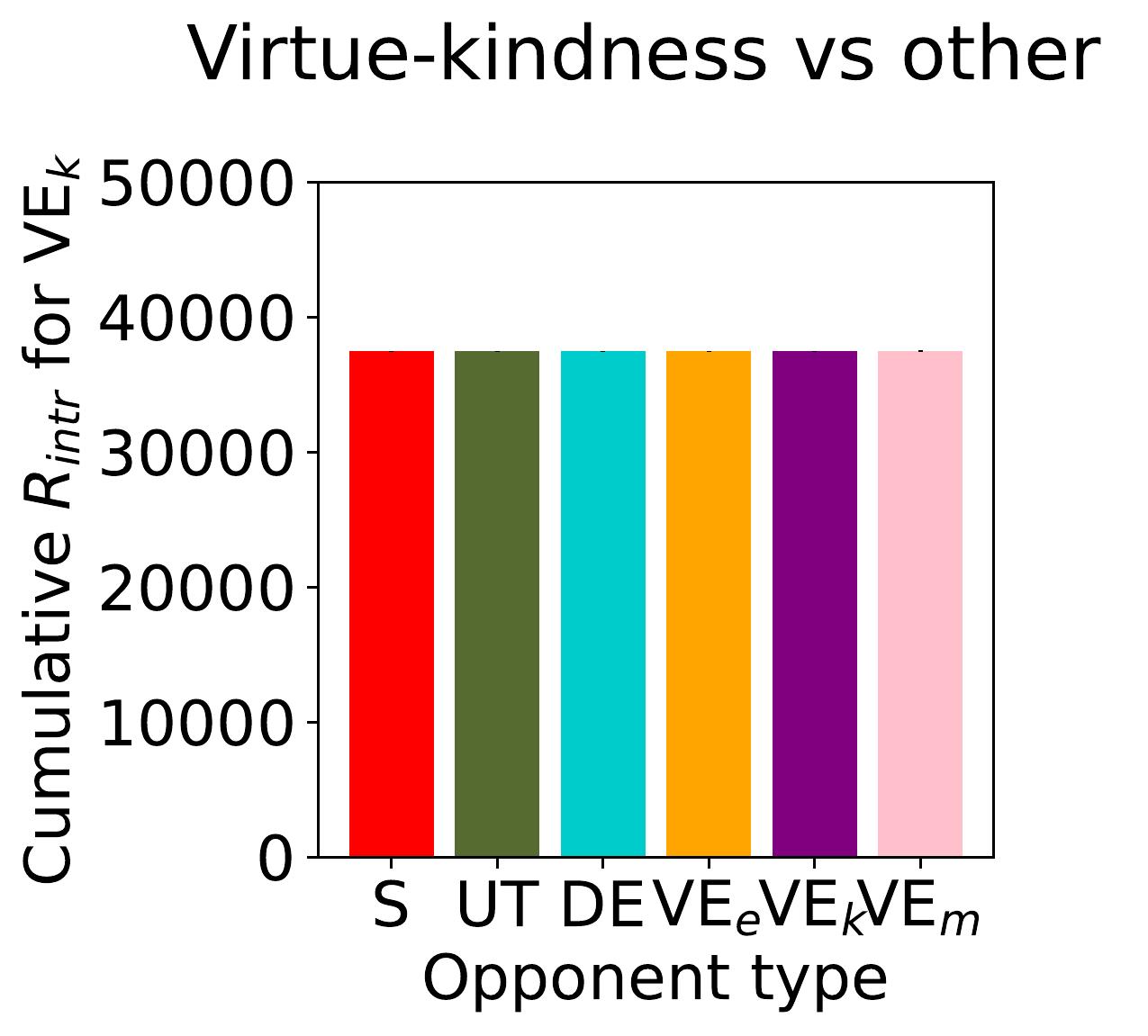}} & \subt{\includegraphics[height=21mm]{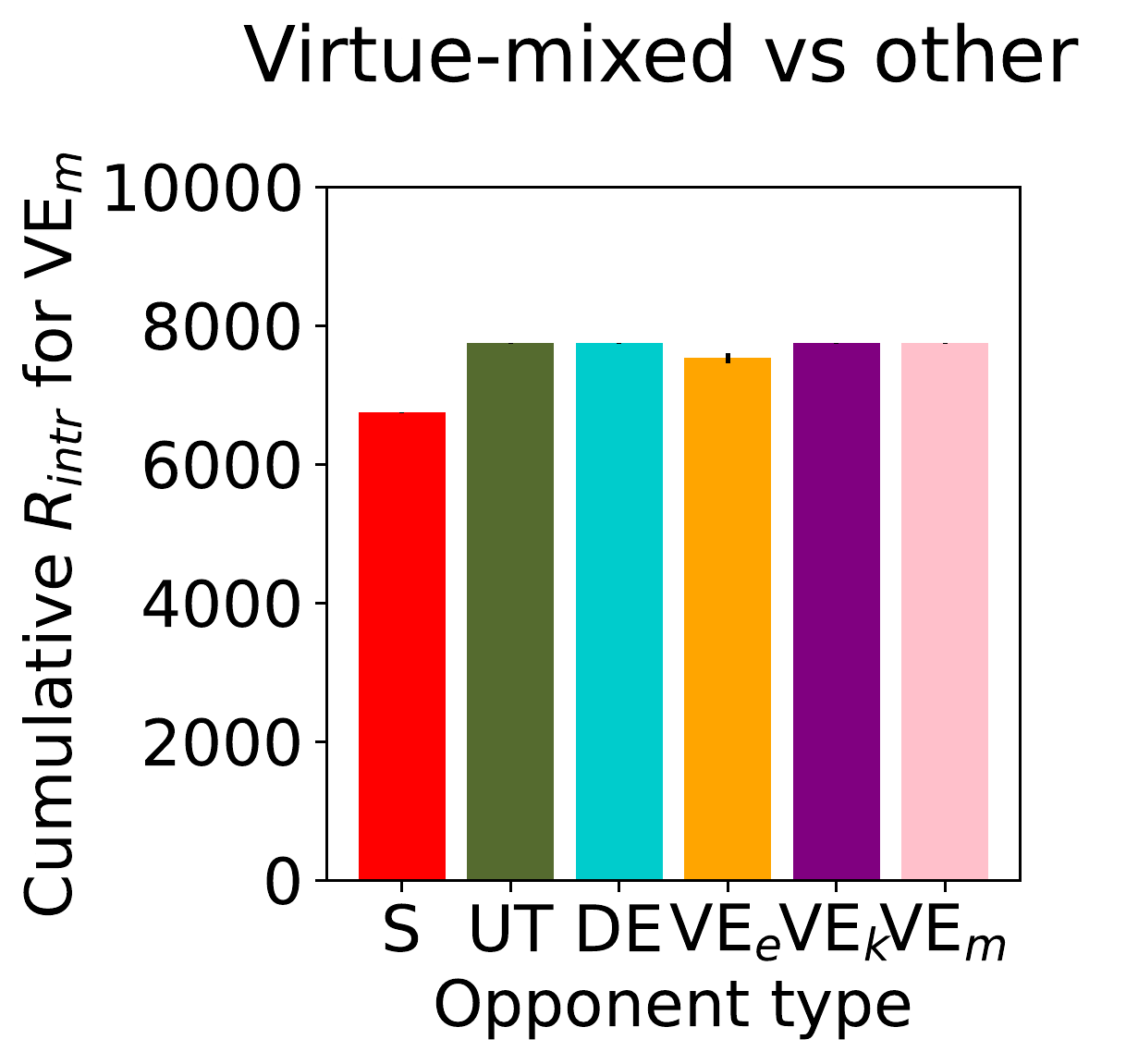}}
\\
\end{tabular}
\\ A. Iterated Prisoner's Dilemma \\

\begin{tabular}[t]{|c|cccccc}
\toprule
\makecell[cc]{\rotatebox[origin=c]{90}{\thead{Game Reward}}} & \subt{\includegraphics[height=21mm]{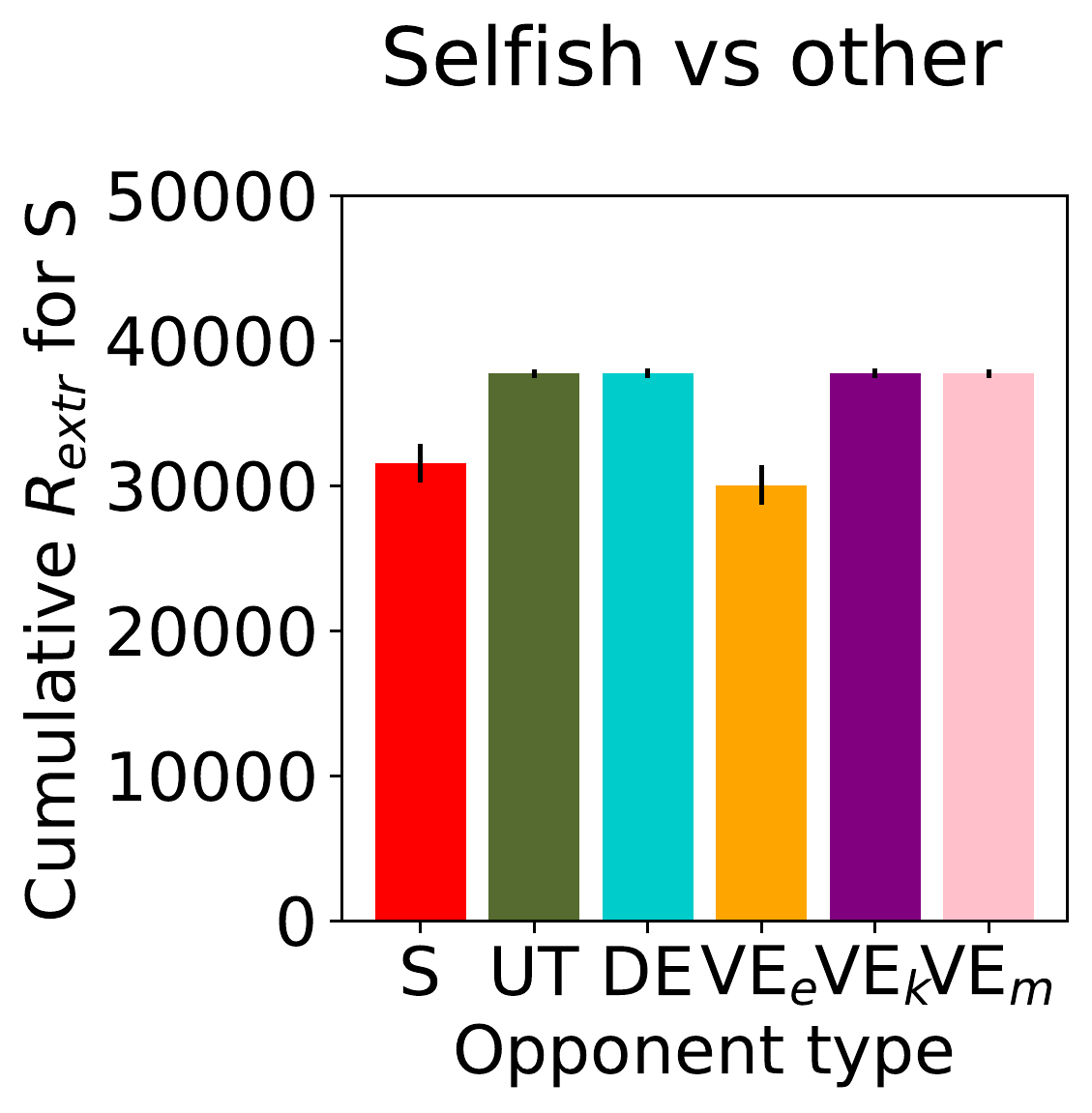}} & \subt{\includegraphics[height=21mm]{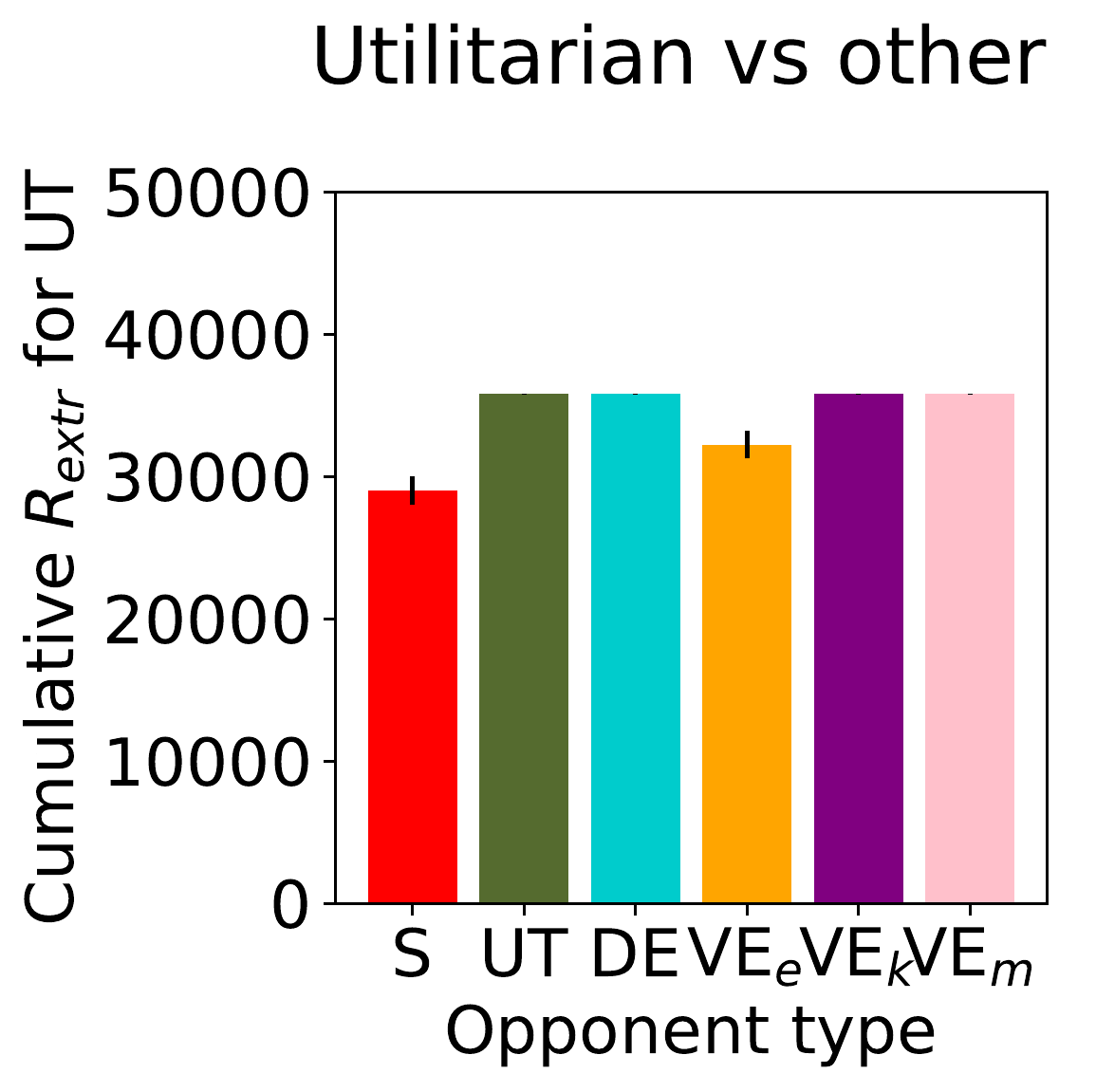}} & \subt{\includegraphics[height=21mm]{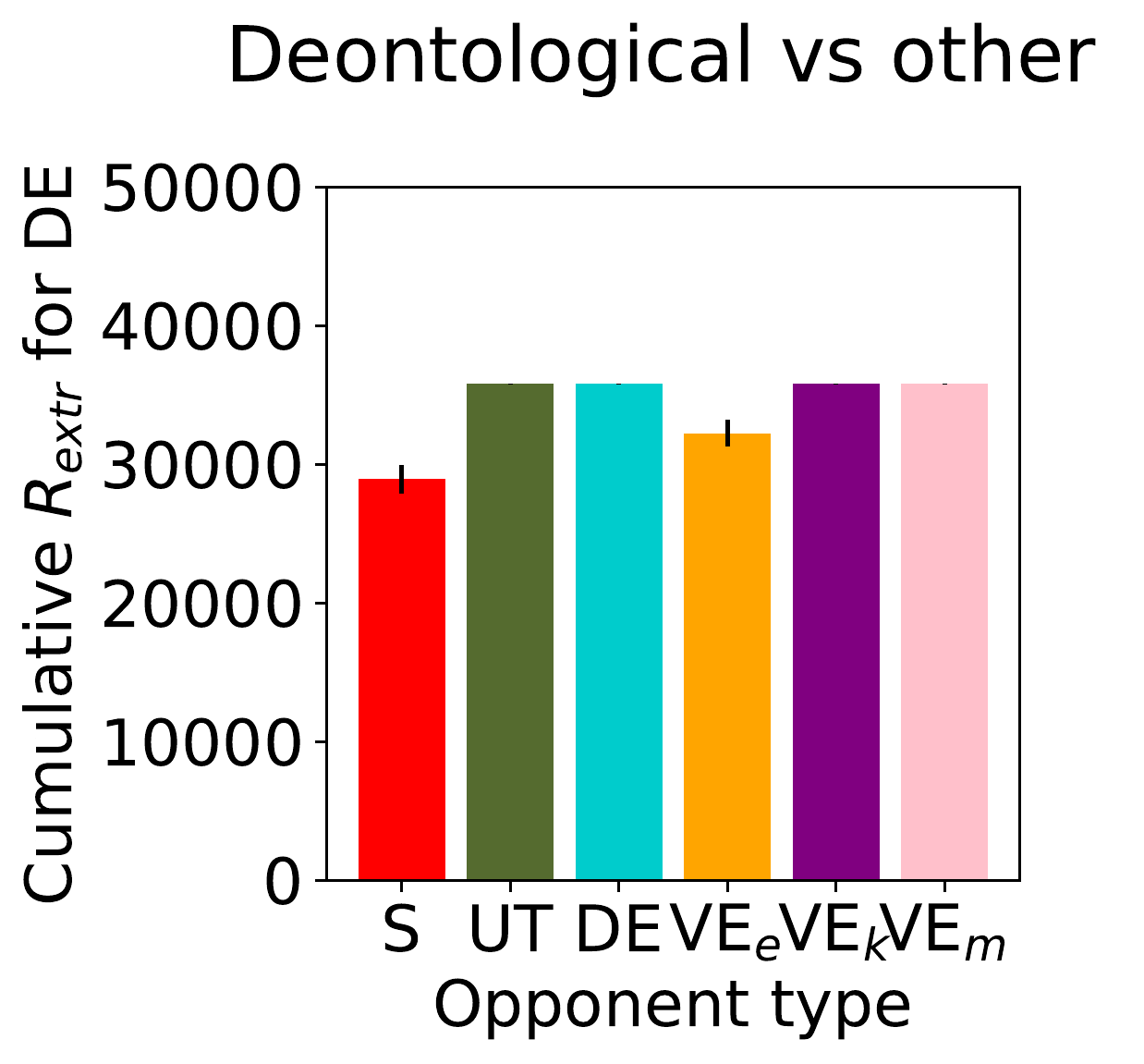}} & \subt{\includegraphics[height=21mm]{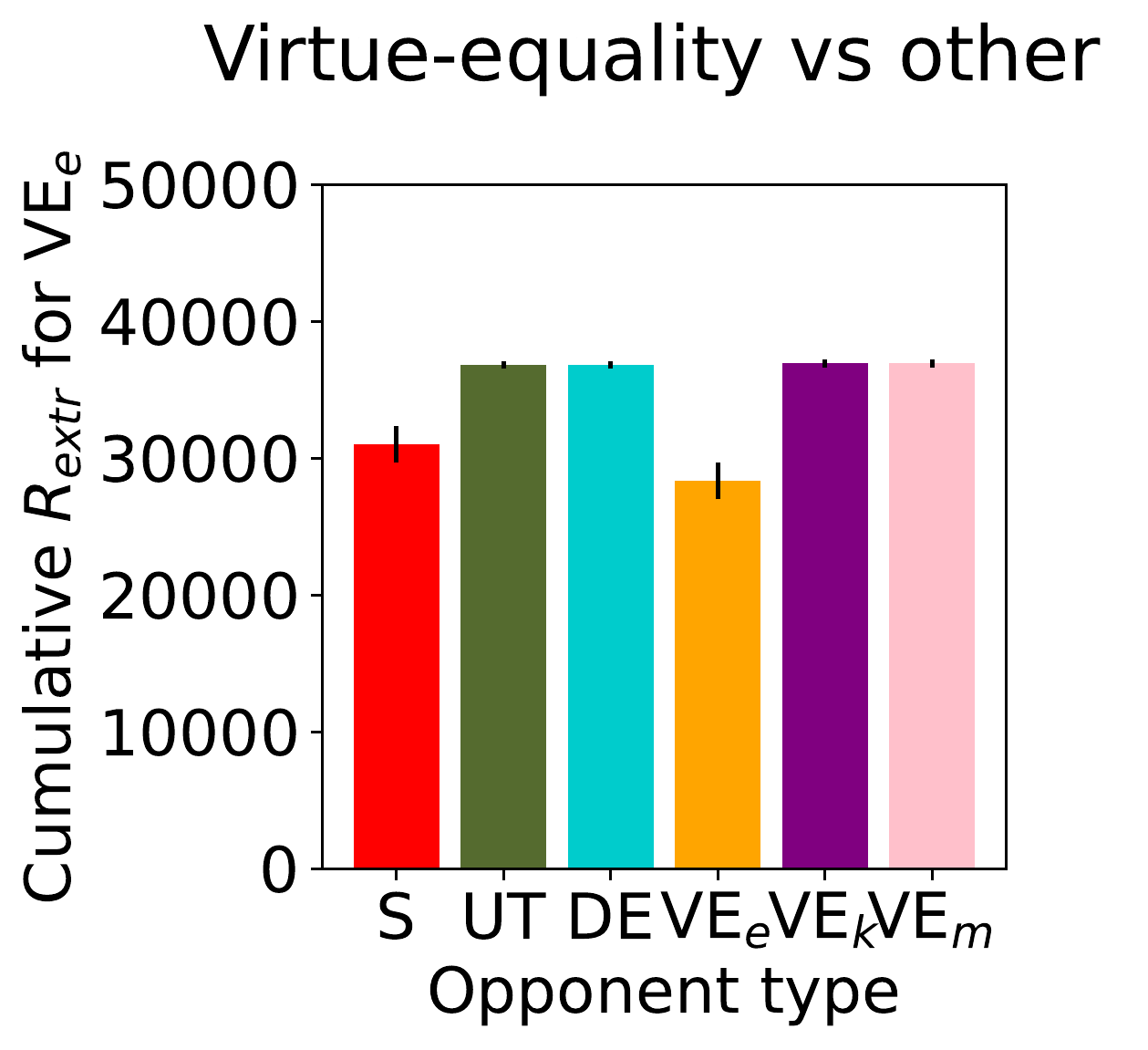}} & \subt{\includegraphics[height=21mm]{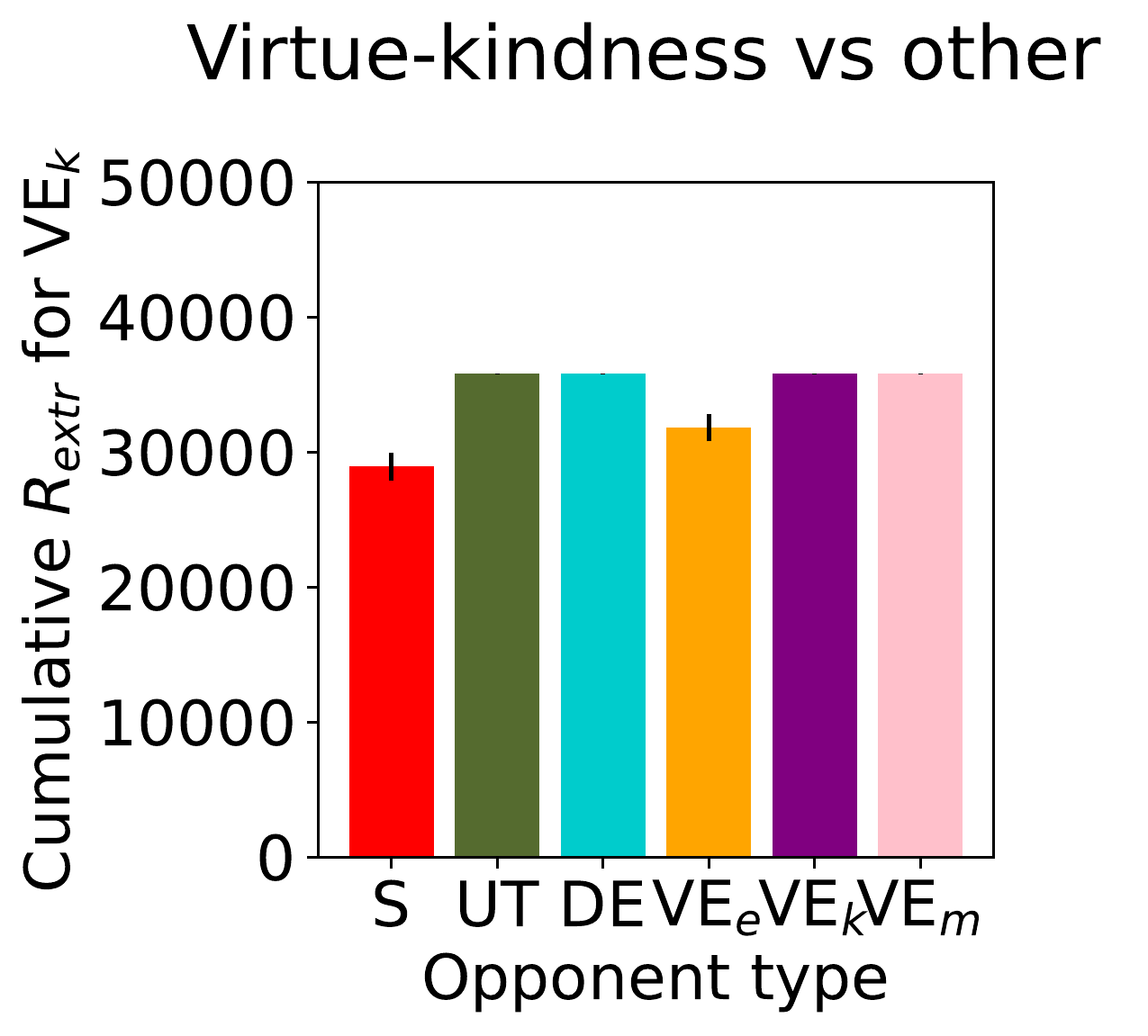}} & \subt{\includegraphics[height=21mm]{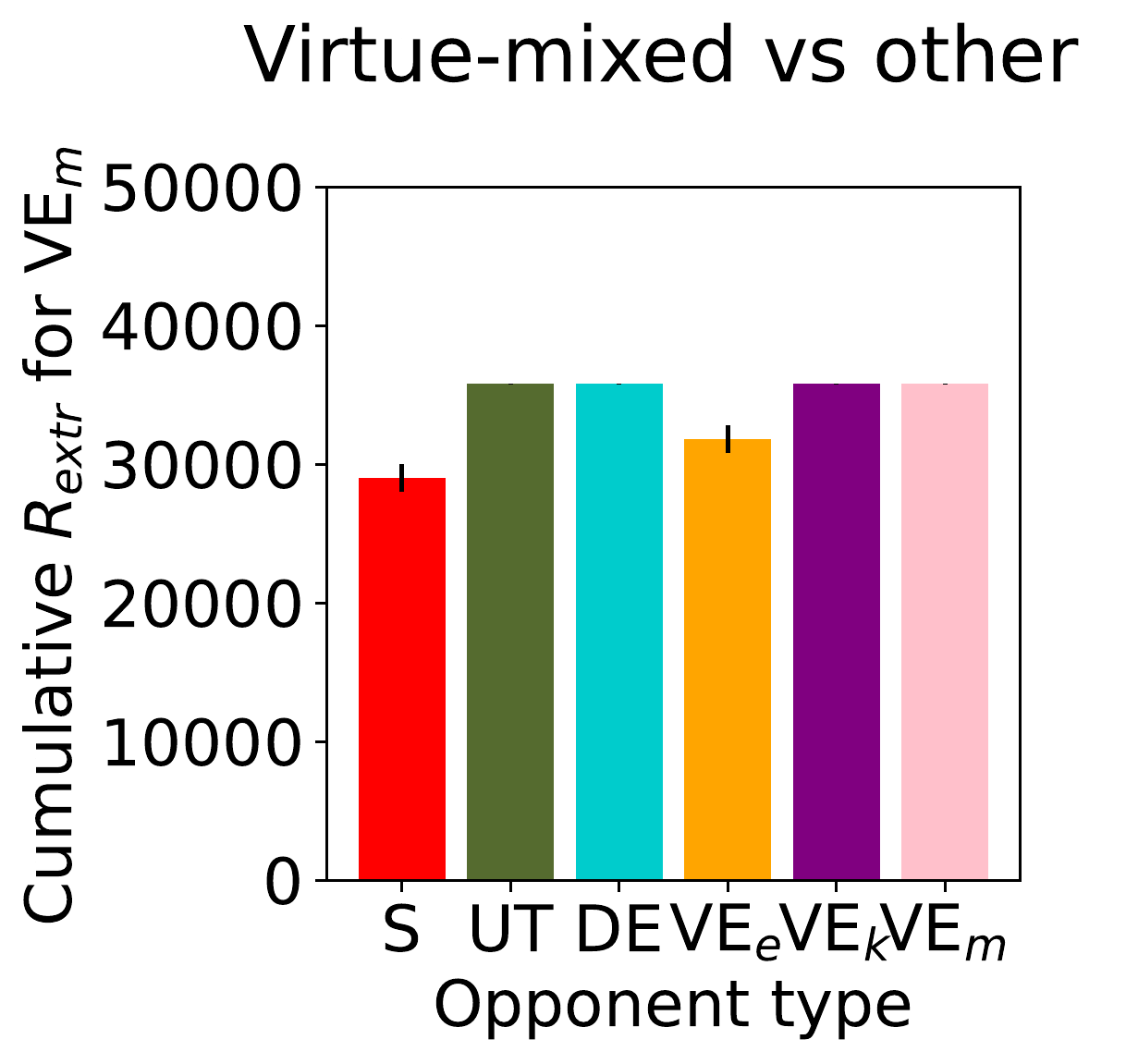}}
\\ 
\makecell[cc]{\rotatebox[origin=c]{90}{\thead{Moral Reward}}} &  & \subt{\includegraphics[height=21mm]{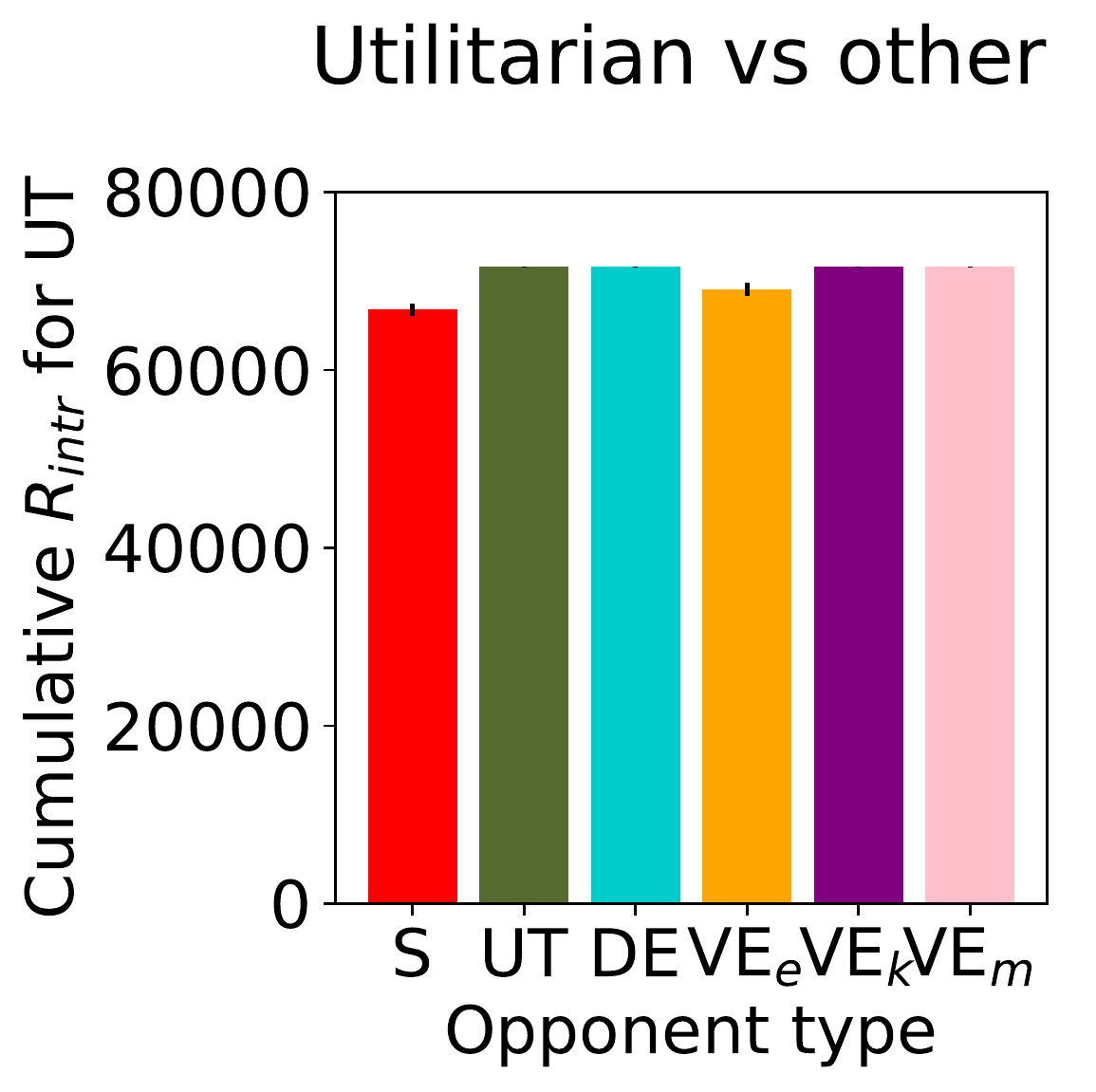}} & \subt{\includegraphics[height=21mm]{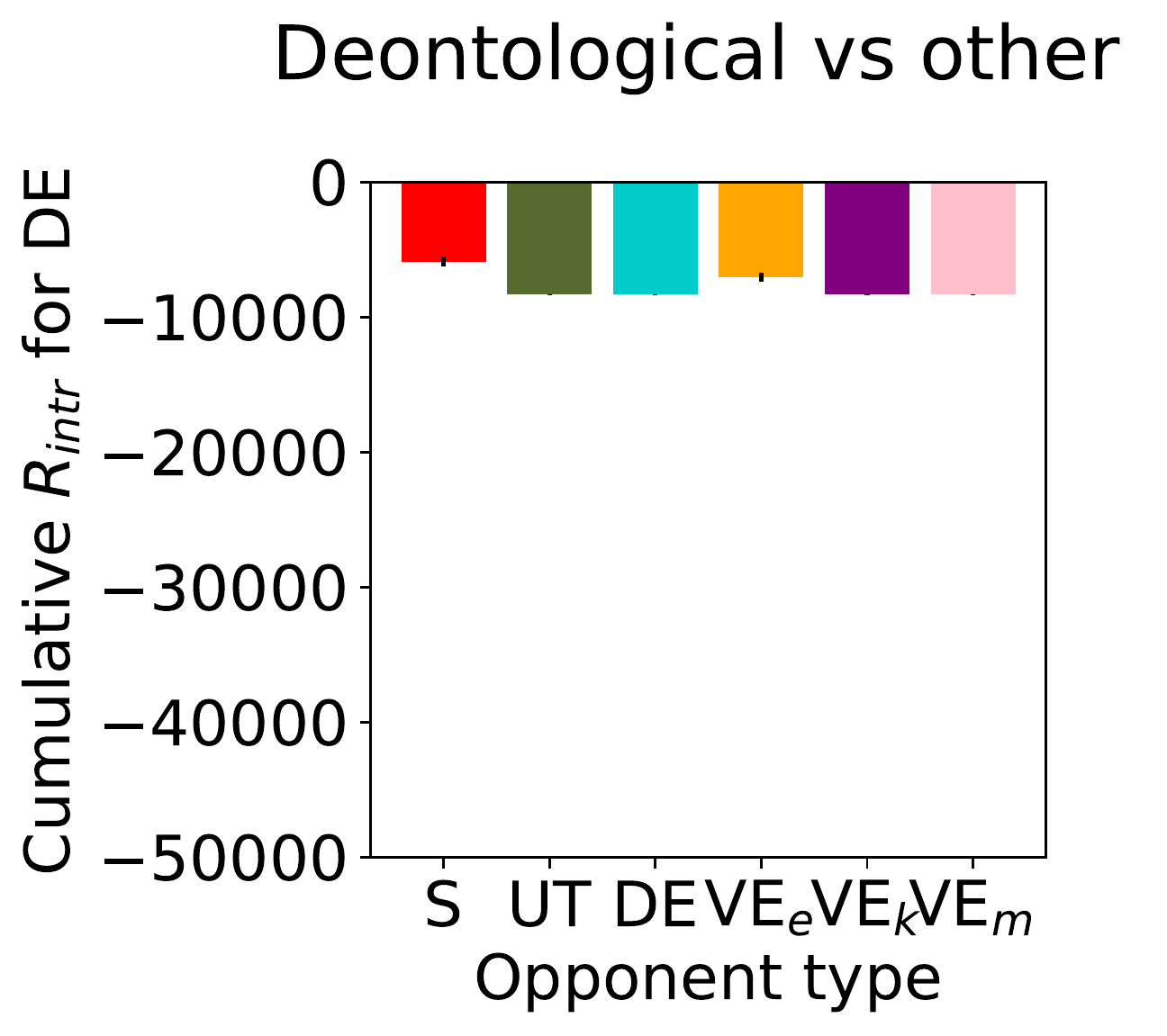}} & \subt{\includegraphics[height=21mm]{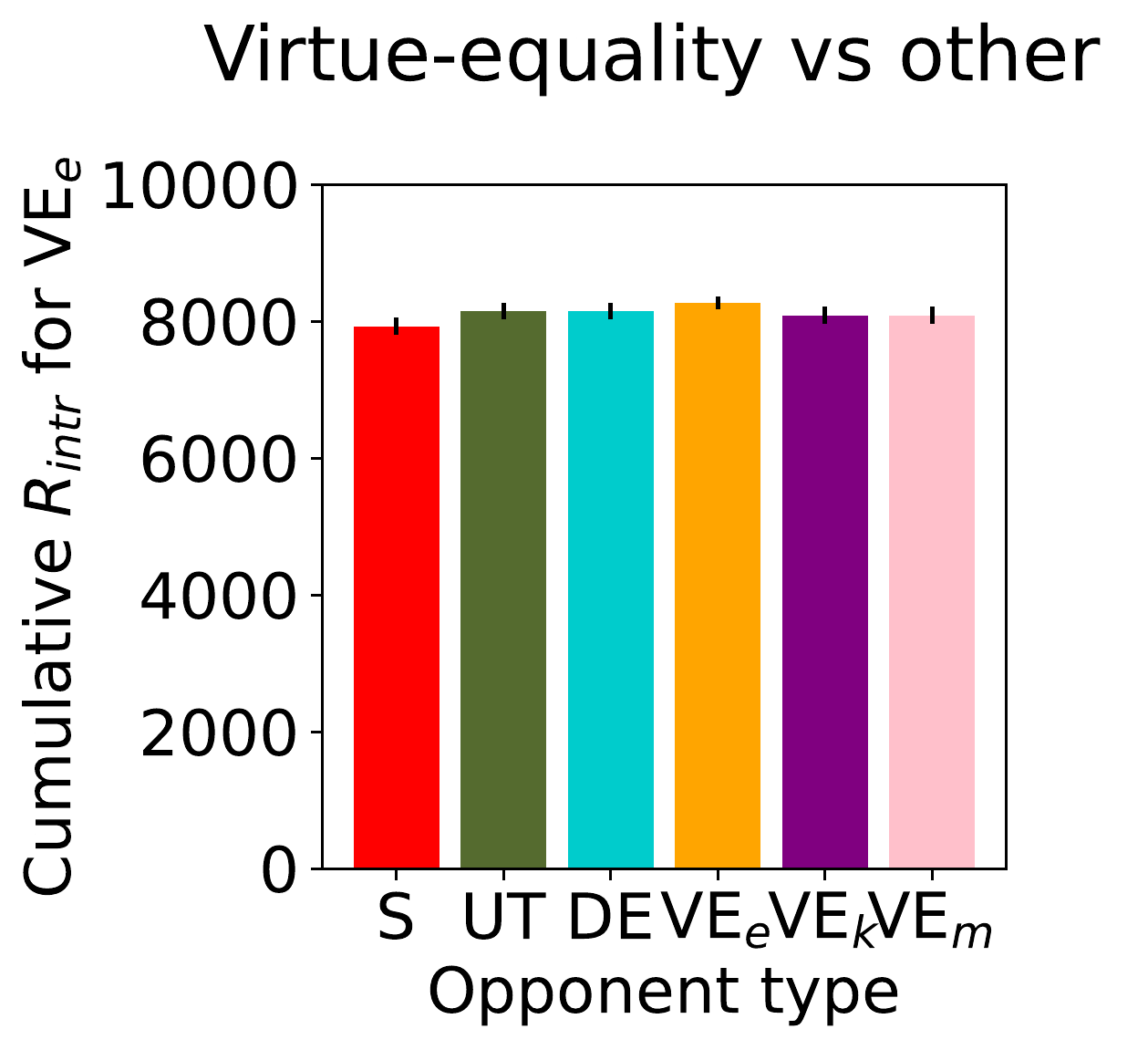}} & \subt{\includegraphics[height=21mm]{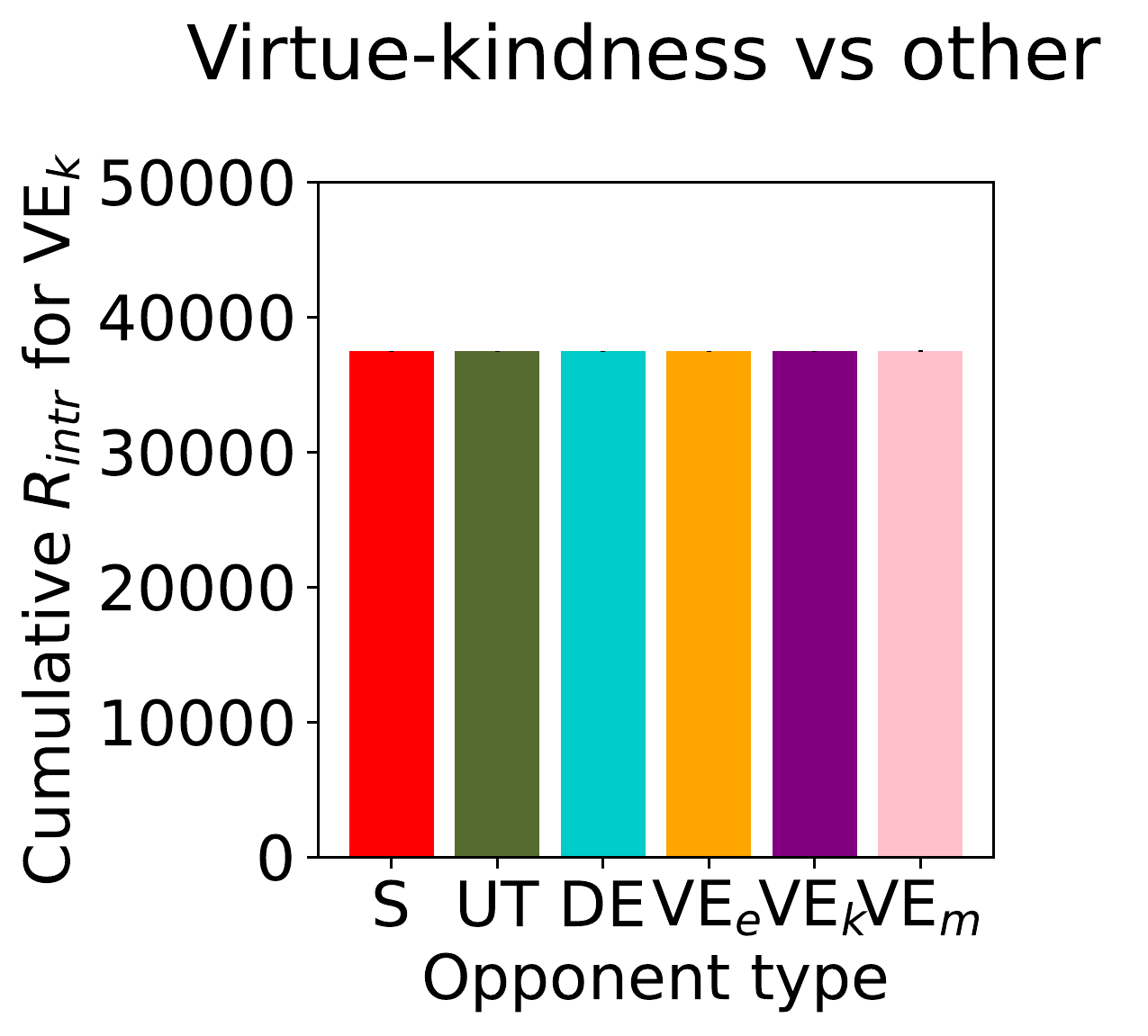}} & \subt{\includegraphics[height=21mm]{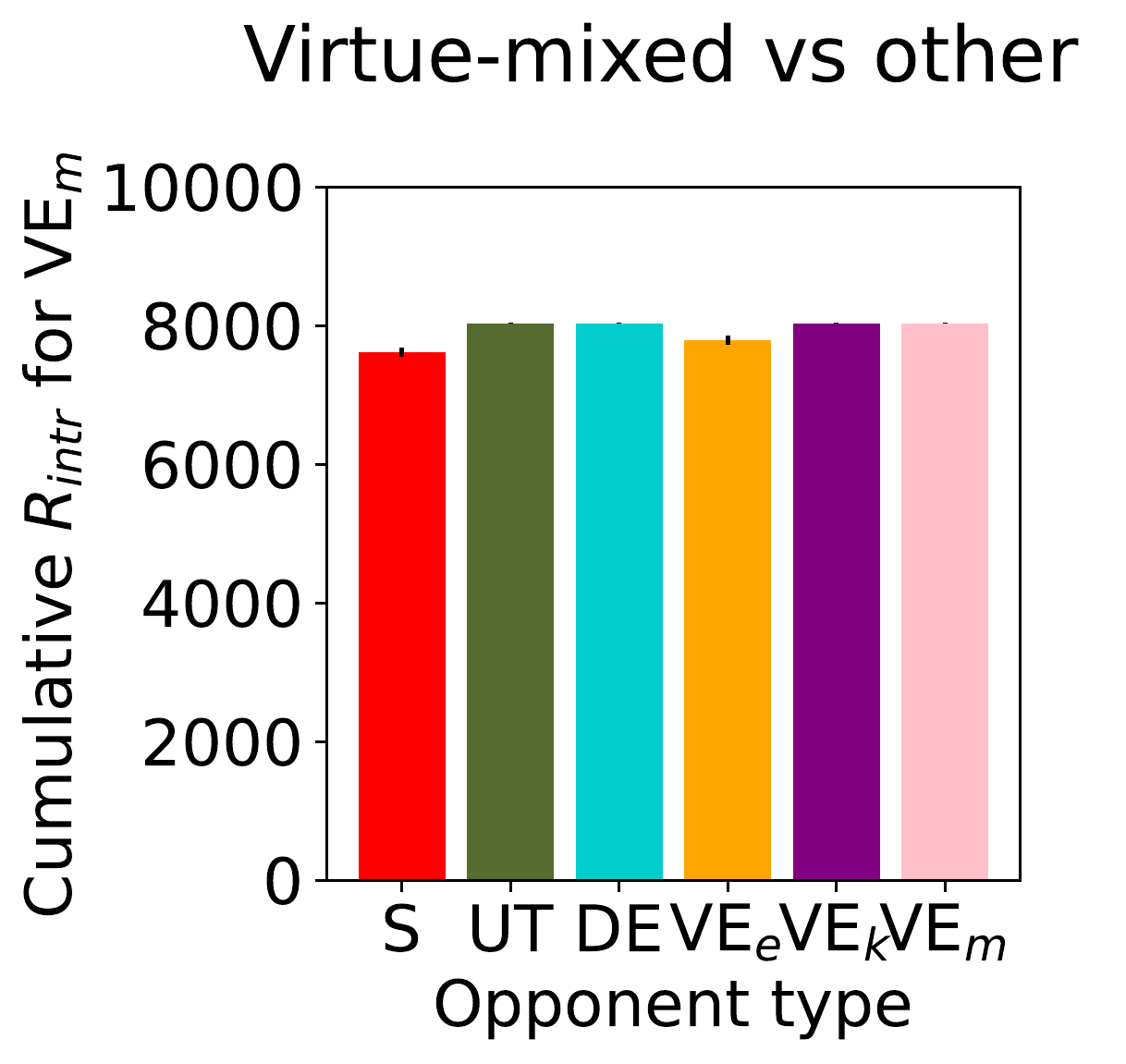}}
\\
\end{tabular}
\\ B. Iterated Volunteer's Dilemma \\

\begin{tabular}[t]{|c|cccccc}
\toprule
\makecell[cc]{\rotatebox[origin=c]{90}{\thead{Game Reward}}} & \subt{\includegraphics[height=21mm]{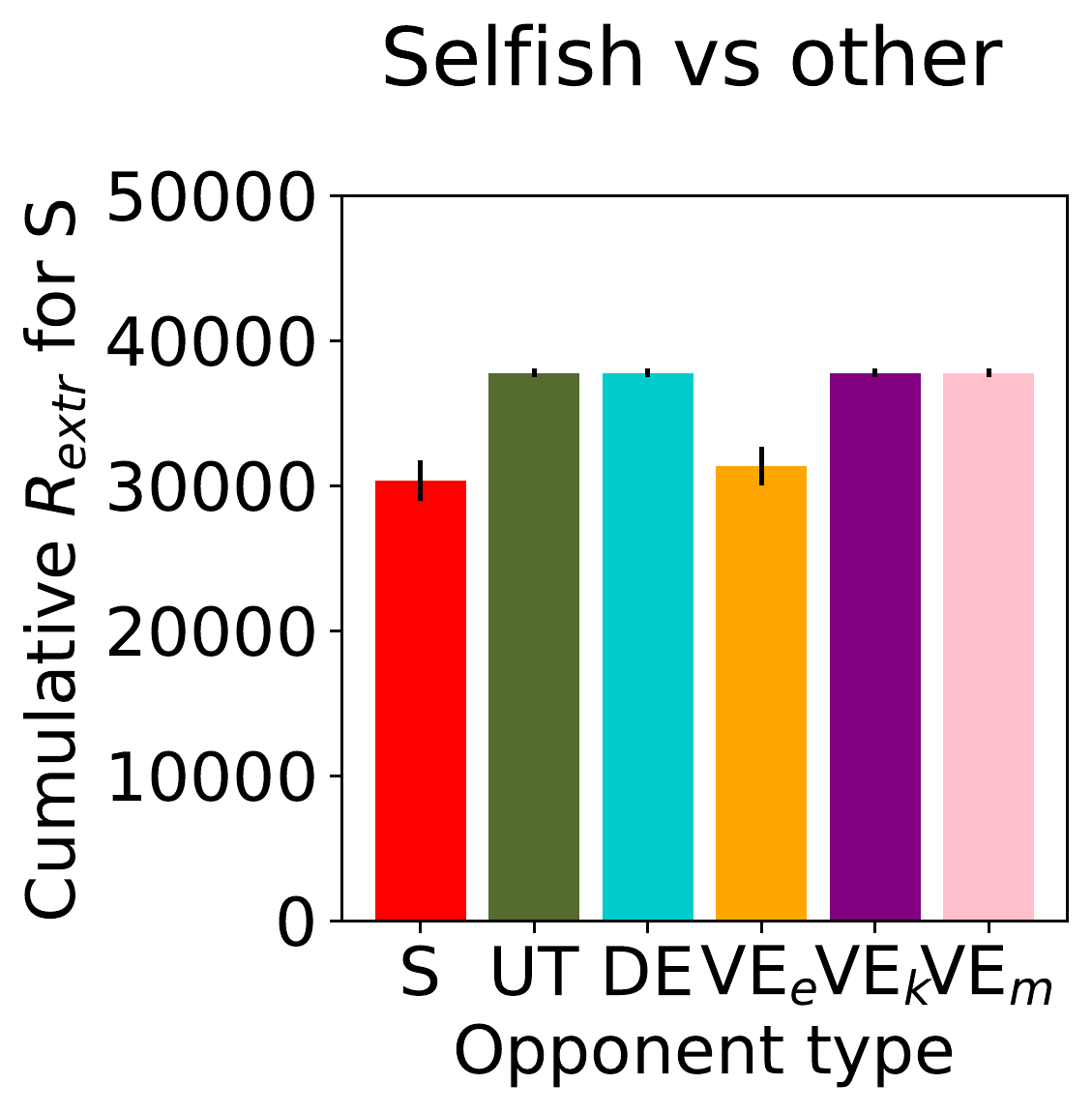}} & \subt{\includegraphics[height=21mm]{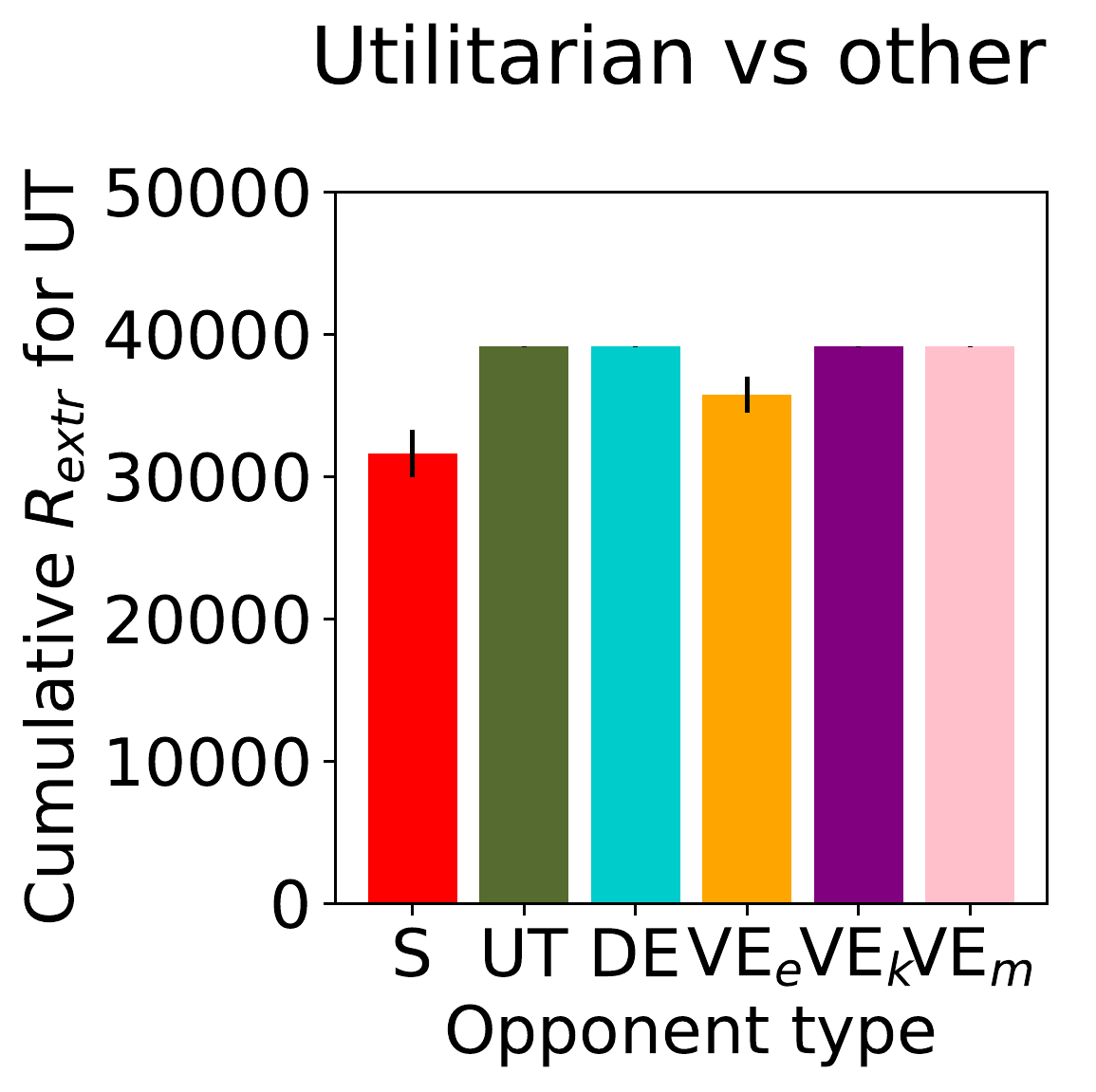}} & \subt{\includegraphics[height=21mm]{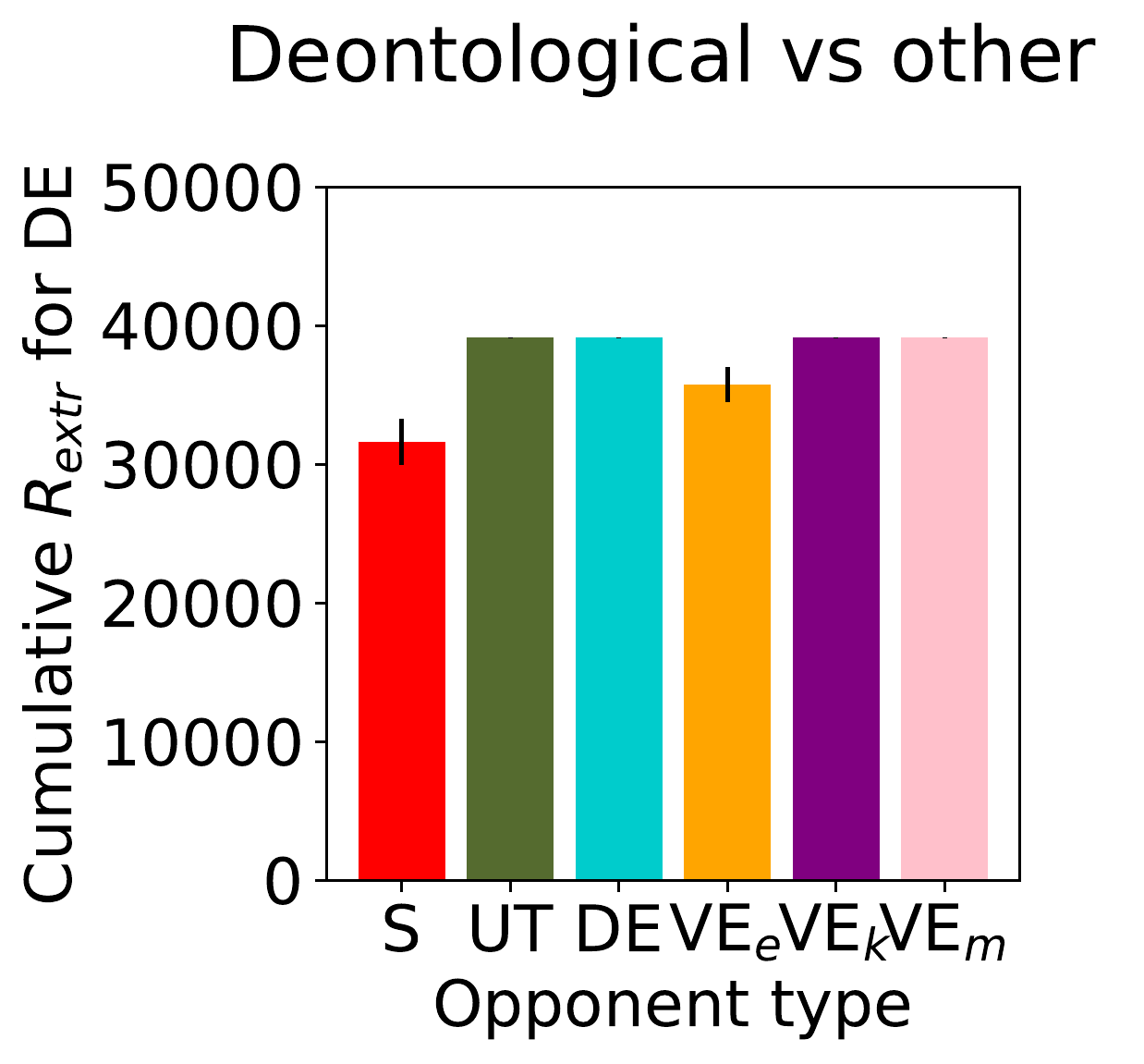}} & \subt{\includegraphics[height=21mm]{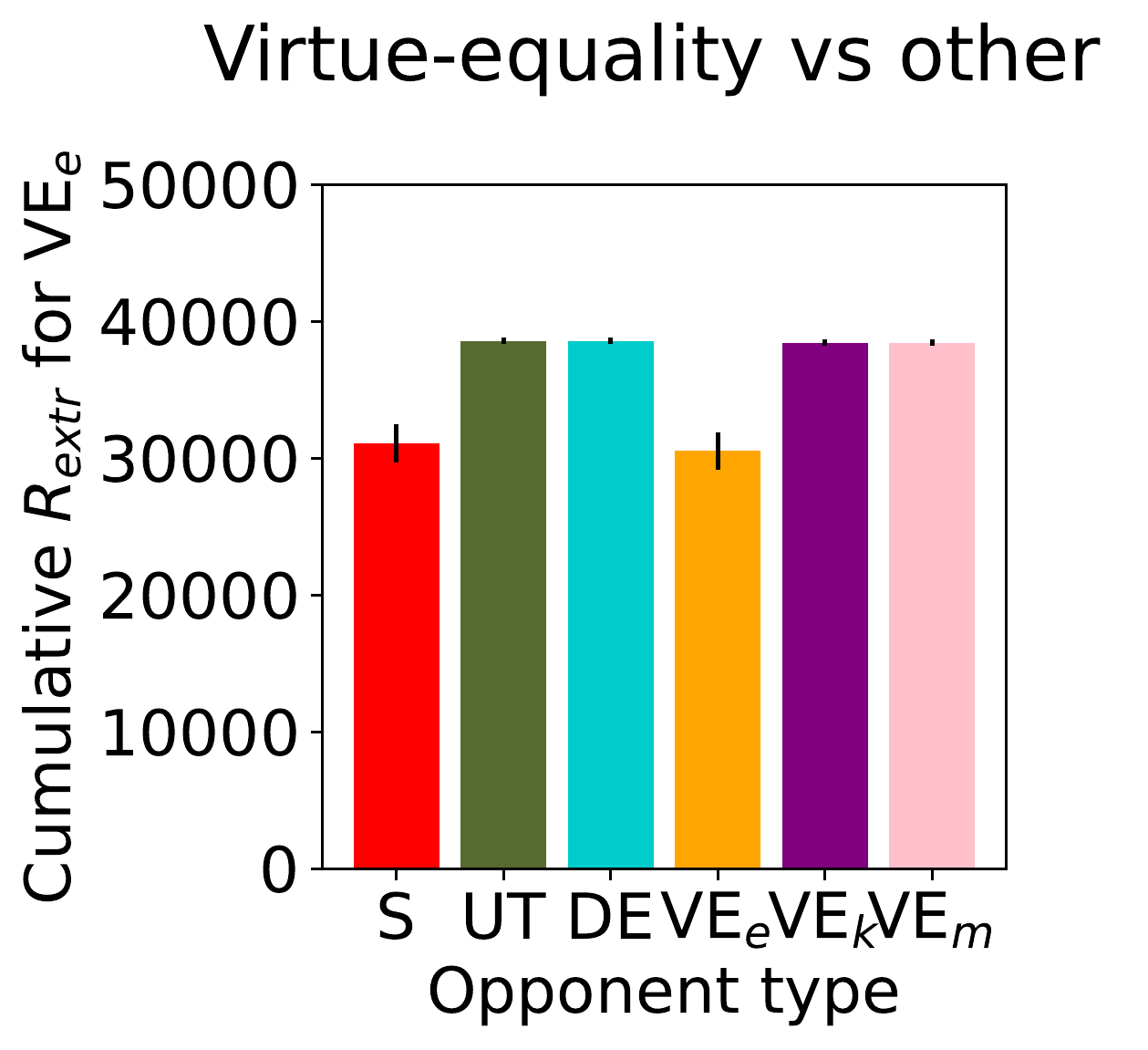}} & \subt{\includegraphics[height=21mm]{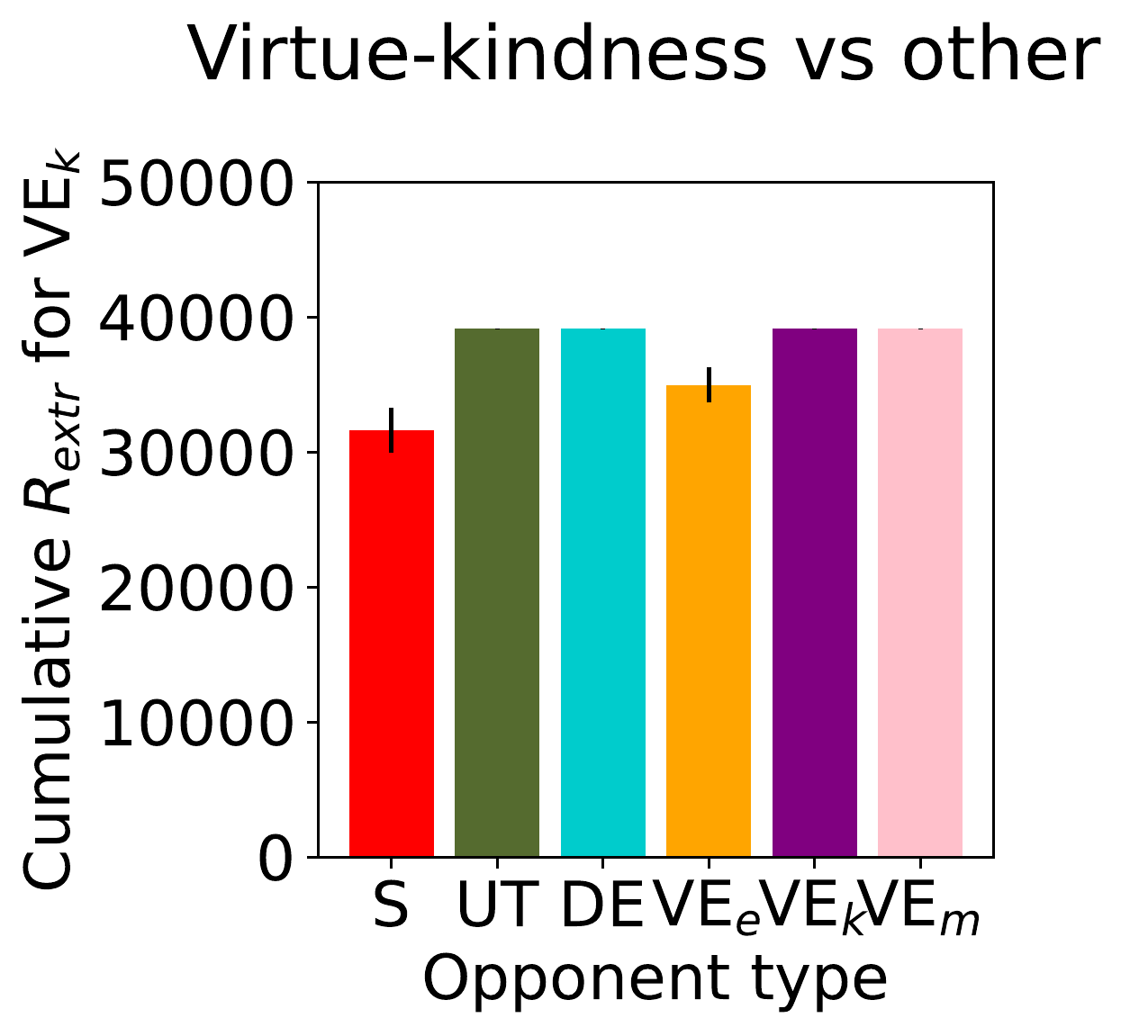}} & \subt{\includegraphics[height=21mm]{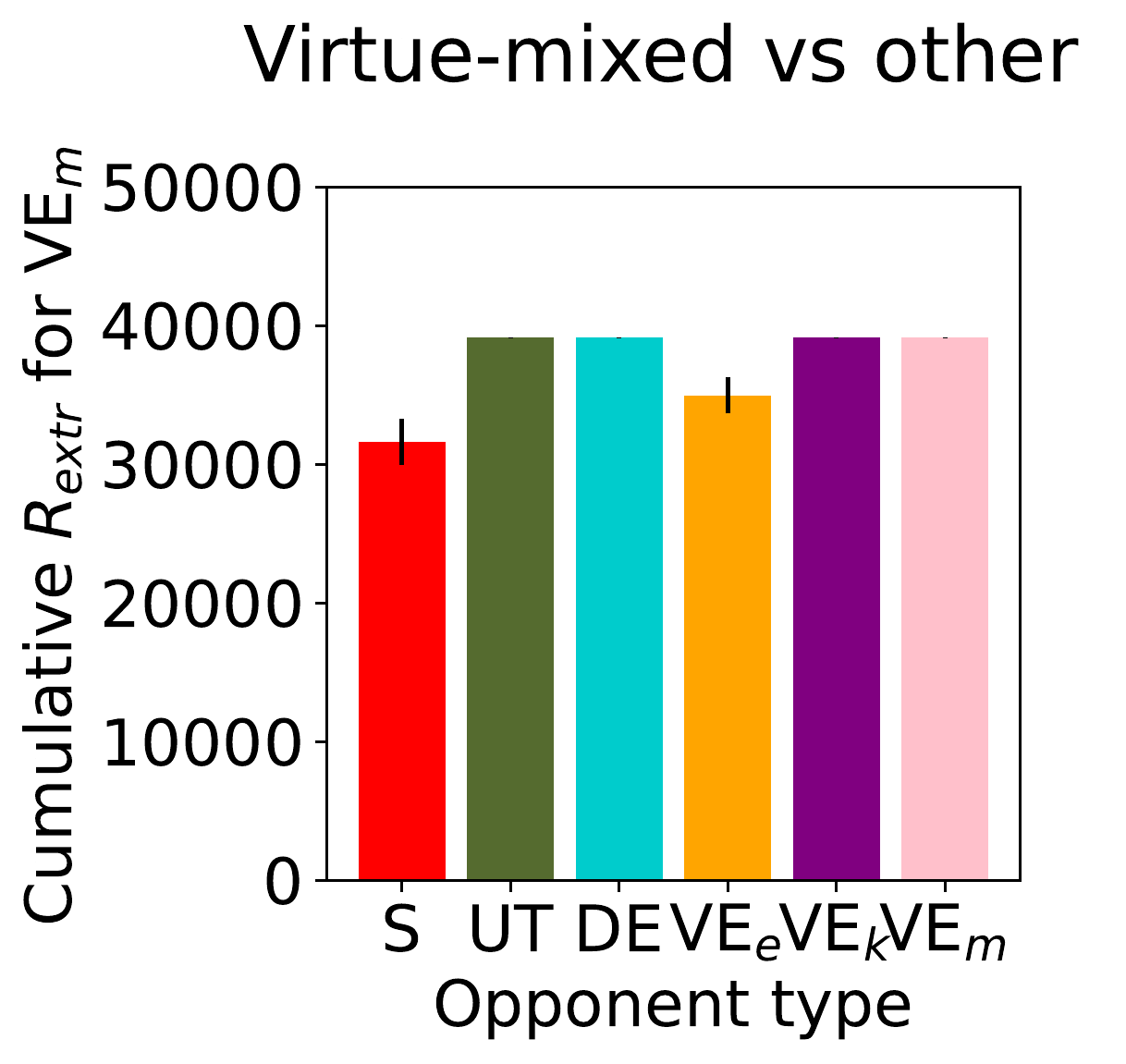}}
\\ 
\makecell[cc]{\rotatebox[origin=c]{90}{\thead{Moral Reward}}} &  & \subt{\includegraphics[height=21mm]{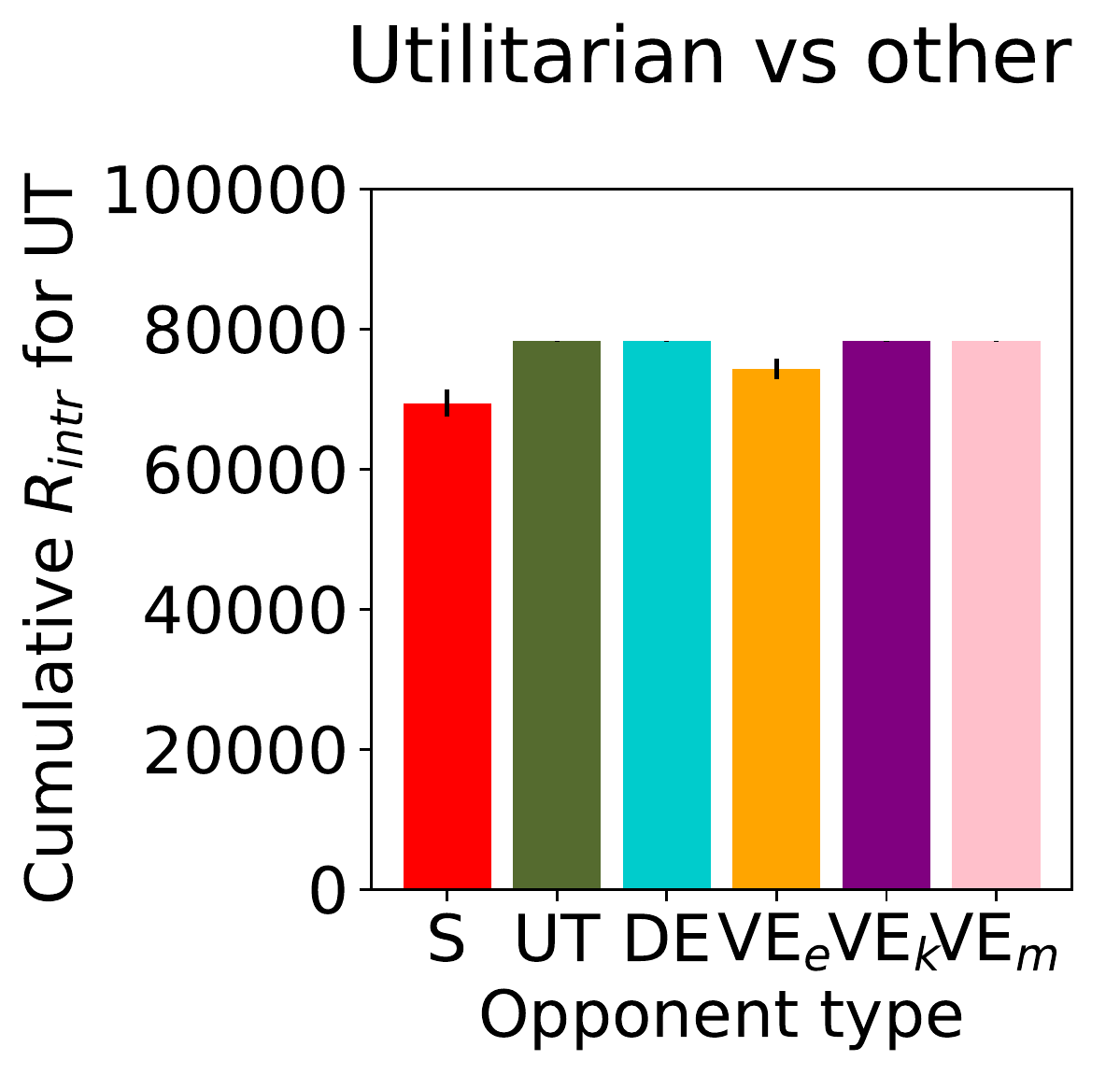}} & \subt{\includegraphics[height=21mm]{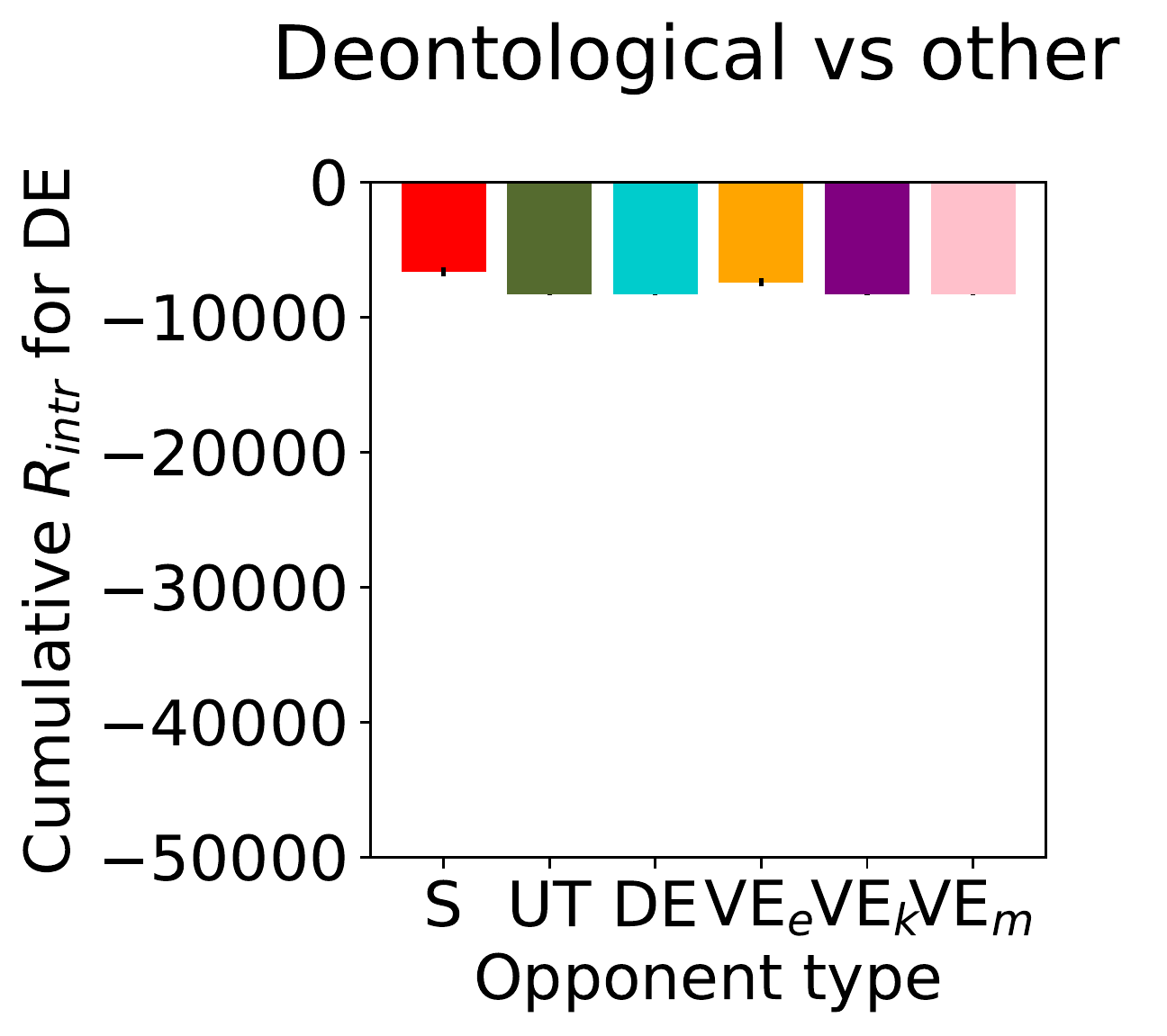}} & \subt{\includegraphics[height=21mm]{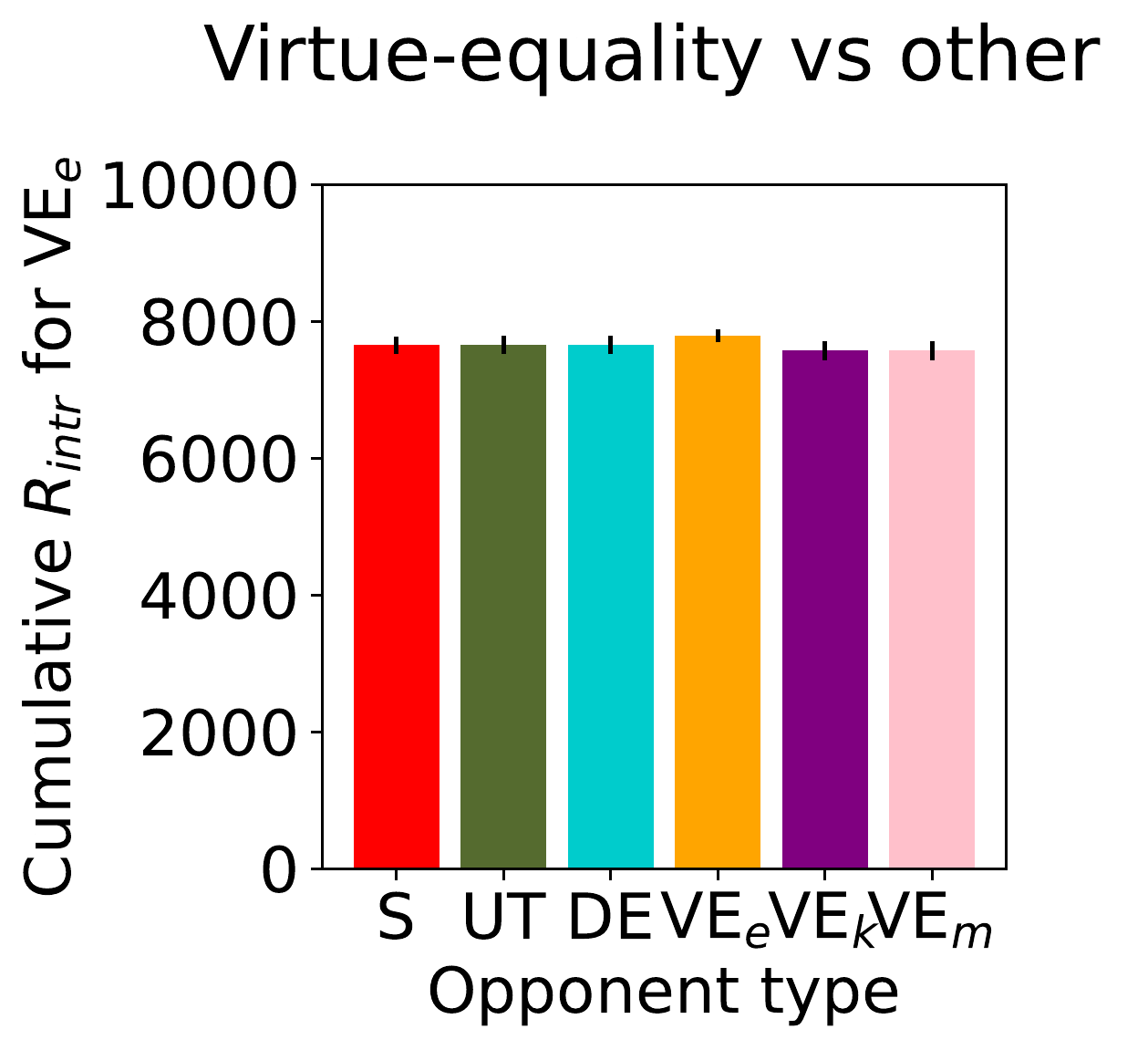}} & \subt{\includegraphics[height=21mm]{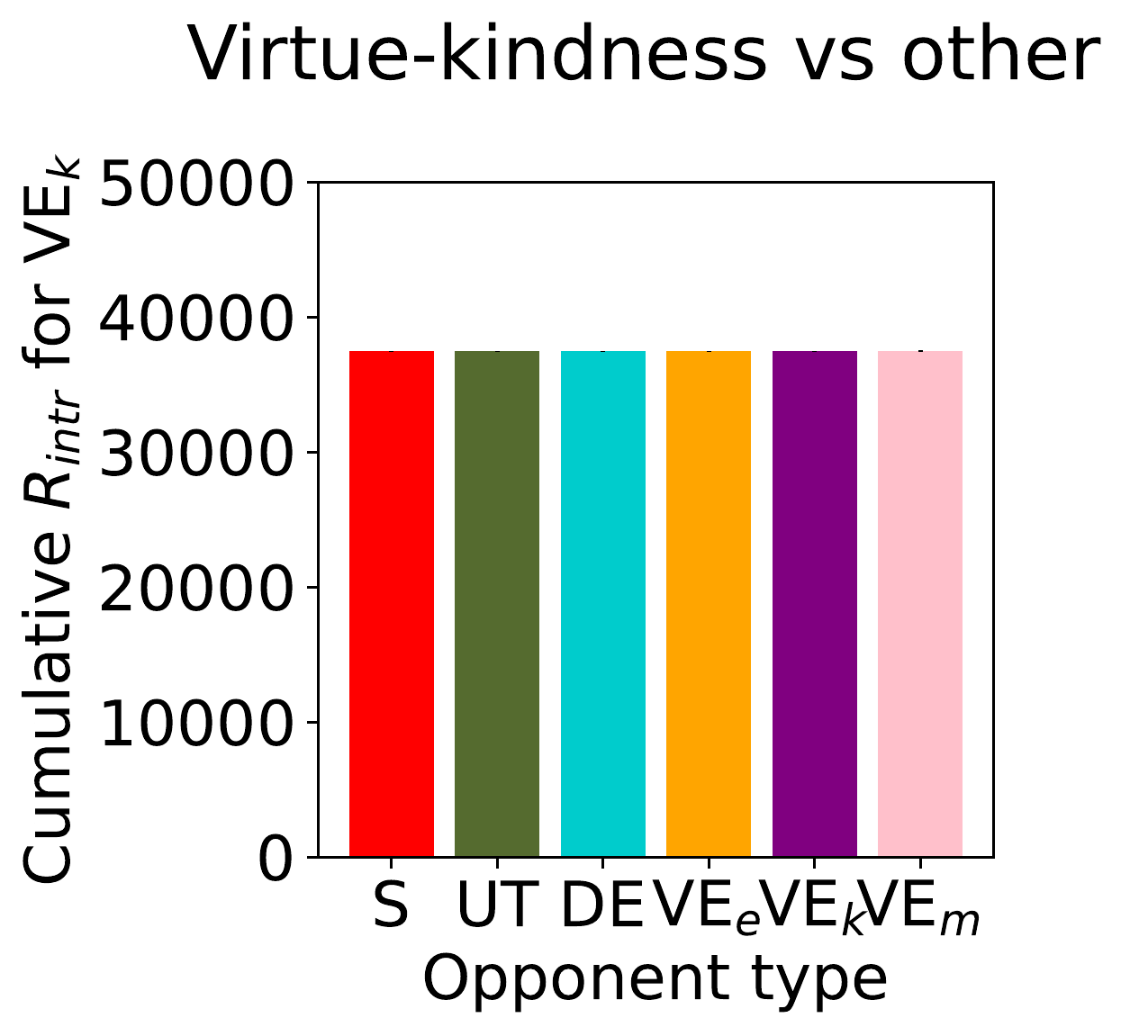}} & \subt{\includegraphics[height=21mm]{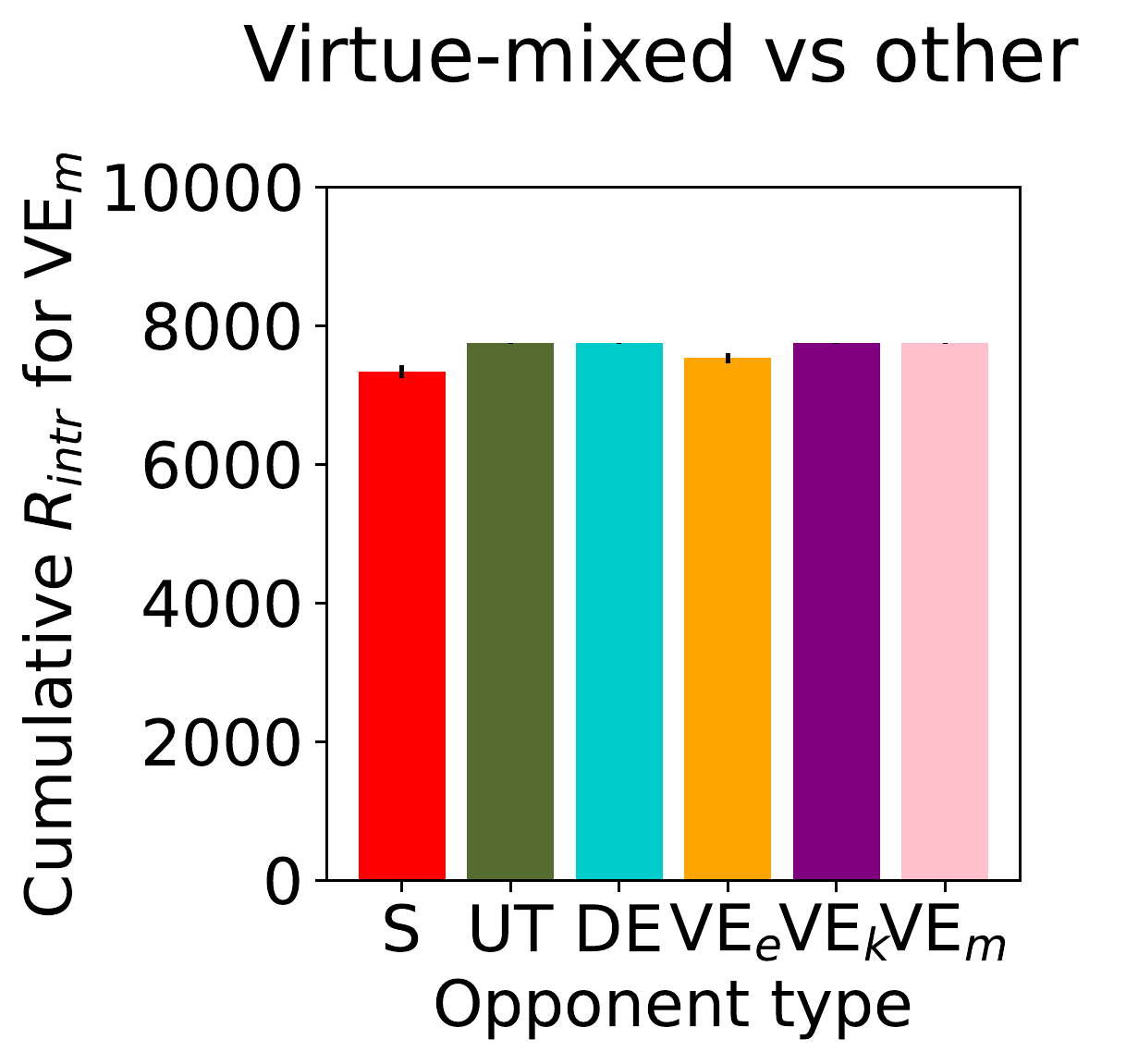}}
\\
\end{tabular}
\\ C. Iterated Stag Hunt \\ 

\caption{Game \& moral reward (cumulative) obtained after 10000 iterations by a given player type $M$ (column) vs. all possible learning opponents $O$ - for all three games (panels A-C). The plots display averages across the 100 runs $\pm$ 95\%CI.}
\label{fig:reward}
\end{figure*}

Next, we consider \textit{game reward} (i.e., in our case, equal to the extrinsic reward) and \textit{moral reward} (i.e., in our case, equal to the intrinsic reward) accumulated by each agent type $M$ against all other agent types. Figure \ref{fig:reward} presents average cumulative rewards across 100 runs, and their Confidence Intervals, for all three games. 

We first analyze extrinsic game reward obtained (row one in panels A,B,C, Figure \ref{fig:reward}). Across all three environments, we observe that, on average, the \textit{Selfish} agent (first column in each panel) shows better performance in games against most non-selfish opponents (due to the exploitation observed), but significantly worse when facing an agent of their same type or the \textit{Virtue-equality} agent (due to the mutual defection observed.
We also see that the \textit{Utilitarian}, \textit{Deontological}, \textit{Virtue-kindness} and \textit{Virtue-mixed} agents similarly obtain the highest game reward when facing another non-selfish agent of this type - because of mutual cooperation (as observed in the pairwise actions). However, against \textit{Selfish} or \textit{Virtue-equality} agents, they do worse on the three games, since exploitation emerges.

Considering the intrinsic moral reward (row two in panels A,B,C, Figure \ref{fig:reward}), we observe that across the three environments the moral agents are broadly able to learn to achieve high intrinsic reward as expected. 
However, it is worth noting some differences that can be observed in the figure. Specifically, the \textit{Utilitarian}, \textit{Deontological} and \textit{Virtue-mixed} agents obtain a smaller moral reward when learning against \textit{Selfish} or \textit{Virtue-equality} opponents. This is a direct consequence of the alignment between the best playing strategies for the game and those that emerge from acting morally. The \textit{Virtue-equality} or \textit{Virtue-kindness} agents achieve stable levels of reward regardless of who they learn against - so exploiting or being exploited by others, as observed in the pairwise action plots, does not get reflected in the average cumulative reward for these agents.

Furthermore, for the \textit{Utilitarian}, \textit{Deontological} and \textit{Virtue-mixed} players (second, third and final column in panels A,B,C, Figure \ref{fig:reward}), we observe an interaction between moral and game reward. We find that playing better morally is associated with better game performance - for example, these agents obtain a smaller moral reward against \textit{Selfish} or \textit{Virtue-equality} opponents, and they also do worse in terms of game reward on these same occasions. No such effects are observed for the \textit{Virtue-equality} and \textit{Virtue-kindness} agents on any of the games.

\subsection{Reward over time}

In the main paper we show cumulative game and moral reward obtained by learning player $M$ vs. all possible learning opponents $O$. Here we present the same rewards over time (over the 10000 iterations), for a consideration of the learning dynamics (Figures 15
-\ref{fig:reward_pairs_learning_STH}). Due to the linear $\epsilon$-decay from 1.0 to 0, we observe a near-linear convergence to the final outcome over time for all types of agents, with only slight variations in the shape. We also observe an overlap in the learning curves of the \textit{Utilitarian, Deontological, Virtue-kindness} and \textit{Virtue-mixed} agents.

\section{Social Outcomes - with Confidence Intervals}

In the paper we present heat-maps summarizing average values (across 100 runs) for collective, Gini and minimum return on all three games. In Figures \ref{fig:outcomes_IPD_CI}
-\ref{fig:outcomes_STH_CI} we present the same data but in bar plots showing 95\% Confidence Intervals. As stated in the paper, a consideration of Confidence Intervals does not change the interpretation of the relative return values. 

\section{Does Exploration Aid Moral Agents' Learning?}

In the main body of the paper we present agents who start off by exploring 100\% of the time and then linearly decay their exploration rate to 0 by the final iteration. This allows the agents to observe all state-action pairs enough times early in the learning process, and they learn to play optimally (i.e. maximizing their moral reward $R_{M_{intr}}$ by the end of the 10000 iterations). It may be of interest to understand how our moral agents might learn without such major exploration at the start. 

In Figure \ref{fig:exploration} we present the impact of implementing a less exploratory agent - one that starts off exploring 5\% of the time and maintains a steady exploration rate instead. For ease of interpretation, we consider the case of each agent learning against its own kind on the Iterated Prisoner's Dilemma (patterns of learning are similar across the three games).  

With smaller exploration (left, $\epsilon=5\%$), the \textit{Selfish}, \textit{Utilitarian} or \textit{Virtue-equality} agents do not learn a consistent strategy across the 100 runs. The \textit{Selfish} agent learns three strategies - each being learned on 1/3 of the runs: mutual defection, to exploit its opponent, or to be exploited. This is sub-optimal learning, as the \textit{Selfish} agent  that is maximizing its own game reward would have gotten a greater payoff from never being exploited. The two consequentialist agents - i.e. \textit{Utilitarian} and \textit{Virtue-equality} - learn one of the four possible strategies strategies on 1/4 of the runs each. We observe that this learning stabilizes early on, and a deeper analysis showed that it is heavily impacted by early experience - i.e. the agents update their Q-values quickly at the start and then remain stuck at a local optimum and unable to learn the optimal strategy over the 10000 iterations. 

The high exploration rate (right, $\epsilon=100\%$ decaying to 0), on the other hand, allows the agents to observe the value of the alternative action and/or state, and to learn more optimally. The \textit{Deontological} or \textit{Virtue-kindness} agents learned to mutually cooperate in either of the settings, so exploration rate has less impact on their learning.

\section{The Effect of Different Weights on the two Virtues in \textit{Virtue-mixed} Agent.}

In the main results, we found that the \textit{Virtue-mixed} agent with equivalent \textit{equality} and \textit{kindness} weights ($\beta=0.5$) learned to be exploited as much as the \textit{Virtue-kindness} agent (equivalent to $\beta=0$), as demonstrated by \textit{Virtue-mixed} learning against all moral opponents (presented in the paper and in pairwise action plots above - Figures \ref{fig:action_pairs_learning_IPD}-\ref{fig:action_pairs_baseline_STH}). To investigate this further, in Figures \ref{fig:extra_QLVM_beta_IPD}-\ref{fig:extra_QLVM_beta_STH} we explore what proportions of \textit{equality} versus \textit{kindness} in the multi-objective reward allow the \textit{Virtue-mixed} agent to defend themselves better against exploitation. Across all three games we find that only very large relative weightings on the \textit{equality} reward ($\beta>0.8$) steer the mixed agent away from being exploited, but as a result the mixed agent essentially behaves as a purely \textit{equality}-driven agent would.

\section{Why Do Equality-focused Agents Learn to Exploit their Opponent?}

We note in the main paper that, on the final iteration, the \textit{Virtue-equality} agent learnt to exploit other non-selfish agents a small proportion of the time (up to 20\%). To investigate this further, we conducted two additional analyses: 
\begin{enumerate}
  \item Using an example pair from the Iterated Prisoner's Dilemma, visualize the last 20 actions performed by every player in the pair (see Figure \ref{fig:QLVE_e-deepdive}) to understand the underlying types of strategies that the equality agent learned. We choose the example case of \textit{Virtue-equality} learning against a \textit{Utilitarian} opponent as the \textit{Utilitarian} was one of the most cooperative ones).
  \item Run the experiments with the Virtue-equality agent for long enough to converge to a stable policy across the 100 runs, and analyse the policy learnt (see Figures \ref{fig:longer_action_pairs_IPD}-\ref{fig:longer_action_pairs_ISH}).
\end{enumerate}

From analysis 1, we can observe that over the 100 runs, the equality agents learn to play a mixed strategy that alternates between D$|$(C,C) and C$|$(C,C) 11\% of the time, and also learn an exploitative Always Defect strategy (D$|$(C,C)) on 9\% of the runs. These runs are not efficient in terms of reward obtained given that the opponent is an always-cooperative \textit{Utilitarian} agent, so the best response for \textit{Virtue-equality} would have been to cooperate against them to get the best equality score. 

Analysis 2 offers an answer as to why this might have happened. We discovered that that the Q-value updates of the \textit{Virtue-equality} agent had not converged on some of the runs over 10000 iterations. In contrast, after running the same set of experiments for long enough for all pairs to converge instead (50000 iterations instead of the 10000 reported in the paper), with the exploration rate $epsilon$ again decaying linearly from 1 to 0, we found that the \textit{Virtue-equality} agent converges to a fully Cooperative policy on 100\% of the runs. Plots of action pairs by the \textit{Virtue-equality} agent versus every type of opponent in each game are presented in Figures \ref{fig:longer_action_pairs_IPD}-\ref{fig:longer_action_pairs_ISH}.

\newpage


\newpage~
\newpage~

\newpage~

\newpage~

\begin{figure*}[!h]
\centering
\begin{tabular}[t]{|c|cccccc}
\toprule
& Selfish & Utilitarian & Deontological & Virtue - equality & Virtue - kindness & Virtue - mixed \\
\midrule
\makecell[cc]{\rotatebox[origin=c]{90}{\thead{Game Reward}}} & 
\subt{\includegraphics[width=20mm]{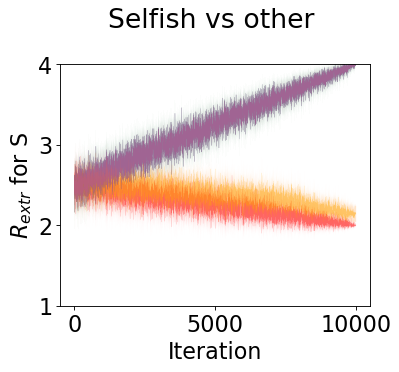}} & \subt{\includegraphics[width=20mm]{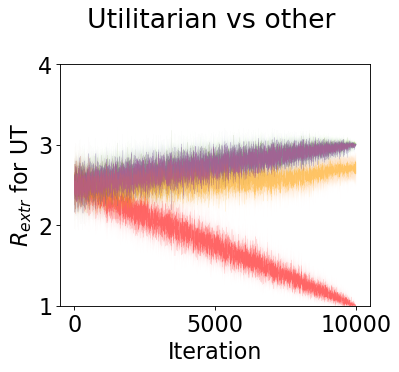}} & \subt{\includegraphics[width=20mm]{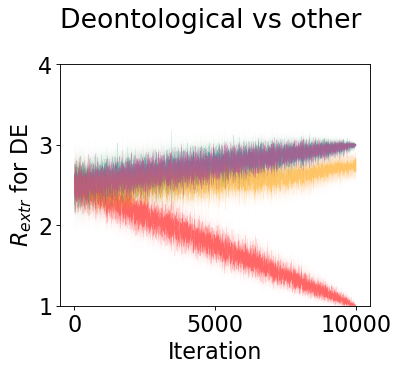}} & \subt{\includegraphics[width=20mm]{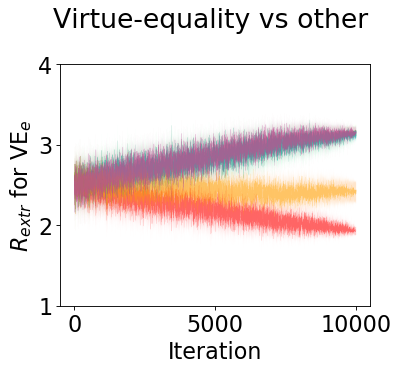}} & \subt{\includegraphics[width=20mm]{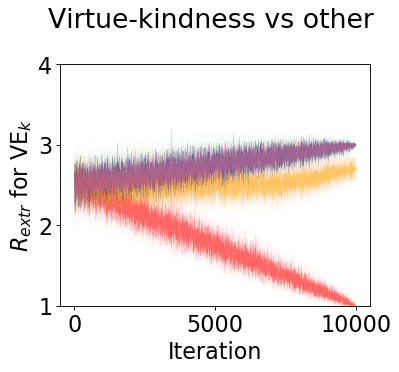}} & \subt{\includegraphics[width=20mm]{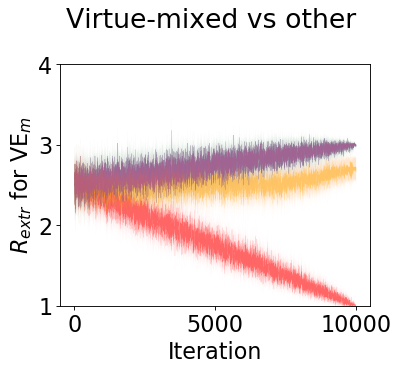}}
\\ \hline
\makecell[cc]{\rotatebox[origin=c]{90}{\thead{Moral Reward}}} & 
\subt{\includegraphics[width=15mm]{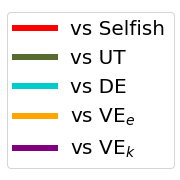}} & \subt{\includegraphics[width=20mm]{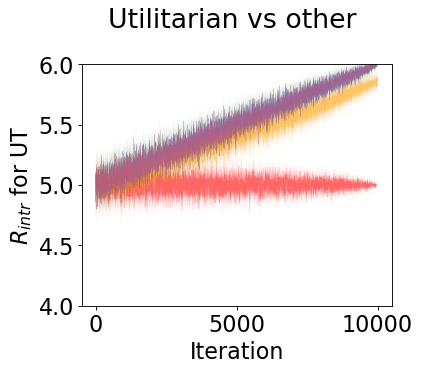}} & \subt{\includegraphics[width=20mm]{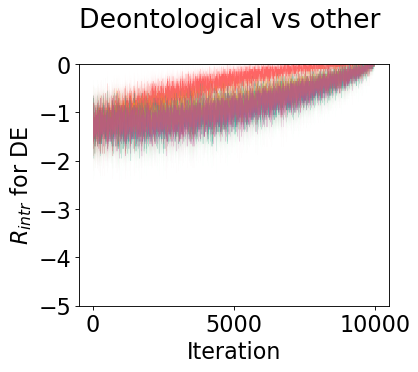}} & \subt{\includegraphics[width=20mm]{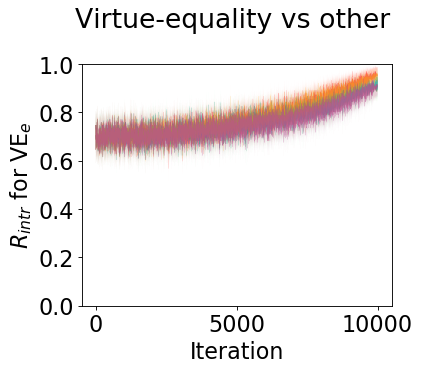}} & \subt{\includegraphics[width=20mm]{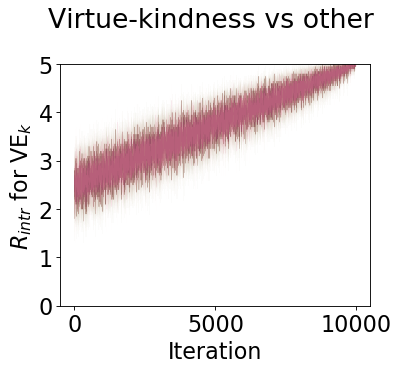}} & \subt{\includegraphics[width=20mm]{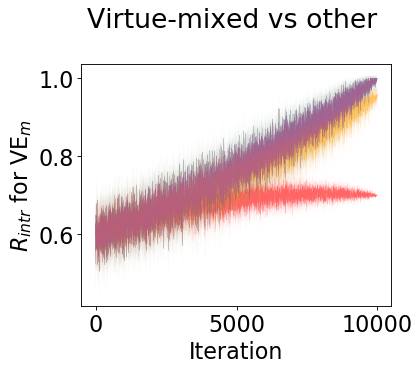}}
\\
\bottomrule
\end{tabular}
\label{fig:reward_p_l_IPD} 
\caption{Iterated Prisoner's Dilemma game. Game \& moral reward (per iteration) obtained by moral learning player type $M$ (row) vs. all possible learning opponents $O$. The plots display average across the 100 runs $\pm$ 95\%CI.}
\end{figure*}

\begin{figure*}[!h]
\centering
\begin{tabular}[t]{|c|cccccc}
\toprule
& Selfish & Utilitarian & Deontological & Virtue - equality & Virtue - kindness & Virtue - mixed \\
\midrule
\makecell[cc]{\rotatebox[origin=c]{90}{\thead{Game Reward}}} & \subt{\includegraphics[width=24mm]{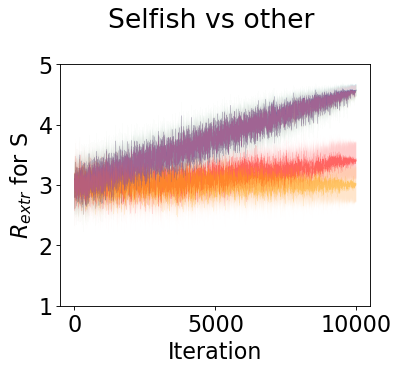}} & \subt{\includegraphics[width=20mm]{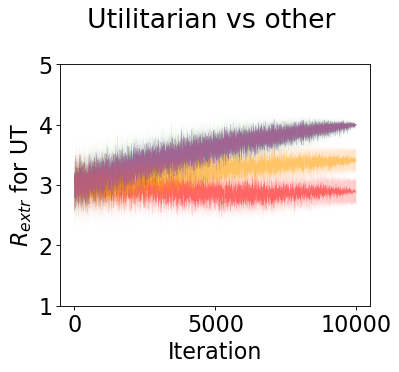}} & \subt{\includegraphics[width=20mm]{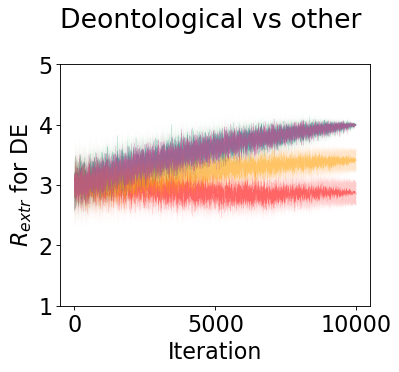}} & \subt{\includegraphics[width=20mm]{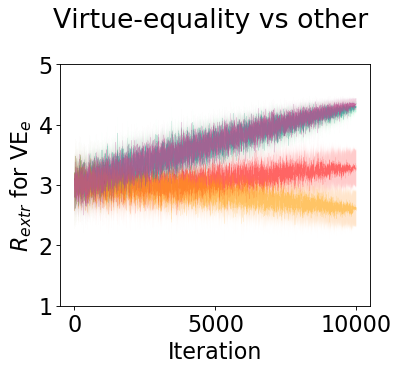}} & \subt{\includegraphics[width=20mm]{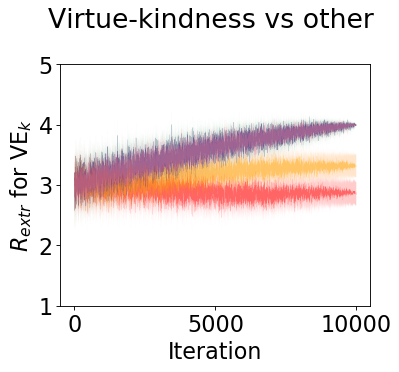}} & \subt{\includegraphics[width=20mm]{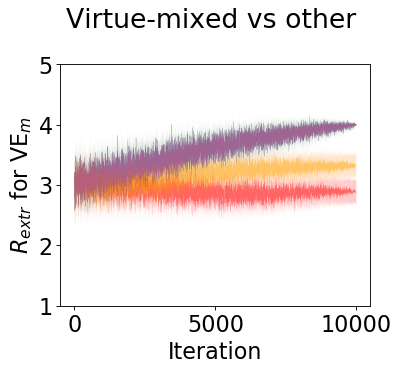}}
\\ \hline
\makecell[cc]{\rotatebox[origin=c]{90}{\thead{Moral Reward}}} & 
\subt{\includegraphics[width=15mm]{IJCAI-2023/IPD_cooperation/legend.png}} & \subt{\includegraphics[width=20mm]{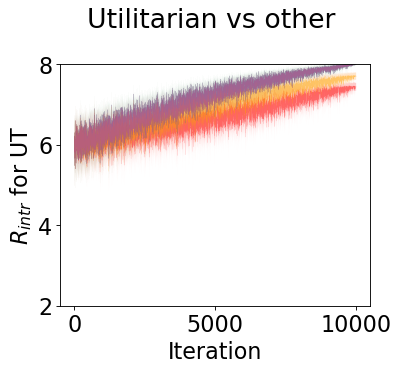}} & \subt{\includegraphics[width=20mm]{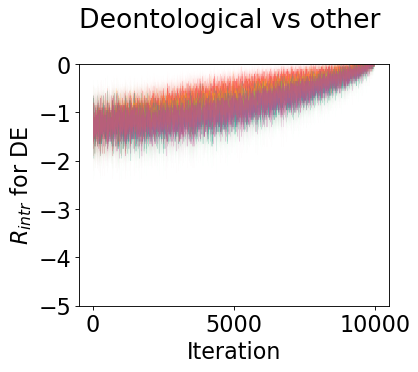}} & \subt{\includegraphics[width=20mm]{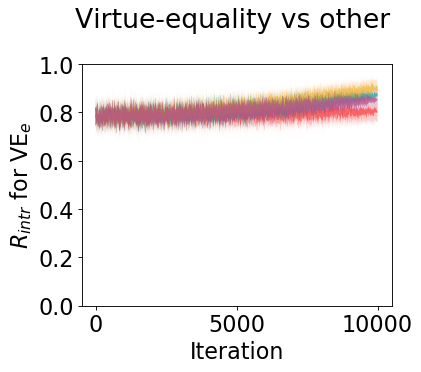}} & \subt{\includegraphics[width=20mm]{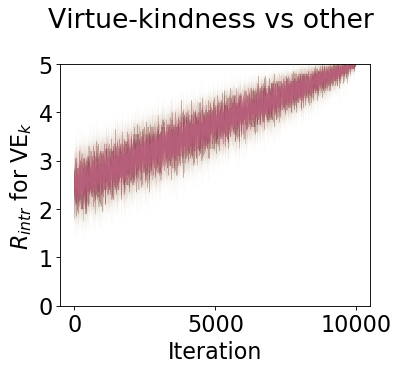}} & \subt{\includegraphics[width=20mm]{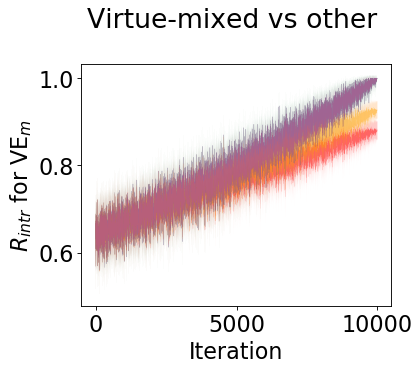}}
\\
\bottomrule
\end{tabular}
\caption{Iterated Volunteer's Dilemma game. Game \& moral reward (per iteration) obtained by moral learning player type $M$ (row) vs. all possible learning opponents $O$. The plots display average across the 100 runs $\pm$ 95\%CI.}
\label{fig:reward_pairs_learning_VOL}
\end{figure*}

\begin{figure*}[!h]
\centering
\begin{tabular}[t]{|c|cccccc}
\toprule
& Selfish & Utilitarian & Deontological & Virtue - equality & Virtue - kindness & Virtue - mixed \\
\midrule
\makecell[cc]{\rotatebox[origin=c]{90}{\thead{Game Reward}}} & \subt{\includegraphics[width=20mm]{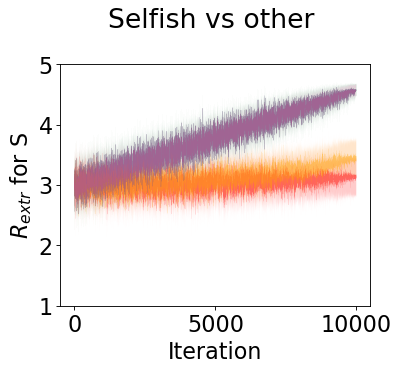}} & \subt{\includegraphics[width=20mm]{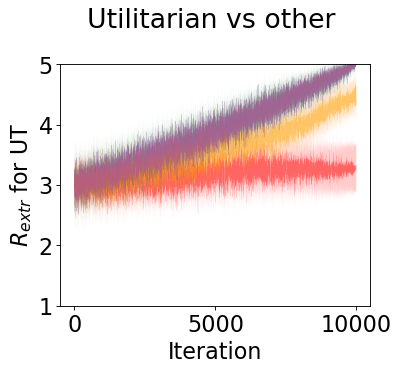}} & \subt{\includegraphics[width=20mm]{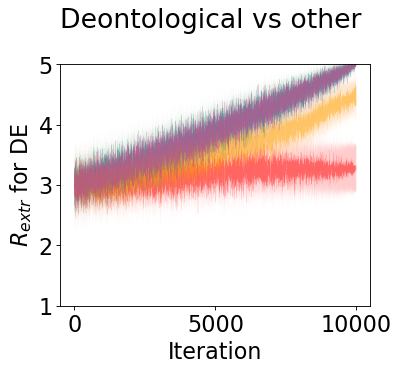}} & \subt{\includegraphics[width=20mm]{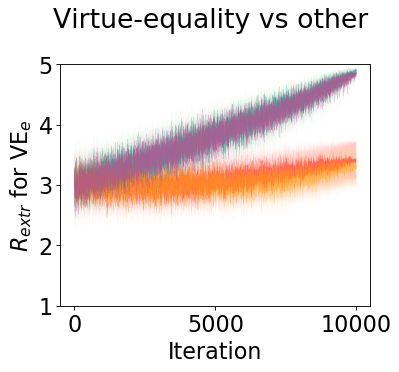}} & \subt{\includegraphics[width=20mm]{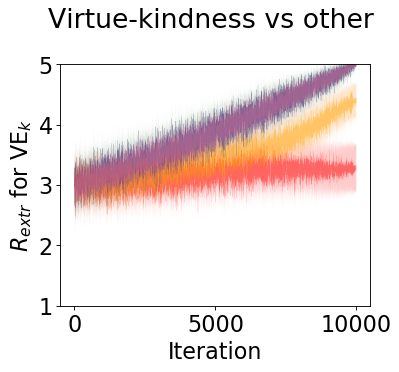}} & \subt{\includegraphics[width=20mm]{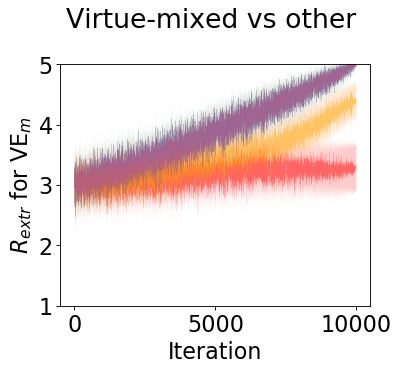}}
\\ \hline
\makecell[cc]{\rotatebox[origin=c]{90}{\thead{Moral Reward}}} & \subt{\includegraphics[width=15mm]{IJCAI-2023/IPD_cooperation/legend.png}} & \subt{\includegraphics[width=20mm]{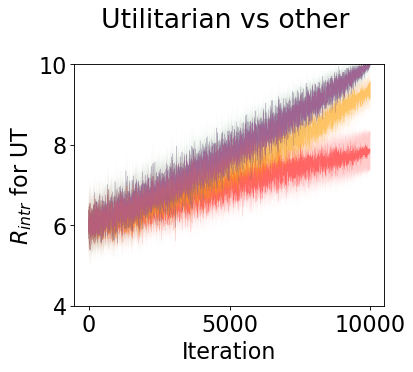}} & \subt{\includegraphics[width=20mm]{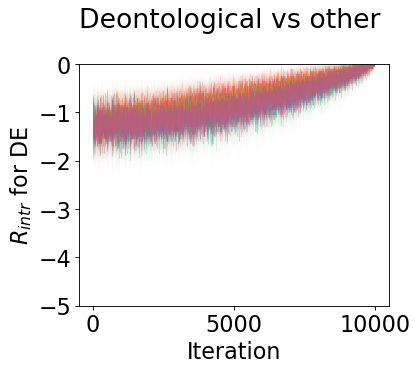}} & \subt{\includegraphics[width=20mm]{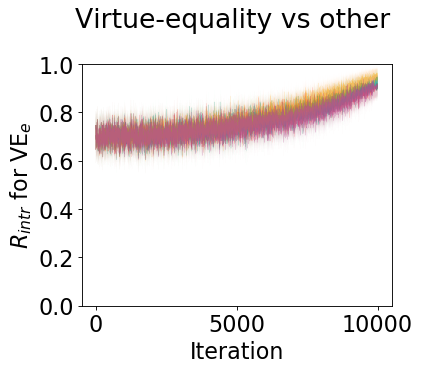}} & \subt{\includegraphics[width=20mm]{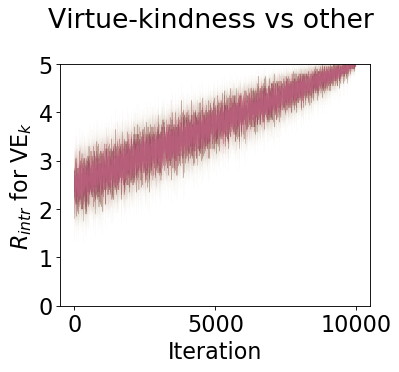}} & \subt{\includegraphics[width=20mm]{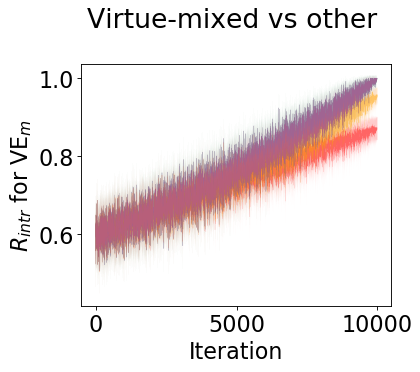}}
\\
\bottomrule
\end{tabular}
\caption{Iterated Stag Hunt game. Game \& moral reward (per iteration) obtained by moral learning player type $M$ (row) vs. all possible learning opponents $O$. The plots display average across the 100 runs $\pm$ 95\%CI.}
\label{fig:reward_pairs_learning_STH}
\end{figure*}

\begin{figure*}[!h]
\centering
\begin{tabular}[t]{|c|cccccc}
\toprule
& Selfish & Utilitarian & Deontological & Virtue - equality & Virtue - kindness & Virtue - mixed \\
\midrule
\makecell[cc]{\rotatebox[origin=c]{90}{\thead{Collective Return}}} & \subt{\includegraphics[height=18mm]{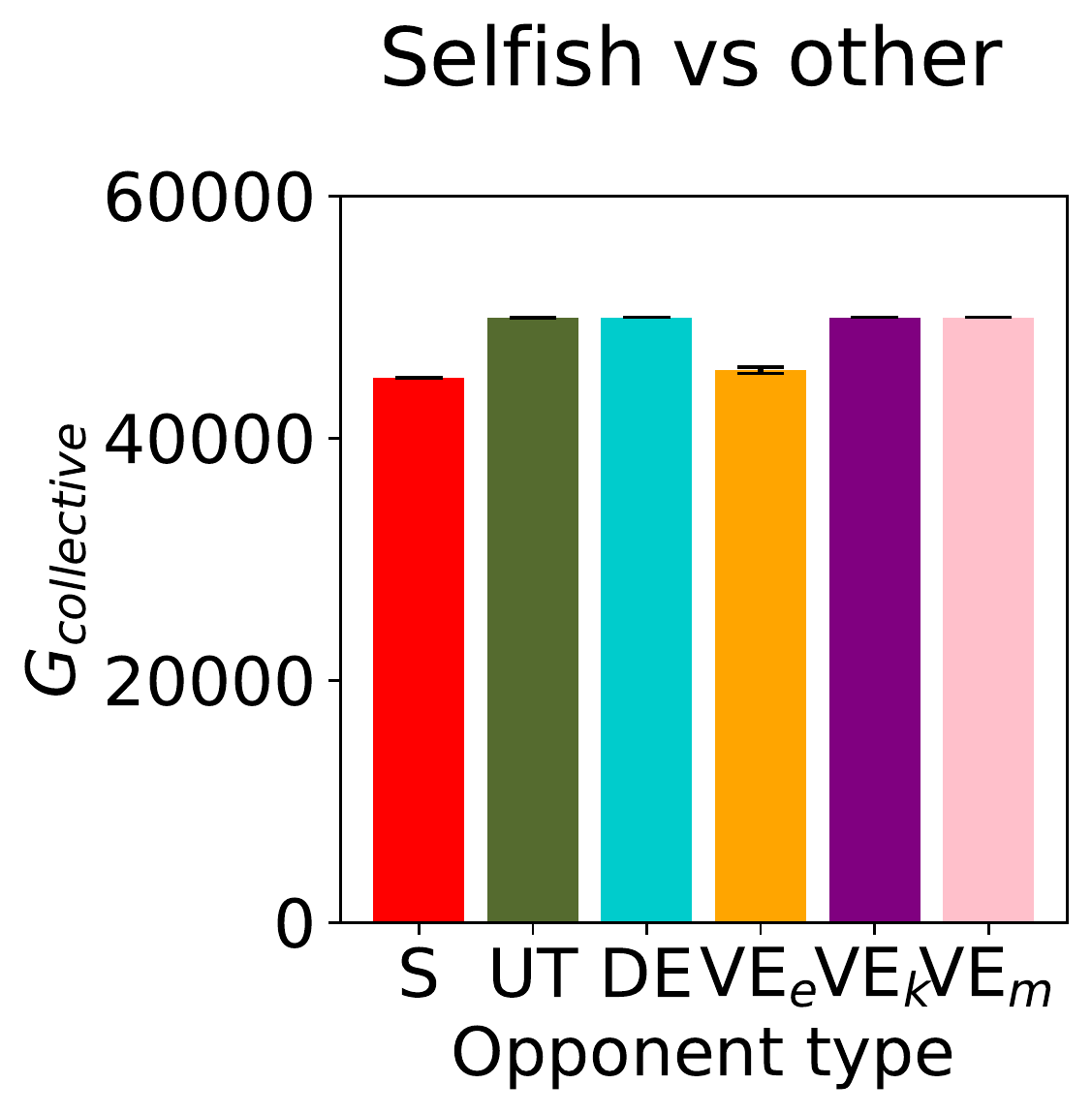}} & \subt{\includegraphics[height=18mm]{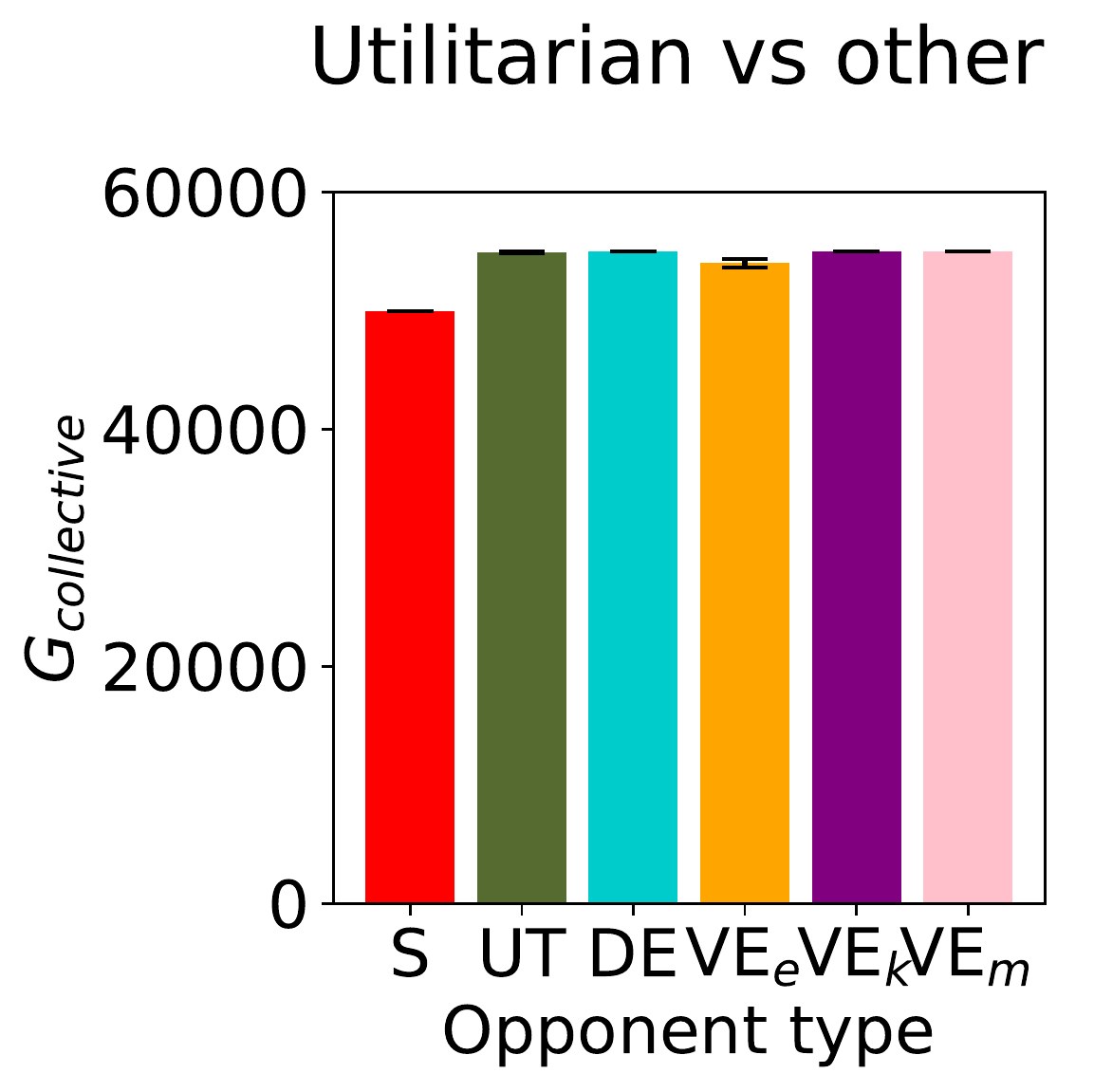}} & \subt{\includegraphics[height=18mm]{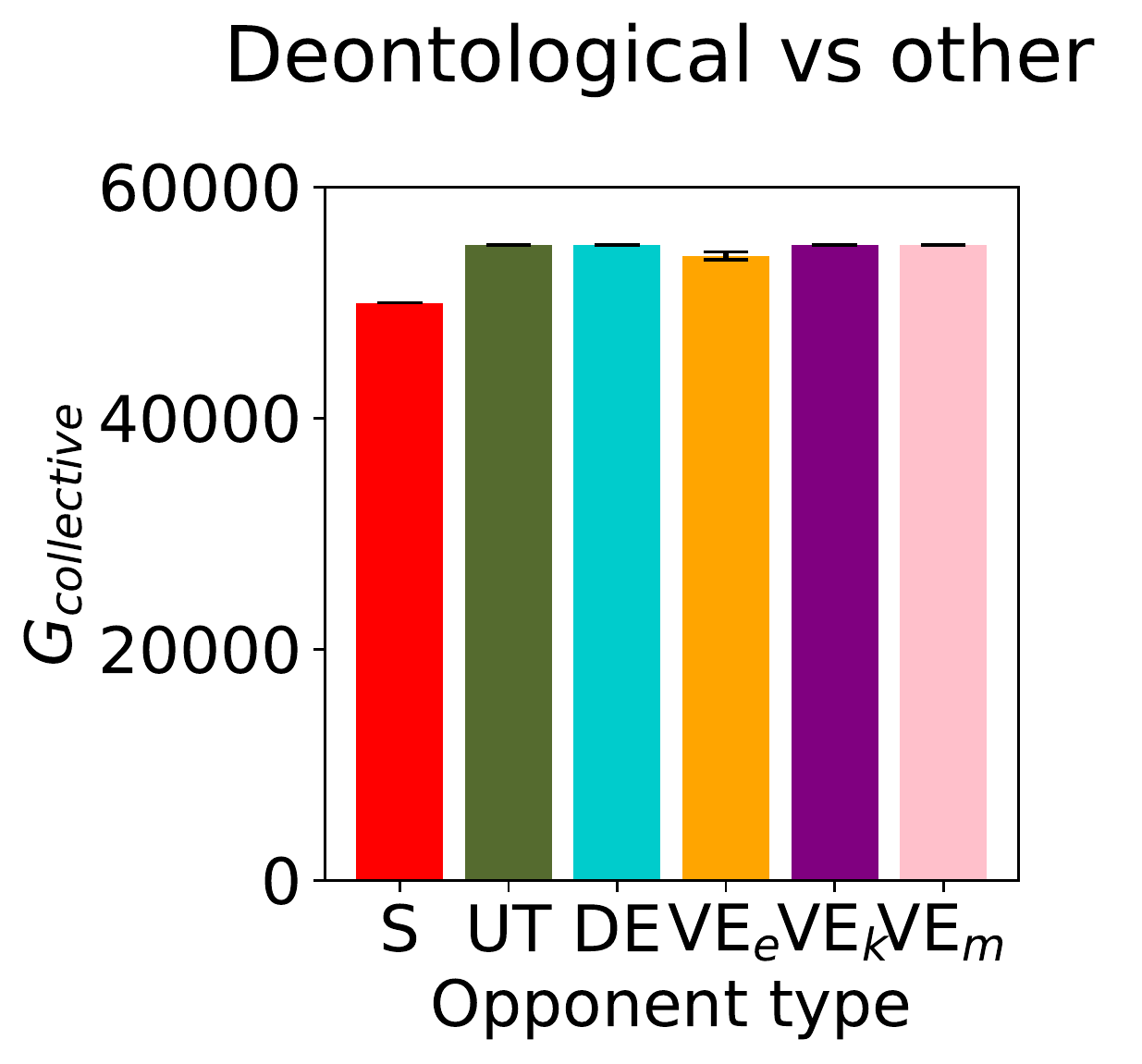}} & \subt{\includegraphics[height=18mm]{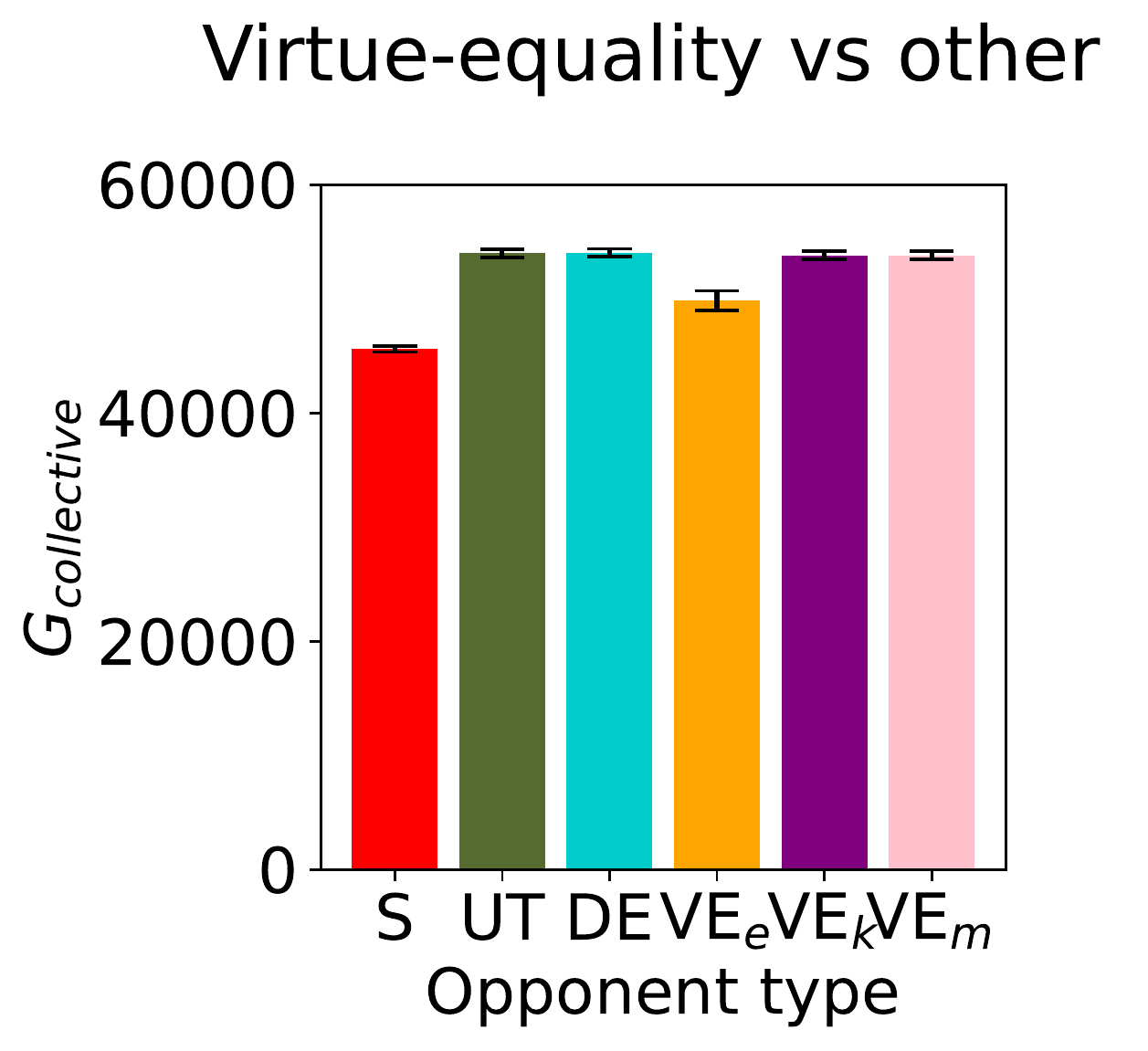}} & \subt{\includegraphics[height=18mm]{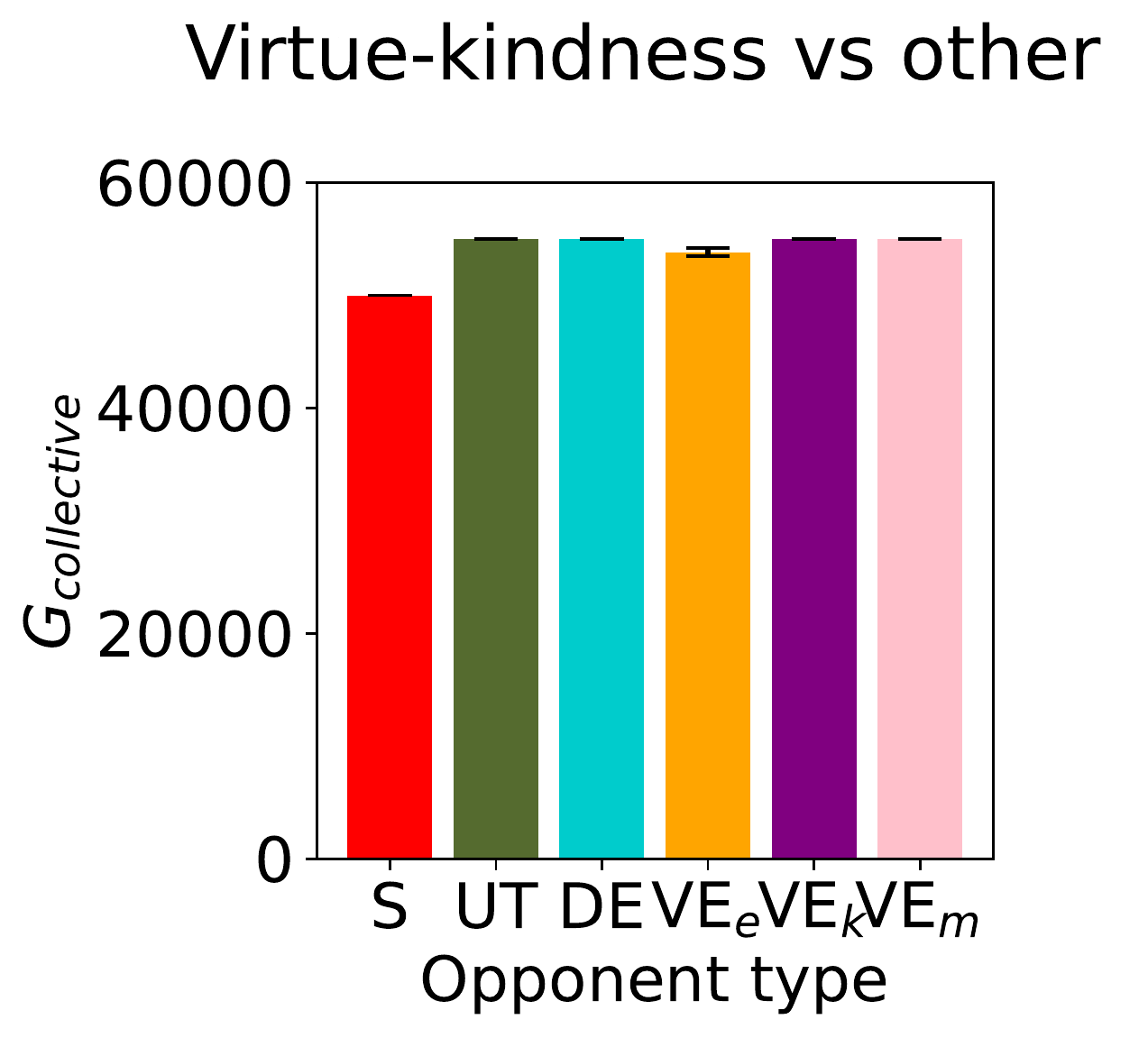}} & \subt{\includegraphics[height=18mm]{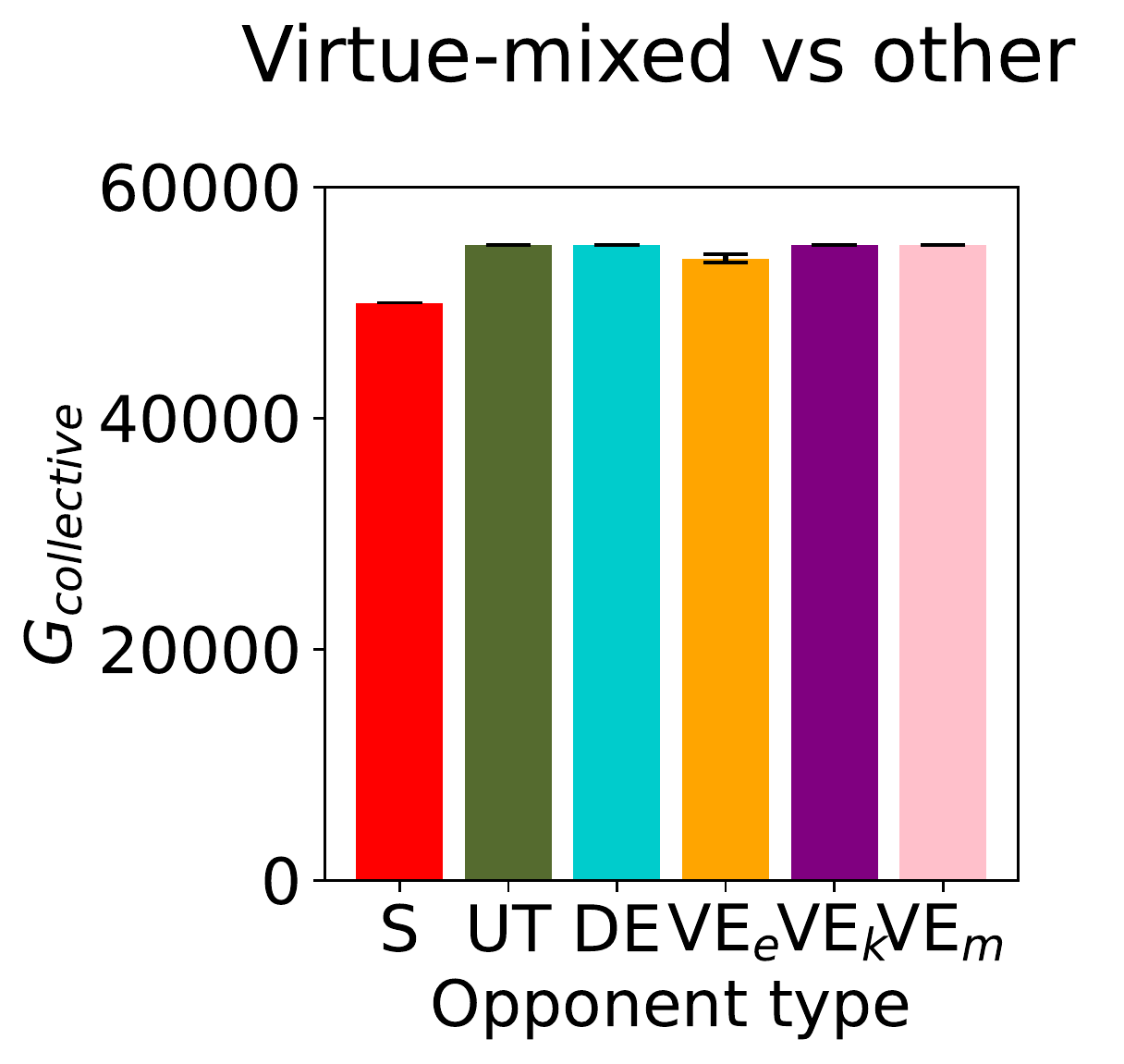}}
\\ 
\makecell[cc]{\rotatebox[origin=c]{90}{\thead{Gini Return}}} & \subt{\includegraphics[height=18mm]{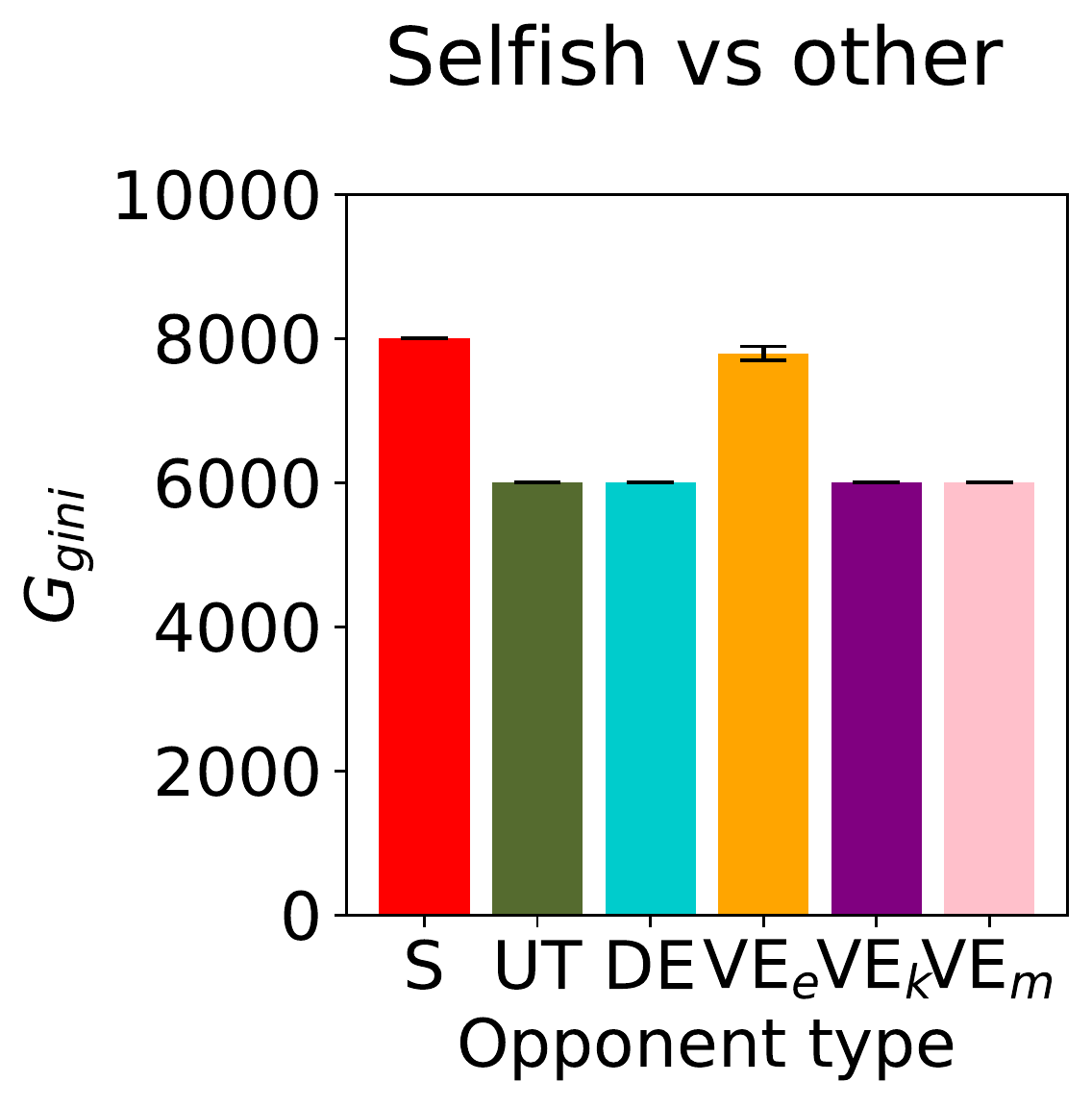}} & \subt{\includegraphics[height=18mm]{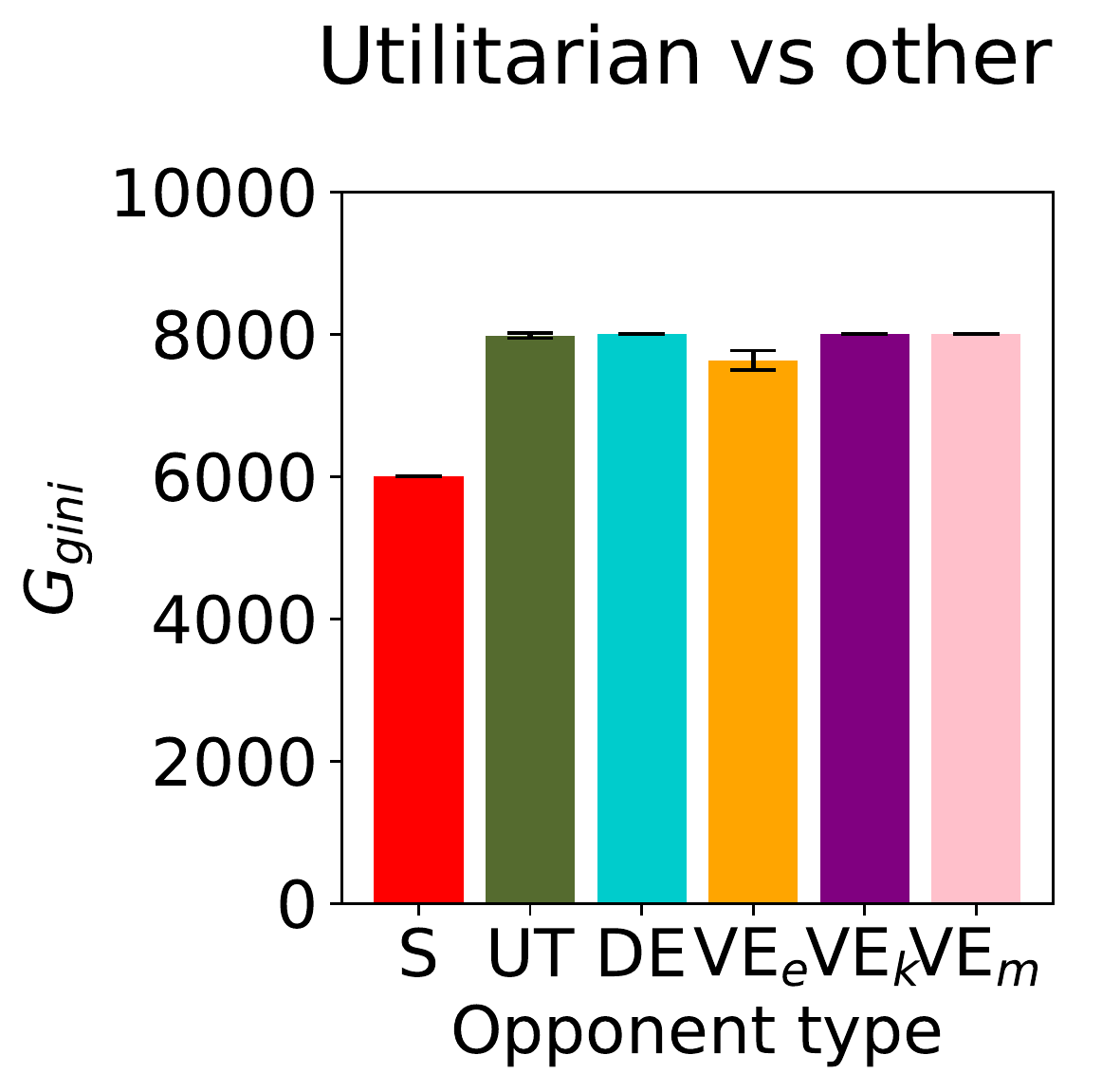}} & \subt{\includegraphics[height=18mm]{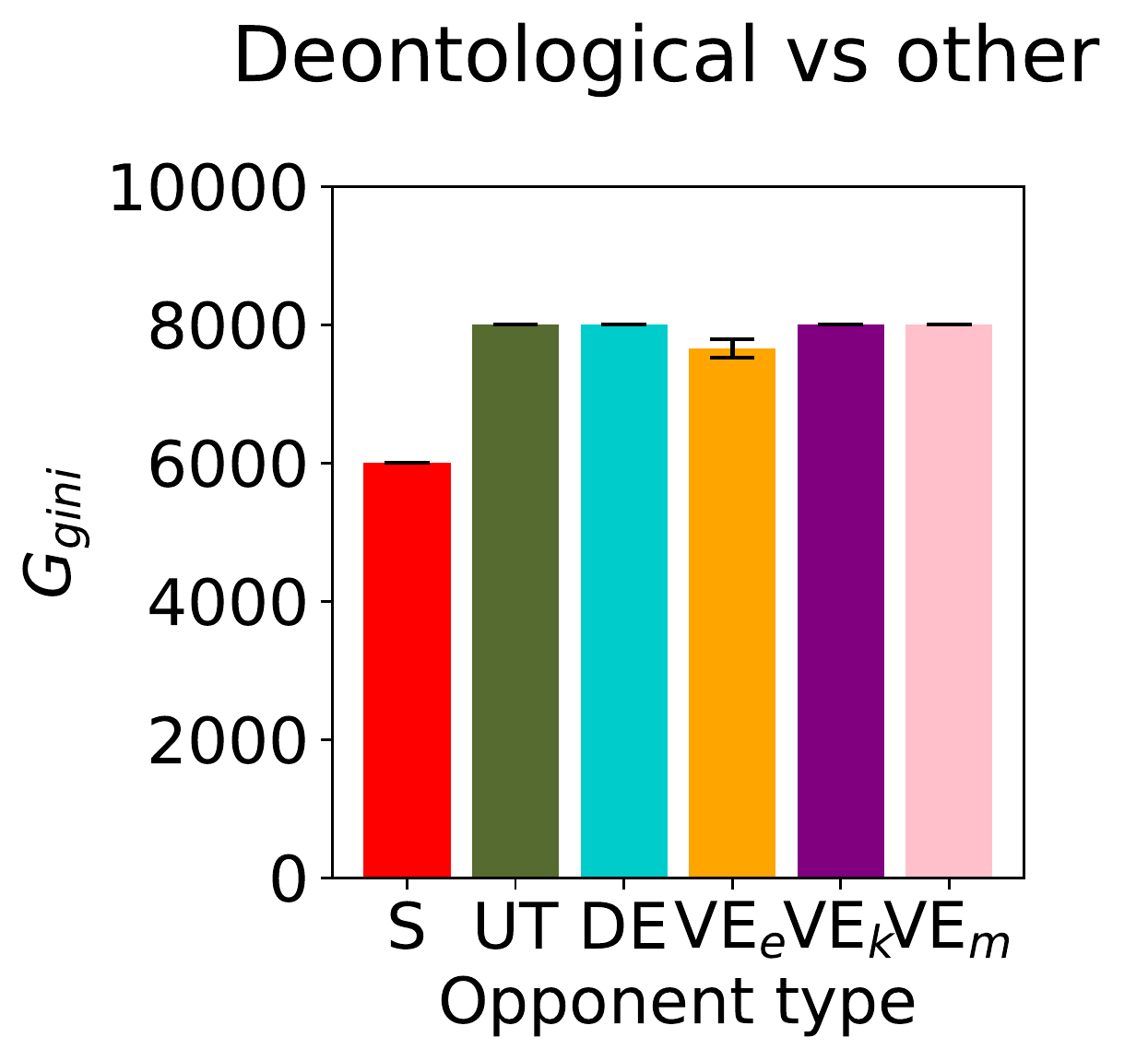}} & \subt{\includegraphics[height=18mm]{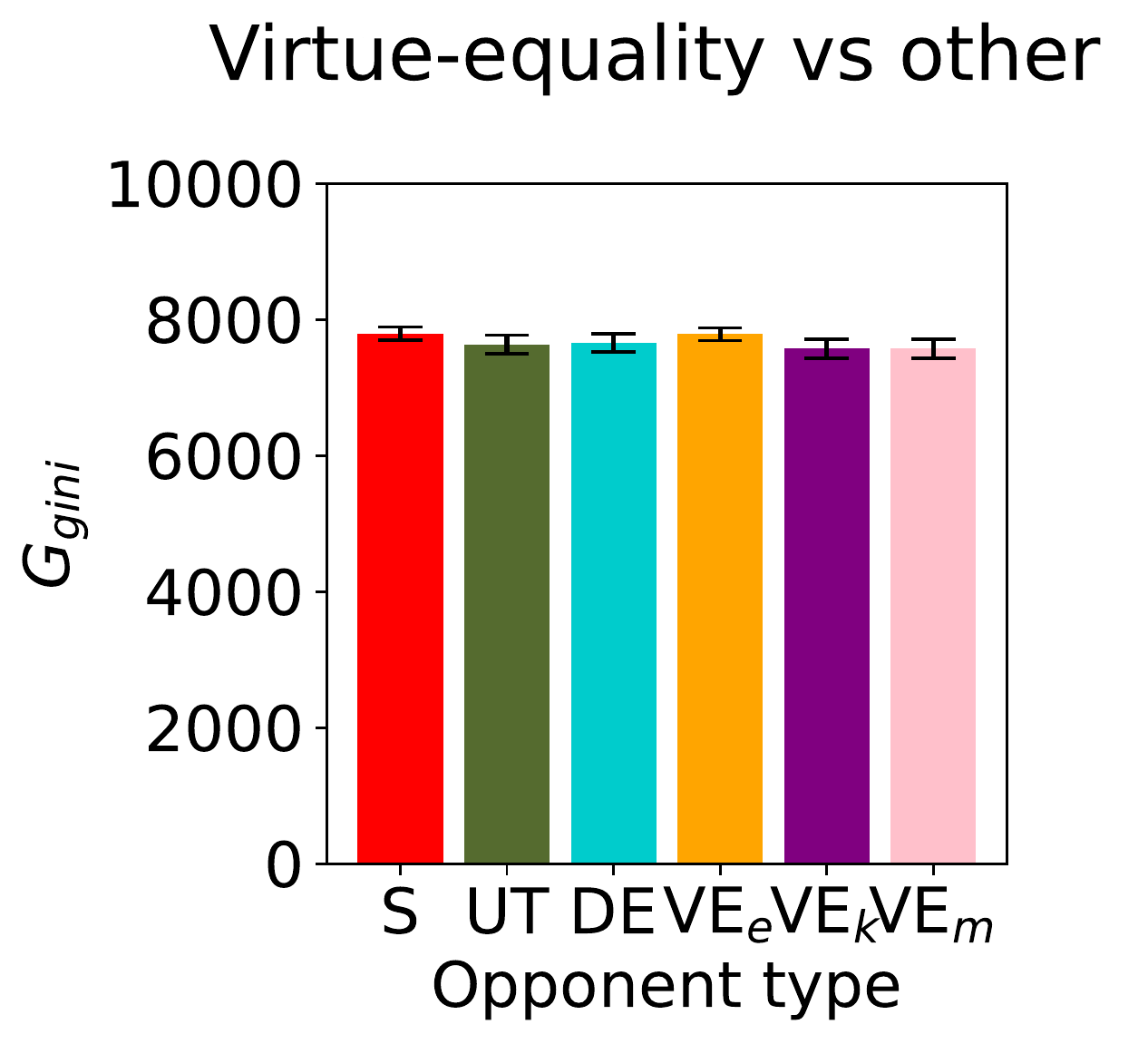}} & \subt{\includegraphics[height=18mm]{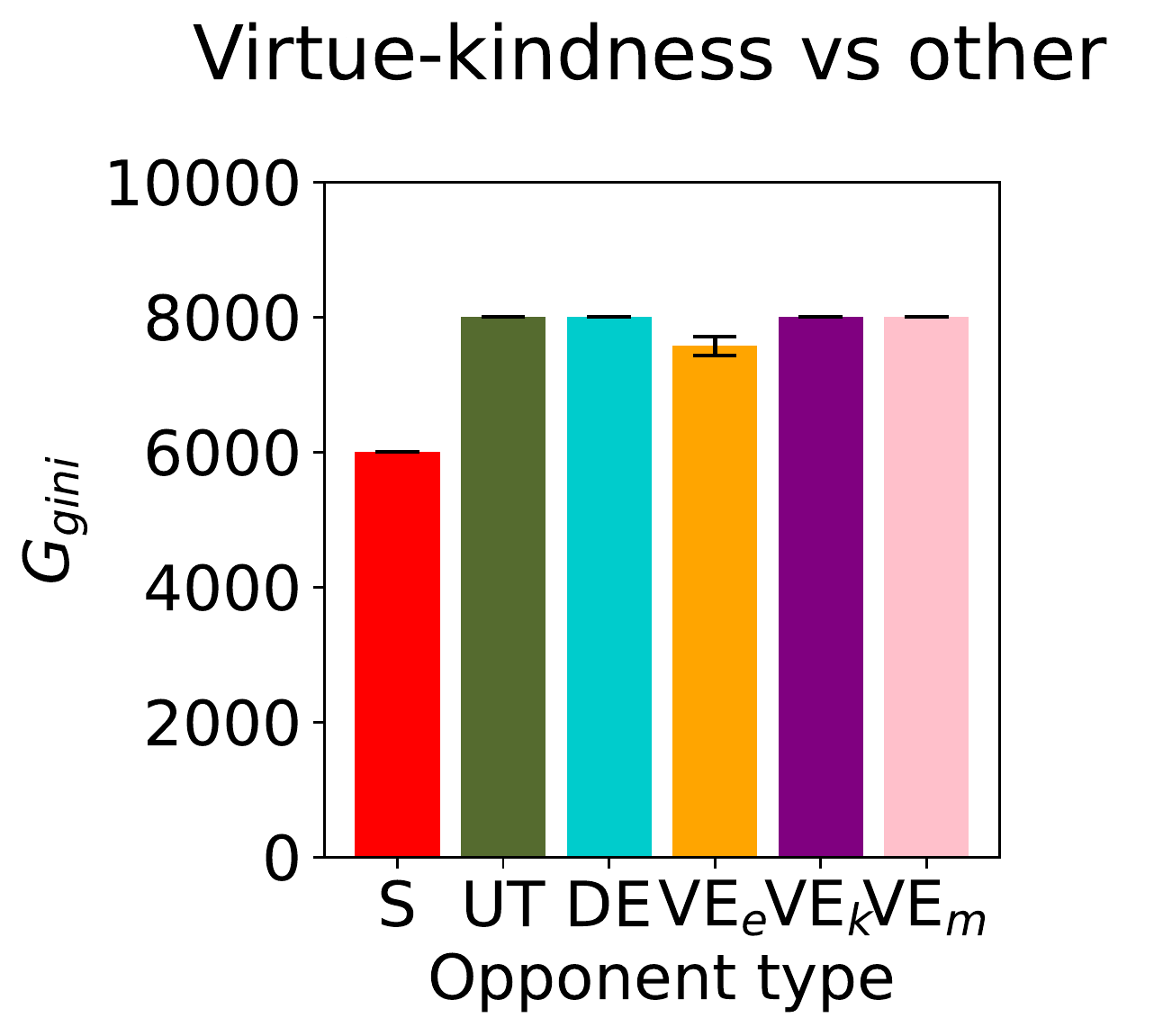}} & \subt{\includegraphics[height=18mm]{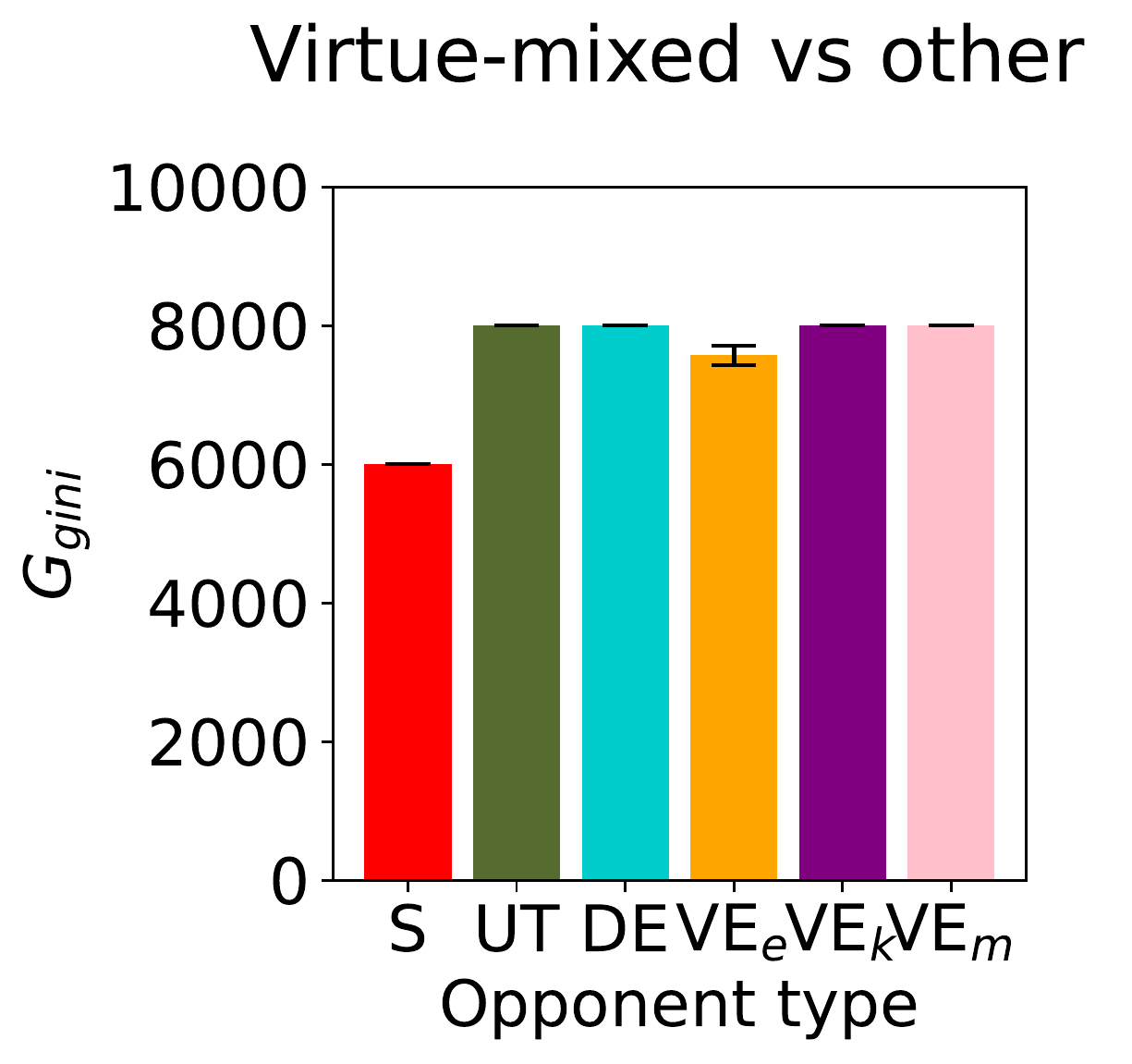}}
\\
\makecell[cc]{\rotatebox[origin=c]{90}{\thead{Min Return}}} & \subt{\includegraphics[height=18mm]{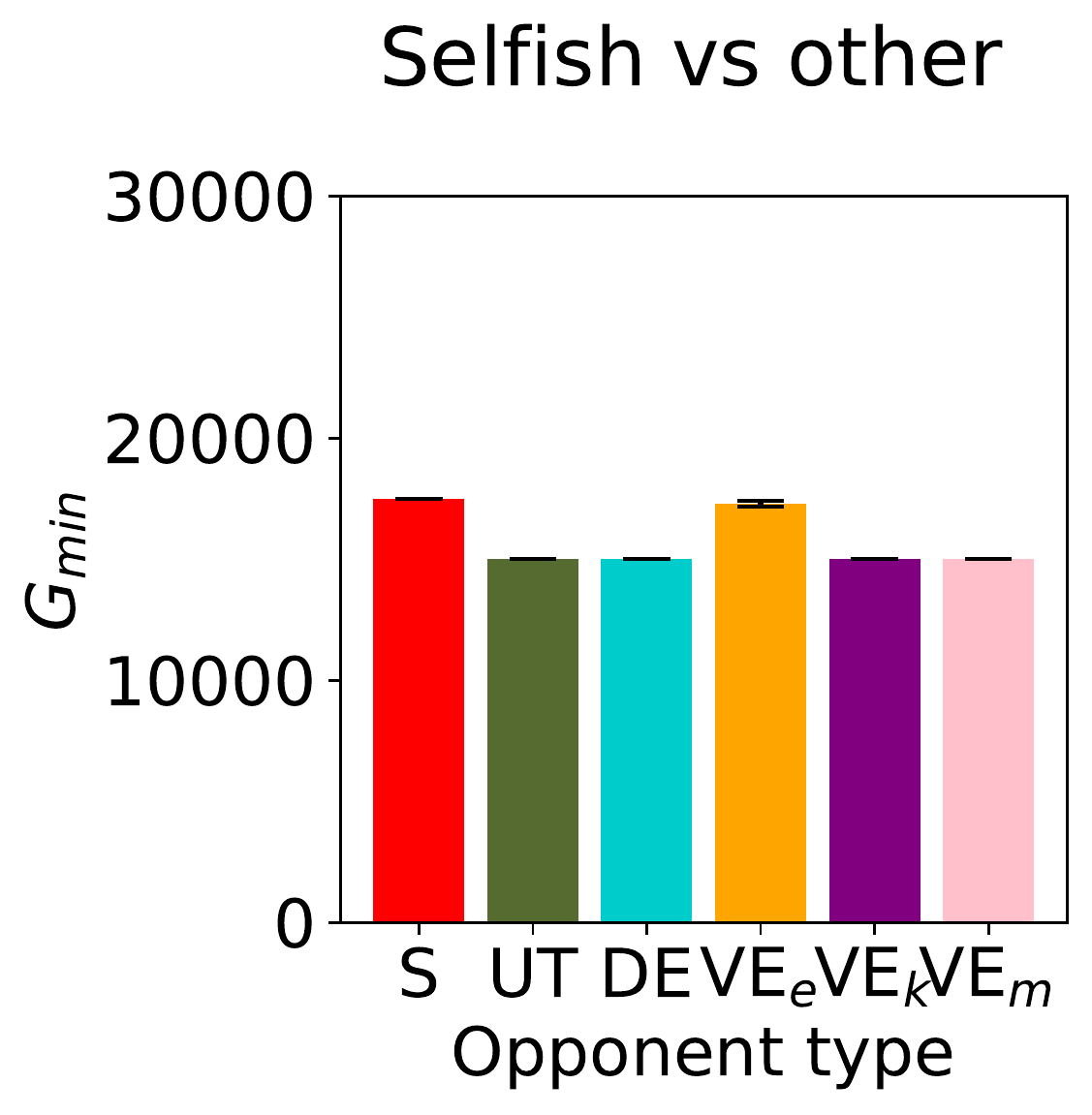}} & \subt{\includegraphics[height=18mm]{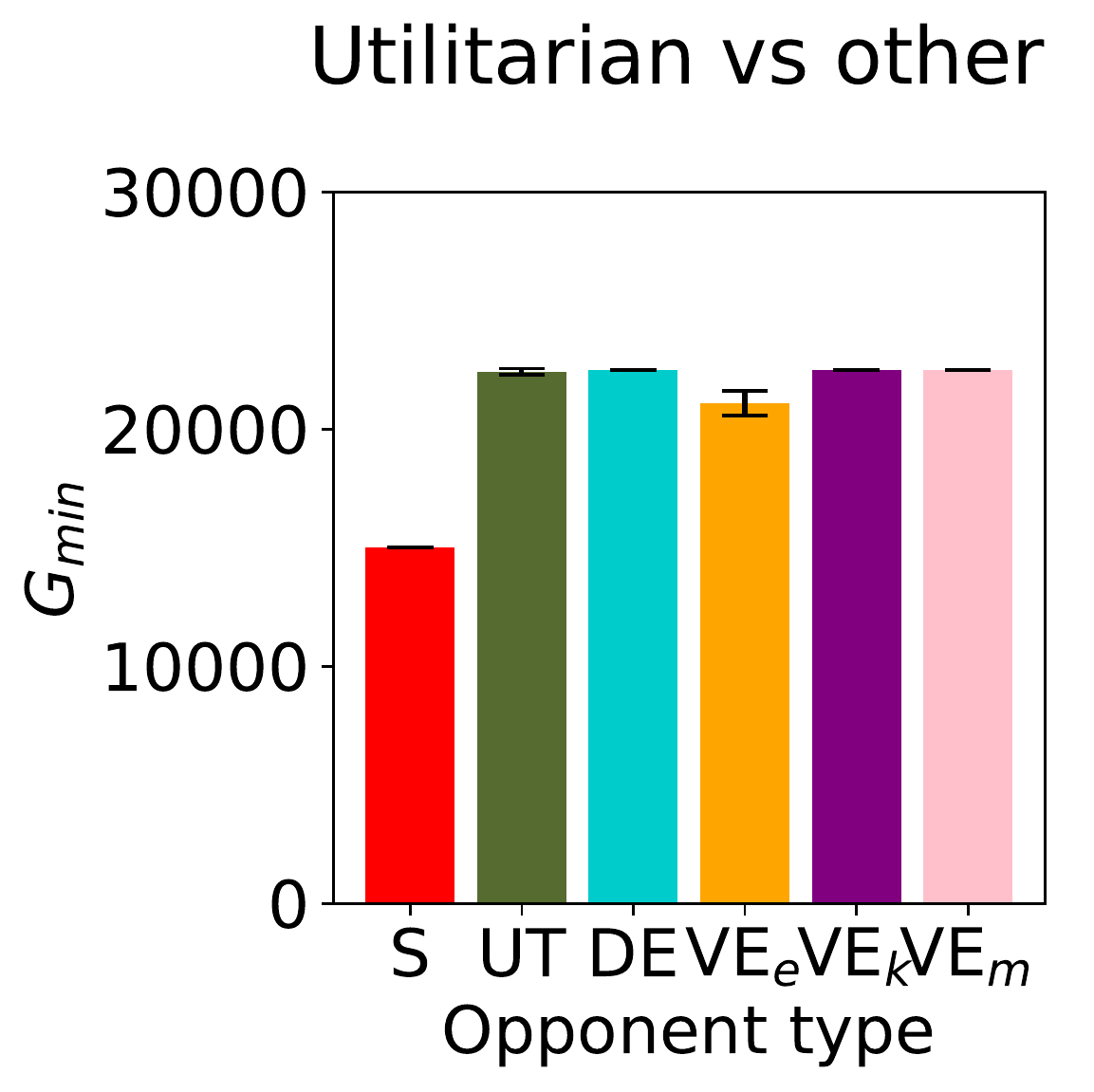}} & \subt{\includegraphics[height=18mm]{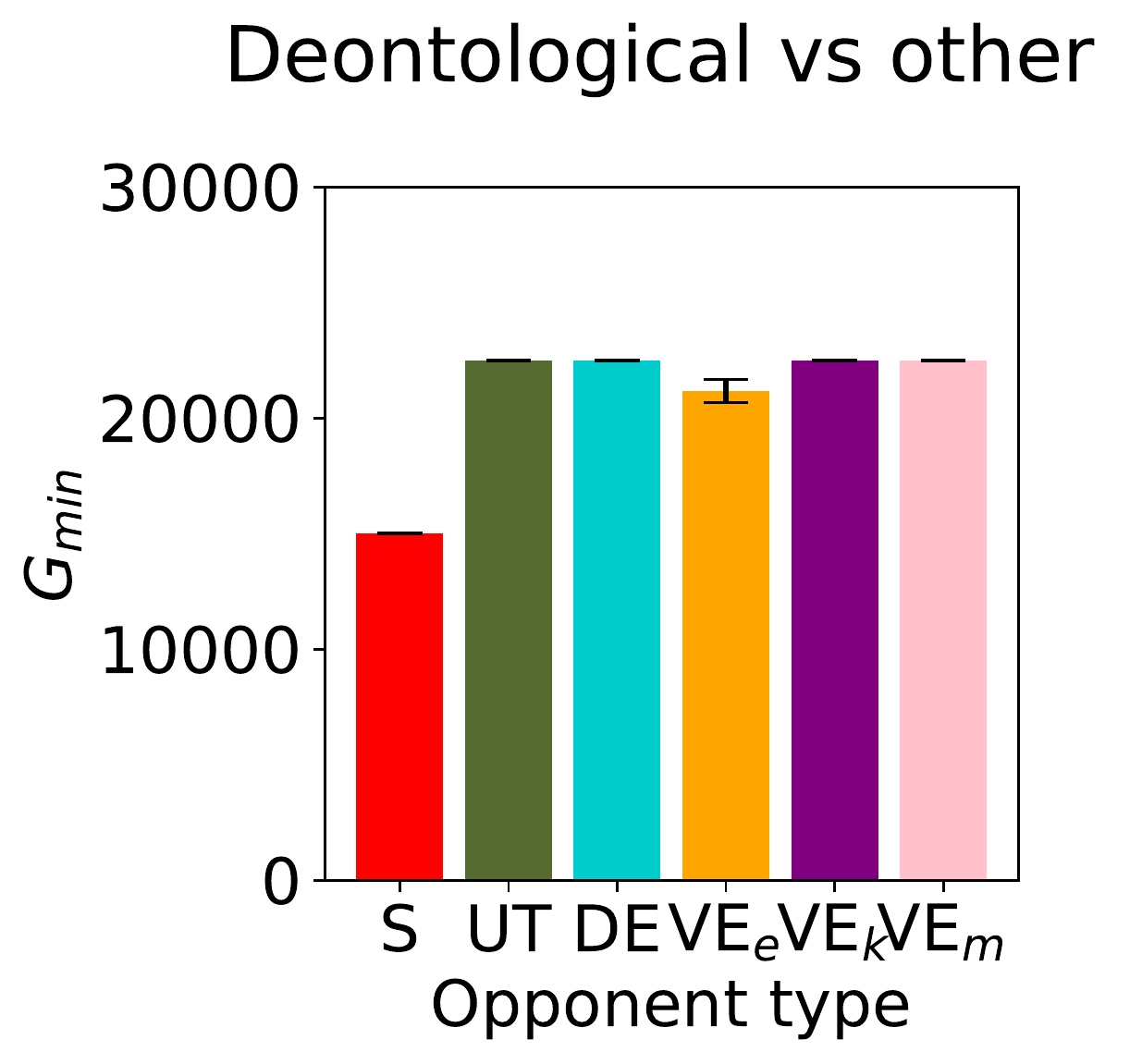}} & \subt{\includegraphics[height=18mm]{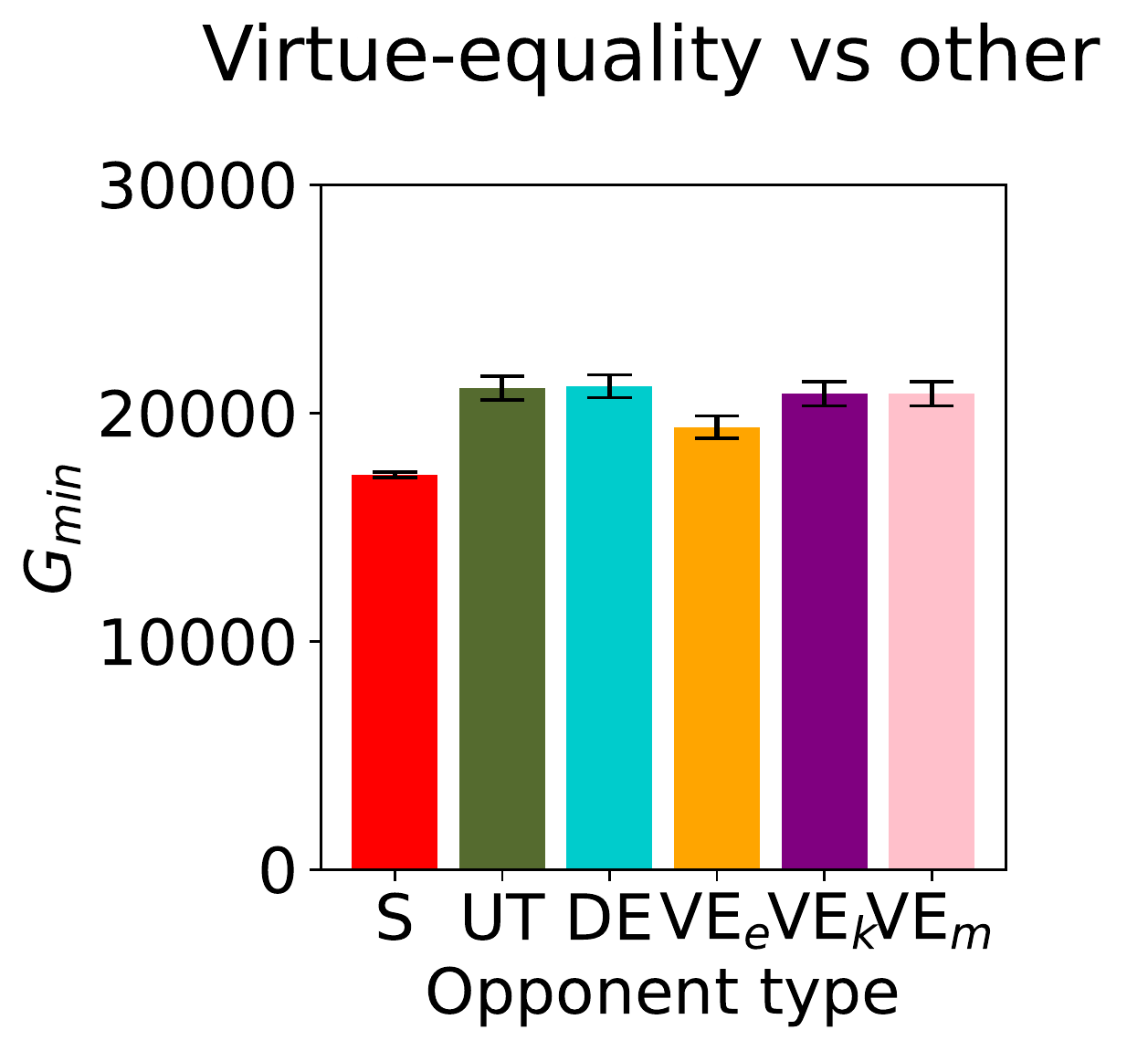}} & \subt{\includegraphics[height=18mm]{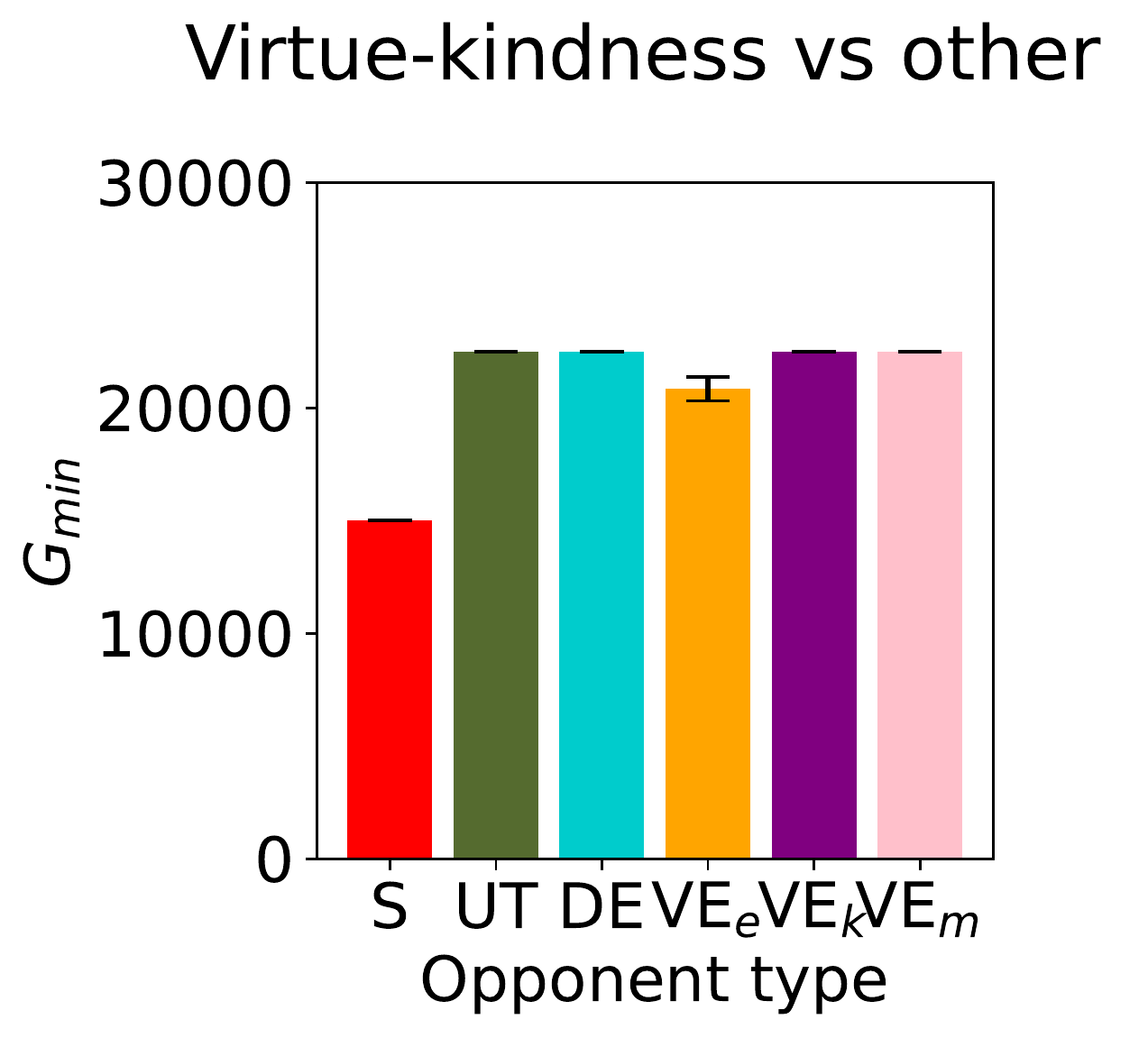}} & \subt{\includegraphics[height=18mm]{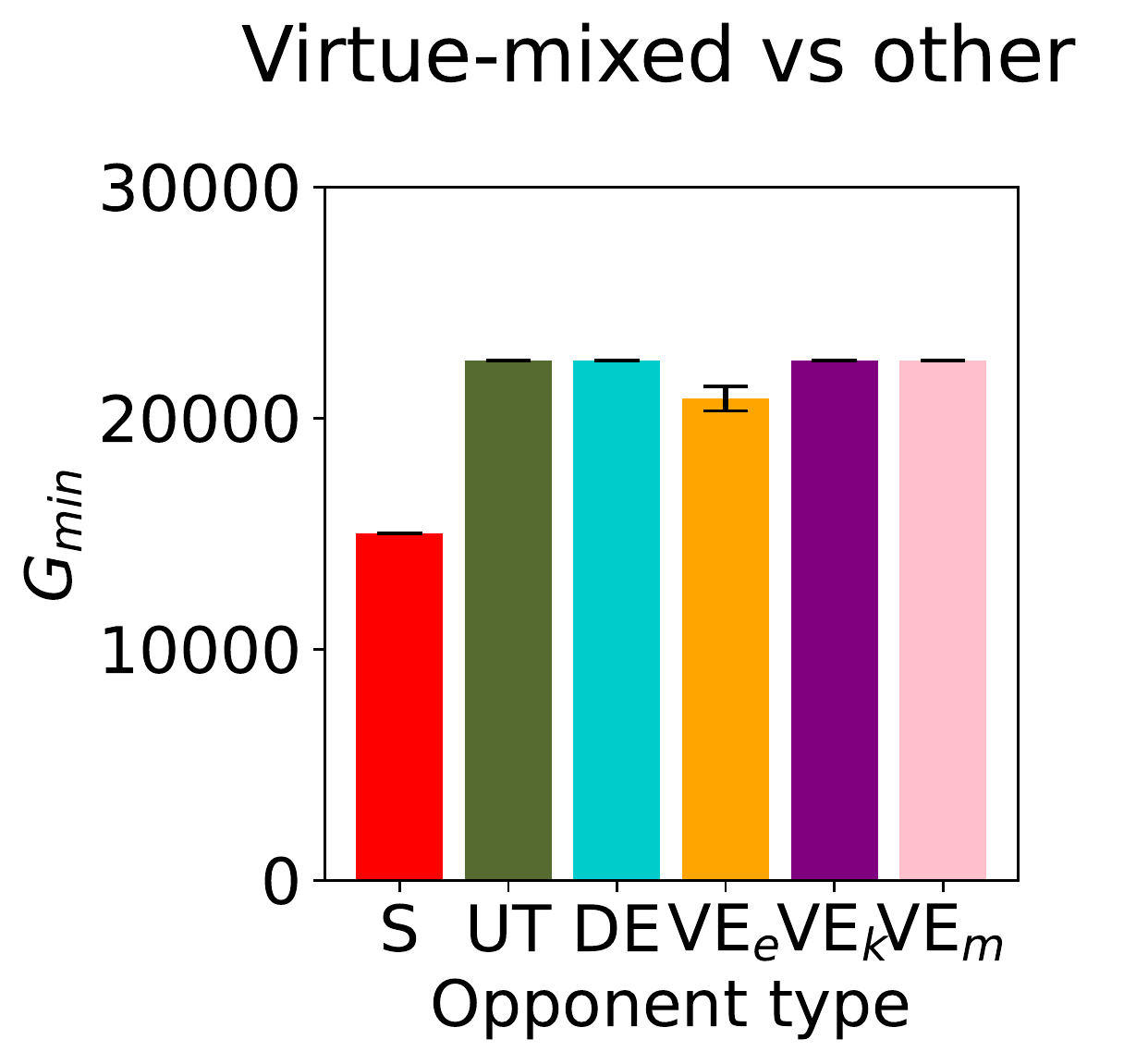}}
\\
\bottomrule
\end{tabular}
\caption{Iterated Prisoner's dilemma game. Relative societal outcomes observed for learning player type $M$ (row) vs. all possible learning opponents $O$. The plots display averages across the 100 runs $\pm$ 95\%CI.}
\label{fig:outcomes_IPD_CI}
\end{figure*}

\begin{figure*}[!h]
\centering
\begin{tabular}[t]{|c|cccccc}
\toprule
& Selfish & Utilitarian & Deontological & Virtue - equality & Virtue - kindness & Virtue - mixed \\
\midrule
\makecell[cc]{\rotatebox[origin=c]{90}{\thead{Collective Return}}} & \subt{\includegraphics[height=18mm]{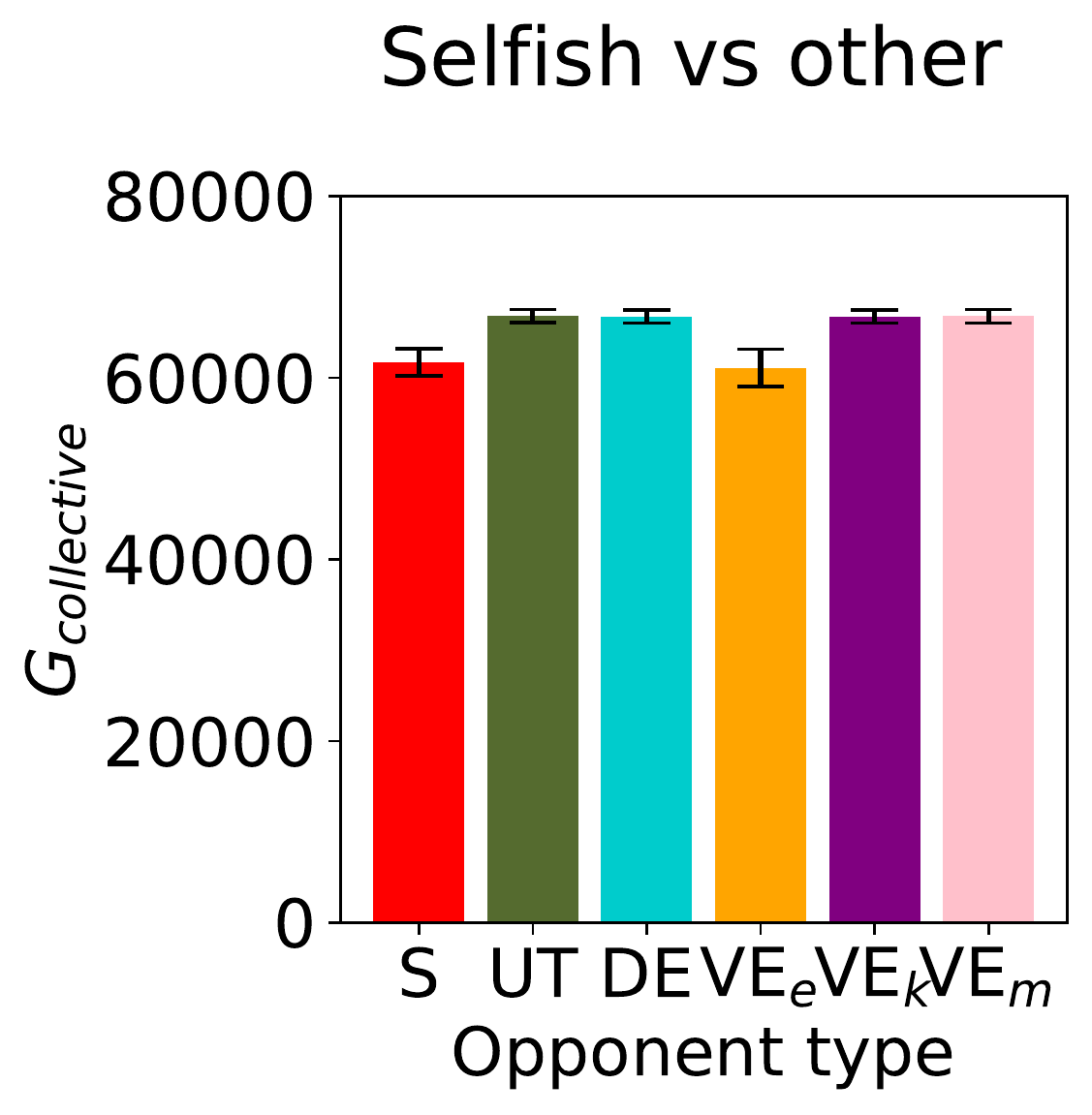}} & \subt{\includegraphics[height=18mm]{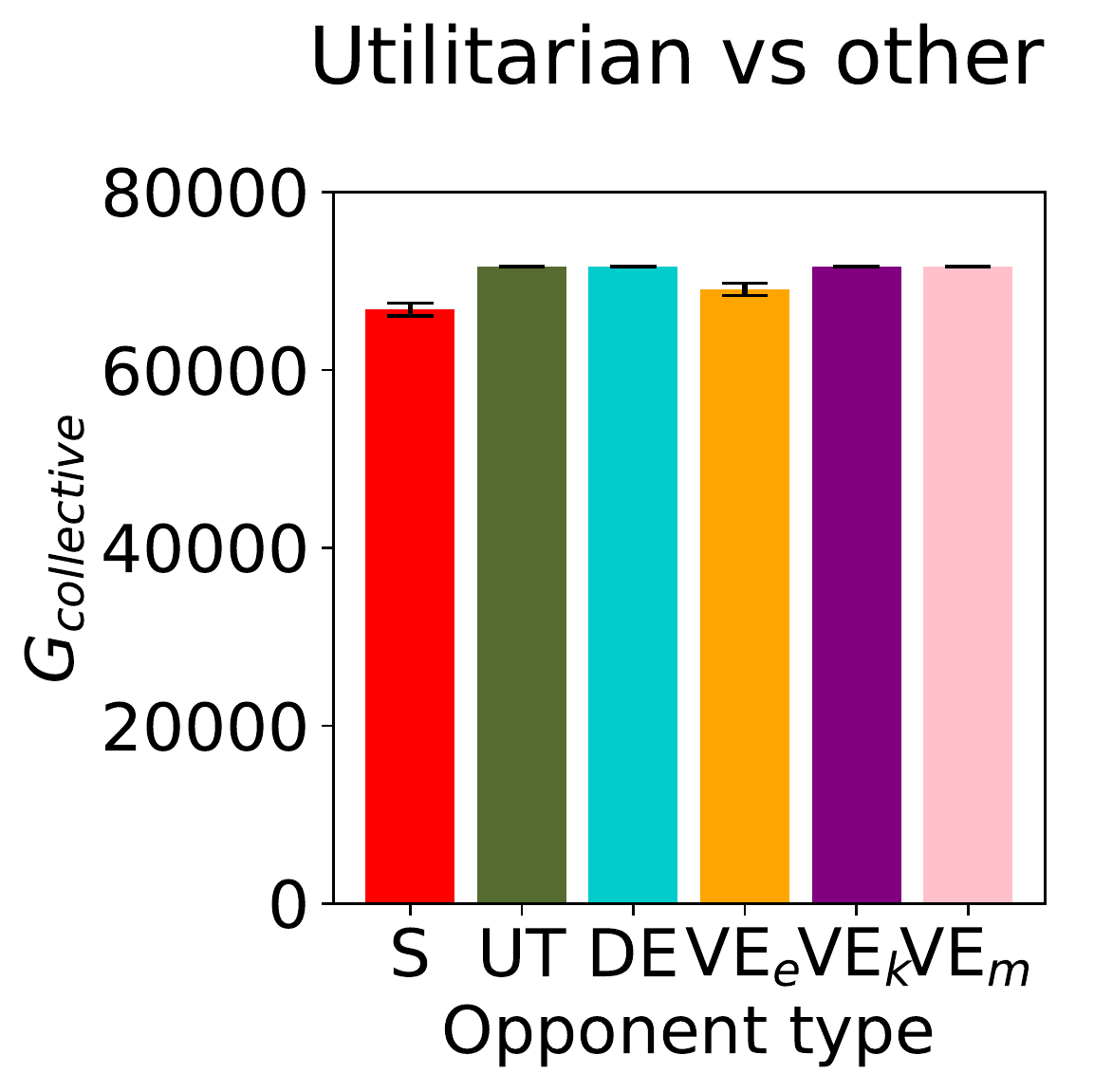}} & \subt{\includegraphics[height=18mm]{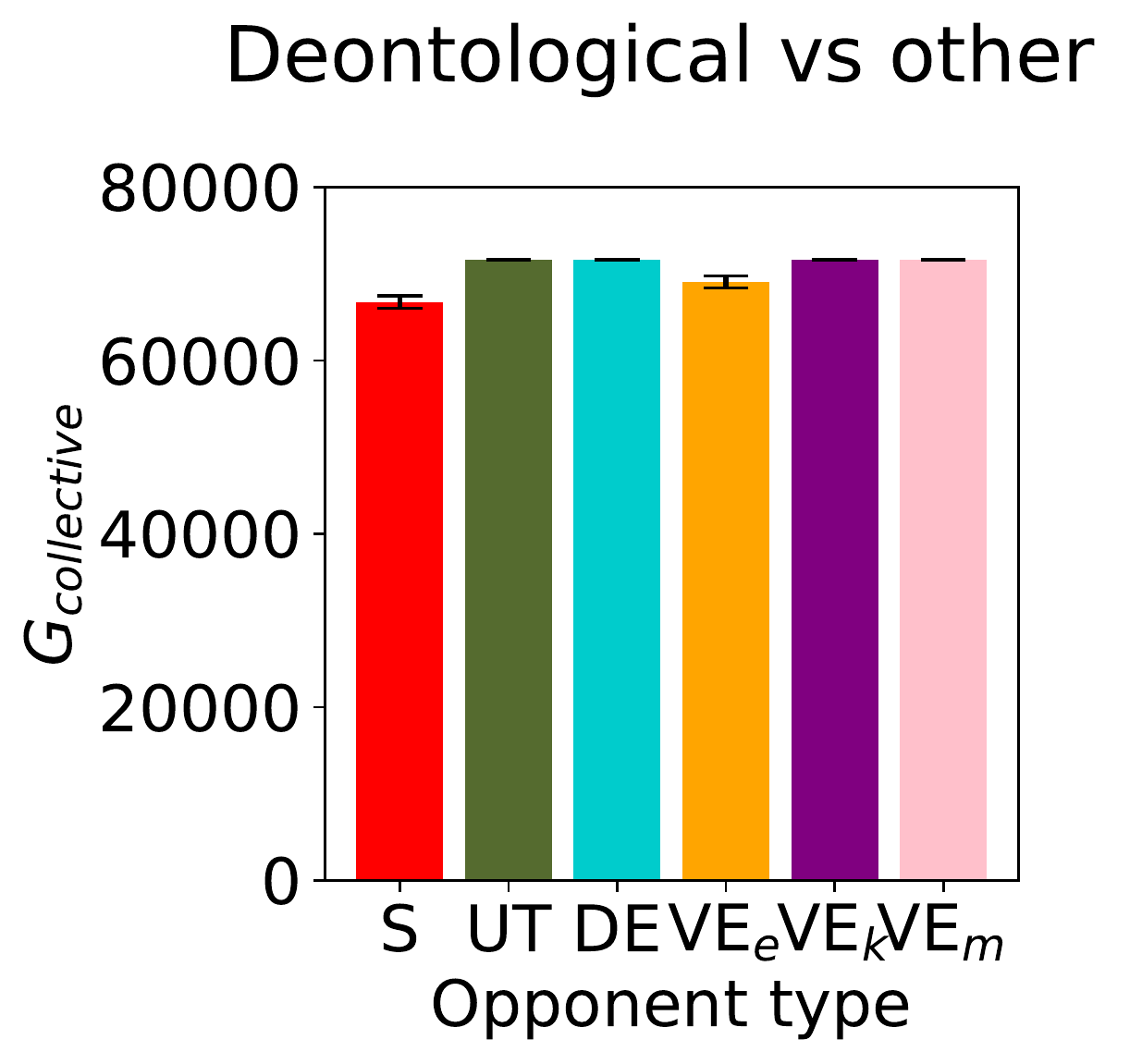}} & \subt{\includegraphics[height=18mm]{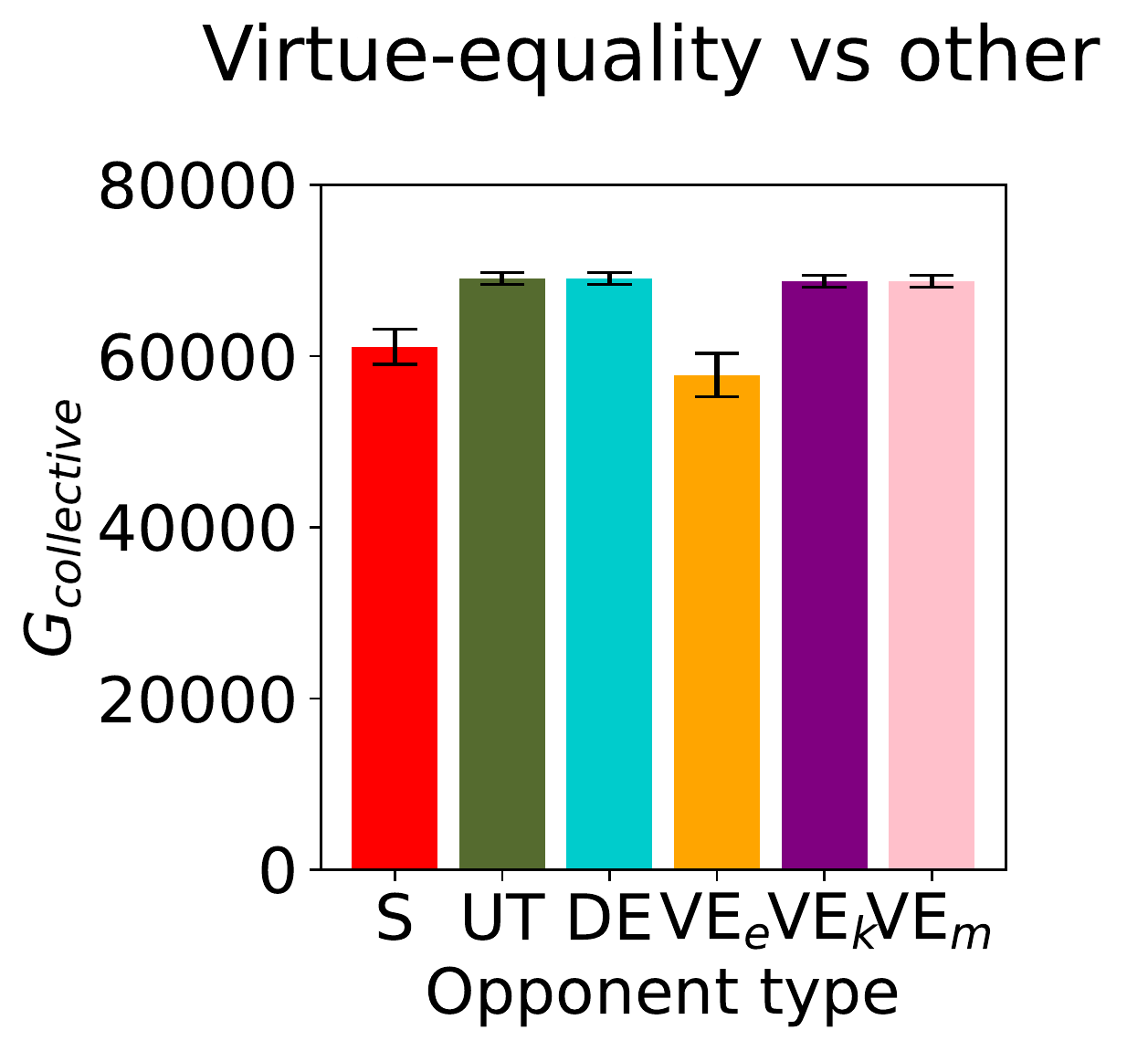}} & \subt{\includegraphics[height=18mm]{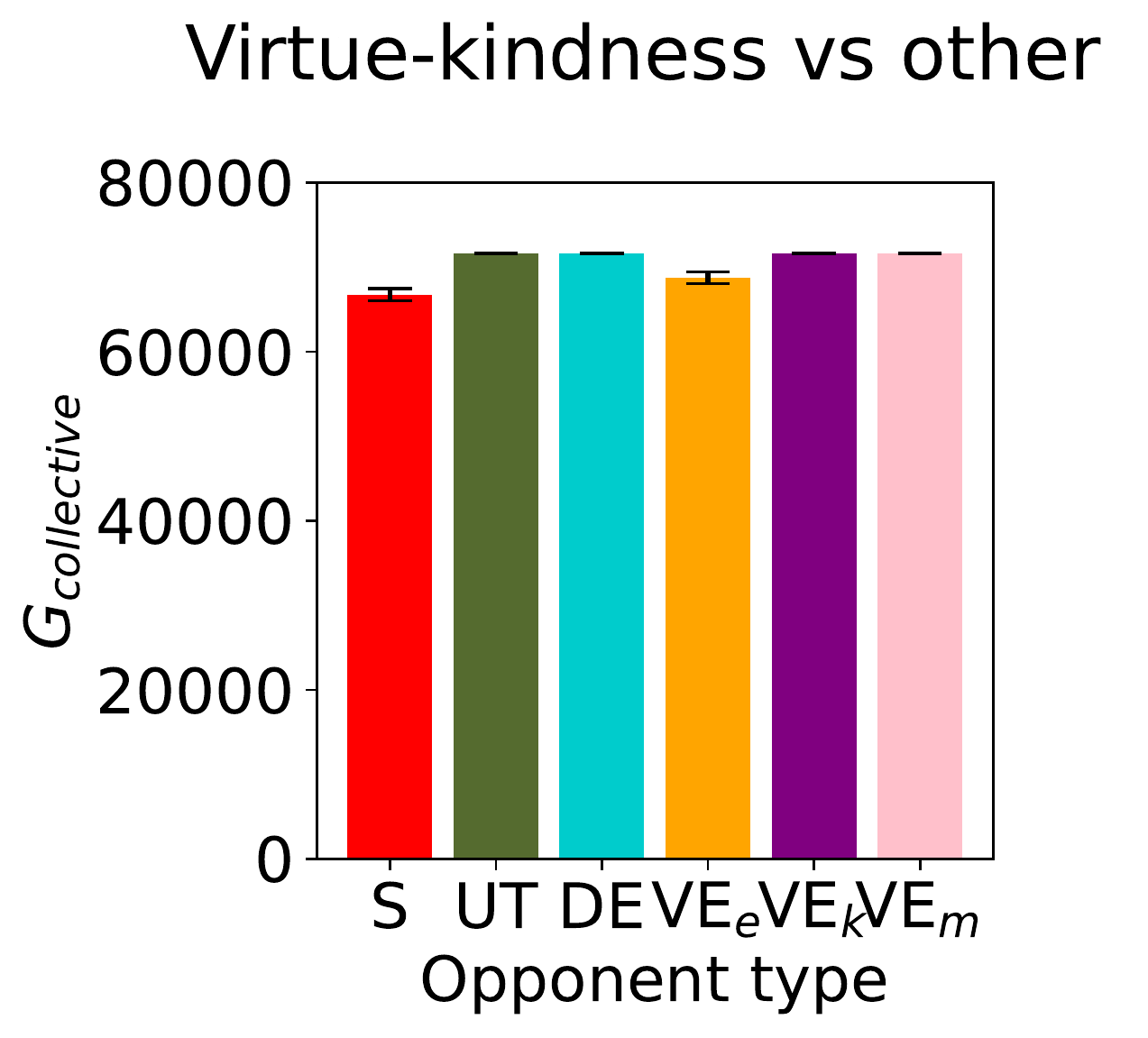}} & \subt{\includegraphics[height=18mm]{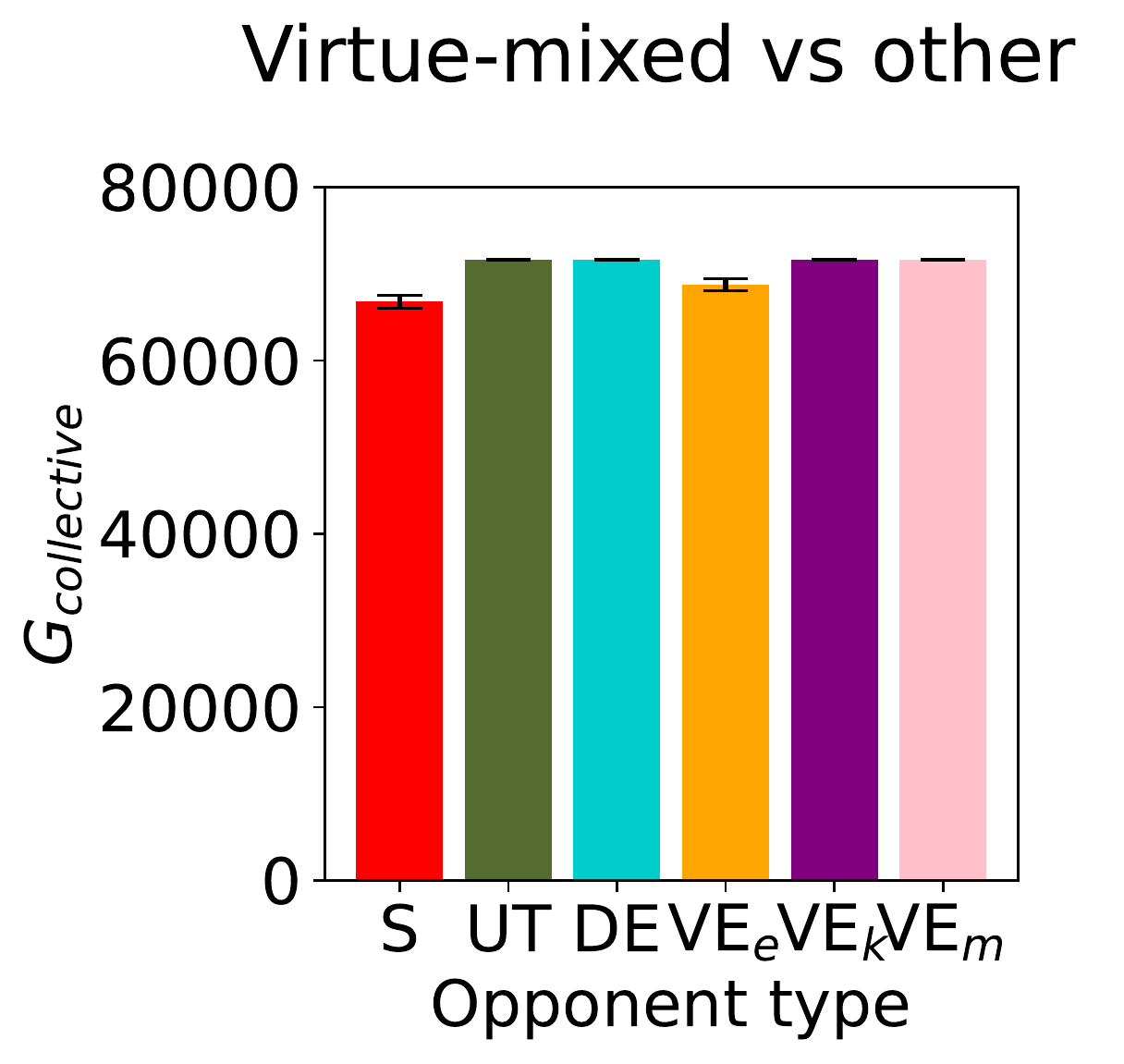}}
\\ 
\makecell[cc]{\rotatebox[origin=c]{90}{\thead{Gini Return}}} & \subt{\includegraphics[height=18mm]{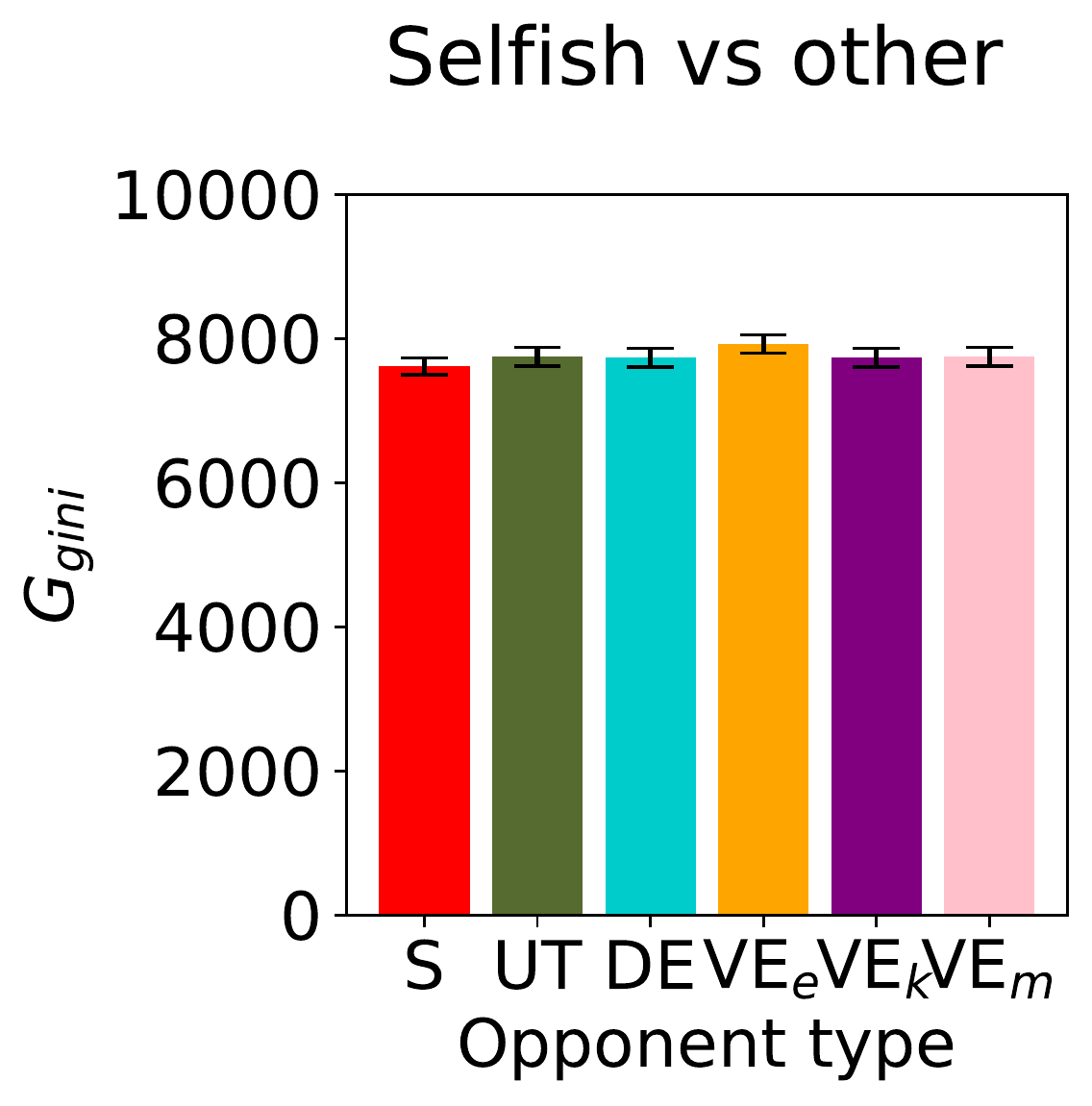}} & \subt{\includegraphics[height=18mm]{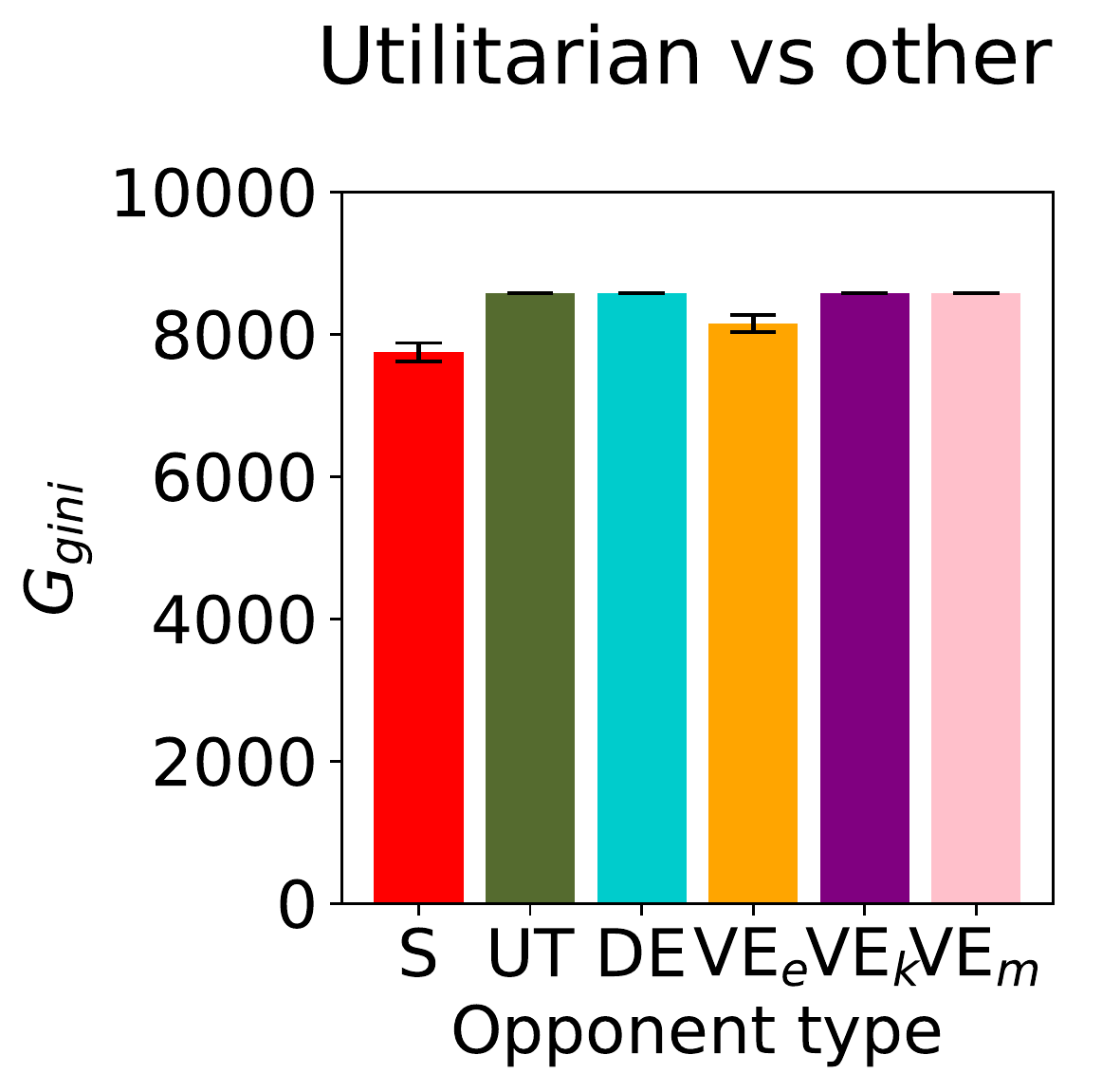}} & \subt{\includegraphics[height=18mm]{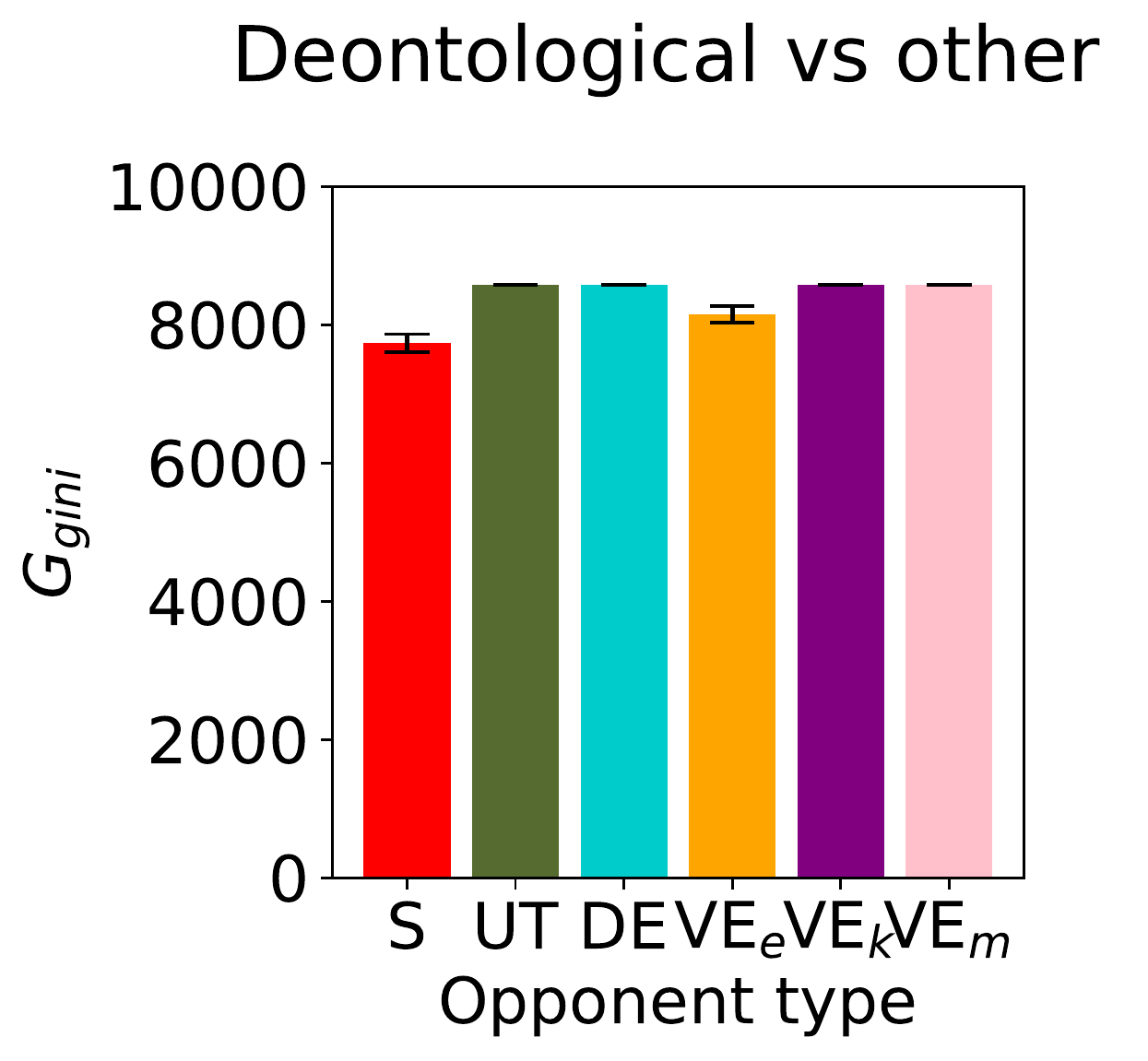}} & \subt{\includegraphics[height=18mm]{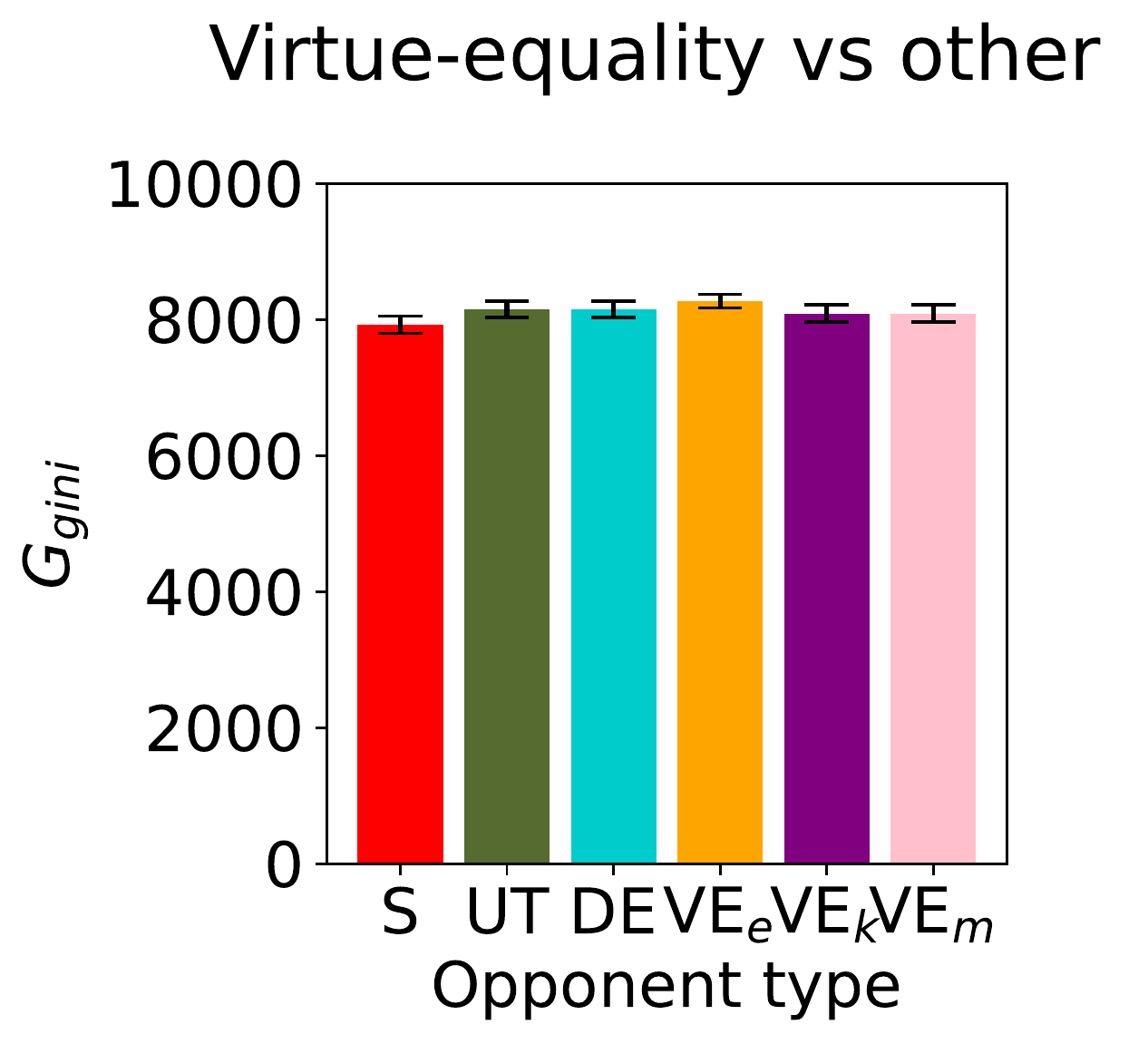}} & \subt{\includegraphics[height=18mm]{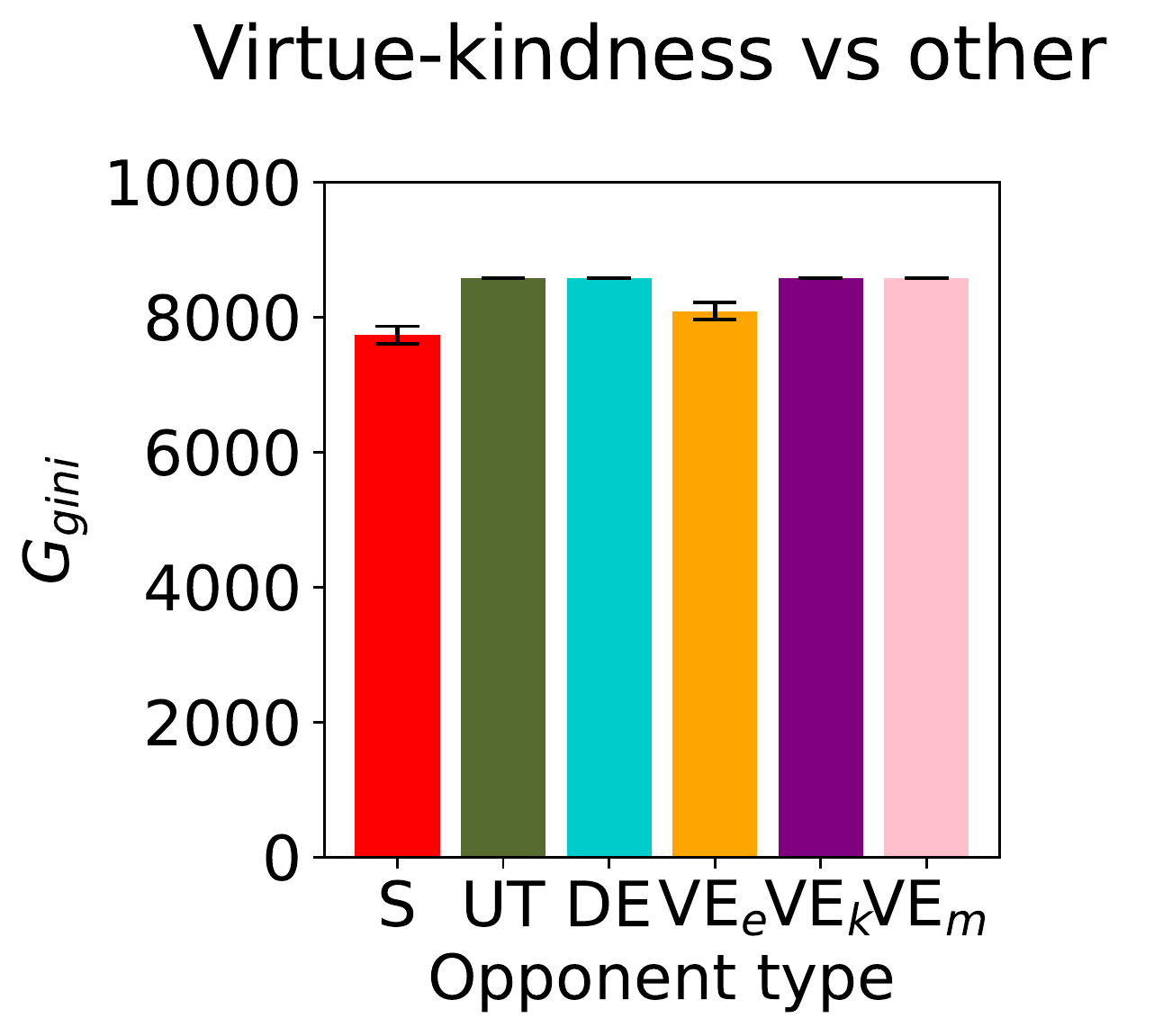}} & \subt{\includegraphics[height=18mm]{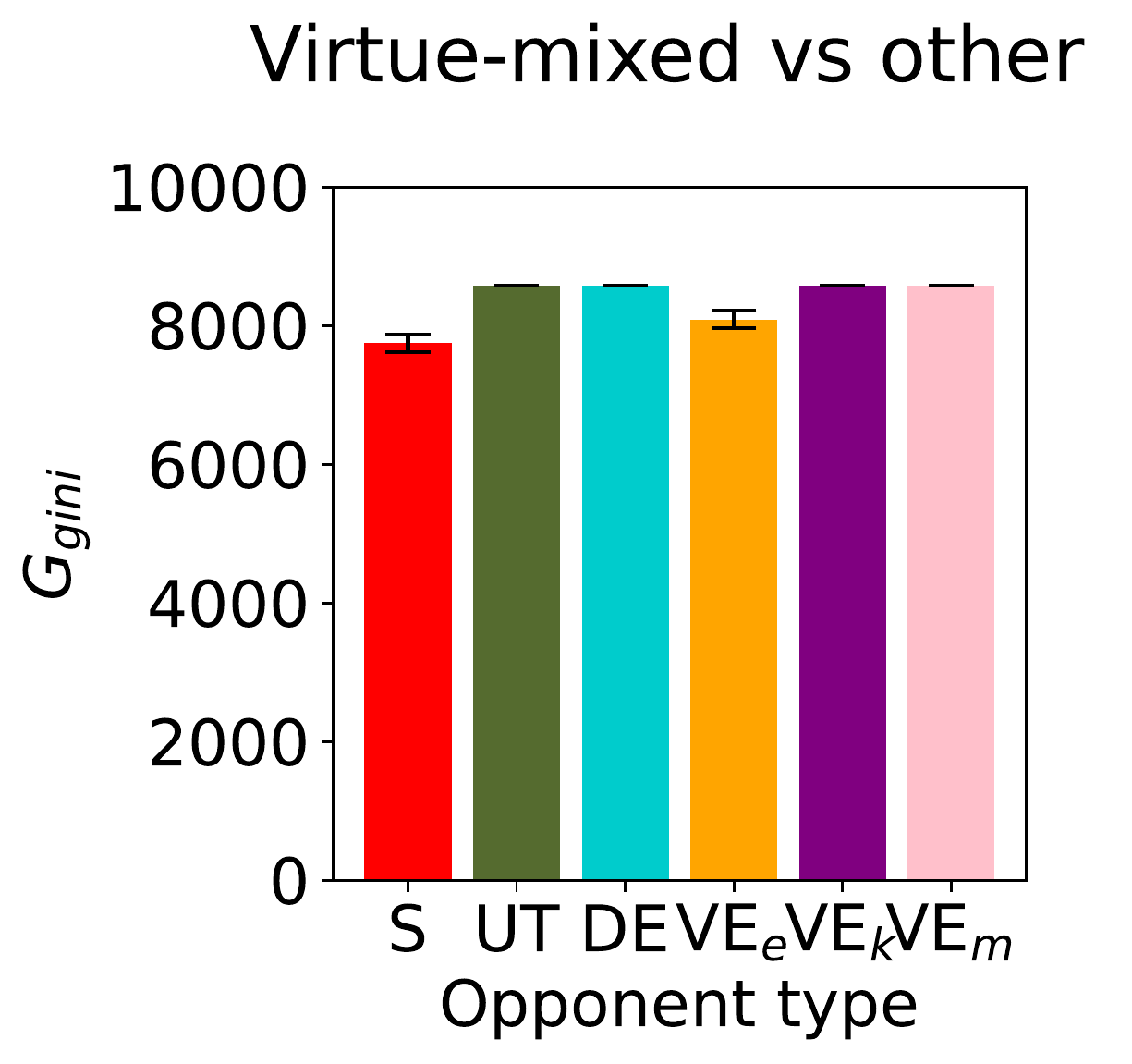}}
\\
\makecell[cc]{\rotatebox[origin=c]{90}{\thead{Min Return}}} & \subt{\includegraphics[height=18mm]{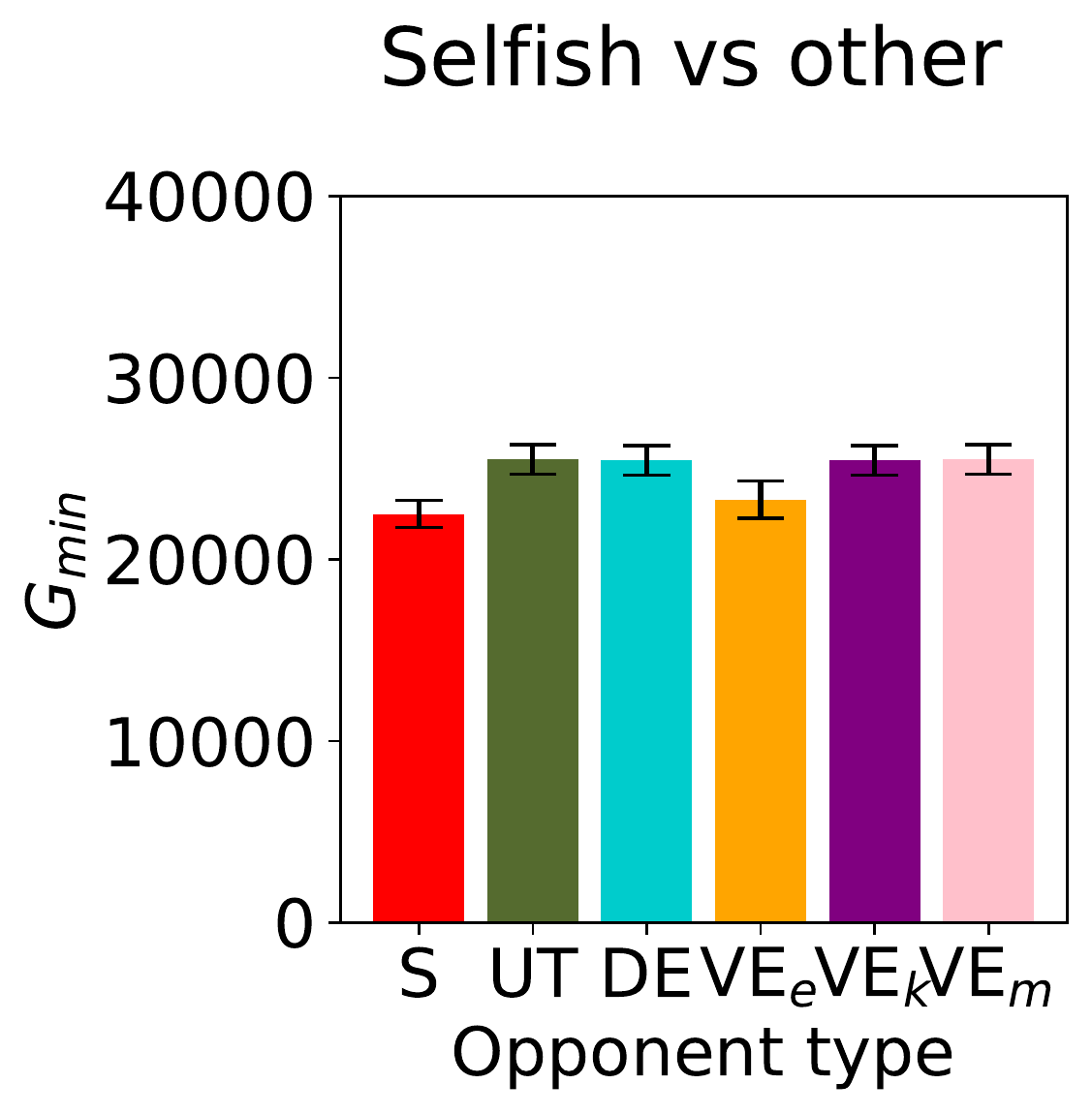}} & \subt{\includegraphics[height=18mm]{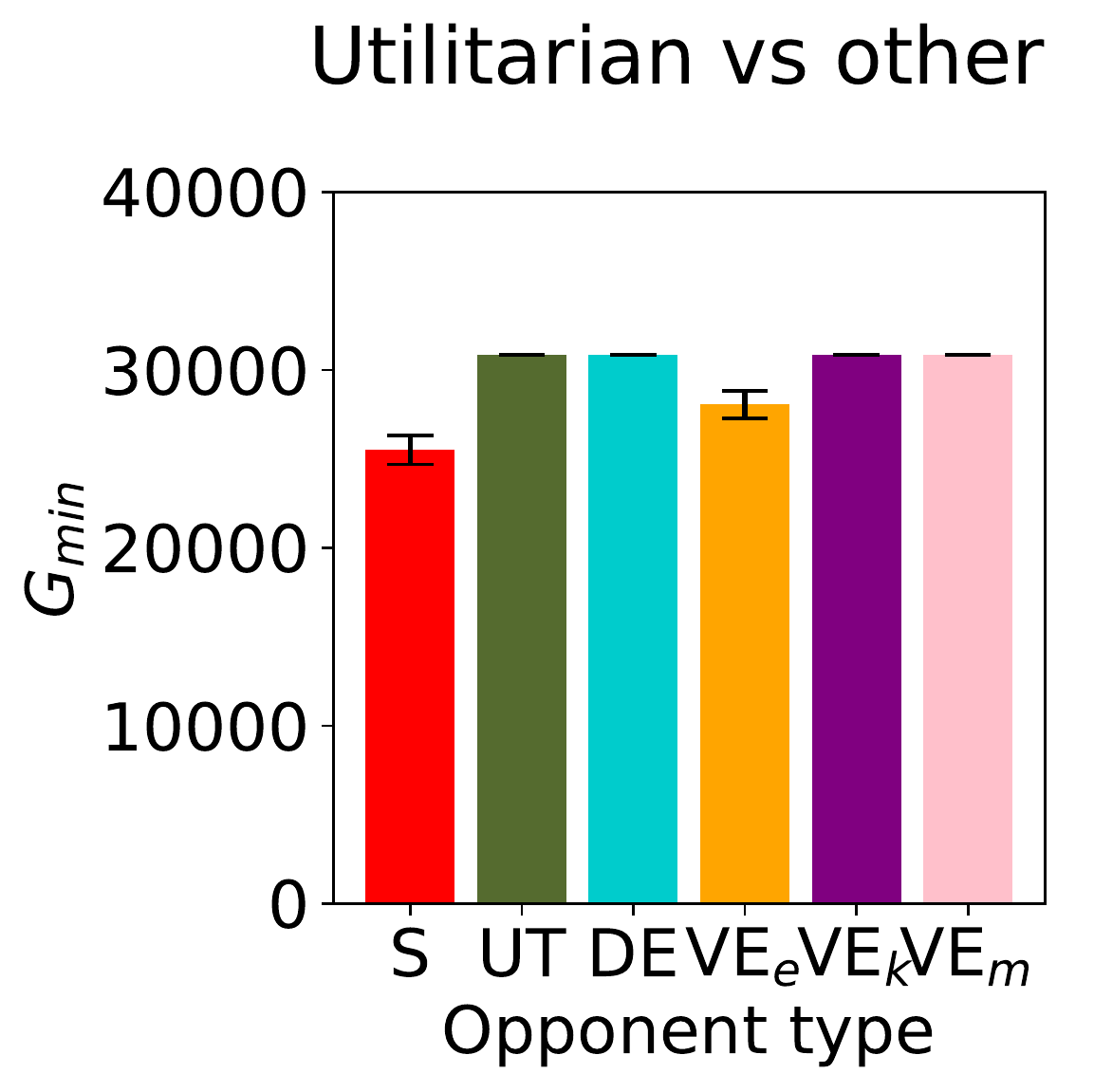}} & \subt{\includegraphics[height=18mm]{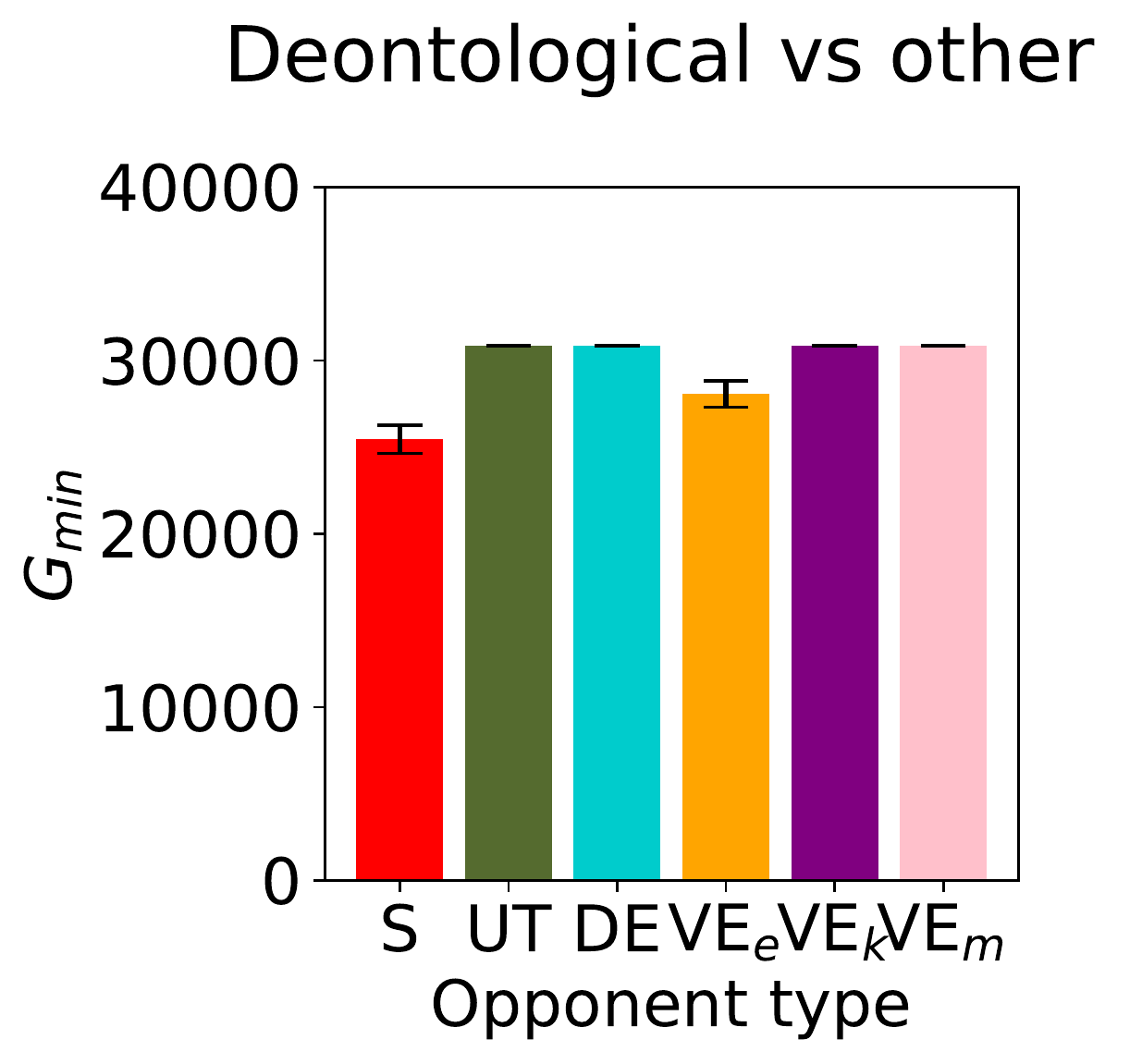}} & \subt{\includegraphics[height=18mm]{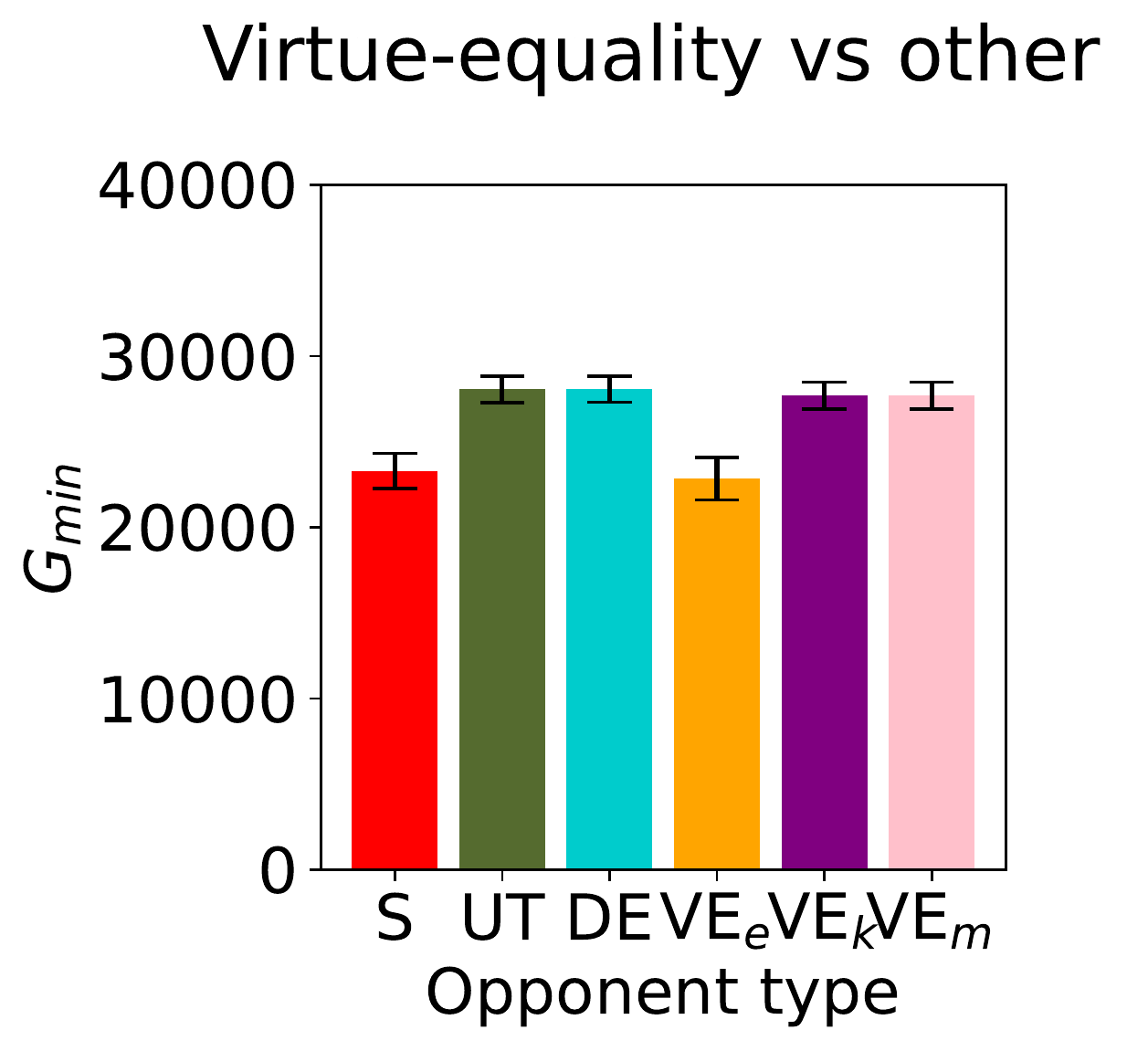}} & \subt{\includegraphics[height=18mm]{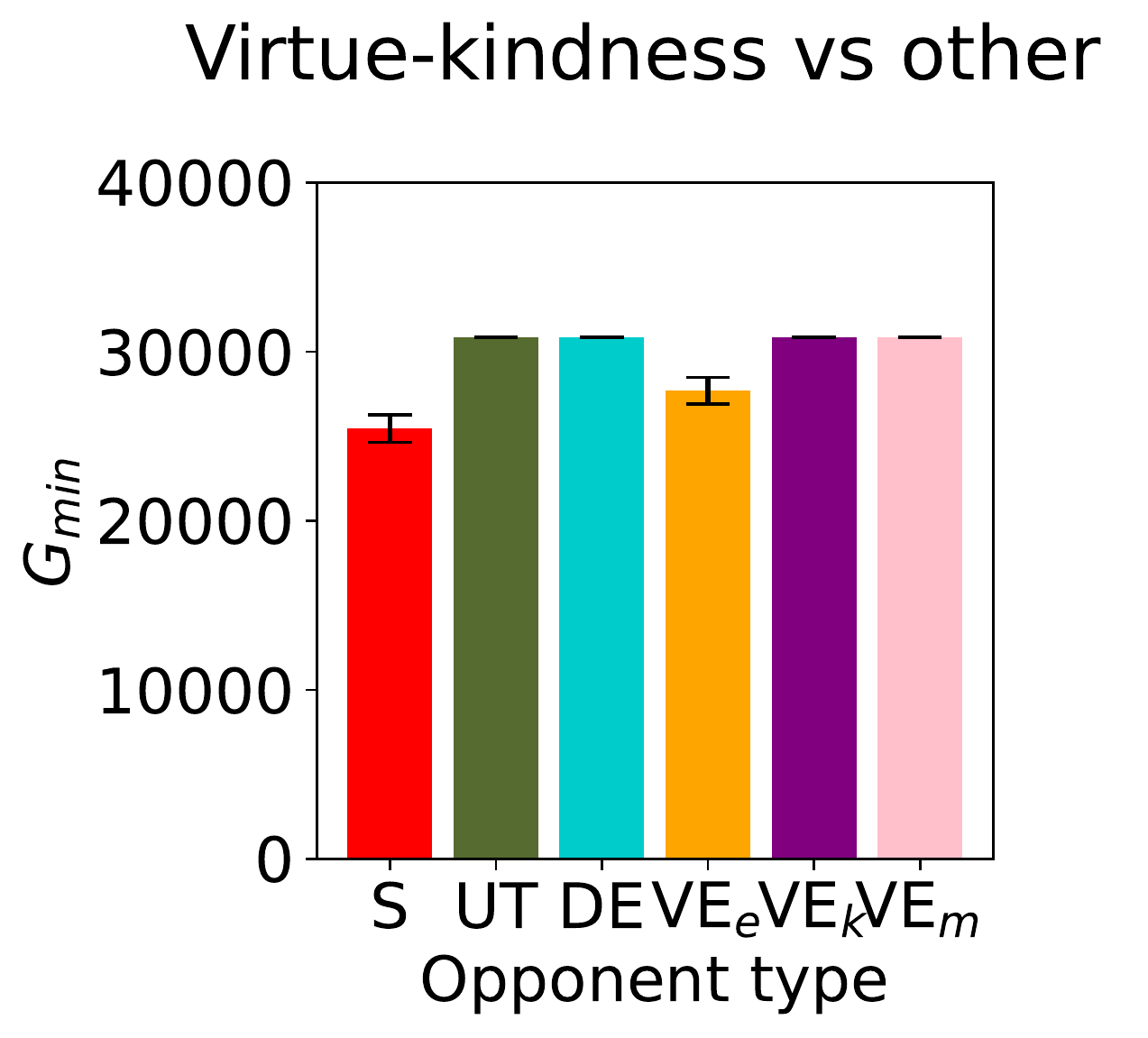}} & \subt{\includegraphics[height=18mm]{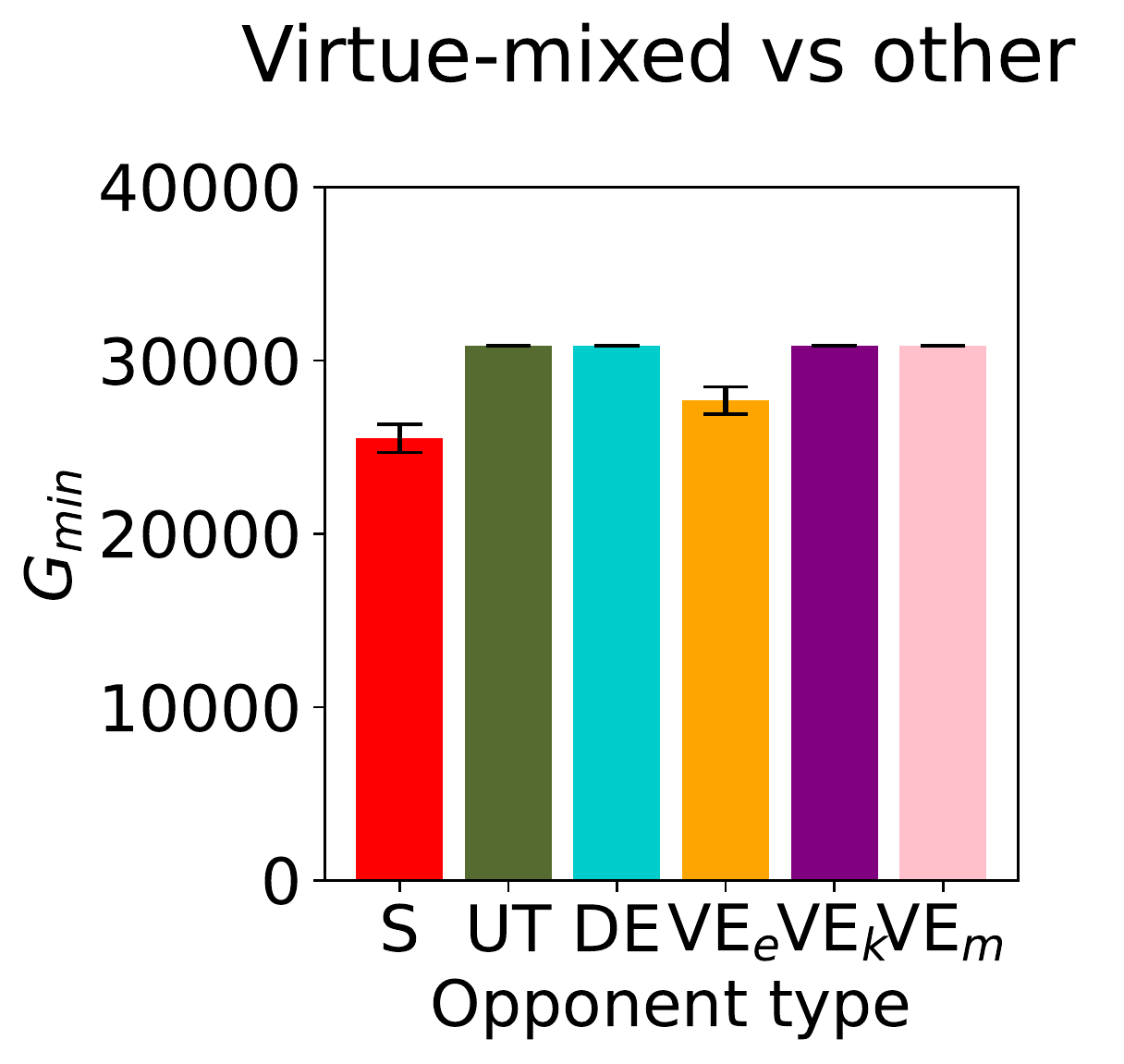}}
\\
\bottomrule
\end{tabular}
\caption{Iterated Volunteer's dilemma game. Relative societal outcomes observed for learning player type $M$ (row) vs. all possible learning opponents $O$. The plots display averages across the 100 runs $\pm$ 95\%CI.}
\label{fig:outcomes_VOL_CI}
\end{figure*}

\begin{figure*}[!h]
\centering
\begin{tabular}[t]{|c|cccccc}
\toprule
& Selfish & Utilitarian & Deontological & Virtue - equality & Virtue - kindness & Virtue - mixed \\
\midrule
\makecell[cc]{\rotatebox[origin=c]{90}{\thead{Collective Return}}} & \subt{\includegraphics[height=18mm]{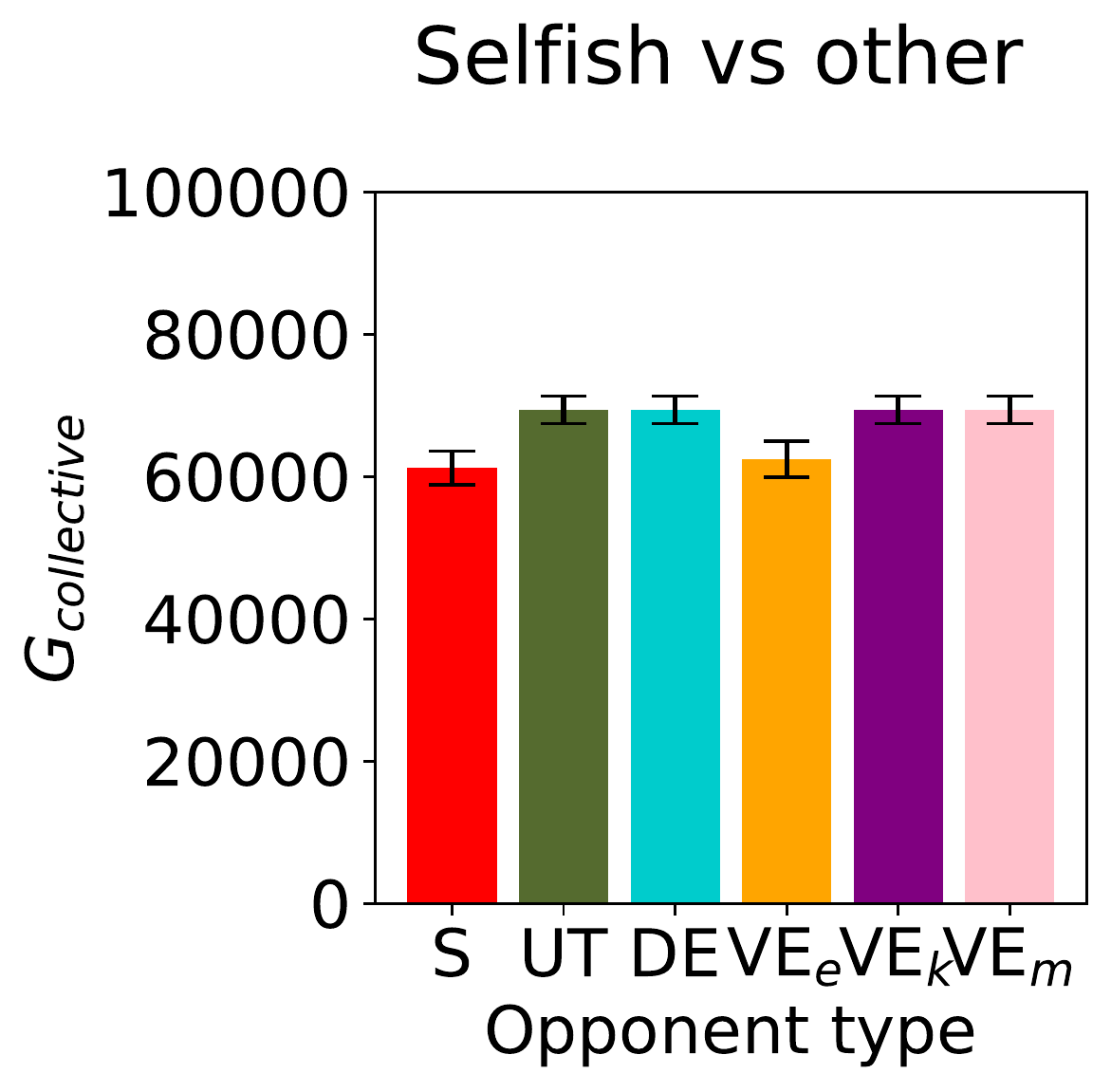}} & \subt{\includegraphics[height=18mm]{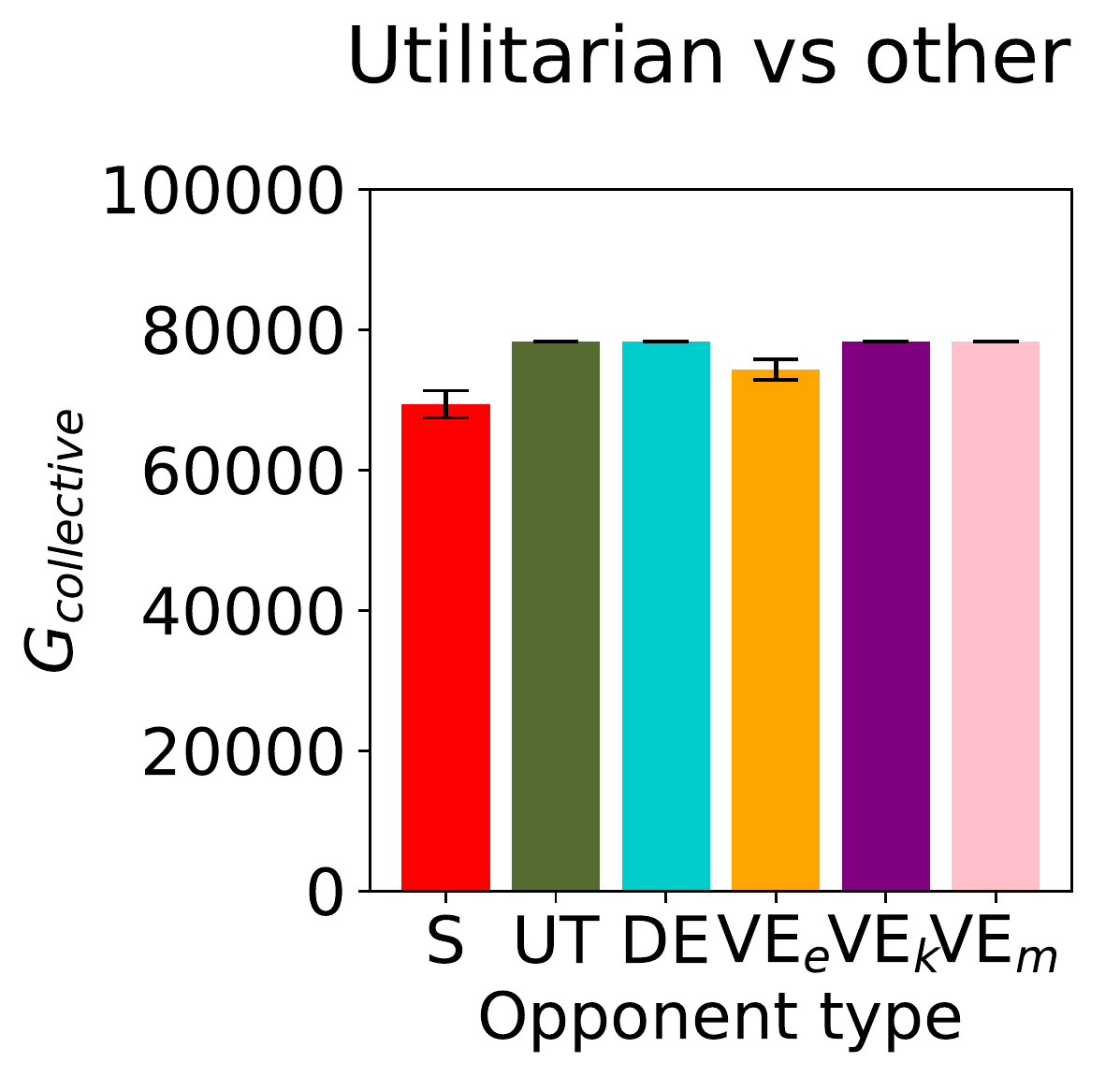}} & \subt{\includegraphics[height=18mm]{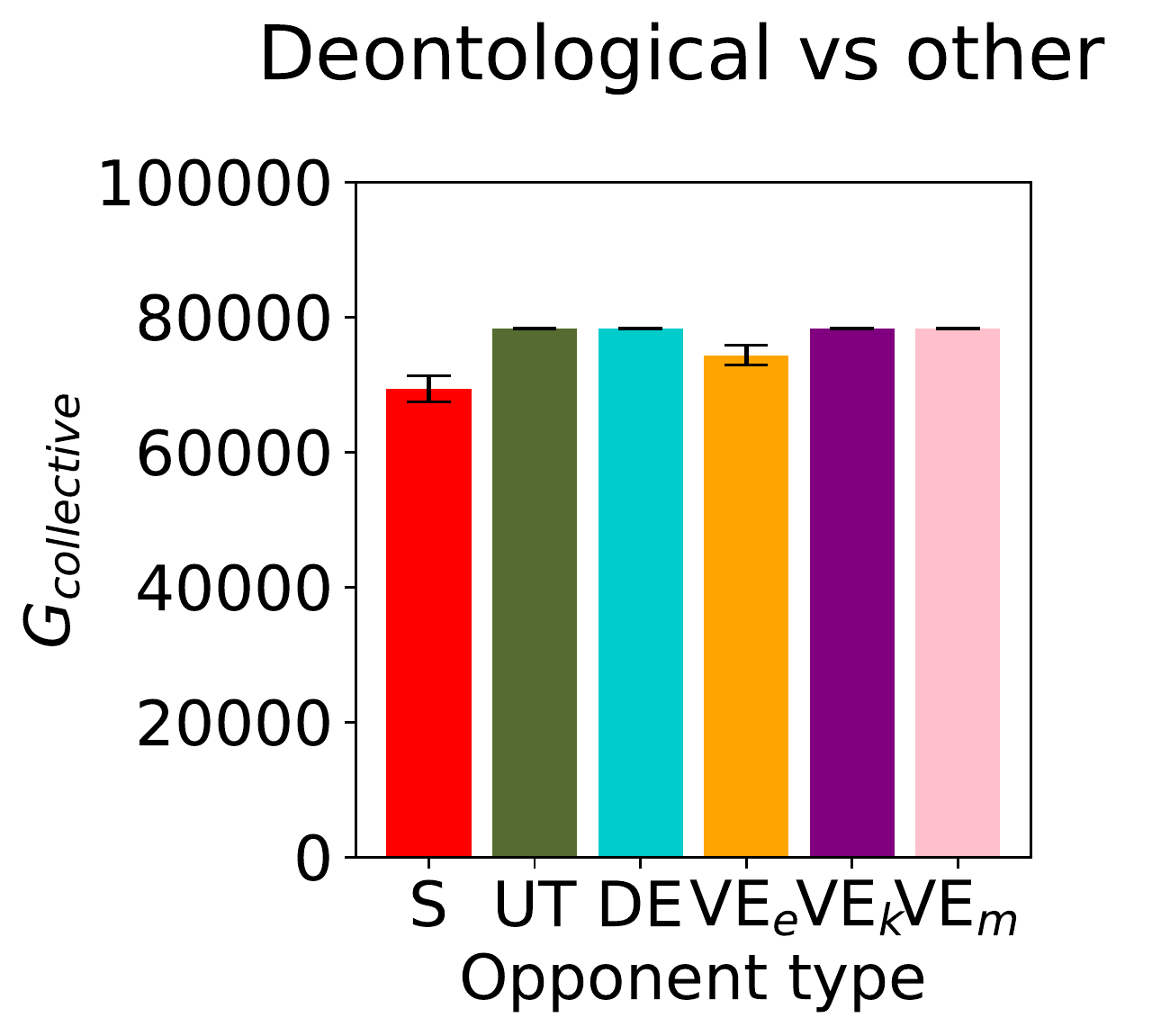}} & \subt{\includegraphics[height=18mm]{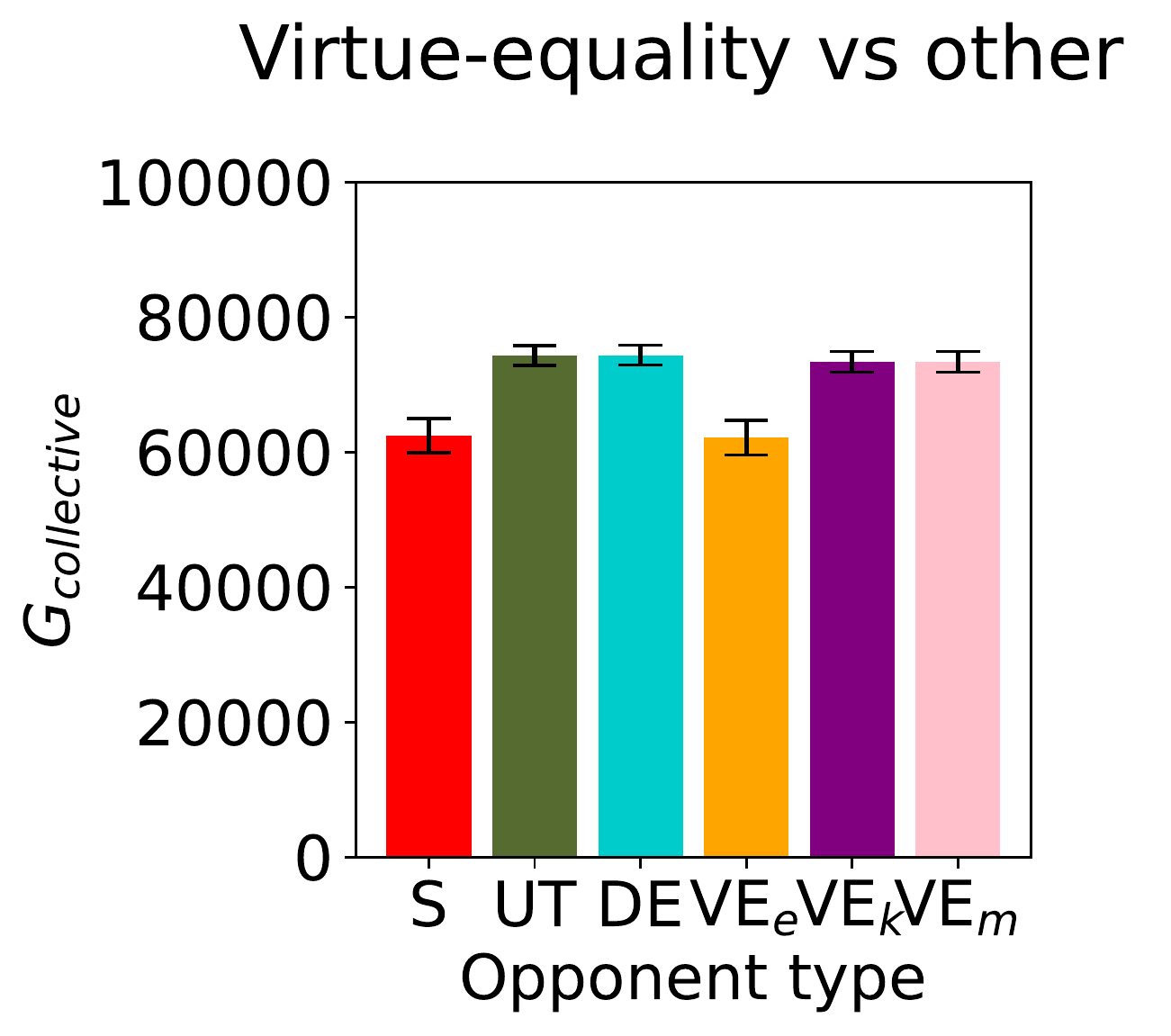}} & \subt{\includegraphics[height=18mm]{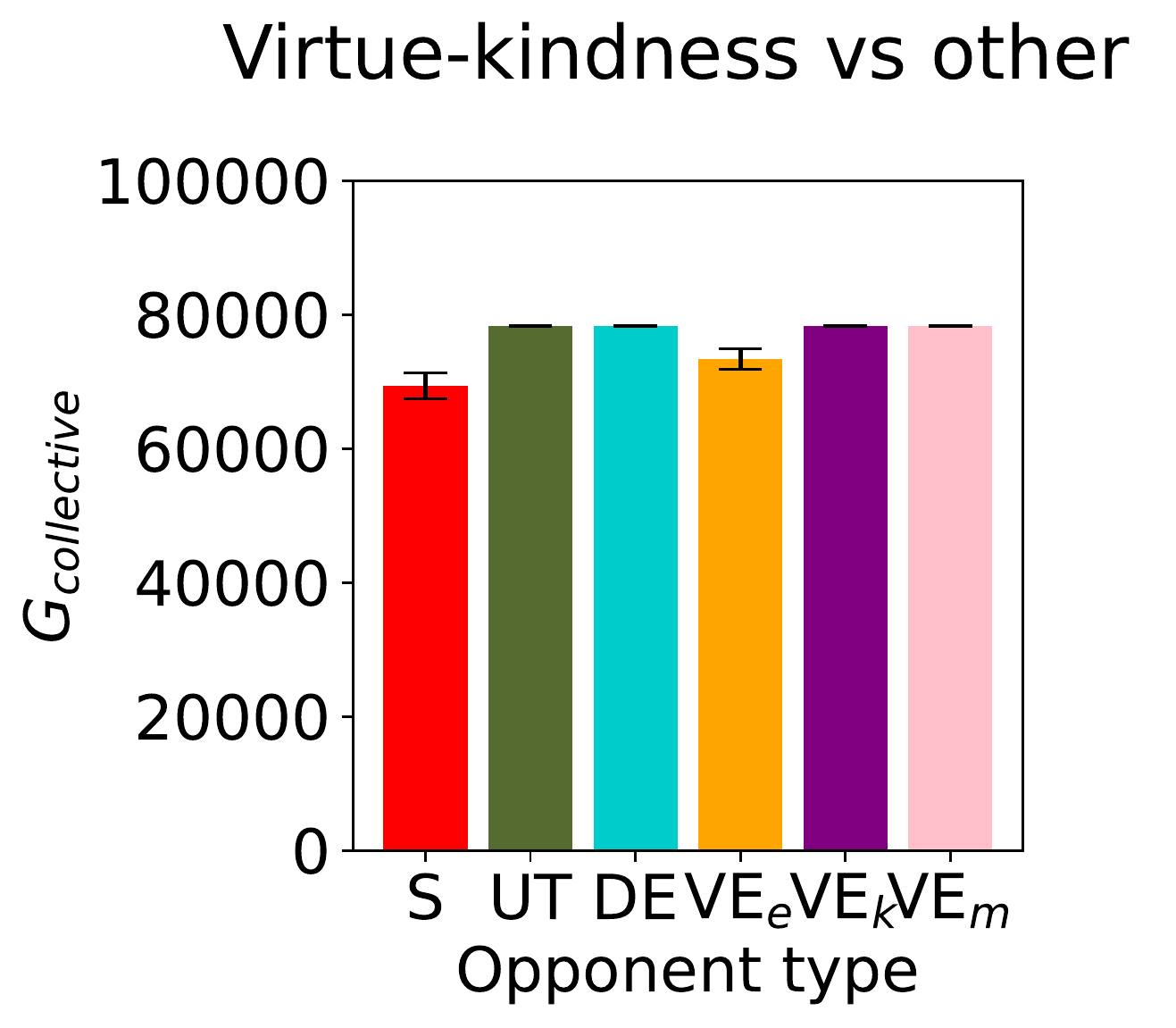}} & \subt{\includegraphics[height=18mm]{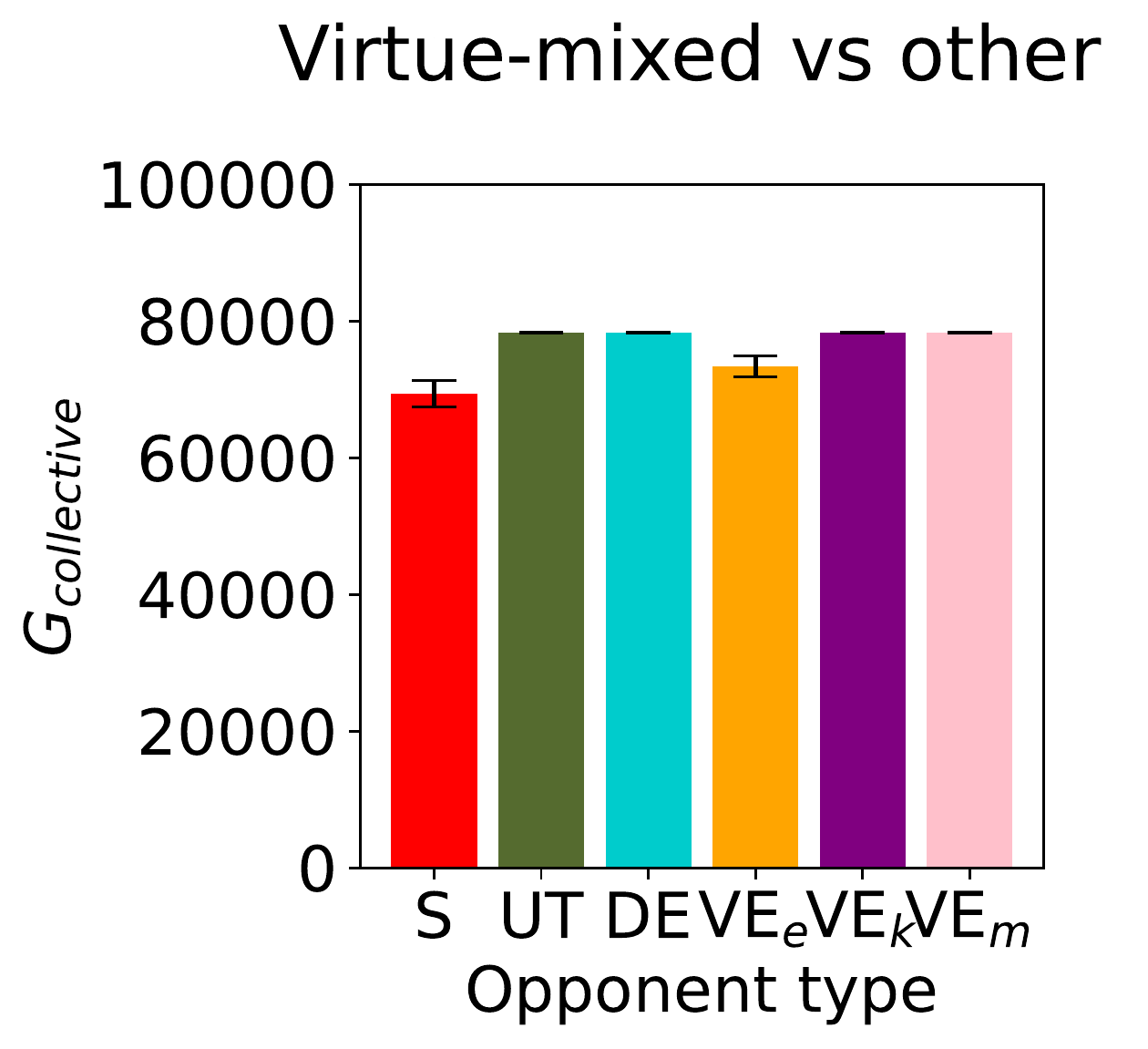}}
\\ 
\makecell[cc]{\rotatebox[origin=c]{90}{\thead{Gini Return}}} & \subt{\includegraphics[height=18mm]{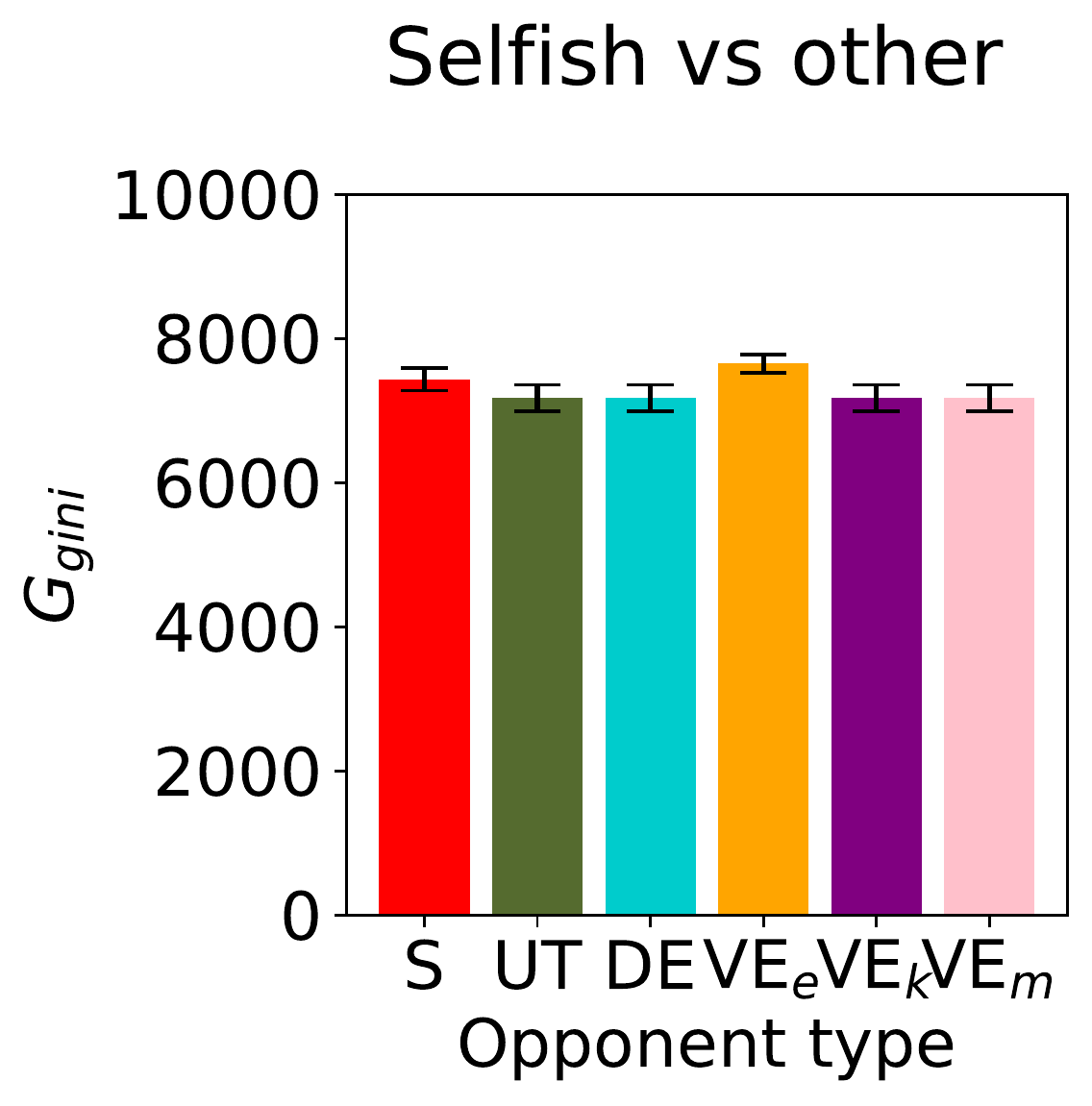}} & \subt{\includegraphics[height=18mm]{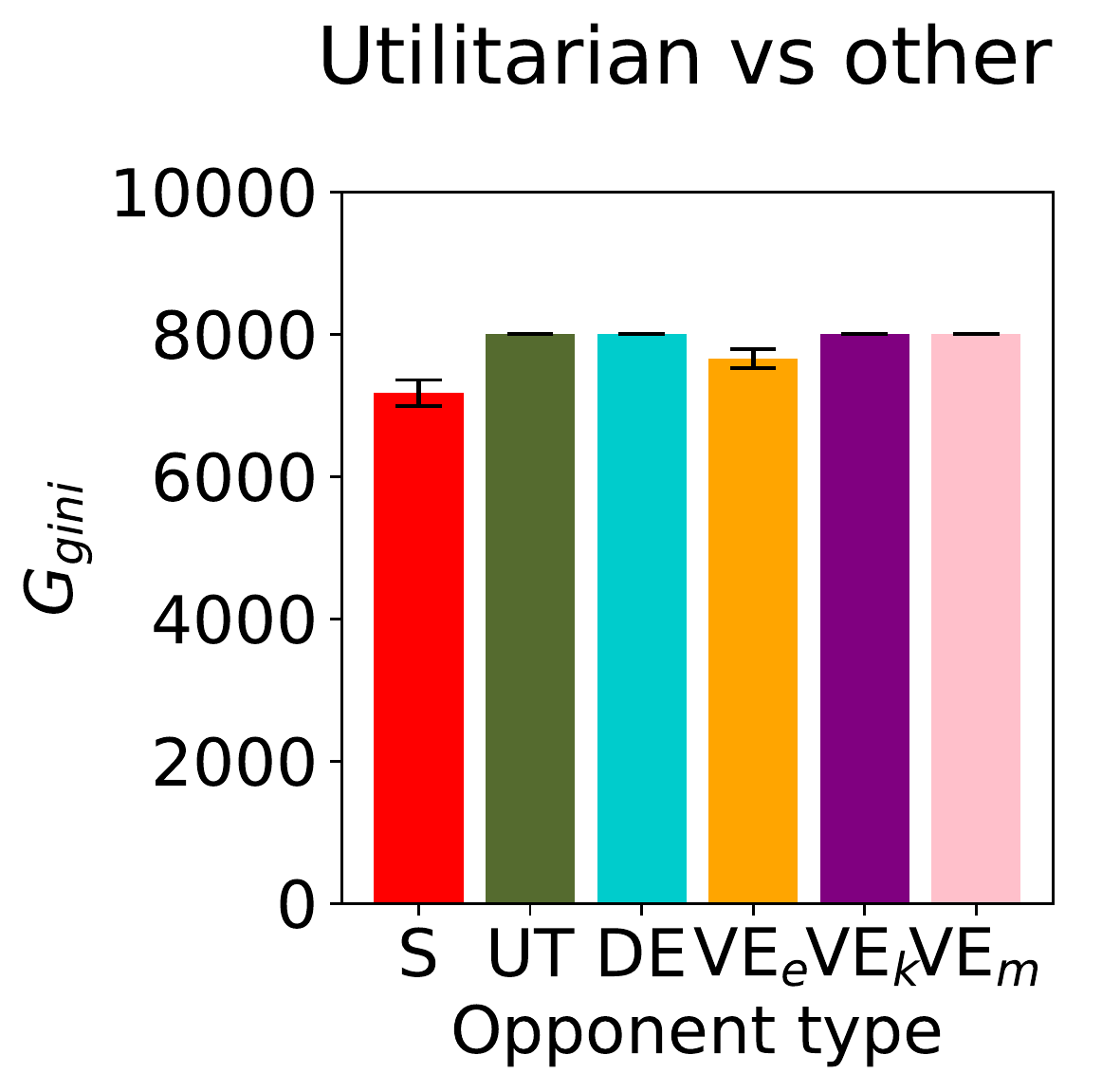}} & \subt{\includegraphics[height=18mm]{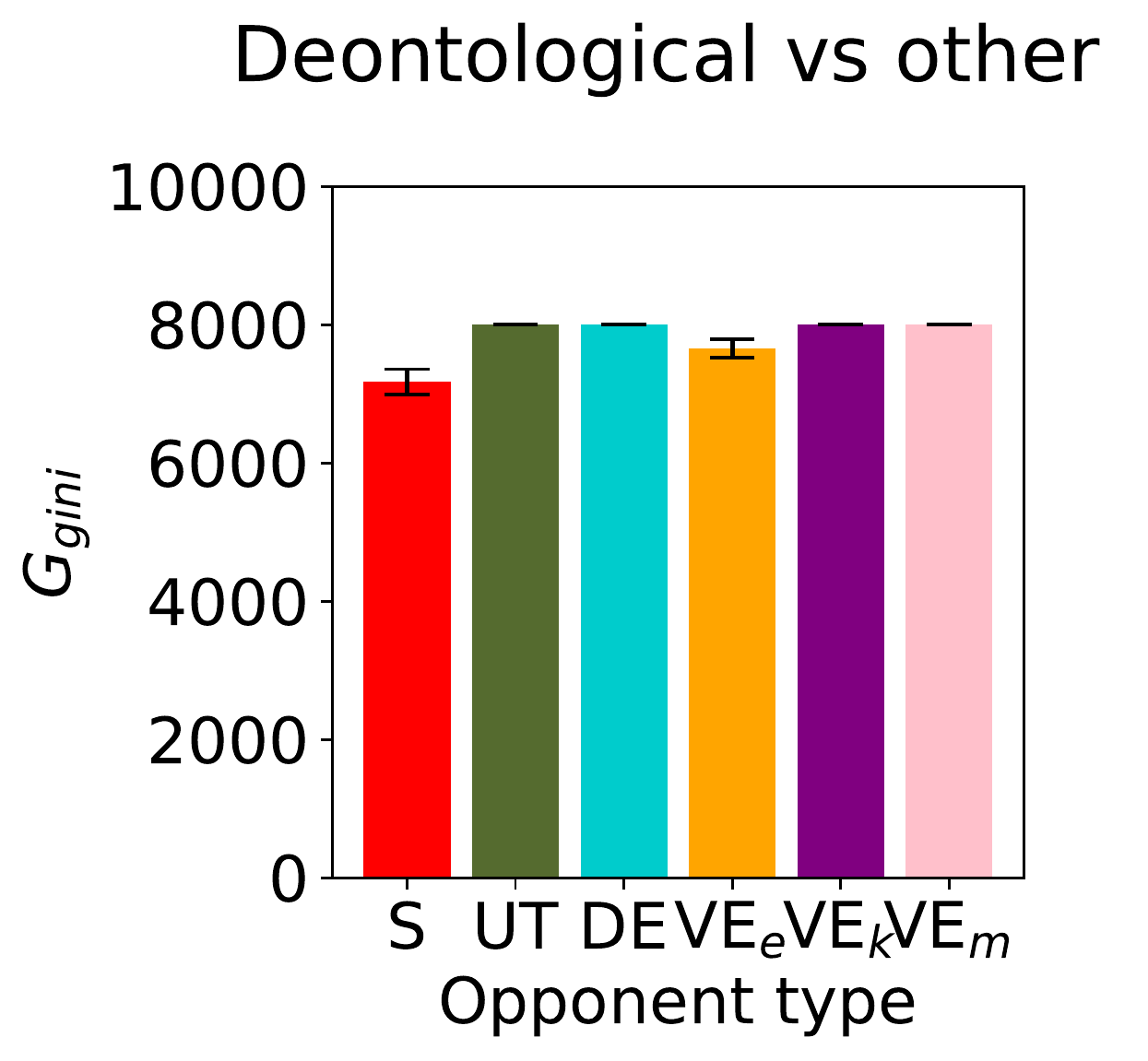}} & \subt{\includegraphics[height=18mm]{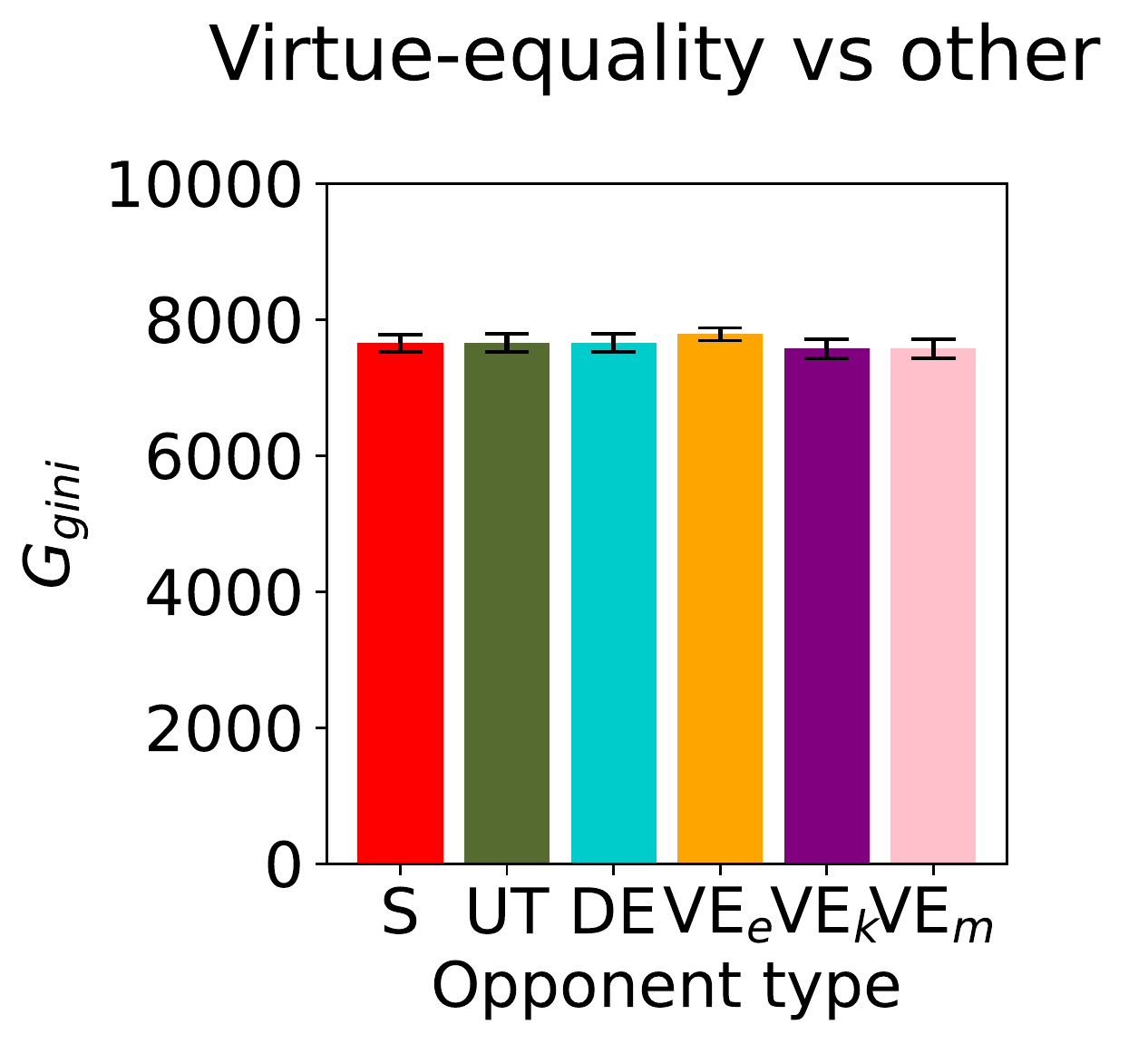}} & \subt{\includegraphics[height=18mm]{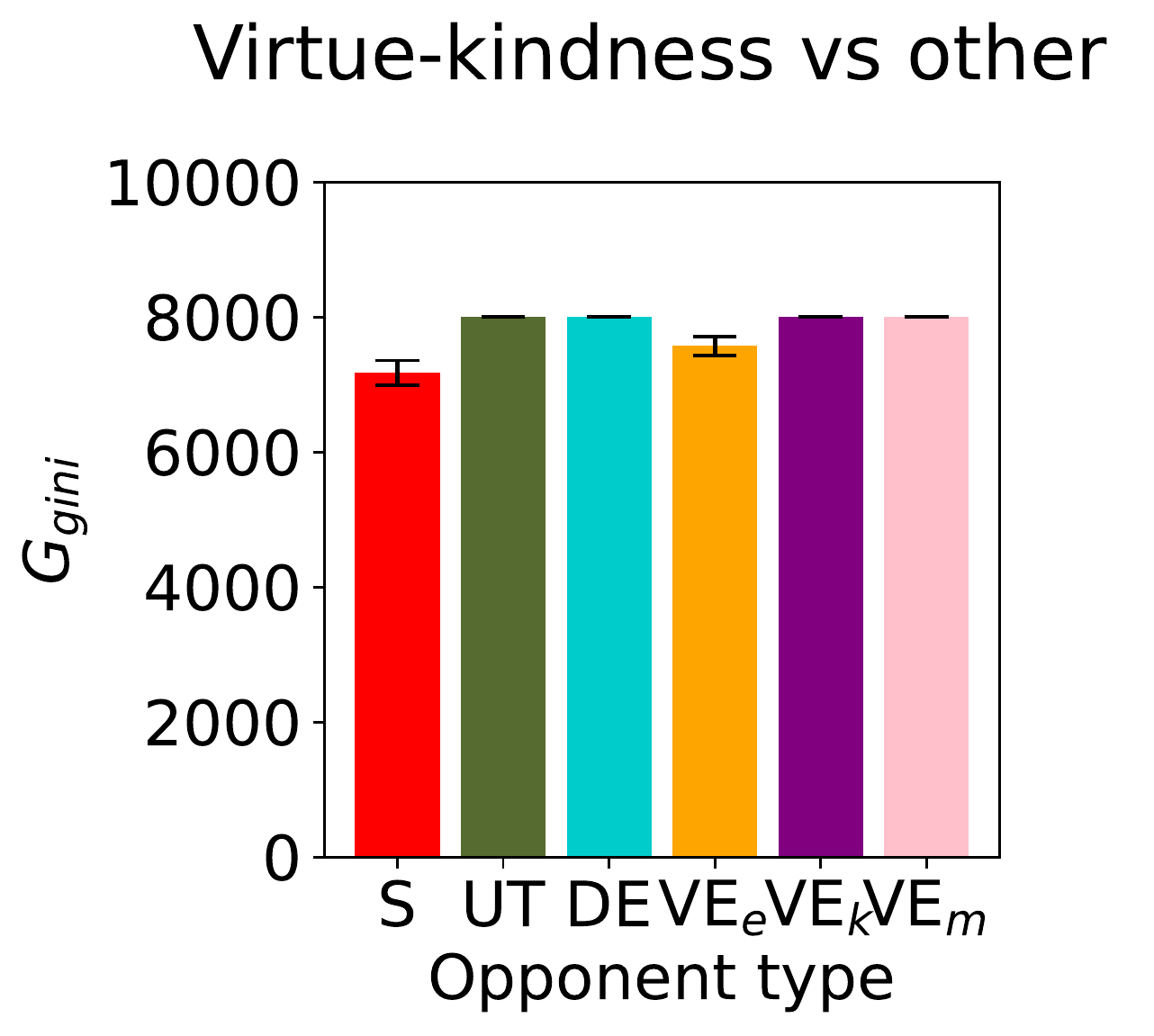}} & \subt{\includegraphics[height=18mm]{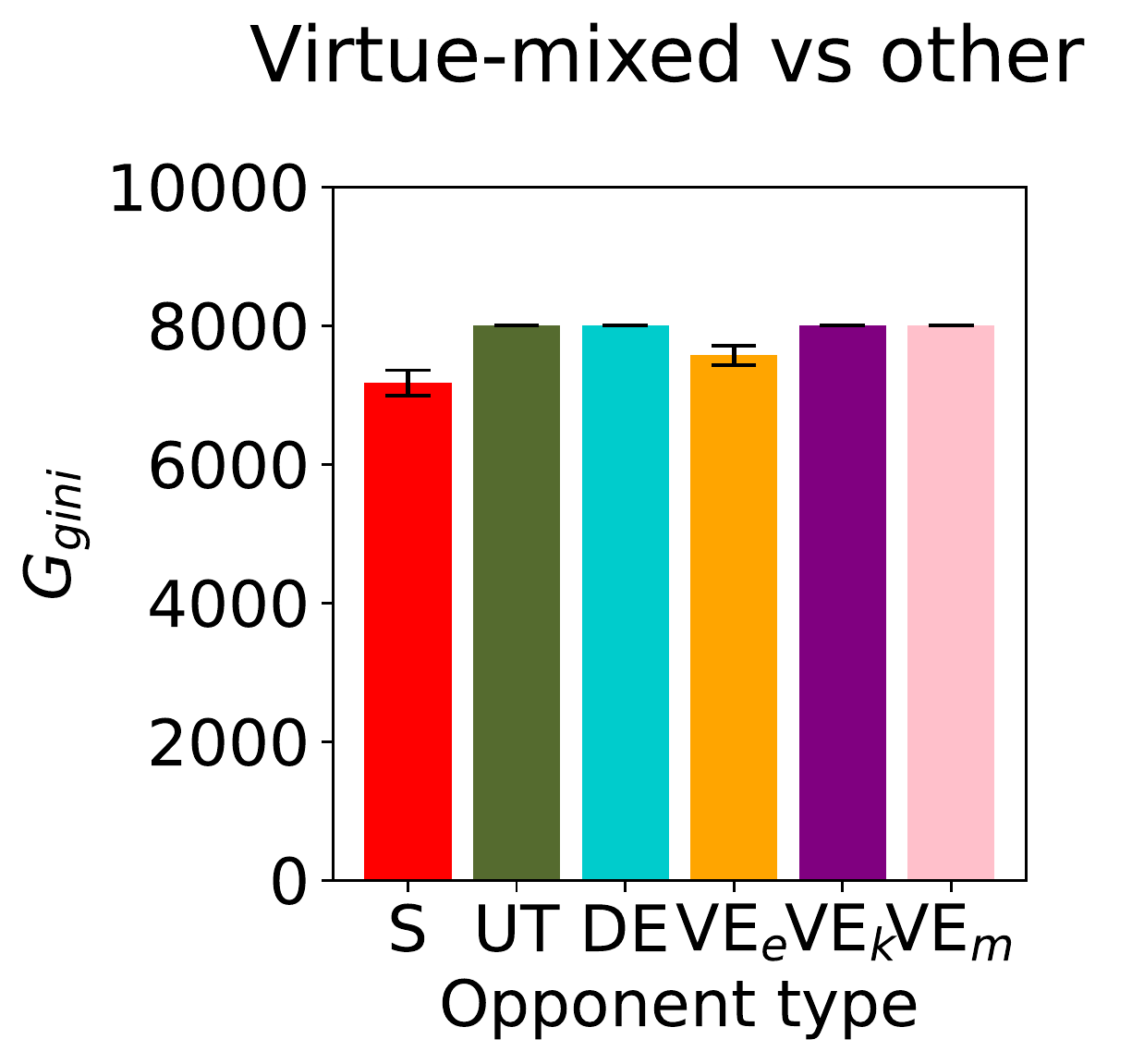}}
\\
\makecell[cc]{\rotatebox[origin=c]{90}{\thead{Min Return}}} & \subt{\includegraphics[height=18mm]{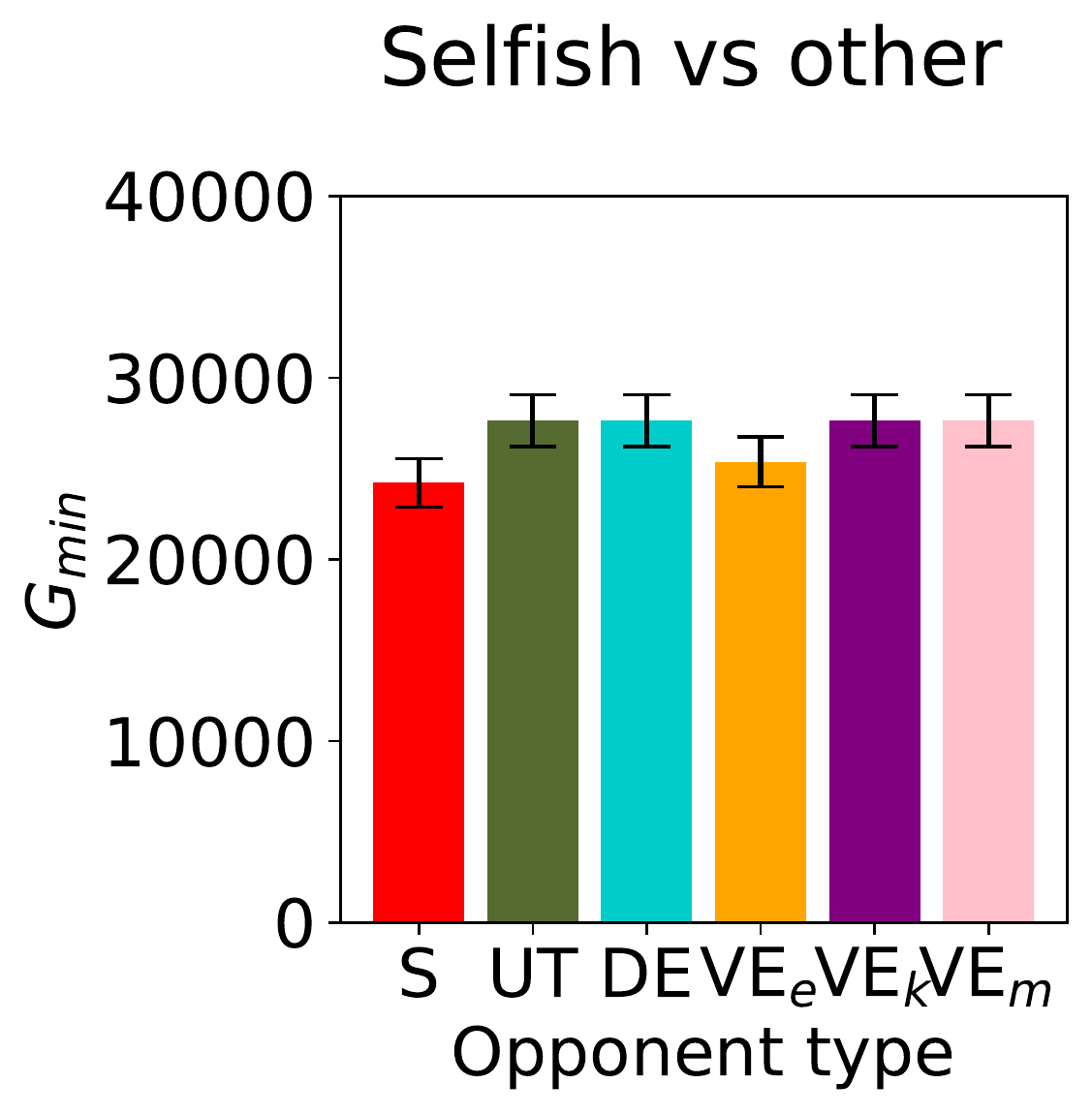}} & \subt{\includegraphics[height=18mm]{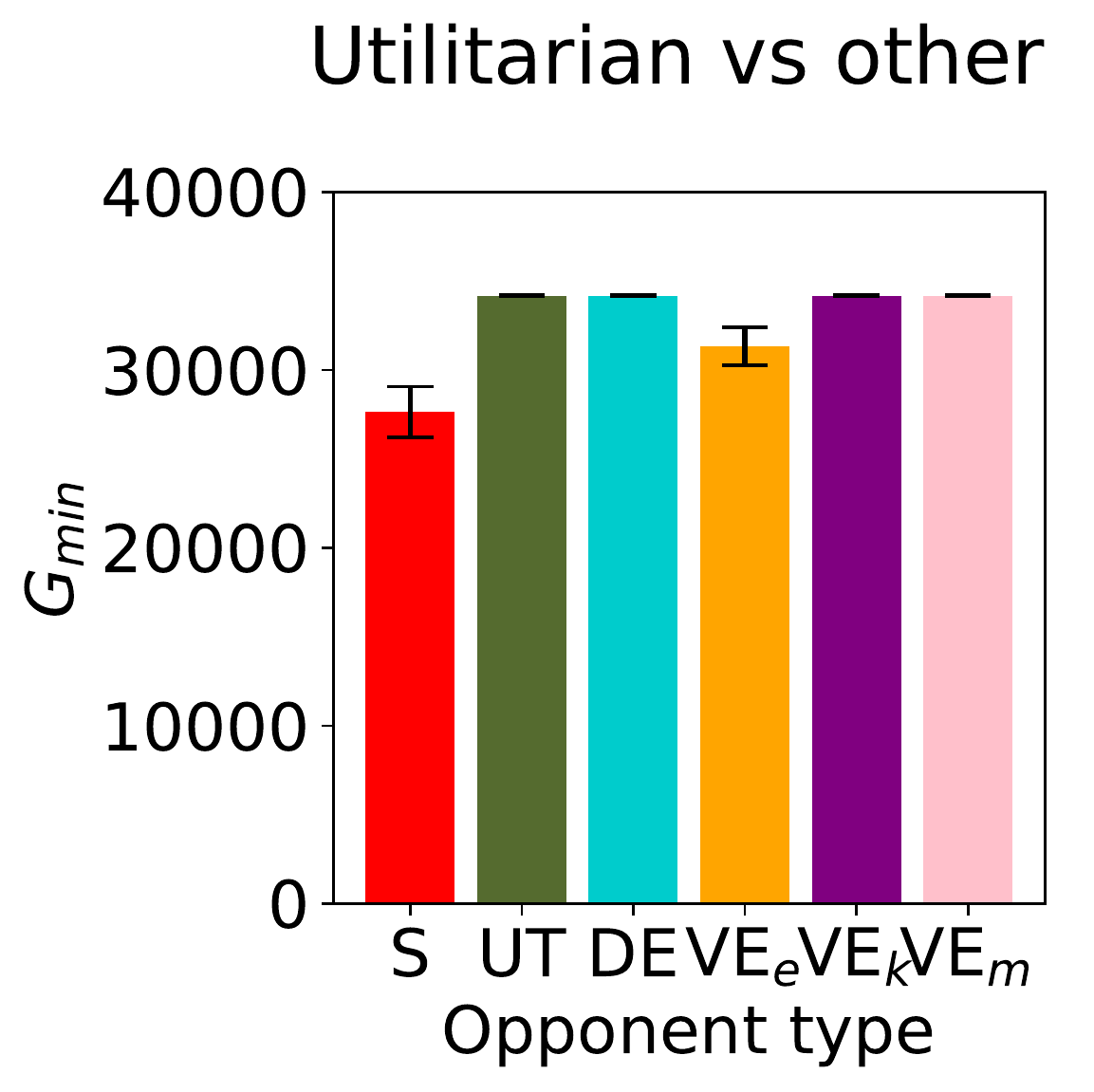}} & \subt{\includegraphics[height=18mm]{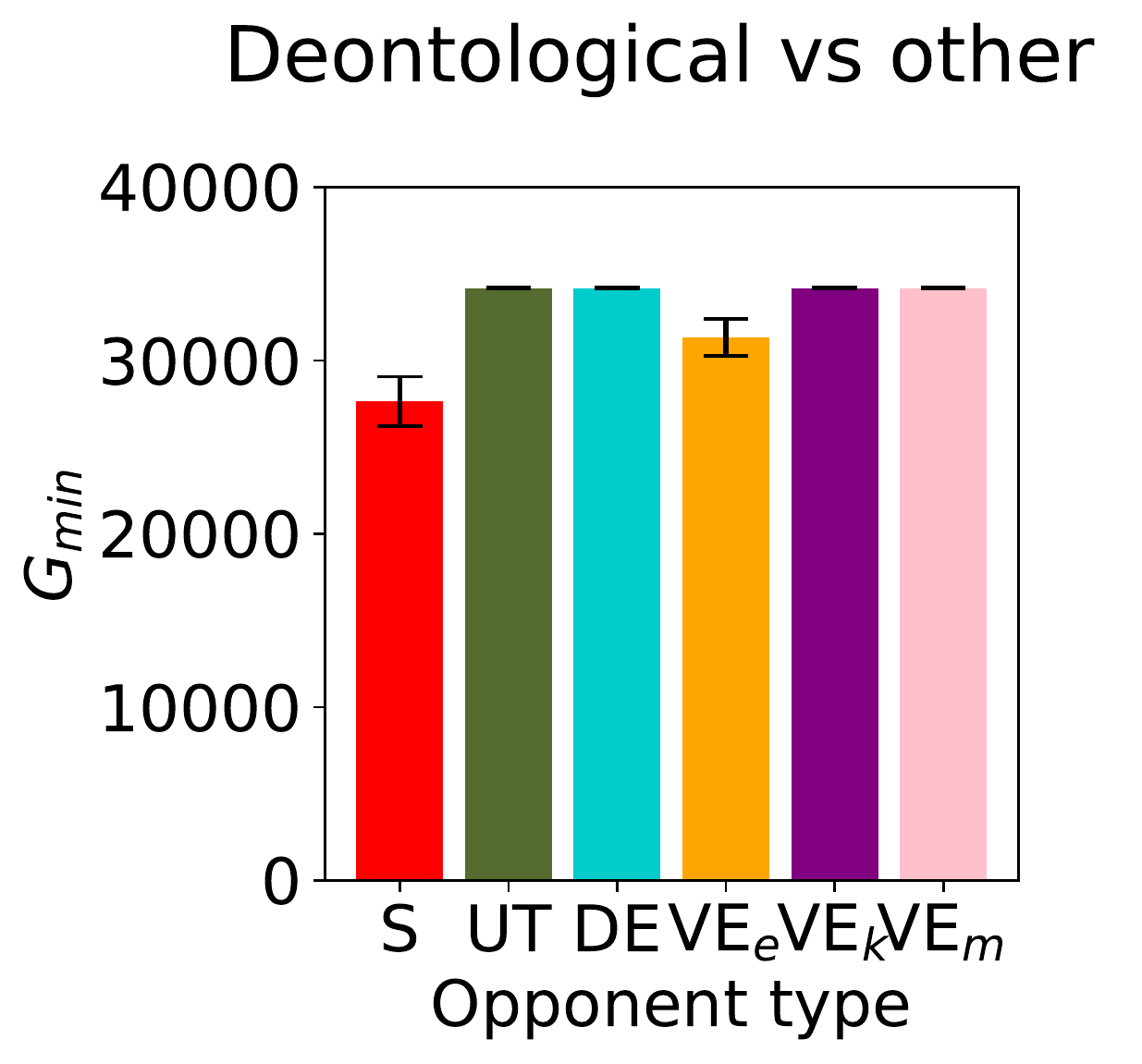}} & \subt{\includegraphics[height=18mm]{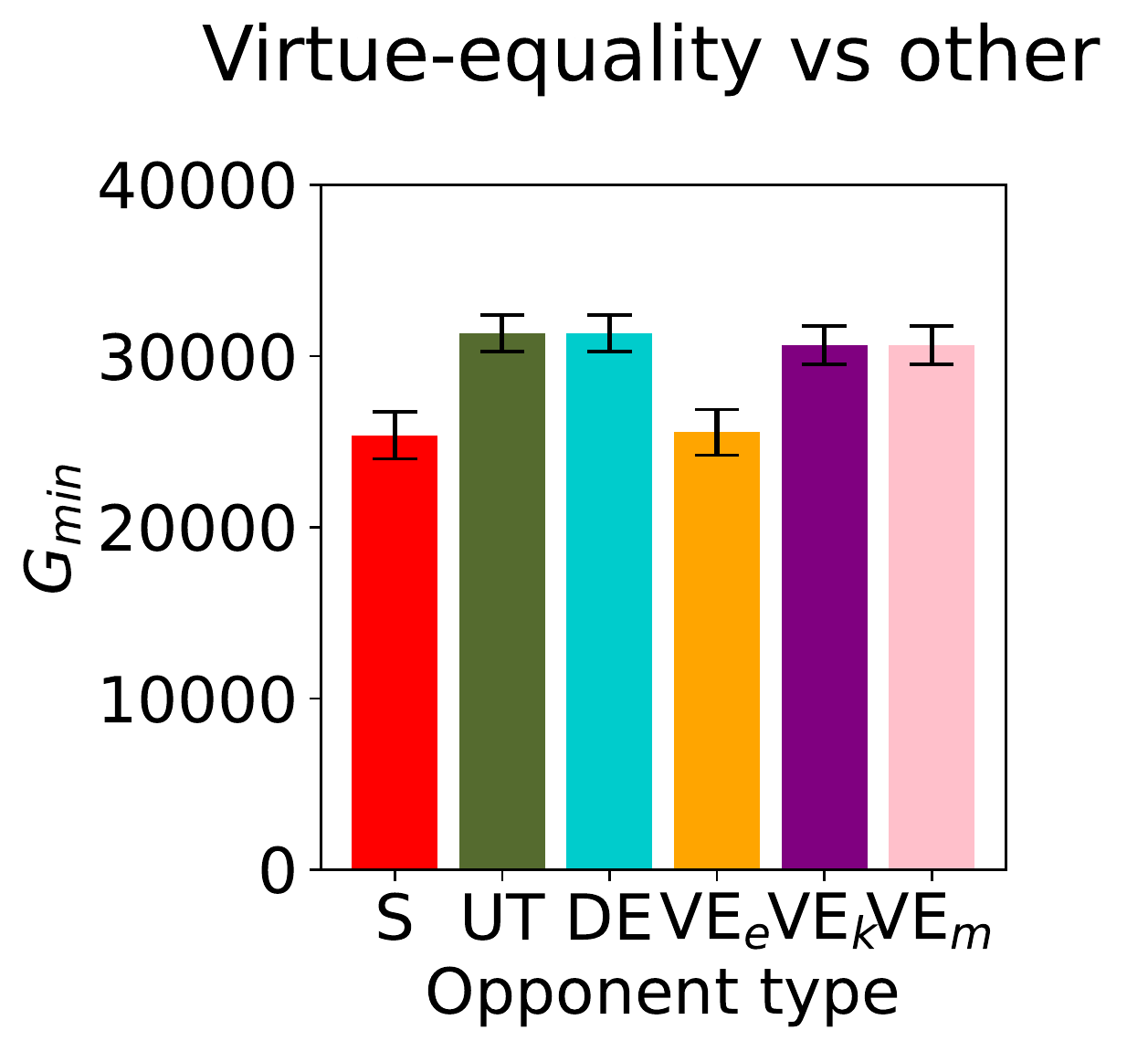}} & \subt{\includegraphics[height=18mm]{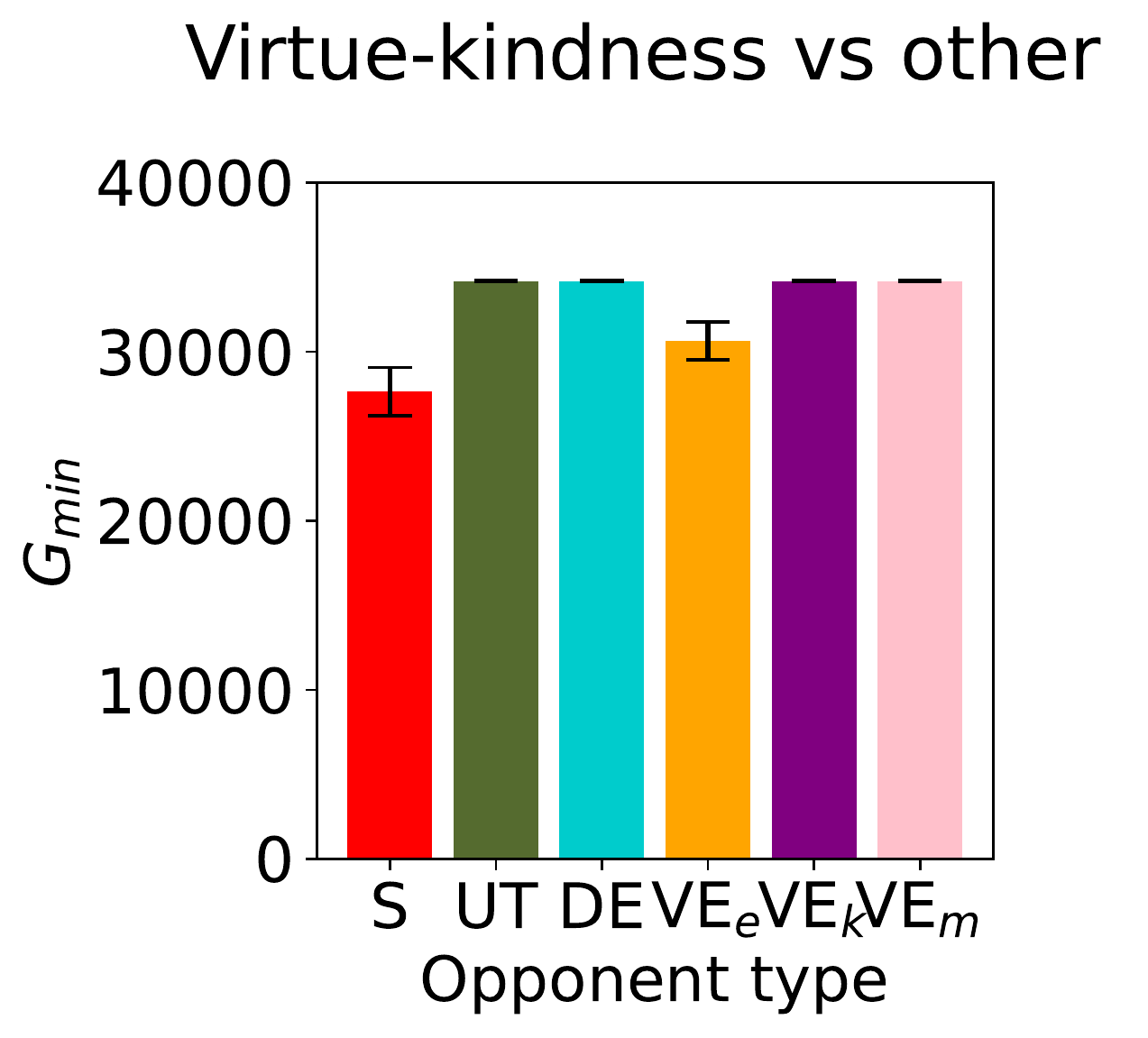}} & \subt{\includegraphics[height=18mm]{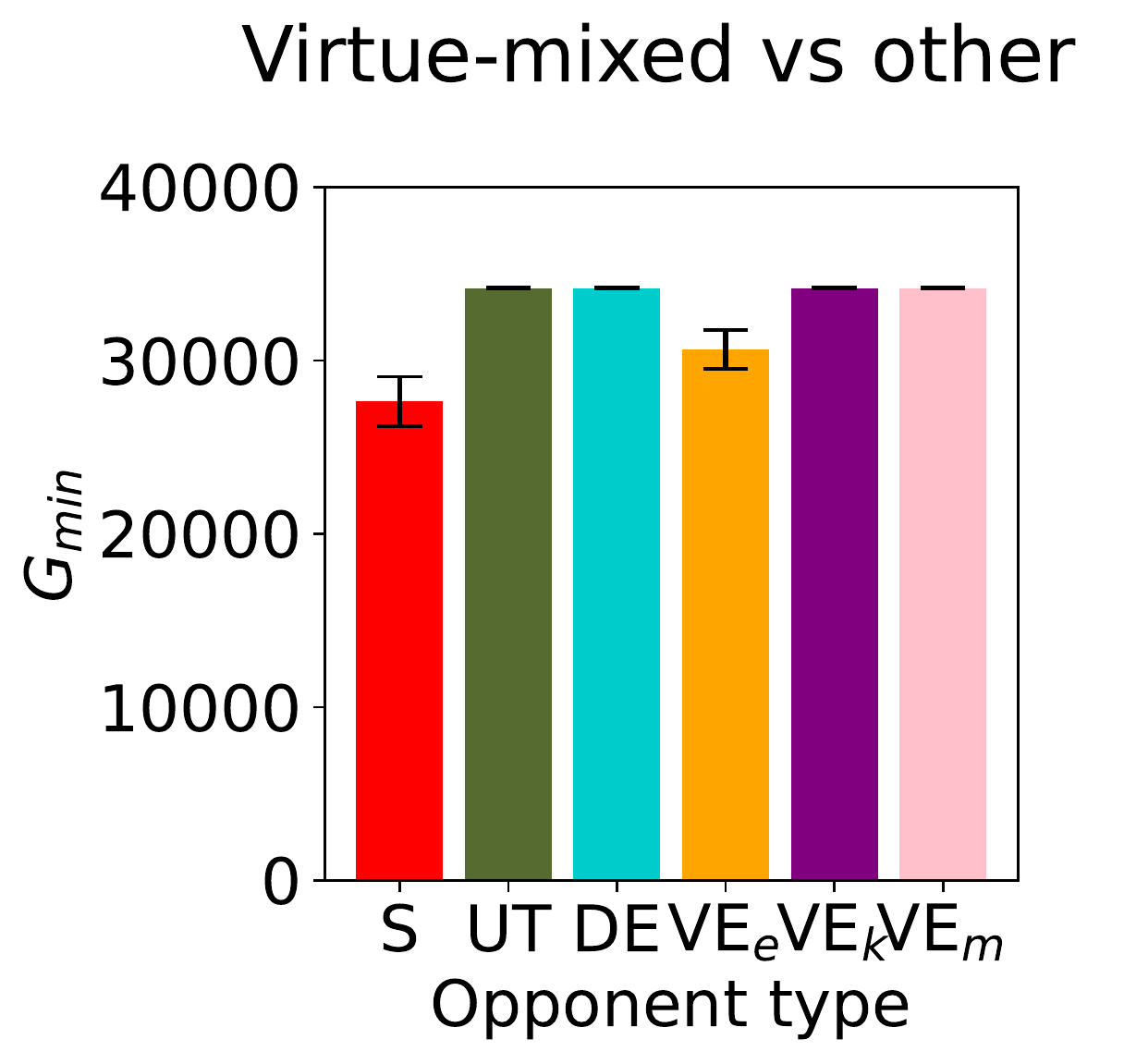}}
\\
\bottomrule
\end{tabular}
\caption{Iterated Stag Hunt game. Relative societal outcomes observed for learning player type $M$ (row) vs. all possible learning opponents $O$. The plots display averages across the 100 runs $\pm$ 95\%CI.}
\label{fig:outcomes_STH_CI}
\end{figure*}

\renewcommand\theadfont{}

\begin{figure}[!h]
\centering
\begin{tabular}{|c|cc}
\toprule
 & \thead{$\epsilon=5\%$, no decay} &  \thead{$\epsilon=100\%$ decaying linearly to $0$ \\ (as reported in the main paper)} \\
\midrule
\makecell[cc]{\rotatebox[origin=c]{90}{Selfish}} & 
\subt{\includegraphics[width=30mm]{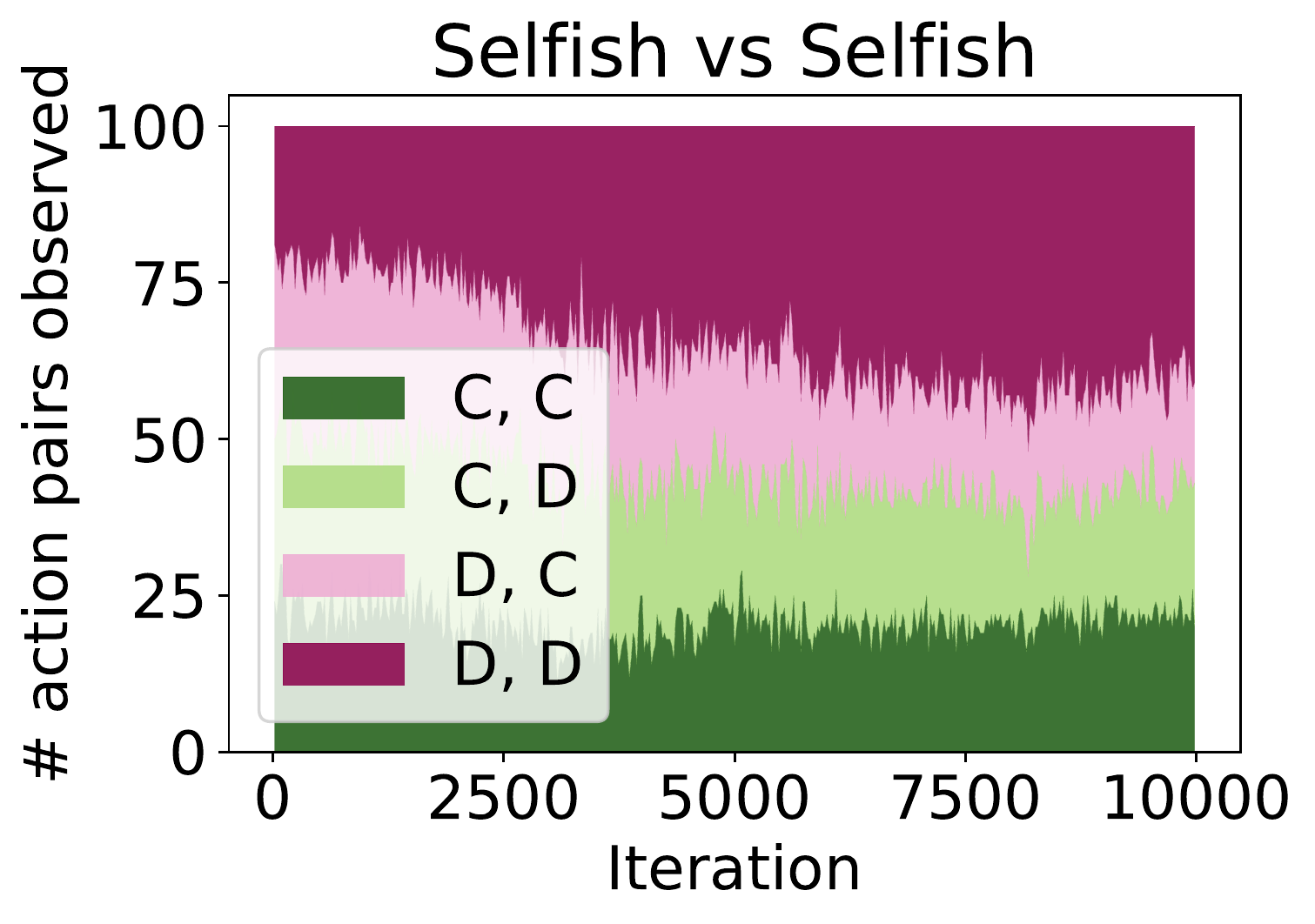}}
& \subt{\includegraphics[width=30mm]{IJCAI-2023/IPD_action_pairs/pairs_QLS_QLS.pdf}}
\\
\makecell[cc]{\rotatebox[origin=c]{90}{ Utilitarian }} & 
\subt{\includegraphics[width=30mm]{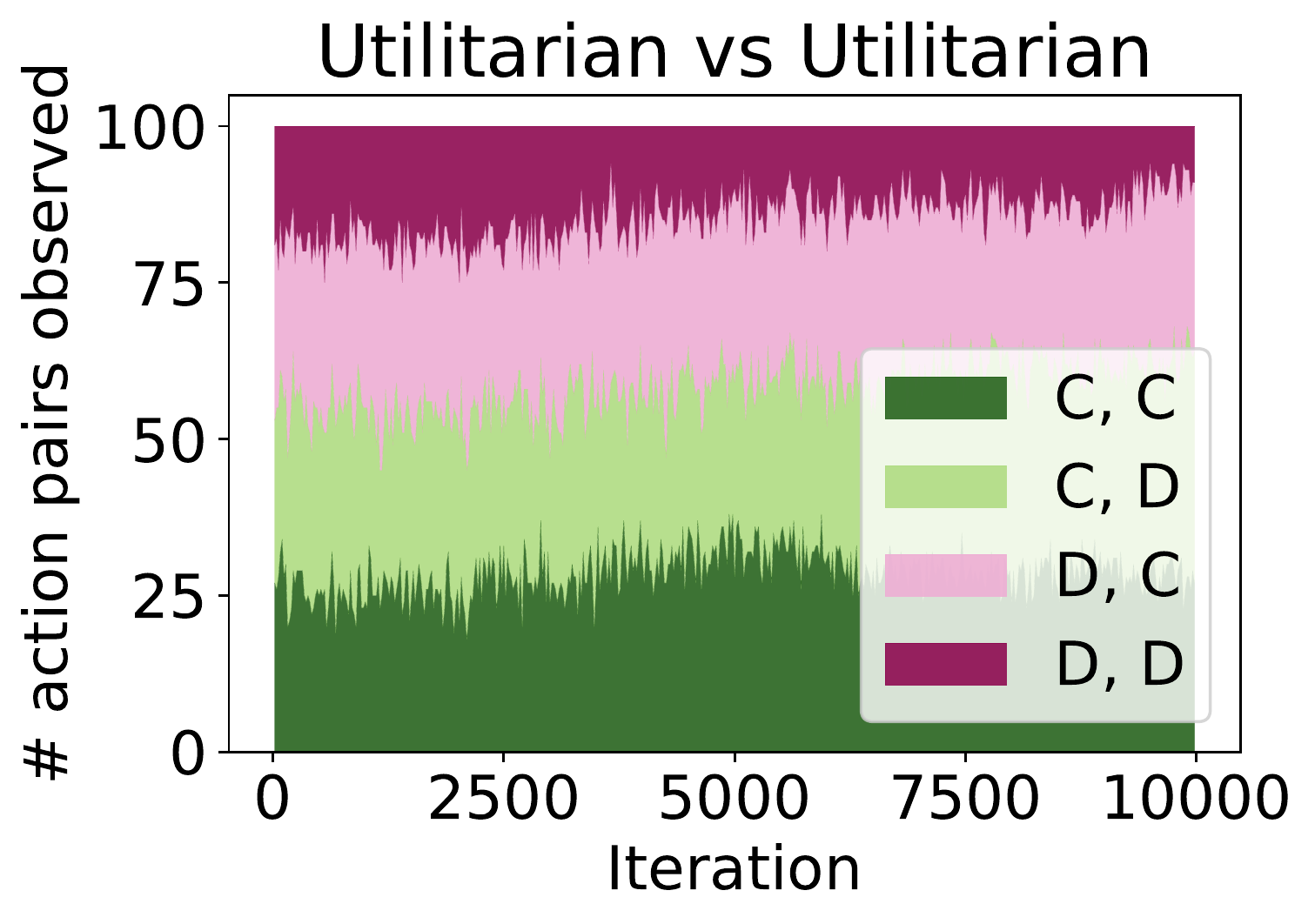}}
&\subt{\includegraphics[width=30mm]{IJCAI-2023/IPD_action_pairs/pairs_QLUT_QLUT.pdf}}
\\
\makecell[cc]{\rotatebox[origin=c]{90}{ Deontological }} &
\subt{\includegraphics[width=30mm]{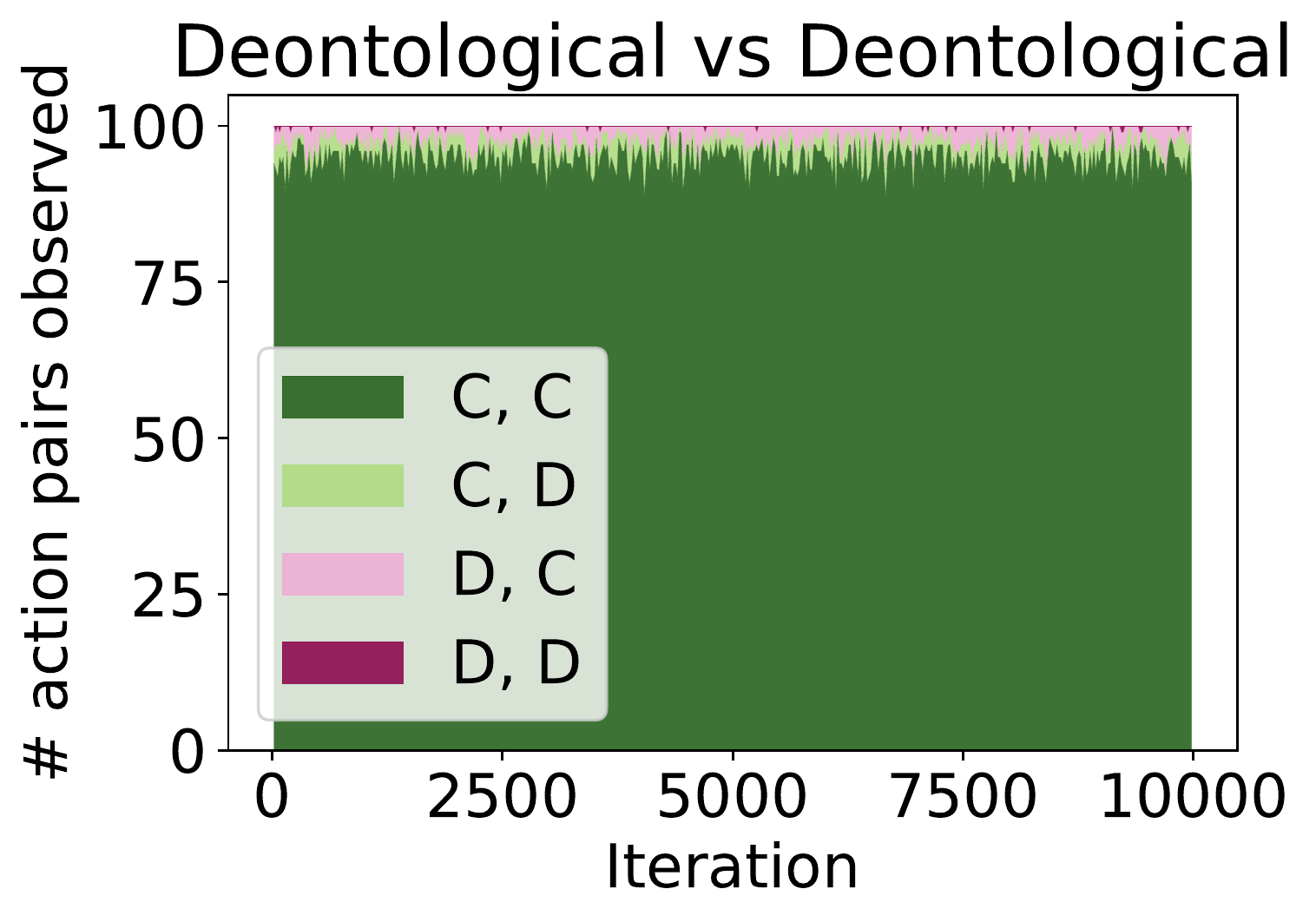}}
&\subt{\includegraphics[width=30mm]{IJCAI-2023/IPD_action_pairs/pairs_QLDE_QLDE.pdf}}

\\
\makecell[cc]{\rotatebox[origin=c]{90}{ Virtue-eq. }} &
\subt{\includegraphics[width=30mm]{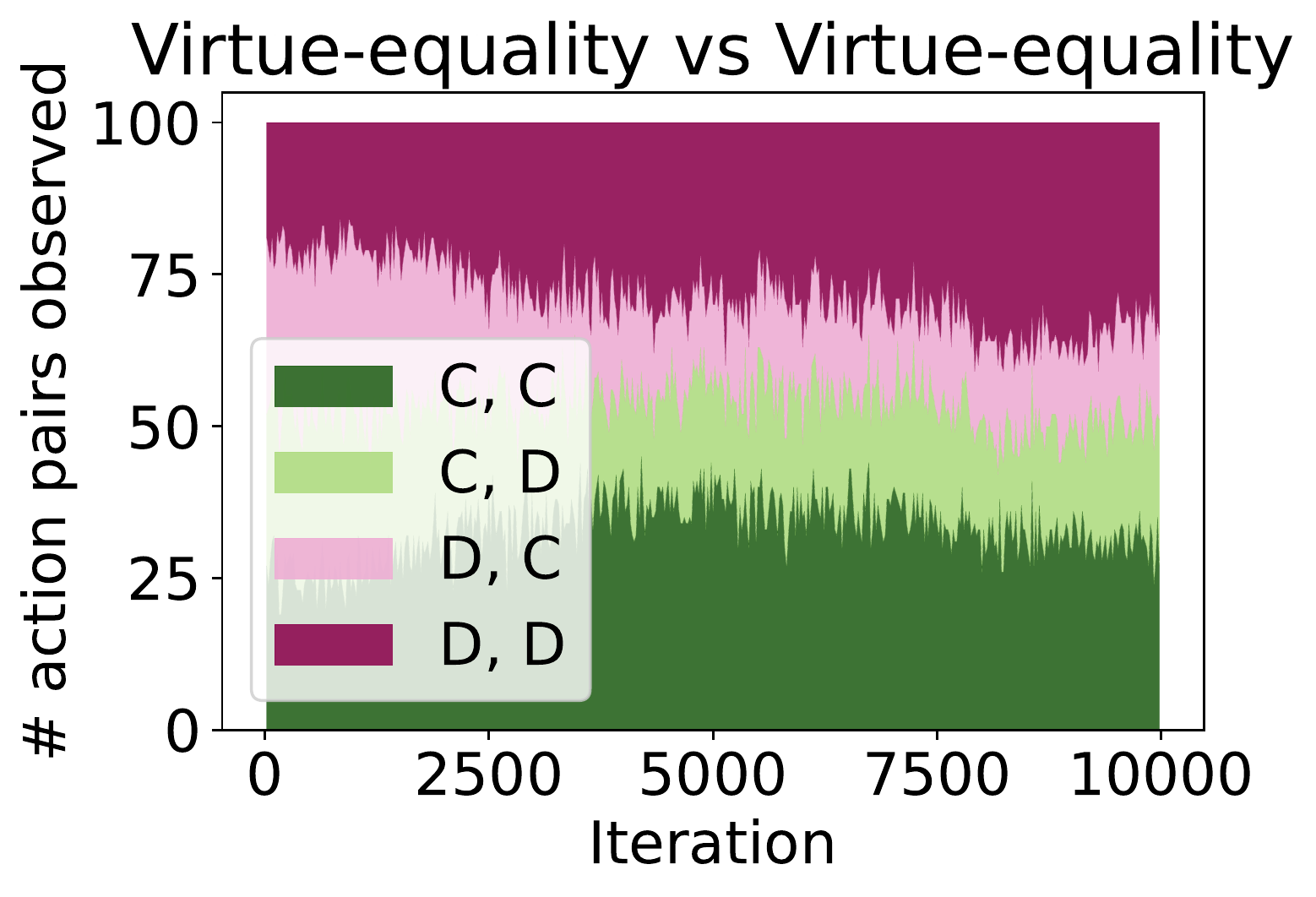}}
&\subt{\includegraphics[width=30mm]{IJCAI-2023/IPD_action_pairs/pairs_QLVE_e_QLVE_e.pdf}}
\\
\makecell[cc]{\rotatebox[origin=c]{90}{ Virtue-kind. }} &
\subt{\includegraphics[width=30mm]{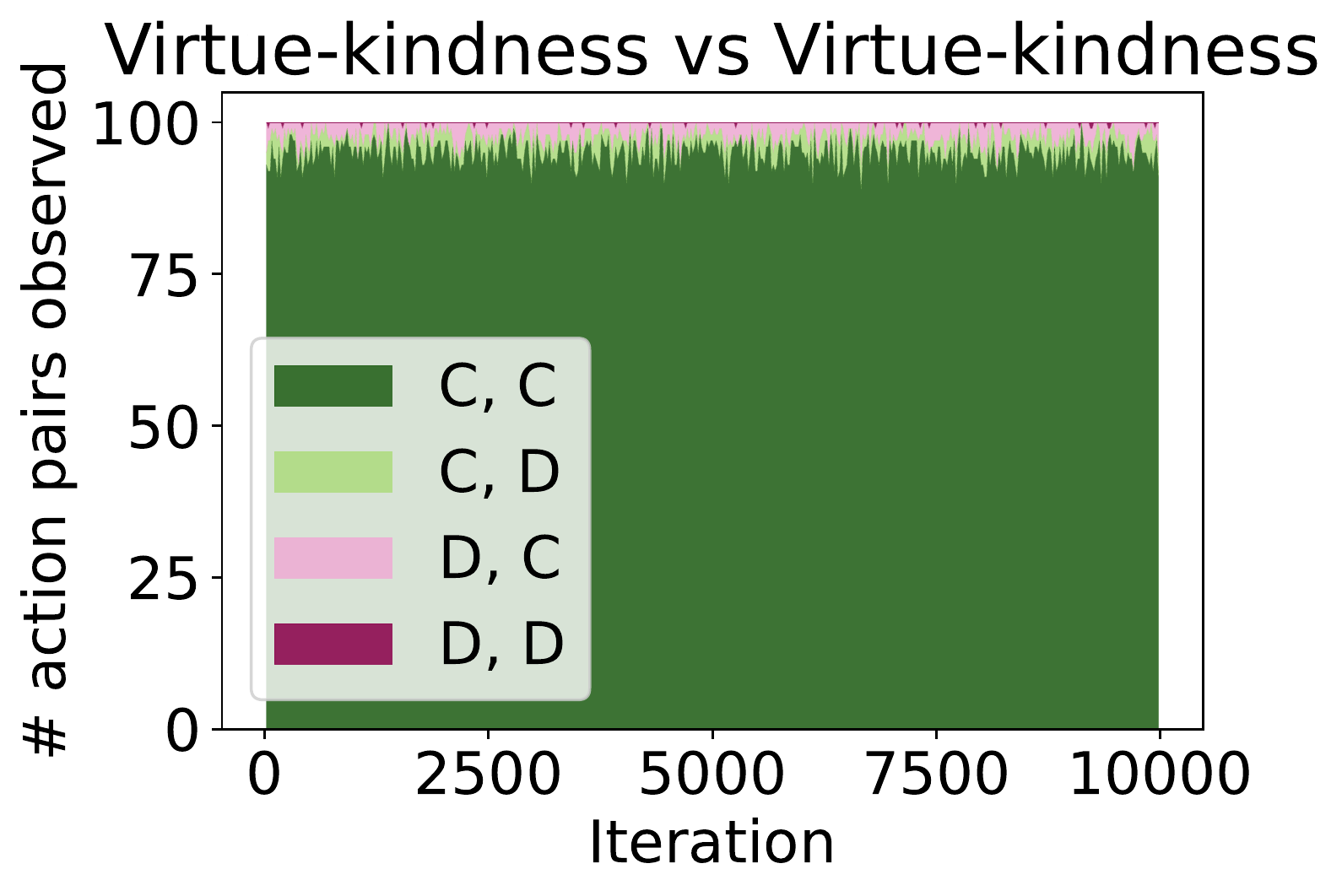}}
&\subt{\includegraphics[width=30mm]{IJCAI-2023/IPD_action_pairs/pairs_QLVE_k_QLVE_k.pdf}}
\\
\bottomrule
\end{tabular}
\caption{Iterated Prisoner's Dilemma game. The simultaneous action plots illustrate the impact of exploration on the learning of moral agents. For simplicity, we show each moral agent learning against its own kind only, and compare learning with a smaller exploration rate (left, $\epsilon=5\%$) versus the large exploration rate (right, $\epsilon=100\%$ decaying to 0), as reported in the paper.}
\label{fig:exploration}
\end{figure}

\renewcommand\theadfont{}

\begin{figure*}[!h]
\centering
\begin{tabular}{|c|cccccc}
\toprule
 & \thead{vs QLS; $\beta=0$ \\(fully 'kind')} & vs QLS; $\beta=0.2$ & vs QLS; $\beta=0.4$ & vs QLS; $\beta=0.6$ & vs QLS; $\beta=0.8$ & \thead{vs QLS; $\beta=1.0$ \\ (fully 'equal')}\\
\midrule
\makecell[cc]{\rotatebox[origin=c]{90}{Virtue-mixed}} &
\subt{\includegraphics[width=20mm]{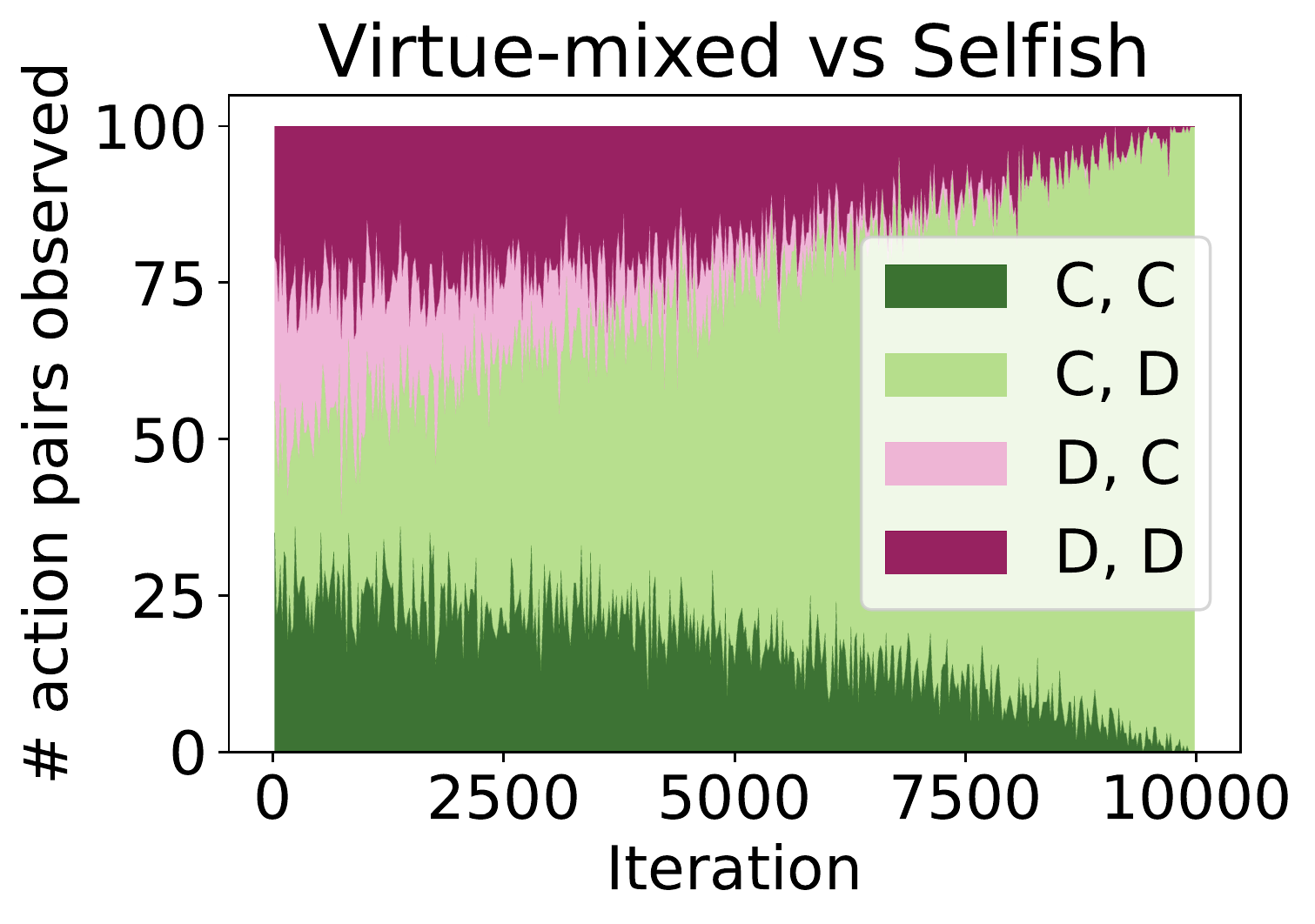}}
&\subt{\includegraphics[width=20mm]{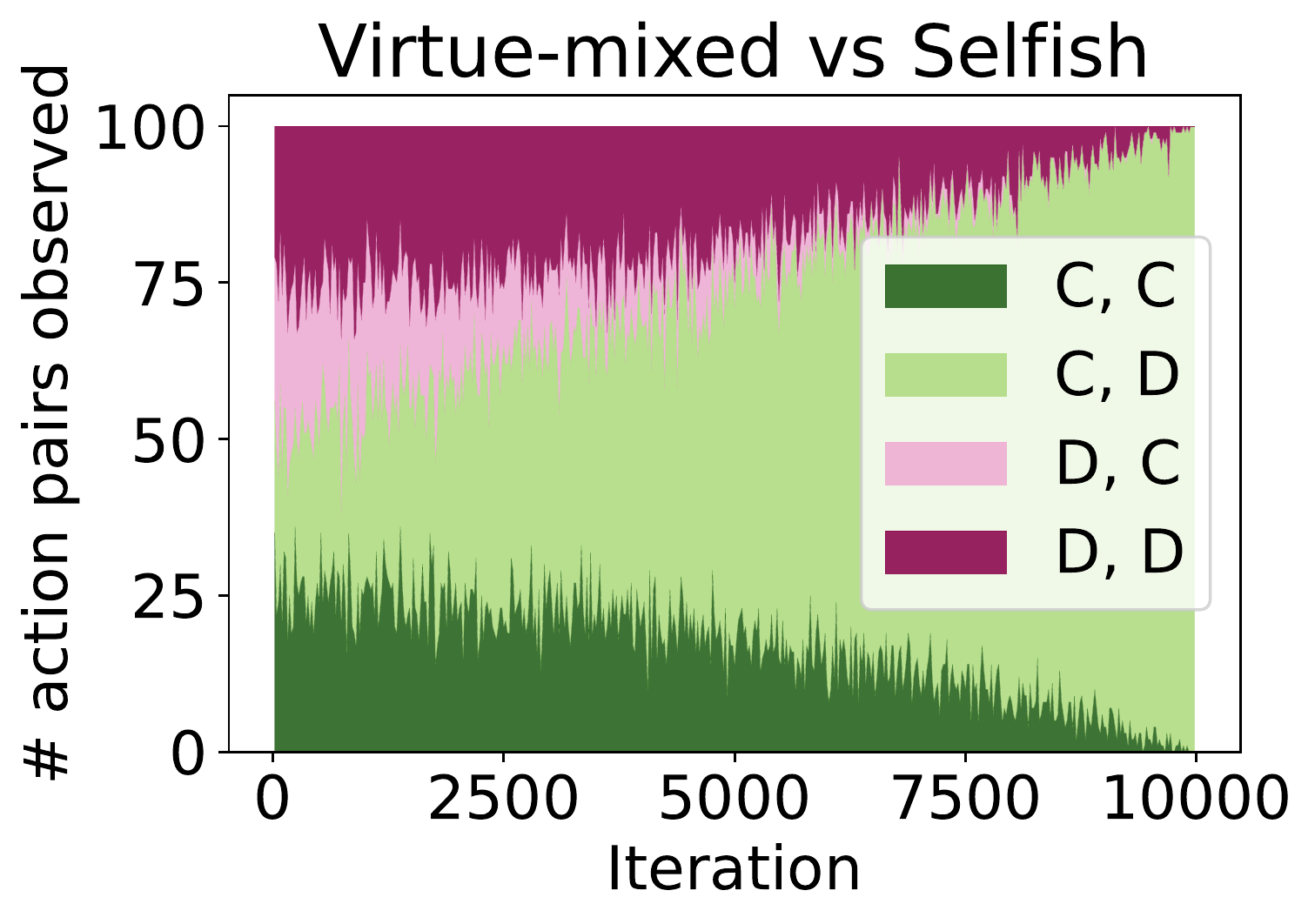}}
&\subt{\includegraphics[width=20mm]{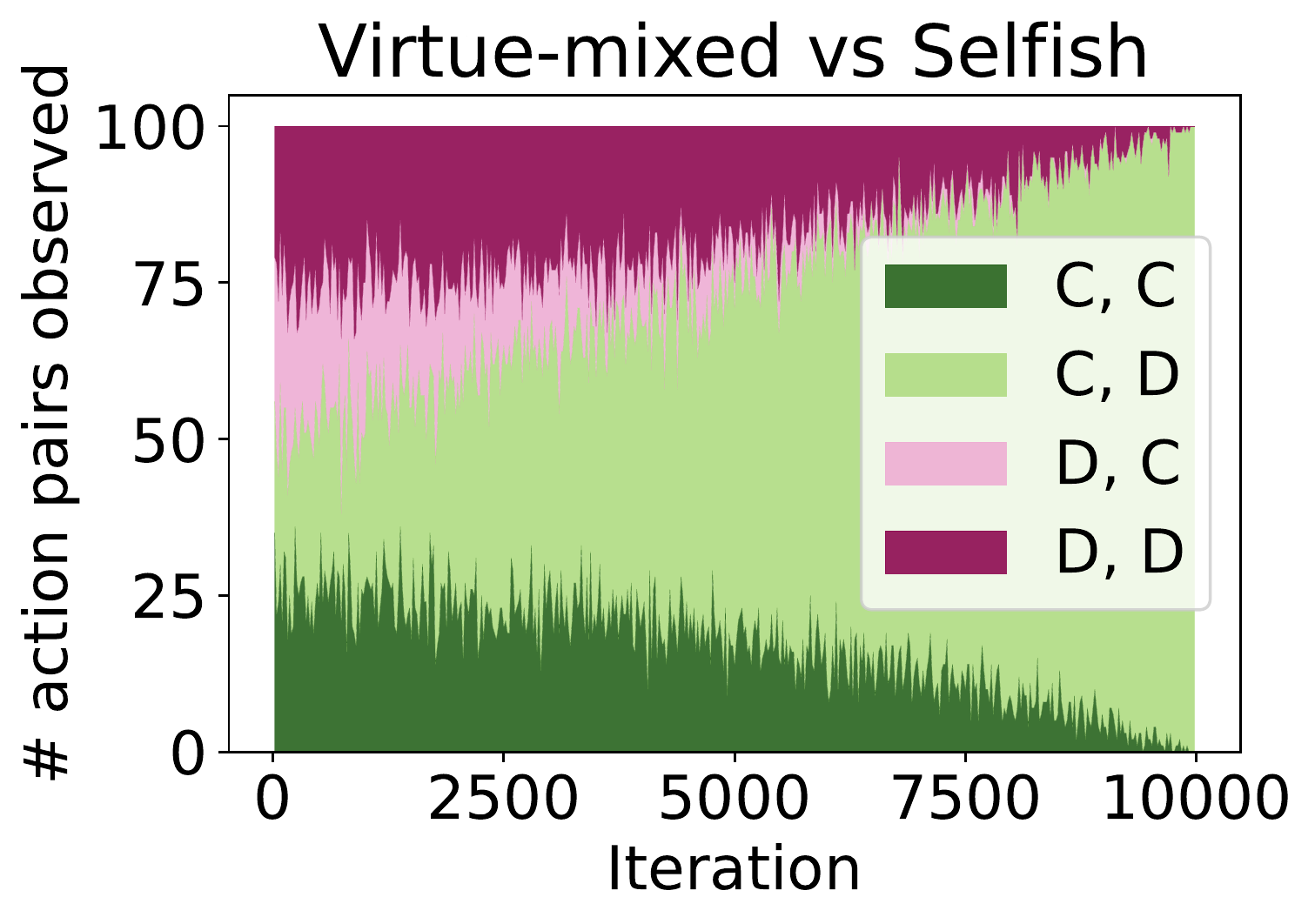}}
&\subt{\includegraphics[width=20mm]{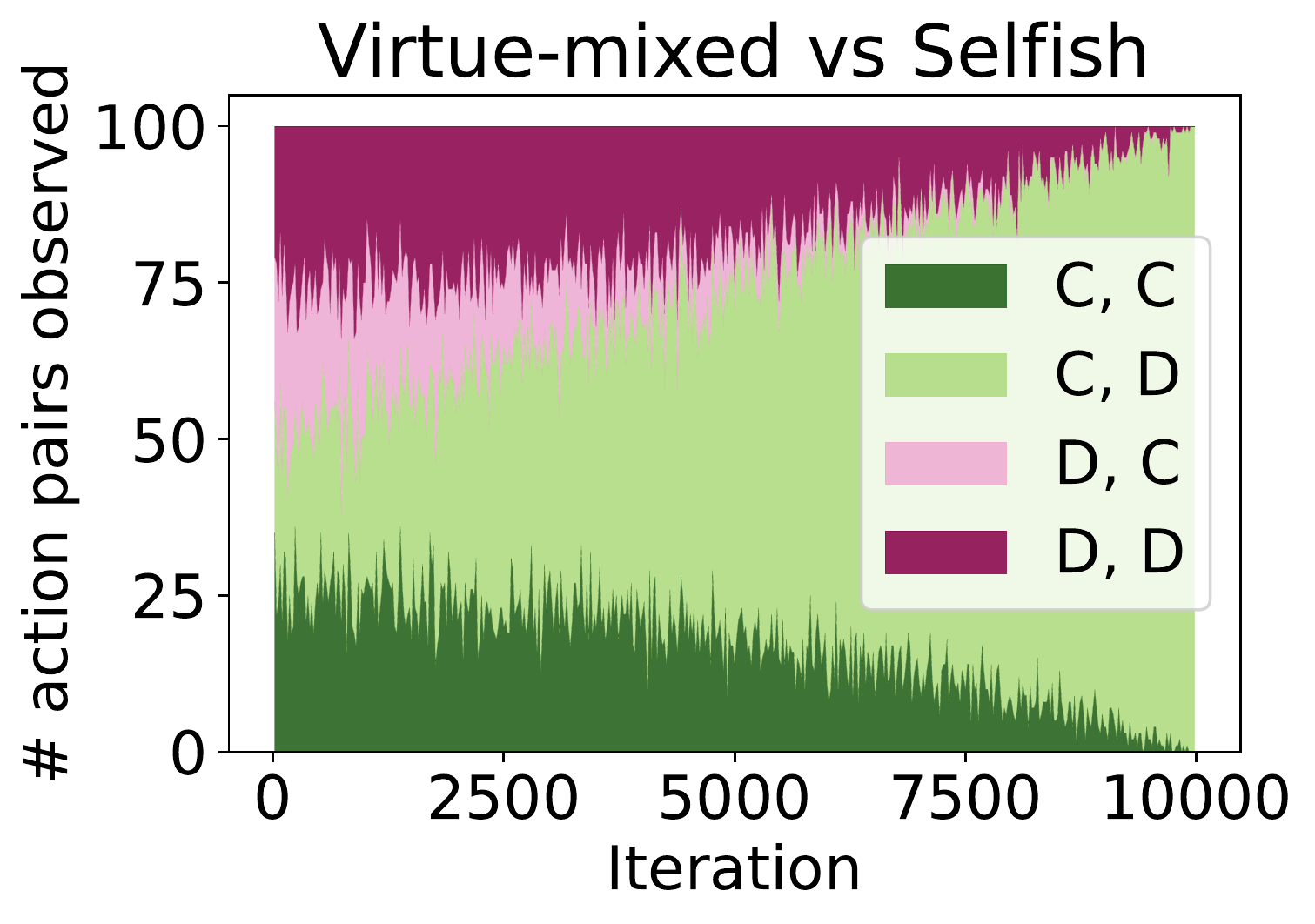}}
&\subt{\includegraphics[width=20mm]{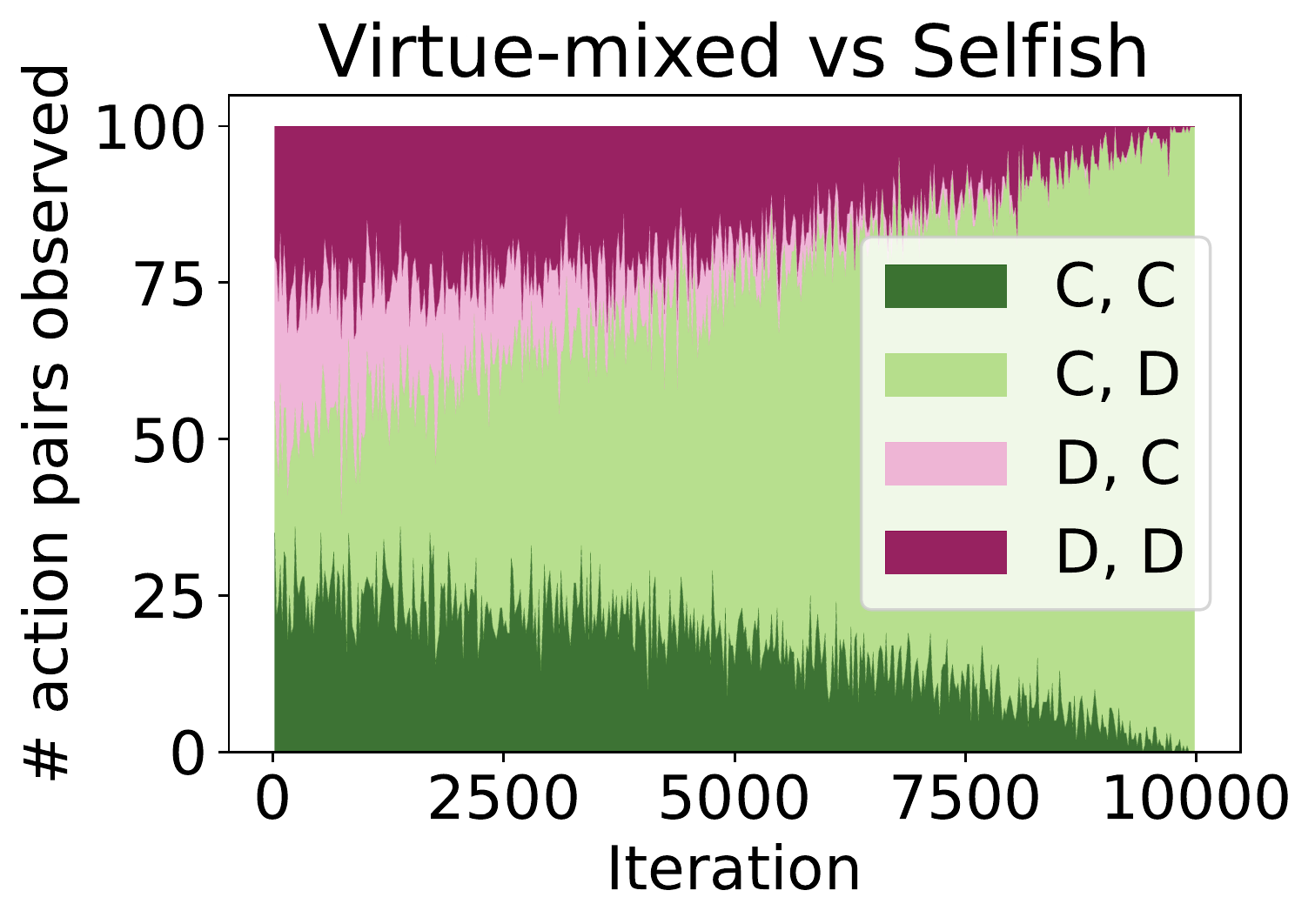}}
&\subt{\includegraphics[width=20mm]{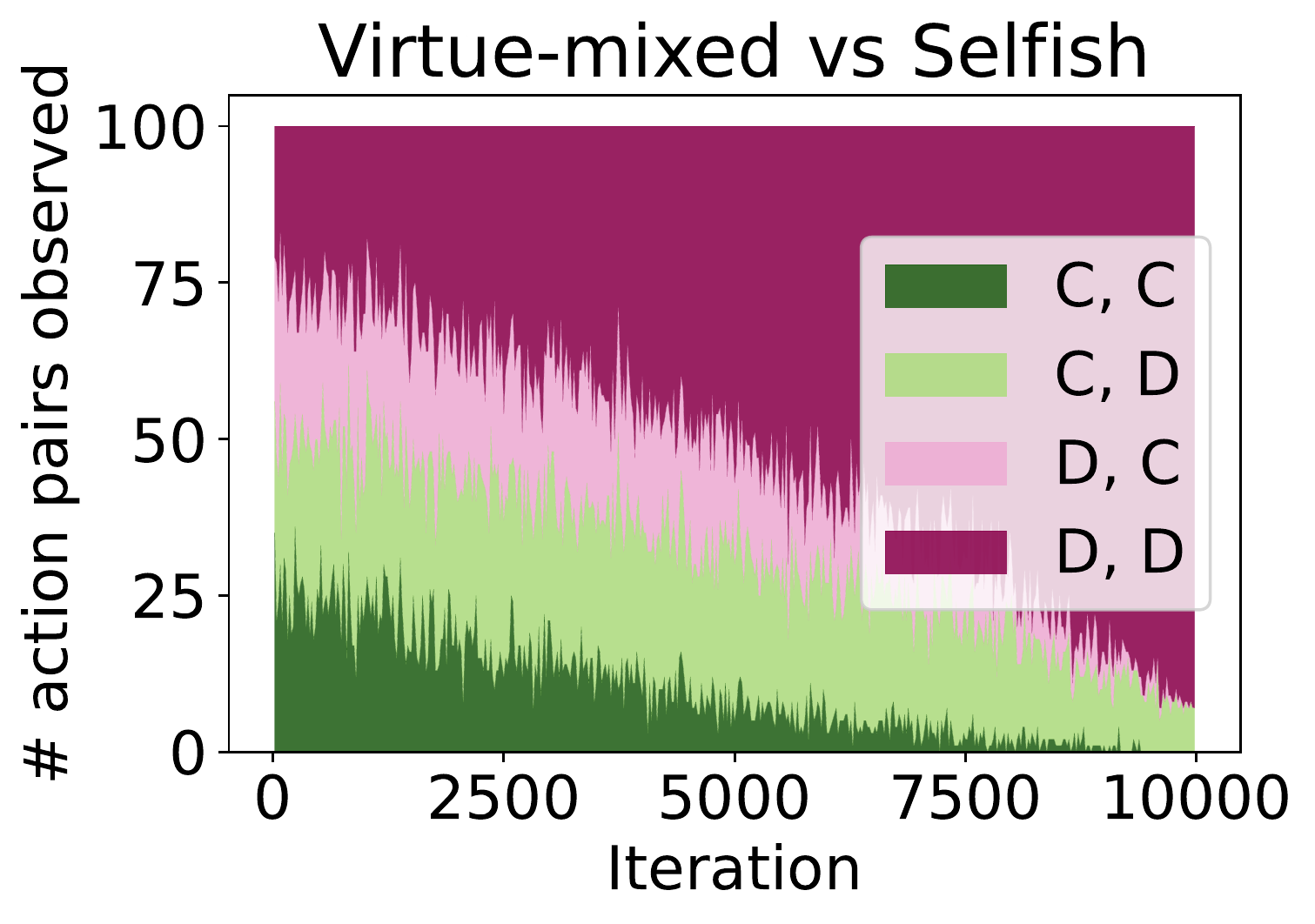}}
\\
\bottomrule
\end{tabular}
\caption{Iterated Prisoner's Dilemma. The actions displayed by \textit{Virtue-mixed} agent defined with different weights $\beta \in (0, 0.2, 0.4, 0.6, 0.8, 1)$ against a \textit{Selfish} opponent.}
\label{fig:extra_QLVM_beta_IPD}
\end{figure*}

\begin{figure*}[!h]
\centering
\begin{tabular}{|c|cccccc}
\toprule
 & \thead{vs QLS; $\beta=0$ \\(fully 'kind')} & vs QLS; $\beta=0.2$ & vs QLS; $\beta=0.4$ & vs QLS; $\beta=0.6$ & vs QLS; $\beta=0.8$ & \thead{vs QLS; $\beta=1.0$ \\ (fully 'equal')}\\
\midrule
\makecell[cc]{\rotatebox[origin=c]{90}{Virtue-mixed}} &
\subt{\includegraphics[width=20mm]{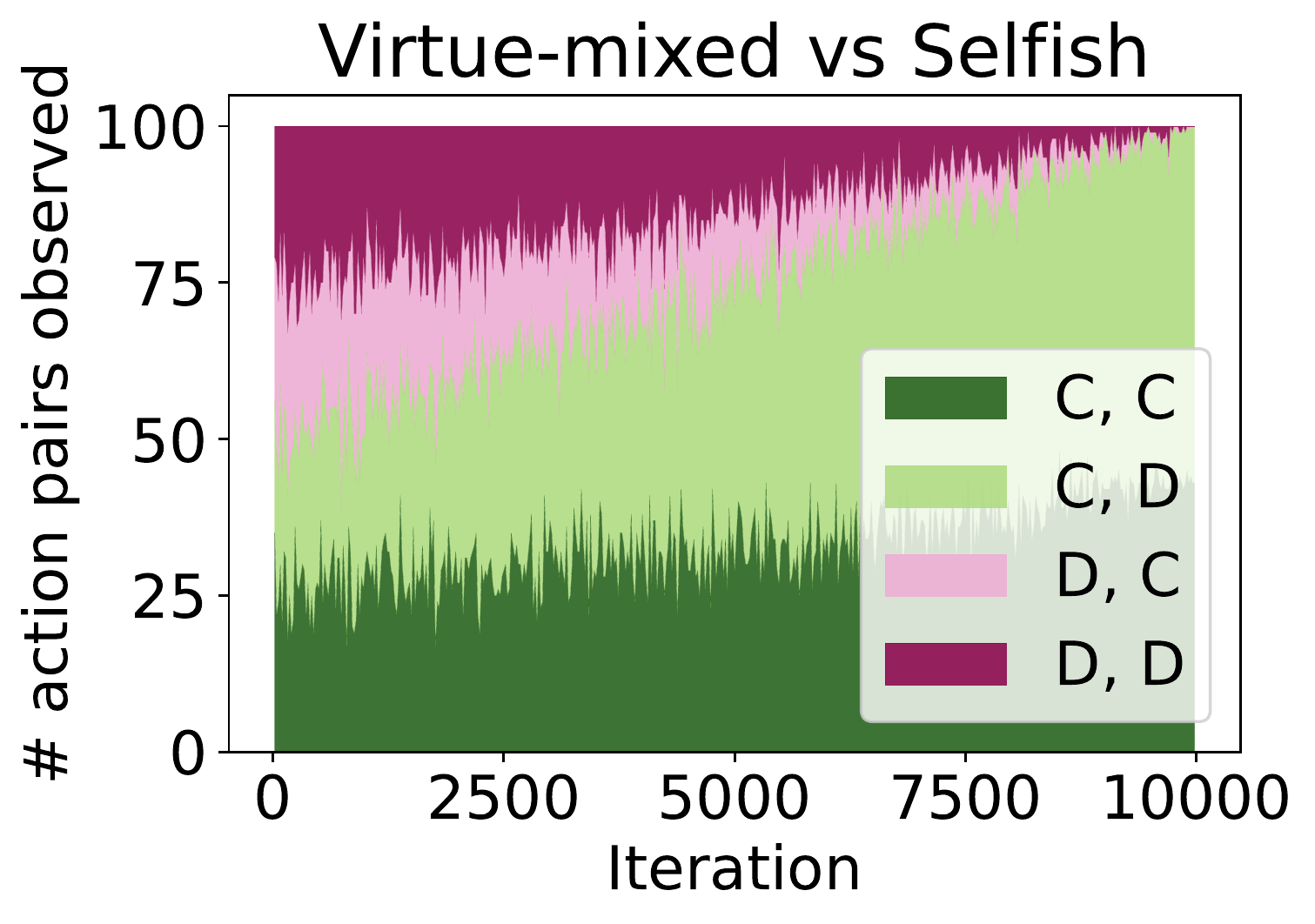}}
&\subt{\includegraphics[width=20mm]{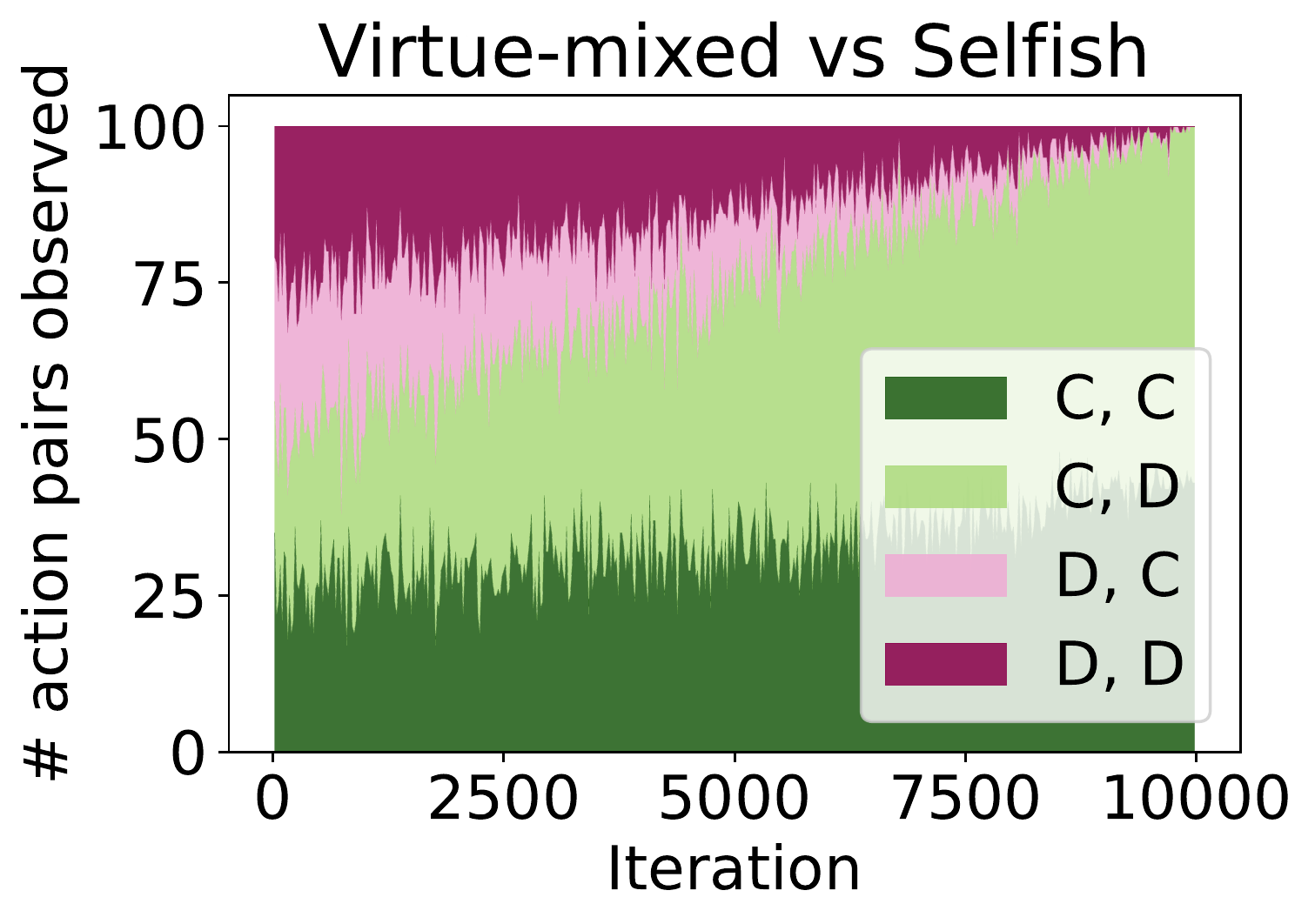}}
&\subt{\includegraphics[width=20mm]{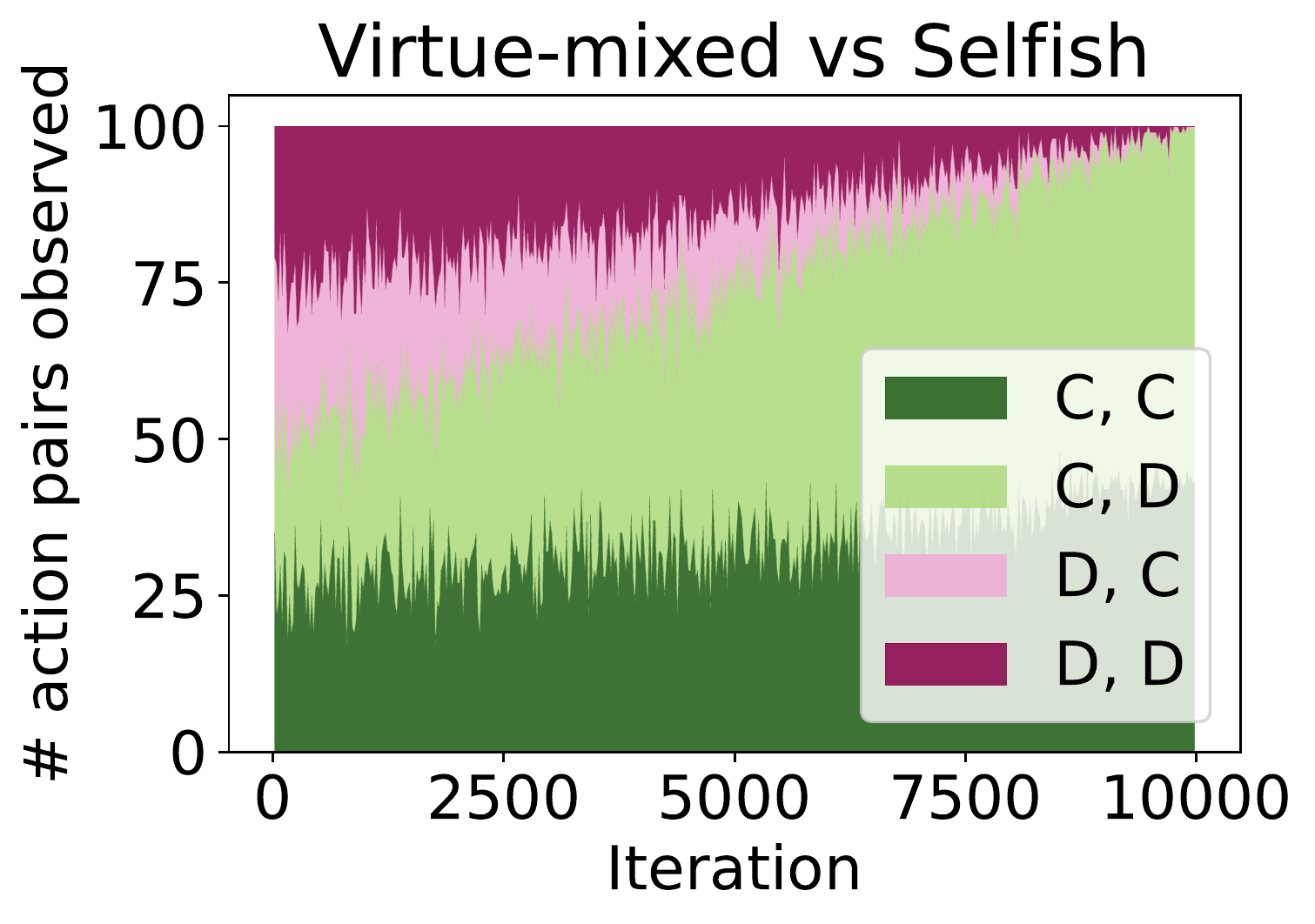}}
&\subt{\includegraphics[width=20mm]{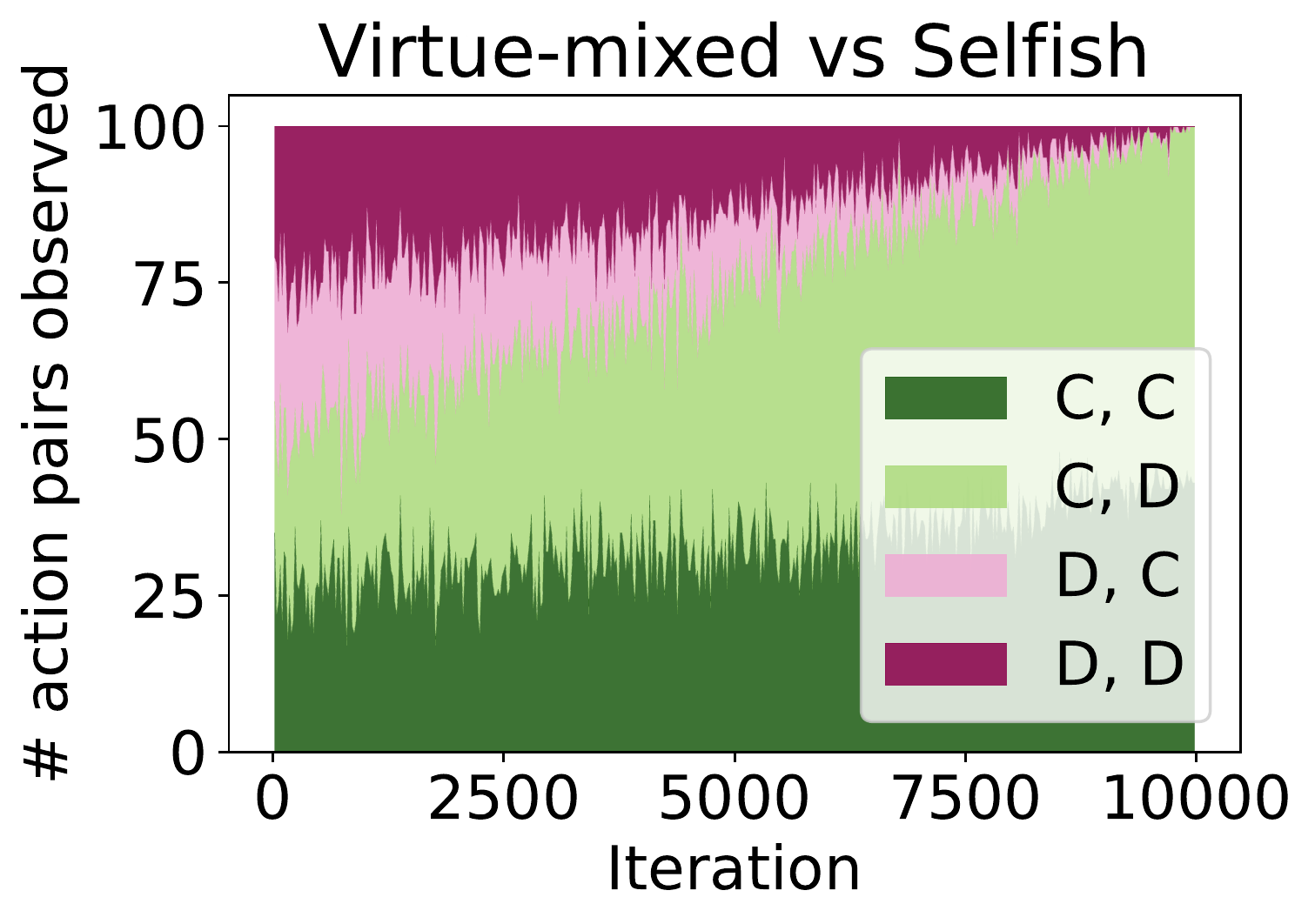}}
&\subt{\includegraphics[width=20mm]{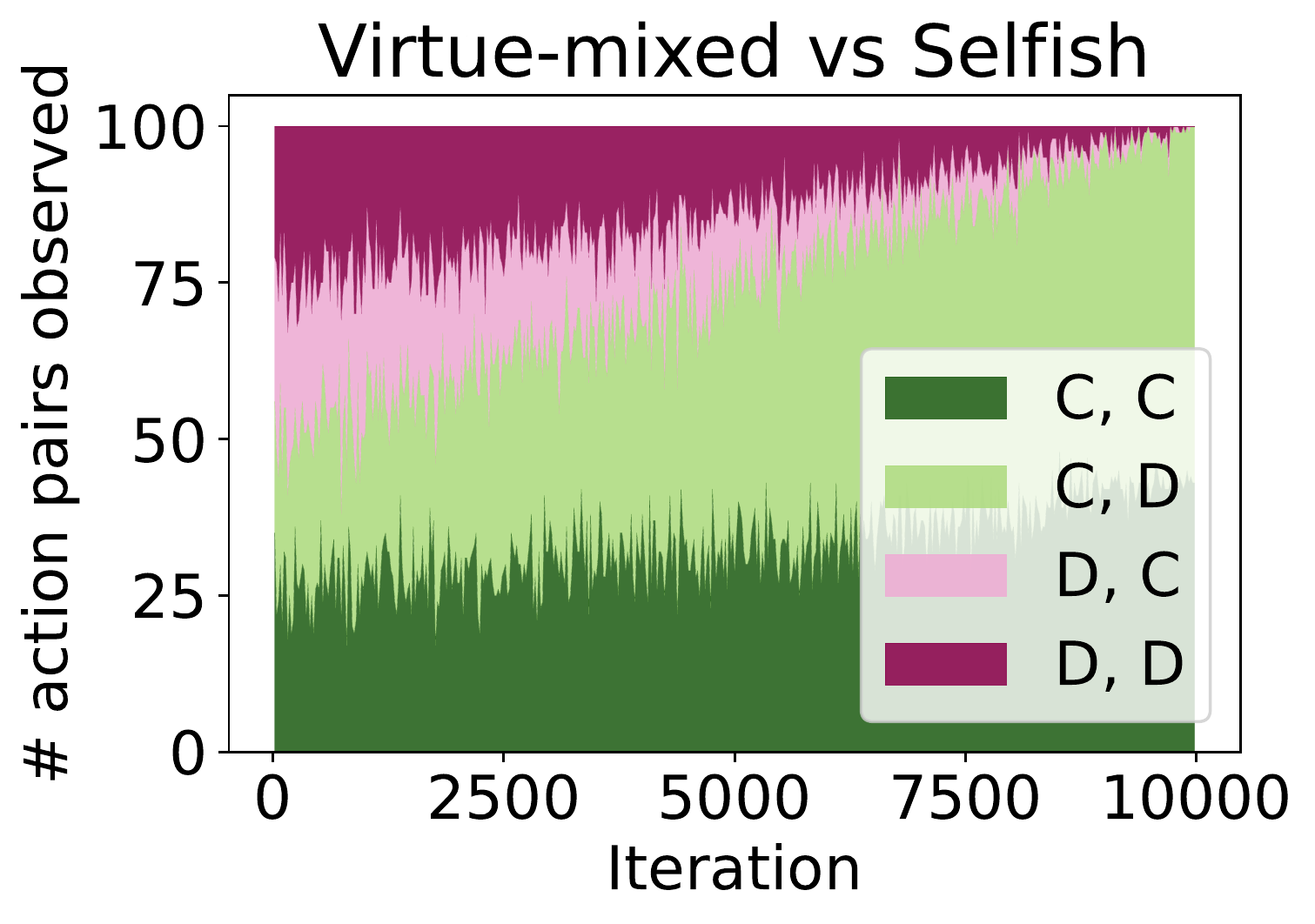}}
&\subt{\includegraphics[width=20mm]{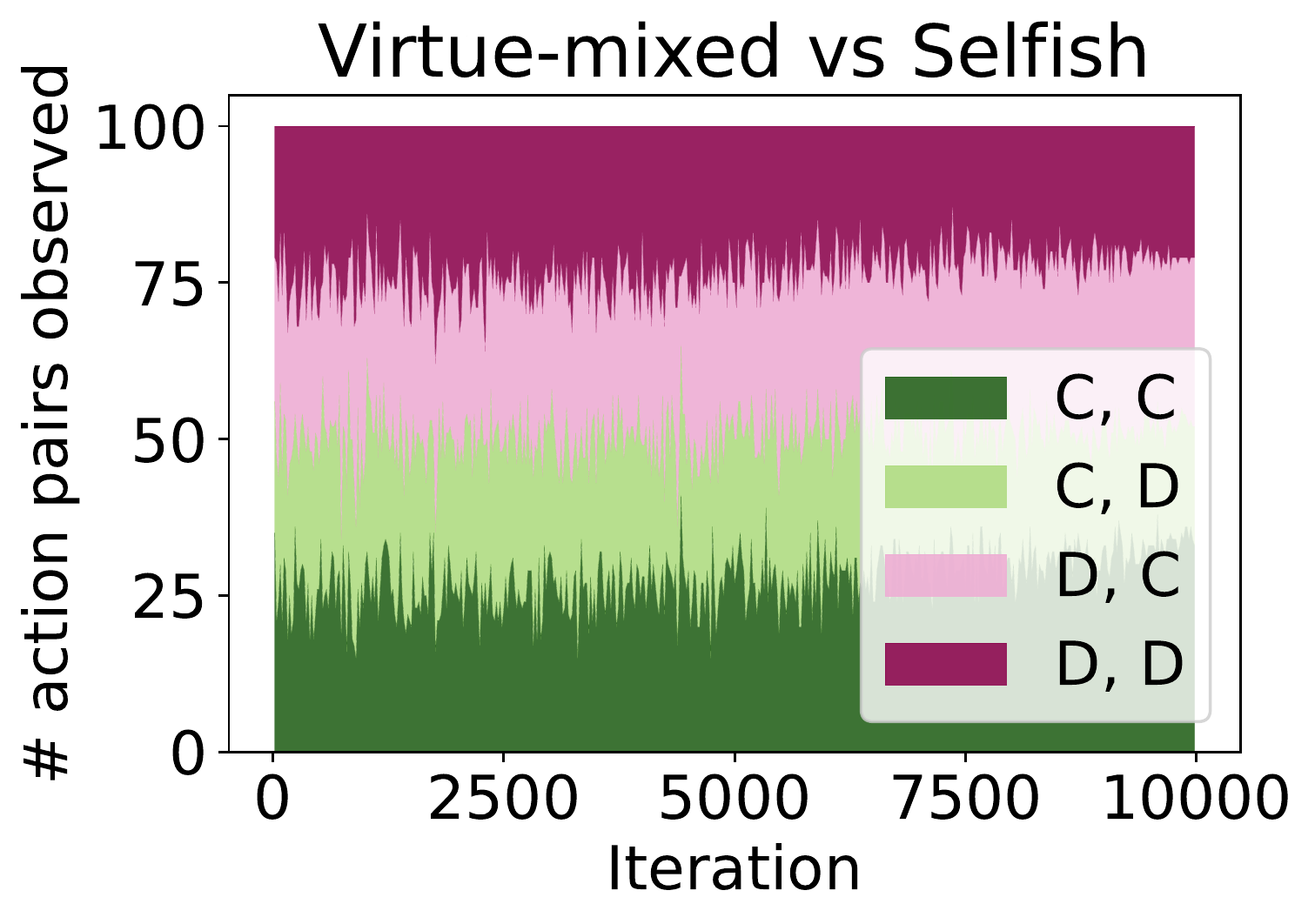}}
\\
\bottomrule
\end{tabular}
\caption{Iterated Volunteer's Dilemma. The actions displayed by \textit{Virtue-mixed} agent defined with different weights $\beta \in (0, 0.2, 0.4, 0.6, 0.8, 1)$ against a \textit{Selfish} opponent.}
\label{fig:extra_QLVM_beta_VOL}
\end{figure*}

\begin{figure*}[!h]
\centering
\begin{tabular}{|c|cccccc}
\toprule
 & \thead{vs QLS; $\beta=0$ \\(fully 'kind')} & vs QLS; $\beta=0.2$ & vs QLS; $\beta=0.4$ & vs QLS; $\beta=0.6$ & vs QLS; $\beta=0.8$ & \thead{vs QLS; $\beta=1.0$ \\ (fully 'equal')}\\
\midrule
\makecell[cc]{\rotatebox[origin=c]{90}{Virtue-mixed}} &
\subt{\includegraphics[width=20mm]{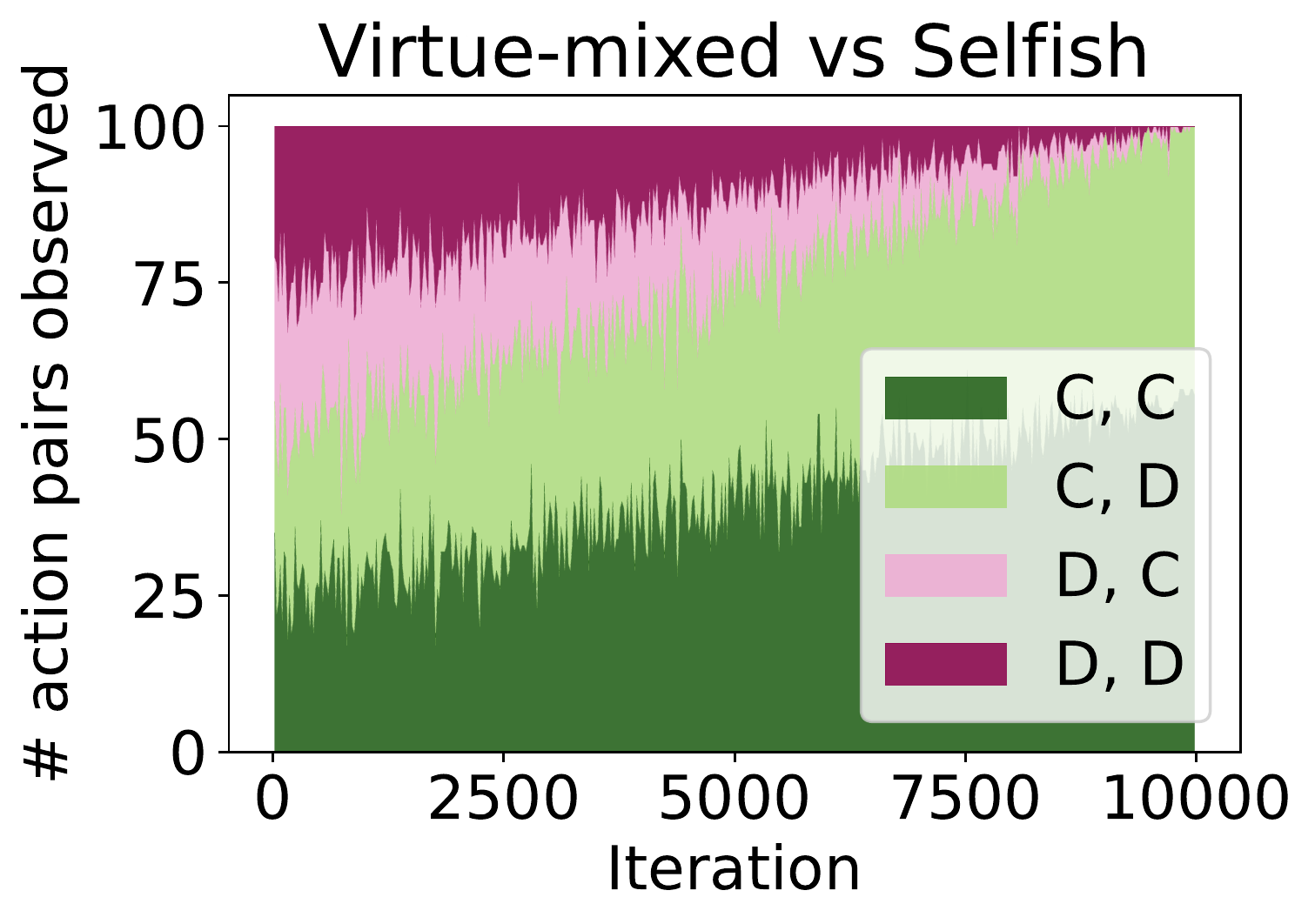}}
&\subt{\includegraphics[width=20mm]{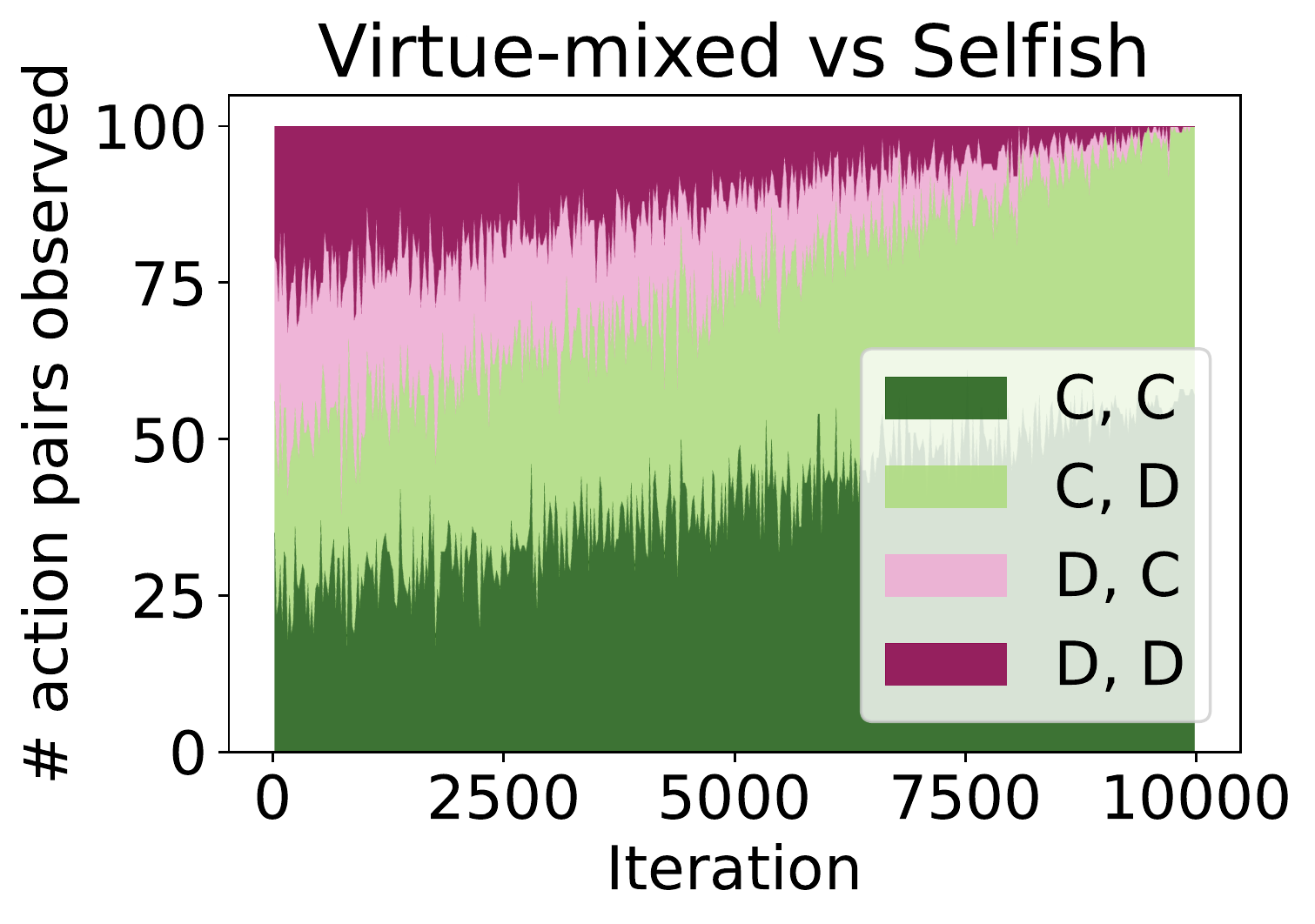}}
&\subt{\includegraphics[width=20mm]{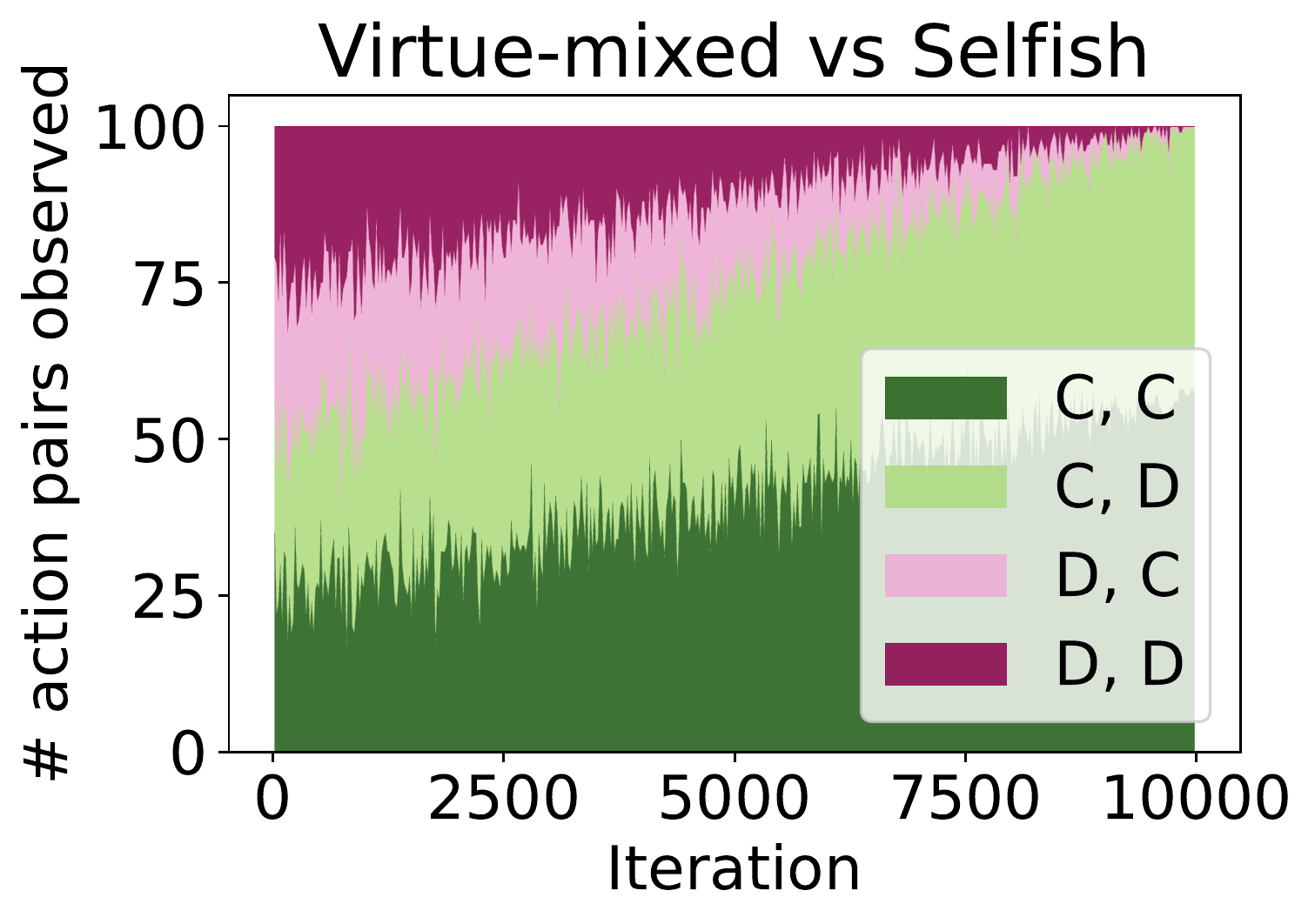}}
&\subt{\includegraphics[width=20mm]{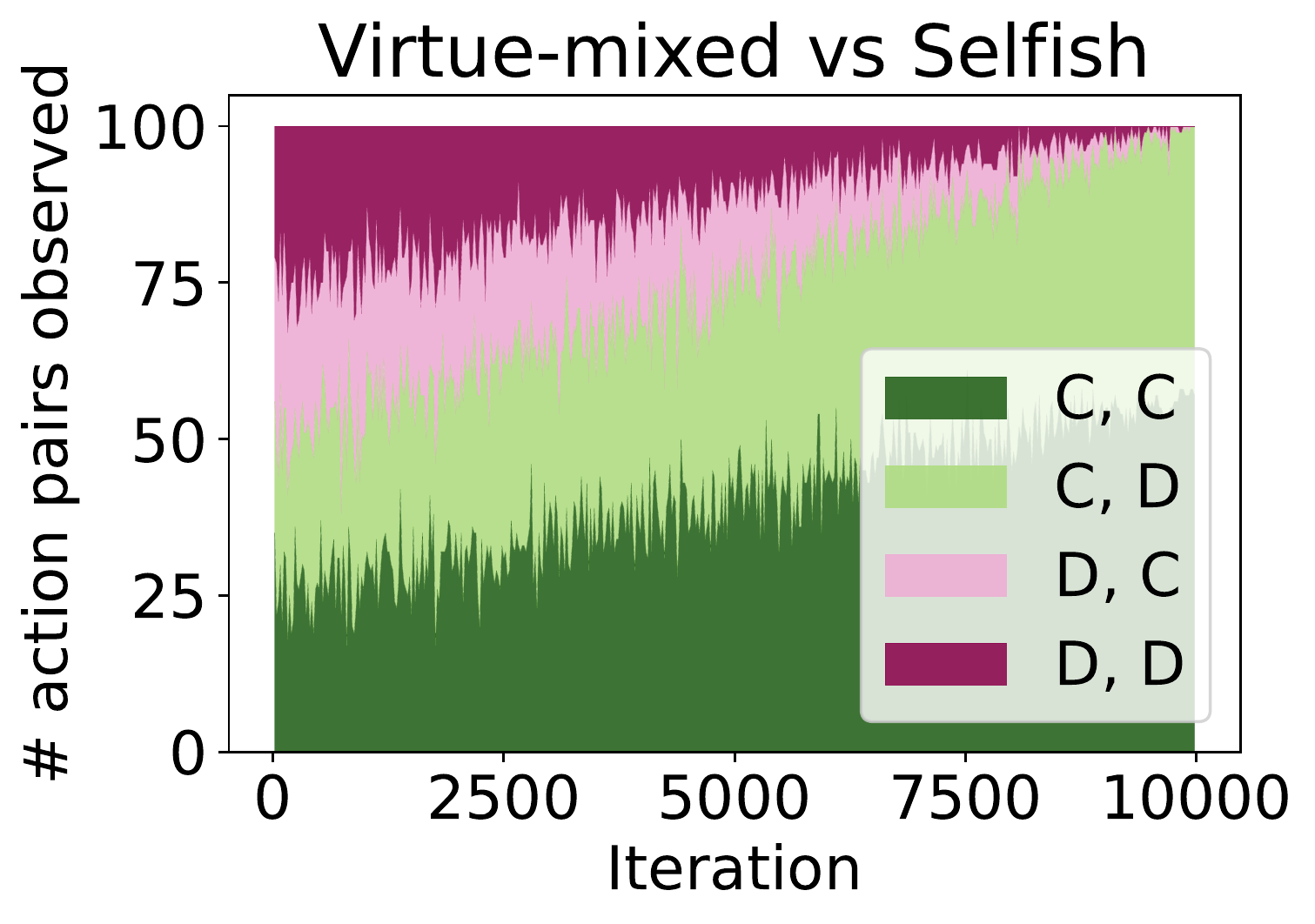}}
&\subt{\includegraphics[width=20mm]{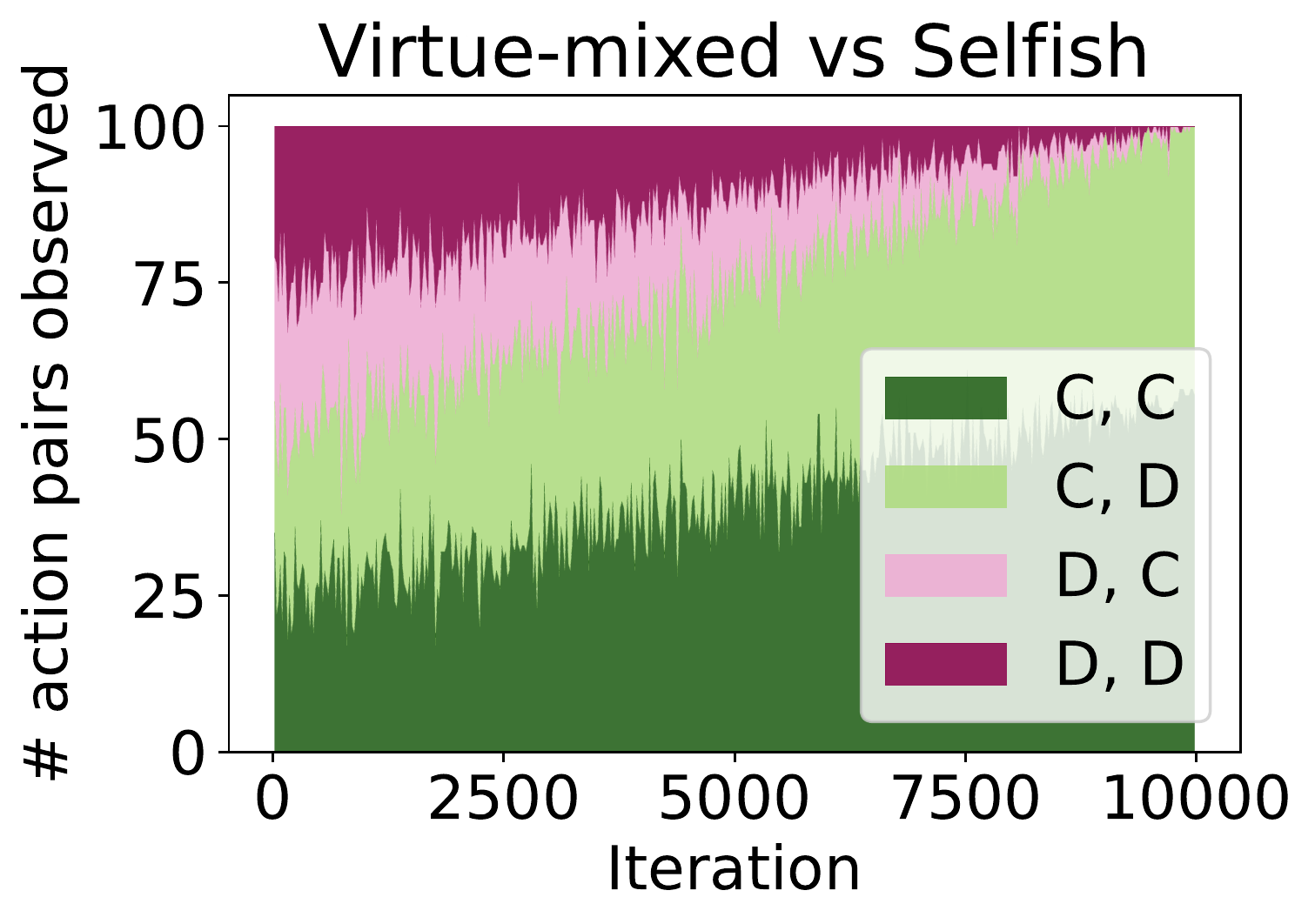}}
&\subt{\includegraphics[width=20mm]{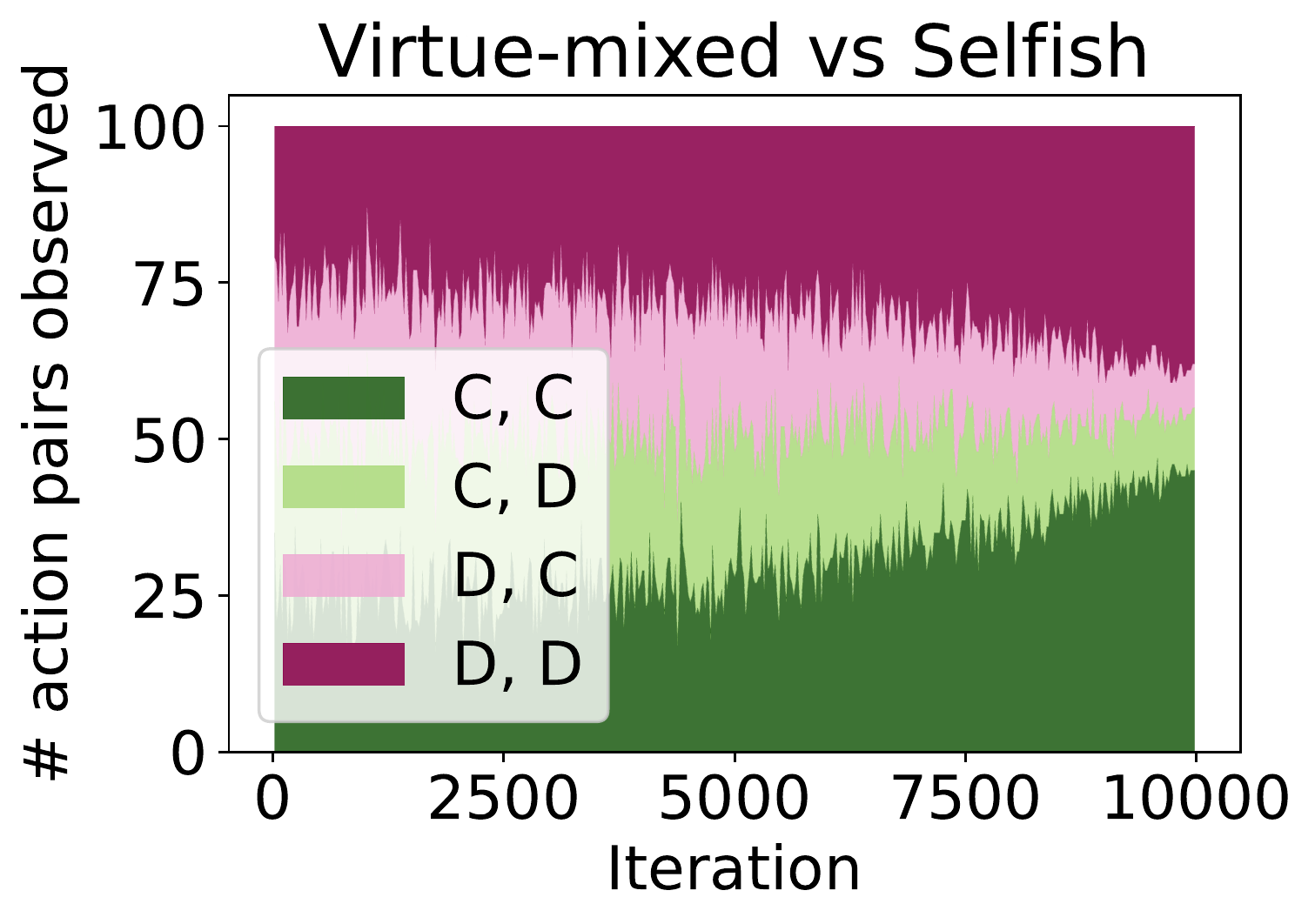}}
\\
\bottomrule
\end{tabular}
\caption{Iterated Stag Hunt. The actions displayed by \textit{Virtue-mixed} agent defined with different weights $\beta \in (0, 0.2, 0.4, 0.6, 0.8, 1)$ against an \textit{Selfish} opponent.}
\label{fig:extra_QLVM_beta_STH}
\end{figure*}

\begin{figure}[!h]
\centering
\includegraphics[width=87mm]{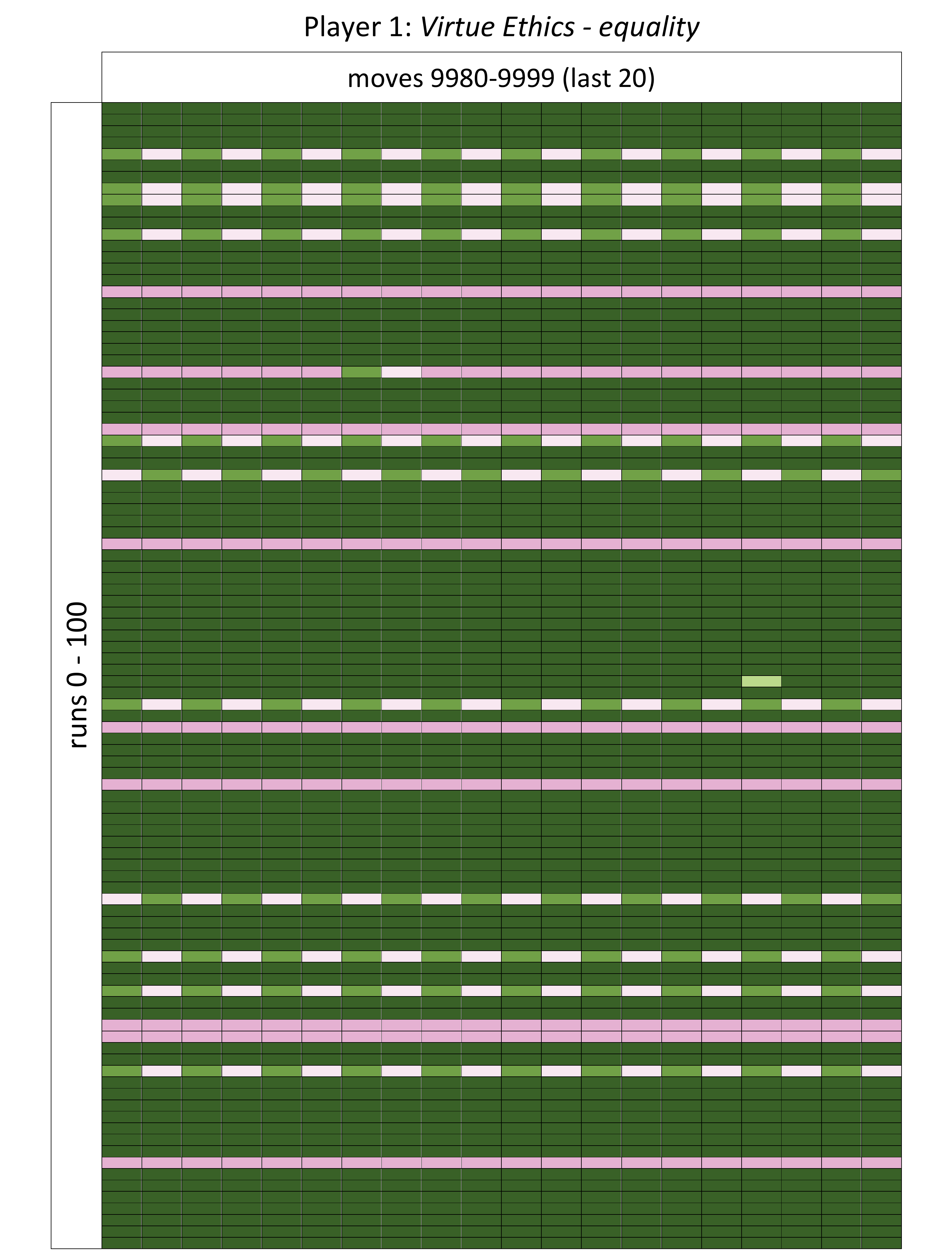}
\includegraphics[width=87mm]{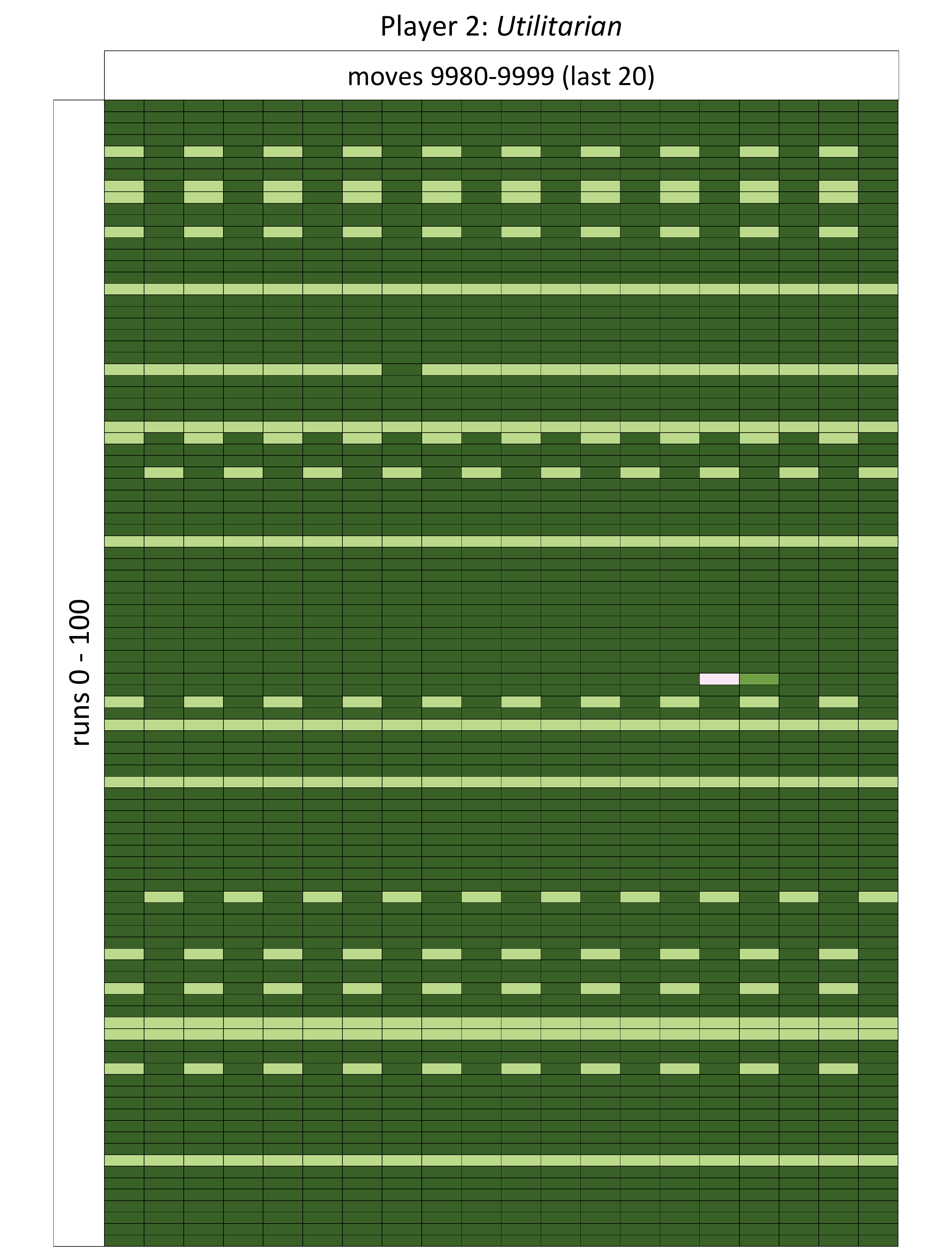}
\includegraphics[width=18mm]{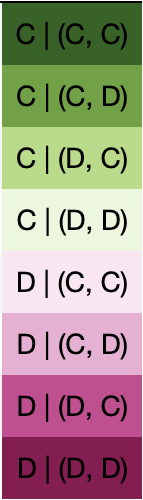}
\caption{Iterated Prisoner's Dilemma. The last 20 actions played by the \textit{Virtue-equality} agent (left) and the \textit{Utilitarian} opponent (right), compared across the 100 runs. For each agent, we display each action given the state that the agent observed (see legend on the right). Each row represents a single run, and the 20 columns represent the last 20 consecutive moves observed (out of 10000).}
\label{fig:QLVE_e-deepdive}
\end{figure}


           
\newpage

\begin{figure}[h!]
\centering
\begin{tabular}{|c|cccccc}
\toprule
IPD & Selfish & Utilitarian & Deontological & Virtue-equality & Virtue-kindness\\
\midrule

\makecell[cc]{\rotatebox[origin=c]{90}{ Virtue-eq. }} &
\subt{\includegraphics[width=28mm]{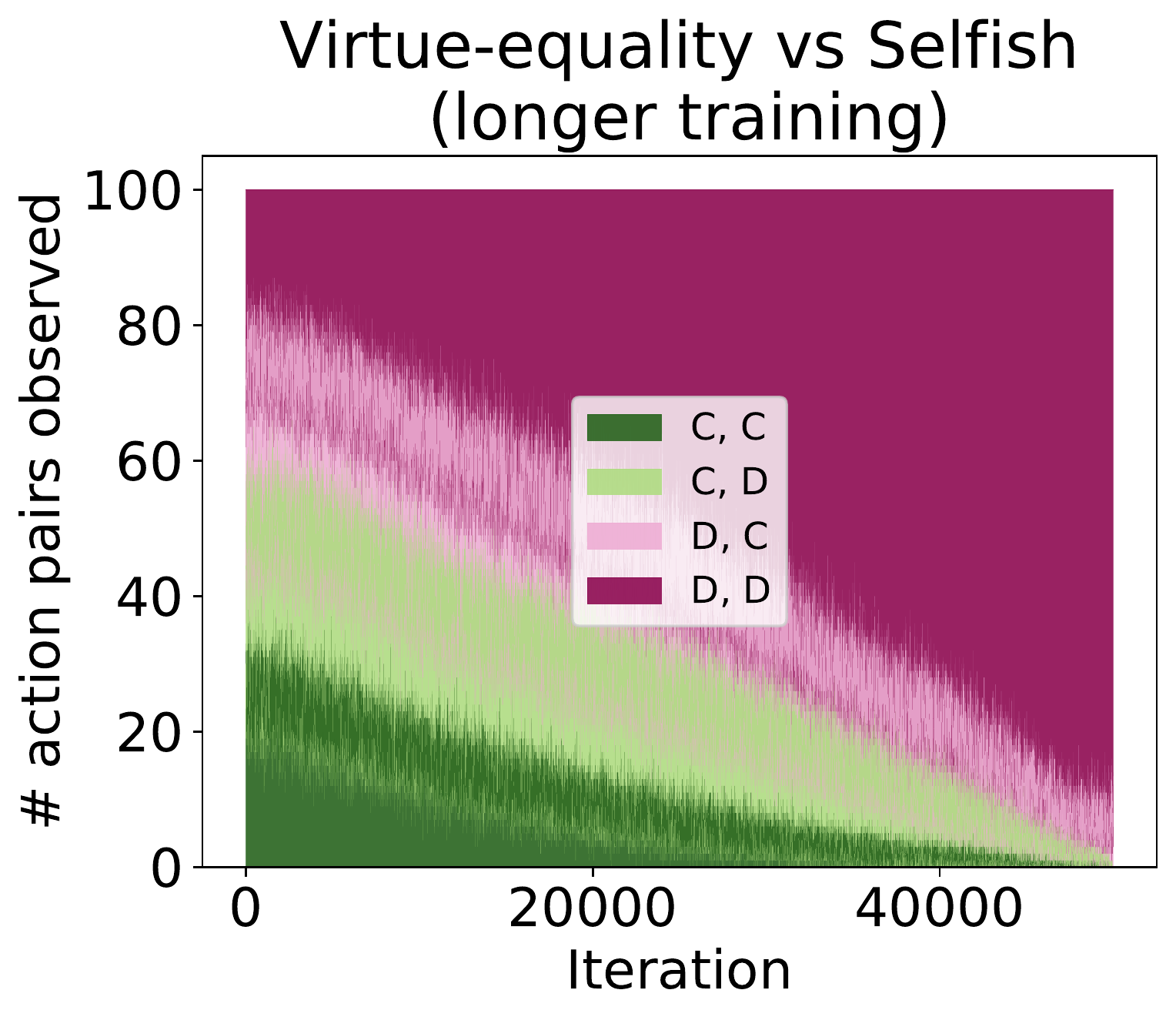}}
&\subt{\includegraphics[width=28mm]{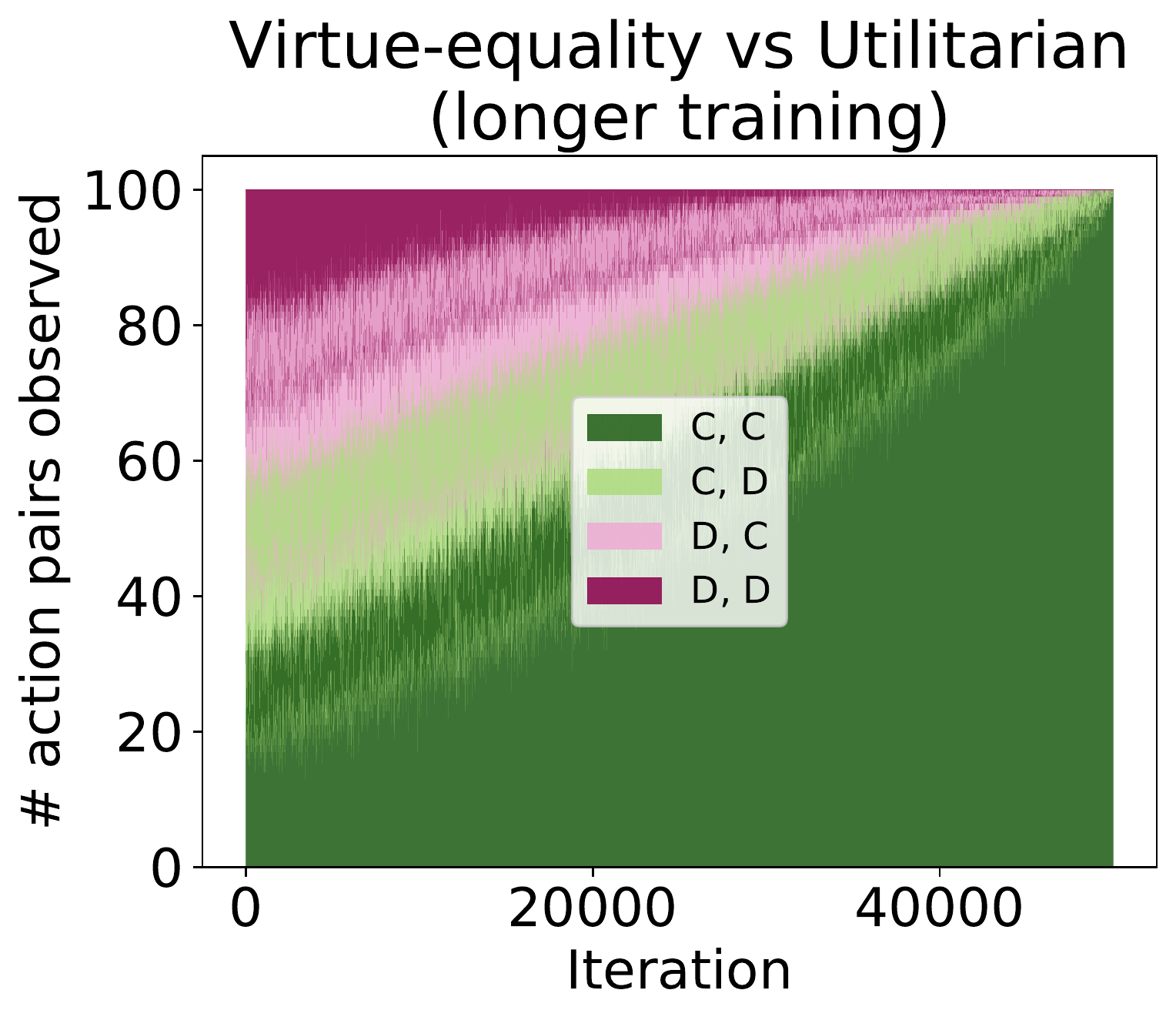}}
&\subt{\includegraphics[width=28mm]{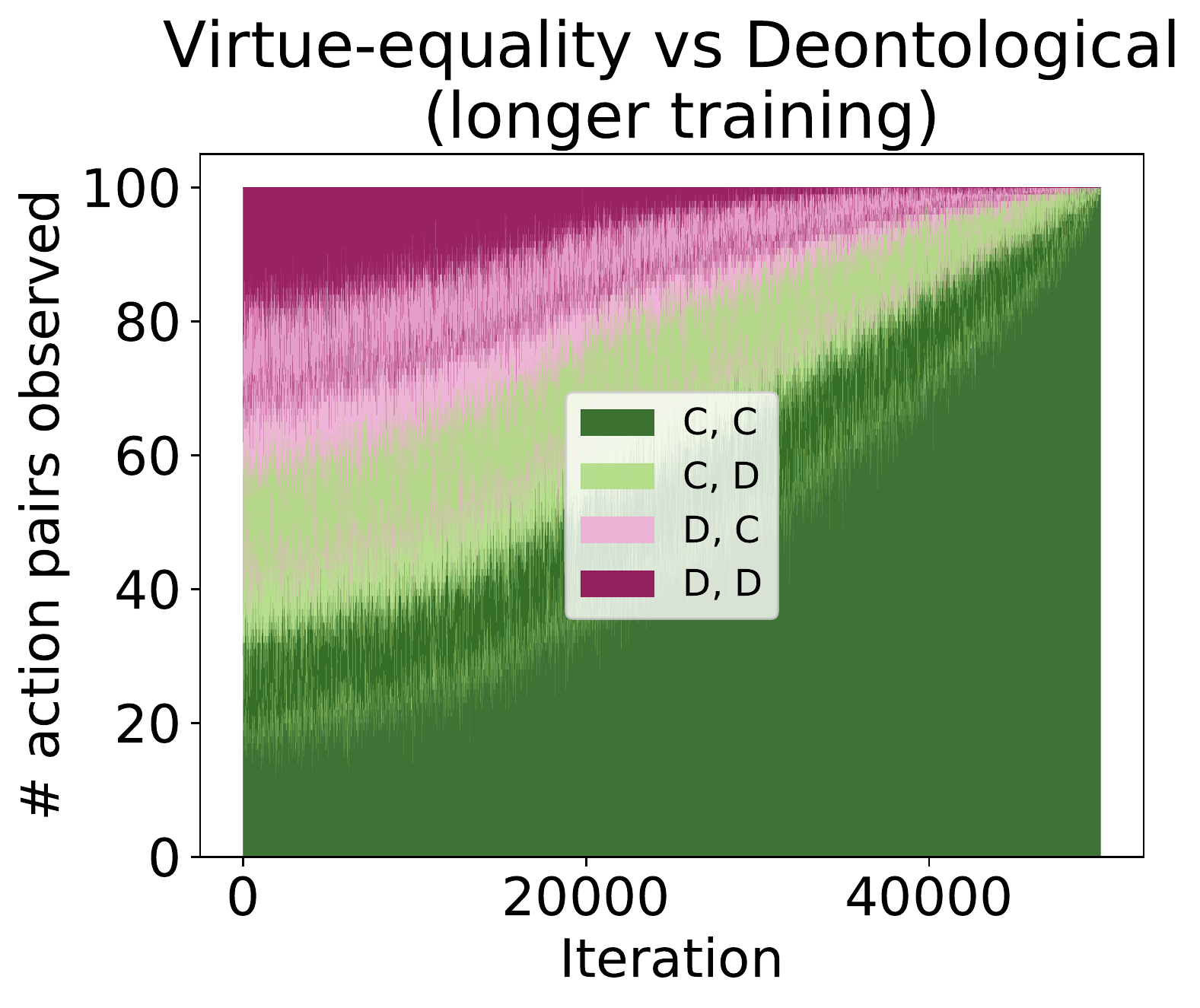}}
&\subt{\includegraphics[width=28mm]{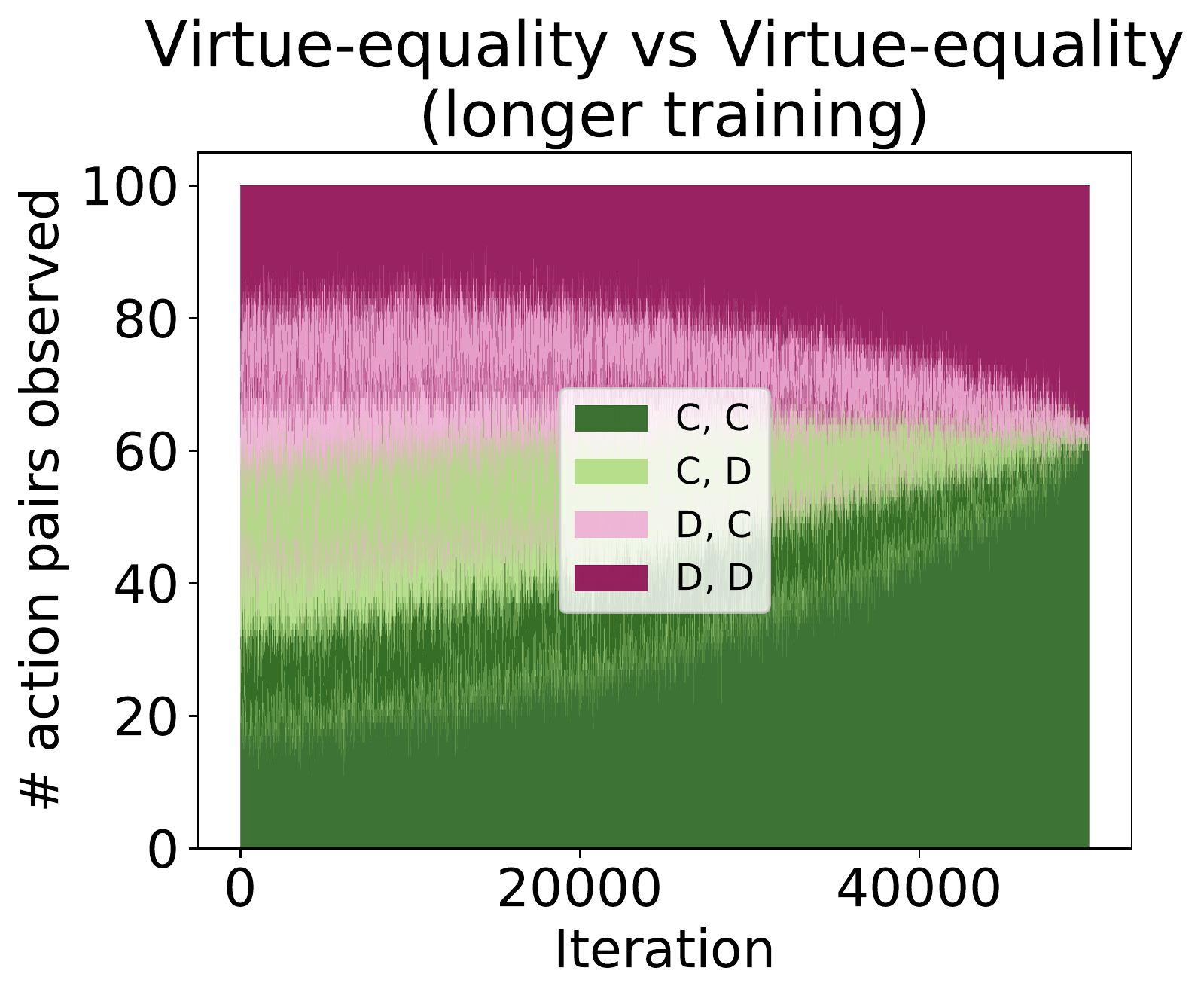}}
&\subt{\includegraphics[width=28mm]{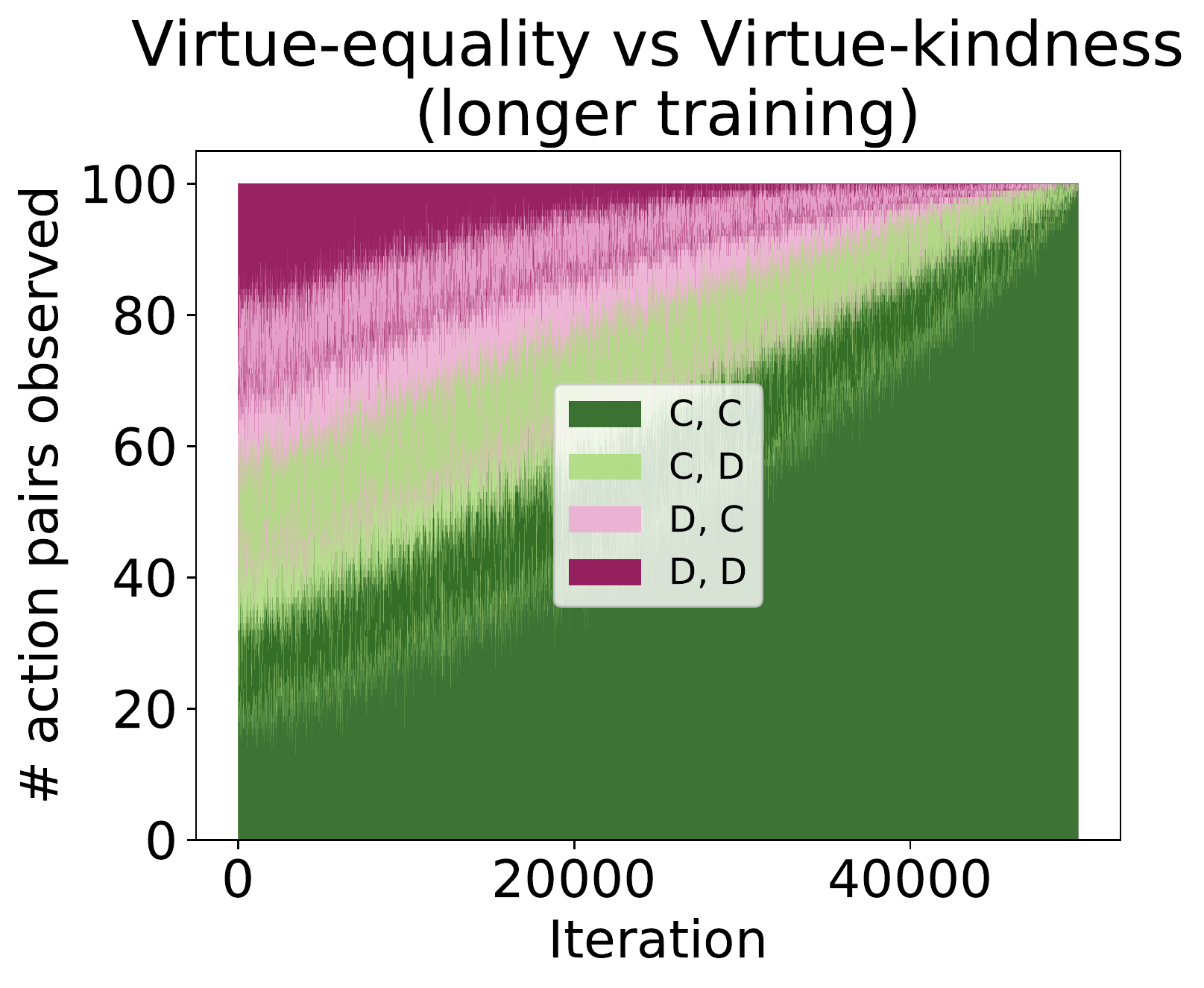}}
\\
\bottomrule
\end{tabular}
\caption{Iterated Prisoner's Dilemma. Simultaneous pairs of actions played by the \textit{Virtue-equality} agent each type of opponent across the 100 runs when training until convergence (50000 iterations instead of the 10000 presented in the main paper). Exploration rate $\epsilon$ decays linearly from 1 to 0 over the training. These results show that, with a longer training period, the \textit{Virtue-equality} agent can converge to a fully cooperative policy against the \textit{Utilitarian, Deontological} and \textit{Virtue-kindness} players.}
\label{fig:longer_action_pairs_IPD}
\end{figure}

\begin{figure}[!h]
\centering
\begin{tabular}{|c|cccccc}
\toprule
IVD & Selfish & Utilitarian & Deontological & Virtue-equality & Virtue-kindness\\
\midrule

\makecell[cc]{\rotatebox[origin=c]{90}{ Virtue-eq. }} &
\subt{\includegraphics[width=28mm]{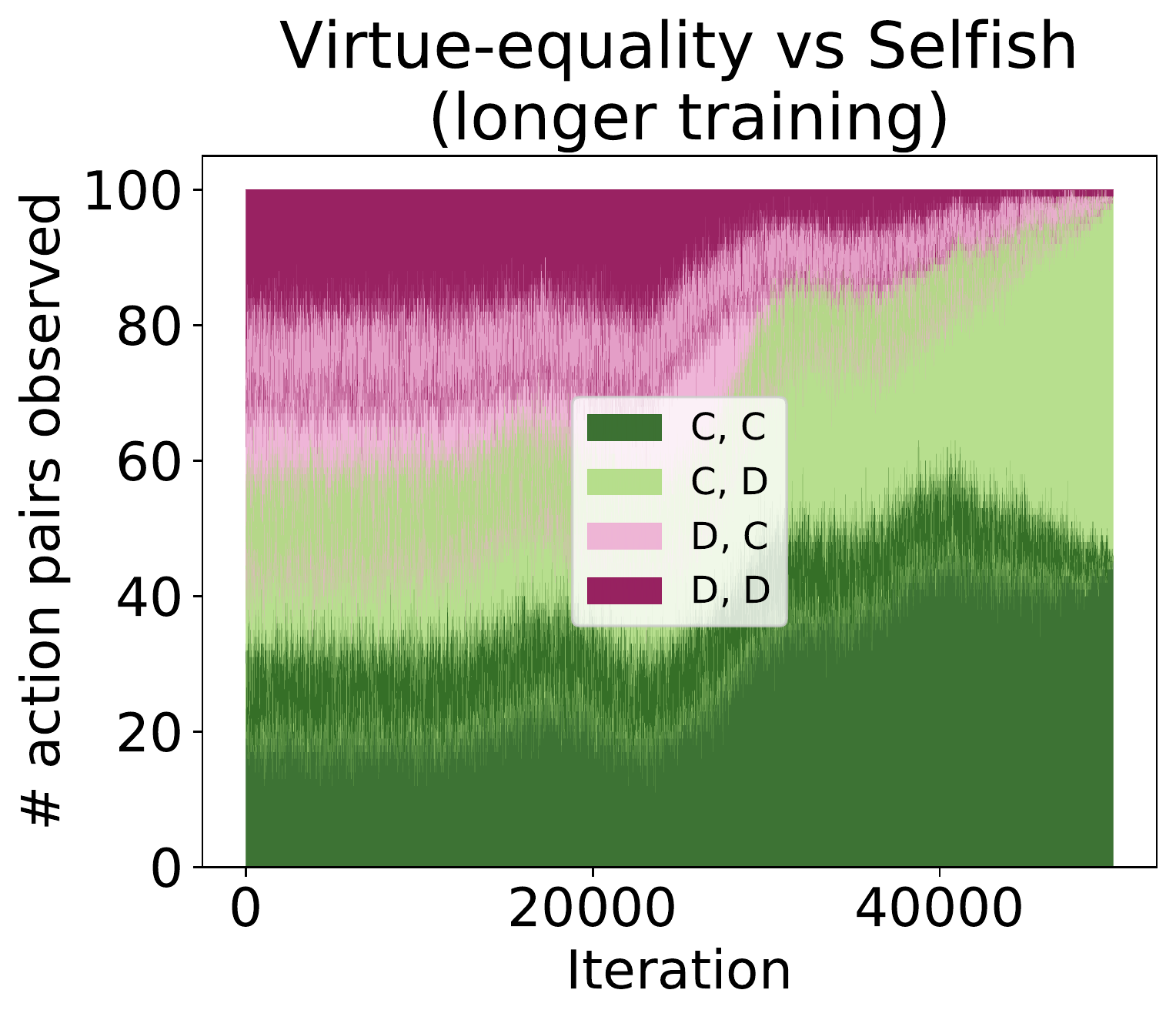}}
&\subt{\includegraphics[width=28mm]{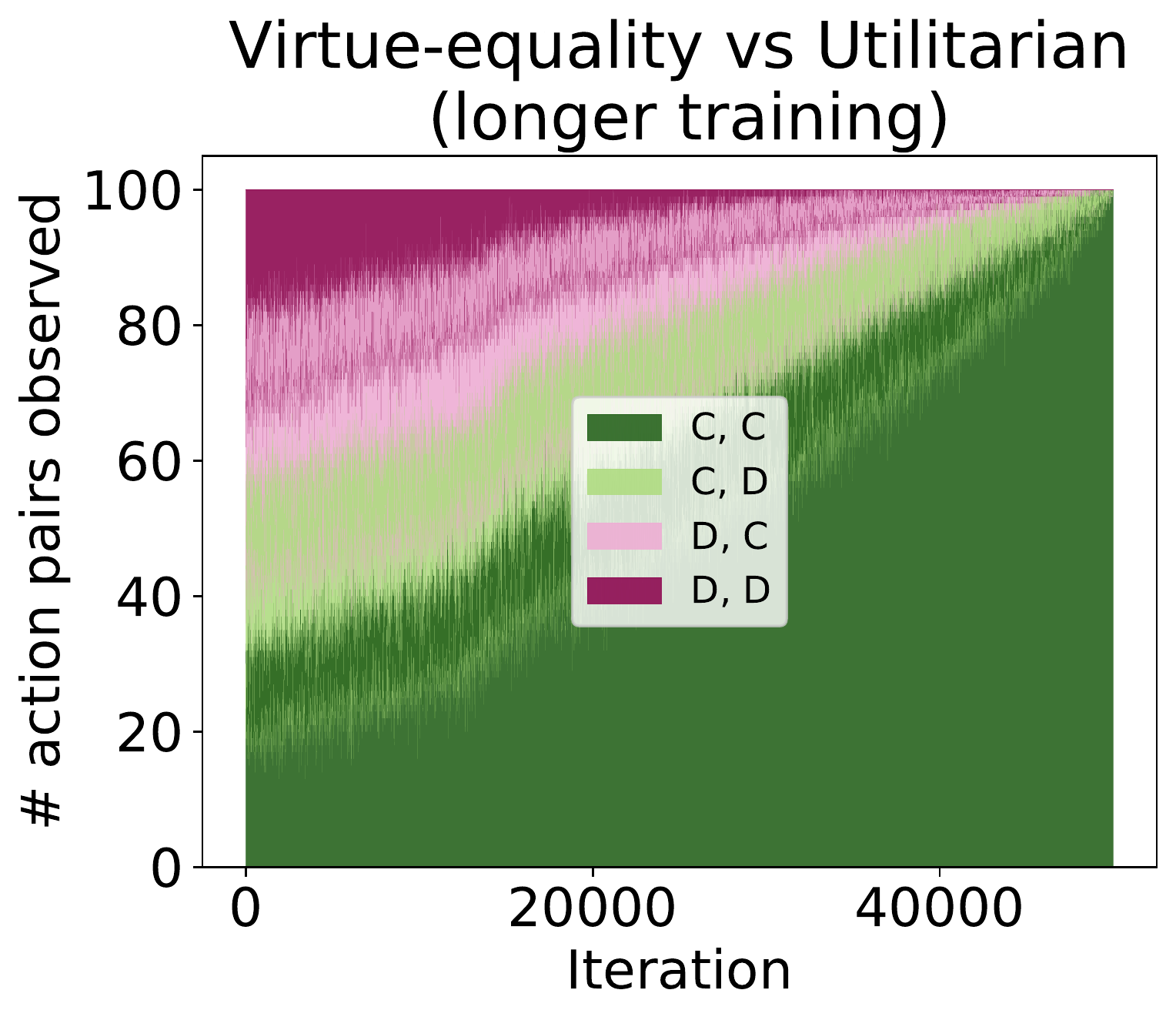}}
&\subt{\includegraphics[width=28mm]{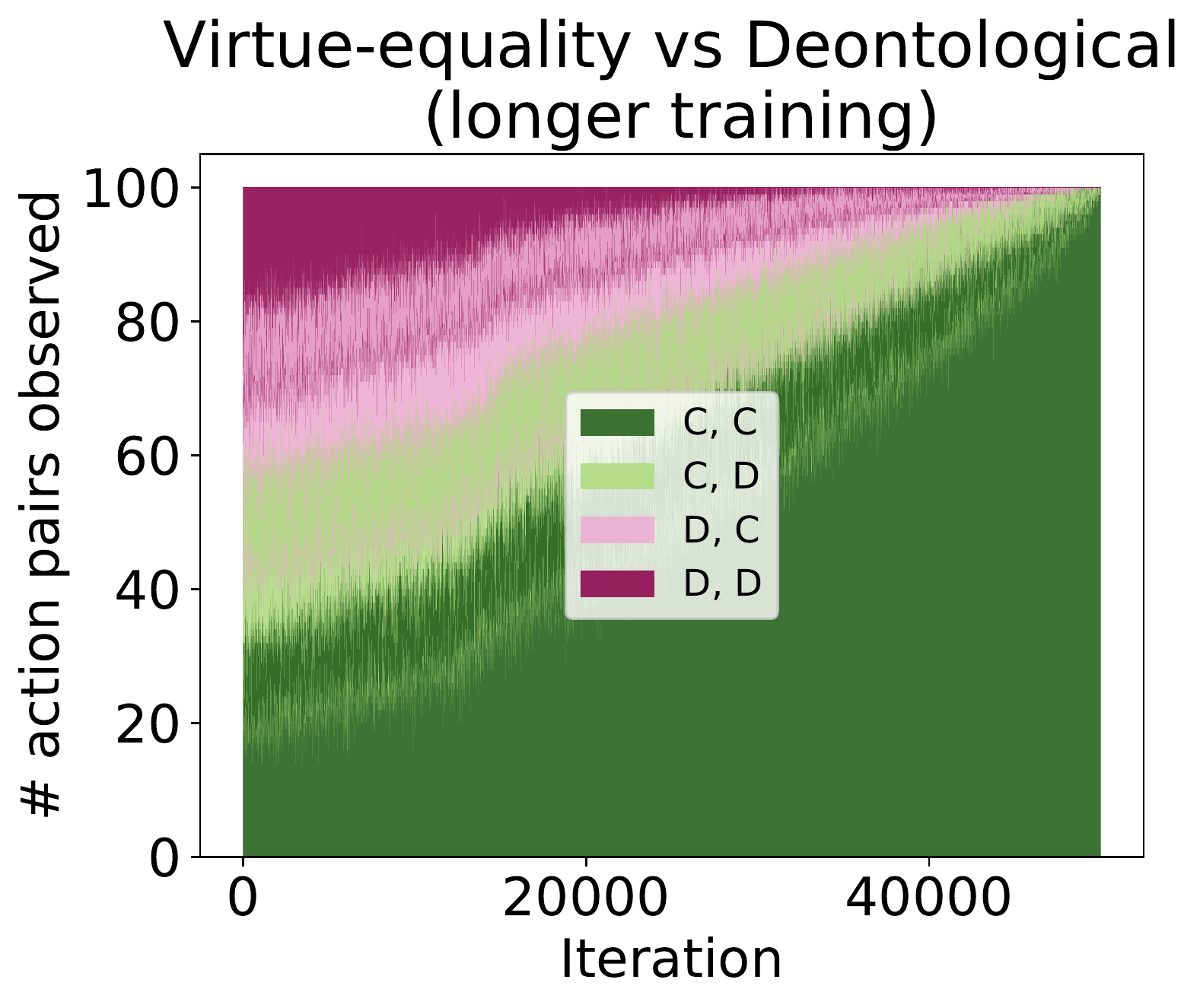}}
&\subt{\includegraphics[width=28mm]{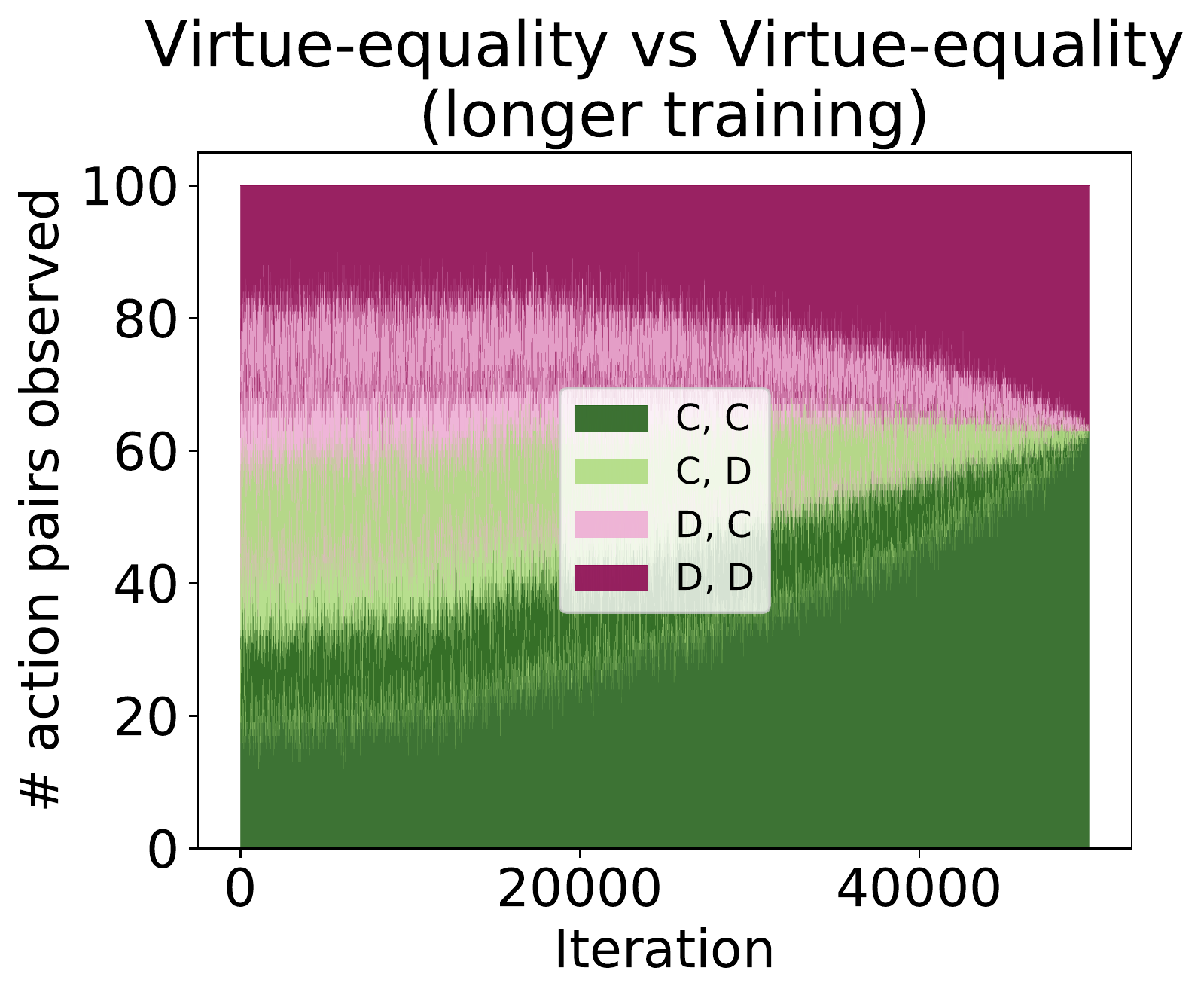}}
&\subt{\includegraphics[width=28mm]{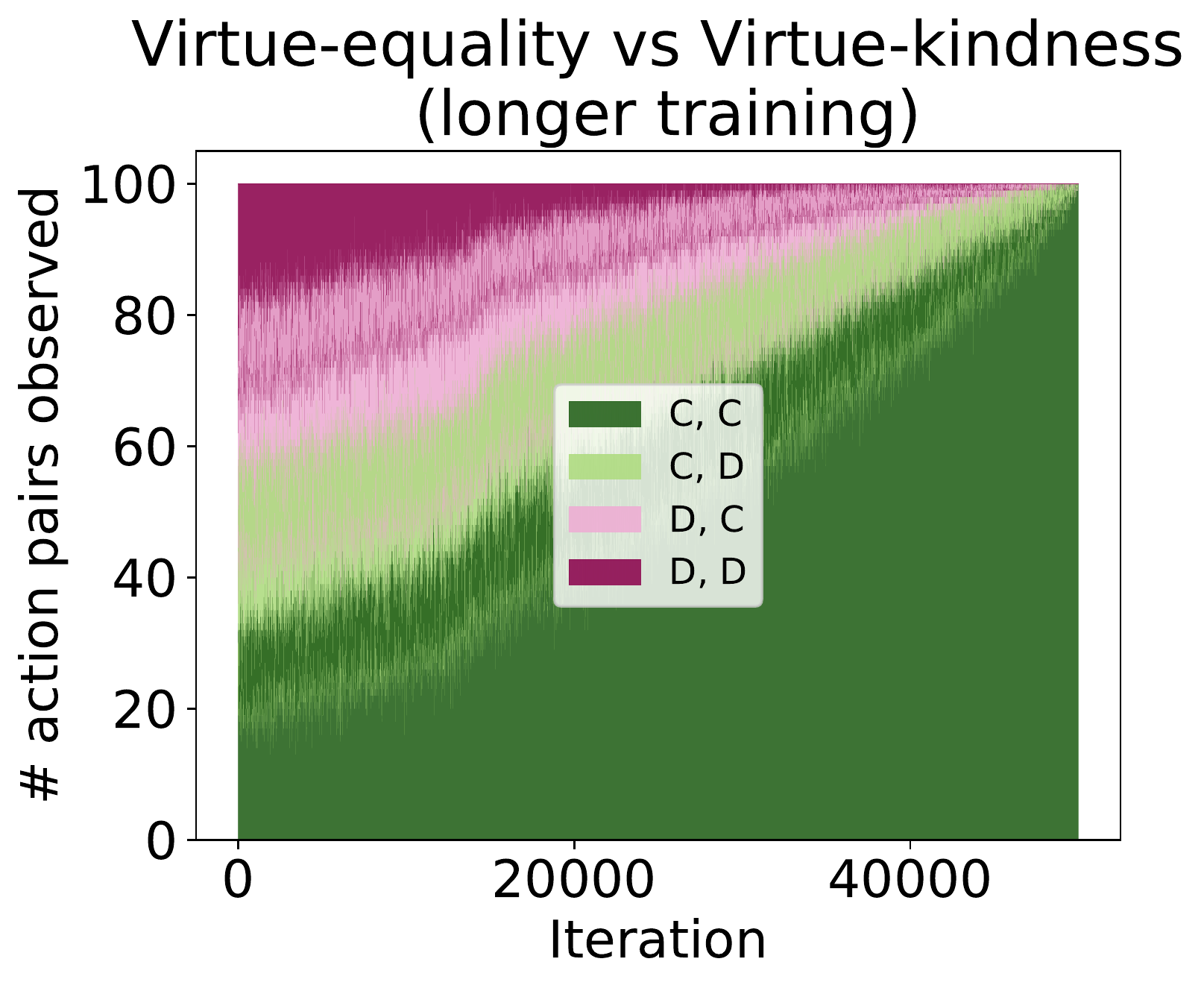}}
\\
\bottomrule
\end{tabular}
\caption{Iterated Volunteer's Dilemma. Simultaneous pairs of actions played by the \textit{Virtue-equality} agent each type of opponent across the 100 runs when training until convergence (50000 iterations instead of the 10000 presented in the main paper). Exploration rate $\epsilon$ decays linearly from 1 to 0 over the training. These results show that, with a longer training period, the \textit{Virtue-equality} agent can converge to a fully cooperative policy against the \textit{Utilitarian, Deontological} and \textit{Virtue-kindness} players.}
\label{fig:longer_action_pairs_IVD}
\end{figure}

\begin{figure}[!h]
\centering
\begin{tabular}{|c|cccccc}
\toprule
ISH & Selfish & Utilitarian & Deontological & Virtue-equality & Virtue-kindness\\
\midrule

\makecell[cc]{\rotatebox[origin=c]{90}{ Virtue-eq. }} &
\subt{\includegraphics[width=28mm]{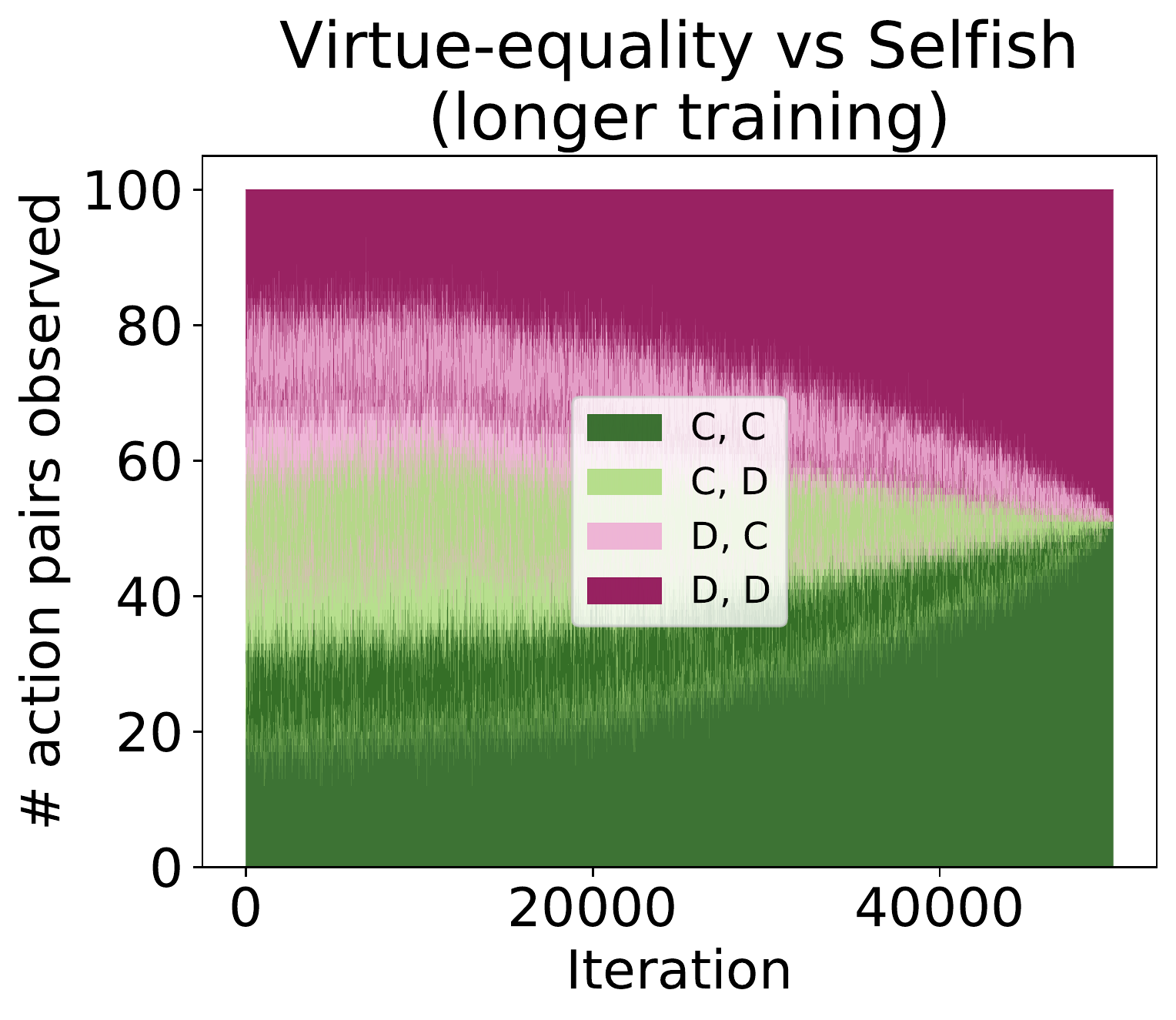}}
&\subt{\includegraphics[width=28mm]{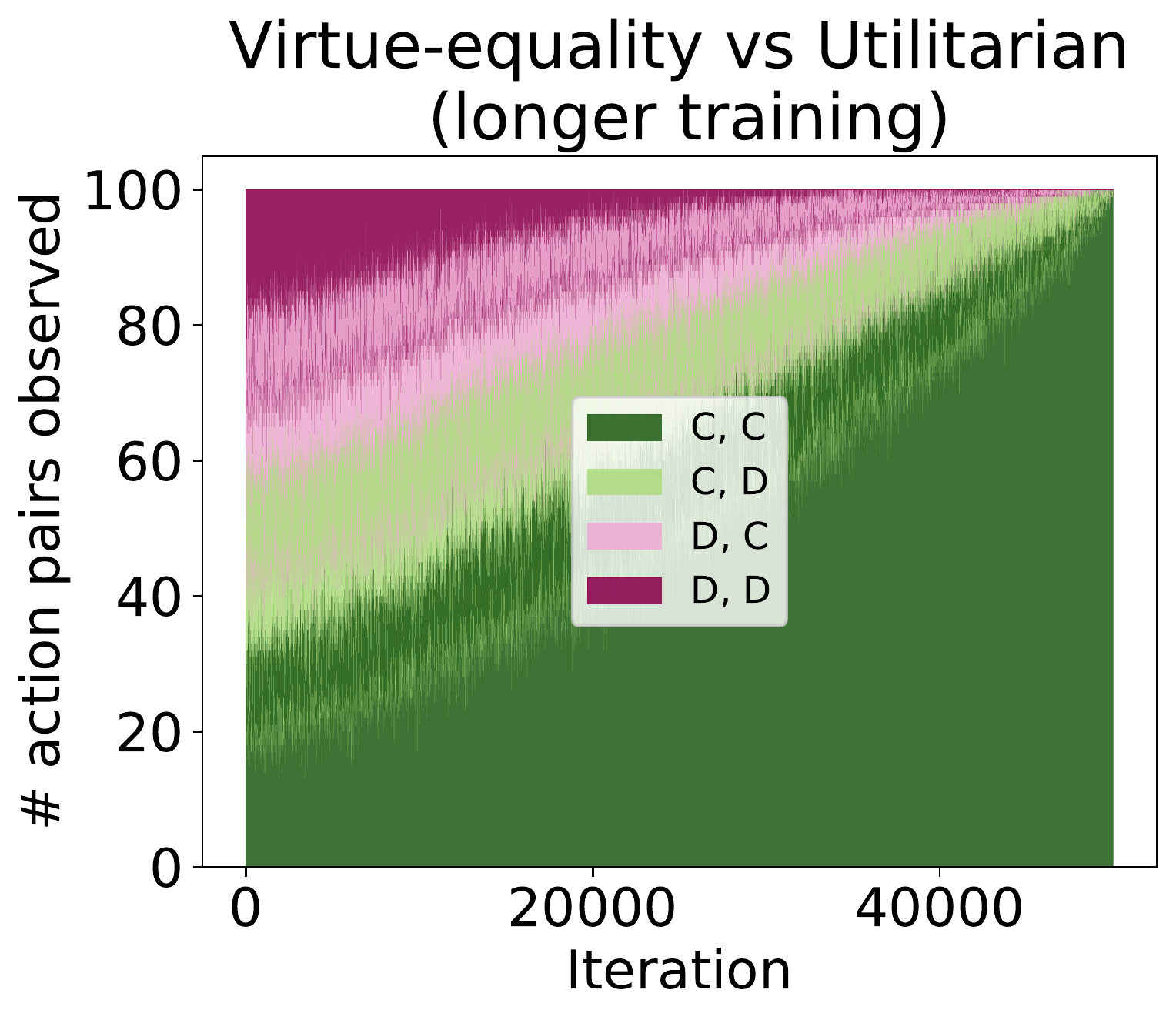}}
&\subt{\includegraphics[width=28mm]{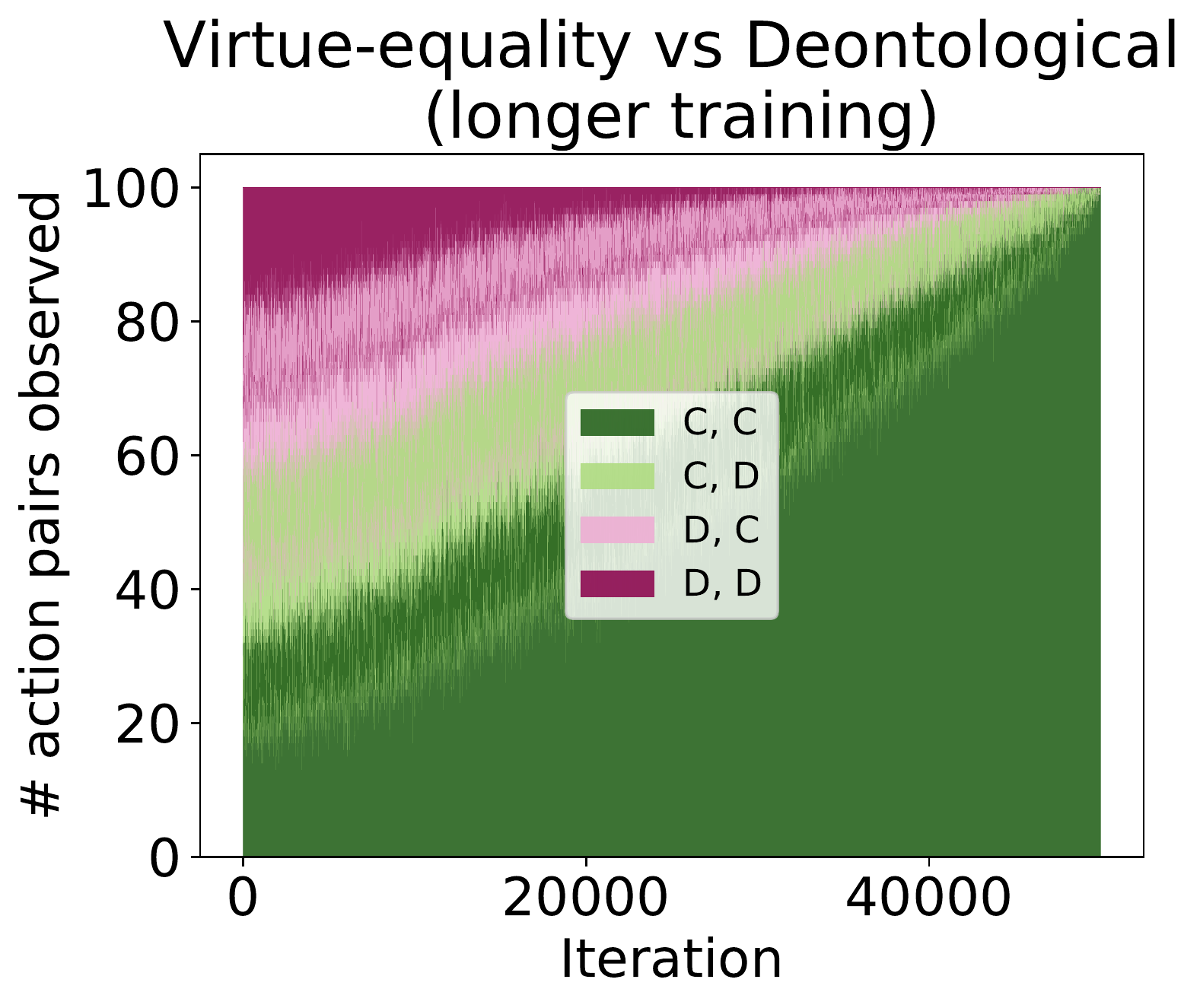}}
&\subt{\includegraphics[width=28mm]{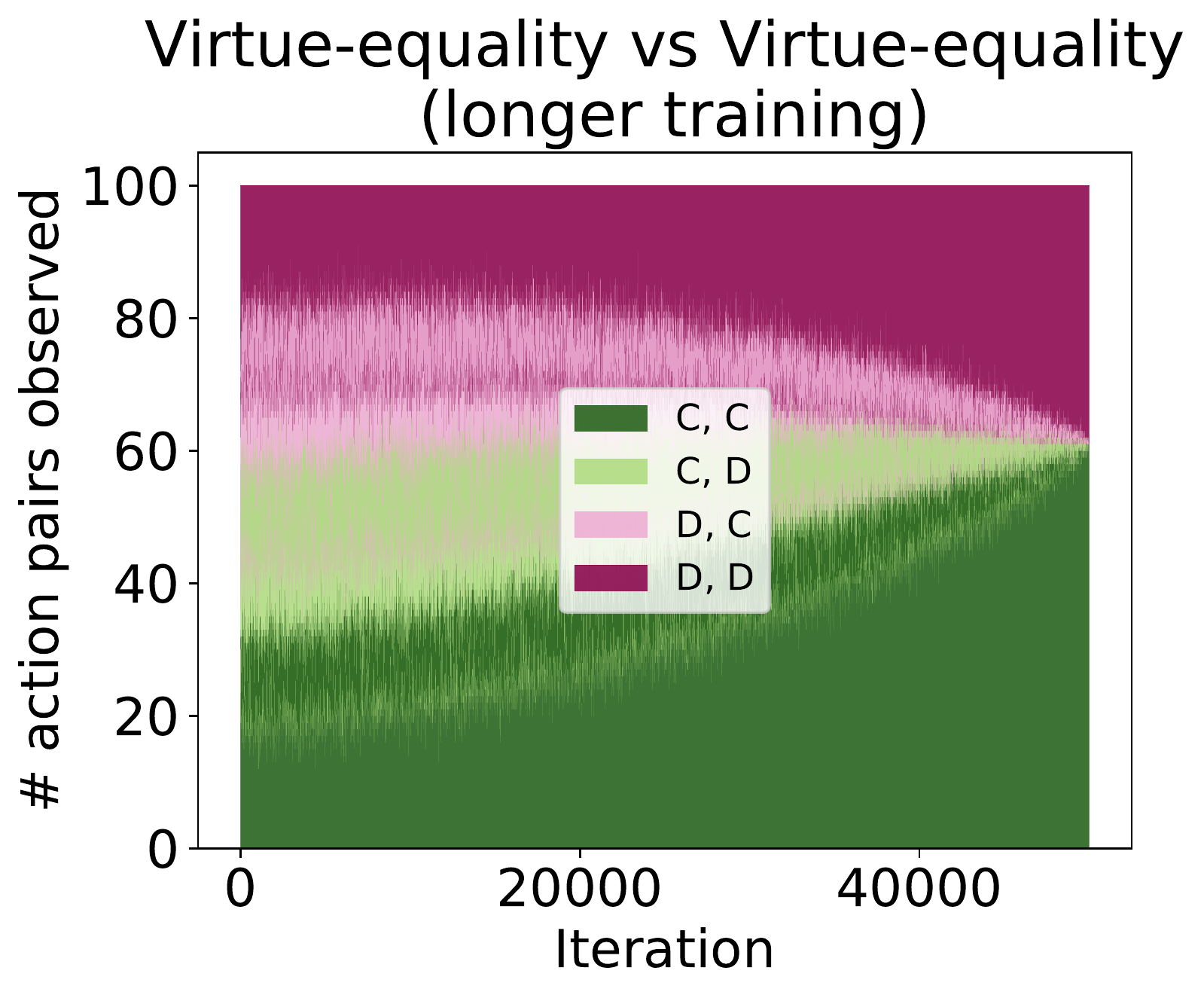}}
&\subt{\includegraphics[width=28mm]{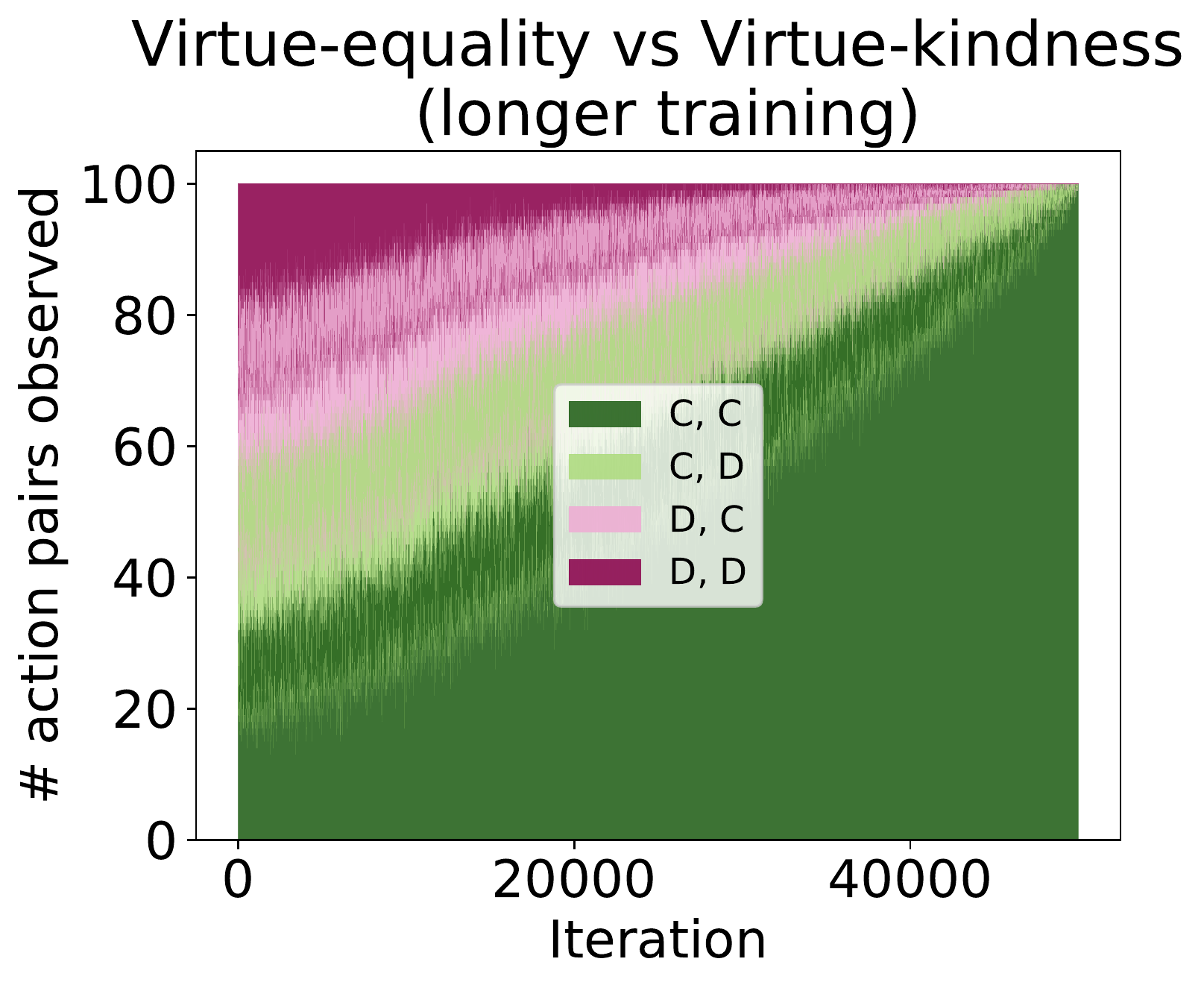}}
\\
\bottomrule
\end{tabular}
\caption{Iterated Stag Hunt. Simultaneous pairs of actions played by the \textit{Virtue-equality} agent each type of opponent across the 100 runs when training until convergence (50000 iterations instead of the 10000 presented in the main paper). Exploration rate $\epsilon$ decays linearly from 1 to 0 over the training. These results show that, with a longer training period, the \textit{Virtue-equality} agent can converge to a fully cooperative policy against the \textit{Utilitarian, Deontological} and \textit{Virtue-kindness} players.}
\label{fig:longer_action_pairs_ISH}
\end{figure}

\end{document}